\documentclass[version=last
    ,DIV=11
    ,BCOR=12mm
    ,twoside=true
    ,twocolumn=false
    ,open=right
    ,headinclude=false
    ,footinclude=false
    ,chapterprefix=true
    ,draft=false
    ,fontsize=10pt
    ,headings=normal
    ,headsepline=true
    ,footsepline=true
    ,titlepage=true
    ,parskip=half+
    ,toc=graduated
    ,cleardoublepage=current
    ,appendixprefix=true
    ,listof=totoc
    ,index=totoc
    ,bibliography=totoc
    ,captions=tableheading
    ,bibliography=oldstyle
]{scrbook}

\usepackage[marginparwidth=0.5cm, 
paperwidth=6.35in, paperheight=9in,
left=15.75mm, right=15.75mm, 
bottom=22.272mm, top=20.272mm, 
footskip=1.13cm, 
verbose]{geometry} 

\usepackage[labelformat=simple,font=footnotesize,format=plain,labelfont=bf]{caption}
\usepackage[font=scriptsize,labelfont=bf]{subfig}
\usepackage{cite}
\usepackage[pdftex]{graphicx}
\usepackage[]{algorithm}
\usepackage[]{algorithmicx}
\usepackage[]{algpseudocode}
\usepackage[utf8]{inputenc}
\usepackage[greek,english]{babel}
\usepackage[T1]{fontenc}
\usepackage{lmodern}
\usepackage{amsfonts}
\usepackage[cmex10]{amsmath}
\usepackage{graphicx}
\usepackage{rotating}
\usepackage[bottom]{footmisc}
\usepackage{multirow}
\usepackage{afterpage}
\usepackage{tablefootnote}
\usepackage{array,booktabs}
\usepackage{tabularx}
\usepackage{enumitem}
\usepackage{xcolor}
\usepackage{threeparttable}
\usepackage{array,multirow}
\usepackage{booktabs, makecell, tabularx}
\usepackage{relsize}
\usepackage{tikz}
\usepackage{upgreek}
\usepackage{oubraces}
\usepackage{listings}
\usepackage{longtable}
\usepackage{IEEEtrantools}

\def\blankpage{%
      \clearpage%
      \thispagestyle{empty}%
      \null%
      \clearpage}

\newenvironment{ChapterAbstract}{\rightskip1in\itshape}{}

\makeatletter
\newcommand{\thickhline}{%
    \noalign {\ifnum 0=`}\fi \hrule height 1.5pt
    \futurelet \reserved@a \@xhline
}
\makeatother

\usepackage{mathtools}

\newcommand{\specialcell}[2][c]{%
  \begin{tabular}[#1]{@{}l@{}}#2\end{tabular}}

\usepackage{url}
\usepackage[export]{adjustbox}

 \deffootnote[1em]{1em}{1em}{\textsuperscript{\thefootnotemark}}

\usepackage[unicode, hyperfootnotes=false]{hyperref}
\hypersetup{final=true,
    colorlinks=true,
    linkcolor=black,
    filecolor=black,
    urlcolor=black,
    citecolor=black,
    pdfauthor={Vasileios K. Leon},
    pdftitle={From Circuits to SoC Processors: Arithmetic Approximation Techniques \& Embedded Computing Methodologies for DSP Acceleration},
    pdfsubject={PhD Thesis, PhD Dissertation, Ph.D., Doctor of Philoshopy, Doctor of Engineering, Electrical and Computer Engineering},
    pdfkeywords={Approximate Computing, Arithmetic Circuits, Hardware Design, Hardware Accelerators, Computer Hardware, ASIC, FPGA, VPU, SoC, Heterogeneous Computing, Embedded Systems, Digital Signal Processing, Computer Vision, Convolutional Neural Networks}
}

\usepackage{tikz}
\usetikzlibrary{shadings,shadows,shapes.arrows}

\usepackage{pifont}

\newcommand\eng[1]{\selectlanguage{english}#1\selectlanguage{greek}}
\usepackage{tabularx}
\usepackage{ltablex}

\renewcommand{\chaptermark}[1]{\markboth{#1}{#1}}
\definecolor{myblue}{rgb}{0, 0.466, 0.709}
\definecolor{myblue2}{rgb}{0.258, 0.521, 0.956}
\definecolor{mygreen}{rgb}{0, 0.815, 0.686}
\definecolor{blue_dlsb}{RGB}{61,61,61}
\definecolor{red_perf}{RGB}{184, 25, 39}
\definecolor{blue_round1}{RGB}{15, 127, 169}
\definecolor{blue_round2}{RGB}{120, 202, 232}

\definecolor{codeblue}{rgb}{0.262, 0.392, 0.937}
\definecolor{codegray}{rgb}{0.5,0.5,0.5}
\definecolor{codered}{rgb}{0.909, 0.270, 0.250}
\definecolor{backcolour}{rgb}{0.991,0.991,0.991}
\lstdefinestyle{mystyle}{
    backgroundcolor=\color{backcolour},   
    commentstyle=\color{codeblue},
    keywordstyle=\color{red},
    numbers=none,
    stringstyle=\color{codered},
    basicstyle=\fontsize{9pt}{9.4pt}\ttfamily,
	breakatwhitespace=false,         
	breaklines=false,                 
	captionpos=b,                    
	keepspaces=true,                 
	showspaces=false,                
	showstringspaces=false,
	showtabs=false,                  
	tabsize=4
}
\renewcommand*{\chapterformat}{%
  \mbox{\chapappifchapterprefix{\nobreakspace}\thechapter
  \IfUsePrefixLine{}{\enskip}}%
}

\addtokomafont{pageheadfoot}{\footnotesize}
\addtokomafont{pagenumber}{\footnotesize}
\addtokomafont{chapterprefix}{\fontsize{23}{24}\selectfont}
\addtokomafont{partprefix}{\fontsize{25}{26}\selectfont}

\usepackage{xpatch}
\makeatletter
\xpretocmd{\@endpart}{%
  \ifx\@abstract\@empty\else
    \bigskip
    \begin{quote}\@abstract\end{quote}
    \global\let\@abstract\@empty
  \fi
}{}{}
\newcommand{\partabstract}[1]{%
  \renewcommand{\@abstract}{#1}%
}
\newcommand{\@abstract}{}
\makeatother

\hypersetup{pdfstartview={XYZ null null null}}

\begin{document}

\newgeometry{marginparwidth=0.0cm,  left=27.82771pt, right=27.82771pt, top=20mm, bottom=20mm} 

\frontmatter
\thispagestyle{empty}

\begin{tabular}{@{}p{\textwidth}@{}}
\vspace{-35pt}
\begin{center}
\includegraphics[scale=0.045]{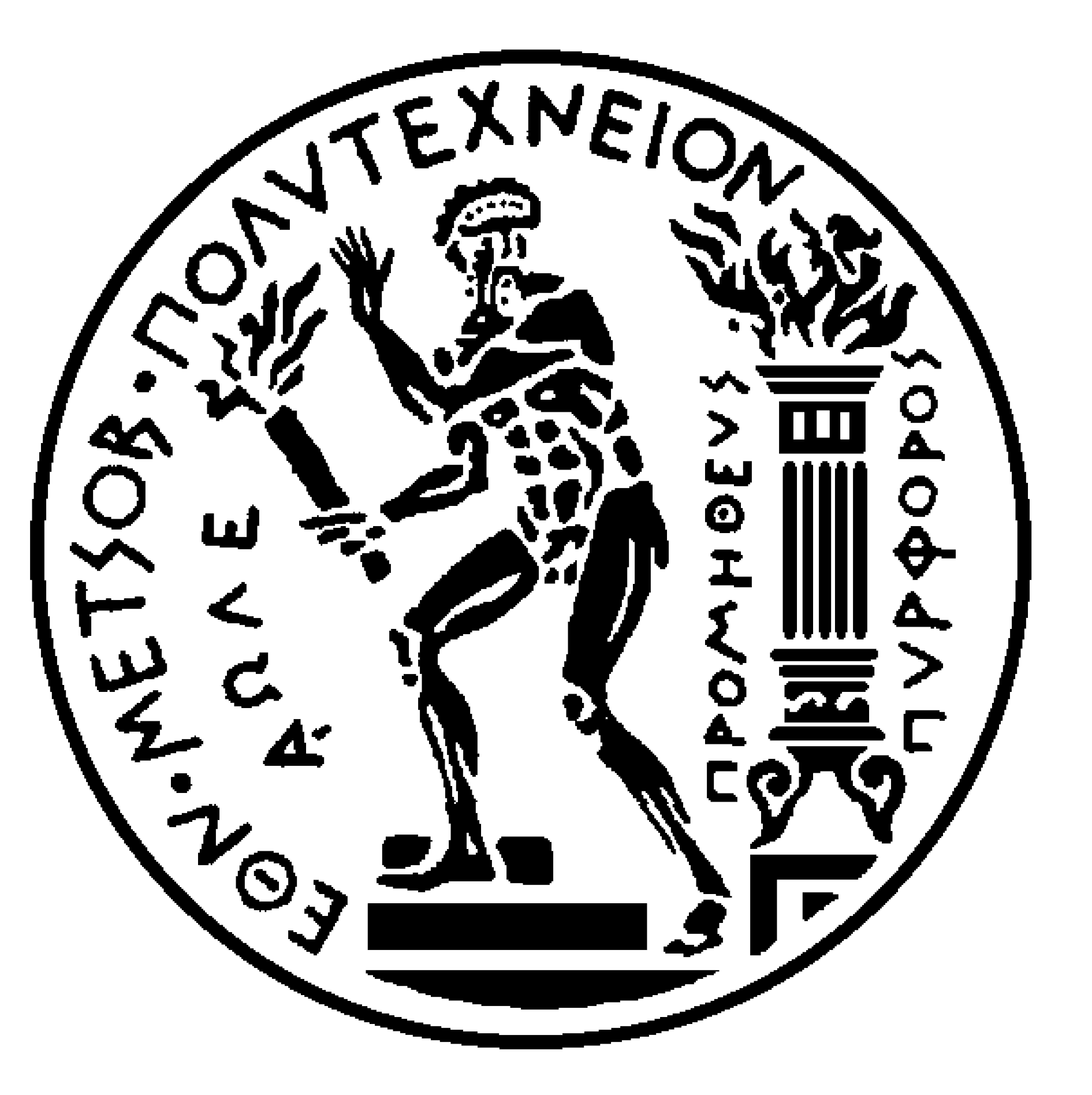}\\ 
{\fontfamily{cmr}\fontsize{15.7}{13}\selectfont\textsc{\textbf{National Technical University of Athens}}}\\[3.5pt]
{\fontfamily{cmr}\fontsize{13}{13}\selectfont\textsc{School of Electrical and Computer Engineering}}\\[3.5pt]
{\fontfamily{cmr}\fontsize{13}{13}\selectfont\textsc{Division of Computer Science}}\\
\end{center}
\end{tabular}

\vspace{35pt}

\begin{tabular}{@{}p{\textwidth}@{}}
\toprule[2pt]
\midrule\\[-20pt]
\begin{center}
{\huge\fontfamily{lmss}\selectfont
    \textbf{From Circuits to SoC Processors:\\
    Arithmetic Approximation Techniques\\
    \hspace{-2pt}\& Embedded Computing Methodologies\\ 
    for DSP Acceleration\\}}
\end{center}
\\[-1pt]
\midrule
\toprule[2pt]
\end{tabular}

\vspace{51pt}

\begin{center}
{\fontsize{16.3}{18}\selectfont Ph.D. Dissertation}\\[3pt]
{\fontsize{18.3}{20}\selectfont\fontfamily{lmss}\selectfont
\textbf{Vasileios K. Leon}\\}
\end{center}

\vspace*{\fill}

\begin{center}
{\fontsize{13.5}{13}\selectfont Athens}\\[4pt]
{\fontsize{13.5}{13}\selectfont October 2022}\\
\end{center}

\blankpage

\newgeometry{marginparwidth=0.0cm,
left=15.75mm, right=15.75mm, 
top=20mm, bottom=20mm} 

\thispagestyle{empty}

\vspace*{85pt}

\begin{flushleft}
{\Large \fontfamily{lmss}\selectfont
\textbf{From Circuits to System-on-Chip Processors:}\\
\textbf{Arithmetic Approximation Techniques}\\
\textbf{\& Embedded Computing Methodologies}\\
\textbf{for Digital Signal Processing Acceleration}\\}
\end{flushleft}

\vspace{35pt}

{\Large \fontfamily{lmss}\selectfont
\textbf{Vasileios Leon}\\}

\vspace{60pt}

{\fontsize{11.3pt}{12pt}\fontfamily{lmr}\selectfont
\underline{Examination Committee}}\\[3pt]
Prof. Kiamal Pekmestzi (Supervisor), \emph{National Technical University of Athens}\\[1.5pt]
Prof. Dimitrios Soudris, \emph{National Technical University of Athens}\\[1.5pt]
Assoc. Prof. Georgios Goumas, \emph{National Technical University of Athens}\\[1.5pt]
Prof. Dionysios Reisis, \emph{National and Kapodistrian University of Athens}\\[1.5pt]
Prof. Apostolos Dollas, \emph{Technical University of Crete}\\[1.5pt]
Prof. Dimitris Gizopoulos, \emph{National and Kapodistrian University of Athens}\\[1.5pt]
Prof. Antonis Paschalis, \emph{National and Kapodistrian University of Athens}\\

\vfill

{Submitted to National Technical University of Athens in partial fulfillment of the requirements for the degree of Doctor of Engineering (PhD in Engineering) in Computer Science.}

\blankpage

\thispagestyle{empty}

\begin{center}
{\fontsize{10.5pt}{12pt}\fontfamily{lmr}\selectfont
\begin{tabular}{c l}
\hspace{-4pt}\raisebox{-.42\height}{\includegraphics[scale=0.03]{MISCs/ntua_logo.png}} & \hspace{-10pt}\specialcell{National Technical University of Athens\\[0.9pt]School of Electrical \& Computer Engineering\\[0.9pt]Division of Computer Science\\[0.9pt]Microprocessors and Digital Systems Laboratory\\[3pt]}
\end{tabular}}
\end{center}

\phantom{a}\\[-22pt] 

\begin{center}
\centering
\begin{tabular}{p{0.78\textwidth}}
\midrule\\[-6pt]
\centering
{\fontsize{14}{15.5}\selectfont
\fontfamily{lmss}\selectfont
\textbf{From Circuits to SoC Processors:}\\
\textbf{Arithmetic Approximation Techniques}\\
\textbf{\& Embedded Computing Methodologies}\\
\textbf{for DSP Acceleration}\\}
\end{tabular}    
\begin{tabular}{p{0.78\textwidth}}
\phantom{a}\\[-23pt]
\midrule\\
\end{tabular}
\end{center}

\begin{center}
{\fontsize{13}{12}\selectfont\fontfamily{lmr}\selectfont
Ph.D. Dissertation\\[4pt]of\\[4pt]
\fontsize{13}{12}\selectfont\fontfamily{lmss}\selectfont
\textbf{Vasileios Leon}\\}
\end{center}

\vspace{12pt}

\begin{tabular}{l l}
\hspace{-9.5pt}
\textbf{Supervising Committee}: 
&  Kiamal Pekmestzi \\
&  Dimitrios Soudris \\
&  Georgios Goumas\\[8pt]
\end{tabular}

{Approved by the Examination Committee on October 10, 2022.\\}

\vspace*{-16pt}

\begin{center}
\begin{tabular}{c c c}
\centering
\includegraphics[scale=0.48]{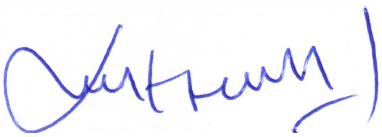} & 
\includegraphics[scale=0.48]{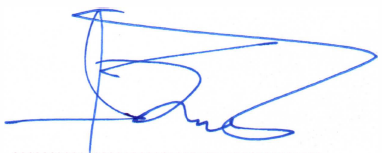} & 
\includegraphics[scale=0.48]{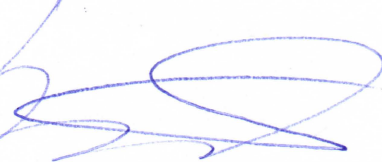} \\[-13pt]
$\dots\dots\dots\dots\dots\dots\dots\dots$ & 
$\dots\dots\dots\dots\dots\dots\dots\dots$ & 
$\dots\dots\dots\dots\dots\dots\dots\dots$ \\ 
Kiamal Pekmestzi & Dimitrios Soudris &  Georgios Goumas\\
Professor NTUA & Professor NTUA & Assoc. Professor NTUA \\[7pt]
\end{tabular}
\begin{tabular}{c@{\hskip 0.5in} c}
\centering
\includegraphics[scale=0.48]{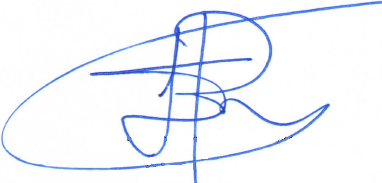} & 
\includegraphics[scale=0.52]{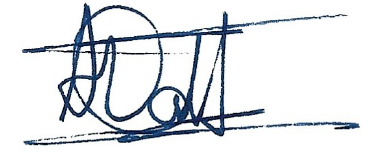} \\[-13pt]
$\dots\dots\dots\dots\dots\dots\dots\dots$ & 
$\dots\dots\dots\dots\dots\dots\dots\dots$ \\ 
Dionysios Reisis &  Apostolos Dollas  \\
Professor NKUA  & Professor TUC \\[7pt]
\includegraphics[scale=0.48]{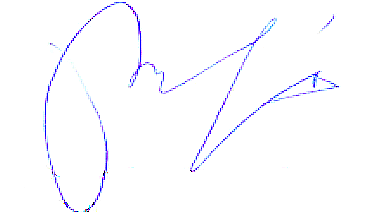} & 
\includegraphics[scale=0.8]{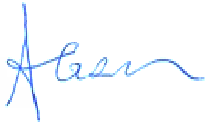} \\[-13pt]
$\dots\dots\dots\dots\dots\dots\dots\dots$ & 
$\dots\dots\dots\dots\dots\dots\dots\dots$ \\ 
 Dimitris Gizopoulos & Antonis Paschalis \\
Professor NKUA & Professor NKUA\\
\end{tabular}
\end{center}

\newpage
\thispagestyle{empty}
~ 
\newpage

\thispagestyle{empty}
\vspace*{\fill}
{\small 
The Ph.D. Dissertation was partially supported by research activities of the European Space Agency (ESA).\\[5pt]

All the reported information and views of the Ph.D. Dissertation lie entirely with the author and do not reflect the official opinion of the National Technical University of Athens. \\[5pt]

Content that is reused from publications that the author has (co-)authored, namely excerpts, figures, and tables, is under copyright with the respective paper publishers (IEEE, ACM, IET, Elsevier). These publications are explicitly stated in the abstract of each Dissertation's chapter. Content that is reused from third-party publications appears with the appropriate copyright note. Reuse of any such content by any interested party requires the publishers' prior consent, according to the applicable copyright policies. Content that has not been published before is copyrighted jointly by the author of the Dissertation and the National Technical University of Athens.\\[5pt]

Copying, storage, and distribution of this work, in whole or part of it, for commercial purposes is prohibited. Reproduction, storage, and distribution for non-profit purposes of educational or research nature is allowed, provided that the source of origin is mentioned and this copyright message is maintained. Questions concerning the use of this work for commercial purposes should be addressed to the author.\\[5pt]

Copyright \copyright \hspace{0.5pt} Vasileios Leon, 2022\\ 
National Technical University of Athens\\
All rights reserved.\\
DOI:  \href{http://dx.doi.org/10.26240/heal.ntua.24738}{\color{blue}{http://dx.doi.org/10.26240/heal.ntua.24738}}
}
\\[22pt]


{\fontsize{11.1pt}{10pt}\fontfamily{lmss}\selectfont \textbf{Vasileios K. Leon}}\\[3pt]
{PhD, Electrical \& Computer Engineering, National Technical University of Athens\\[2pt]
Diploma, Computer Engineering \& Informatics, University of Patras}

\newpage
\thispagestyle{empty}
~ 
\newpage

\newpage

\thispagestyle{empty}

\begin{minipage}{\textwidth}
\vspace*{5cm}
\fontsize{12.5pt}{19pt}\fontfamily{lmr}\selectfont
\raggedleft \emph{\textbf{To my parents}}
\end{minipage}

\newpage
\thispagestyle{empty}
~ 
\newpage

\restoregeometry

\linespread{1.08}    

\chapter{Abstract}
The recent end of Dennard's Scaling and the declining Moore's Law have signified a new era for computing systems. Power efficiency has now become a critical factor for both cloud and edge computing. Concurrently, the rapid growth of compute-intensive applications from the Digital Signal Processing (DSP) and Artificial Intelligence (AI) domains challenges the resources of computing systems. As a result, the computing industry is forced to find alternative design approaches and computing platforms to sustain increased power efficiency, while providing sufficient performance. Among the examined solutions, \emph{Approximate Computing}, \emph{Hardware Acceleration}, and \emph{Heterogeneous Computing} have gained great momentum. Approximate Computing is a novel design paradigm that exploits the inherent error resilience of DSP/AI applications to deliver gains in power, area, and/or performance by reducing the quality of the results. Hardware Acceleration refers to the execution of demanding computational tasks on specialized hardware, such as the Application-Specific Integrated Circuits (ASICs) and the Field-Programmable Gate Arrays (FPGAs), rather than general-purpose processors. Finally, Heterogeneous Computing refers to versatile processing architectures, such as the Vision Processing Units (VPUs), which integrate more than one type of processor and various memory technologies. 

In this Dissertation, we introduce design solutions and methodologies, built on top of the preceding computing paradigms, for the development of energy-efficient DSP and AI accelerators. In particular, we adopt the promising paradigm of Approximate Computing and apply new approximation techniques in the design of arithmetic circuits. Based on our methodology, these arithmetic approximation techniques are then combined with hardware design techniques to implement approximate ASIC- and FPGA-based DSP and AI accelerators. Moreover, we propose methodologies for the efficient mapping of DSP/AI kernels on distinctive embedded devices, such as the new space-grade FPGAs and the heterogeneous VPUs. On the one hand, we cope with the decreased flexibility of space-grade technology and the technical challenges that arise in new FPGA tools and devices. On the other hand, we unlock the full potential of heterogeneity by surpassing the increased hardware complexity and exploiting all the diverse processors and memories.

In more detail, the proposed arithmetic approximation techniques involve bit-level optimizations, inexact operand encodings, and skipping of computations, while they are applied in both fixed- and floating-point arithmetic. To increase the design space and extract the most efficient solutions, we also conduct an extensive exploration on combinations among the approximation techniques. Furthermore, we propose a low-overhead scheme for seamlessly adjusting the approximation degree of our circuits at runtime. In comparison with state-of-the-art designs, the proposed arithmetic circuits feature a very large approximation space, i.e., a wide range of approximation configurations, which allows to maximize the resource gains for a given error constraint. Our techniques induce a mean relative error of up to \raisebox{0.8pt}{$\scriptstyle\sim$}$2\%$, i.e., typical error values for approximate circuits. The most prominent approximate circuits of the Dissertation form a high-resolution Pareto front in a comparative evaluation involving state-of-the-art designs of the literature, and they deliver up to $63\%$ better energy consumption. Finally, our runtime-configurable circuits exhibit a small area overhead of \raisebox{0.8pt}{$\scriptstyle\sim$}$3\%$ compared to the accurate design, and they provide \raisebox{0.8pt}{$\scriptstyle\sim$}$1.5\times$ less energy gains than their respective design-time counterparts with fixed approximation. Nevertheless, they can dynamically change the approximation degree, namely, the accuracy of the calculations, while they still attain remarkable energy gains versus the accurate circuit and state-of-the-art approximate circuits. At the accelerator level, we develop a plethora of approximate kernels for 1D/2D signal processing and Convolutional Neural Networks (CNNs). The experimental results show that we achieve small relative errors for classic DSP calculations and $0\%$--$5\%$ accuracy loss in CNNs for various arithmetic formats while providing up to $70\%$ area and energy savings.

Regarding the DSP acceleration on new space-grade FPGAs, we apply our methodology to efficiently map computer vision algorithms onto the radiation-hardened NanoXplore's FPGAs. In the end, we achieve balanced resource utilization, which is comparable to that of well-established FPGA vendors. Moreover, the throughput is sufficient (e.g., up to $10$ FPS for feature detection on MPixel images), considering the performance requirements of vision-based space applications. In terms of Heterogeneous Computing, we accelerate custom DSP kernels, a sophisticated computer vision pipeline, and a demanding CNN with ResNet-50 backbone on Intel's Myriad VPUs. The proposed methodology and embedded design techniques provide speedups up to $20\times$ for classic DSP on Myriad 2, while the power lies around $1$W. The CNN is accelerated on Myriad X with $2$W, achieving \raisebox{0.8pt}{$\scriptstyle\sim$}$8.5\times$ and \raisebox{0.8pt}{$\scriptstyle\sim$}$1.7\times$ better performance-per-Watt than the ARM CPU and the Jetson Nano GPU, respectively.

\textbf{Keywords:} 
Approximate Computing, 
Approximation Techniques, 
Arithmetic Circuits, 
Computer Arithmetic, 
Hardware Design, 
Hardware Accelerators, 
ASIC, FPGA, VPU, SoC, 
Heterogeneous Computing, 
Embedded Systems,
Space-Grade, 
Digital Signal Processing, 
Computer Vision, 
Convolutional Neural Networks.


\chapter[Greek Abstract]{\selectlanguage{Greek}Περίληψη}
\begin{otherlanguage}{greek}

Το πρόσφατο τέλος της Κλιμάκωσης του \eng{Dennard} και η φθίνουσα πορεία του Νόμου του \eng{Moore} έχουν σηματοδοτήσει μια νέα εποχή για τα υπολογιστικά συστήματα. Η κατανάλωση ισχύος αποτελεί πλέον έναν κρίσιμο παράγοντα, τόσο για το υπολογιστικό νέφος όσο και για υπολογισμούς στην άκρη του δικτύου. Ταυτόχρονα, η ταχεία ανάπτυξη απαιτητικών εφαρμογών από τους τομείς της Ψηφιακής Επεξεργασίας Σήματος (\eng{DSP}) και της Τεχνητής Νοημοσύνης (\eng{AI}) δημιουργεί προκλήσεις στους πόρους των υπολογιστικών συστημάτων. Ως αποτέλεσμα, η βιομηχανία των υπολογιστών υιοθετεί εναλλακτικές μεθόδους σχεδίασης κυκλωμάτων και συστημάτων, ώστε να διατηρήσει χαμηλή κατανάλωση ισχύος, παρέχοντας όμως και επαρκή ταχύτητα. Ανάμεσα στις λύσεις που εξετάζονται, ο \emph{Προσεγγιστικός Υπολογισμός} εκμεταλλεύεται την εγγενή ανθεκτικότητα σε σφάλματα των \eng{DSP/AI} εφαρμογών ώστε να προσφέρει κέρδη σε πόρους μειώνοντας την ποιότητα των αποτελεσμάτων. Η \emph{Επιτάχυνση Υλικού} αναφέρεται στην εκτέλεση απαιτητικών υπολογιστικών εργασιών σε εξειδικευμένο υλικό, όπως τα Ολοκληρωμένα Κυκλώματα Ειδικής Εφαρμογής (\eng{ASICs}) και οι Συστοιχίες Επιτόπια Προγραμματιζόμενων Πυλών (\eng{FPGAs}). Τέλος, ο \emph{Ετερογενής Υπολογισμός} αναφέρεται σε ευέλικτες αρχιτεκτονικές επεξεργασίας με πολλαπλούς τύπους επεξεργαστή και μνήμης, όπως οι Μονάδες Επεξεργασίας Όρασης (\eng{VPUs}). 

Στην παρούσα Διατριβή, εισάγουμε σχεδιαστικές λύσεις και μεθοδολογίες βασισμένες στα προαναφερθέντα πρότυπα σχεδίασης, με στόχο την ανάπτυξη ενεργειακά αποδοτικών επιταχυντών υλικού. Σχετικά με τον Προσεγγιστικό Υπολογισμό, εφαρμόζουμε νέες τεχνικές προσέγγισης στη σχεδίαση αριθμητικών κυκλωμάτων. Οι τεχνικές αυτές συνδυάζονται με βάση τη μεθοδολογία μας με κλασσικές τεχνικές σχεδίασης, ώστε να υλοποιήσουμε προσεγγιστικούς \eng{DSP} και \eng{AI} επιταχυντές σε \eng{ASIC} και \eng{FPGA}. Επιπλέον, προτείνουμε μεθοδολογίες για την αποτελεσματική αποτύπωση \eng{DSP}/\eng{AI} πυρήνων πάνω σε ιδιόμορφες ενσωματωμένες συσκευές, όπως τα νέα \eng{FPGAs} διαστημικού βαθμού και οι ετερογενείς \eng{VPUs}. Όσον αφορά τα \eng{FPGAs}, αντιμετωπίζουμε τις τεχνικές προκλήσεις που προκύπτουν κατά τη χρήση νέων εργαλείων, ενώ για τις \eng{VPUs}, ξεκλειδώνουμε όλες τις δυνατότητες της ετερογένειας, ξεπερνώντας την αυξημένη πολυπλοκότητα υλικού και αξιοποιώντας όλους τους διαφορετικούς πόρους. 

Οι προτεινόμενες τεχνικές αριθμητικής προσέγγισης περιλαμβάνουν βελτιστοποιήσεις σε επίπεδο δυαδικού ψηφίου, μη ακριβείς κωδικοποιήσεις τελεστών, και παράλειψη υπολογισμών, ενώ εφαρμόζονται σε αριθμητική τόσο σταθερής όσο και κινητής υποδιαστολής. Για να αυξηθεί ο χώρος σχεδίασης και να εξάγουμε τις πιο αποτελεσματικές λύσεις, πραγματοποιούμε επίσης μια εκτενή εξερεύνηση πάνω στους συνδυασμούς των τεχνικών. Επιπλέον, προτείνουμε ένα σχήμα χαμηλής επιβάρυνσης για την απρόσκοπτη ρύθμιση του βαθμού προσέγγισης των κυκλωμάτων κατά το χρόνο εκτέλεσης. Σε σύγκριση με σημαντικά κυκλώματα της βιβλιογραφίας, οι προτεινόμενες λύσεις διαθέτουν πολύ μεγαλύτερο χώρο προσέγγισης (ευρύτερο φάσμα προσεγγίσεων), επιτρέποντας τη μεγιστοποίηση των κερδών σε πόρους για έναν δεδομένο περιορισμό σφάλματος. Οι τεχνικές μας προκαλούν ένα μέσο σχετικό σφάλμα έως και \raisebox{0.8pt}{$\scriptstyle\sim$}$2\%$, δηλαδή τυπικές τιμές σφάλματος προσεγγιστικών κυκλωμάτων. Τα πιο εξέχοντα προσεγγιστικά κυκλώματα της Διατριβής σχηματίζουν ένα σύνορο \eng{Pareto} υψηλής ανάλυσης στη συγκριτική αξιολόγηση με σημαντικές εργασίες της βιβλιογραφίας, προσφέροντας έως και $63\%$ καλύτερη κατανάλωση ενέργειας. Τέλος, τα κυκλώματα που μπορούν να ρυθμίσουν δυναμικά την προσέγγιση, έχουν αυξημένη επιφάνεια κατά \raisebox{0.8pt}{$\scriptstyle\sim$}$3\%$ σε σύγκριση με το ακριβές κύκλωμα, και παρέχουν \raisebox{0.8pt}{$\scriptstyle\sim$}1.5$\times$ λιγότερα κέρδη ενέργειας από τα αντίστοιχα κυκλώματα με σταθερή προσέγγιση. Όμως, έχουν τη δυνατότητα να αλλάζουν την ακρίβεια των υπολογισμών, ενώ εξακολουθούν να προσφέρουν αξιοσημείωτα ενεργειακά κέρδη έναντι του ακριβούς κυκλώματος και κυκλωμάτων της βιβλιογραφίας. Σε επίπεδο επιταχυντή, αναπτύσσουμε μια πληθώρα από προσεγγιστικούς πυρήνες για επεξεργασία σημάτων/εικόνων και Συνελικτικά Νευρωνικά Δίκτυα (\eng{CNNs}). Με βάση την πειραματική ανάλυση, τα σφάλματα είναι μικρά σε κλασικούς \eng{DSP} υπολογισμούς και η απώλεια ακρίβειας κυμαίνεται ως $5\%$ στα νευρωνικά δίκτυα, ενώ επιτυγχάνεται έως και $70\%$ εξοικονόμηση επιφάνειας και ενέργειας.

Σχετικά με τα νέα \eng{FPGAs} διαστημικού βαθμού, εφαρμόζουμε τη μεθοδολογία μας για την αποτελεσματική απεικόνιση αλγορίθμων υπολογιστικής όρασης στα ανθεκτικά-σε-ακτινοβολία \eng{FPGAs} της \eng{NanoXplore}. Στο τέλος, επιτυγχάνουμε ισορροπημένη χρήση πόρων, η οποία είναι συγκρίσιμη με αυτή των καθιερωμένων προμηθευτών \eng{FPGAs}. Επιπλέον, η ταχύτητα είναι επαρκής (π.χ., έως και $10$ \eng{FPS} για την ανίχνευση χαρακτηριστικών σε \eng{MPixel} εικόνες), λαμβάνοντας υπόψη τις απαιτήσεις απόδοσης των διαστημικών εφαρμογών. Σχετικά με τον Ετερογενή Υπολογισμό, επιταχύνουμε \eng{DSP} πυρήνες, μια ακολουθία αλγορίθμων υπολογιστικής όρασης, και ένα απαιτητικό \eng{CNN} στις \eng{Myriad VPUs} της \eng{Intel}. Οι προτεινόμενες μεθοδολογίες και τεχνικές ενσωματωμένης σχεδίασης παρέχουν επιτάχυνση έως και $20\times$ σε κλασικούς \eng{DSP} υπολογισμούς στη \eng{Myriad 2} με κατανάλωση ισχύος $1$\eng{W}. Το \eng{CNN} επιταχύνεται στη \eng{Myriad X} με $2$\eng{W}, προσφέροντας \raisebox{0.8pt}{$\scriptstyle\sim$}8.5$\times$ και \raisebox{0.8pt}{$\scriptstyle\sim$}1.7$\times$ καλύτερη απόδοση-ανά-\eng{Watt} από τον επεξεργαστή γενικού-σκοπού \eng{ARM} και τον επεξεργαστή γραφικών \eng{Jetson Nano}, αντίστοιχα.

\textbf{Λέξεις Κλειδιά:} 
Προσεγγιστικός Υπολογισμός, 
Τεχνικές Προσέγγισης, 
Αριθμητικά Κυκλώματα, 
Αριθμητική Υπολογιστών, 
Σχεδίαση Υλικού, 
Επιταχυντές Υλικού,
Ετερογενής Υπολογισμός,
Ενσωματωμένα Συστήματα,
Τεχνολογία Διαστήματος,
Ψηφιακή Επεξεργασία Σήματος, 
Υπολογιστική Όραση, 
Συνελικτικά Νευρωνικά Δίκτυα. 

\end{otherlanguage}

\tableofcontents
\listoffigures
\listoftables
\chapter*{List of Abbreviations}
\addcontentsline{toc}{chapter}{List of Abbreviations}
\chaptermark{List of Abbreviations}

\begin{longtable}{l@{\hspace{30pt}}|@{\hspace{30pt}}l@{\hspace{50pt}}}
\textbf{AI} & Artificial Intelligence\\[1.5pt]
\textbf{ALU} & Arithmetic Logic Unit\\[1.5pt] 
\textbf{ANN} & Artificial Neural Network\\[1.5pt]
\textbf{ASIC} & Application-Specific Integrated Circuit\\[1.5pt]
\textbf{BER}  & Bit Error Rate\\[1.5pt] 
\textbf{BRAVE} & Big Re-programmable Array for Versatile Environments\\[1.5pt]
\textbf{CER} & Correct Edge Ratio\\[1.5pt]
\textbf{CFA} & Circuit Functional Approximation\\[1.5pt] 
\textbf{CIF} & Camera Interface\\[1.5pt]
\textbf{CKG} & Clock Generator\\[1.5pt] 
\textbf{CMB} & Conventional Modified Booth\\[1.5pt]
\textbf{CMX} & Connection Matrix\\[1.5pt]
\textbf{CNN} & Convolutional Neural Network\\[1.5pt]
\textbf{CORDIC} & Coordinate Rotation Digital Computer\\[1.5pt] 
\textbf{COTS} & Commercial-Off-The-Shelf\\[1.5pt]
\textbf{CPU} & Central Processing Unit\\[1.5pt]
\textbf{CUDA} & Compute Unified Device Architecture\\[1.5pt]
\textbf{CV} & Computer Vision\\[1.5pt]
\textbf{CY} & Carry Unit \\[1.5pt]
\textbf{DDR} & Double Data Rate \\[1.5pt]
\textbf{DFF} & Delay Flip-Flop\\[1.5pt]
\textbf{DLSB} & Double Least Significant Bit\\[1.5pt]
\textbf{DMA} & Direct Memory Access\\[1.5pt]
\textbf{DNN} & Deep Neural Network\\[1.5pt]
\textbf{DRAM} & Dynamic Random Access Memory\\[1.5pt]
\textbf{DSE} & Design Space Exploration\\[1.5pt]
\textbf{DSP} & Digital Signal Processing\\[1.5pt]
\textbf{ECR} & Erroneous Classification Ratio\\[1.5pt]
\textbf{EDAC} & Error Detection And Correction\\[1.5pt] 
\textbf{EO} & Earth Observation\\[1.5pt]
\textbf{ESA} & European Space Agency\\[1.5pt]
\textbf{FE} & Functional Element\\[1.5pt]
\textbf{FIFO} & First-In First-Out\\[1.5pt] 
\textbf{FIR}  & Finite Impulse Response\\[1.5pt] 
\textbf{FPGA} & Field-Programmable Gate Array\\[1.5pt]
\textbf{FPS} & Frames Per Second\\[1.5pt]
\textbf{FPU} & Floating-Point Unit\\[1.5pt]
\textbf{GPU} & Graphics Processing Unit\\[1.5pt]
\textbf{HDL} & Hardware Description Language\\[1.5pt]
\textbf{IoT} & Internet of Things\\[1.5pt]
\textbf{LCD} & Liquid Crystal Display\\[1.5pt]
\textbf{LEO} & Low Earth Orbit\\[1.5pt]
\textbf{LLR} & Log-Likelihood-Ratio\\[1.5pt]
\textbf{LNS} & Logarithmic Number System\\[1.5pt]
\textbf{LOCE} & Location Error\\[1.5pt]
\textbf{LSB} & Least Significant Bit\\[1.5pt]
\textbf{LUT} & Look-Up Table\\[1.5pt]
\textbf{MB} & Modified Booth\\[1.5pt]
\textbf{MDK} & Myriad Development Kit\\[1.5pt]
\textbf{ML} & Machine Learning\\[1.5pt]
\textbf{MPDS} & Mega Pixel Disparities per Second\\[1.5pt]
\textbf{MRED} & Mean Relative Error Distance\\[1.5pt]
\textbf{MSB} & Most Significant Bit\\[1.5pt]
\textbf{NaN} & Not-a-Number\\[1.5pt]
\textbf{NCE} & Neural Computer Engine\\[1.5pt]
\textbf{NCS} & Neural Computer Stick\\[1.5pt]
\textbf{OC} & Over-Clocking\\[1.5pt] 
\textbf{ORIE} & Orientation Error\\[1.5pt]
\textbf{PDF} & Probability Density Function\\[1.5pt] 
\textbf{PE} & Processing Element\\[1.5pt]  
\textbf{PLL} & Phase-Locked Loop\\[1.5pt]  
\textbf{PON} & Possibility of Overflow-Normal\\[1.5pt]
\textbf{PP} & Partial Product\\[1.5pt]
\textbf{PRED} & Possibility of Relative Error Distance\\[1.5pt]
\textbf{PSNR} & Peak Signal-to-Noise Ratio\\[1.5pt] 
\textbf{PUN} & Possibility of Underflow-Normal\\[1.5pt]
\textbf{QAM}  & Quadrature Amplitude Modulation\\[1.5pt]
\textbf{QoS} & Quality of Service\\[1.5pt] 
\textbf{RAMB} & Random Access Memory Block\\[1.5pt]
\textbf{RED} & Relative Error Distance\\[1.5pt]
\textbf{RF} & Register File\\[1.5pt]
\textbf{RH} & Radiation Hardened\\[1.5pt]
\textbf{RNS} & Residue Number System\\[1.5pt]
\textbf{ROM} & Read-Only Memory\\[1.5pt]
\textbf{RT} & Radiation Tolerant\\[1.5pt]
\textbf{RTL} & Register-Transfer Level\\[1.5pt]
\textbf{SHAVE} & Streaming Hybrid Architecture Vector Engine\\[1.5pt]
\textbf{SIMD} & Single Instruction Multiple Data\\[1.5pt]
\textbf{SIPP} & Streaming Image Processing Pipeline\\[1.5pt]
\textbf{SNR} & Signal-to-Noise Ratio\\[1.5pt]
\textbf{SoC} & System-on-Chip\\[1.5pt]
\textbf{SoM} & System-on-Module\\[1.5pt]
\textbf{SRAM} & Static Random Access Memory\\[1.5pt]
\textbf{STA} & Static Timing Analysis\\[1.5pt]
\textbf{SSIM} & Structural Similarity Index\\[1.5pt] 
\textbf{SWaP} & Size, Weight and Power\\[1.5pt]
\textbf{TDP} & Thermal Design Power\\[1.5pt] 
\textbf{TMR} & Triple Modular Redundancy\\[1.5pt] 
\textbf{TOPS} & Tera Operations Per Second\\[1.5pt]
\textbf{TPU} & Tensor Processing Unit\\[1.5pt]
\textbf{UART} & Universal Asynchronous Receiver-Transmitter\\[1.5pt]
\textbf{VBN} & Vision-Based Navigation\\[1.5pt]
\textbf{VLIW} & Very Long Instruction Word\\[1.5pt]
\textbf{VOS} & Voltage Over-Scaling\\[1.5pt]
\textbf{VPU} & Vision Processing Unit\\[1.5pt]
\textbf{WFG} & Waveform Generator\\[1.5pt] 
\end{longtable}

\mainmatter
\bstctlcite{IEEEexample:BSTcontrol} 
\chapter{Introduction}
\label{chapter1}

\addtocontents{lof}{\protect\contentsline{chapter}{\protect\numberline{1}Introduction}{}{}}
\addtocontents{lot}{\protect\contentsline{chapter}{\protect\numberline{1}Introduction}{}{}}

\section{The Landscape of Embedded Systems}

\begin{figure}[!t]
\centering
\includegraphics[width=0.92\textwidth]{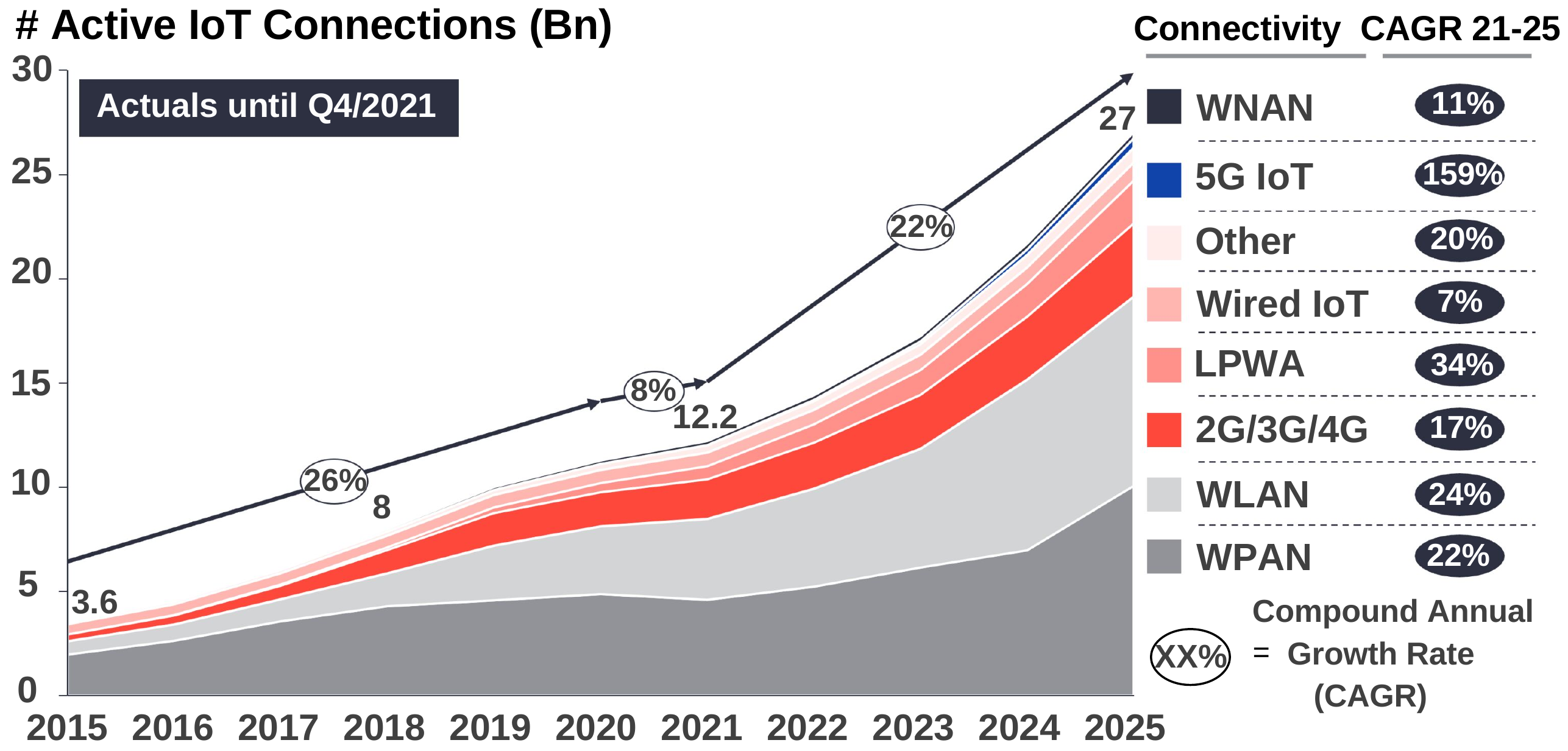}%
\caption[Number of Connected IoT Devices from 2015 to 2025]{Number of connected IoT devices from 2015 to 2025. Source: IoT Analytics,  \url{https://iot-analytics.com/number-connected-iot-devices/}.}%
\label{fig_iott}
\end{figure}

The rapid technological advancements
in processing, communication, storage, and sensing 
have transformed the landscape
of embedded systems.
With the emergence of Internet of Things (IoT) \cite{iot_elsi},
there is a huge increment in the amount of data that are generated,
which imposes technical challenges 
in the typical resource-constrained devices.
On the other hand, 
the transmission of all these data to cloud infrastructures and data centers 
for processing 
creates communication bottlenecks,
does not guarantee real-time response,
while it is often avoided due to safety and privacy issues. 
Due to the ever-growing number of IoT connections,
which is expected to be 27 billion in 2025
(as shown in Figure \ref{fig_iott}),
the original cloud-centric system is already stressed
to meet the runtime requirements.
As a result, 
there is a tendency to
process the data
at the edge of the network, namely,
upon they are generated.
This new computing paradigm is established
by the name Edge Computing \cite{edge_ieee}
and has gained significant momentum over the last years.

Concurrently, 
the massive growth
of demanding applications
and powerful algorithms
from domains such as
Digital Signal Processing (DSP),
Computer Vision (CV),
Artificial Intelligence (AI),
and
Machine Learning (ML),
marks a new era for the 
computing systems at the edge.
Conventional embedded processors,
such as Central Processing Units (CPUs) and microcontrollers,
do not have the computational power
to handle these compute-intensive workloads, 
and thus, 
they are unable to meet the performance requirements \cite{glent, georgis, ai_edge}.
Another limitation is the restricted memory capacity at the edge,
which slows down the processing,
especially in applications with a large amount of I/O data.

It is evident that
the proliferation of data
along with the
emergence of applications with increased complexity 
requires alternative design solutions
to sustain sufficient performance at the edge.
For this reason,
the embedded systems have evolved over the years
and constitute multi-purpose systems that 
sense, act, and communicate with their environment.
In earlier years, 
the embedded systems
consisted of simple microcontrollers
and memories.
In contrast,
contemporary
embedded systems are build
on complex 
System-on-Modules (SoMs)
and
System-on-Chips (SoCs),
i.e., 
they are placed in 
single boards
and single chips,
respectively. 
These systems
integrate CPUs,
novel specialized processors,
memory blocks, 
high-bandwidth peripherals,
and communication interfaces.
Nevertheless,
besides space limitations, 
there is a vital factor
that does not allow to seamlessly
increase the computation resources,
e.g., by adding more devices and/or processing units:
the power consumption. 
Typical embedded systems 
consume some Watts \cite{embed_power},
while 
ultra-low-power embedded systems
(e.g., wearable)
have a power budget of a few milliWatts \cite{embed_power2}.
Consequently,
there is a trade-off
between performance and power:
from the one side,
the applications demand speed and real-time response,
while from the other side,
there are tight power constrains at the edge.

\section{The Evolution of Integrated Circuits}

\begin{figure}[!t]
\centering
\includegraphics[width=0.89\textwidth]{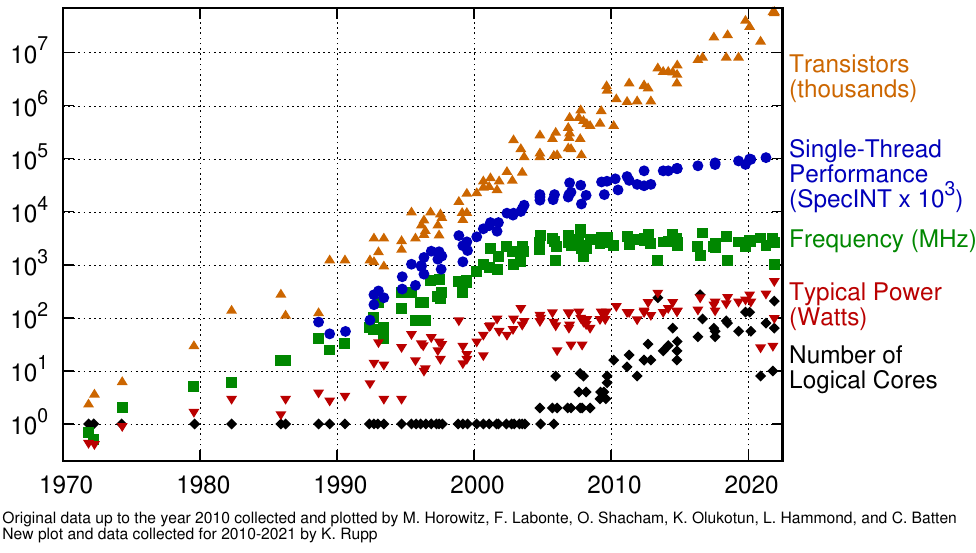}%
\caption[50 Years Trends in Microprocessors]{50-year trends in microprocessors. Source: Karl Rupp,  \url{https://github.com/karlrupp/microprocessor-trend-data}.}%
\label{fig_trend}
\end{figure}

The integrated circuits
constitute the cornerstone of computing systems,
as they inherently impact their performance, power consumption, and area utilization. 
Figure \ref{fig_trend} illustrates the trends for the microprocessors over the last 50 years.
Historically,
the semiconductor technology was driven for more than 40 years 
by two fundamental principles:
Moore's Law \cite{moore}
and 
Dennard's Law \cite{dennard}.
According to Gordon Moore \cite{moore}, 
the number of transistors in
a dense integrated circuit doubles approximately every two years.
As shown in Figure \ref{fig_trend}, 
the transistor scaling is linear until today,
even though several researchers 
forecast that Moore's Law will end soon \cite{moore_future}.
Moore's Law is often quoted together with
the prediction of David House (executive of Intel),
who then said that the overall computing performance would double approximately every 18 months.
On the other hand,
Dennard's Law, 
which expired in the mid-2000s,
is the force behind Moore's Law.
According to Robert Dennard \cite{dennard}, 
the power density 
(power per silicon area) 
remains stable as the transistors get smaller,
so that the power use stays in proportion with the area,
i.e., both voltage and current scale (downward) with the length.
Practically,
for a given area size,
the power consumption of the chip remained 
the same in each transition to a new generation of process technology (technological node). 
Therefore, 
each new process technology 
doubled the number of transistors in a chip 
without increasing the power consumption. 
The combination of the two laws
allowed to scale the supply voltage and the threshold voltage,
resulting in 
lower power per transistor,
and thus,
almost stable power density.

Nevertheless,
the scaling law of Dennard did not consider
the impact of the 
transistor sub-threshold leakage on the total chip power
\cite{dennardend7}.
More specifically, 
in technological nodes of some nanometers,
the decrease of the threshold voltage 
results in an exponential increase of
the leakage power. 
This did not happen in 1970s,
because
the sub-threshold
leakage was small 
and had negligible impact on the total chip power.
As a result,
the threshold voltage can no longer be reduced,
and thus, 
the scaling of the supply voltage stopped
(further scaling could affect the performance).

In summary,
even though
the number of transistors integrated per area is increasing,
the supply voltage is not scaled accordingly,
and thus,
the power density is increased. 
The failure of Dennard's scaling 
in conjunction with other factors
such as the cooling technology
and the natural limits of silicon
has led us to the ``Dark Silicon'' era \cite{dark1, dark2, dark3}.
In this era,
the 
entire circuitry of an integrated circuit cannot be powered-on at the nominal operating voltage for a given Thermal Design Power (TDP) constraint.
All the things considered,
the power efficiency and management is nowadays a critical issue
for computing systems,
either they are placed at the edge (embedded systems)
or on the cloud (data centers). 
Therefore,
the industry of computing systems
is forced to find new 
design approaches 
and computing platforms,
which will improve the power efficiency
while providing the desired performance.

Towards power-efficient, real-time, and high-yielding
computing systems, 
\emph{\textbf{Approximate Computing}} \cite{2013_Han_ETS, 2016_Mittal_ACMsrv, 2016_Xu_IEEEdt, Shafique_2016, 2021_Stanley_ACMsrv},
\emph{\textbf{Hardware Acceleration}} \cite{glent, georgis, kachris, accel1, accel2},
and
\emph{\textbf{Heterogeneous Computing}} \cite{heter1, heterbook, heterog, heter2, heter3}
have attracted much interest from the 
research community.
The Approximate Computing paradigm
exploits the error resilience
of applications from the DSP/AI
domains
to reduce the quality of the results
and deliver in exchange 
gains in power, area, and/or performance.
Hardware Acceleration
refers to the process of
offloading compute-intensive tasks onto specialized hardware,
such as the Application-Specific Integrated Circuits (ASICs) and the Field-Programmable Gate Arrays (FPGAs),
rather than executing them on general-purpose CPUs.
Finally,
Heterogeneous Computing
refers to systems
integrating more than one type of processor,
such as the Graphics Processing Units (GPUs)
and the Vision Processing Units (VPUs).
These design approaches are 
inherently linked
and overlapping each other:
(i) 
practically,
heterogeneous hardware architectures
(i.e., GPUs and VPUs)
belong in the wider range of hardware accelerators, 
and
(ii)
approximations can be applied 
in implementations of all the hardware accelerators.

\section{Approximate Computing}

Approximate computations have been applied 
since the 1960s.
For example,
in one of the first works of the field,
Mitchell proposed the logarithmic-based multiplication/division \cite{1962_Mitchell}).
Nevertheless,
the first systematic efforts to 
define the Approximate Computing paradigm
started in the late 2000s.  
Since then, 
various terms have been used
to describe the process of
generating approximate architectures, programs, and circuits.
Approximate Computing is synonymous or overlaps with them. 
The most well-known terms of the literature are listed as follows:

\begin{itemize}
\setlength\itemsep{-4pt}
\item Chakradhar \emph{et al.} \cite{ChakradharDAC2010}
define ``Best-Effort Computing'' as
\emph{``the approach of designing software/hardware computing systems
with reduced workload, improved parallelization and/or approximate components
towards enhanced efficiency and scalability''}.
\item Carbin \emph{et al.} \cite{2012_Carbin_PLDI} introduce the term ``Relaxed Programming''
to express 
\emph{``the transformation of programs with approximation methods and relaxed semantics
to enable greater flexibility in their execution''}.
\item Chippa \emph{et al.} \cite{2014_Chippa_IEEEtvlsi}
use the term ``Scalable Effort Design'' 
for
\emph{``the systematic approach that embodies the notion of scalable effort into the design process at different levels of abstraction,
involving mechanisms to vary the computational effort and control knobs to achieve the best possible trade-off between energy efficiency and quality of results''}.
\end{itemize}

According to Mittal \cite{2016_Mittal_ACMsrv},
\emph{``Approximate Computing exploits the gap between
the accuracy required by the applications/users and that provided by the computing system
to achieve diverse optimizations''}.
Han and Orshansky \cite{2013_Han_ETS}
distinguish Approximate Computing
from Probabilistic/Stochastic Computing,
stating that 
\emph{``it does not involve assumptions on the stochastic
nature of the underlying processes implementing the system
and employs deterministic designs for producing inaccurate results''}.
Another interesting point of view 
is expressed by Sampson \cite{2015_Sampson_phd},
who claims that 
Approximate Computing 
is based on
\emph{``the idea that we are hindering the efficiency of the computer systems
by demanding too much accuracy from them''}.
In this Dissertation,
we attribute the following definition:

\begin{itemize}[wide=3pt, leftmargin=*]
    \item[$\diamond$]\underline{Approximate Computing}: \emph{A progressive paradigm shift
for the development of systems, circuits \& programs,
build on top of the error-resilient nature of application domains,
and based on disciplined methods to intentionally induce errors 
that will provide 
valuable resource gains
in exchange for 
tunable accuracy loss}.
\end{itemize}

The world of computing systems
is full of workloads 
with intrinsic error tolerance \cite{ChakradharDAC2010}.
These workloads
are significant pillars of 
domains such as
DSP
(e.g, signal/image/video/audio processing),
AI/ML
(e.g., artificial neural networks, clustering),
and Data Analytics
(e.g., web search, data mining).
Approximate Computing exploits this error resilience
across the entire computing stack,
i.e., at both software and hardware levels.
The factors that allow
Approximate Computing to decrease the quality of the results in exchange for resource gains,
originate from \cite{ChakradharDAC2010}:
\begin{itemize}
\setlength\itemsep{-3pt}
    \item the user's intention to accept results of lower quality.
    \item the limited human perception, e.g., in multimedia applications.
    \item the lack of perfect/golden results for validation, e.g., in data mining applications.
    \item the lack of a unique answer/solution, e.g., in machine learning applications.
    \item the application's self-healing property, i.e., the inherent capability to absorb/compensate errors.  
    \item the application's inherent approximate nature, e.g., in probabilistic calculations, iterative algorithms, and learning systems.
    \item the application's analog/noisy real-world input data, e.g., in multimedia/signal processing.
\end{itemize}

There are several challenges in the design of approximate systems.
First of all,
the accuracy is still 
of the utmost importance.
Namely,
even though errors are tolerated,
the accuracy needs to be retained
within the acceptable limits of the application.
This requires the analysis of the application with realistic datasets
and 
the development of accurate models that emulate the approximations.
Moreover,
modern approximate systems
need to adapt their accuracy at runtime
depending on the application's/user's constraints.
Hence,
the approximation methods
should enable
runtime approximation tuning
instead of providing 
a single static
approximation configuration.
Another challenge in Approximate 
Computing
is the development of systems
with cross-layer approximation,
i.e., the synergistic application
of approximation techniques from different layers of the computing stack.
This approach has shown promising results,
however,
sophisticated methodologies 
for the automatic configuration of the approximations in all system's modules
are still missing.

In this \underline{Dissertation},
regarding ``Approximate Computing'',
we report an extensive literature review
of the state-of-the-art software and hardware approximation techniques
and then, we
focus on hardware-level Approximate Computing.
In particular,
we propose new approximation techniques
for the design of power-efficient arithmetic circuits,
and employ 
them along with 
other design techniques
to develop approximate hardware accelerators
from the DSP and AI domains. 

\section{Hardware Acceleration}

\begin{figure}[!t]
\vspace*{-3pt}
\centering
\subfloat[\label{fig_ha1}]{\includegraphics[width=0.79\textwidth]{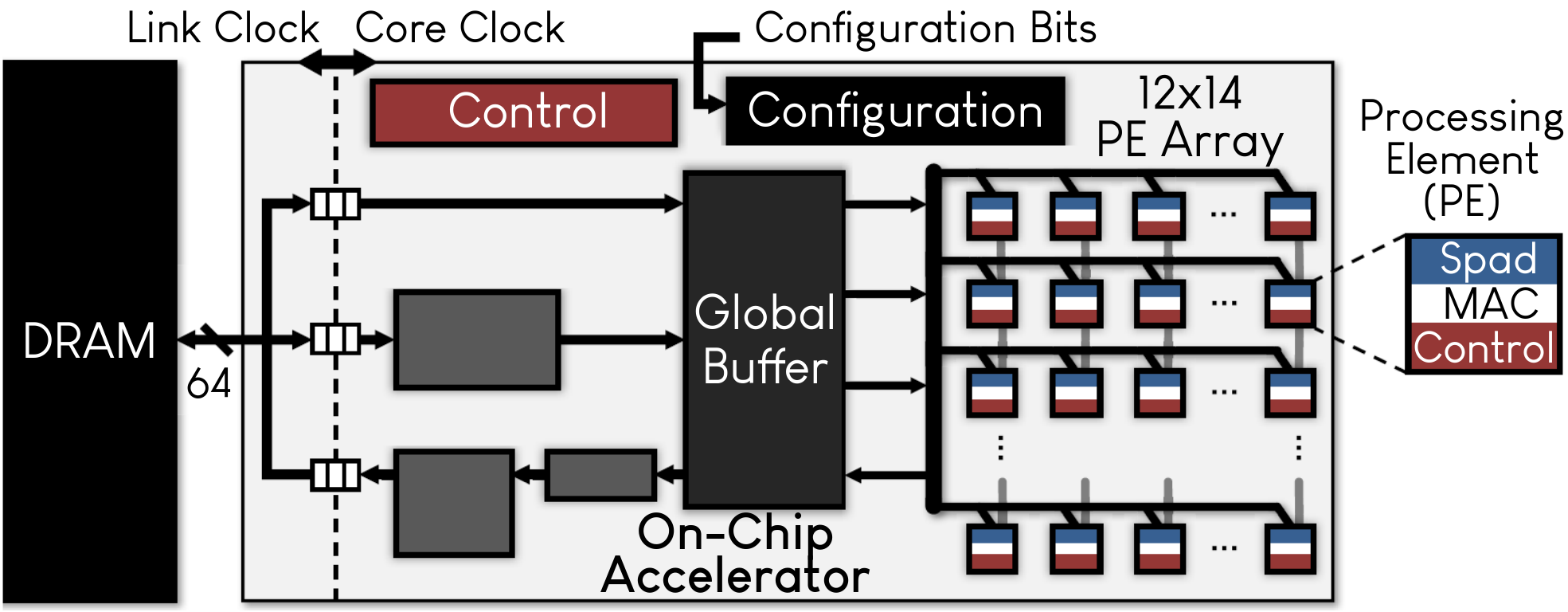}}  %
\\[-0.2pt]
\subfloat[\label{fig_ha2}]{\includegraphics[width=0.79\textwidth]{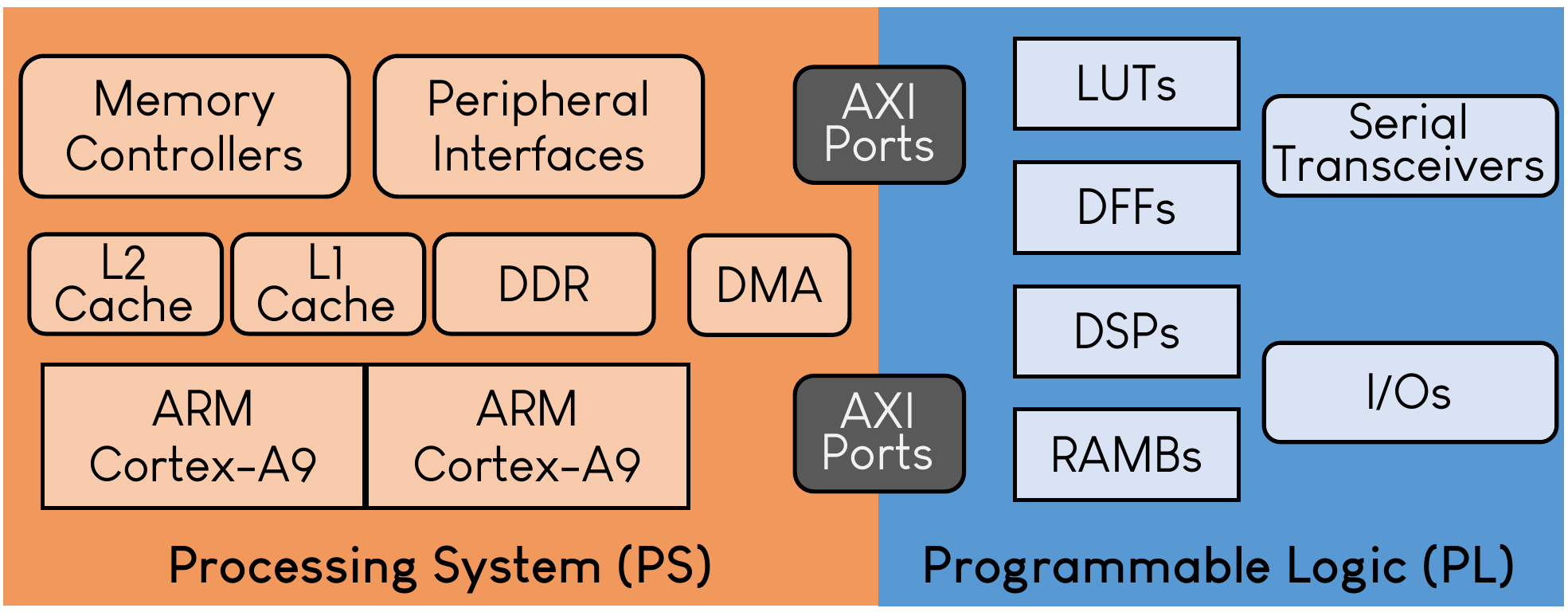}}%
\caption[Architecture of the Eyeriss ASIC and the Xilinx Zynq-7000 SoC FPGA]{High-level architecture of 
\textbf{(a)} the Eyeriss ASIC \cite{eyeriss}
and
\textbf{(b)} the Xilinx Zynq-7000 SoC FPGA \cite{zynqs}.}%
\label{fig_hwaccel}
\vspace*{-7pt}
\end{figure}

The conventional CPU-based computing platforms
do not provide sufficient performance for compute-intensive DSP/AI workloads \cite{glent, georgis, kachris, accel1, accel2}.
As already discussed, 
the computing industry employs novel processors
based on ASICs, FPGAs, GPUs, and VPUs
to cope with the increased demands
of modern DSP/AI workloads.
The execution of high-complexity computing tasks
on customized hardware,
such as the preceding specialized processors,
is widely known as Hardware Acceleration.
In terms of development,
Hardware Acceleration requires
more 
programming effort and time 
than the respective development on general-purpose CPU-based processors.
Nevertheless,
the performance results are very impressive,
providing speedups of orders of magnitude.
We note that
besides the aforementioned accelerators,
the market offers
additional hardware platforms,
such Google's 
Tensor Processing Unit (TPU) \cite{tpu_datac},
which is an AI accelerator.
Below, we introduce in brief the
most common hardware accelerators:
\begin{itemize}
\setlength\itemsep{-3pt}
    \item \underline{ASIC}: an integrated circuit that is customized for a specific application/function. It cannot be reprogrammed or modified after its production. ASICs are used for the efficient implementation of DSP/AI functions and general-purpose tasks of computing systems. 
    \item \underline{FPGA}: an integrated circuit that is manufactured with configurable logic blocks and programmable interconnects. It can be reprogrammed numerous times after its production. FPGAs are mainly used for accelerating heavy DSP/AI functions, as well as for interfacing and prototyping purposes.
    \item \underline{GPU}: a processor integrating multiple specialized small cores. 
    GPUs are used for accelerating graphics rendering, calculations involving massive amount of data, scientific calculations, and CV/AI workloads.
    \item \underline{VPU}: a processor integrating vector cores, image processing filters and AI engines. VPUs excel in imaging/vision tasks and are used for accelerating DSP/AI workloads. 
\end{itemize}
In the remainder of this Dissertation,
we report a plethora of related works and discussions
involving the aforementioned hardware accelerators.
Indicatively, 
in Figure \ref{fig_hwaccel},
we depict
the well-known Eyeriss ASIC for Convolutional Neural Networks (CNNs) \cite{eyeriss} 
and Xilinx's Zynq-7000 SoC FPGA \cite{zynqs}.
Eyeriss is designed for accelerating CNNs with many layers and varying shapes,
and it is
based on a spatial array of 168 Processing Elements (PEs)
and 
a global on-chip buffer.
The data movement is optimized 
by exploiting data reuse and inter-PE communication, 
while data gating and compression 
are used to reduce the power consumption. 
On the other hand,
the Zynq SoC FPGAs have been established in the market
as state-of-the-art devices for Hardware Acceleration.
These devices offer 
the software programmability of an ARM-based processing system 
($2$ ARM Cortex-A9, on-chip memory, caches, memory controllers, and peripherals)
and
the hardware programmability of a traditional FPGA
(programmable resources for logic functions, arithmetic operations, registers, and memories).
All these components are integrated in a single chip
to
provide a
fully scalable SoC platform
for high-performance DSP/AI processing.

In this \underline{Dissertation},
regarding ``Hardware Acceleration'',
we implement approximate arithmetic circuits using standard-cell libraries for ASIC.
Our arithmetic circuits are also 
integrated in hardware DSP/AI functions,
which are accelerated on
standard-cell ASIC and Xilinx's FPGAs.
Moreover,
we implement demanding CV kernels on the new European space-grade FPGAs.
Finally,
we accelerate custom DSP kernels,
a sophisticated CV pipeline,
and CNNs on Intel's VPUs.

\section{Heterogeneous Computing}

\begin{figure}[!t]
\vspace*{-12pt}
\centering
\subfloat[\label{fig_hh1}]{\includegraphics[width=0.5\textwidth]{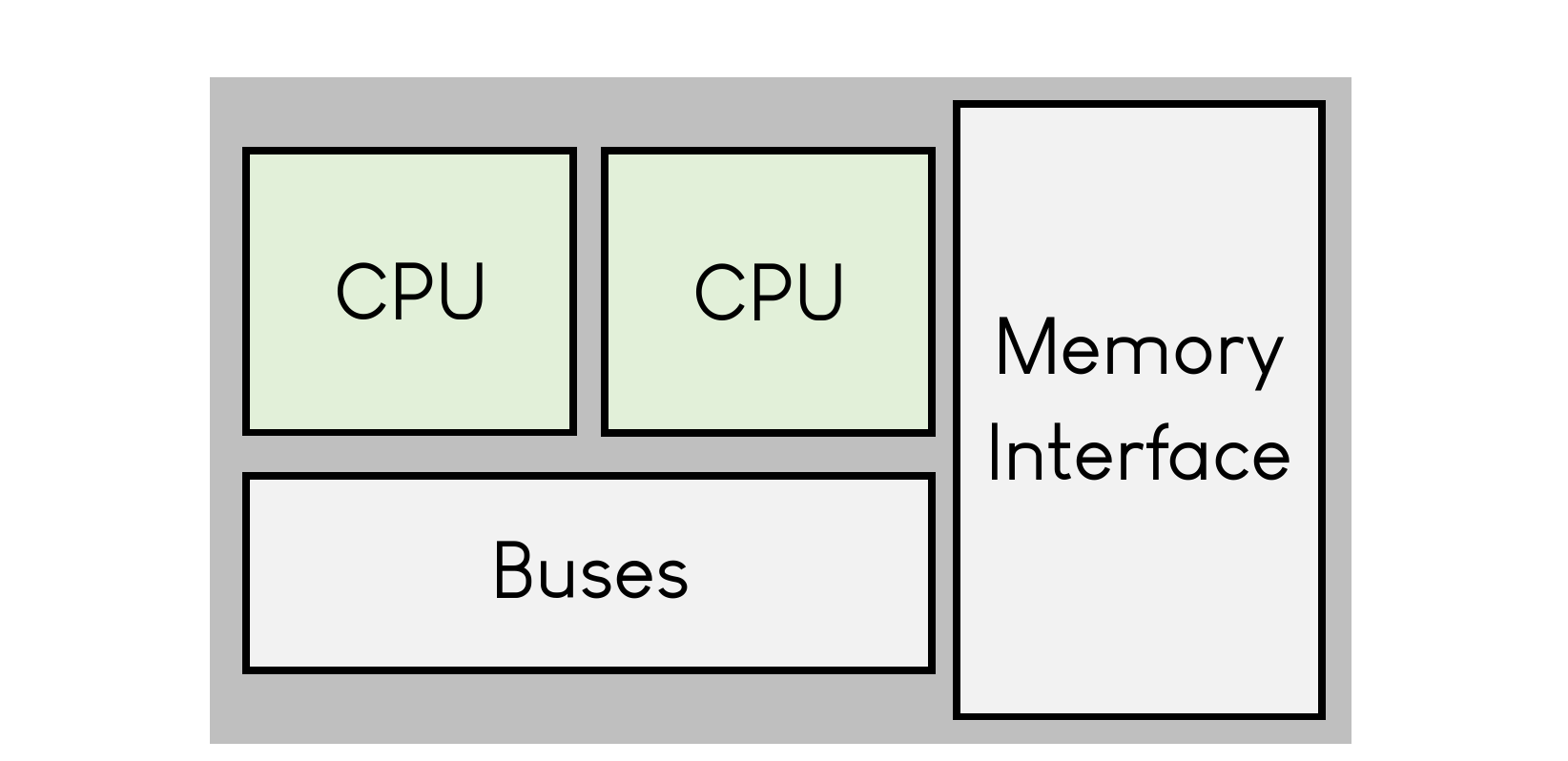}}  %
\hfill
\subfloat[\label{fig_hh2}]{\includegraphics[width=0.5\textwidth]{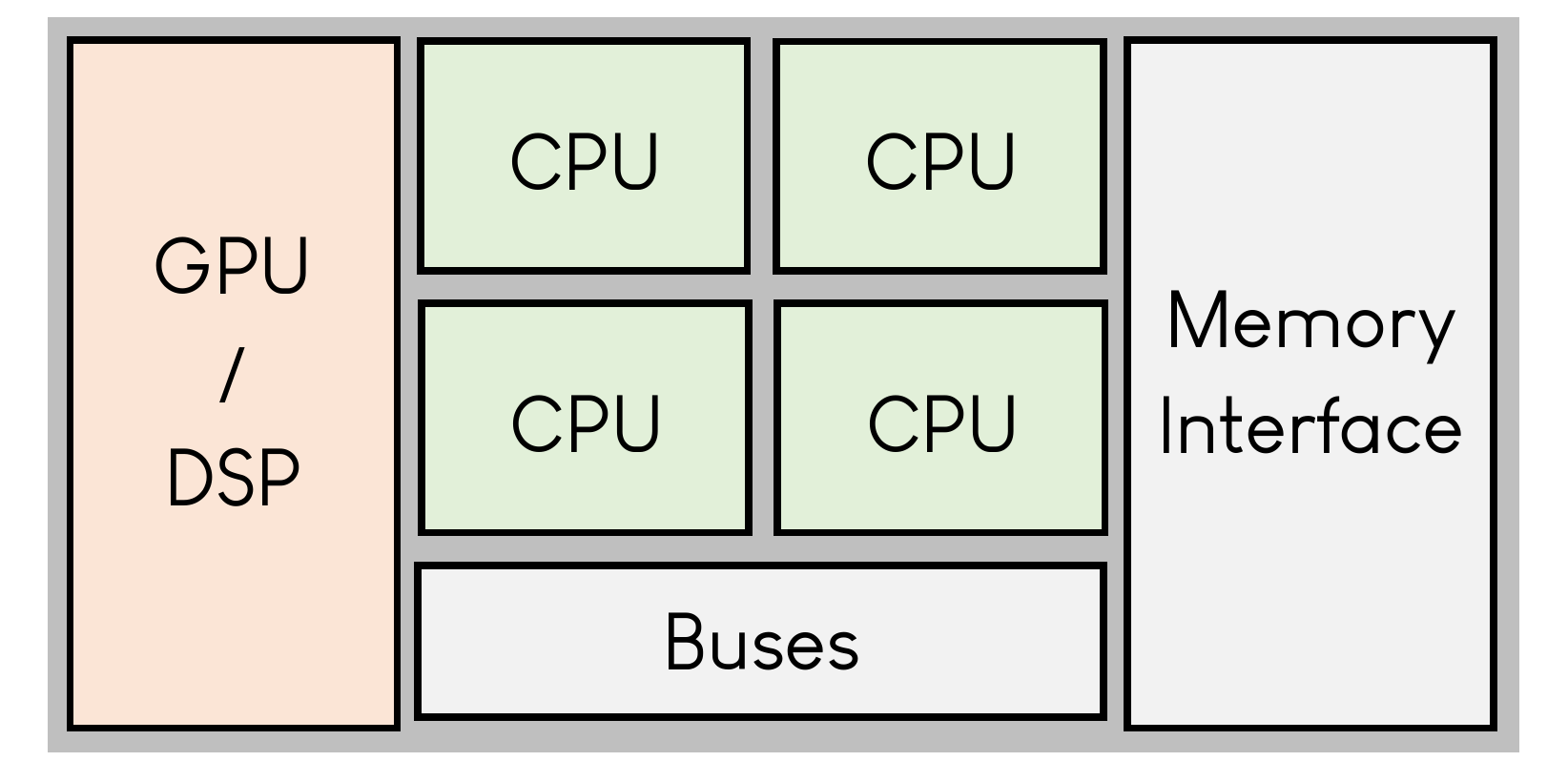}}\\[-6pt]
\subfloat[\label{fig_hh3}]{\includegraphics[width=0.5\textwidth]{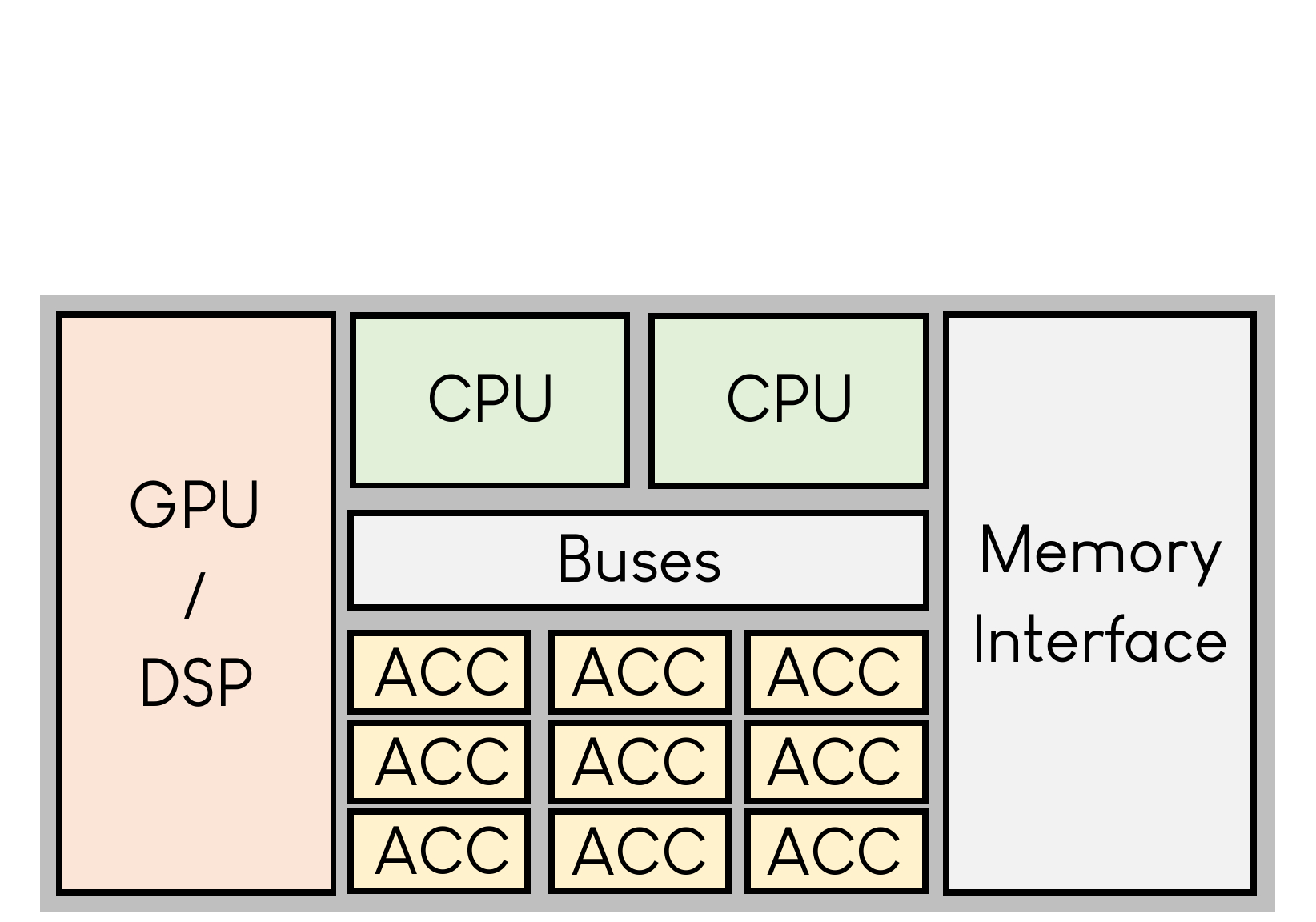}} 
\hfill
\subfloat[\label{fig_hh4}]{\includegraphics[width=0.5\textwidth]{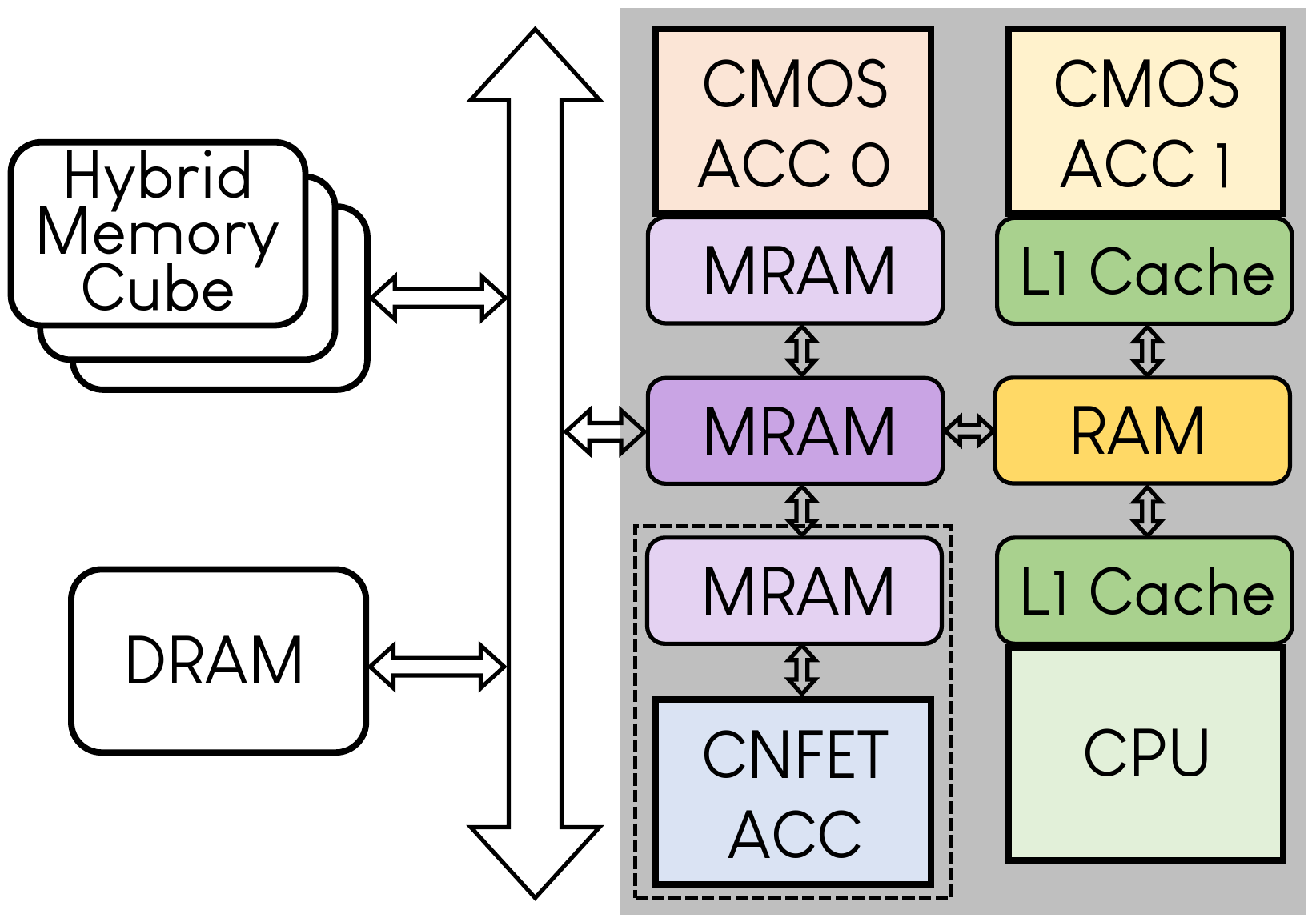}}%
\caption[Evolution of Heterogeneous Computing Architectures]{Evolution of heterogeneous computing architectures \cite{moore_future}:
\textbf{(a)} homogeneous CPU (past),
\textbf{(b)} CPU + GPU/DSP (present), 
\textbf{(c)} CPU + GPU/DSP + accelerators (present),
and 
\textbf{(d)} extreme heterogeneity in processors and memories (future).}%
\label{fig_heter}
\vspace*{-7pt}
\end{figure}

The increased diversity of modern DSP/AI workloads
in I/O, computational, and memory
requirements
has marginalized the use of homogeneous CPU-based computing platforms.
To cope with all these requirements
and provide efficient design solutions,
the computing industry has turned to 
Heterogeneous Computing architectures
\cite{heter1, heterbook, heterog, heter2, heter3}.
These architectures integrate more than one type of processor
and potentially
different memory technologies.
In particular,
contemporary heterogeneous architectures
offer both
general-purpose processors
and
specialized acceleration cores/engines. 
In terms of storage,
the heterogeneity 
usually offers global memory, caches and scratchpad (working) memory.
Figure \ref{fig_heter} shows
the evolution of heterogeneous architectures
according to Shalf \cite{moore_future}. 
Computing architectures with two similar CPUs
or a multi-core CPU
(Figure \ref{fig_hh1})
are now considered homogeneous systems.
Currently,
there are two prevailing kinds
of computing architecture:
\begin{enumerate}
\setlength\itemsep{-3pt}
    \item[(i)]  
    the heterogeneous architecture 
    of Figure \ref{fig_hh2},
    which combines the general-purpose CPU with a GPU/DSP accelerator \cite{heterog}.  
    \item[(ii)] 
    the very heterogeneous architecture of Figure \ref{fig_hh3},
    which 
    additionally includes small accelerators
    (e.g., the VPU SoCs \cite{myriad_vpu}).
\end{enumerate}

The heterogeneity is expected to increase in the future (Figure \ref{fig_hh4}),
integrating accelerators and memories of different technologies. 
The very heterogeneous architecture
of Figure \ref{fig_hh3}
is met in Intel's Myriad VPUs \cite{myriad_vpu},
which constitute embedded SoCs
for Edge Computing.
These VPUs have recently emerged as an attractive solution for accelerating imaging applications
with only $1$--$2$W. 
Compared to the CPU--GPU architectures, 
the VPU SoCs are more heterogeneous,
as they offer 
general-purpose CPUs, 
vector cores, 
hardware filters for image processing,
and a dedicated AI accelerator in the case of Myriad X.

Heterogeneous Computing imposes
several challenges.
From the developer side,
the programming model
involves parallel computing
and mapping to specialized hardware,
and thus,
it is more complex
than the respective one
of conventional computing.
The workloads need to 
be efficiently distributed and parallelized
among the cores and accelerators
to deliver improved performance
and power efficiency. 
Towards this direction,
the developer has to make decisions
regarding 
the application's decomposition into parallel computing tasks,
the selection of the most suitable processor
for each task,
and the identification
of the parts that offer 
limited parallelization opportunities
or do not require acceleration.
In the case of the Myriad VPUs,
there are more technical challenges,
given that these SoCs are mainly build for power efficiency
rather than high performance.
Therefore, the developer needs to
exploit every piece of the VPU heterogeneity
to provide sufficient performance.

In this \underline{Dissertation},
regarding ``Heterogeneous Computing'',
we develop methodologies
for the efficient mapping and scheduling
of DSP/CV kernels and CNNs on the Myriad VPUs.
We introduce several high- and low-level implementation techniques
and evaluate the suitability of the VPUs
as edge processors.

\section{Scope and Contribution of Dissertation}

The \emph{\textbf{scope}} of the Dissertation
is the design of 
\underline{arithmetic circuits}
and
\underline{DSP \& AI} \underline{accelerators}.
In this context,
we propose 
\underline{design solutions and methodologies}
for improving the efficiency of the implementations  
on ASIC/FPGA and multi-core SoCs.
At circuit level,
we adopt the promising design paradigm of Approximate Computing 
and propose new arithmetic approximation techniques, 
which are then used 
to design various approximate hardware accelerators on ASIC/FPGA technology.
At platform level,
we aim to unlock the full potential of new embedded devices,
such as 
the space-grade FPGAs 
and 
the multi-core VPUs,
by surpassing the bottlenecks of the tools 
and 
exploiting the heterogeneity of the SoCs,
respectively,
to accelerate high-performance DSP/AI workloads.

The main \emph{\textbf{differentiation}} of the Dissertation
compared to prior art 
is summarized as follows:
\begin{itemize}
\item At \underline{design technique} level, 
we propose 
approximation methods
that provide a \emph{larger} approximation space,
i.e., multiple approximation configurations,
enabling to maximize the resource gains under a specified error constraint. 
\item At \underline{circuit} level,
we propose 
energy-efficient approximate circuits
that can \emph{seamlessly} 
adjust their approximation configuration at runtime. 
\item At \underline{hardware accelerator} level,
we perform an \emph{extensive} design space exploration 
on approximation techniques,
arithmetic formats, algorithms, and hardware design techniques
to generate approximate ASIC/FPGA-based accelerators. 
\item At \underline{computing platform} level,
we \emph{systematically}
examine the capabilities of the programming tools and exploit the underlying hardware architectures
to accelerate high-performance DSP/AI workloads.
\end{itemize}

The \emph{\textbf{contribution}}
of the Dissertation is summarized as follows:
\begin{itemize}
\item[(i)] In Chapter \ref{chapter2},
we report a comprehensive up-to-date \emph{literature survey} for Approximate Computing,
where we report the basic terminology,
and then, 
classify and analyze the state-of-the-art software and hardware approximation techniques.
\item[(ii)] In Chapter \ref{chapter3},
we highlight the significance of the underlying arithmetic in circuits, 
and show that novel numerical formats
and 
\emph{sophisticated bit-level optimizations} can provide valuable resource gains in hardware. 
\item[(iii)] In Chapter \ref{chapter4},
we address the circuit overheads of the classic high-radix encodings and propose a new approximate hybrid high-radix encoding, 
which is parametric in terms of approximation degree.
This encoding is used to design
the \emph{RAD} family of approximate multipliers. 
\item[(iv)] In Chapter \ref{chapter5},
we introduce a low-overhead dynamic configuration scheme for adjusting the approximation degree of multipliers at runtime.
This technique is applied in fixed- and floating-point arithmetic,
generating the \emph{DyFXU} and \emph{DyFPU}
families of runtime-configurable approximate multipliers.
\item[(v)] In Chapter \ref{chapter6},
we highlight the efficiency of integrating more than one approximation technique in the design of approximate circuits.
In this context,
we combine various state-of-the-art techniques
and provide a very large design space for approximate multiplication.
This extensive exploration
results in the \emph{ROUP} family of approximate multipliers,
which form the state-of-the-art Pareto-front. 
\item[(vi)] In Chapter \ref{chapter7},
we introduce a methodology for designing approximate DSP/AI hardware accelerators. 
Based on this methodology,
we fuse approximation techniques
with various arithmetic formats,
DSP/AI algorithms,
and 
hardware design techniques,
to generate energy-efficient
\emph{hardware accelerators}
for 1D/2D signal processing
and neural networks. 
\item[(vii)] In Chapter \ref{chapter8}, we introduce a methodology for efficiently mapping and accelerating high-performance DSP algorithms on the \emph{new European space-grade FPGAs}.
Based on this methodology,
we apply our tool-level exploration
to surpass issues that arise because
the FPGA vendor is new
and the space-grade FPGAs exhibit decreased flexibility compared to the commercial ones.
\item[(viii)] In Chapter \ref{chapter9}, we introduce a methodology for partitioning, scheduling and optimizing
demanding DSP and AI workloads on the \emph{heterogeneous multi-core VPUs}.
Based on this methodology,
we exploit the full potential of the VPU SoCs
to provide sufficient DSP and AI acceleration with limited power consumption.
\end{itemize}

\section{Structure of Dissertation}

\begin{figure}[!t]
\vspace*{-6pt}
\centering
\includegraphics[width=0.89\textwidth]{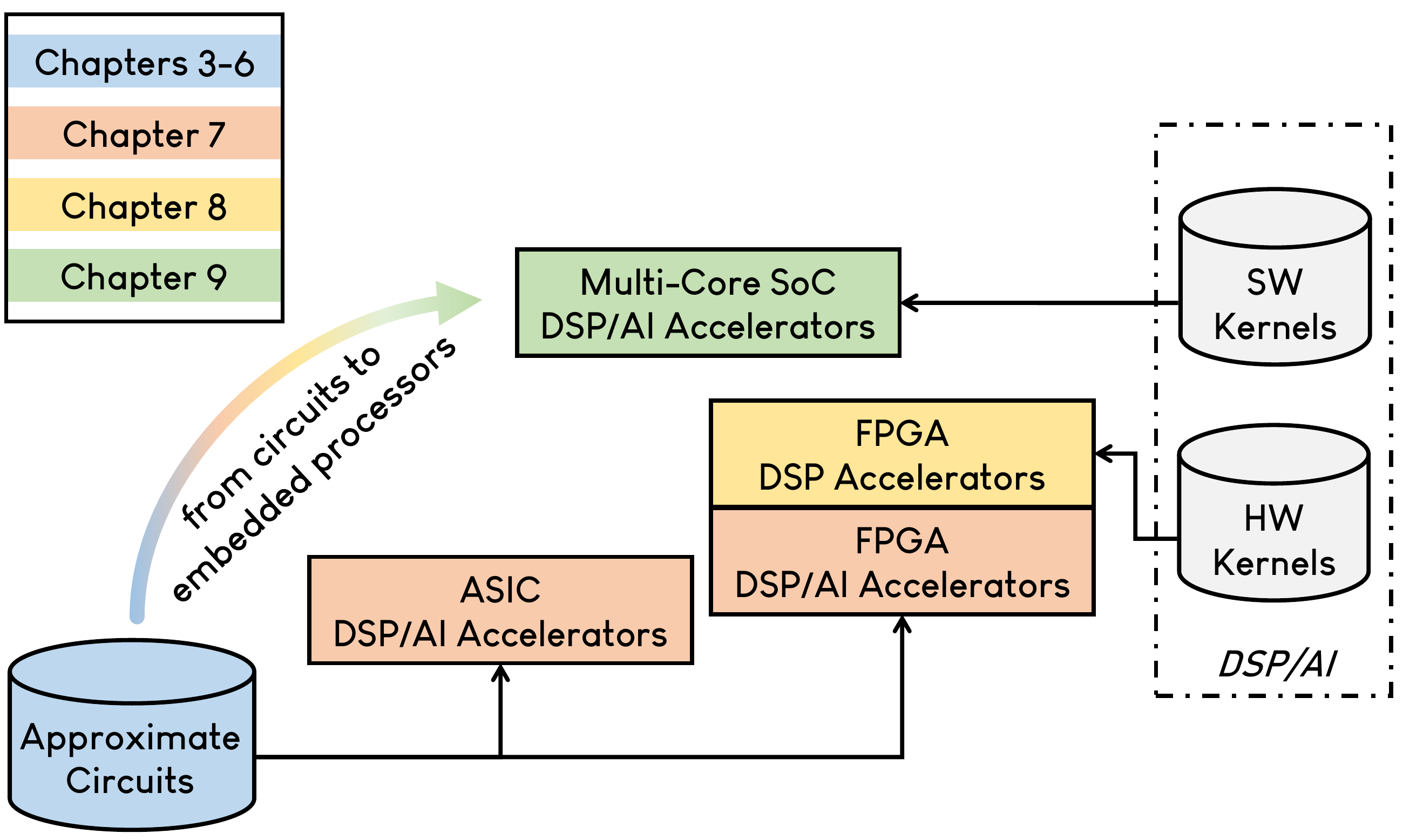}%
\vspace*{-4pt}
\caption[The Structure of the Ph.D. Dissertation]{The structure of the Ph.D. Dissertation.}%
\label{fig_thesis}
\end{figure}

The remainder of the Dissertation
is organized as follows.
Chapter \ref{chapter2}
introduces the Approximate Computing paradigm,
reviewing the terminology
and 
state-of-the-art software \& hardware approximation techniques. 
Chapters \ref{chapter3}--\ref{chapter9}
report the main work of the Dissertation.
Finally,
Chapter \ref{chapter10}
concludes the Dissertation
by summarizing the contributions
and discussing future extensions.
The structure of the main work
is presented in Figure \ref{fig_thesis}
and 
is divided in two parts:
\begin{description}
\item[Part \ref{part1}:] \emph{``Arithmetic Approximation Techniques for Circuit Design''}
\item[Part \ref{part2}:] \emph{``Design Methodologies for Embedded Computing''}
\end{description}

\begin{figure}[!t]
\vspace*{-9pt}
\centering
\includegraphics[width=0.89\textwidth]{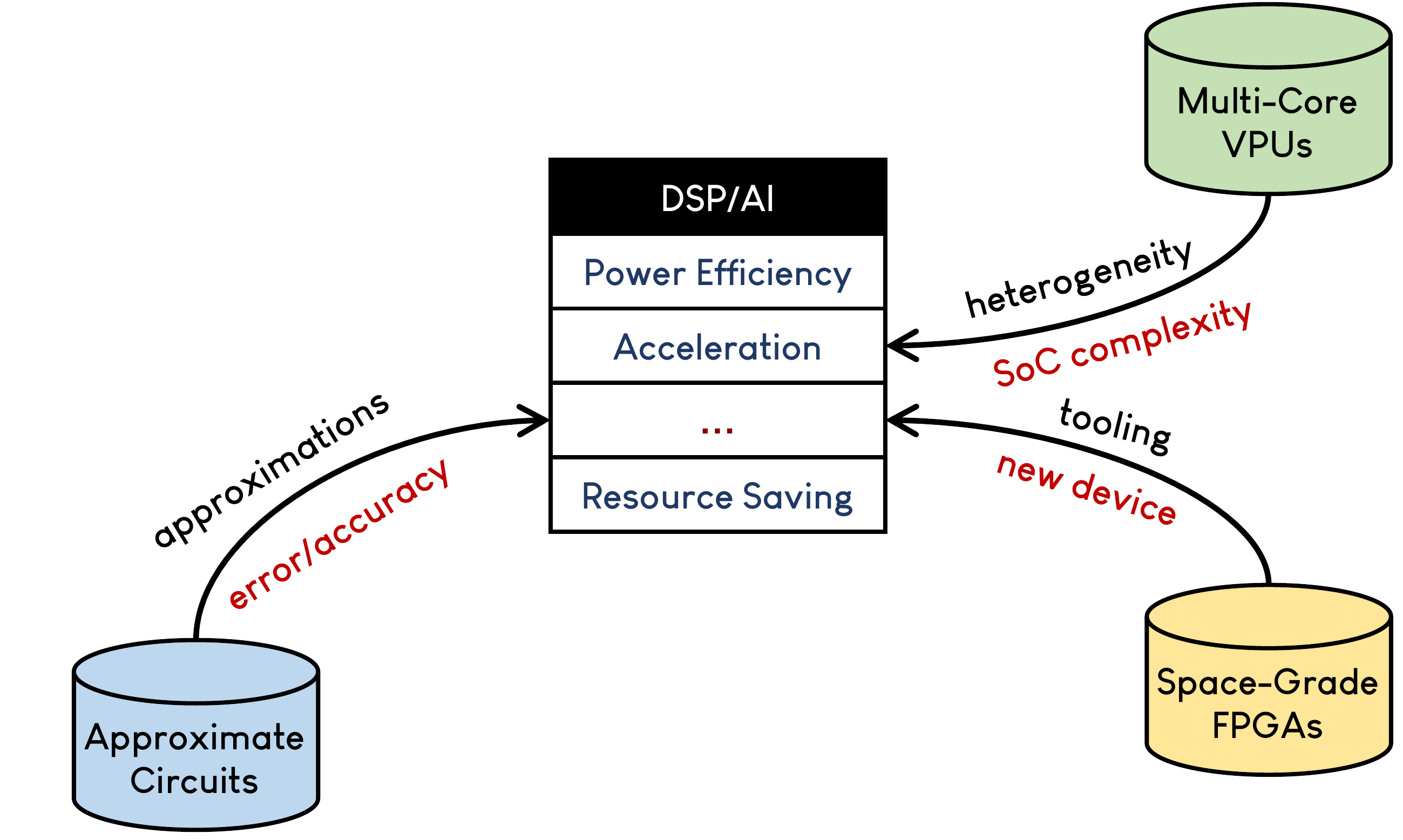}%
\caption[The Goal of the Ph.D. Dissertation]{The goal of the Ph.D. Dissertation achieved through different design layers.}%
\label{fig_thesisgoal}
\vspace*{-6pt}
\end{figure}

Part \ref{part1} includes Chapters \ref{chapter3}--\ref{chapter7}
and focuses on arithmetic approximation techniques
and the design of approximate DSP/AI hardware accelerators. 
Part \ref{part2} includes Chapters \ref{chapter8}--\ref{chapter9}
and focuses on new embedded computing platforms 
(space-grade FPGAs and multi-core VPU SoCs)
and the efficient acceleration of 
DSP/AI kernels.
Both parts have a \emph{\textbf{common goal}}:
the development of energy-efficient DSP/AI accelerators.
Part \ref{part1} reaches this goal
from a lower design abstraction layer,
while 
Part \ref{part2} reaches it
from higher design abstraction layers.
Figure \ref{fig_thesisgoal} depicts
how we achieve this goal and what issues we have to surpass:
(i) by using arithmetic \underline{approximations},
while we have to care about the errors and application's accuracy,
(ii) via extensive and systematic \underline{tooling} on the new space-grade FPGAs,
where we have to surpass the limitations/issues of newly released tools/devices,
(iii) by exploiting the \underline{heterogeneity} of low-power multi-core VPUs,
where we have to cope with resource-constrained edge devices
and the increased complexity of the SoCs. 
Next, we discuss the content of each chapter included in Part \ref{part1} and Part \ref{part2}. 


\textbf{Chapter \ref{chapter3}:} This chapter
acts as introductory to low-level logic optimizations,
targeting to highlight the significance of studying the arithmetic of circuits/accelerators.
For this purpose,
it focuses on the Double Least Significant Bit (DLSB) numerical format,
in which the numbers have an extra least significant bit, 
and it proposes 
sophisticated low-level optimizations.
These bit manipulations
are also used in the next chapters,
and specifically in
the design
of approximate circuits.
More explicitly,
in this chapter,
we 
improve the DLSB multiplication, 
resulting in decreased overheads
versus the straightforward design approach.
Moreover,
as case study,
we demonstrate 
how the proposed optimized circuit
can be used as building block 
in the implementation of large-size multiplications. 

\textbf{Chapter \ref{chapter4}:} This chapter
proposes an approximate hybrid high-radix encoding 
for generating the energy-efficient RAD multipliers.
The proposed encoding scheme
approximately encodes one of the  operands,
using 
the accurate radix-$4$ encoding 
for its most significant part
and an approximate high-radix encoding 
for its least significant part.
The approximation is inserted by 
mapping all the high-radix values to a set of values
including only the $4$ largest powers of two.
The proposed RAD family of approximate multipliers
is configurable,
and can be tuned to achieve the desired energy--accuracy trade-off.

\textbf{Chapter \ref{chapter5}:} This chapter
proposes runtime-configurable approximate multipliers
for fixed- and floating-point arithmetic.
The approximation is inserted by two orthogonal techniques,
i.e., partial product perforation and partial product rounding,
which allow
to integrate a low-overhead scheme
for tuning the approximation at runtime. 
The runtime circuit variants 
DyFXU and DyFPU
deliver negligible overhead
versus their design-time counterparts (AxFXU and AxFPU).
However, 
they still provide energy efficiency
and benefit from
their capability of selecting a different approximation configuration 
(among numerous ones) at runtime.  

\textbf{Chapter \ref{chapter6}:} This chapter
proposes the concept of cooperative approximation,
namely,
the application of more than one arithmetic approximation technique 
in the design of a circuit. 
The goal is twofold: 
(i) to create a very large approximation space
that serves various design scenarios and can handle different 
error constraints or power budgets,
and 
(ii) to identify the most efficient approximation solutions in terms of both accuracy and resources.
Our extensive design space exploration results in $5$ new families of approximate multipliers,
from which,
ROUP is the most prominent,  
as it forms the state-of-the-art Pareto front with increased resolution.

\textbf{Chapter \ref{chapter7}:} This chapter introduces a methodology for the systematic development of approximate DSP/AI hardware accelerators. 
The proposed methodology consists of two stages,
i.e., software-level exploration
and hardware development.
The main feature of the methodology is the combination of
approximation techniques 
with different 
arithmetic formats 
(e.g., fixed/floating-point, quantized integer),
alternative algorithms for the same application,
and classic hardware design techniques.
The goal of this chapter is twofold:
(i) to assess the Dissertation's approximate designs in real-world DSP/AI applications, 
and
(ii) to evaluate the error resilience and quantify the resource gains of approximate DSP/AI accelerators. 

\textbf{Chapter \ref{chapter8}:} This chapter proposes a design methodology
for porting demanding DSP algorithms on the new European space-grade FPGAs. 
The methodology is divided with respect to 
the stages of the typical FPGA design flow. 
Our systematic design approach aids us to
surpass issues that arise when using new tools
or porting designs developed in other FPGA vendors,
as well as confront the decreased flexibility
and lower performance of space-grade FPGAs compared to their commercial counterparts. 
The evaluation is performed 
with hardware kernels for feature detection and stereo vision,
and it includes comparisons to other FPGAs.

\textbf{Chapter \ref{chapter9}:} This chapter proposes a design methodology
for the efficient mapping and acceleration of compute-intensive algorithms on the heterogeneous multi-core VPUs. 
The methodology aims to
highlight the most efficient partitioning and scheduling
schemes in such complex and very heterogeneous SoCs.
In this context, 
we introduce several high- and low-level implementation techniques.
Given that the VPUs are designed
to provide low power consumption,
our methodology and design choices
allow us to deliver sufficient performance
in custom DSP kernels and a sophisticated CV pipeline.
Moreover, we deploy a demanding CNN.
The evaluation includes comparison results with other state-of-the-art embedded devices.

Finally,
even though all chapters are related to each other,
they constitute standalone structures of text.
Namely,
each chapter has its own 
abstract, 
introduction, 
list of contributions,
proposed techniques/designs/methodologies,
experimental evaluation 
and conclusion.

\section{Overview of Technologies, Tools and Devices}

\begin{table}[!b]
\fontsize{9}{10}\selectfont
\renewcommand{\arraystretch}{1.2}
\setlength{\tabcolsep}{6.5pt}
\caption[Overview of Dissertation's Tools and Devices]{Overview of Dissertation's tools and devices.}
\label{tb_tools}
\centering
\begin{tabular}{l|cc} 
\hline
\multicolumn{1}{c|}{\textbf{Tool}} & \textbf{Usage} &   \textbf{Reference} \\
\hline
\hline
Synopsys Design Compiler & standard-cell synthesis & Chapters \ref{chapter3}--\ref{chapter7}  \\
Synopsys PrimeTime & standard-cell power measurement & Chapters \ref{chapter3}--\ref{chapter7}  \\
Siemens QuestaSim & simulation (validation \& power) & Chapters \ref{chapter3}--\ref{chapter8}  \\
Xilinx Vivado & FPGA implementation & Chapters \ref{chapter7}--\ref{chapter8}  \\
Intel Quartus & FPGA implementation & Chapter \ref{chapter8}  \\
Microsemi Libero & FPGA implementation & Chapter \ref{chapter8}  \\
NanoXplore NXmap & FPGA implementation & Chapter \ref{chapter8}  \\
Intel MDK & VPU implementation & Chapter \ref{chapter9}\\
Intel OpenVINO & VPU implementation & Chapter \ref{chapter9} \\
Google TensorFlow & CNN development & Chapters \ref{chapter7},\ref{chapter9} \\
\hline
\end{tabular}
\setlength{\tabcolsep}{23.5pt}
\vspace{4pt}
\begin{tabular}{l|c} 
\hline
\multicolumn{1}{c|}{\textbf{Device/Technology}} &  \textbf{Reference} \\
\hline
\hline
TSMC Standard Cells (65-nm, 45-nm) & Chapters \ref{chapter3}--\ref{chapter7}  \\
Xilinx FPGAs (ZCU106, Zynq-7020) & Chapter \ref{chapter7}  \\
Xilinx FPGA (Virtex-5QV) & Chapter \ref{chapter8}  \\
Intel FPGA (Cyclone III) & Chapter \ref{chapter8}  \\
Microsemi FPGA (RTG4) & Chapter \ref{chapter8}  \\
NanoXplore FPGAs (NG-Medium, NG-Large) & Chapter \ref{chapter8}  \\
Intel VPUs (Myriad 2, Myriad X, NCS2) & Chapter \ref{chapter9}\\
Nvidia  GPU (Jetson Nano) & Chapter \ref{chapter9}\\
\hline
\end{tabular}
\end{table}

In this section,
we report all the industrial-strength programming/development tools, devices/platforms and technologies
that are used in the Dissertation. 
Table \ref{tb_tools} summarizes all the relevant details.
In Chapters \ref{chapter3}--\ref{chapter6},
all the circuits are synthesized
on TSMC standard-cell libraries (65-nm and 45-nm)
with Synopsys' Design Compiler tool.
All the simulations are performed with Siemens' (Mentor Graphics) QuestaSim,
while for the power measurements,
we use Synopsys' PrimeTime. 
The same libraries and tools
are used 
for the synthesis of the ASIC-based accelerators in
Chapter \ref{chapter7}.
Moreover,
this chapter includes
implementation results
for accelerators on 
Xilinx's ZCU106 and Zynq-7020 FPGAs, 
while the associated tool is Vivado.
In Chapter \ref{chapter8},
we report numerous results
from the implementation
of DSP kernels 
on various FPGAs
(Xilinx Virtex-5QV, Intel Cyclone III,
and Microsemi RTG4).
For these implementations,
we use the corresponding software tool
of each FPGA vendor.
Chapter \ref{chapter8} also includes results
for NanoXplore's new space-grade FPGAs
(NG-Medium and NG-Large), 
which are 
generated by the NXmap tool.
Chapter \ref{chapter9}
focuses on Intel's VPUs
(Myriad 2 and Myriad X), 
and all the implementations
are performed
with the associated tools,
i.e., the MDK design suite for custom development
and the OpenVINO framework 
for the deployment of neural networks. 
This chapter also reports
results for Nvidia's Jetson Nano GPU.

\chapter{The Approximate Computing Paradigm}
\label{chapter2}

\addtocontents{lof}{\protect\contentsline{chapter}{\protect\numberline{2}The Approximate Computing Paradigm}{}{}}
\addtocontents{lot}{\protect\contentsline{chapter}{\protect\numberline{2}The Approximate Computing Paradigm}{}{}}

\begin{ChapterAbstract}
The emergence of complex applications
in domains
applying multimedia processing
and machine learning tasks
has transformed
the computing paradigm  
in embedded systems and data centers. 
These applications
involve massive data
and/or high computational complexity.
Consequently, 
there is an emerging need
for computational resources,
which results in increasing more and more the power consumption of the computing systems.
In the past, 
the technology scaling has played 
significant role
towards surpassing these challenges,
however,
its declining efficiency
pushes us to examine new
computing paradigms.
Approximate Computing is such an alternative paradigm,
which trades-off 
accuracy loss and decreased quality of results
for resource gains
(e.g., in power/energy, area, latency, throughput).
This computing paradigm 
is applied to error-resilient applications,
such as those involving multimedia and machine learning,
and it induces errors based on a disciplined approach
to improve the efficiency of the systems/circuits and deliver the desired resource gains.
Approximation opportunities
abound in every layer of the typical computing stack,
i.e., 
from transistors and circuits
to compilers and programming languages.
Therefore, 
there is a great variety
of approximation techniques
at each design abstraction layer,
which study the errors and relax the accuracy from different perspective.
In this chapter, 
we review state-of-the-art research works in Approximate Computing
from all the layers of the computing stack.
At first,
we discuss the terminology of Approximate Computing
and then,
we classify and analyze the state-of-the-art approximation techniques with respect to
their application layer
(software or hardware).
Our taxonomy is fine-grained,
namely we study in-depth the approximation techniques of each layer and
classify them based on their design approach.\\
This chapter is based on our  \textbf{publications} in \textbf{\cite{LeonSURV1, LeonSURV2}}.
\end{ChapterAbstract}
\newpage

\section{Introduction}

Approximate Computing covers the entire computing stack,
i.e., approximation techniques are applied
at all design abstraction layers.
Significant research has been conducted in the field of software approximation techniques,
which involve
approximate programming languages, approximate compilers, approximation frameworks, quality-aware runtime systems, 
as well as approximations via precision scaling,
task skipping and
memoization.
On the other hand,
the most common hardware techniques
target the modification of the circuits and hardware architectures,
i.e., they generate a lossy circuit/architecture
from the nominal accurate one.
There is also wide research on 
the tunable scaling of the circuit's voltage or frequency.

This chapter provides background in Approximate Computing.
In particular,
we introduce the terminology of Approximate Computing
based on our literature search,
and we report an extensive review of the state-of-the-art software and hardware approximation techniques.
Our literature review includes newer works
and more works 
compared to previous well-established surveys in Approximate Computing,
i.e., from 
Mittal \emph{et al.} (2016) \cite{2016_Mittal_ACMsrv},
Xu \emph{et al.} (2016) \cite{2016_Xu_IEEEdt},
and
Shafique \emph{et al.} (2016) \cite{Shafique_2016}.
Furthermore,
it differentiates from the survey of \cite{2021_Stanley_ACMsrv},
as it focuses on implementation details
and analyzes
the approximations of each reviewed technique.

The remainder of this chapter is organized as follows. 
Section \ref{termi} reports the terminology.
Section \ref{clsw} classifies and analyzes the software approximation techniques,
while 
Section \ref{clhw} classifies and analyzes the hardware approximation techniques.

\section{The Terminology of Approximate Computing}
\label{termi}

\begin{table}[!t]
\vspace*{-3pt}
\fontsize{9}{10}\selectfont
\renewcommand{\arraystretch}{1.25}
\setlength{\tabcolsep}{4pt}
\caption[Basic Terminology of Approximate Computing]{Basic terminology of Approximate Computing.}
\begin{tabular}
{>{\raggedright}m{0.24\textwidth}|m{0.71\textwidth}}
\hline  
\makecell[c]{\textbf{Term}} & \makecell[c]{\textbf{Description}}\\
\hline
\hline
\emph{Error-Resilient Application} & The application that tolerates errors and accepts results of lower quality. \\ \hline
\emph{Quality of Service}                   & The quality of the results in terms of errors and accuracy. \\ \hline
\emph{Error Constraints}                     & The quality/accuracy requirements that the results should satisfy. \\ \hline
\emph{Error Threshold}                & The maximum error allowed in the results. \\ \hline
\emph{Golden Result}                        & The result that is obtained from the accurate computations. \\ \hline
\emph{Acceptable Result}                    & The result that satisfies the application's error constraints. \\ \hline
\emph{Variable Accuracy}                    & The capability of providing different levels of accuracy. \\ \hline
\emph{Non-Critical Task}        & The task/computation that can be safely approximated due to its small impact on the quality of the output results. \\ \hline
\emph{Error Analysis}                       & The study involving metrics, mathematics and simulations to examine the range, frequency, scaling, and propagation of errors. \\ \hline
\emph{Approximation Technique}       & The systematic and disciplined approach/method to insert computation errors in exchange for resource gains. \\ \hline 
\emph{Approximation Degree}        & The strength of the approximation technique in terms of computations approximated and errors induced. \\ \hline
\emph{Approximation Configuration}          & An instance of the parameters/settings of the approximation technique. \\ \hline
\emph{Frozen Approximation}                 & The approximation degree is fixed and cannot be re-configured.  \\ \hline
\emph{Dynamic Approximation Tuning}         & The capability of adjusting the approximation degree at runtime to satisfy the desired error constraints. \\ \hline
\emph{Cross-Layer Approximation}            & The approximation that is applied at multiple design abstraction layers (software, hardware, architecture). \\ \hline
\emph{Heterogeneous Approximation}          & The approximation that concurrently applies multiple configurations of different degree within the same system. \\ \hline
\emph{Approximation Space Exploration}        & The study involving error analysis and gain quantification to examine trade-offs and select the suitable approximations. \\ \hline
\emph{Approximation Localization}           & The systematic approach to locate the computations and regions that are offered for approximation. \\ \hline
\emph{Error Modeling}                       & The process of emulating the errors of the approximations. \\ \hline
\emph{Error Prediction}                     & The process of predicting errors before computing the final result. \\ \hline
\emph{Error Detection}                      & The process of identifying an error occurrence. \\ \hline
\emph{Error Compensation}                   & The process of modifying the erroneous result to reduce the error. \\ \hline
\emph{Error Correction}                     & The process of correcting the erroneous result. \\ \hline
\end{tabular}
\label{tb_terms}
\vspace{-8pt}
\end{table}

Table \ref{tb_terms} describes the most frequently used terms in Approximate Computing.
The term \emph{error} is used to 
indicate that the output result
is different from 
the accurate result (produced with conventional computing). 
Error is distinguished from \emph{fault}, 
which refers to an unexpected condition
(e.g., stuck-at-logic in circuits, bit-flips in memories, faults in operating systems) 
that causes the system to unintentionally output erroneous results. 
Another significant term is
\emph{accuracy},
which is defined as 
the distance between the approximate and the 
(nominal) accurate result
and is expressed with various error metrics
(e.g., error rate, mean relative error, mean squared error, classification accuracy).
Accuracy is distinguished from 
\emph{precision},
which expresses the differentiation between nearby discrete values
and does not refer to errors of Approximate Computing but to quantization noise
(inserted by the real-to-digital value mapping). 
More specifically, 
in computer arithmetic, 
the more bits are used for the decimal number part,
the higher the precision,
i.e., there are more bits for the representation
and the numbers are closer to their real value. 
Moreover,
in Approximate Computing,  
the term \emph{Quality-of-Service (QoS)} is used 
to describe the overall quality of the results (in terms of accuracy and errors),
considering the expected/accurate results as baseline
and the application's quality constraints.

\section{Classification of Software Approximation Techniques}
\label{clsw}

\begin{figure}[!b]
    \centering
    \includegraphics[width=1\textwidth]{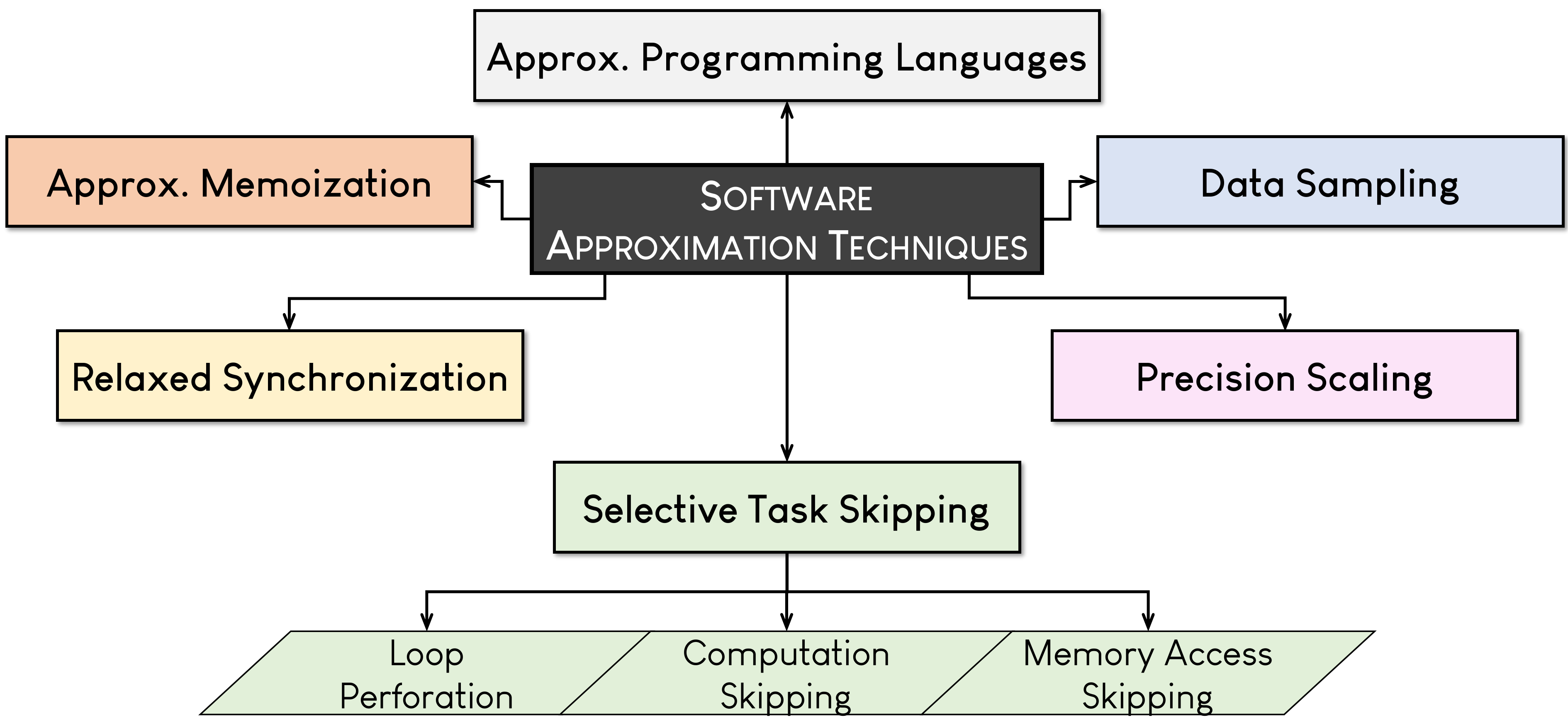}
    \caption[Classification of State-of-the-Art Software Approximation Techniques]{Classification of state-of-the-art software approximation techniques.}
    \label{fig_sw}
\end{figure}

In this section,
we classify and introduce approximation techniques
that are applied at software level,
i.e., the higher level of the design abstraction hierarchy.
The goal of software Approximate Computing
is to improve the execution time of the program
and/or the energy consumption of the system.
The techniques of the literature,
illustrated in Figure \ref{fig_sw},
can be categorized into six classes:
(i) \emph{Selective Task Skipping},
(ii) \emph{Approximate Memoization},
(iii) \emph{Relaxed Synchronization},
(iv) \emph{Precision Scaling},
(v) \emph{Data Sampling},
and (vi) \emph{Approximate Programming Languages}.
Typical techniques
include some of the following features:
approximation libraries/frameworks,
compiler extensions,
accuracy tuning tools,
runtime systems, 
and language annotations. 
Moreover, 
there are numerous techniques
allowing the programmer to 
specify QoS constraints,
provide approximate code variants,
and mark the program's  regions/tasks for approximation.

The remainder of this section
reports representative state-of-the-art works
for software Approximate Computing.
Besides classifying state-of-the-art techniques,
we discuss their approximations and how they are applied. 
Table \ref{tb_sw} reports the references of all the reviewed research works.
We note that the literature also includes entire approximation frameworks,
such as ACCEPT \cite{2015_Sampson_UOW} and OPPROX \cite{2017_Mitra_CGO},
which integrate multiple of these state-of-the-art techniques.
Their goal is to perform an extensive exploration using differing approximation approaches,
identify the best approximation opportunities,
and hence, 
maximize the resource gains while satisfying the application's/user's constraints. 

\begin{table}[!t]
\fontsize{9}{10}\selectfont
\renewcommand{\arraystretch}{1.2}
\setlength{\tabcolsep}{20pt}
\caption[Classification of Software Approximation Techniques]{Classification of software approximation techniques.}
\centering
\label{tb_sw}
\begin{tabular}{l l}
\hline  
\makecell[l]{\textbf{SW Approximation Class}} & \makecell[l]{\textbf{References}}\\
\hline
\hline
Loop Perforation &  \cite{2009_Hoffmann_MIT, 2010_Misailovic_ICSE, 2011_Sidiroglou_FCE, 2015_Shi_IEEEcal, 2017_Omar_ICCD, 2018_Li_ICS, 2020_Baharvand_IEEEtetc, 2015_Tan_ASP-DAC, 2018_Kanduri_DAC} \\
\hline
Computation Skipping & \cite{2009_Meng_IPDPS, 2010_Byna_GPGPU, 2015_Raha_DATE, 2015_Vassiliadis_PPoPP, 2016_Vassiliadis_CGO, 2006_Rinard_ICS, 2007_Rinard_OOPSLA, 2017_Lin_ISCAS, 2018_Akhlaghi_ISCA} \\
\hline
Memory Access Skipping & \cite{2014_Miguel_MICRO, 2016_Yazdanbakhsh_ACMtaco, 2018_Kislal_ELSEclss, 2013_Samadi_MICRO, 2019_Karakoy_ACMmacs, 2015_Zhang_DATE} \\
\hline
Approximate Memoization & \cite{2011_Chaudhuri_FSE, 2014_Samadi_ASPLOS, 2014_Mishra_WACAS, 2017_Brumar_IPDPS, 2015_Keramidas_WAPCO, 2018_Tziantzioulis_IEEEmicro, 2005_Alvarez_IEEEtc, 2013_Rahimi_IEEEtcasii, 2018_Zhang_IEEEcal, 2019_Liu_ISCA} \\
\hline
Relaxed Synchronization & \cite{2012_Renganarayana_RACES, 2012_Misailovic_RACES, 2013_Misailovic_ACMtecs, 2015_Campanoni_CGO, 2020_Stitt_ACMtecs, 2010_Sreeram_IISWC, 2010_Mengt_IPDPS, 2012_Rinard_RACES} \\
\hline
Precision Scaling & \cite{2011_Dinechin_IEEEtc, 2010_Linderman_CGO, 2017_Chiang_POPL, 2017_Darulova_ACMtoplas, 2013_Rubio_SC, 2018_Guo_ISSTA, 2013_Lam_ICS, 2018_Lam_SAGE, 2014_Chiang_PPoPP, 2016_Rubio_ICSE, 2018_Yesil_IEEEmicro, 2015_Tian_GLSVLSI, 2018_Menon_SC, 2020_Brunie_SC, 2019_Laguna_HPC, 2020_Kang_CGO} \\
\hline
Data Sampling & \cite{2012_Laptev_VLDB, 2015_Goiri_ASPLOS, 2019_Hu_MASCOTS, 2017_Quoc_MIDDL, 2016_Krishnan_WWW, 2017_Quoc_ATC, 2018_Wen_ICDCS, 2013_Agarwal_EuroSys, 2016_Kandula_MOD, 2016_Zhang_VLDB, 2016_Anderson_ICDE, 2019_Park_MOD} \\
\hline
Approximate Program. Languages &   \cite{2011_Ansel_CGO, 2007_Sorber_SenSys, 2010_Baek_PLDI, 2015_Boston_OOPSLA, 2012_Carbin_PLDI, 2013_Carbin_OOPSLA, 2014_Misailovic_OOPSLA, 2011_Liu_ASPLOS, 2015_Achour_OOPSLA, 2011_Sampson_PLDI, 2015_Park_FSE, 2014_Park_GIT, 2008_Goodman_UAI, 2014_Mansinghka_CoRR, 2016_Tolpin_IFL, 2014_Bornholt_ASPLOS, 2014_Sampson_PLDI, 2019_Fernando_ACMpapl, 2019_Joshi_ICSE} \\
\hline
\end{tabular}
\vspace*{-5pt}
\end{table}

\subsection{Selective Task Skipping}

\subsubsection{Loop Perforation}
The loop perforation technique 
skips some of the loop iterations in a software program 
to provide performance/energy gains in exchange for
QoS loss.
Subsequently, 
we present several relevant works \cite{2009_Hoffmann_MIT, 2010_Misailovic_ICSE, 2011_Sidiroglou_FCE, 2015_Shi_IEEEcal, 2017_Omar_ICCD, 2018_Li_ICS, 2020_Baharvand_IEEEtetc, 2015_Tan_ASP-DAC, 2018_Kanduri_DAC} involving design space exploration on loop perforation
with programming frameworks and profiling tools.  

Starting with one of the first state-of-the-art works, 
the SpeedPress compiler \cite{2009_Hoffmann_MIT}
supports a wide range of loop perforation types,
i.e., modulo, truncation, and randomized.
It takes as input the original source code,
a set of representative inputs,
as well as a programmer-defined QoS acceptability model,
and it outputs a loop perforated binary.
In the same context, 
Misailovic \emph{et al.} \cite{2010_Misailovic_ICSE}
propose a QoS profiler
to identify computations that can be approximated via
loop perforation. 
The proposed profiling tool
searches the space of loop perforation
and generates results
for multiple perforation configurations.
In \cite{2011_Sidiroglou_FCE}, 
the same authors propose a methodology
to exclude critical loops, 
i.e., whose skipping results in unacceptable QoS,
and they perform exhaustive and greedy design space explorations to 
find the Pareto-optimal perforation configurations
for a given QoS constraint. 

In \cite{2015_Shi_IEEEcal},
the authors propose an architecture 
that employs a profiler 
to identify non-critical loops towards their perforation. 
To protect code segments 
that can be affected by the perforated loops,
the architecture is equipped with
HaRE,
i.e., a hardware resilience mechanism.
Another interesting work is
GraphTune \cite{2017_Omar_ICCD},
which is an input-aware loop perforation scheme
for graph algorithms. 
This approach 
analyzes the input dependence of graphs
to build
a predictive model that finds
near-optimal perforation configurations
for a given accuracy constraint. 
Li \emph{et al.} \cite{2018_Li_ICS}
propose a compiling \& profiling system,
called Sculptor,
to improve the conventional loop perforation,
which skips a static subset of iterations. 
More specifically,
Sculptor dynamically skips a subset of the loop instructions (and not entire iterations)
that do not affect the output accuracy. 
More recently, 
the authors of \cite{2020_Baharvand_IEEEtetc} develop LEXACT,
which is a tool for identifying non-critical code segments
and monitoring the QoS of the program.
LEXACT searches the loop perforation space, 
trying to find perforation configurations that satisfy pre-defined metrics. 

The loop perforation technique has also been used 
in approximation frameworks for heterogeneous multi-core systems combining various approximation mechanisms.  
Tan \emph{et al.} \cite{2015_Tan_ASP-DAC}
propose a task scheduling algorithm,
which employs multiple approximate versions of the tasks
with loops perforated. 
Kanduri \emph{et al.} \cite{2018_Kanduri_DAC}
target applications in which the main computations are continuously repeated,
and they tune the loop perforation at runtime. 

\subsubsection{Computation Skipping}
This technique 
omits the execution of blocks of codes
with respect to 
the acceptable QoS loss,
programmer-defined constraints,
and/or runtime predictions regarding the output accuracy
\cite{2009_Meng_IPDPS, 2010_Byna_GPGPU, 2015_Raha_DATE, 2015_Vassiliadis_PPoPP, 2016_Vassiliadis_CGO, 2006_Rinard_ICS, 2007_Rinard_OOPSLA, 2017_Lin_ISCAS, 2018_Akhlaghi_ISCA}. 
Compared to loop perforation,
these techniques do not focus only on skipping loop iterations,
but also skip higher-level computations/tasks 
e.g., an entire convolution operation. 
Most of the state-of-the-art works perform application-specific computation skipping.

Meng \emph{et al.} \cite{2009_Meng_IPDPS}
introduce a parallel template
to develop 
approximate programs 
for iterative-convergence recognition \& mining algorithms.
The proposed programming template
provides several strategies (implemented as libraries)
for task dropping,
such as convergence-based computation pruning,
computation grouping in stages,
and early termination of iterations. 
Another interesting work involving application-specific computation skipping is presented in \cite{2010_Byna_GPGPU}. 
The authors of this work study 
the error tolerance of
the supervised semantic indexing algorithm
to make approximation decisions.
Regarding their task dropping approach,
they choose to omit 
the processing of common words (e.g., ``the'', ``and'') after the initial iterations, 
as these 
computations have negligible impact on accuracy. 

The authors of
\cite{2015_Raha_DATE}
propose two techniques
to find computations with low impact on the 
QoS of the
Reduce-and-Rank
computation pattern,
targeting to approximate or skip them completely. 
To identify these computations,  
the first technique 
uses intermediate reduction results and ranks,
while the second one 
is based on the spatial or temporal correlation of the input data
(e.g., adjacent image pixels or successive video frames).
Similar to the other state-of-the-art works,
Vassiliadis \emph{et al.} \cite{2015_Vassiliadis_PPoPP, 2016_Vassiliadis_CGO}
propose a programming environment 
that skips (or approximates) computations with respect to programmer-defined QoS constraints.
More specifically, 
the programmer 
expresses the significance of the tasks using pragmas directives,
optionally provides approximate variants of tasks,
and specifies the task percentage to be executed accurately. 
Based on these constraints,
the proposed system
makes decisions at runtime  
regarding the approximation/skipping
of the less significant tasks. 

Rinard \cite{2006_Rinard_ICS} builds probabilistic distortion models 
based on linear regression
to study the impact of computation skipping on accuracy. 
The programmer partitions the computations into tasks, 
which are then marked as 
``critical'' or ``skippable''
through random skip executions. 
The probabilistic models
estimate the output distortion as function of the skip rates of the skippable tasks.
This approach is also applied in parallel programs \cite{2007_Rinard_OOPSLA}, 
where probabilistic distortion models are employed 
to tune the early phase termination at barrier synchronization points,
targeting to keep all the parallel cores busy.

Significant research has also been conducted 
on skipping the computations of 
Convolutional Neural Networks (CNNs).
Lin \emph{et al.} \cite{2017_Lin_ISCAS}
introduce PredictiveNet
to predict the sparse outputs of the nonlinear layers 
and 
skip a large subset of convolutions at runtime. 
The proposed technique,
which does not require 
any modification in the original CNN structure,
examines the most-significant part of the convolution
to predict if the nonlinear layer output is zero,
and then it decides whether
to skip 
the remaining least-significant part computations
or not. 
In the same context, 
Akhlaghi \emph{et al.} \cite{2018_Akhlaghi_ISCA}
propose SnaPEA,
exploiting the convolution--activation algorithmic chain in CNNs (activation inputs the convolution result and outputs zero if it is negative).
This technique
early predicts 
negative convolution results,
based on static re-ordering of the weights 
and monitoring of the partial sums' sign bit,
in order to skip the rest computations.

\subsubsection{Memory Access Skipping}
Another approach 
to improve the execution time and energy consumption
at software level
is the memory access skipping.
Such techniques \cite{2014_Miguel_MICRO, 2016_Yazdanbakhsh_ACMtaco, 2018_Kislal_ELSEclss, 2013_Samadi_MICRO, 2019_Karakoy_ACMmacs, 2015_Zhang_DATE} 
aim to avoid high-latency memory operations,
while as a result, 
they also reduce the number of computations. 

Miguel \emph{et al.} \cite{2014_Miguel_MICRO} exploit the approximate data locality 
to skip the required memory accesses due to L1 cache miss.
In particular,
they employ a load value approximator,
which learns value patterns 
using a global history buffer and an approximator table, 
to estimate the memory data values.
RFVP \cite{2016_Yazdanbakhsh_ACMtaco}
uses value prediction instead of memory accessing.
When selected load operations miss in the cache memory,
RFVP predicts the requested vales
without checking for misprediction
or recovering the values.
As a result,
timing overheads from pipeline flushes
and re-executions
are avoided.
Furthermore,
a tunable rate of cache misses
is dropped
after the value prediction
to eliminate 
long memory stalls. 
The authors of \cite{2018_Kislal_ELSEclss}
propose a framework 
that skips costly last-level cache misses 
according to a programmer-defined error constraint
and an heuristic predicting skipped data. 

To improve the performance of CUDA kernels on GPUs, 
Samadi \emph{et al.} \cite{2013_Samadi_MICRO}
propose a runtime approximation framework,
called SAGE,
which focuses on optimizing the memory operations among other functionalities.
The approximations lie in
skipping selective atomic operations
(used by kernels to write shared variables) 
to avoid conflicts leading to performance decrease.
Furthermore,
SAGE reduces the number of memory accesses
by packing the read-only input arrays, 
and thus, allowing to access more data with fewer requests. 
Karakoy \emph{et al.} \cite{2019_Karakoy_ACMmacs}
propose a slicing-based approach
to identify data (memory) accesses 
that can be skipped
to deliver energy/performance gains
within an acceptable error bound. 
The proposed method applies
backward and forward code slicing
to estimate the gains 
from skipping each output data.
Moreover,
the `$0$' value 
is used for each data access that is not performed. 
The ApproxANN framework \cite{2015_Zhang_DATE},
apart from performing approximate computations,
skips memory accesses 
on neural networks
according to the neuron criticality. 
More specifically, 
a theoretical analysis is adopted
to study the impact of neurons on the output accuracy
and 
characterize their criticality. 
The neuron approximation 
under a given QoS constraint
is tuned
by an iterative algorithm,
which applies the approximations
and
updates the criticality of each neuron (it may change due to approximations in other neurons).

\subsection{Approximate Memoization}
The memoization technique stores
results of previous calculations
or pre-computed values
in memory
to use them instead of performing calculations.
Namely,
this memory functions as a look-up table,
which maps
a set of data identifiers to a set of stored data.
Subsequently,
we focus on approximate memoization techniques \cite{2011_Chaudhuri_FSE, 2014_Samadi_ASPLOS, 2014_Mishra_WACAS, 2017_Brumar_IPDPS, 2015_Keramidas_WAPCO, 2018_Tziantzioulis_IEEEmicro} relying on software frameworks, compilers and programmer's decisions.
Nevertheless,
we note that there are also approaches \cite{2005_Alvarez_IEEEtc, 2013_Rahimi_IEEEtcasii, 2018_Zhang_IEEEcal, 2019_Liu_ISCA}
requiring hardware modification to support memoization.

Chadhuri \emph{et al.} \cite{2011_Chaudhuri_FSE} propose an approximate memoization for computations in loops.
Prior executing an expensive function within a loop,
this technique checks a look-up table
to find
if this computation was previously performed for similar input data.
In this case, 
the cached result is used,
otherwise,
the function is executed 
and the new computation is stored in the look-up table. 
Paraprox \cite{2014_Samadi_ASPLOS} is a software framework
for identifying common patterns in data-parallel programs
and applying tailored approximations.
For the Map \& Scatter/Gather patterns,
Paraprox uses memoization rather than performing computations. 
In particular,
it fills a look-up table with pre-computed data,
which are obtained from 
the execution of the Map \& Scatter/Gather function 
for some representative inputs,
and it performs runtime look-up table queries
instead of the conventional computations. 

iACT \cite{2014_Mishra_WACAS}
is another approximation framework
that applies runtime memoization among other functionalities.
The programmer uses pragmas
to declare the functions for memoization
and specify the error tolerance percentage.
For each function call-site,
the framework creates a global table
to store pairs of function arguments and output results.
In case the function arguments 
are already stored in the table (within an error bound),
the corresponding output results are returned.
Otherwise,
the function is accurately executed
and the new input--output pairs are stored in the table.
The ATM approach \cite{2017_Brumar_IPDPS}
performs runtime task memoization,
relying on hashing functions to store the task inputs
and an adaptive algorithm to automatically decide whether to use memoization or execute the task.
The programmer needs to 
use pragmas to specify the tasks that are suitable for memoization. 
The authors of 
\cite{2015_Keramidas_WAPCO} 
introduce an approximate memoization
mechanism for GPU fragment shading operations,
which reduces the precision of the input parameters
and performs partial matches.
To identify approximate memoization opportunities,
they characterize various fragment shader instructions
in terms of memoization hits and output accuracy. 
Moreover,
runtime policies are proposed
to tune the precision according to the errors introduced.

Contrary to the aforementioned techniques,
TAF-Memo \cite{2018_Tziantzioulis_IEEEmicro}
is an output-based function memoization technique,
i.e., it memoizes function calls based on their output history.
TAF-Memo checks for temporal locality 
by calculating the relative arithmetic difference of two consecutive
output values from the same function call-site.
In case this difference is below the acceptable error constraint,
memoization is applied by returning the last computed output for the following function calls. 

\subsection{Relaxed Synchronization}

The execution of parallel applications
on multi/many-core systems  
requires time-consuming synchronization
to either access shared data
or satisfy data dependencies. 
The literature includes 
various techniques \cite{2012_Renganarayana_RACES, 2012_Misailovic_RACES, 2013_Misailovic_ACMtecs, 2015_Campanoni_CGO, 2020_Stitt_ACMtecs, 2010_Sreeram_IISWC, 2010_Mengt_IPDPS, 2012_Rinard_RACES}
that relax the conventional synchronization requirements guaranteeing error-free execution,
to improve the performance. 

The authors of \cite{2012_Renganarayana_RACES} propose the four-step RaC methodology
to systematically relax synchronization,
while always satisfying a programmer-defined QoS constraint. 
Initially,
the programmer specifies the parallel code segments,
and then applies the RaC methodology.
This methodology 
identifies criteria for quantifying the acceptable QoS,
selects the relaxation points,
modifies the code to enable the execution of both the original and relaxed versions,
and selects the suitable relaxation degree (i.e., which instances to relax for each synchronization point). 
Misailovic \emph{et al.} \cite{2012_Misailovic_RACES}
propose the Dubstep system,
which relaxes the synchronization of parallelized programs
based on a ``find-transform-navigate'' approach.
Dubstep performs a profiling-based analysis of the original program 
to find possible optimizations,
inserts opportunistic synchronization and barriers,
and finally, performs an exploration 
including   
accuracy, performance and safety analysis.  

QuickStep \cite{2013_Misailovic_ACMtecs} is a system for approximately parallelizing sequential programs 
without preserving their semantics 
within statistical accuracy bounds. 
Among other transformations,
QuickStep replicates shared objects to eliminate the bottlenecks of synchronized operations on them. 
HELIX-UP \cite{2015_Campanoni_CGO} is another parallelizing compiler
that selectively relaxes strict adherence to program semantics to tackle runtime performance bottlenecks,
involving profiling and user interaction to tune QoS.
The compiler also offers a synchronization-relaxing knob to decrease the inter-core communication overhead
by synchronizing sequential segments with prior iterations. 
More recently, 
the authors of \cite{2020_Stitt_ACMtecs} introduce PANDORA,
which is an approximate parallelizing framework based on symbolic regression machine learning and sampled outputs of the original function.
To avoid timing bottlenecks, such as data movement and synchronization,
and improve parallelism,
PANDORA eliminates loop-carried dependencies using fitness functions and constraints regarding error and performance. 
In \cite{2010_Sreeram_IISWC},
the authors exploit the concept of 
approximate shared value locality
to reduce synchronization conflicts
in programs using optimistic synchronization.
The reduction of conflicts on approximately local variables,
detected for a given similarity constraint,
is achieved through
an arbitration mechanism that
imprecisely shares the values between threads.
The authors of \cite{2010_Mengt_IPDPS} apply aggressive coarse-grained parallelism on recognition \& mining algorithms
by relaxing or even ignoring 
data dependencies between different iterations.
As a result, the timing overheads are reduced in comparison with the conventional parallel implementation,
which also applies parallelization only within the iteration (iterations are executed serially).
Rinard \cite{2012_Rinard_RACES} introduces synchronization-free updates to shared data structures
by eliminating the conventional use of mutual exclusion 
and dropping array elements at the worst scenario.
Moreover,
the same work 
applies relaxed barrier synchronization,
allowing the threads to pass the barrier without stalling to wait for the other threads. 

\subsection{Precision Scaling}
Precision scaling (tuning) refers to the discipline reduction of the nominal numerical precision,
resulting in improved
calculation speed
and/or 
memory bandwidth \cite{2020_Cherubin_ACMsrv}.
The state-of-the-art software-level works
\cite{2011_Dinechin_IEEEtc, 2010_Linderman_CGO, 2017_Chiang_POPL, 2017_Darulova_ACMtoplas, 2013_Rubio_SC, 2018_Guo_ISSTA, 2013_Lam_ICS, 2018_Lam_SAGE, 2014_Chiang_PPoPP, 2016_Rubio_ICSE, 2018_Yesil_IEEEmicro, 2015_Tian_GLSVLSI, 2018_Menon_SC, 2020_Brunie_SC, 2019_Laguna_HPC, 2020_Kang_CGO}
address several challenges,
such as 
the scaling degree,
scaling automation,
mixed precision,
and dynamic scaling.

Starting with works based on formal methods to
reduce the precision and examine the errors,
the Gappa tool \cite{2011_Dinechin_IEEEtc}
automates the
study of rounding errors in elementary functions and floating-point calculations using interval arithmetic. 
An extended version of this tool is
Gappa++ \cite{2010_Linderman_CGO},
which provides automated analysis of numerical errors in a wide range of computations,
i.e., 
fixed-point, floating-point, linear and non-linear.
This tool integrates several features,
such as
operation rewriting 
to facilitate the isolation of rounding errors,
and affine arithmetic 
to accurately bound linear calculations with correlated errors.
FPTuner \cite{2017_Chiang_POPL} is a tool that 
performs formal error analysis
based on symbolic Taylor expansions
and quadratically constrained quadratic programming.
It searches for precision allocations
that satisfy constraints such as
the number of
operators at a given precision
and the number of type casts.
Rosa \cite{2017_Darulova_ACMtoplas}
is a source-to-source compiler 
that combines satisfiability modulo theories with interval arithmetic
to bound the round-off errors of the fixed- and floating-point formats.

Several works employ heuristics and automated search to scale the precision of floating-point programs.
Precimonious \cite{2013_Rubio_SC} 
searches all the program variables in their order of declaration
using the delta-debugging algorithm,
and it lowers their precision according to an error constraint specified by the programmer.
In the same context, 
HiFPTuner \cite{2018_Guo_ISSTA}
firstly groups dependent variables 
that may require the same precision,
and then it performs a customized hierarchical search. 
Lam \emph{et al.} \cite{2013_Lam_ICS}
introduce a framework
that employs
the breadth-first search algorithm
to identify code regions that
can tolerate lower precision.
Similar to this technique,
CRAFT \cite{2018_Lam_SAGE} performs binary searches 
to initially determine the
required program precision,
and then truncate the results of some of the floating-point instructions.
Towards the detection of large floating-point errors,
the authors of \cite{2014_Chiang_PPoPP} propose S$^3$FP.
This tool is  
based on an heuristic-guided search
to find the inputs causing the largest errors.
The Blame Analysis \cite{2016_Rubio_ICSE}  
combines concrete and shadow execution
to generate a blame set for the program instructions,
which contains the minimum precision requirements under a giver error constraint.
This approach can be also used in cooperation with the previous search-based works,
and specifically,
as pre-processing
to reduce the search space. 
Schkufza \emph{et al.} \cite{2014_Schkufza_PLDI}
treat the scaling of floating-point precision
as a stochastic search problem.
In particular,
they repeatedly 
apply random program transformations
and use a robust search to guarantee the maximum errors.

The concept of dynamic precision scaling,
i.e., the precision tuning at runtime with respect to the input data and error tolerance, 
has been studied in \cite{2018_Yesil_IEEEmicro}.
The dynamic scaling framework of this work integrates
an offline application profiler, 
a runtime monitor to track workload changes,
and an accuracy controller 
to adjust the precision accordingly.
ApproxMA \cite{2015_Tian_GLSVLSI} dynamically scales the precision of
data memory accesses in algorithms
such as mixture model-based clustering.
This framework integrates a runtime precision controller,
which generates custom bit-widths according to the QoS constraints,
and a memory access controller,
which loads the scaled data from memory.
The custom bit-widths are generated
by analyzing a subset of data and intermediate results
and calculating metrics
regarding the error appearance and the number of tolerable errors.

Mixed floating-point precision has also been studied in high-performance computing workloads.
ADAPT \cite{2018_Menon_SC} uses algorithmic differentiation,
i.e., 
a technique for numerically evaluating the derivative of a function corresponding to the program, 
to estimate the output error of high-performance computing workloads. 
It provides a precision sensitivity 
profile to guide the development of mixed-precision programs.  
The authors of \cite{2020_Brunie_SC} 
propose an instruction-based search
that explores information 
about the dynamic program behaviour and the temporal locality.

To enable mixed floating-point precision in GPUs,
the authors of \cite{2019_Laguna_HPC} propose
the GPUMixer tool,
which relies on static analysis to
find code regions where precision scaling improves the performance.
Next,
GPUmixer performs a dynamic analysis
involving shadow computations
to examine if the scaled program configurations satisfy the accuracy constraints.
In the same context, 
PreScaler \cite{2020_Kang_CGO} is an automatic framework
that generates precision-scaled OpenCL programs,
considering both kernel execution and data transfer.
Initially, it employs
a system inspector
to collect
information about precision scaling on the target platform,
and an application profiler 
to identify
memory objects with floating-point elements for potential scaling.
This information is exploited by 
a decision maker,
which finds the best scaling configuration using decision tree search on a minimized space.

\subsection{Data Sampling}
Approximate Computing is also exploited in big data analysis,
in an effort to reduce the increased number of computations
due to the large amount of data. 
The key idea is to perform computations on a representative data sample
instead of the entire dataset.
In this context, 
data sampling methods \cite{2012_Laptev_VLDB, 2015_Goiri_ASPLOS, 2019_Hu_MASCOTS, 2017_Quoc_MIDDL, 2016_Krishnan_WWW, 2017_Quoc_ATC, 2018_Wen_ICDCS, 2013_Agarwal_EuroSys, 2016_Kandula_MOD, 2016_Zhang_VLDB, 2016_Anderson_ICDE, 2019_Park_MOD} 
provide real-time processing with error bounds
in applications involving stream analytics, database search, and model training.  

EARL \cite{2012_Laptev_VLDB}
is an extension of Hadoop
(i.e., 
a software framework that provides distributed storage and big data processing on clusters), 
which delivers early results with reliable error bounds.
It applies statistics-based uniform sampling 
from distributed files.
Goiri \emph{et al.} \cite{2015_Goiri_ASPLOS}
propose the ApproxHadoop framework
to generate approximate MapReduce programs
based on 
task dropping and multi-stage input sampling. 
They also bound the errors
using statistical theories. 
The programmer tunes the approximation
by specifying
either the ratio of task dropping and/or input sampling
or the desired error bound. 
Similarly, 
ApproxSpark \cite{2019_Hu_MASCOTS}
performs sampling at multiple arbitrary points 
of long chains of transformations
to facilitate the aggregation of huge amounts of data.
This framework models 
the clustering information of transformations
as a data provenance tree,
and then it computes the approximate aggregate values as well as error thresholds.
Moreover, 
the sampling rates are dynamically selected
according to programmer-specified error thresholds.

Sampling methods have also been examined
in stream analytics.
StreamApprox \cite{2017_Quoc_MIDDL}
is an approximate stream analytics system
that supports batched and pipelined stream processing.
It employs two sampling techniques 
(stratified and reservoir sampling) 
to approximate the outputs with rigorous error bounds.
IncApprox \cite{2016_Krishnan_WWW} 
combines approximate and incremental computations to 
provide stream analytics with bounded error.
This system executes a stratified sampling algorithm
that selects data for which the results have been memoized from previous runs,
and it adjusts the computations
to produce an incrementally updated output.
On the other hand,
PrivApprox \cite{2017_Quoc_ATC} 
combines sampling and randomized response 
to provide approximate computations and privacy guarantee.
This system integrates a query execution interface
that systematically explores the trade-off
between accuracy and query budget.
ApproxIoT \cite{2018_Wen_ICDCS}
facilitates approximate stream analytics
at the edge
by combining stratified reservoir sampling
and hierarchical processing.

A variety of sampling methods have been employed
in approximate query processing systems for databases.
BlinkDB \cite{2013_Agarwal_EuroSys}
performs approximate distributed query processing,
supporting SQL-based aggregation queries 
with time and error constraints.
It creates stratified samples based on past queries 
and uses an heuristic-based profiler to dynamically
select the sample that meets the query's constraints. 
Another system applying approximate big-data queries 
is Quickr \cite{2016_Kandula_MOD},
which integrates operators 
sampling multiple join inputs
into a query optimizer,
and then it searches for an appropriate 
sampled query plan.
Sapprox \cite{2016_Zhang_VLDB} is a distribution-aware system
that employs the occurrences of sub-datasets
to drive the online sampling.
In particular, 
the exponential number of sub-datasets is reduced 
to a linear one using a probabilistic map, 
and then,
cluster sampling with unequal probability theory 
is applied for sub-dataset sampling. 
Sapprox also determines the optimal sampling unit size 
in relation with approximation costs and accuracy.

Numerous works 
use data sampling to decrease the increased computational cost of model training
for machine learning applications.
Zombie \cite{2016_Anderson_ICDE}
is a two-stage system
that trains approximate models based on clustering and active learning.
The first stage applies offline indexing
to organize the dataset into index groups of similar elements.
Afterwards,
the stage of online querying 
uses the index groups that are likely to output useful features
to create the training subset of data. 
BlinkML \cite{2019_Park_MOD} approximately trains a model on a small sample,
while providing accuracy guarantees.
The sample is obtained through uniform random sampling,
however, in case of very large datasets, 
a memory-efficient algorithm is employed.


\subsection{Approximate Programming Languages}
The high-level approximation of software programs has been examined
through approximate programming languages,
i.e., language extensions 
that allow the programmer to 
systematically declare approximate code regions, variables, loops, and functions,
insert randomness in the program,  
and/or specify error constraints. 
The literature involves numerous works \cite{2011_Ansel_CGO, 2007_Sorber_SenSys, 2010_Baek_PLDI, 2015_Boston_OOPSLA, 2012_Carbin_PLDI, 2013_Carbin_OOPSLA, 2014_Misailovic_OOPSLA, 2011_Liu_ASPLOS, 2015_Achour_OOPSLA, 2011_Sampson_PLDI, 2015_Park_FSE, 2014_Park_GIT, 2008_Goodman_UAI, 2014_Mansinghka_CoRR, 2016_Tolpin_IFL, 2014_Bornholt_ASPLOS, 2014_Sampson_PLDI, 2019_Fernando_ACMpapl, 2019_Joshi_ICSE}
that enable approximate procedural, object-oriented, and probabilistic programming. 

Ansel \emph{et al.}
\cite{2011_Ansel_CGO}
introduce a set of PetaBricks language extensions that allow the programmer to write code of variable accuracy.
These extensions expose
the performance--accuracy trade-off 
to the compiler,
which automatically searches the algorithmic space 
to tune the program according to the programmer's accuracy constraints.
Eon \cite{2007_Sorber_SenSys} is a programming language 
that allows the programmer to annotate program flows (paths) with different energy states.
The Eon runtime system predicts the workload and energy of the system,
and then it adjusts the execution of flows according to the programmer's declarations and the energy constraints. 
In the same context, 
Baek and Chilimbi \cite{2010_Baek_PLDI} propose Green,
which is a two-phase programming framework
providing language extensions 
to approximate expensive functions and loops.
The programmer uses pragma-like annotations to 
specify approximate variants of functions.
In the calibration phase,
Green builds a model to quantify the QoS loss and the performance/energy gains.   
This model is then used in the operational phase
to generate an approximate program satisfying the programmer's QoS constraint. 
DECAF \cite{2015_Boston_OOPSLA} is a type-based approximate programming language
that allows the programmer
to specify the correctness probability 
for some of the program variables. 
The DECAF type system also integrates
solver-aided type inference
to automatically tune the type of the rest variables,
code specialization,
and dynamic typing.
Flikker \cite{2011_Liu_ASPLOS} provides language annotations 
to mark the program variables
and partition the data into critical and non-critical regions 
(stored in unreliable memories).
Topaz \cite{2015_Achour_OOPSLA}
is a task-based language
that maps tasks onto approximate hardware
and uses an outlier detector 
to find and re-execute the computations
producing unacceptable results.

In \cite{2012_Carbin_PLDI},
the authors introduce
language constructs for generating approximate programs
and proof rules for verifying the acceptability properties.
Rely \cite{2013_Carbin_OOPSLA} is an imperative language that allows the programmer 
to introduce quantitative reliability specifications
for generating programs with data stored in approximate memory and inexact arithmetic/logical operations.
Chisel \cite{2014_Misailovic_OOPSLA}
automates the selection of Rely's approximations
while satisfying the programmer-defined reliability and accuracy constraints.
To solve this optimization problem, 
Chisel employs an integer programming solver. 
All these works include safety analysis and program verification for sequential programs.
In contrast, 
Parallely \cite{2019_Fernando_ACMpapl} is a  programming language for approximating parallel programs
through canonical sequentialization, 
i.e., 
a verification method that generates sequential programs capturing the semantics of parallel programs.

Targeting approximations in Java programs,
the authors of \cite{2011_Sampson_PLDI}
propose EnerJ, i.e., a language extension providing type qualifiers to specify data that can be approximately stored or computed.
EnerJ guarantees isolation of the approximate computations.
FlexJava \cite{2015_Park_FSE} offers another
set of language extensions 
to annotate approximate programs.
Using an approximation safety analysis,
FlexJava
automates the approximation of data and operations 
while ensuring safety guarantees. 
ExpAX \cite{2014_Park_GIT}
allows the programmer to explicitly specify error expectations for a subset of Java.
Based on an approximation safety analysis, 
it identifies operations that are candidate for approximation,
and then,
an heuristic-based framework approximates those that statistically satisfy the error expectations.

Significant research has also been conducted on probabilistic programming languages.
Church \cite{2008_Goodman_UAI} is a probabilistic language that inserts randomness on a deterministic function subset using stochastic functions.
The Church semantics are defined in terms of 
evaluation histories and conditional distributions on the latter.
Similarly, 
Venture \cite{2014_Mansinghka_CoRR} is another language
that enables the specification of probabilistic models and inference problems.
The Anglican \cite{2016_Tolpin_IFL} language and runtime system
provides probabilistic evaluation model and functional representations, 
e.g., distributions and sequences of random variables. 

Uncertain\raisebox{0.8pt}{$\scriptstyle <$}T\raisebox{0.8pt}{$\scriptstyle >$} \cite{2014_Bornholt_ASPLOS}
is a language abstraction
that manipulates data as probability distributions.
Random variables are declared as ``uncertain''
and a Bayesian network for representing computations is build,
where nodes correspond to the variables
and edges correspond to conditional variable dependencies.
The Uncertain\raisebox{0.8pt}{$\scriptstyle <$}T\raisebox{0.8pt}{$\scriptstyle >$} runtime system 
performs hypothesis tests and sampling to evaluate the network.
Similarly,  
Sampson \emph{et al.} \cite{2014_Sampson_PLDI}
use probabilistic assertions on random variables.
Their tool,
called MayHap,
performs probabilistic evaluation by
statically building a Bayesian representation network based on the input distribution
and
dynamically interpreting it via sampling.
In the same context,
AxProf \cite{2019_Joshi_ICSE}
is a profiling-based framework for 
analyzing randomized approximate programs.
The programmer specifies probabilistic predicates for the output,
i.e., 
regarding 
the expectation of the output value
and/or
the probability that the output satisfies a condition,
and AxProf generates approximate programs based on statistical tests. 

\vspace*{-5pt}

\section{Classification of Hardware Approximation  Techniques}
\label{clhw}

\begin{figure}[!b]
    \centering
    \includegraphics[width=1\textwidth]{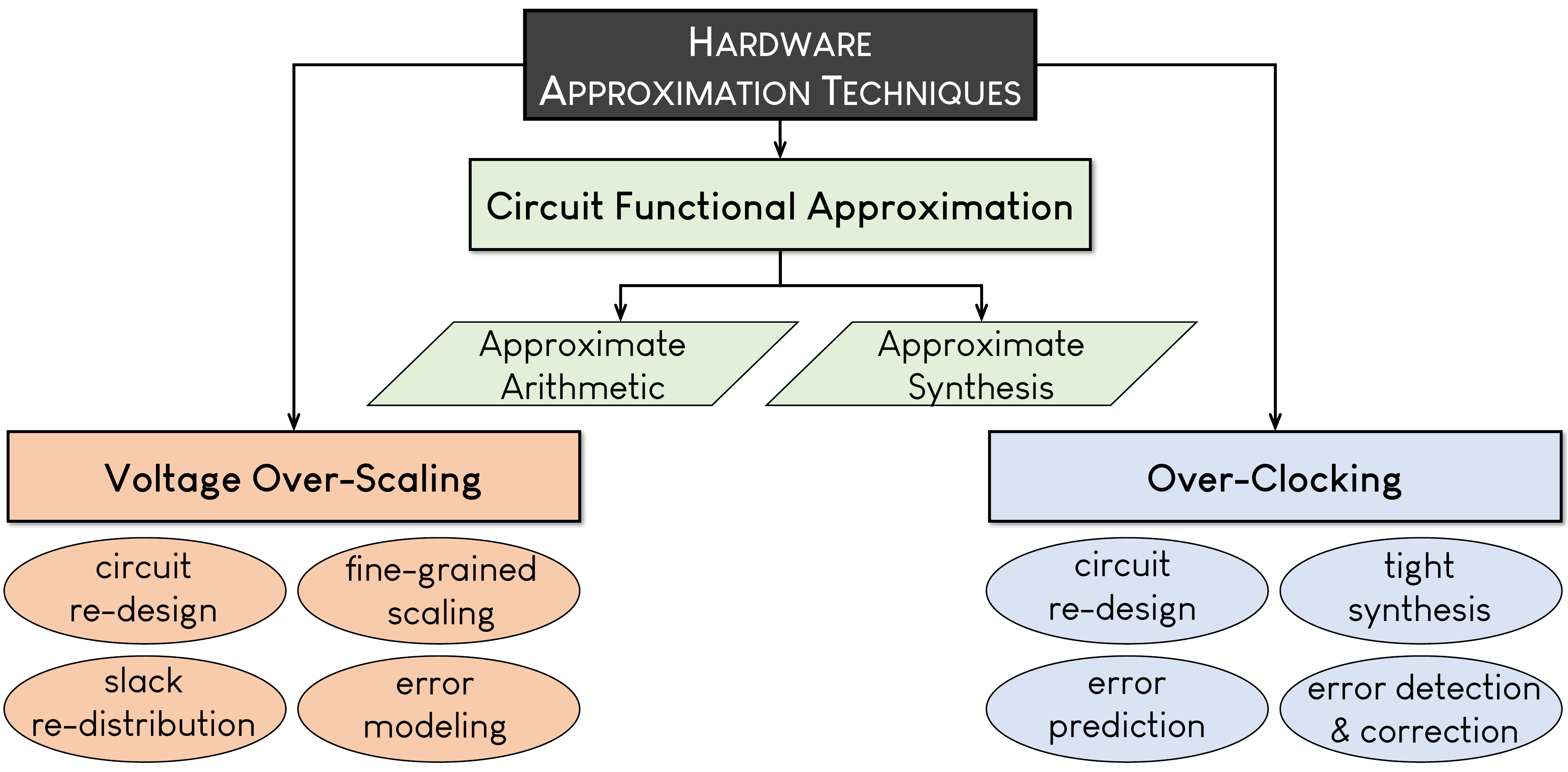}%
    \caption[Classification of State-of-the-Art Hardware Approximation Techniques]{Classification of state-of-the-art hardware approximation techniques.}
    \label{fig_hw}
    \vspace*{-5pt}
\end{figure}

In this section,
we classify and introduce the hardware approximation techniques,
which are applied at the lower level of the design abstraction hierarchy.
These techniques aim to 
improve the area, power consumption, and performance 
of the circuits
i.e., the basic building blocks of accelerators, processors, and computing platforms.
The hardware approximation techniques can be categorized into three classes:
(i) \emph{Circuit Functional Approximation (CFA)},
(ii) \emph{Voltage Over-Scaling (VOS)},
and (iii) \emph{Over-Clocking (OC)}.
In approximate hardware,
we distinguish two types of errors: 
the \emph{functional errors} (produced by CFA)
and the \emph{timing errors} (produced by VOS and OC).
Figure \ref{fig_hw} illustrates the
hardware approximation techniques,
including a further taxonomy
to sub-classes.

In the remainder of this section,
we present state-of-the-art 
works,
organized according to the proposed classification.
Table \ref{tb_hw} summarizes the 
state-of-the-art hardware approximation techniques.
We also note that 
some works belong to more than one sub-class,
however, 
we opt to assign them to the sub-class of their prevalent technique  
and highlight the relevant features.

\begin{table}[!t]
\fontsize{9}{10}\selectfont
\renewcommand{\arraystretch}{1.2}
\setlength{\tabcolsep}{4.5pt}
\caption[Classification of Hardware Approximation Techniques]{Classification of hardware approximation techniques.}
\label{tb_hw}
\centering
\begin{tabular}{l l}
\hline  
\makecell[l]{\textbf{HW Approximation Class}} & \makecell[c]{\textbf{Technique/Approach}}\\
\hline
\hline
\multirow{2}{*}{Adder Approximation} & use of approximate full adder cells \cite{2013_Gupta_IEEEtcad, 2013_Yang_NANO, 2018_Pashaeifar_IEEEtvlsi, 2018_Dalloo_IEEEtvlsi} \\
& segmentation and carry prediction \cite{2013_Kim_ICCAD, 2015_Hu_DATE, 2012_Kahng_DAC, 2013_Ye_ICCAD, 2015_Shafique_DAC, 2018_Akbari_IEEEtcasii, 2018_Xu_IEEEtvlsi, 2020_Ebrahimi_IEEEtcasii} \\
\hline
\multirow{4}{*}{Multiplier Approximation} & truncation and rounding \cite{2015_Hashemi_ICCAD, 2017_Zendegani_IEEEtvlsi, LeonMicro, LeonTECS, 2019_Vahdat_IEEEtvlsi, LeonDAC, 2020_Frustaci_IEEEtcasii} \\
& approximate radix encodings \cite{2017_Liu_IEEEtc, 2019_Venkatachalam_IEEEtc, 2016_Jiang_IEEEtc, 2020_Waris_IEEEtcasii, LeonTVLSI} \\
& use of approximate compressors     \cite{2015_Momeni_IEEEtc, 2017_Akbari_IEEEtvlsi, 2019_Sabetzadeh_IEEEtcasi, 2018_Esposito_IEEEtcasi,  2020_Strollo_IEEEtcasi} \\
& logarithmic approximation \cite{2018_Liu_IEEEtcasi, 2020_Saadat_DATE, 2021_Ansari_IEEEtc, 2021_Pilipovic_IEEEtcasi} \\
\hline
\multirow{3}{*}{Divider Approximation} & bit-width scaling \cite{2016_Hashemi_DAC, 2019_Jiang_IEEEtc} \\
& use of approximate adder/subtractor cells \cite{2015_Chen_GLSVLSI, 2016_Chen_IEEEtc, 2018_Chen_IEEEtmscs, 2020_Adams_IEEEtc} \\
& simplification of computations \cite{2016_Zendegani_DATE, 2017_Vahdat_DATE, 2019_Imani_Date,  2018_Liu_ARITH, 2019_Saadat_DAC} \\
\hline
\multirow{4}{*}{Approximate Synthesis} & structural netlist transformation
\cite{2017_Schlachter_IEEEtvlsi, 2018_Scarabottolo_DATE, 2013_Venkataramani_DATE, 2017_Liu_ICCAD, 2021_Castro_IEEEtcasii} \\
& Boolean rewriting
\cite{2012_Venkataramani_DAC, 2014_Ranjan_DATE, 2013_Miao_ICCAD, 2018_Hashemi_DAC} \\
& high-level approximate description \cite{2015_Yazdanbakhsh_DATE, 2014_Nepal_DATE, 2017_Lee_DATE, 2019_Nepal_IEEEtetc, 2020_Castro_ICCAD} \\
& evolutionary synthesis \cite{2013_Sekanina_ICES, 2015_Vasicek_IEEEtec, 2017_Mrazek_DATE, 2016_Vasicek_FPL, 2019_Vasicek_DATE} \\
\hline
\multirow{4}{*}{Voltage Over-Scaling} & slack re-distribution \cite{2010_Kahng_ASP-DAC} \\
& circuit re-design and architecture modification \cite{2011_Mohapatra_DATE, 2013_Chen_IEEEtvlsi, 2018_Zhang_DAC} \\
& fine-grained scaling \cite{2019_Pandey_DAC, 2020_Wang_IEEEtc, 2019_Zervakis_IEEEtcasii} \\
& error modeling \cite{2010_Liu_IEEEtvlsi, 2012_Jeon_IEEEtcasii, 2017_Ragavan_DATE, 2020_Jiao_IEEEtcad,  2018_Zervakis_IEEEtvlsi} \\
\hline
\multirow{4}{*}{Over-Clocking} & tight synthesis \cite{2018_Alan_IEEEtcad} \\
& circuit re-design and architecture modification \cite{2013_Ramasubramanian_DAC, 2014_Shi_DAC, 2017_Wang_IEEEtvlsi} \\
& error detection \& correction \cite{2014_Choudhury_IEEEtc, 2016_Ragavan_ISVLSI, 2017_Li_IEEEtvlsi} \\
& error prediction \cite{2012_Roy_DAC, 2015_Constantin_DATE, 2016_Jiao_ICCD, 2017_Jiao_DATE} \\
\hline
\end{tabular}
\end{table}

\subsection{Circuit Functional Approximation}
Circuit functional approximation
modifies the original accurate design
by reducing its circuit complexity at logic level.
Typical CFA approaches include:
(i) the modification of the circuit's truth table,
(ii) the use of an approximate version of the initial hardware algorithm,
(iii) the use of small inexact components as building blocks,
and
(iv) approximate circuit synthesis.
The main target of CFA is the arithmetic circuits \cite{2020_Jiang_IEEE},
as they constitute the key processing units of processors and accelerators, 
and thus,
they inherently affect the power efficiency and performance of the entire system.
Interestingly,
the literature provides several open-source libraries of approximate arithmetic circuits,
such as ApproxAdderLib \cite{2015_Shafique_DAC}, EvoApprox8b \cite{2017_Mrazek_DATE} and SMApproxLib \cite{2018_Ullah_DAC}.
In this literature review,
we focus on approximate adders, multipliers, and dividers.
However, 
we note that numerous works design and evaluate other approximate arithmetic operations, 
such as
circuits for 
multiply-accumulate \cite{2019_Chen_IEEEtc, 2019_Gillani_IEEEa}, 
square root \cite{2019_Jiang_IEEEtc},
squaring \cite{2020_Manikanta_IEEEtvlsi},
square-accumulate \cite{2018_Gillani_IEEEa},
and Coordinate Rotation Digital Computer (CORDIC) \cite{2017_Chen_IEEEtscl}.
Moreover,
the literature includes
automated methods
for generating approximate circuits,
which are presented in the context of approximate logic synthesis. 

\subsubsection{Approximate Adders}
Significant research has been conducted on the design of approximate area- and power-efficient adders.
The approximation techniques for inexact adders involve:
(i) \emph{use of approximate full adder cells} \cite{2013_Gupta_IEEEtcad, 2013_Yang_NANO, 2018_Pashaeifar_IEEEtvlsi, 2018_Dalloo_IEEEtvlsi}
and
(ii) \emph{segmentation and carry prediction} \cite{2013_Kim_ICCAD, 2015_Hu_DATE, 2012_Kahng_DAC, 2013_Ye_ICCAD, 2015_Shafique_DAC, 2018_Akbari_IEEEtcasii, 2018_Xu_IEEEtvlsi, 2020_Ebrahimi_IEEEtcasii}.
In the following,
we present representative state-of-the-art works with approximate adders.

The IMPACT adders
are based on inexact full adder cells,
which are approximated
at the transistor level to deliver up to 45\%  
area reduction \cite{2013_Gupta_IEEEtcad}.
Another transistor-level cell approximation is proposed in \cite{2013_Yang_NANO},
where 
the AXA 4-transistor XOR/XNOR-based adders are implemented,
delivering up to $31\%$ gain in dynamic power consumption.
Moreover, 
in \cite{2018_Pashaeifar_IEEEtvlsi}, 
approximate reverse carry-propagate full adders 
are used to build the hybrid RCPA adders.
Targeting higher level approximations, 
the OLOCA adder
splits the addition into accurate and approximate segments \cite{2018_Dalloo_IEEEtvlsi},
and for the latter,
it employs OR gates for the most-significant bit additions
and outputs constant `$1$' for the least-significant ones.
To reduce the worst-case carry propagation delay,
Kim \emph{et al.} \cite{2013_Kim_ICCAD}
propose a carry prediction scheme
leveraging the less-significant input bits,
which
is $2.4\times$ faster than the conventional ripple-carry adder.
Similarly,
Hu \emph{et al.} \cite{2015_Hu_DATE}
introduce 
a carry speculating method
to segment the carry chain
in their design,
which also performs error and sign correction.
Compared to the accurate adder,
the proposed design
is $4.3\times$ faster 
and saves $47\%$ power. 

The quality constraint of applications may vary during runtime,
thus,
research efforts have also focused on 
designing dynamically configurable adders,
which can tune their accuracy. 
In \cite{2012_Kahng_DAC},
the authors propose an accuracy-configurable adder,
called ACA, 
which consists of several sub-adders and an error detection \& correction module.
The design controls the accuracy at runtime 
and can operate in accurate mode. 
Another dynamically configurable adder,
called GDA,
is proposed in \cite{2013_Ye_ICCAD},
where multiplexers select the carry input either from the 
previous sub-adder or the carry prediction unit,
providing a more graceful degradation of the accuracy.
In the same direction,
the GeAr adder
employs multiple sub-adders
of equal length 
to variable approximation modes \cite{2015_Shafique_DAC}. 
This architecture also
supports accurate mode via a
configurable error correction unit. 

More recently,
Akbari \emph{et al.} \cite{2018_Akbari_IEEEtcasii}
introduce the RAP-CLA adder,
which
splits the conventional 
carry look-ahead scheme
into two segments,
i.e., the approximate part and the augmenting part,
supporting approximate and accurate mode.
When operating at the approximate mode,
the augmenting part is power-gated to reduce power consumption.
Another carry-prediction-based approach supporting both modes
is the SARA design \cite{2018_Xu_IEEEtvlsi}.
This adder uses carry ripple sub-adders,
and the carry prediction does not require a dedicated circuitry.
Finally, the BSCA adder,
which is based on a block-based carry speculative approach \cite{2020_Ebrahimi_IEEEtcasii},
integrates an error recovery unit
and
non-overlapped blocks consisting of 
a sub-adder, a carry prediction unit, and a selection unit.

\subsubsection{Approximate Multipliers}
The multiplication circuits have attracted significant interest from the research community.
The literature includes a plethora of inexact multipliers,
which can be categorized according to the prevailing approximation techniques:
(i) \emph{truncation and rounding} \cite{2015_Hashemi_ICCAD, 2017_Zendegani_IEEEtvlsi, LeonMicro, LeonTECS, 2019_Vahdat_IEEEtvlsi, LeonDAC, 2020_Frustaci_IEEEtcasii}, 
(ii) \emph{approximate radix encodings} \cite{2017_Liu_IEEEtc, 2019_Venkatachalam_IEEEtc, 2016_Jiang_IEEEtc, 2020_Waris_IEEEtcasii, LeonTVLSI},
(iii) \emph{use of approximate compressors}     \cite{2015_Momeni_IEEEtc, 2017_Akbari_IEEEtvlsi, 2019_Sabetzadeh_IEEEtcasi, 2018_Esposito_IEEEtcasi,  2020_Strollo_IEEEtcasi},
and (iv) \emph{logarithmic approximation} \cite{2018_Liu_IEEEtcasi, 2020_Saadat_DATE, 2021_Ansari_IEEEtc, 2021_Pilipovic_IEEEtcasi}.
Subsequently,
we introduce the state-of-the-art works from each category.

Starting with the rounding and truncation techniques,
the DRUM multiplier \cite{2015_Hashemi_ICCAD}
dynamically reduces the input bit-width, 
based on the leading `$1$' bits,
to achieve $60\%$ power gain in exchange for mean relative error of $1.47\%$.
Zendegani \emph{et al.} \cite{2017_Zendegani_IEEEtvlsi}
propose
the RoBa multiplier,
which 
rounds the operands to the nearest exponent-of-two 
and performs a shift-based multiplication in segments.
In \cite{LeonMicro},
the PR approximate multiplier
perforates partial products
and applies rounding to the remaining ones,
delivering up to $69\%$ energy gains.
The same approximation technique is integrated 
in the mantissa multiplication of floating-point numbers
to create the AFMU multiplier \cite{LeonTECS}.
Vahdat \emph{et al.} \cite{2019_Vahdat_IEEEtvlsi} 
propose
the TOSAM multiplier, which
truncates the input operands 
according to their leading `$1$' bit.
To decrease the error,
the truncated values are rounded to the nearest odd number.  
In \cite{LeonDAC},
different rounding, perforation and encoding schemes are combined to
extract the most energy-efficient designs.
Finally,
Frustaci \emph{et al.} \cite{2020_Frustaci_IEEEtcasii}
implement an alternative dynamic truncation with correction,
along with an efficient mapping for the remaining partial product bits.

Next,
we present multipliers
generating their partial products
based on approximate radix encodings.
Liu \emph{et al.} \cite{2017_Liu_IEEEtc}
modify the Karnaugh map of the radix-4 encoding
to create approximate encoders for generating the least-significant partial product bits.
A similar approach is followed in \cite{2019_Venkatachalam_IEEEtc},
where approximate radix-4 partial product generators are designed.
Jiang \emph{et al.} \cite{2016_Jiang_IEEEtc}
use an approximate adder to generate the $\pm$3$\times$ multiplicand term in the radix-8 multiplier.
In \cite{2020_Waris_IEEEtcasii},
the authors propose 
a hybrid low-radix encoding,
which encodes the most-significant bits
with the 
accurate radix-4 encoding
and the least-significant bits with the proposed approximate radix-8 encoding. 
Correspondingly,
the hybrid high-radix encoding,
i.e., accurate radix-4 and approximate radix-2$^k$,
is examined in \cite{LeonTVLSI}.

Several works employ approximate compressors for 
the partial product accumulation.
Momeni \emph{et al.} \cite{2015_Momeni_IEEEtc}
modify the truth table of the accurate
4:2 compressor 
to create two simplified designs
and use them in the Dadda multiplier.
The authors in \cite{2017_Akbari_IEEEtvlsi}
design 4:2 compressors,
again for Dadda multipliers,
which can switch
between accurate and approximate mode at runtime,
providing 68\% lower power consumption.
In \cite{2019_Sabetzadeh_IEEEtcasi},
an approximate 4:2 compressor
is implemented in FinFET
based on 
a three-input majority gate,
and then it is used in the Dadda architecture along truncation.
Esposito \emph{et al.} \cite{2018_Esposito_IEEEtcasi}
introduce a new family of approximate compressors 
and assign them to 
each column of the partial product matrix
according to their allocation algorithm.
Another interesting work
is the design of approximate compressors for multipliers using the concept of stacking circuit \cite{2020_Strollo_IEEEtcasi}.

Regarding the approximate logarithmic multipliers,
Liu \emph{et al.} \cite{2018_Liu_IEEEtcasi}
employ
a truncated binary-logarithm converter
and inexact adders for the mantissa addition
to design the ALM family of multipliers.
The logarithmic-based REALM multiplier \cite{2020_Saadat_DATE}
partitions the 
power-of-two intervals
of the input operands
into segments,
and
determines an error compensation factor 
for each one.
The ILM multiplier \cite{2021_Ansari_IEEEtc}
differentiates from the conventional design,
as it 
rounds the input operands to their 
nearest power-of-two
using a nearest `$1$' bit detector.
Pilipovic \emph{et al.} \cite{2021_Pilipovic_IEEEtcasi}
propose
a two-stage trimming 
logarithmic multiplier,
which reduces at first 
the bit-width of
the input operands,
and then  
the bit-width of the
mantissas.

\subsubsection{Approximate Dividers}
The division circuits have received less attention
than the other arithmetic circuits (adders and multipliers),
even though they feature complex calculations. 
Nevertheless,
the literature provides numerous works 
aiming to reduce the large critical paths of the conventional dividers.
The approximation techniques for the division circuits can be categorized as follows:
(i) \emph{bit-width scaling} \cite{2016_Hashemi_DAC, 2019_Jiang_IEEEtc},
(ii) \emph{use of approximate adder/subtractor cells} \cite{2015_Chen_GLSVLSI, 2016_Chen_IEEEtc, 2018_Chen_IEEEtmscs, 2020_Adams_IEEEtc},
and (iii) \emph{simplification of computations} \cite{2016_Zendegani_DATE, 2017_Vahdat_DATE, 2019_Imani_Date,  2018_Liu_ARITH, 2019_Saadat_DAC}.

The first class of approximation techniques
uses exact dividers with reduced bit-width.
The approximate divider of \cite{2016_Hashemi_DAC}
dynamically selects the most relevant bits
from the input operands
and 
performs accurate division at lower bit-width,
providing up to $70\%$ power gains in exchange for $3\%$ average error.
The design makes use of leading `$1$' bit detectors,
priority encoders,
multiplexers,
subtractor
and barrel shifter.
Similarly, 
the AAXD divider of \cite{2019_Jiang_IEEEtc} 
detects the leading `$1$' bits
and 
uses a pruning
scheme to extract the bits
that will be given as input to the divider. 
Additionally, 
the design integrates an error correction unit 
to form the final output.

Regarding the second class of approximation techniques,
Chen \emph{et al.} \cite{2015_Chen_GLSVLSI}
perform the subtraction of the non-restoring array divider with inexact subtractor circuits employing pass transistor logic.
For the proposed divider,
called AXDnr,  
the authors examine different schemes
with regard to which subtractions of the division array to approximate. 
Similarly,
in the AXDr divider of \cite{2016_Chen_IEEEtc},
some of the subtractions of the restoring array divider are performed with inexact subtractor circuits. 
The use of inexact cells has also been examined
in the high radix SRT divider \cite{2018_Chen_IEEEtmscs}.
In this divider,
called HR-AXD,
the inexact cell is 
a signed-digit adder, which is employed  
based on different replacement schemes and 
along with cell truncation and error compensation.
More recently,
Adams \emph{et al.} \cite{2020_Adams_IEEEtc}
introduce two approximate division architectures,
called AXRD-M1 and AXRD-M2, 
which deliver up to 46\% area and 57\% power gains, respectively, 
compared to the exact restoring divider.
The first design replaces 
some of the restoring divider cells with
inexact ones of simplified logic,
while the second one 
involves the elimination of some rows of the divider.

Targeting to perform the division with simplified computations,  
the SEERAD divider \cite{2016_Zendegani_DATE}
rounds the divisor to a specific form 
based on the leading `$1$' bit position,
and thus,
the division is transformed to shift-\&-add multiplication.
Similarly,
Vahdat \emph{et al.} \cite{2017_Vahdat_DATE}
propose the TruncApp divider, 
which
replaces the division
with the multiplication of 
the truncated dividend
by the approximate inverse
divisor.
In the same context, 
the CADE divider of \cite{2019_Imani_Date} 
performs the floating-point division 
by subtracting the input mantissas.
To compensate a large error,
which is estimated by analyzing the 
most-significant input bits,
a pre-computed value is retrieved from memory.
In \cite{2018_Liu_ARITH},
the proposed AXHD divider
approximates the least-significant computations of the division
using a non-iterative logarithmic approach that is 
based on leading `$1$' bit detection 
and subtraction of the logarithmic
mantissas. 
Finally,
Saadat \emph{et al.} \cite{2019_Saadat_DAC}
propose approximate integer and floating-point dividers with near-zero error bias,
called INZeD and FaNZeD, respectively,
by combining an error correction method with the classical approximate logarithmic divider.

\subsubsection{Approximate Synthesis}
An automated approach to
generate inexact circuits
is the approximate logic synthesis. 
This method
provides increased approximation diversity,
i.e., multiple approximate variants of circuits, 
without relying on 
the manual approximation inserted by the designer,
such as in the case of the aforementioned arithmetic approximations.
Another benefit of approximate synthesis 
is that 
several techniques 
generate the approximate variant
that leads  
to the maximum hardware gains
for a given approximation/error constraint. 
The state-of-the-art techniques
can be categorized as follows \cite{2020_Scarabottolo_IEEE}:
(i) \emph{structural netlist transformation}
\cite{2017_Schlachter_IEEEtvlsi, 2018_Scarabottolo_DATE, 2013_Venkataramani_DATE, 2017_Liu_ICCAD, 2021_Castro_IEEEtcasii}, 
(ii) \emph{Boolean rewriting} 
\cite{2012_Venkataramani_DAC, 2014_Ranjan_DATE, 2013_Miao_ICCAD, 2018_Hashemi_DAC}, 
(iii) \emph{high-level approximate description} \cite{2015_Yazdanbakhsh_DATE, 2014_Nepal_DATE, 2017_Lee_DATE, 2019_Nepal_IEEEtetc, 2020_Castro_ICCAD},
and (iv) \emph{evolutionary synthesis} \cite{2013_Sekanina_ICES, 2015_Vasicek_IEEEtec, 2017_Mrazek_DATE, 2016_Vasicek_FPL, 2019_Vasicek_DATE}.

Several works of the literature employ
a direct acyclic graph
to represent the circuit netlist,
where each node corresponds to a gate. 
In this context,
the GLP technique \cite{2017_Schlachter_IEEEtvlsi}
prunes nodes with an iterative greedy approach
according to their impact on
the final output and their toggle activity.
In contrast,
the CC framework \cite{2018_Scarabottolo_DATE}
performs an exhaustive exploration of
all possible
node subsets that can be pruned
without surpassing the error constraint.
Venkataramani \emph{et al.} \cite{2013_Venkataramani_DATE}
propose SASIMI,
which is based on a greedy heuristic
to find signal pairs assuming the same value
and substitute one with the other.
This automatic synthesis framework 
guarantees that the user-defined quality constraint is satisfied 
and generates accuracy configurable circuits.
To apply stochastic netlist transformation,
the SCALS framework \cite{2017_Liu_ICCAD}
maps an initial gate-level network to the targeted technology (standard cell or FPGA),
and then it iteratively extracts sets of sub-netlists
and
inserts random approximations in them.  
These sub-netlists are evaluated
using statistical hypothesis testing. 
More recently,
Castro-Codinez \emph{et al.} \cite{2021_Castro_IEEEtcasii}
propose the AxLS framework,
which converts the Verilog netlist to XML format
and then applies typical transformation techniques, 
e.g., gate pruning,
according to an error threshold.

The second category 
includes techniques
that apply approximations in a formal Boolean representation of the circuit 
before it is synthesized.
The SALSA approach \cite{2012_Venkataramani_DAC} 
encodes the error constraints
into a quality logic function,
which compares the outputs of the accurate and approximate circuits.
Towards logic simplification,
SALSA computes
the ``observability don't cares''
for each output of the approximate circuit,
i.e.,
the set of input values for which 
the output is insensitive.
In the same direction,
but for sequential circuits,
Ranjan \emph{et al.} \cite{2014_Ranjan_DATE} introduce ASLAN.
This framework 
generates several approximate variants
of the 
combinational blocks,
and then it identifies the best
approximations for the entire sequential circuit
based on a gradient-descent approach.
Miao \emph{et al.} \cite{2013_Miao_ICCAD} 
use a two-phase Boolean minimization algorithm
to address the problem of approximate synthesis.
The first phase solves the problem under a given constraint for the error magnitude, 
and the second phase iteratively finds a solution that also
satisfies the error frequency constraint.
In an iterative fashion,
the BLASYS methodology \cite{2018_Hashemi_DAC}
partitions the circuit into smaller circuits,
and for each one,
it generates an approximate truth table 
based on 
Boolean matrix factorization.
The approximate sub-circuits are synthesized 
and the trade-off between error and power/area efficiency
for the entire circuit is evaluated. 

Regarding approximations introduced at the hardware description level,
Yazdanbakhsh \emph{et al.} \cite{2015_Yazdanbakhsh_DATE}
propose the Axilog language annotations,
which provide syntax and semantics for approximate design and reuse in Verilog.
Axilog allows the designer 
to partition the design into 
accurate and approximate segments.
ABACUS \cite{2014_Nepal_DATE} is another interesting work,
which parses
the behavioral/RTL Verilog description of the design
to create its abstract syntax tree.
Next, 
a set of diverse transformations
is applied to the tree
to create approximate variants, 
which are then
written in Verilog.
An expanded version of ABACUS is introduced in
\cite{2019_Nepal_IEEEtetc},
where sorting-based evolutionary algorithms are employed for design space exploration.
Moreover, 
the new ABACUS version focuses on approximations
in critical paths
to facilitate 
the reduction of the supply voltage.
Lee \emph{et al.} \cite{2017_Lee_DATE}  
generate approximate designs in Verilog 
from C accurate descriptions.
The proposed framework
computes 
data statistics and mobility information for
the given design 
and employs an heuristic solver
for optimizing the energy--quality trade-off.
Targeting to high-level synthesis,
the AxHLS approach
\cite{2020_Castro_ICCAD}
performs a design space exploration,
based on analytical models,
to identify the best arithmetic approximations
for a given error constraint.
Starting from a C description,
AxHLS adopts
scheduling and binding
operations
to apply the approximations provided by the exploration
and generate the Verilog code.

The fourth class of techniques for automated synthesis of approximate circuits is based on evolutionary algorithms,
i.e., heuristic-based search algorithms that treat circuit approximation
as multi-objective optimization problem and generate a set of solutions.
In this context,
Sekanina \emph{et al.} \cite{2013_Sekanina_ICES}
use Cartesian genetic programming
to minimize the error in adders
considering the number of logic gates as constraint. 
This approach is extended in \cite{2015_Vasicek_IEEEtec},
where approximate multipliers and median filters
are evolved through randomly seeded Cartesian genetic programming. 
Based on the same utilities, 
the authors of \cite{2017_Mrazek_DATE} propose 
the EvoApprox8b library of approximate adders and multipliers.
This library is generated by examining various trade-offs between accuracy and hardware efficiency,
and it offers different approximation variants and circuit architectures. 
In \cite{2016_Vasicek_FPL},
a search-based technique for evolutionary circuit synthesis for FPGAs is proposed.
In particular, 
this approach represents the circuit as a directed acyclic graph,
and 
re-synthesizes approximate configurations based on Cartesian genetic programming.
Vasicek \emph{et al.} \cite{2019_Vasicek_DATE}
adjust the approximation degree
according to the significance of the inputs.
To do so,
they adopt a weighted error metric 
to determine the significance of each input vector 
and use Cartesian genetic programming to
minimize the circuit's area while satisfying a threshold. 

\subsection{Voltage Over-Scaling}
Voltage over-scaling reduces the circuit's supply voltage
below its nominal value,
while keeping 
the clock frequency constant. 
The circuit operation 
at a lower voltage value 
produces timing errors
due to the failure of the critical paths
to meet the delay constraints.
Nevertheless, 
considering that power consumption
depends on the voltage value,
VOS techniques are continuously examined in the literature.
An exploration and quantification of the benefits and overheads of VOS is presented in \cite{2010_Kurdahi_IEEEtvlsi}.
Research involving VOS can be classified in the following categories:
(i) \emph{slack re-distribution} \cite{2010_Kahng_ASP-DAC},
(ii) \emph{circuit re-design and architecture modification} \cite{2011_Mohapatra_DATE, 2013_Chen_IEEEtvlsi, 2018_Zhang_DAC},
(iii) \emph{fine-grained scaling} \cite{2019_Pandey_DAC, 2020_Wang_IEEEtc, 2019_Zervakis_IEEEtcasii},
and 
(iv) \emph{error modeling} \cite{2010_Liu_IEEEtvlsi, 2012_Jeon_IEEEtcasii, 2017_Ragavan_DATE, 2020_Jiao_IEEEtcad,  2018_Zervakis_IEEEtvlsi}.

Kahng \emph{et al.}
\cite{2010_Kahng_ASP-DAC} 
shift the timing slack
of the frequently executed near-critical paths
through slack redistribution,
and thus,
they reduce the minimum voltage
at which the error rate remains acceptable.
The proposed technique is based on post-layout cell resizing
to deliver the switching activity-aware slack redistribution. 
More specifically,
a heuristic
finds the voltage
satisfying the desired error rate,
and then it increases the transistor width 
of the cells
to optimize the frequently executed paths.

In \cite{2011_Mohapatra_DATE},
the authors
optimize building blocks
for more graceful degradation under VOS,
using two techniques,
i.e., 
dynamic segmentation \& error compensation
and
delay budgeting of chained datapath.
The first technique
bit-slices the datapath of the adder 
and employs a multi-cycle error correction circuitry that tracks the carries.
The second technique 
adds transparent latches 
between chained arithmetic units
to distribute the clock period.
To facilitate VOS,
Chen \emph{et al.} \cite{2013_Chen_IEEEtvlsi}
build their designs
on
the residue number system,
which
provides shorter critical paths
than conventional arithmetic.
They also
employ the reduced precision redundancy scheme
to eliminate the timing errors. 
Another interesting work is Thundervolt \cite{2018_Zhang_DAC}, 
which provides error recovery
in the Multiply-And-Accumulate (MAC) units
of systolic Deep Neural Network (DNN) accelerators.
To detect timing errors,
Thundervolt
employs Razor shadow flip-flops. 
In case an error occurs in a MAC,
a multiplexer 
forwards the previous MAC's accurate partial sum
(stored in the Razor flip-flop)
to the next MAC.

Targeting fine-grained VOS solutions,
i.e., 
different voltages
across the same circuit architecture, 
Pandey \emph{et al.} 
propose GreenTPU \cite{2019_Pandey_DAC}.
This design 
integrates
a timing error control unit 
in each MAC row of 
the systolic array,
which stores 
input sequences producing timing errors.
As a result,
when such an input sequence pattern
is identified,
the voltage of the MAC
is scaled accordingly 
to prevent timing errors. 
In the same context,
the authors of \cite{2020_Wang_IEEEtc}
propose NN-APP.
This framework 
analyzes the error propagation 
in neural networks
to model 
the impact of VOS on accuracy.
Based on this analysis,
as well as an error resilience study for the neurons,
NN-APP uses 
a voltage clustering method
to assign the same voltage
to neurons with similar error resilience.
Another fine-grained VOS approach is proposed in
\cite{2019_Zervakis_IEEEtcasii}.
This framework 
provides voltage heterogeneity
by using
a greedy algorithm to   
solve the optimization problem
of grouping and assigning the voltage of arithmetic units
to different islands. 

The analysis of errors
in circuits under VOS is considered a key factor,
as it guides the aggressiveness of voltage scaling
towards the acceptable error margins.
In \cite{2010_Liu_IEEEtvlsi},
an analytical method to study the errors in voltage over-scaled arithmetic circuits
is proposed.
Similarly, 
the authors of 
\cite{2012_Jeon_IEEEtcasii}
introduce a probabilistic approach to
model the errors
of the critical paths.
In the same category,
we include works relying
on simulations to analyze the errors of VOS.
Ragavan \emph{et al.}
\cite{2017_Ragavan_DATE}
characterize 
arithmetic circuits in terms of energy efficiency and errors
using 
transistor-level SPICE simulation for various voltages. 
Based on this characterization,
they propose a statistical model 
to simulate the behavior of arithmetic operations in VOS systems.
Exploiting the machine learning methods,
Jiao \emph{et al.} \cite{2020_Jiao_IEEEtcad}
propose
LEVAX
to model voltage over-scaled functional units.
This input-aware model
is trained on data from gate-level simulations 
to
predict the timing error rate
for each output bit.
To provide accurate VOS-aware gate-level simulation,
Zervakis \emph{et al.} propose VOSsim \cite{2018_Zervakis_IEEEtvlsi}.
This framework
performs an offline 
characterization of the flip-flop
for timing violations 
and 
calculates the cell delays
for the targeted voltage,
enabling gate-level simulation under VOS.

\subsection{Over-Clocking}
Over-clocking (or frequency over-scaling)
configures the circuit/system at higher clock frequencies
than those that respect the critical paths.
As a result,
timing errors are induced in exchange for increased performance.
A trade-off analysis between accuracy and performance 
when over-clocking FPGA-based designs
is presented in \cite{2013_Shi_FCCM}.
In the same work, 
the authors show that OC 
outperforms the traditional bit truncation for the same error constraint. 
For our analysis,
we consider that 
the state-of-the-art works of the domain focus on the following directions:
(i) \emph{tight synthesis} \cite{2018_Alan_IEEEtcad}, 
(ii) \emph{circuit re-design and architecture modification} \cite{2013_Ramasubramanian_DAC, 2014_Shi_DAC, 2017_Wang_IEEEtvlsi}, 
(iii) \emph{error detection \& correction} \cite{2014_Choudhury_IEEEtc, 2016_Ragavan_ISVLSI, 2017_Li_IEEEtvlsi},
and 
(iv) \emph{error prediction} \cite{2012_Roy_DAC, 2015_Constantin_DATE, 2016_Jiao_ICCD, 2017_Jiao_DATE}.

The first approach aims to reduce the timing errors of OC by 
optimizing the critical paths of the design.
In this context, 
the SlackHammer framework \cite{2018_Alan_IEEEtcad}
synthesizes circuits with tight delay constraints
to 
reduce the number of near-critical paths,
and thus, 
decrease the probability of timing errors when frequency is over-scaled. 
At first, 
SlackHammer isolates the paths
and identifies potential delay optimizations. 
Based on the isolated path analysis,
the framework performs an iterative synthesis
with tighter constraints for the 
primary outputs with negative slack. 

The second class of techniques 
modifies the conventional circuit architecture
to facilitate frequency OC and increase the resilience to timing errors.
The retiming technique \cite{2013_Ramasubramanian_DAC}
re-defines the boundaries of combinational logic
by moving the flip-flops backward or forward between the stages.
Based on this circuit optimization, 
the synthesis is relaxed
by ignoring the 
paths that are bottleneck to minimum period retiming. 
Targeting different circuit architectures, 
Shi \emph{et al.} \cite{2014_Shi_DAC}
adopt an alternative arithmetic,
called online,
and show that online-based circuits are
more resilient to the timing errors of OC
than circuits with traditional arithmetic.
The modification of the initial neural network model
to provide resilience in timing errors
has also attracted research interest. 
In this direction, 
Wang \emph{et al.} \cite{2017_Wang_IEEEtvlsi} 
propose an iterative reclocking-and-retraining framework for  
operating neural network circuits at higher frequencies
under a given accuracy constraint.
The clock frequency is gradually increased
and the network's weights are updated through back-propagation training
until to find the maximum frequency for which
the timing errors are mitigated
and the accuracy constraint is satisfied.

Several works propose circuits for timing error detection \& correction,
enabling the use of over-clocking.
These techniques either 
improve the frequency value of the first failure,
i.e., the first timing error,
or 
reduce the probability of timing errors.  
TIMBER \cite{2014_Choudhury_IEEEtc}
masks timing errors
by borrowing time from successive pipeline stages.
According to this approach,
the use of 
discrete time-borrowing flip-flops
and continuous time-borrowing latches
slows down the appearance of timing errors
with respect to the frequency scaling.
Ragavan \emph{et al.} \cite{2016_Ragavan_ISVLSI}
detect and correct timing errors by
employing a dynamic speculation window 
on the double-sampling scheme.
This technique 
adds an extra register
(called shadow and clocked by a second ``delayed'' clock)  
at the end of the pipelined path
to sample the output data at two different time instances.
The proposed approach also
uses an online slack measurement to adaptively
over-clock the design.
The TEAI approach \cite{2017_Li_IEEEtvlsi}
is based on 
the locality of the timing errors
in software-level instructions,
i.e., the tendency of specific instructions to produce timing errors.
TEAI identifies these instructions at runtime,
and sends error alarms
to hardware,
which is equipped with error detection \& correction circuits.

There is also wide research on the prediction of the timing errors in advance, 
which allows to over-scale the frequency 
according to the acceptable error margins. 
In \cite{2012_Roy_DAC},
the authors introduce an
instruction-level error prediction system
for pipelined micro-processors,
which stalls the pipeline when
critical instructions are detected.
Their method is based on
gate-level simulations to
find the critical paths that are 
sensitized during the program execution.
Similarly,
Constantin \emph{et al.}
\cite{2015_Constantin_DATE}
obtain the maximum delays for 
each arithmetic instruction 
through gate-level simulations,
and 
they dynamically exploit timing margins
to apply frequency over-scaling.

Besides 
instruction-level prediction models,
there are numerous works 
that build models 
based on machine learning 
and simulations of functional units.
A representative work of this approach 
is WILD \cite{2016_Jiao_ICCD},
which builds a 
workload-dependent prediction model
using logistic regression. 
In the same direction,
SLoT \cite{2017_Jiao_DATE} is a supervised learning model that 
predicts timing errors
based on the inputs and the clock frequency.
At first, 
SLoT performs gate-level simulation
to extract timing class labels, 
i.e., ``timing error'' or ``no timing error'', 
for different inputs and frequencies.
These classes are then used, 
along with features extracted from random data pre-processing, 
to train the error prediction model.

\partabstract{\begin{center}{\normalsize\textbf{Prologue}}\\[2pt]\end{center}
{Dissertation's Part \ref{part1} 
focuses on the arithmetic of circuits and accelerators. 
At first, 
it proposes logic-level arithmetic approximation techniques,
and then,
it presents the development of approximate hardware accelerators.
Chapters \ref{chapter3}--\ref{chapter7}
examine different aspects of 
computer arithmetic and approximate circuit design: 
alternative numerical formats in Chapter \ref{chapter3},
hybrid radix encodings in Chapter \ref{chapter4},
dynamic/runtime approximation configuration in Chapter \ref{chapter5},
cooperative approximation in Chapter \ref{chapter6},
and systematic design of approximate accelerators in Chapter \ref{chapter7}. 

Chapter \ref{chapter3}
highlights the benefits 
of applying sophisticated bit-level optimizations. 
Chapter \ref{chapter4}
proposes a hybrid high-radix encoding that is used
to design the RAD family of approximate multipliers.  
Chapter \ref{chapter5}
proposes runtime-configurable approximate multipliers
for fixed-point (DyFXU design family) 
and floating-point arithmetic
(DyFPU design family).  
Chapter \ref{chapter6} 
examines the combination of arithmetic approximation techniques,
resulting in the state-of-the-art 
ROUP family of approximate multipliers.
Finally, 
Chapter \ref{chapter7}
integrates all the proposed approximate circuits
in the design
of ASIC/FPGA accelerators
for DSP and AI applications.

\textbf{Acknowledgements:}
The author of the Ph.D. Dissertation
would like to thank 
Prof. Kiamal Pekmestzi 
for accepting him in the Ph.D. program, 
as well as 
Prof. Dimitrios Soudris
for his support throughout the years.}}

\part{Arithmetic Approximation Techniques~for~Circuit~Design}
\label{part1}

\chapter{Arithmetic Optimization: Double-LSB~Encoding}
\label{chapter3}

\addtocontents{lof}{\protect\contentsline{chapter}{\protect\numberline{3}Arithmetic Optimization: Double-LSB Encoding}{}{}}
\addtocontents{lot}{\protect\contentsline{chapter}{\protect\numberline{3}Arithmetic Optimization: Double-LSB Encoding}{}{}}

\begin{ChapterAbstract}
Computer arithmetic has received significant attention,
as it inherently affects the efficiency and performance
of the hardware accelerators.
In this context,
novel numerical formats are examined,
targeting to provide increased resource gains 
compared to the conventional binary formats
in various application domains, e.g., in Digital Signal Processing (DSP). 
In this chapter,
we adopt the Double Least Significant Bit (DLSB) format,
where the numbers have an extra least significant bit, 
and we apply optimizations in the multiplication,
i.e., one of the most resource-hungry DLSB operations.
The DLSB arithmetic delivers several benefits, 
such as the symmetric representation range, 
the number negation performed only by bitwise inversion, 
the improvement of the residue number circuits,
and the improvement of the rounding process in floating-point calculations.
However, all these advantageous features come with some penalties
in the design of the arithmetic DLSB units.
Towards reducing the DLSB overheads in the multiplication,
we propose an energy-efficient scheme for multiplying $\mathit{2}$'s-complement binary numbers,
which is based on sophisticated bit-level manipulations.
The overhead of the proposed DLSB design
is negligible compared to the conventional design for ordinary $\mathit{2}$'s-complement numbers,
i.e., \raisebox{0.8pt}{$\mathit{\scriptstyle\sim}$}$\mathit{3}$\%
area/energy overhead on average
for different multiplier sizes.
Moreover, 
our design outperforms the straightforward design approach
by providing 
$\mathit{4\times}$--$\mathit{5\times}$
less resource overhead. 
Finally, 
as case study, 
we demonstrate how the DLSB multiplier 
can be effectively used as 
building block for the implementation of larger multiplications, 
delivering $\mathit{31}$\% area and $\mathit{43}$\% energy gains.\\
This chapter is based on our \textbf{publication} in \textbf{\cite{leon_IET}}.
\end{ChapterAbstract}

\newpage 

\section{Introduction}

The integration of novel numerical representation formats 
in the design of circuits 
appears as an effective solution 
for reducing the power/energy consumption, area and delay \cite{Parhami2000}.
In this context, 
several alternative arithmetic formats have been proposed in the literature. 
The Residue Number System (RNS) \cite{Garner1959} and the Logarithmic Number System (LNS) \cite{Swartzlander1975} 
are typical examples of non-standard representation formats 
that aim to improve the hardware efficiency of the arithmetic units. 
Another interesting format is the carry-save representation \cite{Verma2004}, 
which is mainly used to build fast adders with multiple inputs 
and accumulation trees for multipliers.  

In this chapter, 
we focus on the novel \emph{Double Least Significant Bit (DLSB)} arithmetic representation,
which is proposed by Parhami \cite{Parhami2008}. 
In particular, 
the DLSB arithmetic adopts the conventional arithmetic 
and considers an additional Least Significant Bit (LSB),
i.e., an extra bit that has the same weight as the original LSB. 
With this format,  
the number representation range becomes fully symmetric, 
i.e., for $2$'s-complement arithmetic,
it is 
$[-2^{n-1}$, $2^{n-1}]$
instead of $[-2^{n-1}$, $2^{n-1}-1]$. 
Other benefits of 
the DLSB arithmetic 
are
the 
simpler number inversion, 
the simpler rounding process in floating-point calculations, 
and the increased efficiency in residue number operations. 

The symmetry in the representation range allows to perform negation with only bitwise inversion,
hence,
the addition of '$1$'
(performed in the conventional $2$'s-complement arithmetic) 
is eliminated. 
The latter is an important feature, especially in algorithms where the sign change is not followed by another addition \cite{Parhami2008}.
Moreover, the symmetry of DLSB guarantees that all the numbers can be complemented without resulting in overflow. 

In floating-point calculations, 
if the result has more bits than those supported by the format, 
then the extra bits are discarded 
and the remaining ones are adjusted with rounding. 
This process requires addition 
involving carry propagation, 
which is a task that can add significant delay to the floating-point arithmetic units. 
In the case of DLSB format, 
the rounding algorithm needs only to determine the value of the extra LSB in the rounded result \cite{Parhami2008}. 
Specifically, instead of performing the required addition of the regular rounding, 
the extra LSB of the result is set to `$1$' for rounding up,
avoiding in this way the carry propagation. 

The DLSB format has also been used in RNS operations
to improve the design efficiency.
In this direction, 
the DLSB encoding of the $2^n+1$ residues 
is employed to design the modulo $2^n+1$ adder \cite{Jaberipur2006} and multiplier \cite{Jaberipur2010}.
The representation range of the unsigned numbers is $[0$, $2^n]$, 
and thus, 
the end-around carry is stored 
and there is no need for a post-increment operation.
Furthermore, 
the authors of \cite{Vassalos2011} 
propose reverse converters for two 4-moduli sets
that are based on DLSB encodings.
The derived results show that both the area and delay of the DLSB-based converters 
are comparable to that of the converters using the conventional binary encoding.

These advantageous features of the DLSB arithmetic 
come with some delay and area overheads, 
considering the corresponding conventional circuits as baseline.  
In \cite{Parhami2008}, the delay and area theoretical penalties are reported for a variety of DLSB circuits 
(e.g., adders, different types of multipliers, dividers), 
showing that complex arithmetic units
such as the multipliers
exhibit the largest overheads. 
Furthermore, these theoretical overheads have not been actually evaluated with industrial synthesis tool-flows and technology libraries.  
In this chapter,
motivated by the benefits of the DLSB format,
and targeting to eliminate the penalties that arise,
we propose an improved algorithm 
for the multiplication of 
two $2$'s-complement DLSB numbers.
We provide both 
theoretical and robust experimental evaluations on industrial tools,
showing that sophisticated bit-level manipulations
provide significant gains
and decrease the penalties. 

We further motivate the applicability and effectiveness of the proposed DLSB technique
by examining a realistic case study design scenario.
In particular, we show that large-size multiplications
can be efficiently implemented 
using small DLSB multipliers as building blocks.
The multiplication of large operands is a key component in various applications, 
e.g., cryptographic schemes and scientific calculations.
To improve the hardware efficiency, 
significant research has been conducted on the implementation of large-size multiplications \cite{Wang2013, Rafferty2017}.
The mapping of large-size multiplications
in Field-Programmable Gate Arrays (FPGAs)
is also examined in the literature \cite{Dinechin2009}. 
In FPGAs, 
such multiplications are implemented
by segmenting the input operands based on the bit-width of the hardwired multipliers 
that are integrated in the DSP blocks. 

The \textbf{contribution} of this chapter is summarized as follows:
\begin{itemize}[]
\item[(i)] We highlight the significance of computer arithmetic and show that novel numerical formats can provide valuable gains in hardware.
\item[(ii)] We propose an improved algorithm for the multiplication of two $2$'s-complement DLSB numbers, which is strictly defined by a rigorous theoretical analysis involving sophisticated bit-level manipulations. 
\item[(iii)] We show that the proposed DLSB circuit 
outperforms its unoptimized counterpart,
and also,
it can be effectively used as building block for improving the
implementation of large-size multiplications.
\end{itemize}

The remainder of this chapter is organized as follows. 
Section \ref{s3_2} introduces the DLSB arithmetic format,
Section \ref{s3_3} presents the proposed DLSB multiplier,
and
Section \ref{s3_4} includes the theoretical and experimental evaluation.
Finally, 
Section \ref{s3_5} draws the conclusions.

\section{The Double Least Significant Bit Format}
\label{s3_2}

This section includes an introduction in the $2$'s-complement DLSB format \cite{Parhami2008}. 
Let $X = \langle x_{n-1} x_{n-2} \cdots x_0\rangle_{2\text{'s}}$
be a $n$-bit $2$'s-complement number. 
$X$ is converted to a $2$'s-complement DLSB number by attaching an extra LSB ($x_{0+}$) next to the original LSB,
as shown in Eq. \eqref{eq_dlsb}.

\vspace{-12pt}

\begin{equation}
X^+ =  
X + x_{0+} = 
\langle x_{n-1} x_{n-2} \cdots x_0 \rangle_{2\text{'s}} + x_{0+}
\label{eq_dlsb}
\end{equation}

\vspace{-5pt}

For the rest of the discussion, 
we assume that $X^+$ is a $n$-bit $2$'s-complement DLSB number 
(the extra LSB is not calculated in the bit-width). 
Moreover, the following notations are used: 
$\langle \dots \rangle_{2\text{'s}}$ denotes 
a $2$'s-complement number, 
and $\langle \dots \rangle_{\text{u}}$ denotes an unsigned number.
Subsequently, we present some numerical examples.

\noindent \textbf{Example 1.}
The DLSB number $\langle 0111 \rangle_{2\text{'s}} + 1$
is the number $8_{10}$ in the decimal system, 
while $\langle 1010\rangle_{ 2\text{'s}} + 0$ is $-6_{10}$.

\noindent \textbf{Example 2.}
The DLSB number $\langle 0111\rangle_{ 2\text{'s}} + 1  =  8_{10}$
is complemented by inverting all its bits:
$\langle 1000\rangle_{ 2\text{'s}} + 0  =  -8_{10}$.

\begin{figure}[!t]
\centering
\includegraphics[width=0.48\textwidth]{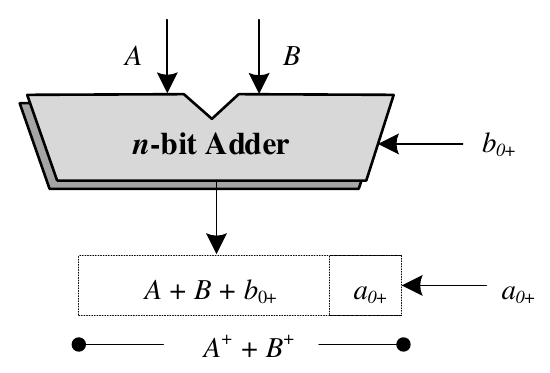}%
\caption[DLSB Addition and Subtraction]{The addition of two $n$-bit DLSB numbers is performed using a conventional adder. 
The subtraction is performed with the same architecture, but with the inverted bits of $B^+$.}%
\label{fig_dlsbadd}
\end{figure}

The addition of two $n$-bit DLSB numbers is performed using a conventional $n$-bit adder \cite{Parhami2008}.
The extra LSB of one of the operands 
is given to the carry input of the adder, 
while the extra LSB of the other 
is attached to the sum as its extra LSB. 
The operation of the addition is illustrated in Figure \ref{fig_dlsbadd}.
The subtraction is performed in the same way,
however,
the subtrahend is firstly complemented, 
namely 
all its bits (including the extra LSB) are logically inverted.  

The multiplication of a DLSB number by a power of two, i.e., $2^e$,  
is performed like in the conventional arithmetic, i.e., via $e$ left shifts. 
Regarding the multiplication of two DLSB numbers, 
well-established architectures and algorithms
can be employed with some extra overheads, as discussed in \cite{Parhami2008}.
This arithmetic operation is exhaustively examined in the rest of the chapter.

\section{Design of DLSB Multiplication Circuits}
\label{s3_3}

In this section,
we design the DLSB multiplier in two flavours,
i.e., based on the straightforward approach 
and a more sophisticated improved approach. 
As baseline multiplication algorithm,
we adopt the Modified Booth (MB) method \cite{Parhami2000},
which facilitates low-level optimizations
and also outperforms several well-established multiplication methods \cite{leon_IET}.

\subsection{Straightforward DLSB Design}

Let $A^+ = 
\langle a_{n-1} a_{n-2} \cdots a_0 \rangle_{2\text{'s}} + a_{0+}$ 
and $B^+ = 
\langle b_{n-1} b_{n-2} \cdots b_0 \rangle_{2\text{'s}} + b_{0+}$ 
be two $n$-bit $2$'s-complement DLSB numbers. 
We form their product as shown in Eq. \eqref{eq_dlsbprod}.

\begin{equation}
A^+ \times B^+ =  
\big( \langle a_{n-1} a_{n-2} \cdots a_0 \rangle_{ 2\text{'s}} + a_{0+} \big) \cdot B^+ \\[4pt]
\label{eq_dlsbprod}
\end{equation}

Following the Modified Booth multiplication algorithm,
$B^+$ is encoded with its extra LSB $b_{0+}$ 
included in the least significant Modified Booth digit,
i.e., 
$b_{-1} = b_{0+}$ instead of $b_{-1} = 0$ (conventional encoding), 
as shown in Eq. \eqref{eq_dlsbopB}--\eqref{eq_dlsbbth}.

\vspace*{-4pt}

\begin{equation}
B^+ = 
-2^{n-1}b_{n-1} + 
\mathlarger{\sum}_{\substack{i=0}}^{\substack{n-2}} 2^{i}b_{i} + b_{0+} = 
\mathlarger{\sum}_{\substack{j=0}}^{\substack{n\text{/}2-1}} 4^{j}b_{j}^{\text{\scriptsize\emph{MB}}}  \label{eq_dlsbopB}
\end{equation}
\begin{equation}
\text{where }  \;
b_{j}^{\text{\scriptsize\emph{MB}}} = -2b_{2j+1} + b_{2j} + b_{2j-1}  \;\; \implies \;\; b_{j}^{\text{\scriptsize\emph{MB}}} \in \{0, \pm 1, \pm 2\}\\[3pt]
\label{eq_dlsbbth}
\end{equation}

The Modified Booth digits 
$b_{j}^{\text{\scriptsize\emph{MB}}}$ $\in$ $\{0,$ ${\pm} 1,$ ${\pm} 2\}$ 
are functions of three consecutive bits of $B^+$ 
($b_{2j+1}$, $b_{2j}$, $b_{2j-1}$), 
and they are formed according Table \ref{tb_bth}.
Each digit is represented by three $1$-bit signals that define its sign ($s_j$) and absolute value ($two_j$, $one_j$). 
Using these encoding signals, 
the Modified Booth digits are calculated
by Eq. \eqref{eq_mb}.

\vspace*{-12pt}

\begin{equation}
b_{j}^{\text{\scriptsize\emph{MB}}} = (-1)^{s_j} \cdot (2two_j + one_j), \;\; 
  j = 0, 1, \cdots n/2-1 \\[4pt]
\label{eq_mb}
\end{equation}

Using Eq. \eqref{eq_dlsbprod} and \eqref{eq_dlsbopB}, 
the multiplication of the DLSB numbers $A^+$ and $B^+$
is formed as shown in Eq. \eqref{eq_dlsbold}.

\vspace*{-8pt}

\begin{eqnarray}
 & A^{+} \times B^{+}
    = \langle a_{n-1} a_{n-2} \cdots a_0 \rangle_{2\text{'s}} \cdot  B^{+} + a_{0+} \cdot B^{+} = \nonumber \\[8pt]
 &  = 
 \langle a_{n-1} a_{n-2} \cdots a_0 \rangle_{2\text{'s}} \cdot  
 \mathlarger{\sum}_{\substack{j=0}}^{\substack{n\text{/}2-1}} 4^j b_{j}^{\text{{\scriptsize\emph{MB}}}}  
 + a_{0+} \cdot \big(\langle b_{n-1} b_{n-2}\cdots  b_0\rangle_{2\text{'s}} + b_{0+}\big)  
 \label{eq_dlsbold} 
\end{eqnarray}

The implementation of the DLSB multiplier
based on Eq. \eqref{eq_dlsbold}
requires a conventional Modified Booth multiplier
to calculate the first term of the final expression,
as well as considerable overhead
to calculate the extra term 
$a_{0+} \cdot \big(\langle b_{n-1} b_{n-2}\cdots  b_0\rangle_{2\text{'s}} + b_{0+}\big)$.

\begin{table}[!t]
\fontsize{9}{10}\selectfont
\renewcommand{\arraystretch}{1.2}
\setlength{\tabcolsep}{9pt}
\caption[Modified Booth Encoding]{Modified Booth encoding.}
\label{tb_bth}
\centering
\begin{tabular}{ccccccc}
\hline
\multicolumn{3}{c}{\textbf{Input}} & \multicolumn{1}{c}{\textbf{MB Digit}} & \multicolumn{3}{c}{\textbf{Output}}\\
\cmidrule(lr){1-3} \cmidrule(lr){4-4} \cmidrule(lr){5-7}
$b_{2j+1}$ & $b_{2j}$ & $b_{2j-1}$ & $b_{j}^{\text{\scriptsize\emph{MB}}}$ & $s_j$ & $two_j$ & $one_j$\\[1pt]
\hline 
\hline 
$0$ & $0$ & $0$ & $0$ & $0$ & $0$ & $0$\\
$0$ & $0$ & $1$ & $1$ & $0$ & $0$ & $1$\\
$0$ & $1$ & $0$ & $1$ & $0$ & $0$ & $1$\\
$0$ & $1$ & $1$ & $2$ & $0$ & $1$ & $0$\\
$1$ & $0$ & $0$ & $-2$ & $1$ & $1$ & $0$\\
$1$ & $0$ & $1$ & $-1$ & $1$ & $0$ & $1$\\
$1$ & $1$ & $0$ & $-1$ & $1$ & $0$ & $1$\\
$1$ & $1$ & $1$ & $0$ & $1$ & $0$ & $0$\\
\hline
\end{tabular}
\vspace*{5pt}
\end{table}

\subsection{Sophisticated DLSB Design}

To reduce the overhead derived by the calculation of the extra term in the straightforward approach, 
we consider an alternative representation for the operand $A^+$. 
Firstly,
we examine how $A^+$ is formed
with respect to the value of $a_{0+}$:

\begin{itemize}

\item if $a_{0+} = 0$, we get:
\begin{equation}
A^+ = 
\langle a_{n-1} a_{n-2} \cdots a_0\rangle_{2\text{'s}} + 0 = 
\langle a_{n-1} a_{n-2} \cdots a_0\rangle_{2\text{'s}}
\label{eq_opA1}
\end{equation}
\vspace*{-16pt}
\item if $a_{0+} = 1$, and using the expression $a_i = - \bar{a}_i + 1$ 
($\bar{a}_i$: inverted $a_i$), we get:
\begin{eqnarray}
 A^{+} & = &
 \langle a_{n-1} a_{n-2} \cdots a_0\rangle_{2\text{'s}} + 1 
 = 
-2^{n-1}a_{n-1} + \mathlarger{\sum}_{\substack{i=0}}^{\substack{n-2}} 2^{i}a_{i} + 1 = \nonumber \\[0pt]
& & \hspace*{-17pt} = -2^{n-1}(-\bar{a}_{n-1} + 1) + \mathlarger{\sum}_{\substack{i=0}}^{\substack{n-2}} 2^{i}(-\bar{a}_{i} + 1) + 1 = \nonumber \\[0pt]
& & \hspace*{-17pt} = -(-2^{n-1}\bar{a}_{n-1} + 2^{n-1}) - \mathlarger{\sum}_{\substack{i=0}}^{\substack{n-2}} 2^{i}\bar{a}_{i}  +  \mathlarger{\sum}_{\substack{i=0}}^{\substack{n-2}} 2^{i} + 1 = \nonumber \\[0pt]
& & \hspace*{-17pt} = -(-2^{n-1}\bar{a}_{n-1} + \mathlarger{\sum}_{\substack{i=0}}^{\substack{n-2}} 2^{i}\bar{a}_{i}) - 2^{n-1} + \mathlarger{\sum}_{\substack{i=0}}^{\substack{n-2}} 2^{i} + 1 = \nonumber \\[0pt]
& & \hspace*{-17pt} = -(-2^{n-1}\bar{a}_{n-1} + \mathlarger{\sum}_{\substack{i=0}}^{\substack{n-2}} 2^{i}\bar{a}_{i}) = -\langle \bar{a}_{n-1} \bar{a}_{n-2} \cdots \bar{a}_0\rangle_{2\text{'s}} 
\label{eq_opA2}
\end{eqnarray}
\end{itemize}

Next, 
we encode $A^+$ using the expression 
$a'_{i} = a_{i} \oplus a_{0+}$. 
Namely, 
each bit of $A = \langle a_{n-1} a_{n-2} \cdots a_0\rangle_{2\text{'s}}$ is driven to a XOR gate 
along with the extra LSB $a_{0+}$
to form 
$A' = \langle a'_{n-1} a'_{n-2} \cdots  a'_0 \rangle_{2\text{'s}}$. 
Hence, 
$A^+$ is encoded as shown in Eq. \eqref{eq_optA}.

\vspace*{-12pt}

\begin{equation}
A^+ = (-1)^{a_{0+}} \cdot (-2^{n-1}a'_{n-1} +
\mathlarger{\sum}_{\substack{i=0}}^{\substack{n-2}} 2^{i}a'_{i}) 
 = (-1)^{a_{0+}} \cdot A', \;\;\;\;\;\;
 a'_{i} = a_{i} \oplus a_{0+}
\label{eq_optA} 
\end{equation}

This encoding is equivalent 
to the initial representation of $A^+$,
as verified by assigning either $0$ or $1$ to $a_{0+}$
in Eq. \eqref{eq_optA}:

\begin{itemize}
    \item for $a_{0+} = 0$, $A^+$ is equal to the expression of Eq. \eqref{eq_opA1}.
    \item for $a_{0+} = 1$, $A^+$ is equal to the expression of Eq. \eqref{eq_opA2}.
\end{itemize}

To perform the DLSB multiplication,
we use Eq. \eqref{eq_dlsbopB} for $B^{+}$
and Eq. \eqref{eq_optA} for $A^{+}$.
Their product is defined as shown in Eq. \eqref{eq_Prr}.

\begin{equation}
A^+ \times B^+ = 
(-1)^{a_{0+}} \cdot A' \cdot \mathlarger{\sum}_{\substack{j=0}}^{\substack{n\text{/}2-1}}  4^{j} b_{j}^{\text{\scriptsize\emph{MB}}}  =  A' \cdot \mathlarger{\sum}_{\substack{j=0}}^{\substack{n\text{/}2-1}} (-1)^{a_{0+}} \cdot 4^{j} b_{j}^{\text{\scriptsize\emph{MB}}} 
\label{eq_Prr} 
\end{equation}

The next step is to integrate the term $(-1)^{a_{0+}}$
in the calculation of $b_{j}^{\text{\scriptsize\emph{MB}}}$,
i.e., the Modified Booth digits,
which are calculated by Eq. \eqref{eq_mb}.
Considering that this term affects the sign
(depending on the value of $a_{0+}$),
we drive $s_j$,
i.e., the sign of the Modified Booth digits,
to a XOR gate along with $a_{0+}$.
As a result,
the absolute value
of the new Modified Booth digits,
labeled as
$b_{j}^{\text{{\scriptsize\emph{MB}}+}}$, 
is the same with that of 
$b_{j}^{\text{{\scriptsize\emph{MB}}}}$,
however, 
their sign depends on $s_j$ and $a_{0+}$.
In particular, 
the new expression for the sign is given by Eq. \eqref{eq_nsign}.

\vspace{-13pt}

\begin{equation}
    s'_{j} = s_{j} \oplus a_{0+}
    \label{eq_nsign}
\end{equation}

Based on the above analysis,
Eq. \eqref{eq_Prr} is transformed to Eq. \eqref{eq_Prr2},
and the calculation of the new Modified Booth digits is performed as shown in Eq. \eqref{eq_bthnew2}.


\begin{equation}
A^+ \times B^+ = A' \cdot \mathlarger{\sum}_{\substack{j=0}}^{\substack{n\text{/}2-1}}  4^{j} b_{j}^{\text{{\scriptsize\emph{MB}}+}}  \\[2pt]
\label{eq_Prr2} 
\end{equation}
\begin{equation}
\text{where }  \;\;
b_{j}^{\text{{\scriptsize\emph{MB}}+}} = 
(-1)^{s'_j} \cdot (2two_j + one_j)\\[3pt]
\label{eq_bthnew2}
\end{equation}

\begin{figure}[!t]
\vspace{-6pt}
\centering
\hspace*{-4pt}
\subfloat[\label{fig_dt1}]{\includegraphics[width=0.37\textwidth]{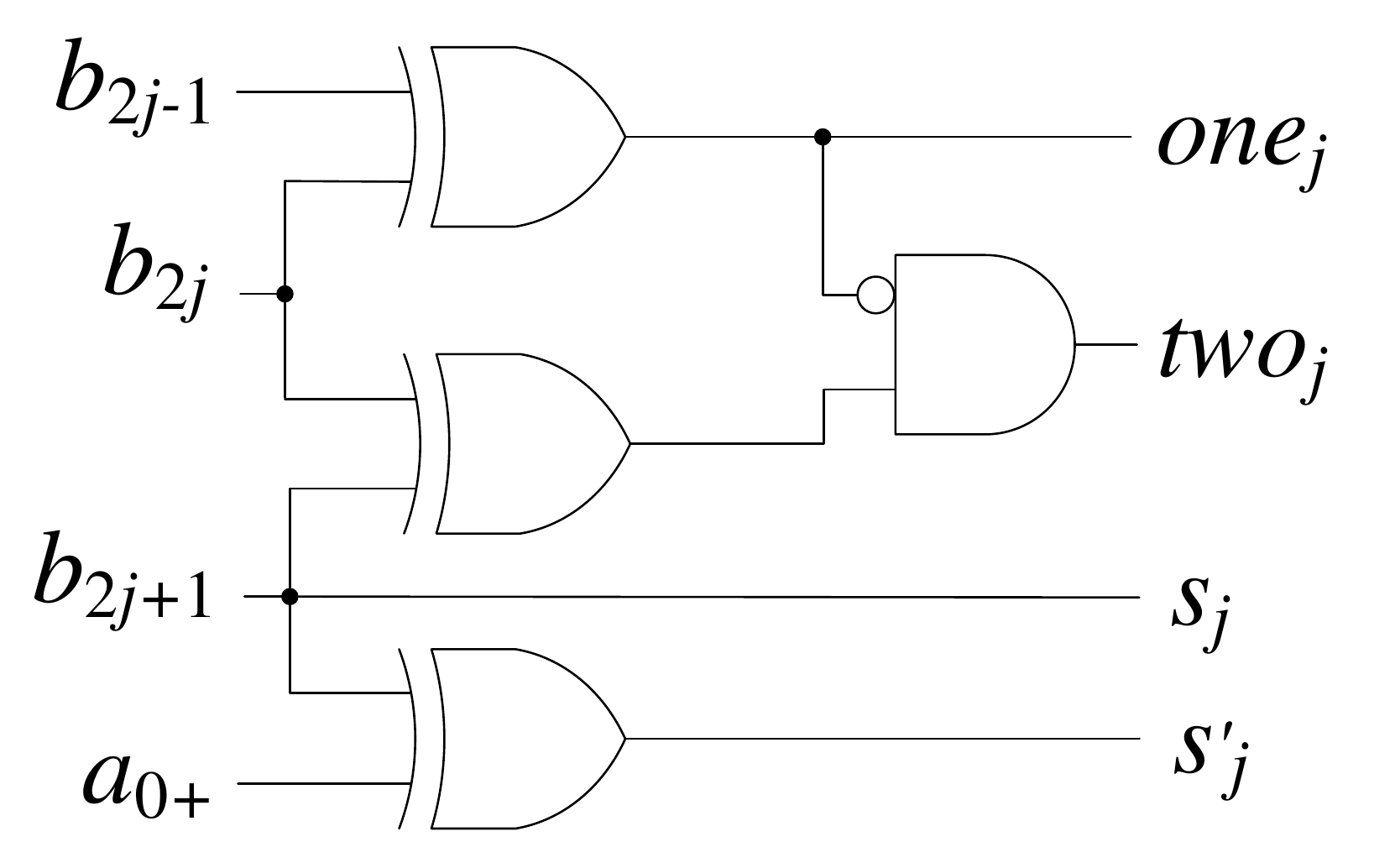}}%
\hspace*{-10pt}
\subfloat[\label{fig_dt2}]{\includegraphics[width=0.37\textwidth]{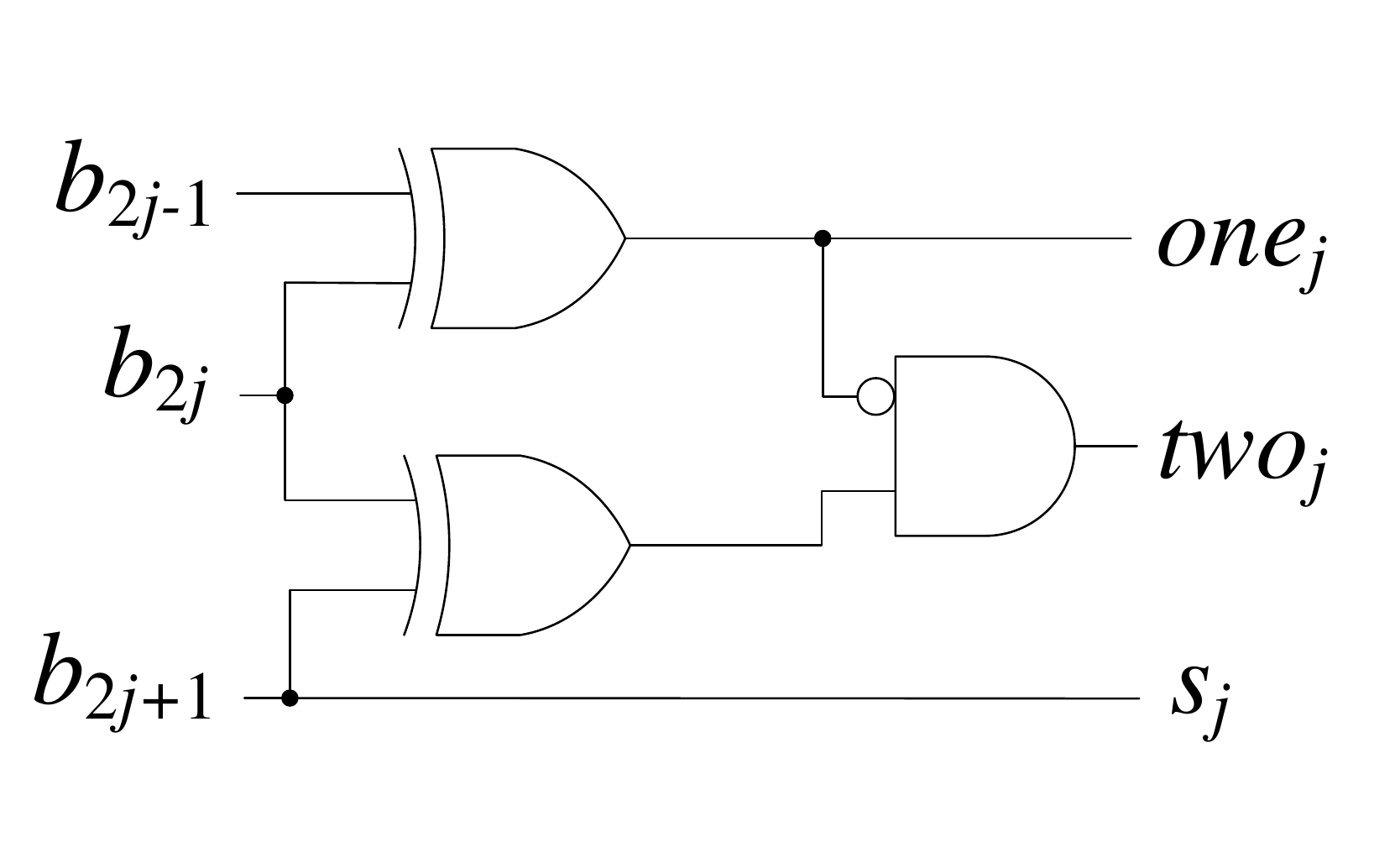}}%
\hspace*{-14pt}
\subfloat[\label{fig_dt3}]{\includegraphics[width=0.32\textwidth]{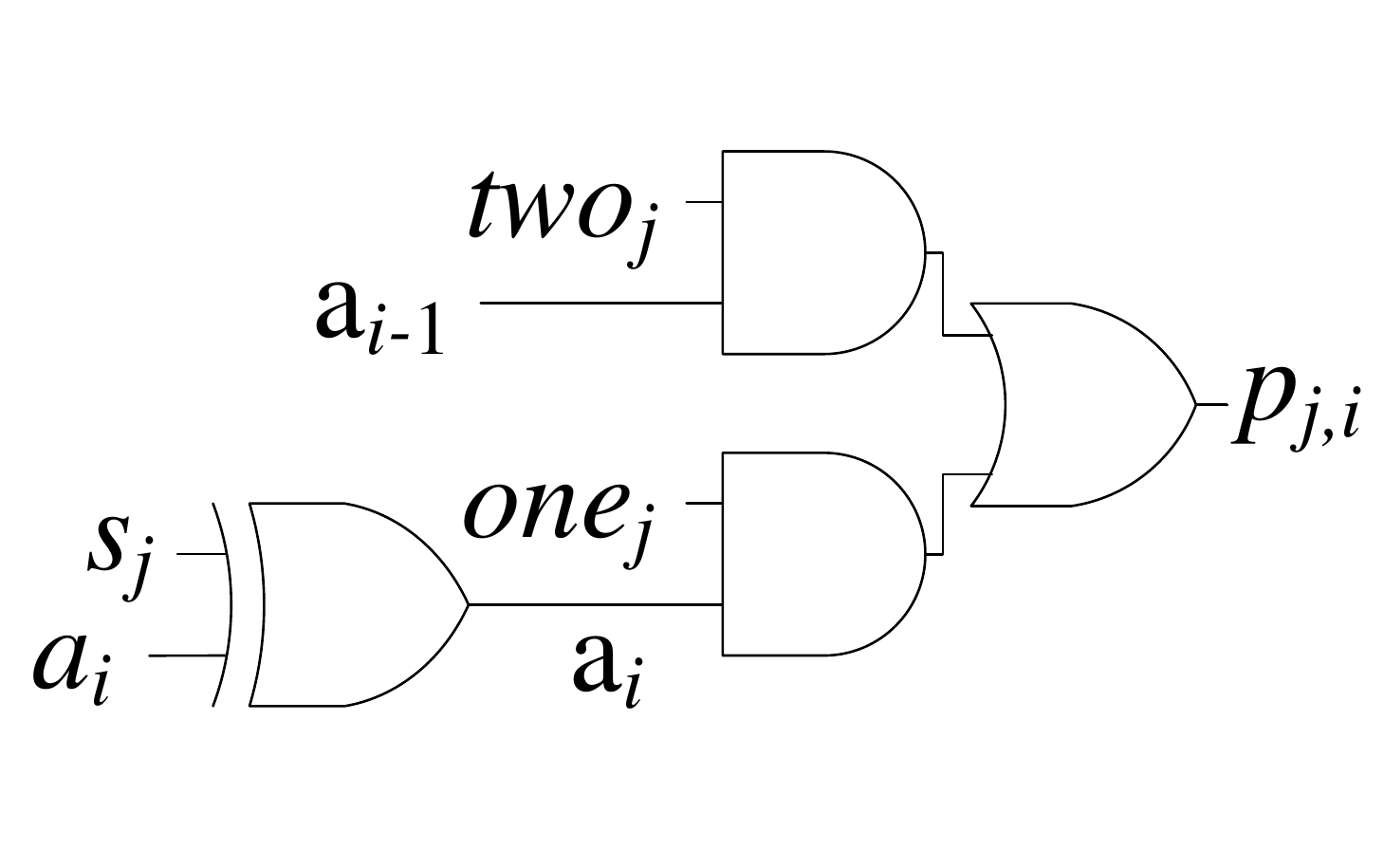}}%
\caption[Circuits of the DLSB Multipliers]{Circuits used in the DLSB multiplication:
\textbf{(a)} DLSB Modified Booth encoder
(used in the sophisticated approach),
\textbf{(b)} conventional Modified Booth encoder
(used in the straightforward approach), 
\textbf{(c)} conventional partial product generator 
($a_i$: $i$-bit of $A$, a$_i = $ $s_j \oplus a_i$, 
used in both approaches).}
\label{fig_tech}
\end{figure}

The logic equations of the encoding signals
$s_j$, $one_j$, and $two_j$
are derived from Table \ref{tb_bth}. 
The proposed encoding circuit
is illustrated in Figure \ref{fig_dt1}.
Compared to the conventional Modified Booth encoder that is illustrated in Figure \ref{fig_dt2},
the proposed encoder has an extra
XOR-2 gate for calculating $s'_{j}$,
which includes the extra LSB of $A^+$.

For the calculation of the product defined in \eqref{eq_Prr2}, 
the generation of $n/2$ partial products $P\!P_j$ is required,
as shown in Eq. \eqref{eq_finprod} and \eqref{eq_finprod2}. 

\begin{equation}
A^+ \times B^+ = A' \cdot \mathlarger{\sum}_{\substack{j=0}}^{\substack{n\text{/}2-1}}  4^{j} b_{j}^{\text{{\scriptsize\emph{MB}}+}}  = \\
\mathlarger{\sum}_{\substack{j=0}}^{\substack{n\text{/}2-1}}  4^{j} P\!P_j 
\label{eq_finprod}
\end{equation}
\begin{equation}
\text{where }  \;\;
    P\!P_j = A' \cdot b_{j}^{\text{{\scriptsize\emph{MB}}+}} = 2^n\,\overline{p}_{j,n} + \mathlarger{\sum}_{\substack{i=0}}^{\substack{n-1}} 2^i p_{j,i}, \;\; 
  j = 0, 1, \cdots n/2-1
  \label{eq_finprod2}
\end{equation}

The generation of the $i$-th bit of the partial product $P\!P_j$ is illustrated at gate level in Figure \ref{fig_dt3}. 
The circuit of the DLSB partial product generator
is the same with that employed in the conventional Modified Booth algorithm.
To calculate the Most Significant Bit (MSB) of each partial product, 
we consider $a'_n = a'_{n-1}$, 
and thus, 
$a_n = a_{n-1}$. 
Their LSB is calculated considering $a'_{-1} = 0$, and thus, 
$a_{-1} = a_{0+}$. 
In comparison with the conventional Modified Booth multiplier, 
the generation of the partial product bits is intact, 
except for the LSBs, 
where $s'_{j} = s_{j} \oplus a_{0+}$ is used instead of $s_j$.

The generated partial products are accumulated, 
properly weighted, 
using a Wallace tree \cite{Weste2010} along with the constant terms ('$1$'s) and the correction terms ($c'_{j}$).
Finally, 
the carry-save output of the Wallace tree is driven to a fast prefix adder \cite{Weste2010} to form the final result. 
Compared to the conventional Modified Booth multiplier, 
again the only difference is the use of
$s'_{j}$ instead of $s_j$ in the calculation of the correction terms,
as shown in its logic function in Eq. \eqref{eq_corr}.

\vspace{-9pt}

\begin{equation}
c'_{j} = s'_{j} \cdot (one_j + two_j)  =
(s_{j} \oplus a_{0+}) \cdot (one_j + two_j)
\label{eq_corr}
\end{equation}

\vspace{-3pt}

Overall, 
the proposed DLSB multiplier incorporates the typical stages of the Modified Booth multiplication circuit: 
(i) encoding, 
(ii) partial product generation, 
(iii) partial product accumulation,
and
(iv) final addition. 
As shown, 
with careful design and bit-level manipulations
we reduce the circuit overheads of the straightforward approach: 
the product $a_{0+} \cdot B^+$ of Eq. \eqref{eq_dlsbold} is eliminated 
and the only overhead regards the encoding stage, 
where an extra signal ($s'_{j}$) is generated.

\begin{figure}[!t]
\centering
\subfloat[\label{fig_dtr1}]{\includegraphics[width=0.49\textwidth]{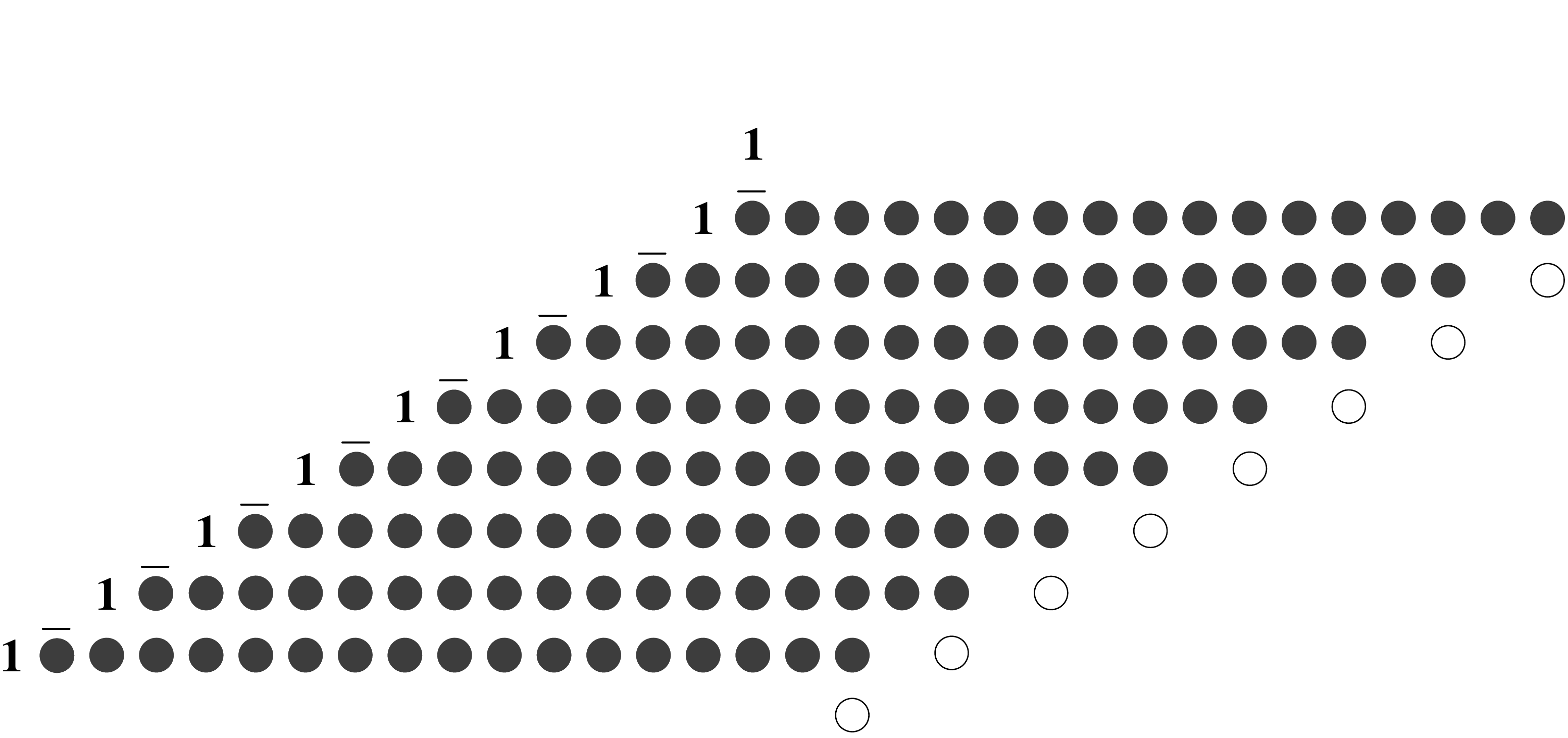}}\\[-6pt]
\subfloat[\label{fig_dtr2}]{\includegraphics[width=0.49\textwidth]{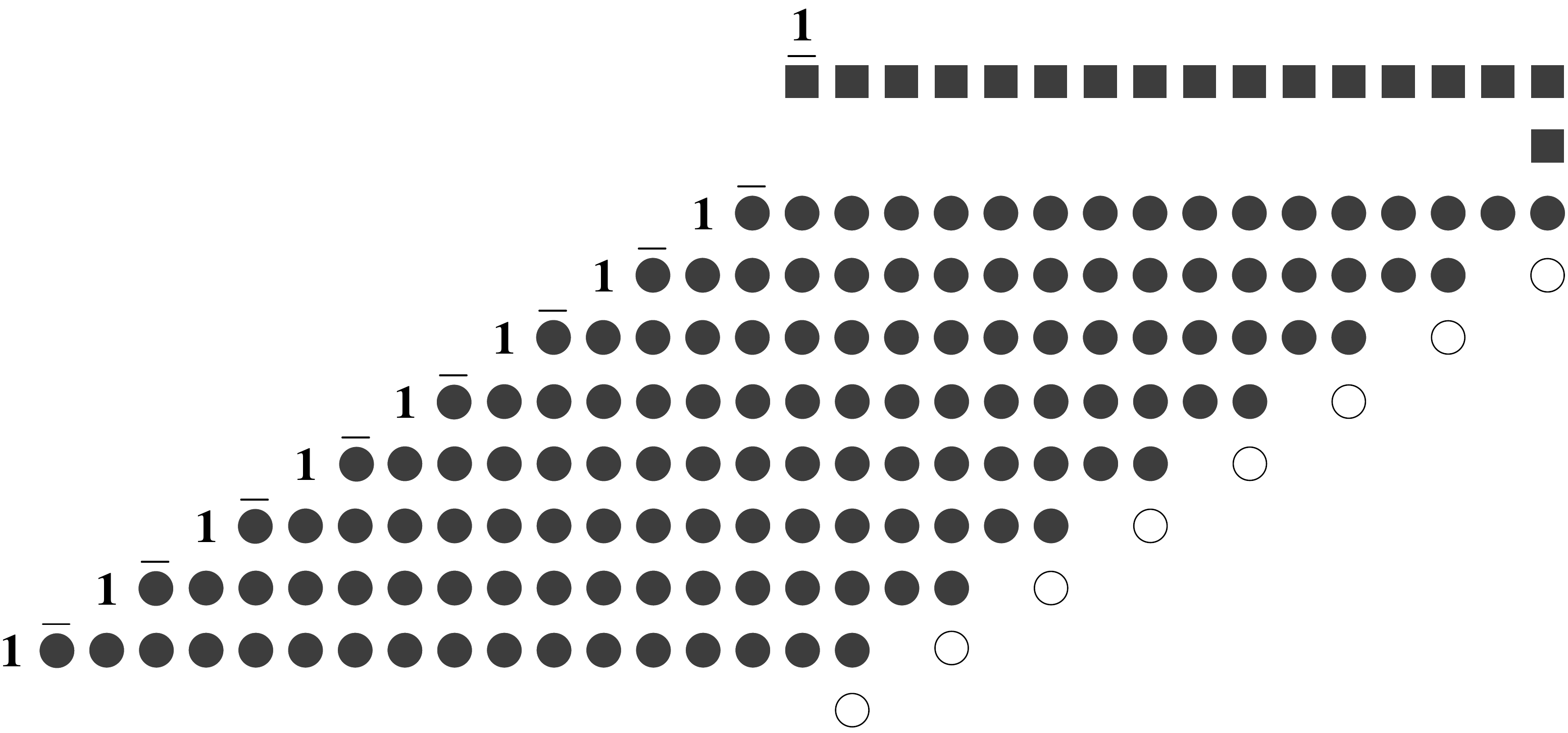}}%
\subfloat[\label{fig_dtr3}]{\includegraphics[width=0.49\textwidth]{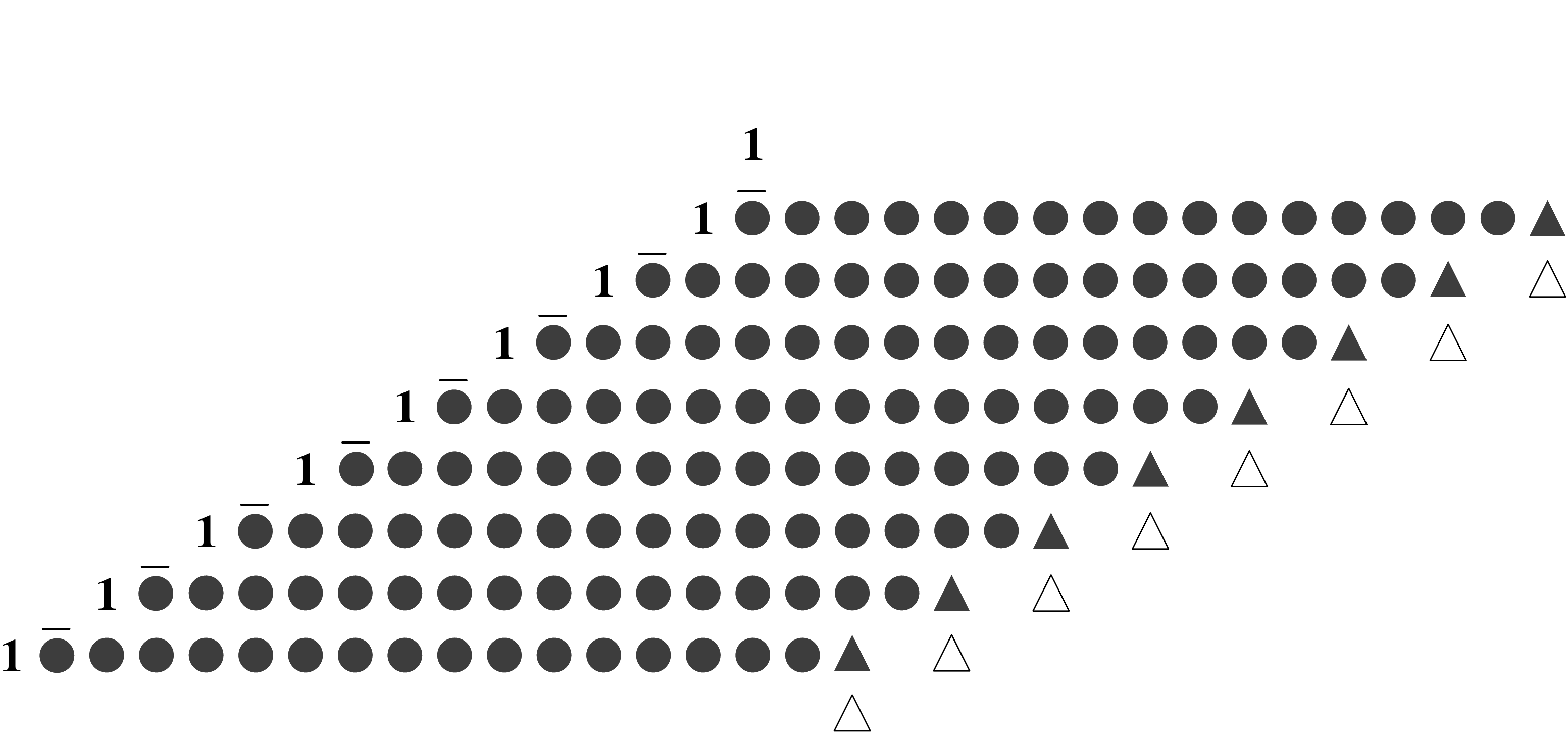}}%
\caption[Partial Product Matrices of the Conventional and DLSB Multipliers]{Partial product matrices of $16$-bit multipliers:
\textbf{(a)} conventional Modified Booth multiplier
(ordinary $2$'s-complement numbers),
\textbf{(b)} straightforward DLSB Modified Booth multiplier 
of Eq. \eqref{eq_dlsbold},
\textbf{(c)} sophisticated DLSB Modified Booth multiplier
of Eq. \eqref{eq_finprod}.\\
Symbols: 
\protect\tikz \protect\fill[blue_dlsb] (1ex,1ex) 
circle (0.63ex);: $p_{j,i}$ \hspace{2pt} 
\protect\tikz \protect\filldraw[fill=white, draw=black] (1ex,1ex) 
circle (0.5ex);: $c_j$ \hspace{2pt}
\protect\tikz \protect\fill[blue_dlsb] (0.77ex,0.77ex) rectangle (0.22,0.22);: $a_{0+} \cdot b_i$ \hspace{2pt}
\protect\tikz \protect\fill[blue_dlsb] (0,0) -- (0.75ex,1.4ex) -- (1.5ex,0) -- (0,0);: $p'_{j,0}$ \hspace{2pt}
\protect\tikz \protect\filldraw[fill=white, draw=black] (0,0) -- (0.65ex,1.25ex) -- (1.2ex,0) -- (0,0);: $c'_j$} 
\label{fig_dtrees}
\end{figure}

For comparison purposes, 
Figure \ref{fig_dtrees} illustrates 
the partial product matrices 
of the $16$-bit examined multipliers,
i.e., 
the conventional Modified Booth multiplier (Figure \ref{fig_dtr1}),
the straightforward DLSB Modified Booth multiplier (Figure \ref{fig_dtr2}),
and
the sophisticated DLSB Modified Booth multiplier (Figure \ref{fig_dtr3}).
All the necessary bits 
(partial product bits, constant terms, and correction terms)
that are required for the partial product accumulation are included. 
As shown,
the straightforward DLSB multiplier 
implements the conventional multiplier plus extra logic
for the product $a_{0+} \cdot B^+$. 
In contrast,
the sophisticated design does not implement this logic, 
it has the same depth of accumulation tree with the conventional multiplier,
and its only difference compared to the latter
is the calculation of $s'_j$.
This signal 
is used instead of $s_j$
for generating the LSB of the partial products ($p'_{j,0}$)
and the correction terms ($c'_j$). 

\section{Evaluation}
\label{s3_4}

In this section,
firstly, 
we evaluate the DLSB multipliers,
and then,
we demonstrate how they can be used to improve the hardware efficiency of large-size multiplications.
We conduct both theoretical 
and experimental analysis.
We consider the following notations for the examined multipliers:
\begin{itemize}
    \item CMB: conventional Modified Booth multiplier for ordinary $2$'s-complement numbers (Figure \ref{fig_dtr1}).
    \item DLSB1: straightforward Modified Booth multiplier for  $2$'s-complement DLSB numbers (Figure \ref{fig_dtr2}).
    \item DLSB2: sophisticated Modified Booth multiplier for $2$'s-complement DLSB numbers (Figure \ref{fig_dtr3}).
\end{itemize}

\subsection{Theoretical Analysis}

As already discussed in the previous section, 
DLSB1 implements CMB 
and requires additional logic ($n + 1$ AND-2 gates) 
for calculating the extra product, 
as well as considerable overhead  
for accumulating it with the rest partial products.
On the other hand,
DLSB2 implements CMB
with a small overhead of $n/2$ XOR-2 gates for calculating $s'_j$.
Namely,
it uses the encoder of Figure \ref{fig_dt1}
instead of that of Figure \ref{fig_dt2}.
Subsequently, we evaluate the area complexity of the DLSB multipliers based on the unit gate model used in \cite{Tsoumanis2014}. 
According to this model, the primitive gates AND-2 and OR-2 are equal to $1$ unit gate, the NOT gate counts as $0.5$ unit gates, and the XOR-2 gate is equal to $2$ unit gates. 
Moreover, the area of a full adder and a half adder is 
$7$ and $3$ unit gates, respectively. 
The gate equivalence of the main components used in the partial product generation and accumulation of the multipliers 
is summarized in Table \ref{tb_ug}.
In the next paragraphs,
we examine separately 
the stages of
the partial product generation
and accumulation,
and then we quantify the resources of the entire multiplication architectures. 

Regarding the partial product generation, 
the $n$-bit CMB multiplier generates 
$n/2$ $(n+1)$-bit partial products,
Therefore, 
it requires 
$n/2$ MB encoders of Figure \ref{fig_dt2}, 
$n/2$ $\!\times\!$ $(n+1)$ 1-bit partial product generators of Figure \ref{fig_dt3}, 
and $n/2$ correction terms generators (signals $c_{j}$). 
Additionally, 
$n/2$ NOT gates are needed for inverting the MSB of each partial product.
DLSB1 requires the same resources,
and also employs 
$n+1$ AND-2 and $1$ NOT gates for calculating the product 
$a_{0+} \cdot B^+$.
Compared to CMB,
DLSB2 uses only a different encoder, 
i.e., the DLSB encoder of Figure \ref{fig_dt1}. 

Next, 
we analyze the resources needed for the partial product accumulation.
CMB and DLSB2 accumulate
$n/2 + 1$ $n$-bit vectors,
i.e., 
$n/2$ partial products and $1$ vector containing the constant and correction terms. 
This multi-operand addition is performed in carry-save form \cite{Weste2010}, 
thus, 
$(n/2-1)$ $\times$ $n$ full adders 
are required in total.
In DLSB1, one more number must be added, 
and thus, $n/2$ $\times$ $n$ full adders are required.
Finally, in all multipliers, the $2n$-bit carry-save output of the Wallace tree is driven to a fast adder \cite{Weste2010}, 
which consists of 
$2n$ half adders, 
$n \log_2 2n$ propagate group circuits (each one is $3$ unit gates), 
and $2n$ XOR-2 gates.

Based on our theoretical analysis about the resources of each design,
Table \ref{tb_overh} reports 
the overhead in unit gates of the DLSB multipliers 
for different bit-widths, 
i.e., $n=8,16,32$. 
This overhead is calculated 
by comparing the unit gates of the DLSB multipliers 
with those of CMB.
As expected,
DLSB1,
which is based on the straightforward design approach,
exhibits larger overhead compared to 
DLSB2.
On the other hand,
the overhead of 
DLSB2 is negligible,
lying in the range $0.5\%$--$1.4\%$. 
The unit gate model is a simplified model, 
but it gives a rough estimation about the circuit area. 
In the next section,
we provide experimental analysis based on industrial-strength synthesis of the circuits, 
in which it is shown that theoretical outcomes and trends 
follow in a fine manner the experimental results.

\begin{table}[!t]
\fontsize{9}{10}\selectfont
\renewcommand{\arraystretch}{1.2}
\setlength{\tabcolsep}{4pt}
\caption[Components of the Conventional and DLSB Multipliers]{Unit gates per component
and components of the $n$-bit multipliers.}
\label{tb_ug}
\centering
\begin{tabular}{l|c|ccc}
\hline
\textbf{Component}  & 
\textbf{Unit Gates} & 
\textbf{\# in CMB}  & 
\textbf{\# in DLSB1} & 
\textbf{\# in DLSB2} \\
\hline \hline
MB Encoder              & $5.5$ & $n/2$ & $n/2$ & --\\
DLSB MB Encoder         & $7.5$ & --    & --    & $n/2$ \\ 
MB PP Generator & $5$   & $n/2$ $\!\times\!$ $(n+1)$                 & $n/2$ $\!\times\!$ $(n+1)$ & $n/2$ $\!\times\!$ $(n+1)$ \\
AND PP Generator         & $1$ & --    & $n+1$    & -- \\ 
Corr. Term Generator    & $2$   &  $n/2$ & $n/2$ & $n/2$\\
Full Adder & $7$ & $(n/2-1)$ $\!\times\!$ $n$ & $n/2$ $\!\times\!$ $n$ & $(n/2-1)$ $\!\times\!$ $n$ \\
Half Adder & $3$ & $2n$ & $2n$ & $2n$ \\
Propagate Group& $3$ & $n \log_2 2n$ & $n \log_2 2n$ & $n \log_2 2n$ \\
XOR-2 Gate & $2$ & $2n$ & $2n$ & $2n$  \\
\hline
\end{tabular}
\end{table}

\begin{table}[!t]
\fontsize{9}{10}\selectfont
\renewcommand{\arraystretch}{1.2}
\setlength{\tabcolsep}{17pt}
\caption[Unit Gate Overhead of the DLSB Multipliers]{Unit gate overhead of the $n$-bit DLSB multipliers.}
\label{tb_overh}
\centering
\begin{tabular}{l|ccc}
\hline 
\textbf{Multiplier} & \textbf{\emph{n $\mathbf{=8}$}} & \textbf{\emph{n $\mathbf{=16}$}} & \textbf{\emph{n $\mathbf{=32}$}}\\
 \hline \hline 
DLSB1 & $11.8 \%$ & $6.7 \%$ & $3.7 \%$ \\ 
DLSB2 & $1.4 \%$ & $0.8 \%$ & $0.5 \%$ \\
\hline
\end{tabular}
\end{table}

\subsection{Experimental Results} 

This section evaluates the multiplication circuits 
in terms of delay, area, and energy consumption, 
using industrial-strength tools. 
All the multipliers
are implemented in Verilog 
and synthesized with the Synopsys Design Compiler 
and the TSMC 45-nm standard-cell library.
The simulations are performed with Mentor Graphics QuestaSim.
Both synthesis and simulation are performed at $1$V, i.e., the nominal supply voltage.
Moreover, the designs are configured 
to operate at their critical path delay,
i.e., at maximum frequency. 
The critical path delay and the area are reported by Synopsys Design Compiler, 
while the power consumption is measured with Synopsys PrimeTime.
The energy consumption is also calculated by the product of delay and power. 

Table \ref{tb_dlsbres} 
presents the experimental results 
from the synthesis of the multipliers for various sizes, i.e., $n=8,16,32$.
Furthermore, the energy and area overheads compared to CMB are reported.
In terms of critical paths, 
as expected, 
the DLSB multipliers deliver almost the same delays with CMB.
DLSB2 exhibits similar delay even for the smallest bit-width,
i.e., $0.38$ns versus $0.37$ns. 
This is justified by the fact that
the depth of DLSB2's partial product tree has not increased. 
In comparison with DLSB1, 
the sophisticated DLSB2 design
achieves higher efficiency for all the resources. 
On average, DLSB2 imposes small area and energy overheads 
(i.e., $3.1\%$ and $3.3\%$, respectively), 
while the corresponding overheads for 
DLSB1 are $11.9\%$ and $15.4\%$.  
Regarding the bit-width scaling, 
the derived results show that as the multiplier's size increases, 
the impact of the extra LSB is becoming smaller in both area and energy. 
This observation is normal,
as by increasing the bit-width, 
the rate of the overheads induced by the DLSB
increases slower 
than the rate of the area added due to the bit-width scaling.

\begin{table}[!t]
\fontsize{9}{10}\selectfont
\renewcommand{\arraystretch}{1.2}
\setlength{\tabcolsep}{12pt}
\caption[Synthesis Results of the DLSB Multipliers on TSMC 45-nm Standard-Cell]{Synthesis results of the $n$-bit DLSB multipliers on TSMC 45-nm standard-cell.}
\label{tb_dlsbres}
\centering
\begin{threeparttable}
\begin{tabular}{cl|c|cc|cc}
\hline 
\multicolumn{2}{c|}{\multirow{2}{*}{\textbf{Design}}} &
\textbf{Delay}      & 
\multicolumn{2}{c|}{\textbf{Area}} & 
\multicolumn{2}{c}{\textbf{Energy}} \\[-2pt]
& & 
(ns) & 
($\upmu$m$^2$) & 
($\%$)\footnotemark\setcounter{footnote}{0} & 
($\upmu$W$\cdot$ns) & 
($\%$)\footnotemark\setcounter{footnote}{0} \\ 
\hline \hline 
\parbox[t]{8mm}{\multirow{3}{*}{\rotatebox[origin=c]{45}{\textbf{\emph{n $\mathbf{=8}$}}}}} & CMB   & $0.37$ & $485$  & $-$    & $384$  & $-$ \\
& DLSB1 & $0.40$ & $572$  & $+18$  & $480$  & $+25$ \\
& DLSB2 & $0.38$ & $510$  & $+5.2$ & $406$  & $+5.7$ \\
\hline 
\parbox[t]{8mm}{\multirow{3}{*}{\rotatebox[origin=c]{45}{\textbf{\emph{n $\mathbf{=16}$}}}}} & CMB   & $0.51$ & $1519$ & $-$    & $1186$ & $-$ \\
& DLSB1 & $0.52$ & $1701$ & $+12$  & $1344$ & $+13.3$ \\
& DLSB2 & $0.51$ & $1562$ & $+2.8$ & $1217$ & $+2.6$ \\
\hline 
\parbox[t]{8mm}{\multirow{3}{*}{\rotatebox[origin=c]{45}{\textbf{\emph{n $\mathbf{=32}$}}}}} & CMB   & $0.67$ & $5189$ & $-$    & $4340$ & $-$ \\
& DLSB1 & $0.67$ & $5491$ & $+5.8$ & $4685$ & $+7.9$ \\
& DLSB2 & $0.67$ & $5256$ & $+1.3$ & $4412$ & $+1.7$ \\
\hline
\end{tabular}
\begin{tablenotes}
   \item[1]{\fontsize{7.7}{8.8}\selectfont Refers to $\%$ area/energy overhead (relative increase) in comparison with CMB.}
   \end{tablenotes}
 \end{threeparttable}
\end{table}

\subsection{Case Study: DLSB for Large-Size Multiplication}

In this section, 
we demonstrate how the sophisticated DLSB multiplier 
can be effectively used 
as building block 
for implementing multiplications
with large input operands.
For the rest of the analysis, 
we consider $2$'s-complement arithmetic  
and small fixed-size multiplication circuits, 
i.e., 
there is the constraint of hardwired arithmetic units 
with specific small bit-width, 
like in embedded devices \cite{Dinechin2009}. 

We follow the well-established approach of decomposing the large bit-width operands to smaller sizes \cite{Dinechin2009}. 
The reduction of the operand bit-width
to match the desired bit-width (imposed by the constraint)
is a recursive procedure.
Nevertheless, 
to ease the description and without loss of generality, 
we focus our analysis on the first recursion, 
assuming that the bit-width constraint 
exposed by the available arithmetic unit is satisfied. 
Next, we discuss how the large-size multiplications are partitioned
in the conventional arithmetic,
and then, 
we present a more efficient DLSB-based partition.

\textbf{Conventional $\mathbf{2}$'s-Complement Arithmetic:}
Let $A$ and $B$ be two $k$-bit $2$'s-complement numbers to multiply. 
Without loss of generality, 
we consider $k = 2n$, 
where $n$ is an integer, 
and we split $A$ and $B$ in two $n$-bit words,
as shown in Eq. \eqref{eq_dlsbpart1}.
The product $A \times B$ is formed as shown in Eq. \eqref{eq_dlsbpart12}.

\vspace*{-4pt}

\begin{equation}
\hspace*{-25pt}
 A =
\langle a_{2n-1} a_{2n-2} \cdots a_{n} \rangle_{2\text{'s}} 
\cdot 2^n 
+ 
\langle a_{n-1} a_{n-2} \cdots a_{0} \rangle_{\text{u}} 
= A_s \cdot 2^n + A_u \nonumber \\[2pt]
\end{equation}
\begin{equation}
\hspace*{-5.5pt} B =
\langle b_{2n-1} b_{2n-2} \cdots b_{n} \rangle_{2\text{'s}} 
\cdot 2^n 
+ 
\langle b_{n-1} b_{n-2} \cdots b_{0} \rangle_{\text{u}} 
= B_s \cdot 2^n + B_u 
\label{eq_dlsbpart1}
\end{equation}
\begin{align}
A \times B & =  (A_s \cdot 2^n + A_u)\cdot(B_s \cdot 2^n + B_u) = \nonumber\\
 & 
= A_s \cdot B_s \cdot 2^{2n} + (A_s \cdot B_u + A_u \cdot B_s)\cdot 2^{n} + A_u \cdot B_u  
  \label{eq_dlsbpart12} 
\end{align}

The above product requires the execution of $4$ multiplications:
(i)   $2$'s-complement $\times$ $2$'s-complement ($A_s \cdot B_s$),
(ii)  unsigned $\times$ unsigned                 ($A_u \cdot B_u$),
(iii) unsigned $\times$ $2$'s-complement         ($A_u \cdot B_s$),
(iv)  $2$'s-complement $\times$ unsigned         ($A_s \cdot B_u$).
We note that the multiplications involving unsigned numbers
must employ an extra bit for the sign extension.
As a result,
to cover all the bit-width cases 
and calculate the $4$ sub-products
with the same module,
the $(n+1)$-bit CMB multiplier needs to be used.

\textbf{$\mathbf{2}$'s-Complement DLSB Arithmetic:} 
We propose an alternative operand partition that 
aims to create DLSB-like bit vectors.
As shown in Eq. \eqref{eq_dlsbpart123a}--\eqref{eq_dlsbpart123b}, 
we divide the operands in two $2$'s-complement segments,
and attach the MSB of their least significant part,
i.e., $a_{n-1}$ and $b_{n-1}$, 
as an extra LSB in their most significant part.
Assuming an extra LSB equal to $0$ for the least significant parts, 
the $k$-bit operands are now formed as 
functions of two $n$-bit $2$'s-complement DLSB numbers. 

\vspace*{-19pt}

\begin{align}
A & = \big(\langle a_{2n-1} a_{2n-2} \cdots a_{n} \rangle_{2\text{'s}} + a_{n-1} \big) 
\cdot 2^n + \big(\langle a_{n-1} a_{n-2} \cdots a_{0} \rangle_{2\text{'s}} + 0\big) = \nonumber\\
 & 
= A_1^+ \cdot 2^n + A_2^+  
  \label{eq_dlsbpart123a} 
\end{align}
\begin{align}
 B & = \big(\langle b_{2n-1} b_{2n-2} \cdots b_{n} \rangle_{2\text{'s}} + b_{n-1} \big) 
\cdot 2^n + \big(\langle b_{n-1} b_{n-2} \cdots b_{0} \rangle_{2\text{'s}} + 0\big) = \nonumber\\
 & 
 = B_1^+ \cdot 2^n + B_2^+ 
\label{eq_dlsbpart123b} 
\end{align}

\begin{figure}[!t]
\centering
\includegraphics[width=0.805\textwidth]{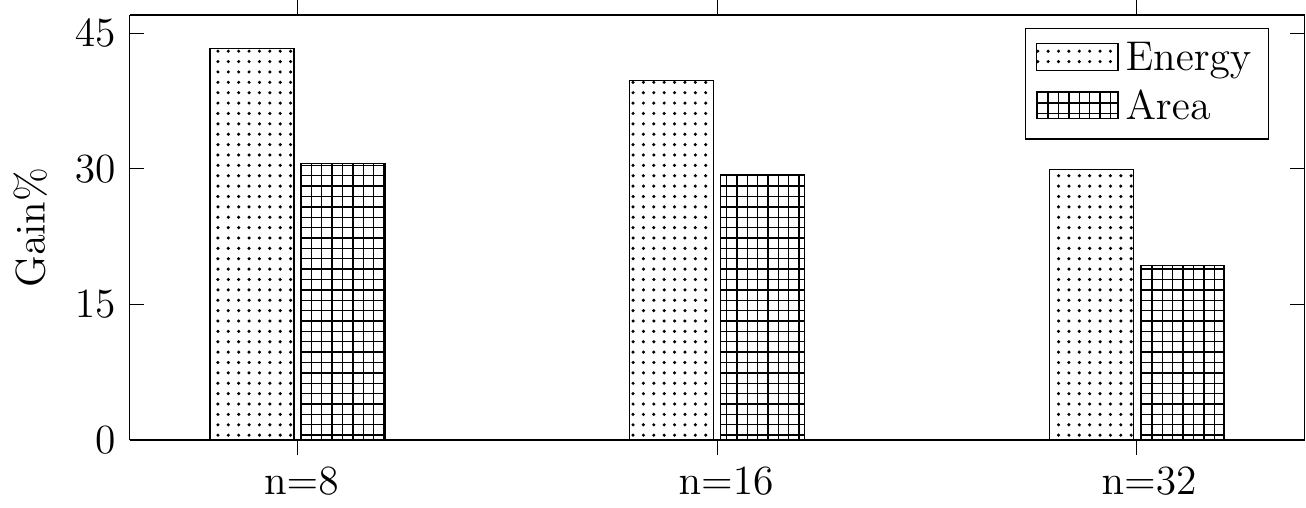}%
\caption[Resource Gains of the DLSB Multiplier]{Energy and area gains provided by $n$-bit DLSB2 compared to $(n + 1)$-bit CMB (targeting the use of DLSB2 instead of CMB as building block in large-size multiplications).}%
\label{fig_dlsbgains}
\end{figure}

Considering the proposed DLSB-based partition,
the $4$ sub-products, 
derived by applying again the distributive property, 
can be calculated by the $n$-bit $2$'s-complement DLSB multiplier,
which can effectively replace the $(n + 1)$-bit CMB as building block.
To evaluate the efficiency of the proposed scheme, 
we compare the $n$-bit DLSB2 with the $(n + 1)$-bit CMB for various bit-widths. 
The experimental setup is the same as described in the previous section.
Figure \ref{fig_dlsbgains} illustrates the energy and area gains achieved by the proposed DLSB2 compared to CMB.
In particular,
DLSB2 provides significant gains, especially for small bit-width (i.e., $30.6 \%$ and $43.3 \%$ area and energy gains, respectively, for $n = 8$).
The inefficiency of CMB is justified by the extra bit in the multiplier's bit-width,
which results in the generation of one more partial product
and the addition of an extra bit in each partial product.

\section{Conclusion}
\label{s3_5}

The goal of this chapter was to highlight
the benefits of using novel numerical formats
and applying low-level arithmetic optimizations and sophisticated bit-level manipulations.
In particular,
we adopted the DLSB arithmetic,
which 
assumes an extra LSB in the number representation 
and provides 
advantageous features, 
such as the representation symmetry
and
the simplification of rounding in floating-point operations. 
Targeting to decrease the overheads induced by the extra LSBs,
we introduced a new and effective algorithm 
for multiplying two $2$'s-complement DLSB numbers. 
Through an alternative representation, 
we optimized the Modified Booth multiplication scheme
to decrease the overheads of the straightforward DLSB multiplication.
According to our experimental evaluation for different multiplication bit-widths,
our DLSB design provides 
average overheads of \raisebox{0.8pt}{$\scriptstyle\sim$}$3\%$
in area and energy.
In contrast,
the DLSB design that  
does not adopt our sophisticated low-level optimizations,
suffers from 
average overheads of \raisebox{0.8pt}{$\scriptstyle\sim$}$12\%$
and \raisebox{0.8pt}{$\scriptstyle\sim$}$15\%$
in area and energy,
respectively.
Finally, 
we demonstrated how the proposed DLSB multiplier can be efficiently used as building block in
large-size multiplications, replacing the conventional multiplier.
In this case, 
both the area and energy consumption are significantly improved,
considering that the DLSB multiplier outperforms the conventional design
that is needed to implement the sub-products, 
by up to $30.6\%$ and $43.3\%$ in area and energy,
respectively. 
\chapter{Arithmetic Approximation: Hybrid~High-Radix~Encoding}
\label{chapter4}

\addtocontents{lof}{\protect\contentsline{chapter}{\protect\numberline{4}Arithmetic Approximation: Hybrid High-Radix Encoding}{}{}}
\addtocontents{lot}{\protect\contentsline{chapter}{\protect\numberline{4}Arithmetic Approximation: Hybrid High-Radix Encoding}{}{}}

\begin{ChapterAbstract}
Approximate Computing forms a design alternative that exploits the intrinsic error resilience of various applications and produces energy-efficient circuits with small accuracy loss.
In this chapter, 
we examine the impact of
applying
low-level optimizations and disciplined approximations
in the design of arithmetic circuits. 
Towards this direction,
we propose an approximate hybrid high-radix encoding 
for generating the energy-efficient RAD multipliers.
The proposed encoding scheme
approximately encodes one of the operands,
using 
the accurate radix-$\mathit{4}$ encoding 
for its most significant part
and an approximate higher radix encoding 
for its least significant part.
In the high-radix encoding,
the approximations are applied 
by mapping all the high-radix values to a set of values
that includes only the $\mathit{4}$ largest powers of two.
The proposed hybrid encoding is configurable 
and can be tuned to achieve the desired energy--accuracy trade-off.
Another important feature of the proposed design
is that the error induced by the approximations
is bounded by a Gaussian distribution with near-zero average.
In addition, 
the mean relative error of the multiplication 
depends only on the approximately encoded operand,
and thus,
the calculation of the corresponding error metrics 
is accelerated,
eliminating the need for exhaustive hardware simulations.
In terms of resources,
the proposed multipliers deliver up to $\mathit{56}$\% energy and $\mathit{55}$\% area gains
compared to the accurate radix-$\mathit{4}$ multiplier, 
when operating at the same frequency.
Moreover, 
our designs outperform state-of-the-art multipliers 
by up to $\mathit{40}$\% in energy consumption for similar error values.
Finally, 
we examine the scalability of our technique,
showing that 
as the multiplier's size increases, 
our designs
achieve larger gains in energy consumption and critical path delay,
i.e., up to $\mathit{64}$\% and $\mathit{22}$\%, respectively, 
while the error remains constant. \\
This chapter is based on our  \textbf{publication} in \textbf{\cite{LeonTVLSI}}.
\end{ChapterAbstract}

\newpage

\section{Introduction}

In modern embedded systems and data centers, 
power efficiency and performance are critical design concerns.
Considering that various application domains 
exhibit an intrinsic error resilience 
(e.g., digital signal processing,
data analytics, and data mining \cite{ChakradharDAC2010, ChippaDAC2013}),   
Approximate Computing 
\cite{2016_Mittal_ACMsrv, 2016_Xu_IEEEdt, 2021_Stanley_ACMsrv} 
appears as an effective solution to provide remarkable power gains and/or speed improvements.
In Approximate Computing, 
error is viewed as a commodity 
that can be traded for significant gains in
resources 
(e.g., area, power, energy, latency, or throughput) \cite{LingamneniCICC2014}, 
and thus, 
it composes a promising design paradigm
to generate power-efficient systems and circuits.
In particular, 
Approximate Computing exploits the innate error tolerance of the applications 
and deliberately relaxes the correctness of some computations
to decrease their power consumption 
and/or accelerate their execution.

To take advantage of the benefits
provided by Approximate Computing, 
massive research has been conducted in the field of approximate hardware and circuits.
In this field, 
automated synthesis techniques \cite{ShinDATE2010} 
and hardware description language annotations \cite{2015_Yazdanbakhsh_DATE} have been proposed to facilitate  
the design of approximate hardware. 
Significant research has also been reported in approximate processors that employ quality programmable vectors \cite{Venkataramani}, neural networks \cite{Esmaeilzadeh} and approximate custom instructions \cite{Kamal2014}.
Circuit-level approximations involve voltage over-scaling 
\cite{2010_Kurdahi_IEEEtvlsi, 2017_Ragavan_DATE} and over-clocking \cite{2015_Constantin_DATE, 2013_Shi_FCCM} techniques.
Voltage over-scaling lowers the supply voltage of the circuit below the nominal value, and thus, the power consumption is decreased in exchange for erroneous outputs due to the critical paths' failure to meet the timing constraints \cite{2010_Kurdahi_IEEEtvlsi}. 
Over-clocking inserts timing violations, reducing expensive timing guard-bands and providing performance improvement \cite{2013_Shi_FCCM}.
Another approach is to apply
logic-level approximations,
i.e., modify the truth table, employ inexact components, or prune nodes of the circuit.

The main target of logic-level Approximate Computing is the arithmetic circuits \cite{2020_Jiang_IEEE},
which are key computation units in 
general-purpose processors and custom hardware accelerators.
Extensive research is reported in approximate adders \cite{GuptaTCADICS2013, 2013_Gupta_IEEEtcad, ZhuTVLSI2010, MahdianiTCSI2010, VermaDATE2008}, 
which 
provide significant gains in critical path delays and power consumption.
On the other hand, 
research activities on the approximate multipliers \cite{KulkarniICVLSI2011, 2015_Momeni_IEEEtc, LinICCD2013, KyawEDSSC2010, LiuDATE2014, NarayanamoorthyTVLSI2015, 2015_Hashemi_ICCAD, ZervakisTVLSI2016, 2016_Jiang_IEEEtc, 2017_Liu_IEEEtc, BabicMM2011, 2017_Zendegani_IEEEtvlsi} 
are less comprehensive compared to the respective on approximate adders.
In the multiplication circuit, 
approximations can be applied in the partial product generation \cite{ZervakisTVLSI2016, 2016_Jiang_IEEEtc, 2017_Liu_IEEEtc}
and the partial product accumulation 
\cite{KulkarniICVLSI2011, LinICCD2013, 2015_Momeni_IEEEtc, LiuDATE2014}.
Both types of approximation are synergistic and can be applied in collaboration to achieve higher power reduction \cite{2016_Jiang_IEEEtc, 2017_Liu_IEEEtc}.
Although significant research regards the partial product accumulation, 
approximation techniques in the partial product generation have received less attention.
Another limitation of the existing approximate multipliers is that the majority of them (e.g., \cite{2015_Momeni_IEEEtc, KulkarniICVLSI2011, KyawEDSSC2010, BabicMM2011}) does not examine signed multiplication.

In this chapter, 
motivated by the resource gains provided by arithmetic circuits,
we present a novel \emph{hybrid high-radix encoding} for approximate multipliers.
Our technique addresses some of the issues of prior designs,
such as the signed multiplication and the reduction of the critical paths, 
and it also provides better results than state-of-the-art works. 
The proposed encoding is hybrid,
i.e., it splits the number and encodes it with two different schemes
(accurate and approximate).
More specifically, 
the Most Significant Bits (MSBs) 
are encoded using the accurate radix-$4$ encoding, 
whereas the $k$ Least Significant Bits (LSBs) 
are encoded using the approximate high-radix-$2^{k}$ (where $k \geq 4$).
To simplify the increased complexity 
induced by the conventional high-radix encodings, 
we alter their truth tables 
and generate their approximate variants. 
Using this approximate hybrid encoding,
the number of the partial products is decreased,
and simpler tree architectures are used for the partial product accumulation.

The \textbf{contribution} of this chapter is summarized as follows:

\begin{itemize}[]
\item[(i)] We highlight the efficiency of applying disciplined low-level approximations, which deliver valuable resource gains
while keeping the accuracy in acceptable levels. 
\item[(ii)] We address the circuit overheads of the classic high-radix encodings, e.g., radix-$64$, radix-$256$, radix-$1024$,
which constitute them inefficient for use in multiplication circuits. \item[(iii)] We propose a new hybrid encoding for approximate partial product generation, which is parametric in terms of approximation degree and can be combined with other techniques, e.g., with approximate partial product accumulation.    
\item[(iv)] We show that the proposed technique outperforms related state-of-the-art approximation techniques,
providing remarkable area and energy gains for comparable error values.
\end{itemize}

The remainder of this chapter is organized as follows. 
Section \ref{s4_2} introduces the proposed approximate hybrid high-radix encoding and its application in the design of approximate radix-based multipliers. 
Section \ref{s4_3} performs the evaluation of the proposed design,
including theoretical resource analysis, error analysis, and comparative experimental results.
Finally, 
Section \ref{s4_4} draws the conclusions.

\section{Design of Approximate High-Radix Encodings and Multipliers}
\label{s4_2}

Conventional high-radix operand encodings
reduce the total number of partial products in multipliers,
and thus,
the partial product accumulation is performed with simpler adder tree architectures.
However,
the high-radix encoders,
as well as the high-radix-based partial product generators,
are characterized by increased logic complexity,
negating thus the benefits of the partial product reduction.
Therefore, 
the prevailing radix encoding for multiplication circuits
is radix-$4$,
outperforming its high-radix counterparts.

In this section, 
we propose an approximate hybrid encoding,
which applies simultaneously the conventional (accurate) radix-$4$ encoding
and a new (approximate) high-radix-$2^{k}$ encoding. 
In more detail,
we present the functionality of the hybrid encoding,
we introduce the proposed 
approximations that simplify the complexity of the conventional high-radix encoding,
and finally,
we design inexact multipliers based on the proposed approximate encoding.

\subsection{Approximate Operand Encoding}
Let $B$ be a $n$-bit $2$'s-complement number
and $k$ be an even number belonging in the interval
$[4$, $n-2]$, i.e.,
$k = 2m \! \! : m \in \mathbb{Z}$ and $2 \leq m \leq (n-2)/2$.
$B$ is divided into two segments with respect to $k$:
the most significant part containing its $n-k+1$ MSBs
and the least significant part containing its $k$ LSBs
(there is a shared bit between the two segments).
The $n-k+1$ MSBs are encoded
with the radix-$4$ encoding, 
while the $k$ LSBs are 
encoded with the radix-$2^{k}$ encoding,
as shown in Eq. \eqref{eq_rad1}--\eqref{eq_rad3}.
\begin{equation}
B = \langle b_{n-1} b_{n-2} \cdots b_0\rangle_{2\text{'s}} 
= -2^{n-1}b_{n-1} +
\mathlarger{\sum}_{\substack{i=0}}^{\substack{n-2}} 
2^{i}b_{i} =
\mathlarger{\sum}_{\substack{j=k/2}}^{\substack{n/2-1}} 
 4^{j} y_{j}^{R4} + y^{R2^{k}}_0 \\[3pt]
\label{eq_rad1}
\end{equation}
\begin{equation}
 y_{j}^{R4} = -2 b_{2j+1} + b_{2j} + b_{2j-1}  
\;\; \implies \;\; y_{j}^{R4} \in \{0, \pm 1, \pm 2\} \\[6pt]
\label{eq_rad2}
\end{equation}
\begin{equation} 
 \!\!\!\! y^{R2^{k}}_0 \!=\! -2^{k-1}b_{k-1} + 2^{k-2}b_{k-2} + \dots + b_{0}  
\!  \implies \!
y^{R2^{k}}_0 \!\!
\in \! \{0, \pm 1, \dots, \pm (2^{k - 1} \!\! -  1), -2^{k - 1}\} \\[5pt]
\label{eq_rad3}
\end{equation}

Eq. \eqref{eq_rad2} refers to the radix-$4$ encoded digits,
while
Eq. \eqref{eq_rad3} refers to the radix-$2^{k}$ encoded digit.
In more detail, 
the radix-$4$ encoding creates 
$(n - k)/2$ digits $y_{j}^{R4}$
$\in$ $\{0,$ $\pm 1,$ $\pm 2\}$, 
while 
the radix-$2^{k}$ encoding creates 
the digit $y^{R2^{k}}_0$ 
$\in$ $\{0,$ $\pm 1,$ $\pm 2,$ $\pm 3,$ $\dots,$ $\pm (2^{k - 1} - 1),$ $-2^{k - 1}\}$.
In total, $B$ is encoded with $(n - k)/2$ + $1$ digits. 

The encoding circuit for 
the above accurate hybrid encoding 
features increased logic due to
the large number of $y^{R2^{k}}_0$ values
and the values that are not power of two. 
Therefore,
we propose an alternative approximate variant of the high-radix encoding,
while 
we keep the low-complexity radix-$4$ encoding of the MSBs
to avoid huge errors. 
To approximate $y^{R2^{k}}_0$, 
we map 
all the values that are not power of two, 
as well as the $k-4$ smallest powers of two,   
to their nearest of the $4$ largest powers of two or $0$.
In this way, the approximate digit $\hat{y}^{R2^{k}}_0$ takes values from a smaller set
that includes only $4$ absolute values plus $0$. 
In addition, the sum of these values is $0$, like in the set of values of $y_{j}^{R4}$.
We choose to keep only the $4$ largest powers of two, 
so that the approximate radix-$2^k$ encoder 
requires about double the 
logic of the $(n - k)/2$ accurate radix-$4$ encoders. 

Table \ref{tb_rad1} presents the accurate radix-$4$ encoding. 
The signal $sign_j$
indicates the sign of $y_{j}^{R4}$,
i.e., it is activated when the digit is negative. 
The signals
$\times1_j$ and $\times2_j$
regard the magnitude of $y_{j}^{R4}$, 
i.e.,
they are activated 
when its magnitude is $1$ and $2$,
respectively. 
The logic functions of these signals are 
provided in Eq. \eqref{eq_r4log1}--\eqref{eq_r4log3}.

\vspace*{-25pt}

\begin{eqnarray}
& sign_j = b_{2j+1}                         & \label{eq_r4log1}  \\[3pt]
& \times1_j = b_{2j-1} \oplus b_{2j}        & \label{eq_r4log2} \\[3pt]
& \times2_j = (b_{2j+1} \oplus b_{2j}) 
  \cdot \overline{(b_{2j-1} \oplus b_{2j})} & \label{eq_r4log3}
\end{eqnarray}

Table \ref{tb_rad2} presents the approximate radix-$2^k$ encoding. 
The first two columns show the proposed mapping to create the approximate set of values ($y^{R2^{k}}_0 \rightarrow \hat{y}^{R2^{k}}_0$). 
For example, 
if the value of $y^{R2^{k}}_0$,
as originally calculated by Eq. \eqref{eq_rad3},
belongs in 
$[2^{k-5}, \, 2^{k-4}+2^{k-5})$, 
it is mapped to $2^{k-4}$.
Similar to the radix-$4$ encoding,
the encoding signals are activated 
to indicate the value of $\hat{y}^{R2^{k}}_0$.
The logic functions of the approximate radix-$2^k$ encoding signals are 
provided in Eq. \eqref{eq_rklog1}--\eqref{eq_rklog5}.
We note that for $k=4$,
we consider an extra bit $b_{-1} = 0$.

\vspace*{-15pt}

\begin{eqnarray}
& sign = b_{k-1} & \label{eq_rklog1}  \\[3pt]
& \times2^{k-4} = (\overline{b}_{k-2} \cdot \overline{b}_{k-3} \cdot \overline{b}_{k-4} + b_{k-2} \cdot b_{k-3}  \cdot b_{k-4}) 
\cdot (b_{k-4} \oplus b_{k-5}) & \label{eq_rklog2} \\[3pt]
& \times2^{k-3} = \overline{b}_{k-1} \cdot \overline{b}_{k-2} \cdot (\overline{b}_{k-3} \cdot b_{k-4} \cdot b_{k-5} + b_{k-3}  \cdot \overline{b}_{k-4}) \; + & \nonumber \\
& \hspace{37pt} +\;  b_{k-1} \cdot b_{k-2} \cdot (b_{k-3}  \cdot \overline{b}_{k-4} \cdot \overline{b}_{k-5} + \overline{b}_{k-3} \cdot b_{k-4}) & \label{eq_rklog3}  \\[3pt]
& \times2^{k-2} = \overline{b}_{k-2} \cdot b_{k-3} \cdot (b_{k-1} + b_{k-4})
+ b_{k-2}  \cdot \overline{b}_{k-3} \cdot (\overline{b}_{k-1}  + \overline{b}_{k-4}) & \label{eq_rklog4}  \\[3pt]
&\times2^{k-1} = \overline{b}_{k-1} \cdot b_{k-2} \cdot b_{k-3} + b_{k-1} \cdot \overline{b}_{k-2}  \cdot \overline{b}_{k-3} & \label{eq_rklog5} 
\end{eqnarray}

\begin{table}[!t]
\fontsize{9}{10}\selectfont
\renewcommand{\arraystretch}{1.2}
\setlength{\tabcolsep}{9pt}
\caption[Accurate Radix-$4$ Encoding]{Accurate radix-$4$ encoding.}
\label{tb_rad1}
\centering
\begin{tabular}{ccccccc}
\hline
\multicolumn{3}{c}{\textbf{Operand Bits}} & \multicolumn{1}{c}{\textbf{R4 Digit}} & \multicolumn{3}{c}{\textbf{Encoding Signals}}\\
\cmidrule(lr){1-3} \cmidrule(lr){4-4} \cmidrule(lr){5-7}
$b_{2j+1}$ & $b_{2j}$ & $b_{2j-1}$ & $y_{j}^{R4}$ & $sign_j$ & $\times2_j$ & $\times1_j$\\
\hline \hline 
$0$ & $0$ & $0$ & $0$ & $0$ & $0$ & $0$\\
$0$ & $0$ & $1$ & $1$ & $0$ & $0$ & $1$\\
$0$ & $1$ & $0$ & $1$ & $0$ & $0$ & $1$\\
$0$ & $1$ & $1$ & $2$ & $0$ & $1$ & $0$\\
$1$ & $0$ & $0$ & $-2$ & $1$ & $1$ & $0$\\
$1$ & $0$ & $1$ & $-1$ & $1$ & $0$ & $1$\\
$1$ & $1$ & $0$ & $-1$ & $1$ & $0$ & $1$\\
$1$ & $1$ & $1$ & $0$ & $1$ & $0$ & $0$\\
\hline
\end{tabular}
\end{table}

\begin{table}[!t]
\fontsize{9}{10}\selectfont
\renewcommand{\arraystretch}{1.2}
\setlength{\tabcolsep}{3.5pt}
\caption[Approximate High-Radix-$2^k$ Encoding]{Approximate high-radix-$2^k$ encoding.}
\label{tb_rad2}
\centering
\begin{tabular}{lcccccc}
\hline
\multicolumn{1}{c}{\textbf{Accurate Digit}} & 
\textbf{Approx. Digit}  &
\multicolumn{5}{c}{\textbf{Encoding Signals}}\\
\cmidrule(lr){1-1} \cmidrule(lr){2-2} \cmidrule(lr){3-7}
\multicolumn{1}{c}{$y_{0}^{R2^k}$} & $\hat{y}^{R2^k}_0$ & $sign$ & $\times2^{k-1}$ & $\times2^{k-2}$ & $\times2^{k-3}$ & $\times2^{k-4}$\\
\hline \hline 
 $[0, \, 2^{k-5})$     & $0$    & $0$ & $0$ & $0$ & $0$ & $0$\\
 $[2^{k-5}, \, 2^{k-4}+2^{k-5})$   & $2^{k-4}$  & $0$ & $0$ & $0$ & $0$ & $1$\\
 $[2^{k-4}+2^{k-5}, \, 2^{k-3}+2^{k-4})$  & $2^{k-3}$  & $0$ & $0$ & $0$ & $1$ & $0$\\
 $[2^{k-3}+2^{k-4}, \, 2^{k-2}+2^{k-3})$  & $2^{k-2}$  & $0$ & $0$ & $1$ & $0$ & $0$\\
 $[2^{k-2}+2^{k-3}, \, 2^{k-1})$ & $2^{k-1}$ & $0$ & $1$ & $0$ & $0$ & $0$\\
 $[-2^{k-1}, \, -2^{k-2}-2^{k-3})$ & $-2^{k-1}$ & $1$ & $1$ & $0$ & $0$ & $0$\\
 $[-2^{k-2}-2^{k-3}, \, -2^{k-3}-2^{k-4})$  & $-2^{k-2}$  & $1$ & $0$ & $1$ & $0$ & $0$\\
 $[-2^{k-3}-2^{k-4}, \, -2^{k-4}-2^{k-5})$  & $-2^{k-3}$  & $1$ & $0$ & $0$ & $1$ & $0$\\
 $[-2^{k-4}-2^{k-5}, \, -2^{k-5})$   & $-2^{k-4}$  & $1$ & $0$ & $0$ & $0$ & $1$\\
 $[-2^{k-5}, \, 0)$    & $0$    & $1$ & $0$ & $0$ & $0$ & $0$\\
\hline
\end{tabular}
\end{table}

Overall, 
starting from the accurate hybrid encoding of Eq. \eqref{eq_rad1}--\eqref{eq_rad3},
we approximate the high-radix encoding of the LSBs
through the mapping of Table \ref{tb_rad2}, 
and thus, 
$B$ is approximately encoded to $\tilde{B}$ as shown in Eq.
\eqref{eq_brad1}--\eqref{eq_brad3}. 

\vspace{-5pt}

\begin{equation}
 \tilde{B} = 
 \mathlarger{\sum}_{\substack{j=k/2}}^{\substack{n/2-1}} 4^{j} y_{j}^{R4} + \hat{y}^{R2^{k}}_0
 \label{eq_brad1} 
\end{equation}
\begin{eqnarray}
& &  \text{where } \; y_{j}^{R4} \in \{0, \;  \pm 1, \;  \pm 2\}  \label{eq_brad2} \\[4pt]
& &  \text{and } \; \hat{y}^{R2^{k}}_0 \in \{0, \;  \pm 2^{k-4}, \;  \pm 2^{k-3}, \;  \pm 2^{k-2}, \;  \pm 2^{k-1}\} 
\label{eq_brad3}
\end{eqnarray}

Next, 
we present some examples to show how the proposed 
hybrid encoding is applied.
We consider $n=16$,
i.e., $16$-bit numbers 
and $k=6,10$, i.e., 
the LSBs are encoded with radix-$64$ and radix-$1024$, respectively.  
In the hybrid radix-$64$ encoding, 
the bits of $B$ are grouped as shown in Eq. \eqref{eq_rad64b}.
Regarding the approximation of the radix-$64$ encoding, 
the $4$ largest powers of two
are $\pm4$, $\pm8$, $\pm16$ and $\pm32$.
Thus,
we map all the rest values 
of the original digit $y^{R64}_0$ 
to their nearest value from the set 
$\{0, \;  \pm 4, \;  \pm 8, \;  \pm 16, \;  \pm 32\}$
according to Table \ref{tb_rad2} (for $k=6$). 
\begin{equation}
\overunderbraces{&&&\br{3}{y_{6}^{R4}} &&\br{3}{y_{4}^{R4}}&&\br{6}{y^{R64}_0}}%
{&b_{15}&b_{14}&b_{13}&b_{12}&b_{11}&b_{10}&b_{9}&b_{8}&b_{7}&b_{6}&b_{5}&b_{4}&b_{3}&b_{2}&b_{1}&b_{0}}%
{&\br{3}{y_{7}^{R4}} &&\br{3}{y_{5}^{R4}} &&\br{3}{y_{3}^{R4}}}
\label{eq_rad64b}
\end{equation}

In the hybrid radix-$1024$ encoding, 
the bits of $B$ are grouped as shown in Eq. \eqref{eq_rad1024b}.
Similarly, 
the values of the digit $y^{R1024}_0$
are mapped to 
$\{0, \;  \pm 64, \;  \pm 128, \;  \pm 256, \;  \pm 512\}$ 
according to Table \ref{tb_rad2} (for $k=10$). 
\begin{equation}
\overunderbraces{&&&\br{3}{y_{6}^{R4}} &&\br{10}{y_{0}^{R1024}}}%
{&b_{15}&b_{14}&b_{13}&b_{12}&b_{11}&b_{10}&b_{9}&b_{8}&b_{7}&b_{6}&b_{5}&b_{4}&b_{3}&b_{2}&b_{1}&b_{0}}%
{&\br{3}{y_{7}^{R4}} &&\br{3}{y_{5}^{R4}} }
\label{eq_rad1024b}
\end{equation}

For the implementation of the hybrid high-radix encoding,
the designer needs only to 
use the logic functions of Eq. \eqref{eq_r4log1}--\eqref{eq_r4log3}
to generate the $(n-k)/2$ accurate radix-$4$ encoders,
as well as 
the logic functions of Eq. \eqref{eq_rklog1}--\eqref{eq_rklog5}
to generate the single approximate radix-$2^k$ encoder. 
We note that 
an important feature of this encoding
is that 
the logic resources of the 
approximate radix-$2^k$ encoder
are fixed and independent of $k$. 

\subsection{RAD: Approximate High-Radix Multiplier}

In this section, 
we present how the proposed hybrid encoding can be used
to design approximate high-radix multipliers.
We consider the notation 
RAD$2^k$
for the multiplier
that implements the hybrid high-radix-$2^k$ encoding. 
Let $A \cdot B$ be the accurate multiplication of 
two $n$-bit $2$'s-complement numbers.
We encode the multiplicand $B$
with the proposed approximate high-radix encoding 
and perform the approximate multiplication $A \cdot \tilde{B}$. 

\begin{figure}[!t]
\centering
\includegraphics[width=0.91\textwidth]{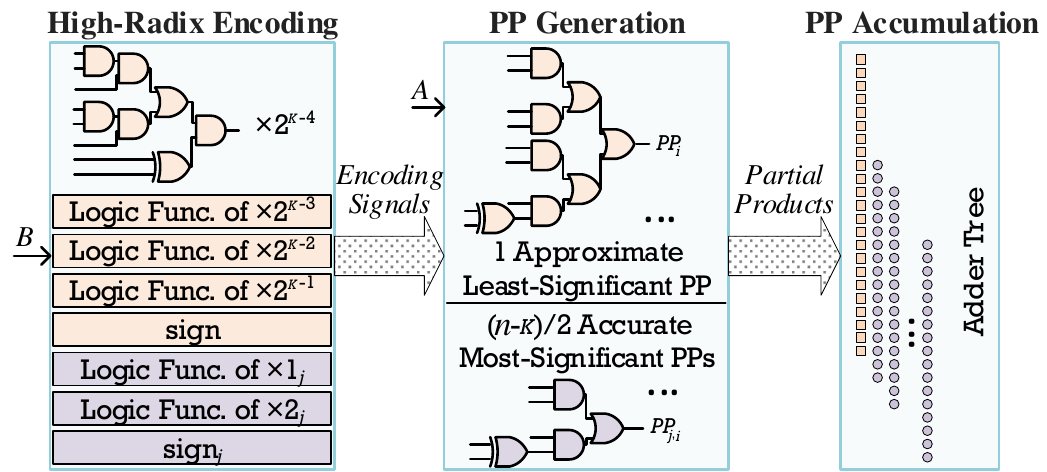}%
\caption[Architecture of the Approximate High-Radix Multiplier]{The architecture of the RAD$2^k$ multiplier that is based on the approximate hybrid high-radix encoding.}%
\label{fig_radmul}
\vspace*{3pt}
\end{figure}

The approximate architecture of the RAD$2^k$ multiplier 
is illustrated in Figure \ref{fig_radmul}.
It consists of three stages: 
operand encoding, partial product generation and partial product accumulation.
In the first stage, 
$B$ is encoded as discussed in the previous section.
The second stage inputs $A$
and the encoding signals of $\tilde{B}$,
and it generates
$(n-k)/2$ accurate partial products based on the radix-4 encoding (digits $y_j^{R4}$) 
and 1 approximate partial product based on the radix-2$^k$ encoding (digit $\hat{y}^{R2^{k}}_0$).
The approximate partial product practically 
substitutes the $k/2$ least significant partial products of the accurate radix-$4$ multiplier. 
Namely, we achieve 
a reduction of $k/2-1$ partial products,
as the accurate radix-$4$ multiplier generates $n/2$ partial products,
while the proposed one generates $(n-k)/2+1$.   
Table \ref{tb_radprods}
summarizes 
the partial products 
that can be generated by 
the accurate radix-$4$ encoding
and the 
approximate radix-$2^k$
encodings ($k=6,8,10$).
All the possible partial products
express the multiplication of 
$A$ with the encoded digits of $B$,
which take the values discussed in the previous section. 
In the last stage, 
the partial products 
are accumulated to form the final approximate product.
We note that the accumulation stage is orthogonal to the 
approximate encoding and partial product generation.
Therefore, 
the designer can choose any method for the accumulation,
either accurate or approximate. 

\begin{table}[!t]
\fontsize{9}{10}\selectfont
\renewcommand{\arraystretch}{1.2}
\setlength{\tabcolsep}{24pt}
\caption[Partial Products Generated by the Radix Encodings]{Partial products generated by the radix encodings ($k=6,8,10$).}
\label{tb_radprods}
\centering
\begin{tabular}{l|c}
\hline
\textbf{Encoding} & \textbf{Partial Products}\\
\hline \hline
Radix-$4$  & $0$, $\pm A$, $\pm2A$\\
Radix-$64$   & $0$, $\pm4A$, $\pm8A$, $\pm16A$, $\pm32A$ \\
Radix-$256$  & $0$, $\pm16A$, $\pm32A$, $\pm64A$, $\pm128A$ \\
Radix-$1024$ & $0$, $\pm64A$, $\pm128A$, $\pm256A$, $\pm512A$ \\
\hline
\end{tabular}
\end{table}

\begin{figure}[!t]
\centering
\subfloat[\label{fig_radd64}]{\includegraphics[width=0.48\textwidth]{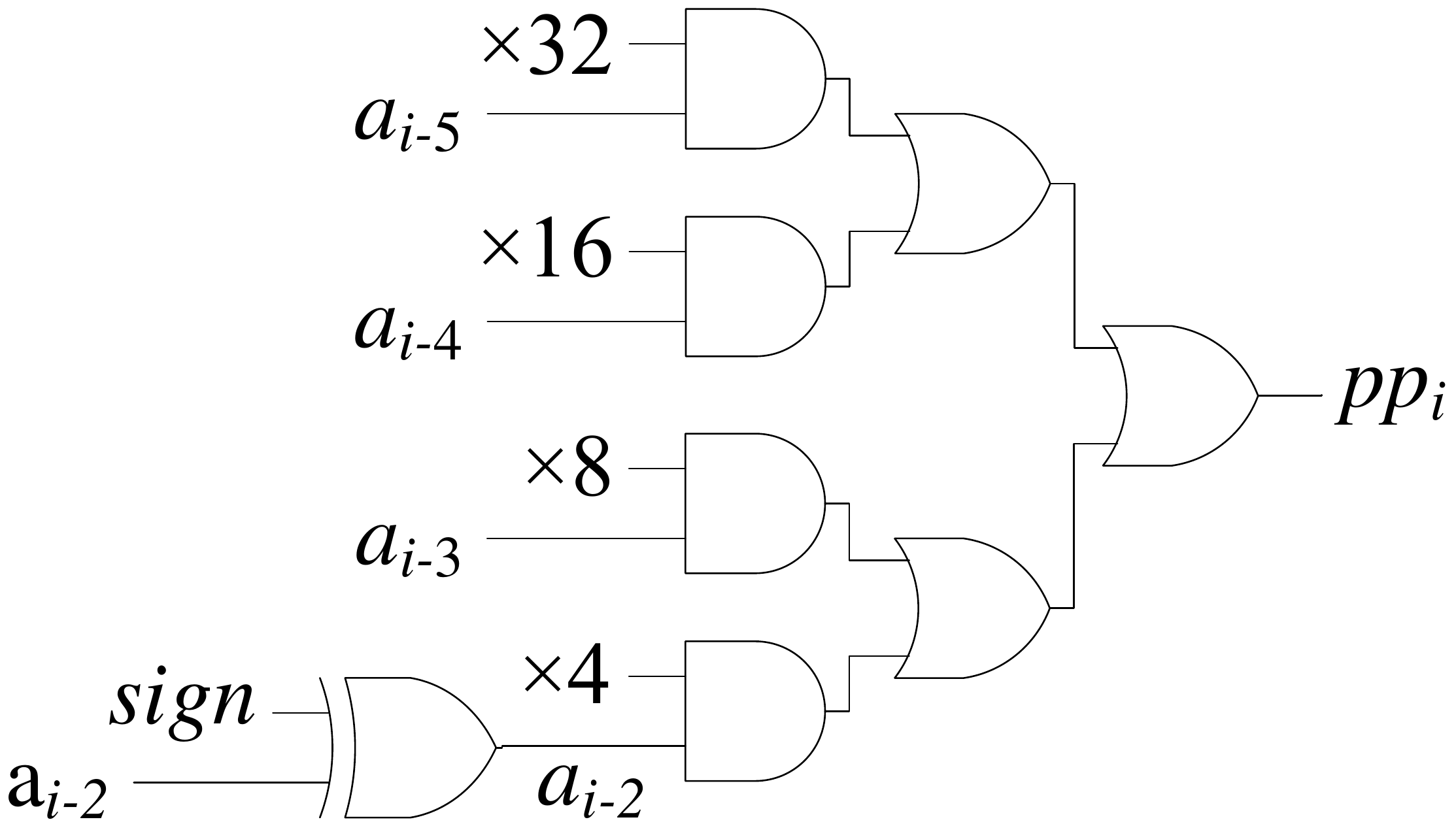}}%
\\[-9pt]
\subfloat[\label{fig_radd256}]{\includegraphics[width=0.48\textwidth]{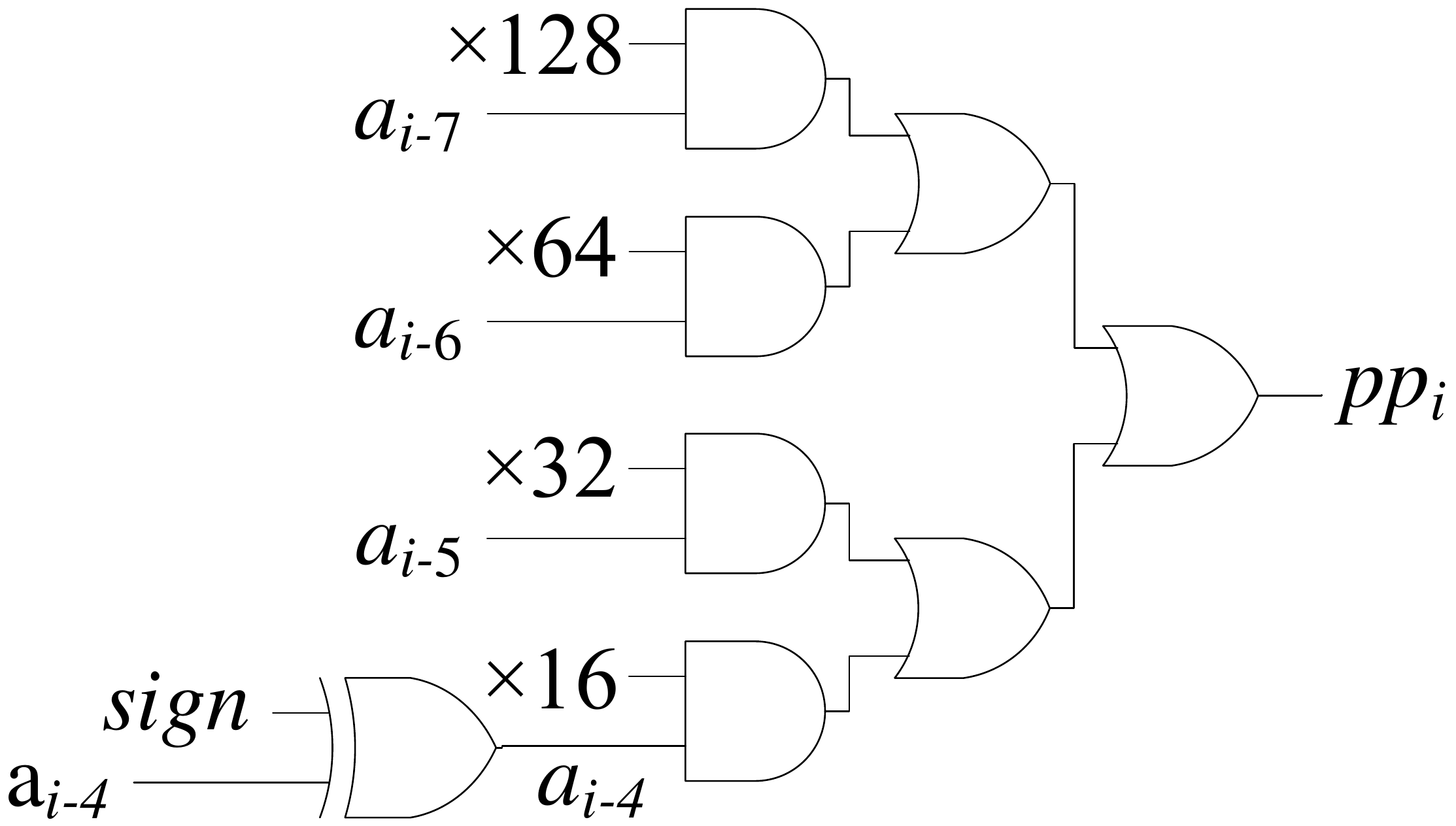}}%
\hfill
\subfloat[\label{fig_radd1024}]{\includegraphics[width=0.48\textwidth]{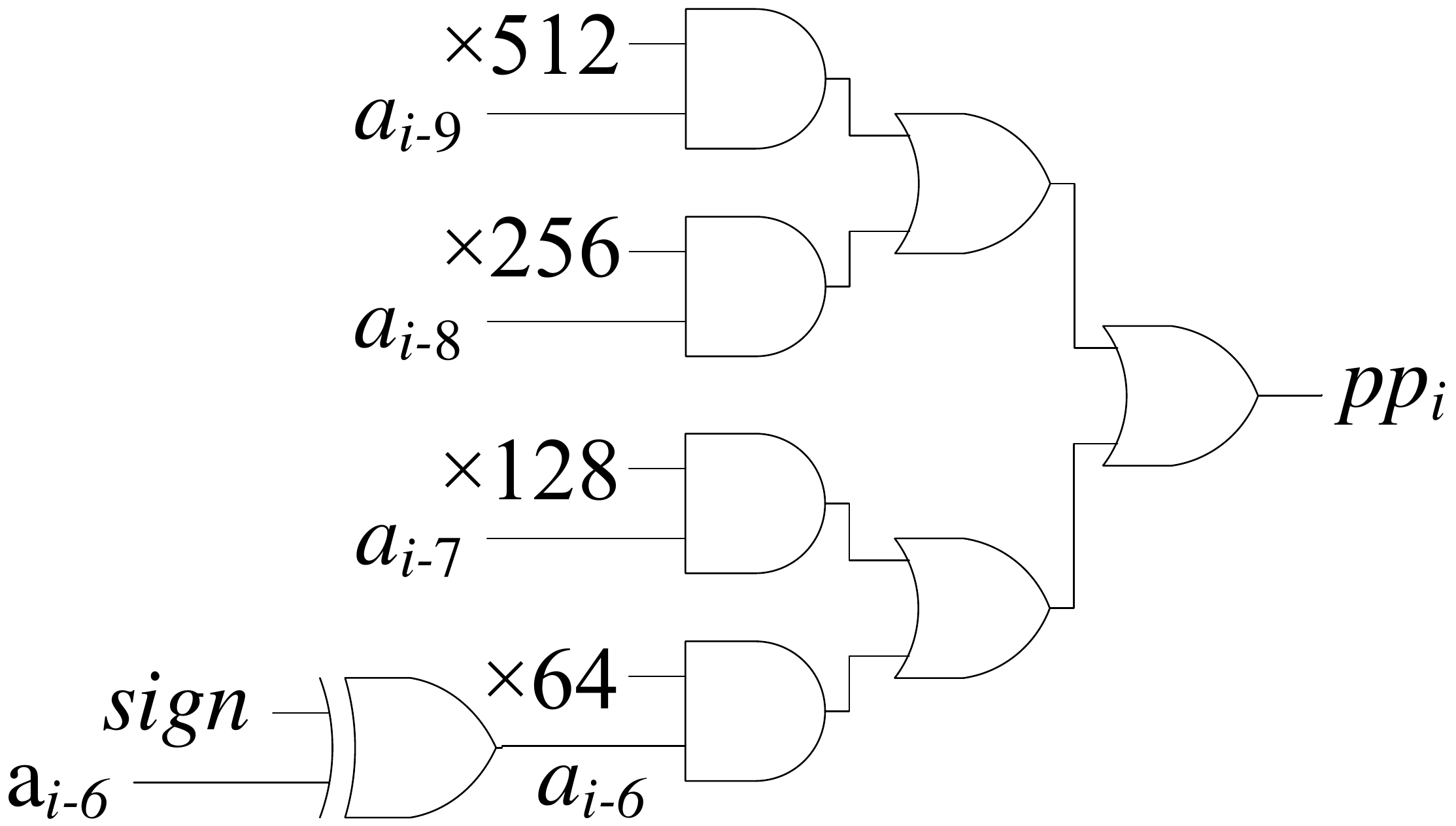}}%
\caption[Partial Product Generators of the Approximate High-Radix Multipliers]{Approximate $i$-bit partial product generators based on the high-radix encodings:
\textbf{(a)} radix-$64$ ($k=6$), 
\textbf{(b)} radix-$256$ ($k=8$), 
\textbf{(c)} radix-$1024$ ($k=10$).\\ 
Notations: $a_i$: $i$-bit of $A$, a$_i = $ $s_j \oplus a_i$}
\label{fig_pprad}
\end{figure}

The circuit of the $i$-bit partial product generator
based on radix-$4$ encoding
is the one discussed in Chapter \ref{chapter3}
and illustrated in Figure \ref{fig_dt3}. 
It inputs the signals $sign_j$, $\times 1_j$, $\times 2_j$
and the bits of $A$,
and it calculates the partial product bits.
In Figure \ref{fig_pprad},
we illustrate
the $i$-bit partial product generators 
based on the approximate radix-$64$, radix-$256$, and radix-$1024$
encodings.
As shown,
similar to the encoders, 
the area of the partial product generator for $1$ bit is fixed
and independent of the encoding
(i.e., the value of $k$). 

\begin{figure}[!t]
\centering
\subfloat[\label{fig_tr64}]{\includegraphics[width=0.49\textwidth]{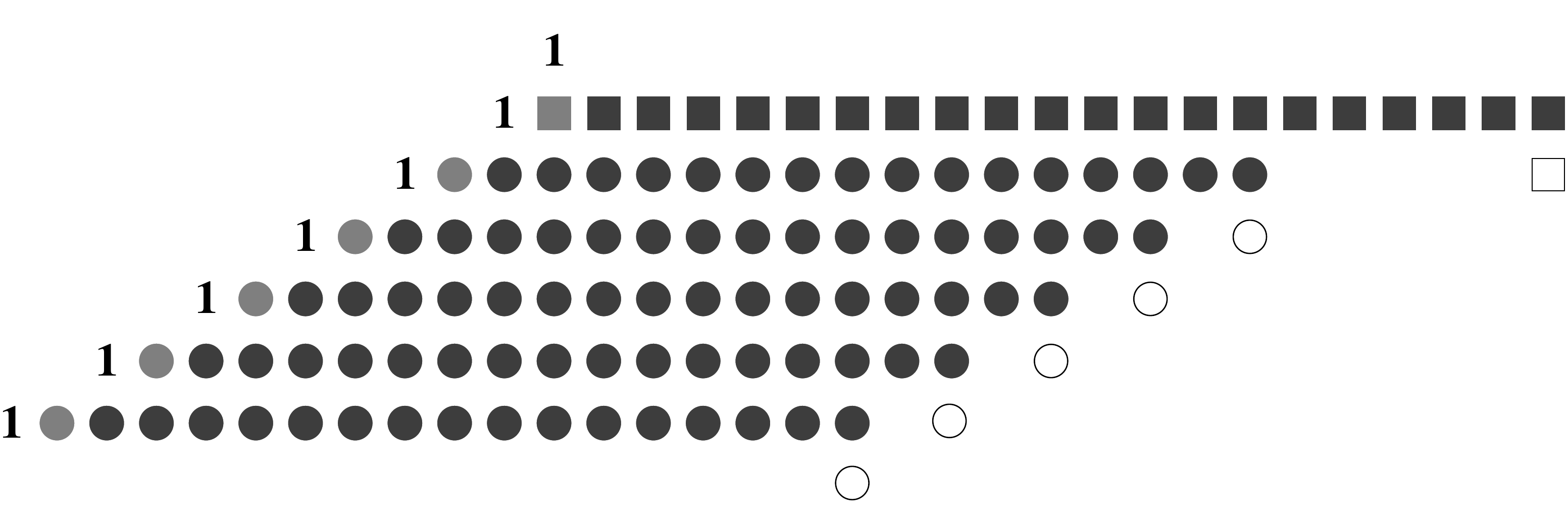}}\\
\subfloat[\label{fig_tr256}]{\includegraphics[width=0.49\textwidth]{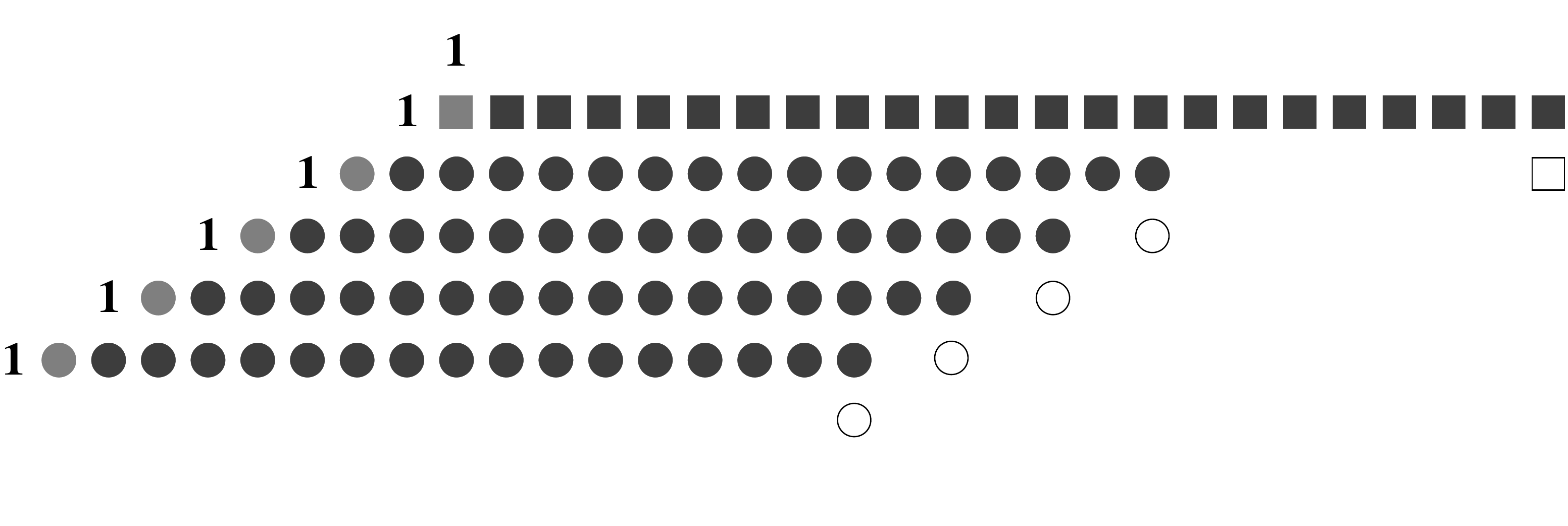}}%
\hspace{1pt}
\subfloat[\label{fig_tr1024}]{\includegraphics[width=0.49\textwidth]{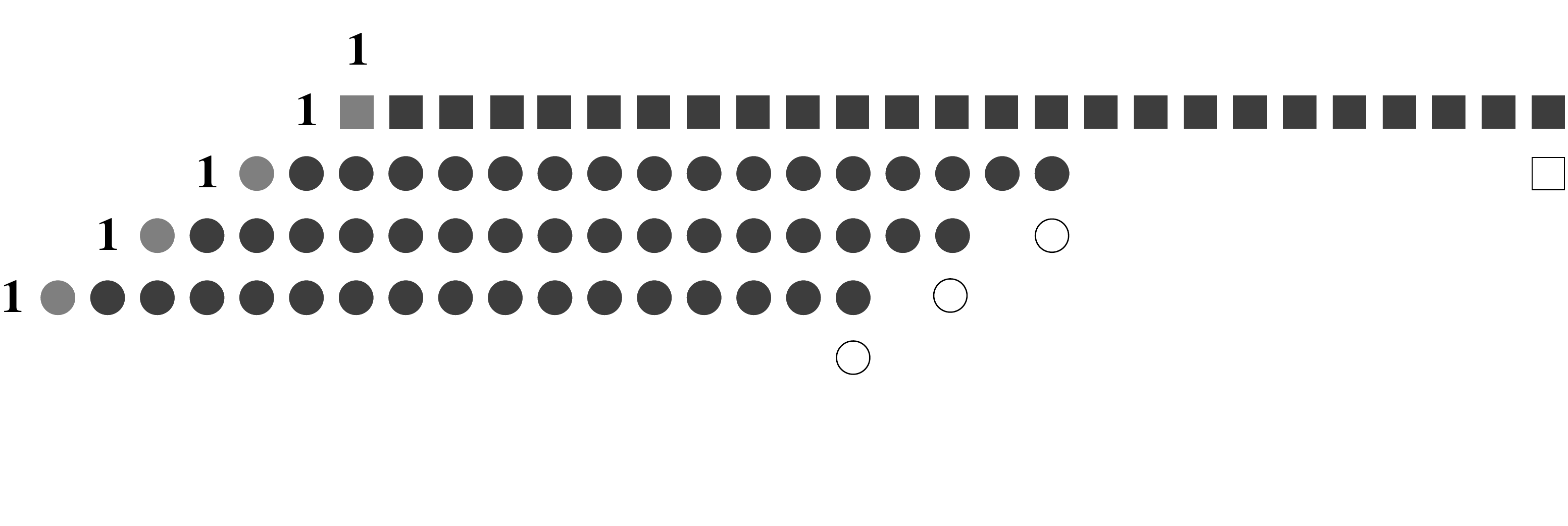}}%
\caption[Partial Product Matrices of the Approximate High-Radix Multipliers]{Partial product matrices of the $16$-bit approximate multipliers using 
the accurate radix-$4$
and an approximate high-radix-$2^k$ encoding:
\textbf{(a)} radix-$64$ ($k=6$), 
\textbf{(b)} radix-$256$ ($k=8$), 
\textbf{(c)} radix-$1024$ ($k=10$).\\ 
Symbols: 
\protect\tikz \protect\fill[blue_dlsb] (1ex,1ex) 
circle (0.63ex);: 
partial product bits from radix-$4$ 
\hspace{2pt} 
\protect\tikz \protect\fill[blue_dlsb] (0.77ex,0.77ex) rectangle (0.22,0.22);: 
partial product bits from radix-$2^k$ 
\hspace{2pt}
\protect\tikz \protect\filldraw[fill=gray, draw=gray] (1.2ex,1.2ex) circle (0.6ex);, \protect\tikz \protect\filldraw[fill=gray, draw=gray] (0.77ex,0.77ex) rectangle (0.22,0.22);:
inverted partial product MSBs
\hspace{2pt}
\protect\tikz \protect\filldraw[fill=white, draw=black] (1.2ex,1.2ex) circle (0.6ex);, \protect\tikz \protect\filldraw[fill=white, draw=black] (0.77ex,0.77ex) rectangle (0.22,0.22);:
correction terms}
\label{fig_radtrees}
\end{figure}

Figure \ref{fig_radtrees} illustrates the partial product matrices
of the $16$-bit approximate multipliers
for $k=6,8,10$.
The matrices also include constant and corrections terms,
as those discussed in Section \ref{s3_3}.
As shown,
each increment of the $k$ parameter
omits an accurate radix-$4$ partial product,
which is encoded in the larger approximate radix-$2^k$
partial product. 
The bit-width of each accurate partial product is $n+1$,
i.e., there are 
$n$ generated partial product bits (black circles) 
plus the inverted MSB of the partial product (gray circle).
Correspondingly,
the bit-width of the approximate partial product
is $n+k-1$.

\section{Evaluation}
\label{s4_3}

In this section,
we evaluate our approximate high-radix multipliers.
Firstly, we report a theoretical analysis
on the resources of our approximate designs,
then we examine the error introduced by the approximations,
and finally,
we attach experimental results
including comparisons with state-of-the-art multipliers.

\subsection{Theoretical Analysis}

The advantage of the RAD$2^k$ multipliers
is their decreased logic compared to 
the accurate radix-$4$ multiplier 
(labeled as ACCR4),
which results in faster operation and area/power gains.
RAD$2^k$ generates a larger least significant partial product,
however,
this product substitutes $k/2$ radix-$4$ partial products,
and also,
the corresponding circuits for its generation
do not impose significant overheads. 
To provide a theoretical evaluation,
we employ the unit gate model used in \cite{Tsoumanis2014}: 
the AND-2/OR-2 gate is equal to $1$ unit gate, 
the NOT gate is equal to $0.5$ unit gate,
the XOR-2 gate counts as $2$ unit gates,
and the unit gates of the full adder and half adder are $7$ and $3$, respectively. 
Based on this model, 
Table \ref{tb_rug1} reports the unit gates of each component used in the RAD$2^k$ multipliers. 

For the partial product generation,
ACCR4 uses 
$n/2$ radix-$4$ encoders and 
$(n/2) \times n$ radix-$4$ partial product generators 
to produce the $n/2$ $n$-bit partial products.
Correspondingly,
RAD$2^k$ employs 
$(n-k)/2$ radix-$4$ encoders, 
$1$ radix-$2^k$ encoder, 
$((n-k)/2) \times n$ radix-$4$ partial product generators, and
$n+k-2$ radix-$2^k$ partial product generators. 
For the partial product accumulation,
we choose the Wallace tree and a final fast adder,
like in the design of Chapter \ref{chapter3}.
ACCR4
accumulates 
$n/2  +  1$ operands,
i.e., $n/2$ partial products and $1$ operand with the correction and constant terms,
while 
RAD$2^k$ accumulates 
$(n-k)/2 + 2$ operands.
Considering the logic required for the
multi-operand accumulation of 
$n$-bit numbers in carry-save form \cite{Weste2010},
as well as our analysis in Section \ref{s3_4},
ACCR4 
requires $7n(n - 2)/2$ unit gates,
while 
RAD$2^k$ requires 
$7n(n  -  k)/2$ unit gates. 
Finally,
regarding the final $2n$-bit fast addition,
which is performed in both ACCR4 and RAD$2^k$,
it requires 
$2n$ half adders, 
$n \log_2 2n$ propagate group circuits (each one is $3$ unit gates), 
and $2n$ XOR-2 gates.

\begin{table}[!t]
\vspace*{3pt}
\fontsize{9}{10}\selectfont
\renewcommand{\arraystretch}{1.2}
\setlength{\tabcolsep}{18pt}
\caption[Unit Gates per Component of the Approximate High-Radix Multipliers]{Unit gates per component of RAD-$2^k$ multipliers.}
\label{tb_rug1}
\centering
\begin{tabular}{l|c|c}
\hline
\textbf{Component} & \textbf{Reference} & \textbf{Unit Gates} \\
\hline \hline
Radix-$4$ Encoder & Eq. \eqref{eq_r4log1}--\eqref{eq_r4log3}                       & $5.5$ \\
Radix-$2^k$ Encoder & Eq. \eqref{eq_rklog1}--\eqref{eq_rklog5}                       & $41.5$ \\
Radix-$4$ PP Generator & Fig. \ref{fig_dt3}    & $5$  \\
Radix-$2^k$ PP Generator & Fig. \ref{fig_pprad}  & $9$  \\
\hline
\end{tabular}
\vspace*{9pt}
\end{table}

Based on the above analysis, 
Table \ref{tb_rug2} includes 
the per-stage and total number of unit gates
for ACCR4 and RAD$2^k$ ($k=6,8,10$).
As shown, 
even though the entire encoding component of the  
RAD$2^k$ multipliers has
\raisebox{0.8pt}{$\scriptstyle\sim$}$1.5\times$ 
more unit gates than the encoder of ACCR,
their partial product generation and accumulation
requires less resources than that of ACCR. 
More specifically,
the RAD$2^k$ partial product generation and accumulation
are  
up to $29\%$ and $57\%$ better,
respectively. 
In total,
the logic reduction achieved by RAD$2^k$
is up to $33\%$. 
We note that
the unit gate model is a simplified model 
(e.g., it does not take into account the interconnections complexity), 
and it only gives a rough estimation for the area reduction. 
The exact resource gains are presented in Section \ref{s4_3_3}.

\begin{table}[!t]
\fontsize{9}{10}\selectfont
\renewcommand{\arraystretch}{1.2}
\setlength{\tabcolsep}{10pt}
\caption[Unit Gates of the Approximate High-Radix Multipliers]{Unit gates of $16$-bit RAD$2^k$ multipliers ($k=6,8,10$).}
\label{tb_rug2}
\centering
\begin{tabular}{l|cccc}
\hline
\textbf{Multiplier Stage} &   \textbf{ACCR4} & \textbf{RAD64} & \textbf{RAD256} &  \textbf{RAD1024} \\
\hline \hline
Radix-$4$ Encoding        & $44$  & $25$    & $20$   & $15$   \\
Radix-$2^k$ Encoding      & $-$   & $41.5$  & $41.5$ & $41.5$   \\
Radix-$4$ PP Generation       & $640$ & $400$   & $320$  & $240$ \\
Radix-$2^k$ PP Generation     & $-$   & $180$   & $198$  & $216$ \\
PP Accumulation           & $784$ & $560$   & $448$  & $336$ \\
Final Addition            & $400$ & $400$   & $400$  & $400$ \\
\hline
\textbf{Total Unit Gates} & $1868$       & $1606.5$ & $1427.5$ & $1248.5$ \\
\textbf{Reduction}   & $-$ & $14\%$  & $24\%$  & $33\%$ \\
\hline
\end{tabular}
\end{table}

\subsection{Error Analysis}

The quality of the results calculated by approximate circuits is of utmost importance,
thus, 
the analysis of the errors induced by the approximations
constitutes a separate object of study.
In this context, we evaluate the accuracy of the 
RAD$2^k$ multipliers
based on the execution of their software models emulating the logic-level approximations. 
Because our approximations can be emulated at software level,
we can provide fast and accurate error analysis.
This is an important feature of our approximate designs,
eliminating the need for time-consuming hardware simulations to obtain the approximate results.  
In our analysis, we employ error metrics
that are widely used in the field of Approximate Computing \cite{LiangTC2013, 2016_Jiang_IEEEtc, 2017_Liu_IEEEtc}.
These metrics aim to evaluate the significance and the frequency  of the errors in approximate arithmetic circuits.

To evaluate the mean error of our designs,
we calculate the Mean Relative Error Distance (MRED),
which is the average of 
the Relative Error Distance (RED) values 
for a set of inputs.
RED refers to the arithmetic difference between the accurate and the approximate result, 
divided by the accurate result. 
In our case,
let $A \cdot B$ be the accurate multiplication
and 
$A \cdot \tilde{B}$ be the approximate multiplication 
(calculated by the RAD$2^k$ multipliers).
$B$ is accurately encoded as in Eq. \eqref{eq_rad1}
and $\tilde{B}$ is approximately encoded as in Eq. \eqref{eq_brad1}.
Considering these encodings,
the RED of the multiplication is calculated by 
Eq. \eqref{eq_redrad1}.

\vspace{-6pt}

\begin{equation}
 \text{RED}_{AB} = \dfrac{\lvert A \cdot B - A \cdot \tilde{B} \rvert}{\lvert A \cdot B \rvert} = \dfrac{\lvert B-\tilde{B} \rvert}{\lvert B \rvert}  
 = \dfrac{\lvert y^{R2^{k}}_0 -\hat{y}^{R2^{k}}_0 \rvert}{\biggl| \mathlarger{\sum}_{\substack{j=k/2}}^{\substack{n/2{-}1}}  4^{j} y_{j}^{R4} + y^{R2^{k}}_0 \biggr|} 
 = \text{RED}_B
 \label{eq_redrad1}
\end{equation}
Hence, 
the RED of the RAD$2^k$ multipliers 
depends only on $B$,
i.e., $\text{RED}_{AB} = \text{RED}_{B}$. 
Let $p_A$ and $p_B$ be the Probability Density Functions (PDFs) of $A$ and $B$, respectively. 
In this case, MRED is calculated by Eq. \eqref{eq_mredrad}.

\vspace{-4pt}

\begin{eqnarray}
 &  \text{MRED} = 
 \mathlarger{\sum}\limits_{\substack{\forall A,B}} p_A(A) \cdot p_B(B) \cdot \text{RED}_{AB} =
 \mathlarger{\sum}\limits_{\substack{\forall A}} p_A(A) \cdot \mathlarger{\sum}\limits_{\substack{\forall B}} p_B(B) \cdot \text{RED}_{B} = & \nonumber \\[4pt] 
 &  \hspace*{-7pt} = \mathlarger{\sum}\limits_{\substack{\forall B}} p_B(B) \cdot \text{RED}_{B} =
       \mathlarger{\sum}\limits_{\substack{\forall B}} p_B(B) \cdot \dfrac{\lvert y^{R2^{k}}_0 - \hat{y}^{R2^{k}}_0 \rvert}{\biggl| \mathlarger{\sum}_{\substack{j=k/2}}^{\substack{n/2-1}} 4^{j} y_{j}^{R4} + y^{R2^{k}}_0 \biggr|} & \label{eq_mredrad}
\end{eqnarray}

Therefore, the MRED of the RAD$2^k$ multipliers 
for a set of input pairs $\{A,B\}$
can be calculated
using only the values of $B$ 
and their PDF values (e.g., for uniform input distribution:
$p_B(B) = 1/2^{n}$).

The possibility of RED $ > M\%$
(PRED$_M$)
is another widely used error metric \cite{2016_Jiang_IEEEtc, 2017_Liu_IEEEtc}.
More specifically, this metric is defined 
as shown in Eq. \eqref{eq_pred}.

\begin{equation}
\text{PRED}_M = p(\text{RED}_{AB} \geq M\%)
=  \dfrac{\# A \cdot B|_{\text{RED}_{AB} \geq M\%}}{\# A \cdot B}
\label{eq_pred}
\end{equation}

\begin{figure}[!t]
\centering
\subfloat[\label{fig_pdf1}]{\includegraphics[width=0.44\textwidth]{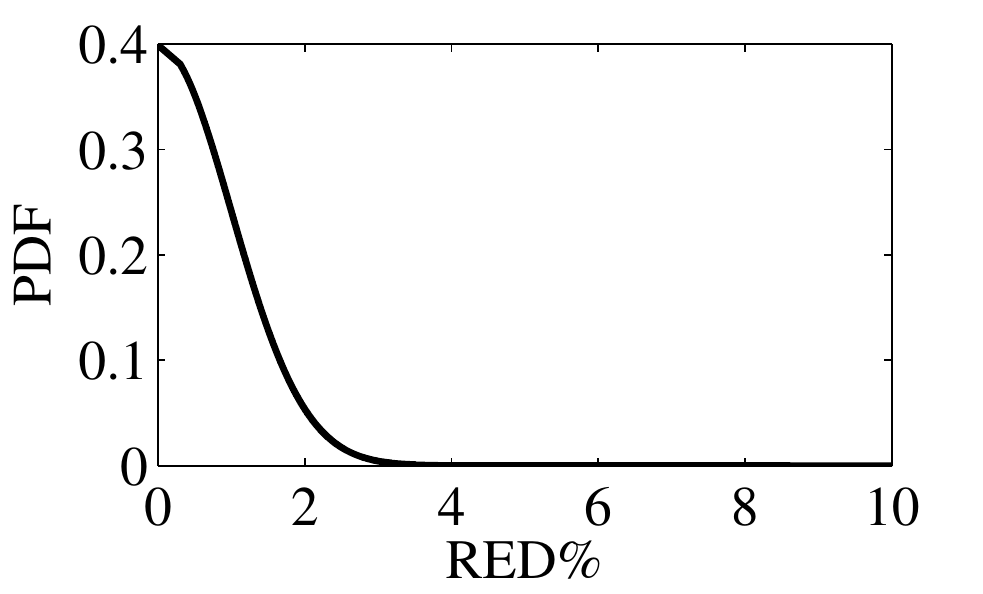}} \hspace{38pt} %
\subfloat[\label{fig_pdf2}]{\includegraphics[width=0.44\textwidth]{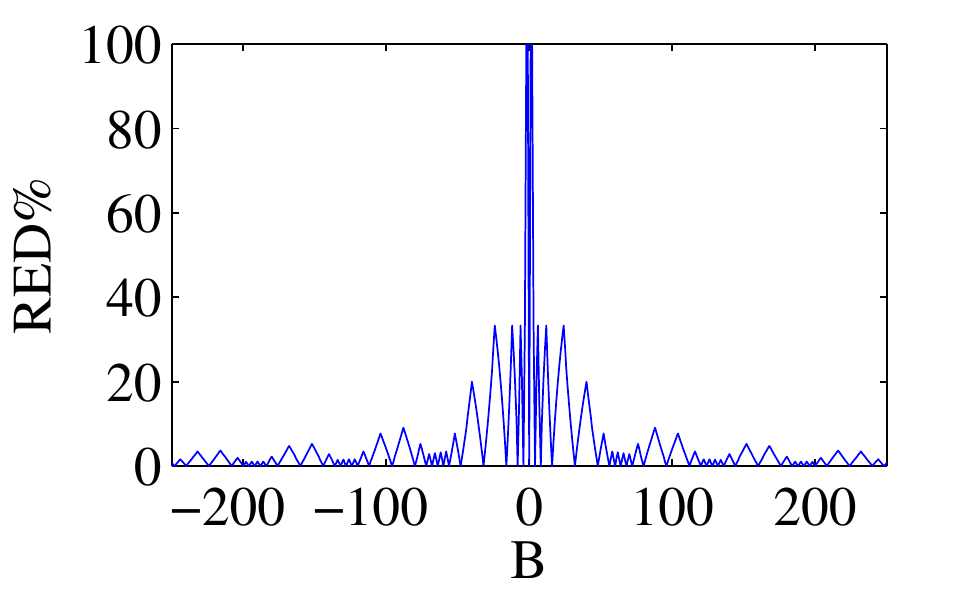}}\\[-3pt]%
\subfloat[\label{fig_pdf3}]{\includegraphics[width=0.44\textwidth]{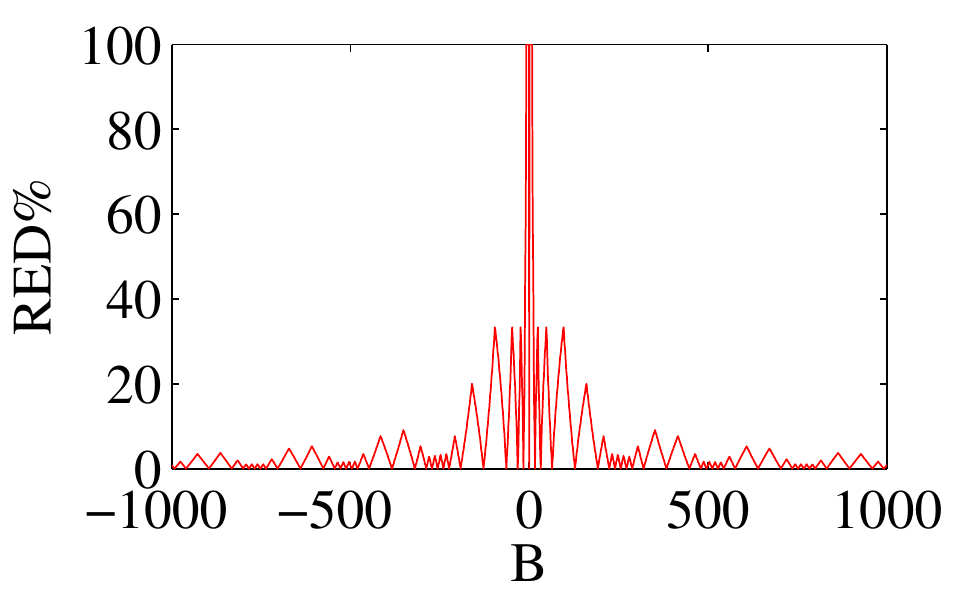}}\hspace{38pt} 
\subfloat[\label{fig_pdf4}]{\includegraphics[width=0.44\textwidth]{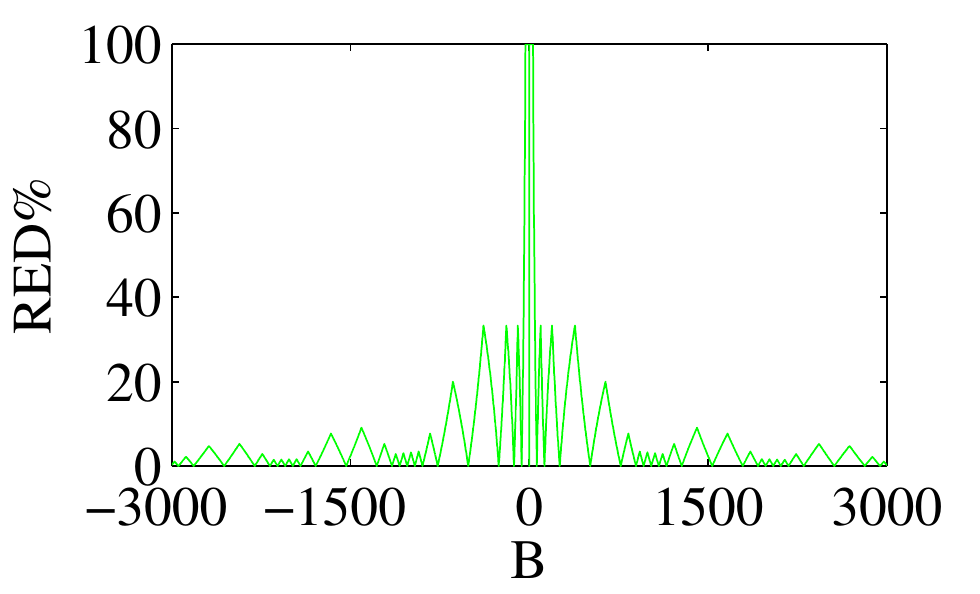}}%
\caption[Error Distribution of the Approximate High-Radix Multipliers]{\textbf{(a)} Probability density function of RED for RAD64.
The RED distribution with respect to the values of $B$ for
\textbf{(b)} RAD64, 
\textbf{(c)} RAD256,
and \textbf{(c)} RAD1024.}
\label{fig_pdfs}
\end{figure}

In our error analysis,
we consider $16$-bit multiplication.  
In Figure \ref{fig_pdf1},
we illustrate the PDFs of RED for the RAD64 multiplier.
RAD256 and RAD1024 deliver a similar curve, 
proving that the probability of having small RED is high.
In Figures \ref{fig_pdf2}--\ref{fig_pdf4},
we plot the RED distribution for RAD$2^{k}$ ($k = 6,8,10$)
with respect to the encoded operand $B$. 
As shown, 
the error follows a Gaussian distribution with bounds for maximum error, i.e., the error is large only for a
small near-zero interval.
For the rest values of $B$ (either positive or negative),
RED $\in [0\%, 1\%]$.
In particular, 
RAD64 exhibits RED larger than $10\%$ 
only if $B\in [-100, 100]$. 
The respective intervals for RAD256 and RAD1024 are $[-400, 400]$ and $[-1500, 1500]$. 
These intervals represent a very small fraction of all the possible values of $B$ for $16$-bit arithmetic.
As a result, 
PRED$_{10}$ is $0.001\%$, $0.004\%$, and $0.02\%$ for RAD64, RAD256, and RAD1024, respectively.

Overall, the proposed $16$-bit RAD$2^{k}$ multipliers
introduce small errors for the majority of the multiplications.
Even RAD1024, which applies the most aggressive approximations, 
features a MRED of $0.93\%$,
while its PRED$_{2}$ and PRED$_{10}$ are $6.74\%$ and $0.02\%$, respectively.
We note that the accuracy loss of approximate designs
should be examined along with the respective gains in resources.
Thus, our analysis in the next section 
examines the trade-off
between accuracy and resources. 

An advantage of the RAD$2^{k}$ multipliers
compared to other state-of-the-art designs is that 
their error, 
as shown in Eq. \eqref{eq_redrad1},
depends only on the operand that is encoded ($B$).
This feature enables the fast calculation of the RED-based error metrics.
In case RED was a function of both $A$ and $B$,
its calculation would require to
use a larger input dataset (to cover different pair combinations)
and also
execute the approximate multiplication.

\subsection{Experimental Results} 
\label{s4_3_3}

This section includes the evaluation of the RAD$2^{k}$ multipliers in terms of resources 
(delay, area, power and energy).
It also examines the accuracy of the approximate designs
in parallel with the provided resource gains.
The evaluation consists of two stages, i.e., 
the comparison of RAD$2^{k}$ with state-of-the-art approximate multipliers
and the exploration of RAD$2^{k}$'s bit-width scaling.

The designs are implemented in Verilog,
synthesized with Synopsys Design Compiler 
and the TSMC 65-nm standard-cell library,
and simulated with Mentor Graphics QuestaSim. 
Both synthesis and simulation are performed at $1$V, 
i.e., the nominal supply voltage.
The critical path delay and the area of the circuits
are reported by Synopsys Design Compiler, 
while the power consumption is measured with Synopsys PrimeTime.
Energy is defined as the product of power and delay. 
Moreover, 
we define the gain of the approximate design 
as the relative resource reduction from the accurate design.

\subsubsection{Comparative State-of-the-Art Evaluation}

For comparison, 
we implement 
the accurate radix-$4$ (ACCR4) 
and radix-$8$ (ACCR8) multipliers, 
as well as
relevant state-of-the-art approximate multipliers
\cite{2016_Jiang_IEEEtc, 2017_Liu_IEEEtc, 2015_Hashemi_ICCAD}. 
R8ABM1 and R8ABM2-15 \cite{2016_Jiang_IEEEtc} 
employ the radix-$8$ encoding 
and calculate the product $3A$ with an adder operating approximately for the $8$ LSBs. 
R8ABM2-15 extends the approximation by 
truncating the $15$ LSBs of the partial products. 
R4ABM1-14, R4ABM1-16, R4ABM2-14, and R4ABM2-16 \cite{2017_Liu_IEEEtc} 
use the accurate radix-$4$ encoder for producing the MSBs of the partial product matrix 
and an approximate radix-$4$ encoder for the $14$/$16$ LSBs. 
The difference between R4ABM1 and R4ABM2
is that the latter design 
performs more aggressive approximations.
For the radix multipliers of \cite{2016_Jiang_IEEEtc} and \cite{2017_Liu_IEEEtc},
the partial product accumulation is performed accurately
with a Wallace tree and a fast adder, like in RAD$2^{k}$. 
Finally, 
we implement DRUM6 \cite{2015_Hashemi_ICCAD},
which selects a $6$-bit segment, 
starting from the leading non-zero bit of the input operands, 
and sets the LSB of the truncated values to `$1$'. 

Table \ref{tb_radhw} presents the results
from the synthesis of the multipliers
for $16$-bit input operands. 
The circuits are configured 
to operate at their critical path delay,
i.e., at maximum frequency. 
We note that 
Table \ref{tb_radhw} also reports error metrics 
(MRED and PRED$_2$), 
which have been obtained by performing simulations over all the possible input combinations.
At first,
we notice the inefficiency of ACCR8
compared to ACCR4, 
which proves that the conventional accurate high-radix encodings
impose significant overheads.
Next, 
we compare the accuracy and the hardware efficiency of the examined multipliers.  

RAD64 gives the best MRED among all the designs, 
as approximations are performed only in the $6$ LSBs of $B$. 
In addition, the approximated values have small absolute distance from the accurate ones.
In R4ABM1-14, 
a very small error is introduced, 
considering that only four entries of the radix-$4$ encoder's K-Map are modified \cite{2017_Liu_IEEEtc}, while R8ABM1 has a small MRED because the approximate adder that calculates $3A$ 
involves both accurate and approximate parts. 
However, the errors of R8ABM1 are of greater significance, 
namely there are large absolute distances from the accurate results. 
In R8ABM2-15, the error does not increase significantly, 
although the $15$ LSBs of the partial products are truncated. 
This is due to the correction `$1$' added to the $17$-th bit that compensates the error generated by the truncated lower part \cite{2016_Jiang_IEEEtc}. 
Finally, DRUM6 delivers the worst MRED among all the multipliers with $1.47\%$, 
although its error distribution is bounded. 
Our RAD$2^{k}$ multipliers deliver MRED smaller than $1\%$ as standalone circuits, meaning that they can be used in real-world applications that tolerate mean errors of $5\%$ or $10\%$ \cite{ParkASPLOS16}.

\begin{table}[!t]
\fontsize{9}{10}\selectfont
\renewcommand{\arraystretch}{1.2}
\setlength{\tabcolsep}{8pt}
\caption[Experimental Results of Approximate Multipliers on TSMC 65-nm Standard-Cell]{Experimental results of the $16$-bit approximate multipliers on TSMC 65-nm standard-cell.}
\label{tb_radhw}
\centering
\begin{tabular}{l| c c c c| c c}
\hline
\multicolumn{1}{c|}{\multirow{2}{*}{\textbf{Design}}} & \textbf{Delay} & \textbf{Power} & \textbf{Area} & \textbf{Energy} & \textbf{MRED}  & \textbf{PRED$_\mathbf{2}$} \\[-2pt]
& 
(ns) & 
($\upmu$W) & 
($\upmu$m$^2$) & 
($\upmu$W$\cdot$ns) &  
$(\%)$ &
$(\%)$ \\
\hline \hline 
ACCR4                         & $0.75$ & $4998$ & $4153$ & $3749$ & $-$    & $-$ \\
ACCR8                         & $0.80$ & $5343$ & $4639$ & $4274$ & $-$    & $-$ \\
RAD64                         & $0.72$ & $4497$ & $3489$ & $3238$ & $0.08$ & $0.42$ \\
RAD256                        & $0.69$ & $3493$ & $2769$ & $2410$ & $0.28$ & $1.69$ \\
RAD1024                       & $0.65$ & $3422$ & $2624$ & $2224$ & $0.93$ & $6.74$ \\
R8ABM1 \cite{2016_Jiang_IEEEtc}     & $0.77$ & $5058$ & $4210$ & $3895$ & $0.15$ & $1.05$ \\
R8ABM2-15 \cite{2016_Jiang_IEEEtc}  & $0.74$ & $3377$ & $2926$ & $2499$ & $0.61$ & $2.70$ \\
R4ABM1-14 \cite{2017_Liu_IEEEtc}    & $0.74$ & $4676$ & $3958$ & $3460$ & $0.12$ & $0.41$ \\
R4ABM1-16 \cite{2017_Liu_IEEEtc}    & $0.73$ & $4447$ & $3725$ & $3246$ & $0.49$ & $1.49$ \\
R4ABM2-14 \cite{2017_Liu_IEEEtc}    & $0.73$ & $4648$ & $3732$ & $3393$ & $0.24$ & $0.68$ \\
R4ABM2-16 \cite{2017_Liu_IEEEtc}    & $0.72$ & $4307$ & $3467$ & $3101$ & $1.18$ & $2.50$ \\
DRUM6 \cite{2015_Hashemi_ICCAD} & $1.07$ & $2148$ & $3993$ & $2298$ & $1.47$ & $28.85$ \\
\hline
\end{tabular}
\end{table}

DRUM6 has the worst critical path delay, 
as it initially calculates the absolute value of the products \cite{2015_Hashemi_ICCAD}, 
while our designs  
RAD64, RAD256, and RAD1024 are the fastest circuits, 
with delays of $0.72$ns, $0.69$ns, and $0.65$ns, respectively. 
Regarding energy, 
R8ABM1 delivers the worst consumption among all the multipliers, as it may use an approximate adder to calculate $3A$, but it contains a $7$-bit precise adder to avoid very large errors \cite{2016_Jiang_IEEEtc}. 
The energy is improved in R8ABM2-15 due to the truncated partial product bits. 
However, as our multipliers feature small MRED, 
the truncation technique can also be applied to them 
to provide even better resource gains. 
The radix-$4$ multipliers of \cite{2017_Liu_IEEEtc} exhibit similar results, 
with R4ABM2-16 delivering the best  area and energy gains 
from their design family. 
RAD1024 features the best energy consumption, with DRUM6 being second. 
However, the latter delivers significantly larger MRED than all the other multipliers.
The same applies to 
its PRED$_2$, which is increased ($28.85\%$),
while the other multipliers 
have PRED$_2$ smaller than $3\%$ (apart from RAD1024 that has PRED$_2$ equal to $6.74\%$).

\begin{figure}[!t]
\centering
\includegraphics[width=0.83\textwidth]{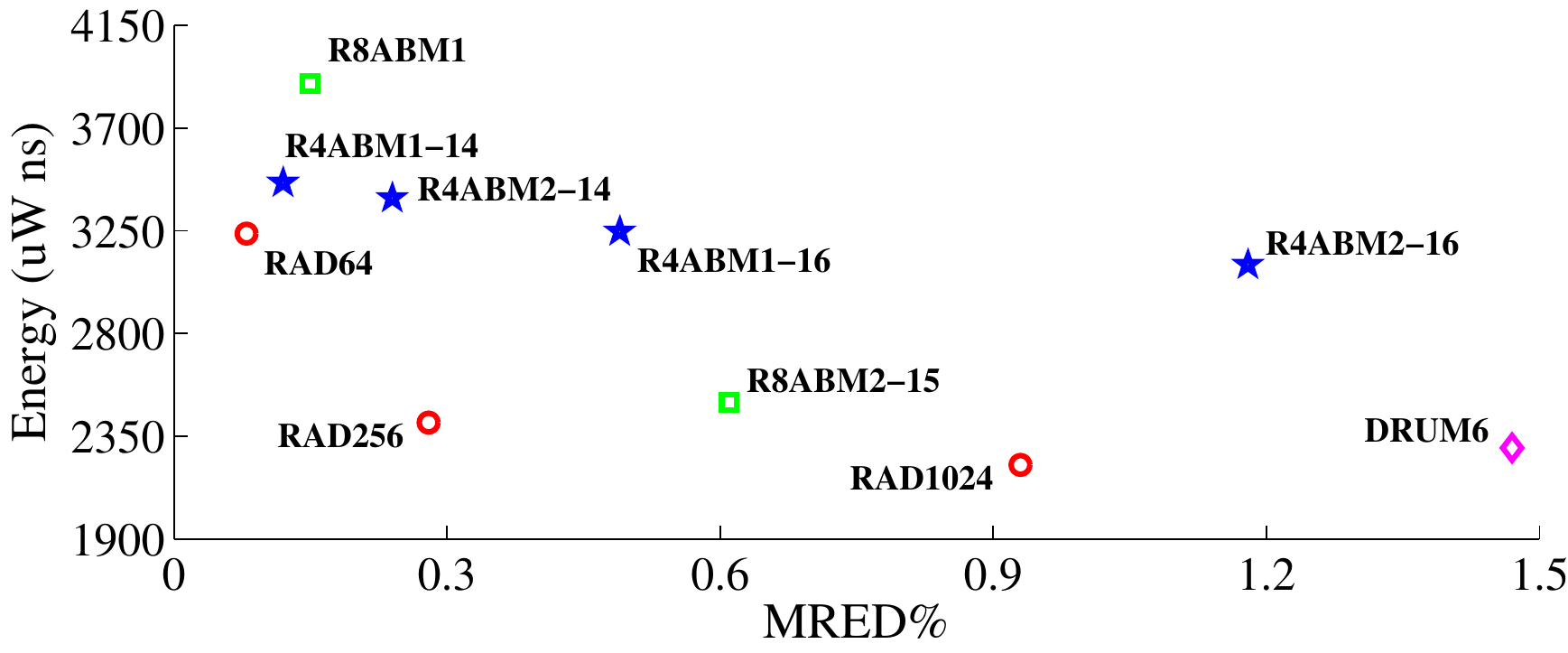}%
\caption[Comparative Pareto Analysis for Approximate Multipliers]{Comparative Pareto analysis for the approximate multipliers considering MRED and energy.}%
\label{fig_radpareto}
\vspace*{-10pt}
\end{figure}

Figure \ref{fig_radpareto} illustrates the scatter plot 
of the examined multipliers for MRED and energy consumption. 
The purpose of this plot
is to highlight the most prominent designs
when considering both error and energy consumption. 
As shown, 
the Pareto Front is formed exclusively by RAD$2^{k}$ multipliers.
Namely, 
RAD$2^{k}$ constitute the most efficient approximate multiplier alternatives 
compared to all the examined state-of-the-art approximate multipliers, 
as they exhibit the best energy--MRED trade-off.
RAD64 attains the smallest MRED value 
and should be preferred when error is of high importance, while RAD1024 delivers the highest energy reduction. 
Finally, 
RAD256 features significant energy reduction for a very small error value. 

\begin{figure}[!t]
\centering
\includegraphics[width=0.83\textwidth]{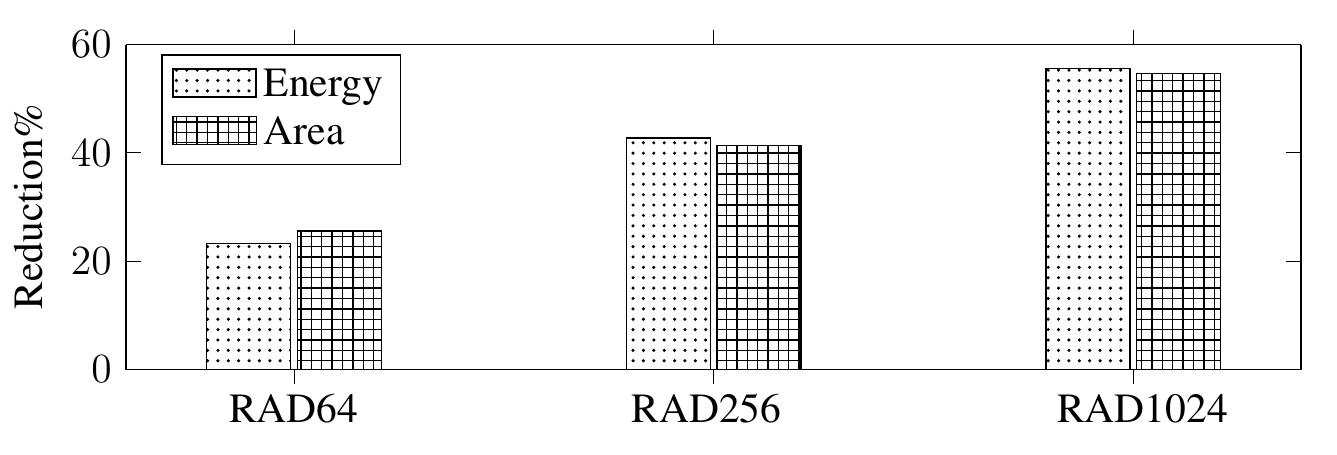}%
\caption[Area and Energy Gains of the Approximate High-Radix Multipliers]{Area and energy gains of RAD$2^{k}$ compared to the accurate radix-$4$ multiplier.}%
\label{fig_radgains}
\vspace*{-5pt}
\end{figure}

Next, 
we evaluate the RAD$2^{k}$ multipliers 
by exploring the energy and area gains compared to the accurate radix-$4$ multiplier. 
The main target of Approximate Computing is to trade accuracy for energy gains, 
i.e., generate good enough results at comparable performance and lower energy consumption. 
Thus, 
in this evaluation, we leverage the delay slack between the RAD$2^{k}$ multipliers and the accurate one, and targeting energy efficiency, 
we synthesize and simulate the RAD$2^{k}$ designs at the critical path delay of ACCR4.
Figure \ref{fig_radgains} reports the delivered gains of RAD$2^{k}$ multipliers compared to ACCR4, 
when they operate at the same frequency.
Remarkable reduction in area and energy
is attained by the RAD$2^{k}$ multipliers, 
and it is shown that the gains become larger 
when using higher radix encoding, because less partial products are generated. 
In these designs, 
the number of the partial products is reduced from $8$ to $6$, $5$ and $4$, 
and thus, 
the area and the depth of the accumulation tree is reduced by up to $50\%$. 
As expected, RAD1024 delivers the largest gains 
($56\%$ in energy consumption and $55\%$ in area).


The presented analysis regards fixed-point arithmetic, 
but it can be also extended for floating-point multiplication.
In the latter case, 
the proposed RAD$2^{k}$ multipliers can replace the fixed-point multiplication of the mantissas,
which is executed in floating-point multiplication.
Considering that the error of RAD$2^{k}$ is very small, 
it will affect only the mantissa's accuracy,
whereas the exponent value (calculated by an accurate adder) 
will remain the same as in the accurate floating-point multiplication.
Therefore, 
similar error values are expected in RAD$2^{k}$-based
floating-point multiplication.
In terms of energy gains, 
they depend on the floating-point representation.
For higher floating-point precision, 
similar or even larger energy gains are expected.

\subsubsection{Evaluation of Bit-width Scaling}

Finally, 
we examine the efficiency of RAD$2^{k}$
when increasing the multiplier's bit-width,
i.e., to $24$ and $32$ bits.
We consider ACCR4 as the baseline design
and an MRED of $0.28\%$ as a quality constraint
that should be satisfied. 
We select $0.28\%$ as error constraint
because it is 
the MRED of the $16$-bit RAD256 multiplier,
which is the most efficient design
when taking into account both the provided energy gains 
and the error values.
Namely, as shown in Figure \ref{fig_radpareto}, 
RAD256 attains a very efficient energy-error trade-off, 
i.e., significant error reduction for very small error.

\begin{figure}[!t]
\centering
\includegraphics[width=0.77\textwidth]{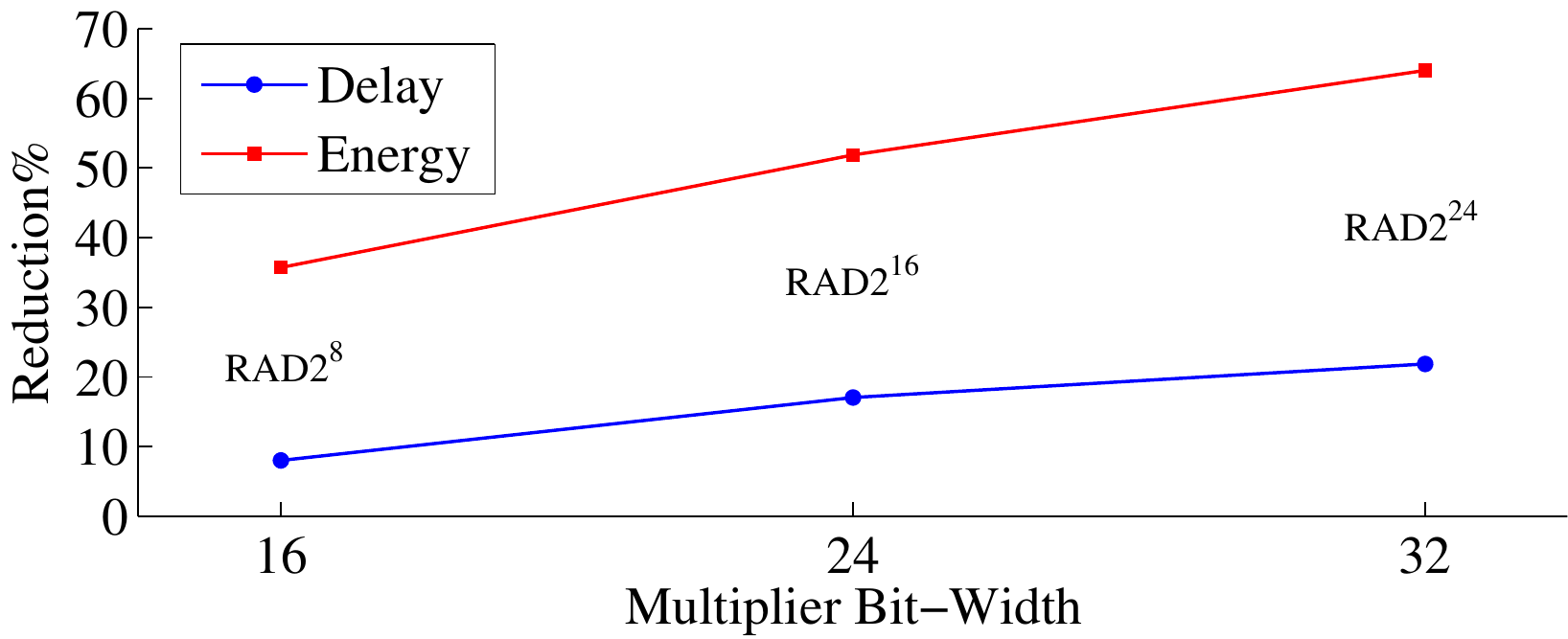}%
\caption[Energy and Delay Gains of the Approximate High-Radix Multipliers for Scaled Bit-width]{Energy and delay gains of RAD$2^{k}$ for scaled bit-width and MRED $ = 0.28\%$ (error constraint).}%
\label{fig_radscale}
\end{figure}

Figure \ref{fig_radscale} presents
the scaling of the gains  
in delay and energy 
with respect to the multiplier's size.
For the $16$-bit arithmetic,
we implement
the high-radix-$2^8$ multiplier (RAD256),
while for $24$-bit and $32$-bit arithmetic, 
we implement 
the high-radix-$2^{16}$ and high-radix-$2^{24}$ multipliers, respectively, 
in order to deliver the same MRED of $0.28\%$.
The reductions in energy consumption and delay scale up to $64\%$ and $22\%$, respectively, 
while the quality constraint is satisfied. 
Thus, 
the results show that as the multiplier's size increases, 
the proposed hybrid high-radix encoding 
achieves larger gains in critical path delay and energy consumption for the same error.
The scaling behavior is theoretically confirmed, 
as the RAD$2^{k}$ designs of Figure \ref{fig_radscale} have an MRED of $0.28\%$, 
while generating $37\%$, $58\%$, and $60\%$ less partial products for $16$, $24$, and $32$ bits, respectively. 
More specifically, 
each one generates $5$ partial products in total
($1$ approximate and $4$ accurate).  
In contrast, 
ACCR4 generates 
$8$, $12$, and $16$ partial products,
respectively. 

\section{Conclusion}
\label{s4_4}

In this chapter, 
we introduced an approximate hybrid high-radix encoding for generating the partial products of a signed multiplier. 
The most significant bits of the multiplicand are encoded with the accurate radix-$4$ encoding, 
while its $k$ least significant bits 
are encoded with an approximate high-radix-$2^k$ encoding.
The parameter $k$ determines the approximate high-radix encoding,
and thus the approximation degree,
and it can be tuned to 
provide the desired trade-off between accuracy and resources.
Our approximation approach 
maps all the high-radix values
to a set including only the $4$ largest powers of two.
Therefore, 
it surpasses the bottlenecks of the conventional high-radix encodings,
which provide partial product reduction,
however,
they suffer from increased encoding logic.
The error of the RAD$2^{k}$ multipliers
follows a Gaussian distribution with near-zero average.
The mean relative error of our designs 
lies in the range $0.08\%$--$0.93\%$,
and also,
it depends only on the approximately encoded operand,
allowing the fast calculation of the error metrics
and eliminating the need for exhaustive circuit simulations.
In terms of resource savings, 
our designs achieve 
up to $56\%$ energy and $55\%$ area gains
compared to the accurate radix-$4$ multiplier, 
when operating at the same frequency.
Compared to state-of-the-art approximate multipliers, 
RAD$2^{k}$ constitute better approximate design alternative, 
as they form the Pareto front in our analysis
involving energy and error.
Finally, 
our approximation technique is scalable, 
delivering larger resource gains 
as the multiplier's size increases,
while keeping the error constant.  
More specifically,
for $32$-bit multiplication,
the gains reach to 
$64\%$ and $22\%$ in
energy and delay,
respectively,
while the mean relative error is retained at $0.28\%$. 

\chapter{Dynamic Approximation: Runtime-Configurable Arithmetic Circuits}
\label{chapter5}

\addtocontents{lof}{\protect\contentsline{chapter}{\protect\numberline{5}Dynamic Approximation: Runtime-Configurable Arithmetic Circuits}{}{}}
\addtocontents{lot}{\protect\contentsline{chapter}{\protect\numberline{5}Dynamic Approximation: Runtime-Configurable Arithmetic Circuits}{}{}}

\begin{ChapterAbstract}
The challenging deployment of 
Digital Signal Processing (DSP) and Artificial Intelligence (AI)
algorithms
pushes the community to examine alternative design approaches,
such as Approximate Computing.
This novel design paradigm 
provides valuable resource gains 
by exploiting the error tolerance of the DSP/AI applications.
Nevertheless,
approximate designs with fixed approximation configuration
provide limited flexibility 
and cannot accommodate different workloads 
or tune the quality of the results
with respect to the given accuracy and energy constraints.
As a result,
there is a growing need
for approximate circuits and systems that 
support multiple approximation configurations
and can seamlessly change among them at runtime.
In this context,
we design runtime-configurable approximate multipliers
for integer/fixed-point and floating-point arithmetic.
We employ two orthogonal approximation techniques,
i.e., partial product perforation and partial product rounding,
to enlarge the approximation space,
and 
we provide a low-overhead configuration scheme
for tuning the approximation at runtime. 
The evaluation is performed
for both the design-time (static) 
and runtime (dynamic) approximate variants,  
and it involves an in-depth error analysis 
and diverse experimental results. 
The error analysis shows that our designs
feature slow error scaling with multiple values,
which allows them to satisfy various accuracy constraints.
According to the experimental results,
the design-time variants
outperform all the examined state-of-the-art multipliers, 
considering either error or resource gains
as target. 
The runtime variants 
provide negligible area overhead 
(e.g., $\mathit{4}$\% and $\mathit{2}$\% for half and
single floating-point precision, respectively) 
and smaller energy gains,
however,
they still deliver remarkable gains versus the accurate multiplier
and other state-of-the-art design-time multipliers.
In more detail, 
the fixed-point runtime variant consumes only
up to
$\mathit{1.2\times}$ and $\mathit{1.7\times}$
more energy than 
its design-time variant
for low-strength and more aggressive approximation, respectively. 
Correspondingly, 
the floating-point runtime variant
provides 
\raisebox{0.8pt}{$\scriptstyle\sim$}$\mathit{1.4\times}$
and \raisebox{0.8pt}{$\scriptstyle\sim$}$\mathit{1.6\times}$
less energy gains
in half and single precision, respectively. \\
This chapter is based on our
\textbf{publications} in 
\textbf{\cite{LeonMicro, LeonTECS}}.
\end{ChapterAbstract}

\newpage 

\section{Introduction}

The proliferation of compute-intensive workloads
from domains such as Digital Signal Processing (DSP) and Artificial Intelligence (AI)
is changing the landscape 
in both cloud and embedded computing.
The efficient deployment of DSP/AI applications
is a first-class concern 
due to their
increased computational and memory demands, 
as well as the resource constraints
imposed by the computing systems 
(e.g., specific energy budget 
or limited number of available processing units).
For this reason,
the research community explores 
new design alternatives towards 
power-efficient and high-performance computing.  

One of the most attractive and well-established 
solutions is the emerging paradigm of 
Approximate Computing 
\cite{2016_Mittal_ACMsrv, 2016_Xu_IEEEdt, 2021_Stanley_ACMsrv}, 
which exploits the 
inherent approximate nature
and error resilience of the DSP/AI applications \cite{ChakradharDAC2010, ChippaDAC2013}.
This design approach 
trades accuracy loss for resource gains, 
e.g., in power, area, or throughput.
In particular,
errors are inserted in the computations based on a systematic and disciplined approach,
which aims to provide resource gains
while retaining the quality of the results at acceptable levels.
Approximation techniques are applied at all layers of the computing pyramid,
i.e., 
from algorithms to software and down to circuits and transistors. 

The increased diversity of the 
real-world error-resilient applications 
demands flexible approximate designs
that offer various approximation configurations.
More explicitly, 
the approximate system/circuit/architecture  
needs to be capable of adjusting the approximation degree
in order to satisfy the end-to-end application-specific accuracy, 
while providing the desired performance 
within the constrained power envelope. 
Even for the same application,
different input distributions may require 
a new approximation configuration
to provide the acceptable quality of results 
(e.g., consider an approximate video processor
that inputs frames with different content). 
As a result,
there is a growing need 
for approximate designs
that can:
(i) provide multiple approximation configurations (namely different levels of accuracy), 
and
(ii) dynamically configure their approximation degree, 
either offline 
(e.g., before starting the execution of a new application)
or at runtime, 
with respect to the given accuracy/power constraints.

In this chapter,
motivated by the demand for dynamic approximation configuration,
we target the arithmetic circuits  
and design approximate multipliers that can configure their 
\emph{approximation degree at runtime}. 
Contrary to the RAD$2^k$ circuits of Chapter \ref{chapter3},
which are configured at design-time 
to implement a fixed (frozen) approximation,
the designs of this chapter 
provide a larger approximation space 
and seamless dynamic configuration.
The proposed family of runtime-configurable approximate multipliers
can be integrated in processors
and custom hardware accelerators 
that need to tune the approximation degree. 
We note that we 
neither examine which approximation configuration should be selected
nor we apply automatic approximation tuning.
Our goal is to implement 
circuits that can change approximation
on an efficient way
(without significant area overhead and timing penalties). 

Additionally, 
compared to Chapter \ref{chapter3} and Chapter \ref{chapter4}, 
we examine floating-point arithmetic,
i.e., 
we design both runtime-configurable fixed- and floating-point multipliers.
Arithmetic computations impose a trade-off in range, precision, and hardware resources.
Range is the capability of representing small/large numbers, while precision is the differentiation between nearby values.
Fixed-point arithmetic delivers hardware-friendly designs, 
but it sacrifices range and offers limited precision.
On the other hand, floating-point arithmetic provides a larger range of values and higher precision for the same word-length, but it suffers from increased hardware cost.
Furthermore, 
the floating-point format 
offers a simplified programming model, contrary to fixed-point, which needs to compensate for the quantization noise.
Below, we discuss the significance of the floating-point arithmetic
and analyze 
what pushed us to design floating-point multipliers. 

Numerous compute-intensive algorithms use a wide range of values and require high precision.
Therefore, floating-point arithmetic is favored in applications from 
domains such as DSP, computer graphics, scientific computing, and speech recognition, 
which handle real numbers and produce results with unpredictable range.
However, the increased hardware cost of floating-point calculations results in using fixed-point arithmetic or alternative data formats that are more hardware-efficient, 
but they 
do not provide the benefits of floating-point.
For this reason, 
there is also limited integration of 
Floating-Point Units (FPUs) in embedded devices, 
e.g., the commercial Field-Programmable Gate Arrays (FPGAs)
do not have hardwired FPU blocks \cite{Langhammer2015}.
The power inefficiency of FPUs is also proven by a recent study on Graphics Processing Units (GPUs) \cite{ZhangDAC2014}.
This study
revealed that the portion of the power consumed for arithmetic operations in compute-intensive benchmarks reaches more than the $70\%$ of the total power, 
with the FPU being the most power-hungry unit.
Specifically for floating-point multiplication,
it is widely used in operations such as convolution, matrix multiplication, and Fourier transform, 
and thus, its efficiency inherently affects the entire application. 
Recent studies on general OpenCL applications from the AMD APP SDK showed that over $85\%$ of the floating-point arithmetic involved multiplication \cite{ImaniDAC2019}.
Although floating-point multipliers are more expensive in power and area, they have received less research attention 
than their fixed-point counterparts 
for disciplined approximations 
\cite{KulkarniICVLSI2011, LiuDATE2014, 2015_Momeni_IEEEtc, 2017_Liu_IEEEtc, 2017_Akbari_IEEEtvlsi,  JiangTCASI2019, 2019_Venkatachalam_IEEEtc, NarayanamoorthyTVLSI2015, 2015_Hashemi_ICCAD, 2017_Zendegani_IEEEtvlsi, 2019_Vahdat_IEEEtvlsi}.

Regarding the technical details of our work,
we employ two orthogonal approximation techniques,
i.e., \emph{partial product perforation} 
and \emph{partial product rounding}. 
The first technique perforates entire partial products,
while the second technique rounds them to a smaller bit-width.
These two techniques 
are applied in collaboration 
to generate a large approximation space,
satisfying the requirement for multiple approximation configurations
and varying levels of accuracy. 
More specifically,
at first, 
we discard partial products starting from the least significant,
and then,
we apply rounding to the remaining ones.
Besides offering a large approximation space,
we select these two techniques
because their configuration is associated with the input operands, 
and thus,
we can easily change it at runtime with negligible area overhead.
This overhead is $2n$ AND gates
and two $n$-bit control signals (one per operand), 
where $n$ is the operand bit-width.
The approximation configuration
is determined by simply setting the bits of the control signals
to either `$1$' or `$0$'.

The \textbf{contribution} of this chapter is summarized as follows:

\begin{itemize}[]
\item[(i)] We highlight the significance of dynamic approximation configuration and integrate this attractive feature in the design of approximate multipliers.
\item[(ii)] We extend our approximation techniques to floating-point arithmetic, which imposes increased hardware cost compared to integer/fixed-point arithmetic.
\item[(iii)] We combine two orthogonal approximation techniques
to generate a large approximation space,
which can accommodate numerous accuracy constraints
and explore the accuracy--energy trade-off to provide the most efficient solution.
\item[(iv)] We introduce a low-overhead dynamic configuration scheme for adjusting the approximation degree at runtime.
\item[(v)] We show that the proposed solution outperforms related state-of-the-art designs 
in both fixed- and floating-point arithmetic, 
providing remarkable area and energy gains for comparable error values and slow error scaling. 
\end{itemize}

The remainder of this chapter is organized as follows. 
Section \ref{s5_2} introduces the proposed approximation techniques
and the design-time and runtime variants of our approximate fixed- and floating-point multipliers. 
Section \ref{s5_3} includes the evaluation of the proposed designs,
including error analysis and comparative experimental results.
Finally, 
Section \ref{s5_4} draws the conclusions.

\section{Design of Runtime-Configurable Approximate Multipliers}
\label{s5_2}

To facilitate the dynamic configuration of the approximation degree,
we target to approximate partial product generation,
i.e., the multiplication stage that is associated with the input operands.
Moreover, 
to provide multiple approximation configurations, 
we employ two orthogonal approximation techniques,
namely, 
the partial product perforation 
and the partial product rounding.
The approximation degree of each technique is tuned independently 
and can be easily tailored to satisfy 
various design constraints, 
e.g., maximum error bound or specific power budget.
Regarding the multiplication scheme, 
we select the radix-$4$ encoding for generating the partial products, 
as it outperforms other well-established algorithms \cite{leon_IET}.
Firstly,
we present our approximate design for integer/fixed-point arithmetic,
which is also the main approximate component of our floating-point unit.
Subsequently, 
we present the extension of our design to support
dynamic approximation configuration.

\subsection{AxFXU: Approximate Fixed-Point Multiplier}
Let $A = \langle a_{n-1} a_{n-2} \cdots a_0\rangle_{2\text{'s}}$
and
$B = \langle b_{n-1} b_{n-2} \cdots b_0\rangle_{2\text{'s}}$
be two $n$-bit $2$'s-complement numbers. 
The accurate radix-$4$ multiplication $A \times B$
is performed by generating and accumulating $n/2$ partial products.
The application of partial product perforation
omits the generation of
$P$ successive partial products,  
starting from the least significant ones.
Afterwards, 
partial product rounding is applied 
to round the remaining partial products
to a smaller bit-width,
i.e., it truncates their $R-1$ Least Significant Bits (LSBs) 
and add their $(R-1)$-th bit to the most significant part.
We name this family of approximate circuits as
AxFXU$|_{P,R}$,
which denotes approximate fixed-point multiplication
with $P$ and $R$ configuration of perforation and rounding, 
respectively. 

Using the accurate radix-$4$ encoding for $B$,
as presented in Chapter \ref{chapter4},
the approximate multiplication in AxFXU
is performed as shown in Eq. \eqref{eq_pr1}.

\begin{equation}
A \times B|_{P,R} = 
\mathlarger{\sum}_{\substack{j=P}}^{\substack{n\text{/}2-1}} 4^{j} \tilde{P\!P}_j  =
\mathlarger{\sum}_{\substack{j=P}}^{\substack{n\text{/}2-1}} 4^{j} A_R \cdot  y_{j}^{R4} 
\label{eq_pr1}
\end{equation}
\begin{align}
\text{where } \; \; \; &  P \in [0, n/2-1), \;  \; R \in [0, n-1) \\[4pt]
&  y_{j}^{R4} = -2 b_{2j+1} + b_{2j} + b_{2j-1}  
\;\; \implies \;\; y_{j}^{R4} \in \{0, \pm 1, \pm 2\}  \\[4pt]
&  A_R = \langle a_{n-1} a_{n-2} \cdots a_R\rangle_{2\text{'s}} + a_{R-1}
\label{eq_pr2} 
\end{align}

\begin{figure}[!t]
\vspace*{6pt}
\centering
\includegraphics[width=0.77\textwidth]{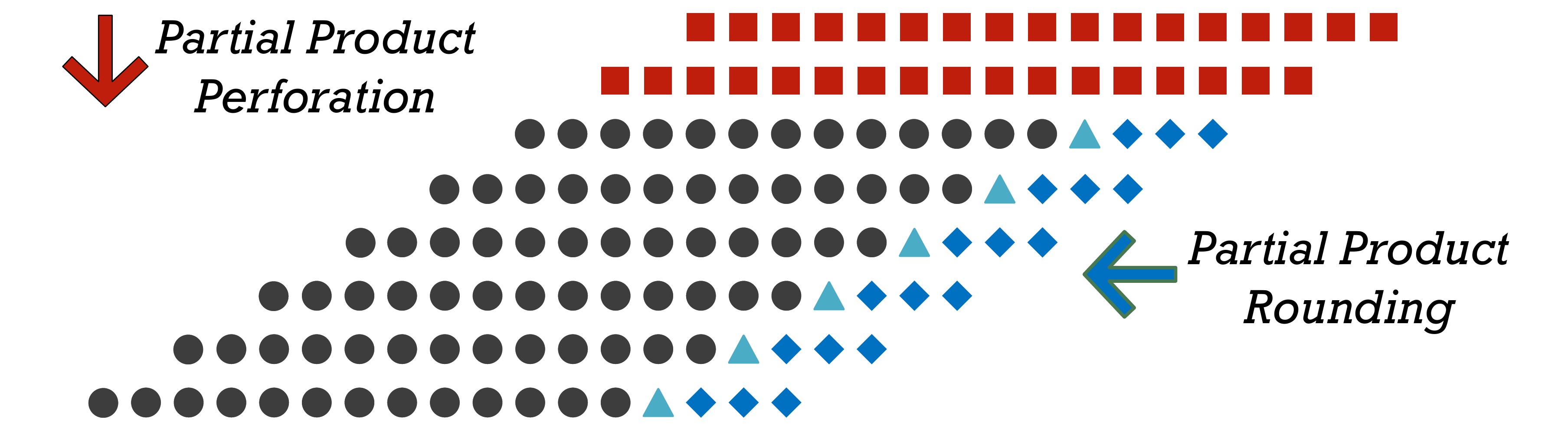}%
\caption[Partial Product Matrix of the Approximate Perforation-\&-Rounding Multiplier]{Approximate partial product matrix that is generated
using partial product perforation ($P=2$) and rounding ($R=4$). \\ 
Symbols: 
\protect\tikz \protect\fill[red_perf] (0.77ex,0.77ex) rectangle (0.22,0.22);: perforated bits
\hspace{2pt}
\protect\tikz \protect\fill[blue_round1] 
(0,0.7ex) -- (0.75ex,1.4ex) -- (1.5ex,0.7ex) -- (0.75ex,0ex) ;: truncated bits
\hspace{2pt}
\protect\tikz \protect\fill[blue_round2] (0,0) -- (0.75ex,1.4ex) -- (1.5ex,0) -- (0,0);: bits added to the remaining words for rounding
\hspace{2pt}
\protect\tikz \protect\fill[blue_dlsb] (1ex,1ex) 
circle (0.63ex);: accurately generated (remaining) bits}%
\label{fig_pr}
\end{figure}

Therefore, 
the final approximate product of AxFXU
is calculated by
accumulating the $n/2 - P$ most significant rounded partial products $\tilde{P\!P}_j$. 
The application of the two approximation techniques is independent, 
as illustrated in Figure \ref{fig_pr}.
As shown, the partial product matrix is reduced vertically
by perforation and horizontally by rounding.

The partial product generation technique
removes $P$ radix-$4$ encoders
and $P \times n$ $1$-bit partial product generators
from the accurate multiplier
(see Section \ref{s4_2} of Chapter \ref{chapter4}).
These components are required to implement
the perforated partial product bits
(red rectangles in Figure \ref{fig_pr}).
To improve the efficiency of rounding
and avoid possible penalties
due to the addition of $a_{R-1}$ 
(blue rhombuses in Figure \ref{fig_pr}), 
we adopt our bit-level manipulations
for DLSB arithmetic 
(see Section \ref{s3_4} of Chapter \ref{chapter3}),
assuming $a_{0+} = a_{R-1}$.
As a result, 
the remaining partial product bits are generated 
like in the accurate multiplier,
and only one XOR gate is added in the calculation of each correction term,
i.e., $sign_j \oplus a_{R-1}$ is used instead of $sign_j$.

\subsection{AxFPU: Approximate Floating-Point Multiplier}

Next, 
we introduce the family of our approximate floating-point multipliers,
which are 
named AxFPU$|_{P,R}$.
This design
makes use of AxFXU 
for multiplying the mantissas,
it employs the logic of the conventional floating-point multiplier
for the rest operations (e.g., exponent addition),
and also, 
it integrates some extra functionalities 
to handle the approximations. 
Before introducing AxFPU, 
we make a brief introduction in floating-point arithmetic 
and the representation of the floating-point data.
This information is required
to understand the design of the floating-point multiplier,
but mainly, we include it 
as background to the error analysis of AxFPU.

\subsubsection{The Floating-Point Arithmetic}
Floating-point arithmetic performs a systematic approximate mapping of the real arithmetic.
More specifically, the floating-point format is an established encoding for representing a finite subset of the continuum of real numbers. 
A floating-point datum can be 
a finite non-zero number, 
a signed zero ($\pm0$), 
a signed infinity ($\pm\infty$), or 
a Not-a-Number (NaN).

In base (radix) $b \geq 2$, 
the finite non-zero numbers and the signed zero 
are represented approximately 
with a fixed number of significant digits 
(called mantissa), 
and they are scaled by raising the base to an integer
(called exponent).
In addition, they include a sign that indicates 
if they are positive or negative.
These three parameters
are defined in more detail as follows:

\vspace{-7pt}

\begin{itemize}[noitemsep]
    \item the sign is `$0$' (positive) or `$1$' (negative).  
    \item the exponent is any integer that belongs in the interval $[e_{min},e_{max}]$, where $e_{min} = 1 - e_{max}$.
    \item the mantissa is any number that belongs in the interval $[0,b)$, and it is represented as $\langle d_0 \cdot d_1 d_2 \dots d_{m-1} \rangle$, where $0 \leq d_i < b$.
\end{itemize}

\vspace{-7pt}

The smallest normal floating-point magnitude is $b^{e_{min}}$.
All the non-zero floating-point numbers with magnitude less than $b^{e_{min}}$ are called subnormal, because they lie between zero and the smallest normal magnitude.

The IEEE-754 standard \cite{ieee754} defines various encodings 
for the binary floating-point format ($b=2$).
The most widely used binary formats 
are presented in the left side of Table \ref{tb_fpieee}.
The Most Significant Bit (MSB) of a floating-point datum is the sign ($S$).
The exponent is encoded using an offset, 
referred as exponent bias in the IEEE-754 standard, 
which is equal to $e_{max}$.
Therefore, the biased exponent $E$ is equal to $e + e_{max}$, where $e$ is the true exponent, 
and it is stored as a $w$-bit unsigned integer.
Regarding the mantissa, 
its leading bit ($d_0$) is implicitly encoded with the biased exponent. 
Thus, 
only $m-1$ bits of mantissa are stored, 
i.e., $M = \langle d_1 d_2 \dots d_{m-1} \rangle$, 
even though the total precision is $m$ bits.

In case the biased exponent $E$ 
belongs in the interval $[1, 2^w-2]$, 
the floating-point datum 
represents the normal value of Eq. \eqref{eq_normal}.

\vspace*{-5pt}

\begin{equation}
(-1)^{S} \cdot 2^{E-e_{max}} \cdot \Bigg(1 + \sum_{i=1}^{m-1} 2^{-i}d_i \Bigg) 
\label{eq_normal}    
\end{equation}

In case the biased exponent $E$ is zero 
and the mantissa $M$ is non-zero, 
the floating-point datum is a subnormal value.
Additionally, 
special numbers ($\pm0$, $\pm\infty$, NaN) are defined depending on the values of exponent and mantissa. 
All the possible floating-point data types 
are summarized in the right side of Table \ref{tb_fpieee}.

\subsubsection{Design of Accurate Architecture}

\begin{table}[!t]
\fontsize{9}{10}\selectfont
\setlength{\tabcolsep}{2.4pt}
\renewcommand{\arraystretch}{1.2}
\caption[IEEE Floating-Point Formats and Data Types]{The IEEE floating-point formats and data types \cite{ieee754}.}
\label{tb_fpieee}
\centering
\begin{tabular}{lccc}

\hline
\multicolumn{1}{c}{\multirow{3}{*}{\textbf{Format Properties}}}
 & \multicolumn{3}{c}{\textbf{Precision}}
 \\
\cmidrule(lr){2-4}
 & \textbf{Half} & \textbf{Single} & \textbf{Double} \\
\hline\hline
Total Bits      & $16$ & $32$  & $64$   \\
- Sign Bit     & $1$  & $1$   & $1$           \\ 
- Exponent Bit  ($w$) & $5$  & $8$   & $11$     \\
- Mantissa Bits ($m-1$) & $10$ & $23$  & $52$   \\
Exponent Bias ($e_{max}$) & $15$ & $127$ & $1023$    \\
\hline
\end{tabular}
\hspace{-1pt}
\begin{tabular}{ccc}
\hline
 \multicolumn{1}{c}{\multirow{3}{*}{\textbf{Datum Type}}}  & \multicolumn{2}{c}{\textbf{Case}}
 \\
\cmidrule(lr){2-3}
 & \textbf{Exponent} & \textbf{Mantissa}  \\
\hline\hline
 Normal & $\in[1, 2^w-2]$  & don't care   \\
 Subnormal &   $=0$             & $\neq0$       \\ 
 $\pm0$ &   $=0$             & $=0$           \\
  NaN  & $=2^w-1$         & $\neq0$     \\
 $\pm\infty$ & $=2^w-1$         & $=0$         \\
\hline
\end{tabular}
\end{table}

Before introducing AxFPU, 
we present the baseline 
accurate floating-point multiplication architecture.
Let $A_{f}$ and $B_{f}$ be two $n$-bit floating-point normal numbers.
The product $A_{f} \times B_{f}$ is calculated by Eq. \eqref{eq_fpmul}.

\vspace*{-5pt}

\begin{equation}
A_{f} \times B_{f} = (-1)^{(S_A+S_B)} \cdot 2^{(E_A+E_B-2e_{max})} \cdot
    \underbrace{\Bigg(1 + \sum_{i=1}^{m-1} 2^{-i} a_i \Bigg)}_{A=1.M_A} \cdot
    \underbrace{\Bigg(1 + \sum_{i=1}^{m-1} 2^{-i} b_i \Bigg)}_{B=1.M_B}  
    \label{eq_fpmul}
\end{equation}

The above multiplication is performed in the following 
basic steps: 
(i) calculation of the result's sign,
(ii) addition of the exponents, 
(iii) multiplication of the mantissas, 
(iv) normalization of the mantissa product, 
(v) update of the result's exponent, 
(vi) rounding of the mantissa product,
and
(vii) handling of special cases. 
Below, we discuss the technical details of each step. 

The sign of the result is calculated by the XOR of the input signs, i.e., $S_R = S_A \oplus S_B$.
Regarding the result's exponent, the input exponents ($E_A$, $E_B$) are biased, thus, the bias ($e_{max}$) is removed before their addition.
Afterwards, 
the bias is added to the result's exponent, 
i.e., in total $E_R = E_A + E_B - e_{max}$.  
For the mantissa multiplication, 
all the $m$ mantissa bits are employed.
The MSB of each mantissa string is `$1$', 
as we assume normal floating-point numbers.
Therefore, 
the multiplication of 
$1.M_A = \langle 1 a_1 a_2 \dots a_{m-1} \rangle$ and 
$1.M_B = \langle 1 b_1 b_2 \dots b_{m-1} \rangle$ is performed.
Due to multiplying normal floating-point numbers, 
the result can be in one of the following forms: 
(a) $01.xx \dots x$, 
(b) $10.xx \dots x$,  
or 
(c) $11.xx \dots x$.
The next step is to normalize the product 
in case it is in the form (b) or (c), 
so that there is only one leading `$1$' before the radix point.
To do so, the radix point is moved one place to the left and the intermediate exponent $E_R$ is increased by $1$.
Otherwise, the exponent and the already-normalized mantissa remain intact.
Finally, only the $m-1$ MSBs that are placed after the radix point can be stored in the mantissa field, thus, bit rounding is applied.

Our architecture 
also handles special cases that may arise due to the exponent value.
Specifically, 
if the exponent is too small/large   
to be represented,
then 
underflow/overflow occurs.
To consider these special cases, 
an underflow/overflow detector is employed 
after the component that 
performs the exponent update in case of normalization. 
This detector checks the value of the exponent
and decides if the result is normal number, underflow or overflow.
If the exponent lies in the interval $[1, 2^w-2]$, 
then the result is a normal number.
In contrast, 
if the result is smaller than $1$, 
it is marked as underflow, 
while if it is bigger than $2^w-2$, 
it is marked as overflow. 
We note that if underflow occurs, 
the exponent is stored as $0$ 
and the product $A_{f} \times B_{f}$ is 
either a subnormal number or $\pm0$, 
depending on the value of the mantissa product 
(see Table \ref{tb_fpieee}).
Similarly, 
if overflow occurs, 
the exponent is stored as $2^w-1$ 
and the product $A_{f} \times B_{f} $ is either NaN or $\pm\infty$.

\subsubsection{Design of Approximate Architecture}

The most costly component of the floating-point unit 
is the mantissa multiplier \cite{TongTVLSI2000}, 
as the rest of the circuits are comparators, 
small adders,
and multiplexers.
Thus, 
to improve its efficiency, 
we use AxFXU to multiply the mantissas.

Considering the accurate floating-point multiplication of Eq. \eqref{eq_fpmul} 
and the approximate AxFXU multiplication of Eq. \eqref{eq_pr1},
the multiplication in 
AxFPU$|_{P,R}$ is calculated by Eq. \eqref{eq_axfpu}.

\vspace*{-6pt}

\begin{equation}
   A_{f} \times B_{f}|_{P,R} =
   (-1)^{(S_A+S_B)} \cdot 2^{(E_A+E_B-2e_{max})} \cdot 
    \underbrace{A_R \cdot \sum_{j=P}^{m/2-1} 4^{j}y_{j}^{R4}}_{A \times B|_{P,R}}  
    \label{eq_axfpu}
\end{equation}

\begin{figure}[!t]
\centering
\includegraphics[width=0.85\textwidth]{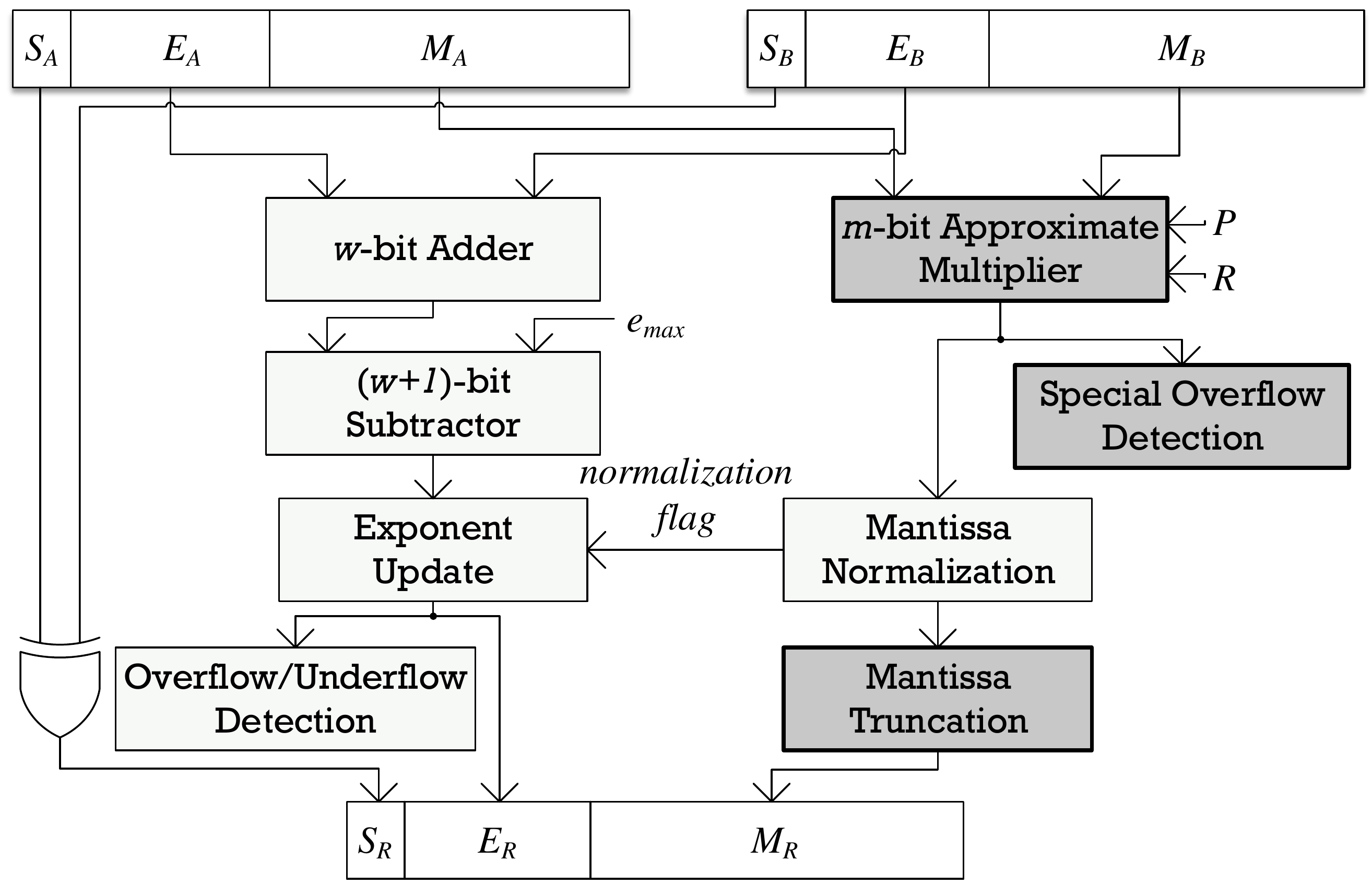}%
\caption[Architecture of the Approximate Perforation-\&-Rounding Floating-Point Multiplier]{The approximate floating-point multiplication architecture of AxFPU$|_{P,R}$. 
The configuration parameters $P$ and $R$ adjust the approximation degree of perforation and rounding, respectively.
The highlighted blocks are modified/added 
compared to the conventional accurate design.}%
\label{fig_blafp}
\vspace*{-5pt}
\end{figure}

Figure \ref{fig_blafp} illustrates the block diagram of AxFPU.
In comparison with the baseline accurate design, 
an additional detection component is implemented,
because due to the approximations, 
the mantissa multiplication of very large operands 
may produce a result in the form $00.xx \dots x$.
We remind that the result of a conventional floating-point multiplication is always in the form
$01.xx \dots x$, 
$10.xx \dots x$, 
or 
$11.xx \dots x$.
In AxFPU, 
this may happen 
due to discarding negative partial products.
In such a case, 
the absence of these products results 
in not decreasing
the value of the final product
(as in the accurate multiplication),
and thus, 
the floating-point result appears in the aforementioned form.
This case is handled as overflow, 
however, 
we note that according to our experimental results, 
the possibility of having 
either the conventional or 
this special overflow is around $0.35\%$ on average for half and single floating-point precision.
Another differentiation of AxFPU 
compared to the accurate design regards the rounding process.
Considering that 
the mantissa product is now approximate, 
its rounding to the $m-1$ MSBs,
which is typically applied after normalization, 
offers negligible accuracy gain.
Therefore, 
we eliminate the rounding unit, i.e., the second most power-consuming component of the floating-point multiplier \cite{TongTVLSI2000}, 
and we simply truncate the leftover LSBs. 
We note that this design choice is optional  
and the designer can apply rounding instead of truncation.

\subsection{Dynamic Configuration of the Approximation Degree}

The AxFXU$|_{P,R}$ and AxFPU$|_{P,R}$ circuits apply static approximation, namely, the approximation parameters $P$ and $R$ are configured at the design time and cannot change after the implementation. 
In this section, we introduce their
runtime variants (DyFXU and DyFPU),  
which can dynamically configure their approximations degree,
i.e.,
adjust the parameters $P$ and $R$ at runtime.
These designs employ the accurate radix-$4$ multiplier,
thus, 
they also support operation in fully-accurate mode.
The two approximation techniques that are applied
(partial product perforation and partial product rounding) 
in conjunction with the selected radix-$4$ multiplication algorithm 
facilitate the design of a low-overhead scheme for dynamic configuration because:
\begin{itemize}
    \item[(i)] the $2P-1$ LSBs of $B$ are used only for the generation of the $P$ least significant partial products.
    \item[(ii)] the $R-1$ LSBs of $A$ are used only for the generation of the $R-1$ LSBs of each partial product.
\end{itemize}
From the first ascertainment, 
we conclude that the truncation of 
the $2P-1$ LSBs of $B$ 
is equivalent to configuring perforation to $P$.
From the second ascertainment, 
we conclude that the truncation of 
the $R-1$ LSBs of $A$, 
along with the addition of $a_{R-1}$ to the most significant part of $A$, 
is equivalent to configuring rounding to $R$.
Therefore, 
to dynamically configure the parameters $P$ and $R$,
we drive the $i$-th bit of each operand
to an AND gate
along with an input signal $s_i$ ($i=1,2,\dots n$).
The value of $s_i$
determines whether the $i$-th bit of the operand
will be ``virtually'' truncated ($s_i=0$) or not ($s_i = 1$).  

\begin{figure}[!t]
\centering
\includegraphics[width=0.85\textwidth]{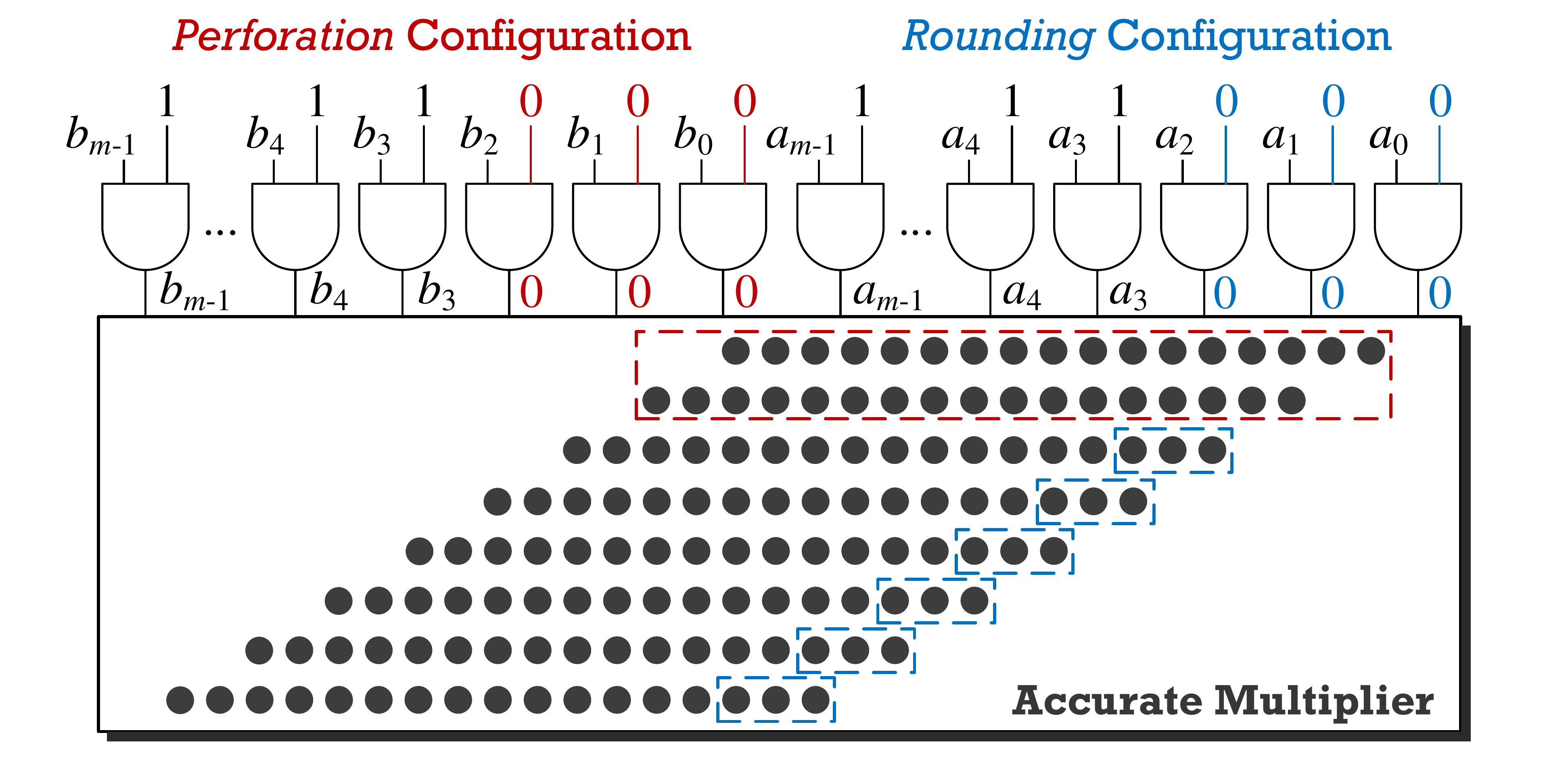}%
\caption[Dynamic Approximation Configuration in the Approximate Perforation-\&-Rounding Multiplier]{Dynamic configuration of the approximation degree by adjusting perforation and rounding via control signals. 
The ``virtual'' truncation of $3$ LSBs in each operand sets the highlighted partial product bits to $0$, activating $P=2$ (perforation configuration) and $R=4$ (rounding configuration).}%
\label{fig_dpr}
\end{figure}

Figure \ref{fig_dpr} illustrates the dynamic configuration of the approximation degree to $P=2$ and $R=4$. 
The control signal $s_i$ is set to `$0$' in the AND gates driven by:
\begin{itemize}
    \item[(i)] the $3$ LSBs of $B$, and thus, the $2$ least significant partial products are not calculated (they are $0$) $\; \rightarrow \;$ perforation is configured to $P = 2$.
    \item[(ii)] the $3$ LSBs of $A$, and thus, the $3$ LSBs of each partial product are not calculated (they are $0$) $\; \rightarrow \;$ rounding is configured to $R = 4$.
\end{itemize}
In the rest AND gates, $s_i$ is set to `$1$', 
and the corresponding partial products bits are accurately generated.

It is obvious that the runtime variants
do not provide area gains, 
as they implement
the entire $n$-bit multiplier
plus 
$2n$ AND gates and $2n$ $1$-bit input signals.
However, they still deliver significant energy gains 
while offering
accurate operation mode 
and 
multiple levels of accuracy,
which can be configured at runtime.  
In terms of performance, 
the critical paths are not reduced, 
and thus, 
the clock frequencies are set at their nominal value, 
i.e., that of the accurate designs, 
to satisfy the accurate operation mode.
To provide performance gains, 
the frequencies can be dynamically 
scaled to satisfy the critical paths
of the respective design-time multipliers,  
e.g., 
operate DyFXU$|_{2,4}$ at the maximum  frequency of AxFXU$|_{2,4}$.

The control signals that tune the approximation degree are directly exposed to the system. 
Namely, they can be seamlessly set to $0$/$1$ at runtime to increase/decrease the approximation degree.
The tuning can be performed by a high-level policy
(e.g., in an approximate custom processor/accelerator)
that sets the control signals
with respect to
pre-stored error values 
and
an error constraint defined 
by the application/user.
For instance, 
the mean error for all the combinations of the approximation parameters $P$ and $R$ can be calculated offline
(e.g., for an application-specific input distribution or for all the possible inputs),
in order to be stored and accessed by the high-level policy.
In this scenario,
given an error constraint
(that may change at runtime), 
the high-level policy selects the 
most resource-efficient approximation configuration that satisfies it.
A similar approach is proposed in \cite{turk_cpu}, 
where the authors use runtime-configurable circuits in an approximate RISC-based processor.

\section{Evaluation}
\label{s5_3}

In this section,
we evaluate the
design-time and runtime variants
of our
fixed-point and floating-point multipliers.
The evaluation begins
with the error analysis,
which examines the accuracy of our designs,
and continues with the experimental results from the synthesis of the circuits,
involving comparisons with state-of-the-art approximate designs.

\subsection{Error Analysis}

In the design of approximate circuits, 
the error inserted in the computations 
is considered a critical issue, 
and thus, 
its impact on the final result is studied 
using either circuit simulations or rigorous error expressions and models. 
To evaluate our approximate designs in terms of accuracy,
we execute the respective software models 
emulating the logic-level approximations. 
Our analysis is based on the calculation of well-established error metrics,
as well as new error metrics that are tailored to the designs,
such as in the case of floating-point approximations. 

\subsubsection{Study of AxFXU Accuracy}

Firstly,
we study the error of AxFXU, 
namely, our approximate fixed-point multiplier. 
We consider the error metrics used in the error analysis of Chapter \ref{chapter4}.
In brief, 
these metrics are defined as follows:
\begin{itemize}
    \item RED$_{AB}$: the relative error distance between the approximate multiplication and the accurate multiplication for a given operand pair $A$ and $B$.
    \item MRED: the average of all relative error distances for a given set of operand pairs.  
    \item PRED$_2$: the possibility of having a relative error distance larger than $2\%$. 
\end{itemize}
We note that the approximation space is large,
i.e., there are numerous approximation configurations
(combinations of the $P$ and $R$ parameters).
Therefore, 
for each examined configuration,
we consider 
$200$K different input pairs,
which are uniformly distributed over the multiplication bit-width.
The uniform input distribution forms a stressed scenario for  AxFXU, 
because more narrow distributions would lead to biased MRED values,
which could be efficiently handled with a limited set
of approximation configurations. 
Nevertheless,
we note that 
AxFXU tunes the approximation degree with two independent parameters, 
and as a result, 
it provides the flexibility to handle different input distributions.

\begin{figure}[!t]
\centering
\subfloat[\label{fig_mred16}]{\includegraphics[width=0.49\textwidth]{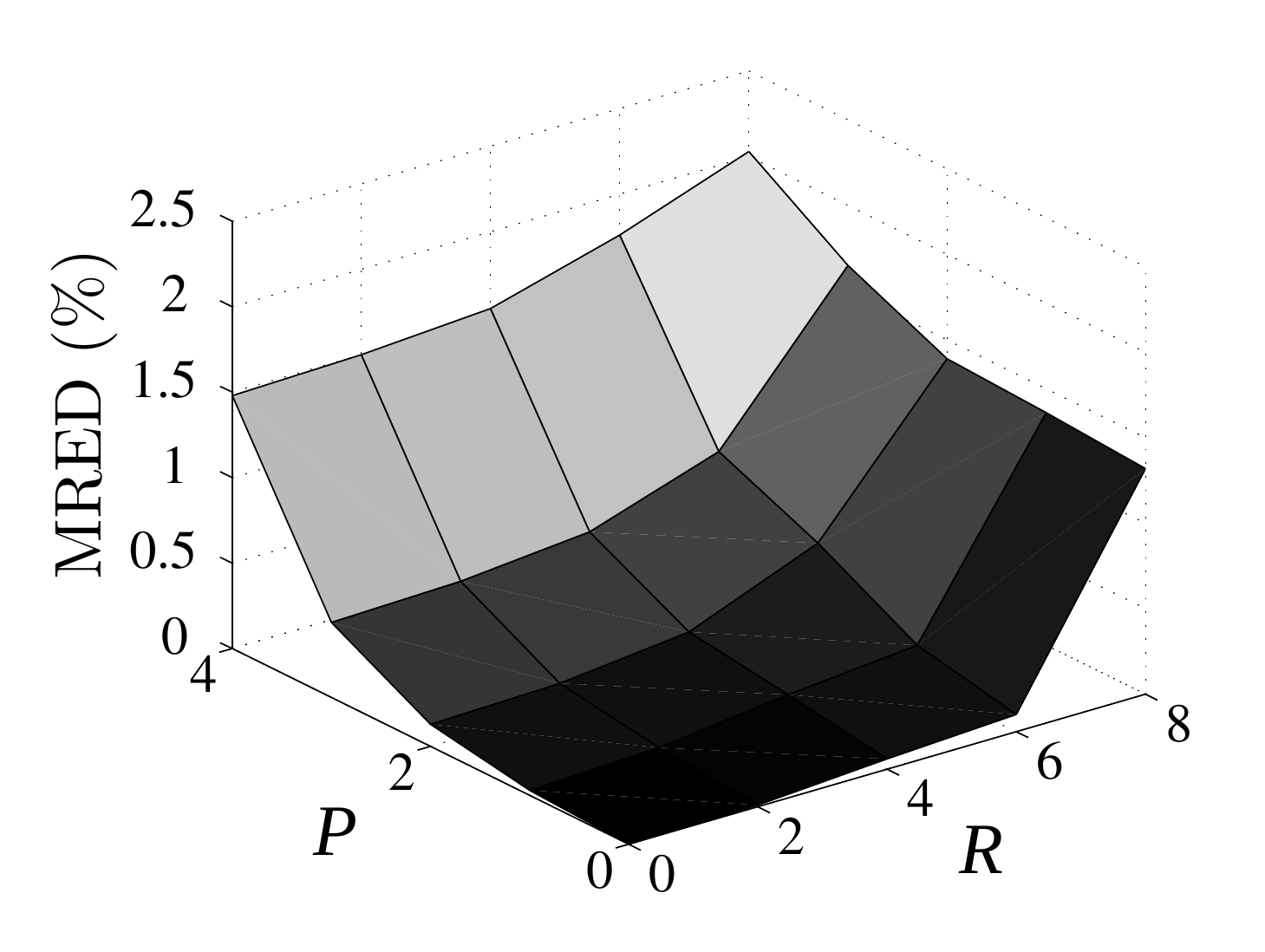}}%
\\[-7pt]
\subfloat[\label{fig_mred24}]{\includegraphics[width=0.49\textwidth]{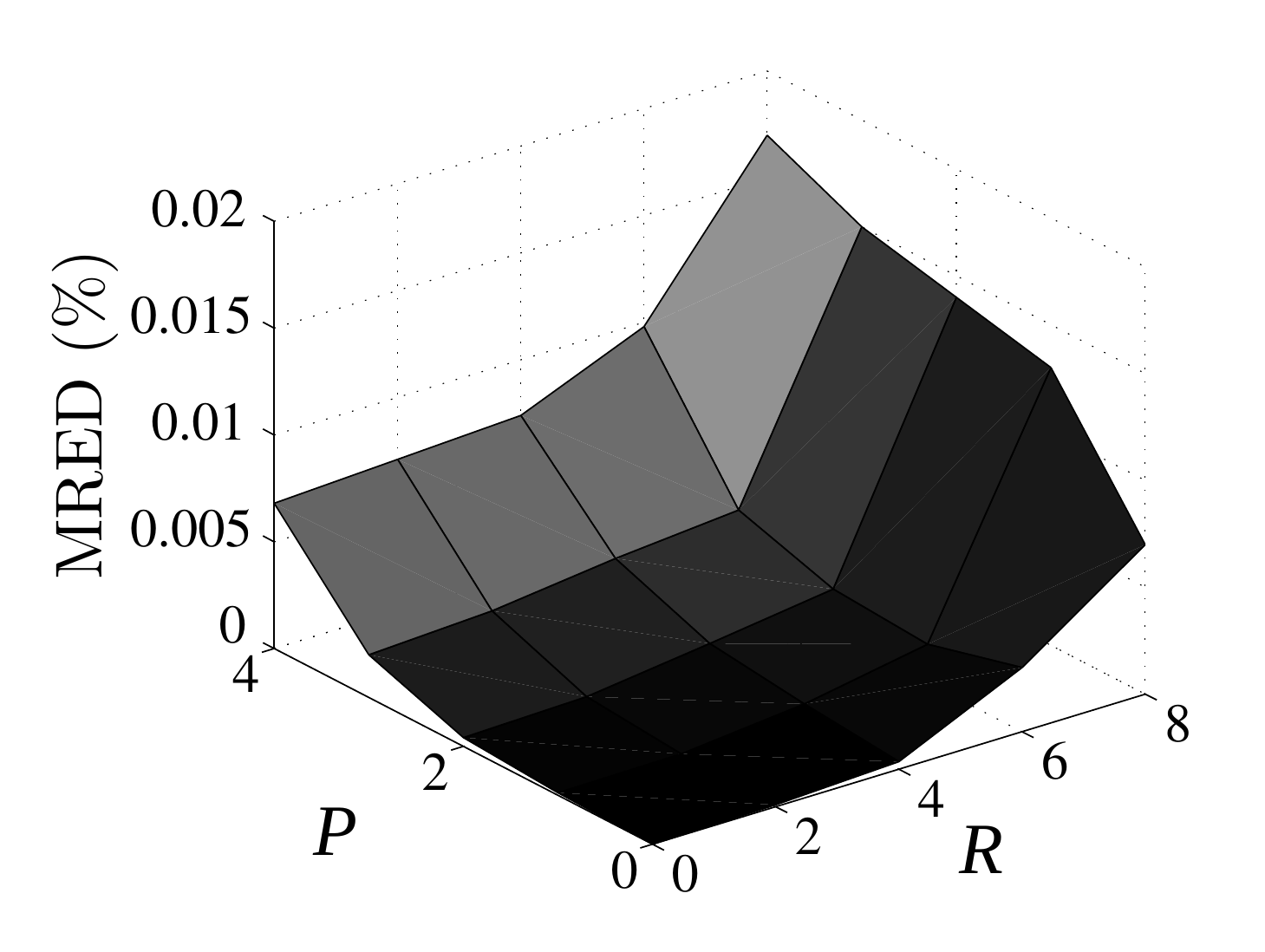}}%
\hspace{3pt}
\subfloat[\label{fig_mred32}]{\includegraphics[width=0.49\textwidth]{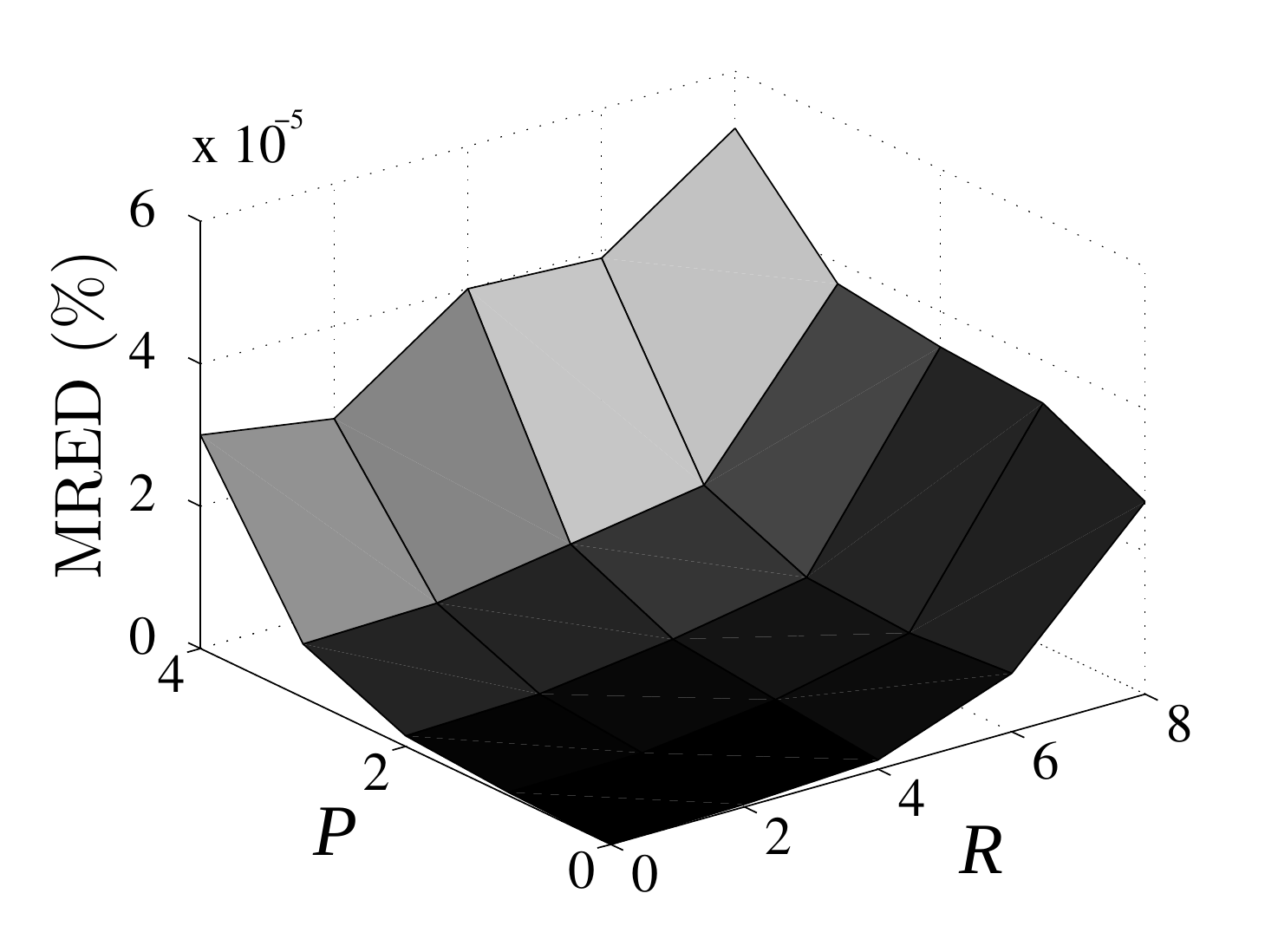}}%
\caption[Error Variation of the Approximate Perforation-\&-Rounding Fixed-Point Multiplier]{MRED variation of AxFXU with respect to the
approximation configuration $P$ (perforation) and $R$ (rounding) for multiplication bit-width:
\textbf{(a)} $n=16$, 
\textbf{(b)} $n=24$, 
and 
\textbf{(c)} $n=32$.}%
\label{fig_mredfx}
\end{figure}

Figure \ref{fig_mredfx} 
presents how MRED is affected by the 
approximation configuration
for multiplication bit-width 
$n = 16, 24, 32$. 
As shown, the MRED values increase linearly
with the approximation degree, 
i.e., as more approximations are applied. 
Perforation introduces larger error than rounding, 
due to the significance of the bits that are pruned 
(entire partial products versus partial product bits).
Moreover, 
as the bit-width increases, 
MRED is less affected by the approximations.
Specifically for 
$n = 24, 32$, 
MRED is up to $0.02\%$ and $0.00005\%$, respectively. 
This feature is  
an advantage of our designs,
because it provides the flexibility to perform more aggressive approximations in large-sized multipliers, 
resulting in increased energy and area gains. 
Correspondingly, 
as the multiplier’s size increases, 
a smaller error is introduced 
to retain the same energy budget and area. 

Regarding the significance of the errors,
which is examined with the metric PRED$_2$,
the large approximation space of AxFXU
generates designs with varying PRED$_2$ values.
Namely, there are designs with zero or near-zero 
PRED$_2$, as well as 
designs with PRED$_2$ of $5\%$ or $10\%$,
which, however, 
deliver remarkable resource gains.
Concluding, 
the set of approximations of AxFXU
can support various design scenarios
involving different error and resource constraints.

\subsubsection{Study of AxFPU Accuracy}

Next,
we study the error of AxFPU, 
namely, our approximate floating-point multiplier. 
We consider the same error metrics with AxFXU,
however, 
we also employ new error metrics to 
evaluate the special cases of floating-point arithmetic. 

Let $A_{f} \times B_{f}$ be
the accurate floating-point multiplication 
of Eq. \eqref{eq_fpmul}
and $A_{f} \times B_{f}|_{P,R}$ be
the approximate floating-point multiplication 
of Eq. \eqref{eq_axfpu}.
The RED of AxFPU is formed as shown in Eq. \eqref{eq_fpred}.

\begin{equation}
\text{RED}_{A_{f}B_{f}}  = 
\dfrac {\bigl\lvert A_{f} \cdot B_{f} - A_{f} \cdot B_{f}|_{P,R} \bigr\rvert} {\bigl\lvert A_{f} \cdot B_{f} \bigr\rvert}
= 
\dfrac {\bigl\lvert A \cdot B - A \cdot B |_{P,R} \bigr\rvert} {\bigl\lvert A \cdot B  \bigr\rvert} =
\text{RED}_{AB} \\[4pt]
\label{eq_fpred}
\end{equation}

Hence, 
the RED of AxFPU
is equal to the RED of the approximate mantissa multiplier
(implemented with AxFXU),
i.e., 
$\text{RED}_{A_{f}B_{f}} = \text{RED}_{AB}$.
It is important to mention that 
this formula assumes that
possible normalization in the accurate mantissa multiplication, 
which results in increasing the exponent, 
also appears in the approximate multiplication.
Additionally, 
we assume that AxFPU does not produce erroneous need for normalization.
Regarding MRED and PRED$_2$,
they are 
calculated as defined in Chapter \ref{chapter4}. 

Considering that in floating-point multiplication overflow or underflow may occur, 
RED is involved in the calculation of MRED and PRED$_2$ 
only in three cases,
depending on the floating-point data type of the approximate and the respective accurate product: 
\begin{itemize}
    \item[(i)] if both products are normal numbers, where RED is calculated by Eq. \eqref{eq_fpred}.
    \item[(ii)] if both products are overflow, where RED is considered $0$.
    \item[(iii)] if both products are underflow, where RED is considered $0$.
\end{itemize}
To evaluate the possibility
of having one of the rest combinations,
e.g., normal accurate product and underflow approximate product, 
two more error metrics are introduced: 
\begin{itemize}
    \item PON: the possibility of overflow approximate product and normal accurate product, or vice versa.
    \item PUN: the possibility of underflow approximate product and normal accurate product, or vice versa.   
\end{itemize}

We note that the appearance of unexpected floating-point data type
in the approximate result, 
which occurs due to the applied approximations, 
is not examined in prior works of approximate floating-point multipliers.
The PON and PUN metrics are calculated by Eq. \eqref{eq_pon}--\eqref{eq_pun}.

\vspace{-10pt}

\begin{align}
\text{PON} =  
p(&
A_{f} \times B_{f}|_{P,R}: \text{overflow} \;\; \& \;\; 
A_{f} \times B_{f}: \text{normal} \;\;\; | \nonumber\\
& 
A_{f} \times B_{f}|_{P,R}: \text{normal} \hspace{-0.4pt} \;\; \;\; \& \;\;
A_{f} \times B_{f}: \text{overflow})
\label{eq_pon}
\end{align}
\begin{align}
\text{PUN} =  p(&
A_{f} \times B_{f}|_{P,R}: \text{underflow} \;\; \& \;\;A_{f} \times B_{f}: \text{normal} \;\;\; | \nonumber\\
 & 
A_{f} \times B_{f}|_{P,R}: \text{normal} \hspace{1pt} \;\;\;\;\;\; \& \;\; A_{f} \times B_{f}: \text{underflow})\label{eq_pun}
\end{align}
 
Table \ref{tb_fppossib} summarizes all the possible accurate and approximate products and reports 
which error metric is employed for each combination.
According to the analysis of our approximations
as well as exhaustive simulations,
there are two impossible combinations. 
Specifically, AxFPU cannot produce overflow/underflow
in case the accurate result is underflow/overflow.

\begin{table}[!t]
\fontsize{9}{10}\selectfont
\renewcommand{\arraystretch}{1.2}
\setlength{\tabcolsep}{4pt}
\caption[Error Metrics for Approximate Floating-Point Multipliers]{Error metrics for approximate floating-point multipliers.}
\label{tb_fppossib}
\centering
\begin{tabular}{cc c |c}
\hline
\textbf{Acc. Product} & \textbf{Appr. Product} & \textbf{Possible} & \textbf{Error Metrics} \\
\hline
\hline
normal    & normal    & \checkmark & MRED, PRED$_2$: w/ RED of Eq. \eqref{eq_fpred} \\
normal    & underflow & \checkmark & PUN  \\
normal    & overflow  & \checkmark & PON \\
underflow & normal    & \checkmark & PUN \\
underflow & underflow & \checkmark & MRED, PRED$_2$: w/ RED$=0$ \\
underflow & overflow  & \ding{53}  & --  \\
overflow  & normal    & \checkmark & PON \\
overflow  & underflow & \ding{53}  & --  \\
overflow  & overflow  & \checkmark & MRED, PRED$_2$: w/ RED$=0$ \\
\hline
\end{tabular}
\end{table}
 
\begin{figure}[!t]
\vspace{-20pt}
\centering
\subfloat[\label{fig_fpmred16}]{\includegraphics[width=0.47\textwidth]{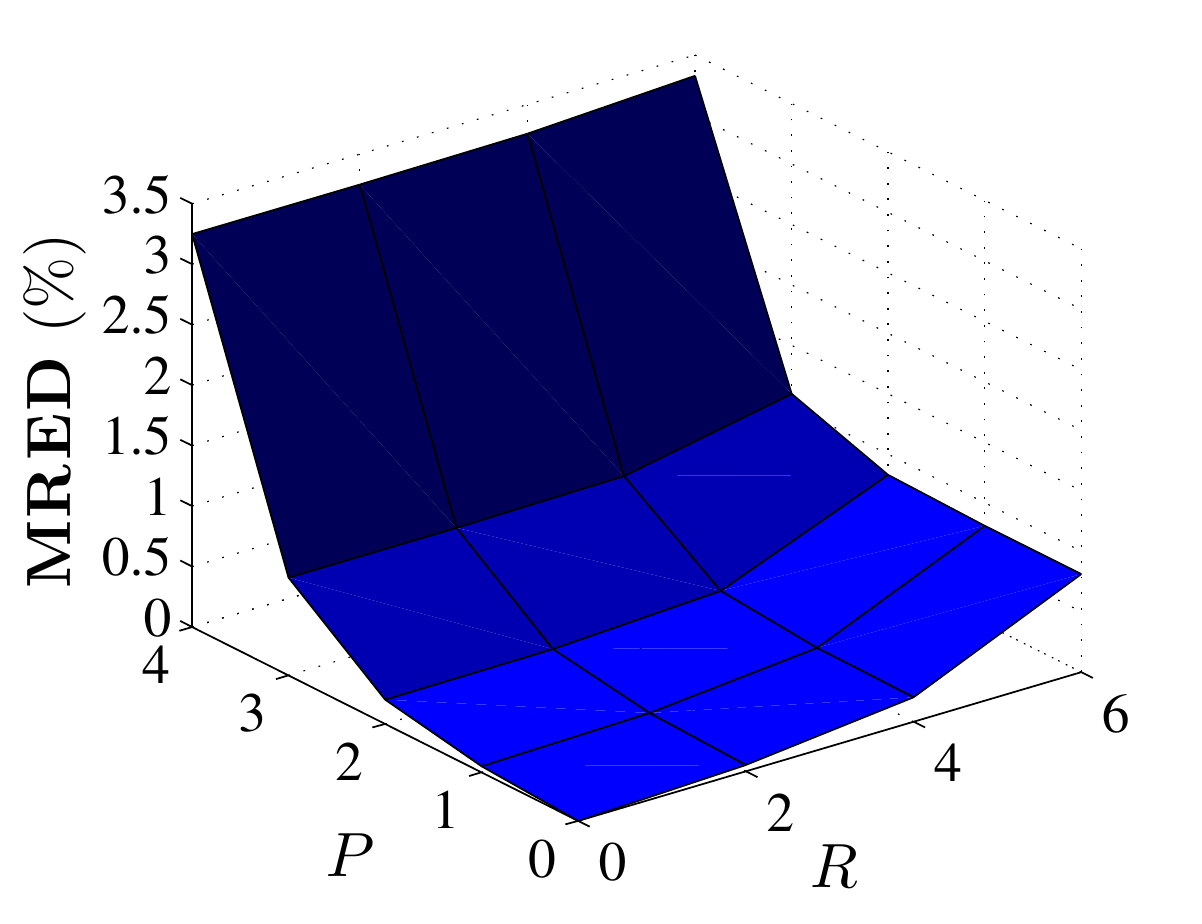}} \hspace{8pt} %
\subfloat[\label{fig_pred16}]{\includegraphics[width=0.47\textwidth]{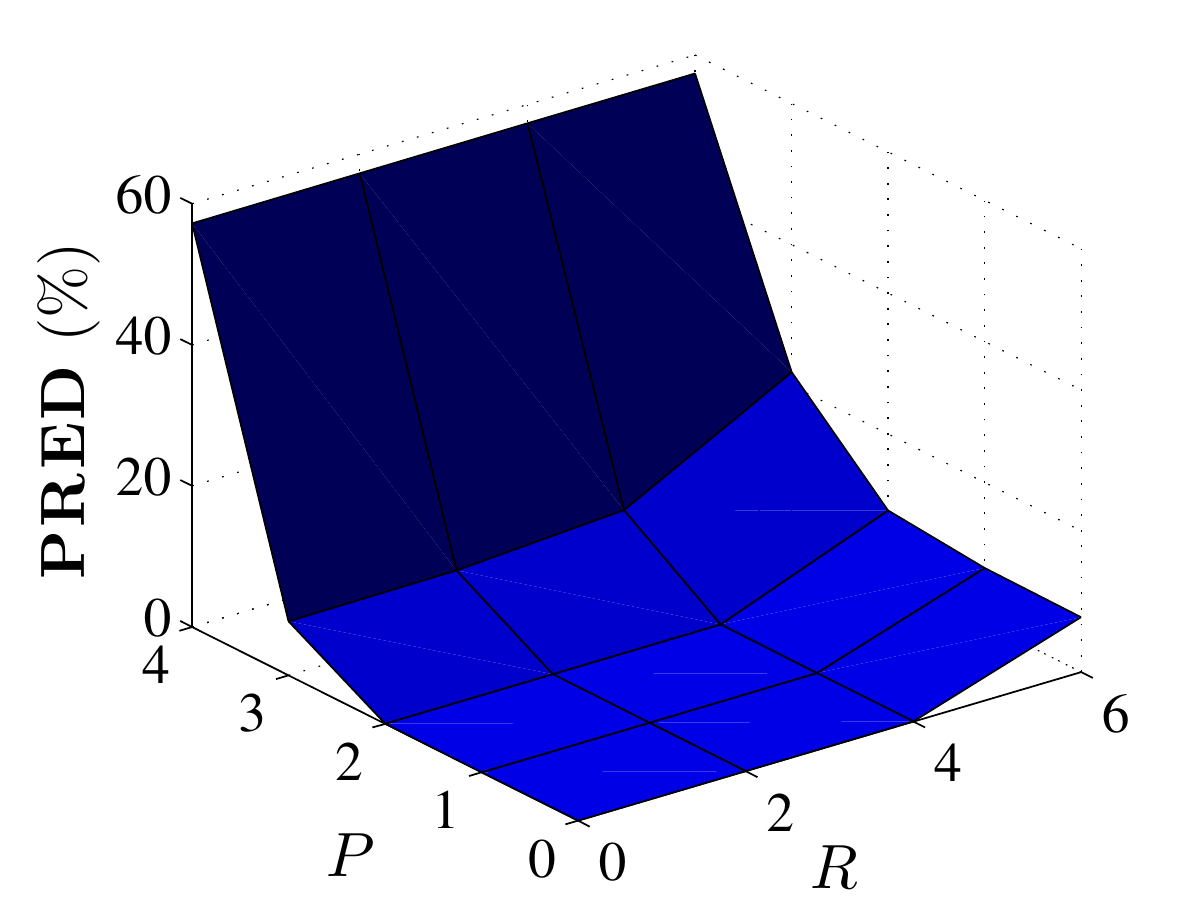}}\\[-10pt]
\subfloat[\label{fig_pon16}]{\includegraphics[width=0.47\textwidth]{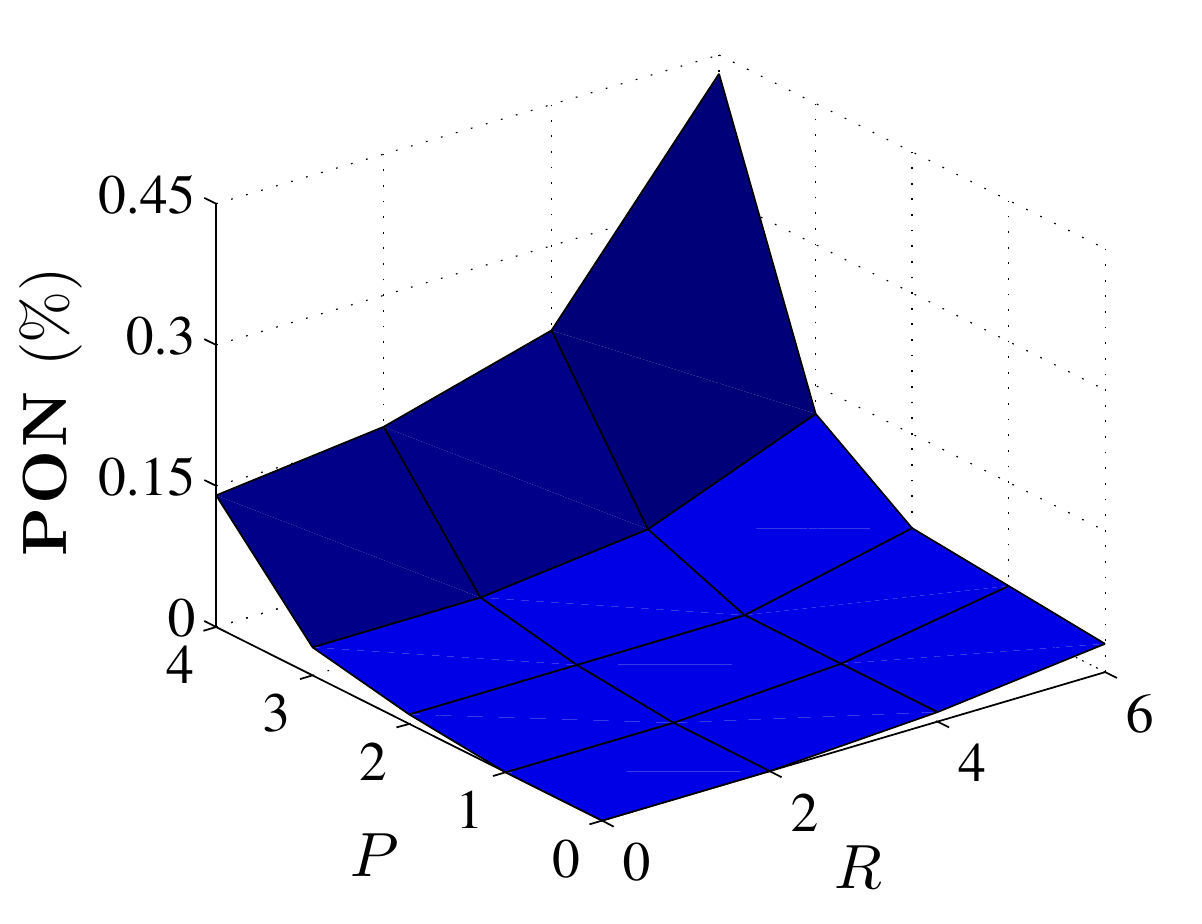}}\hspace{8pt} 
\subfloat[\label{fig_pun16}]{\includegraphics[width=0.47\textwidth]{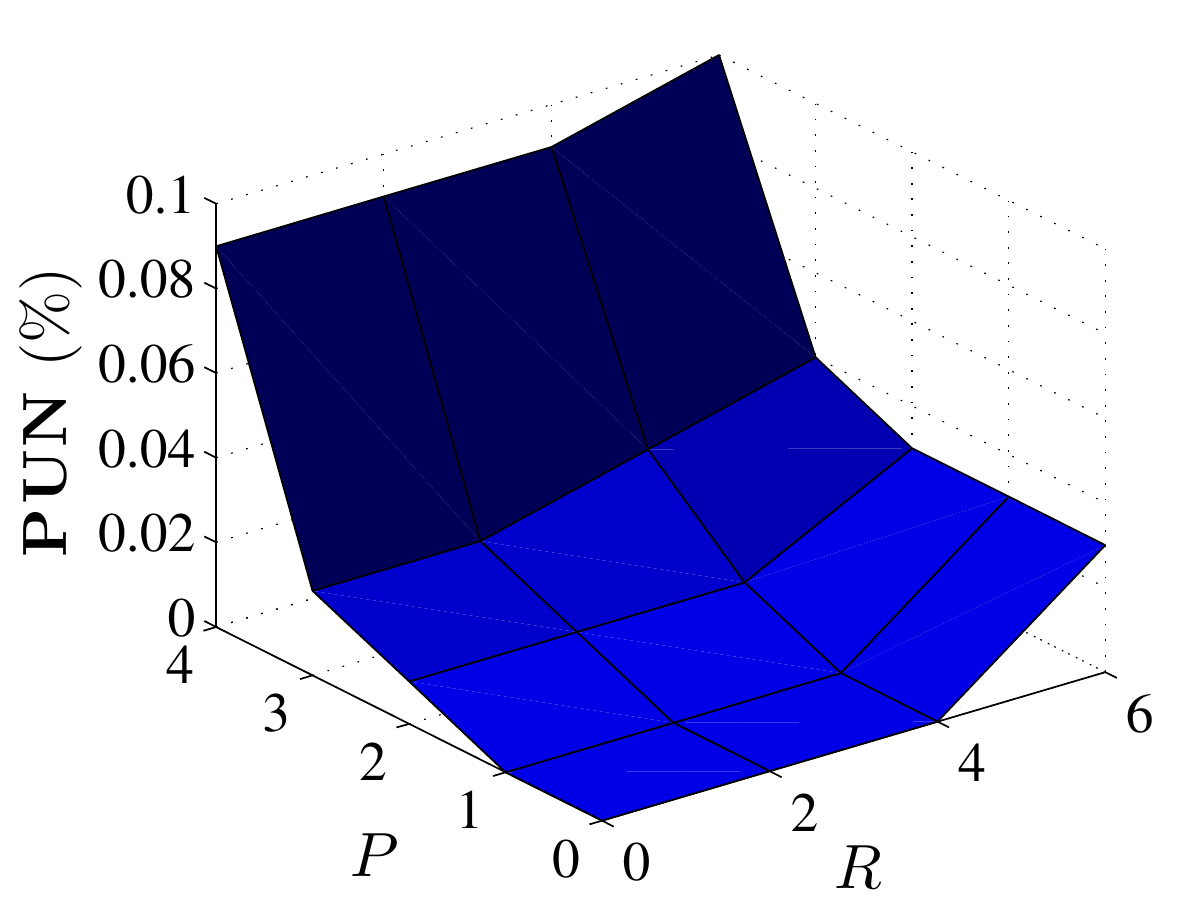}}%
\caption[Error Variation of the Approximate Perforation-\&-Rounding Half-Precision Floating-Point Multiplier]{Variation of error metrics for half-precision AxFPU with respect to the
approximation configuration $P$ (perforation) and $R$ (rounding):
\textbf{(a)} MRED, 
\textbf{(b)} PRED$_2$, 
\textbf{(c)} PON, 
\textbf{(d)} PUN.}%
\label{fig_fp16}
\end{figure}

\begin{figure}[!t]
\vspace{-20pt}
\centering
\subfloat[\label{fig_fpmred32}]{\includegraphics[width=0.47\textwidth]{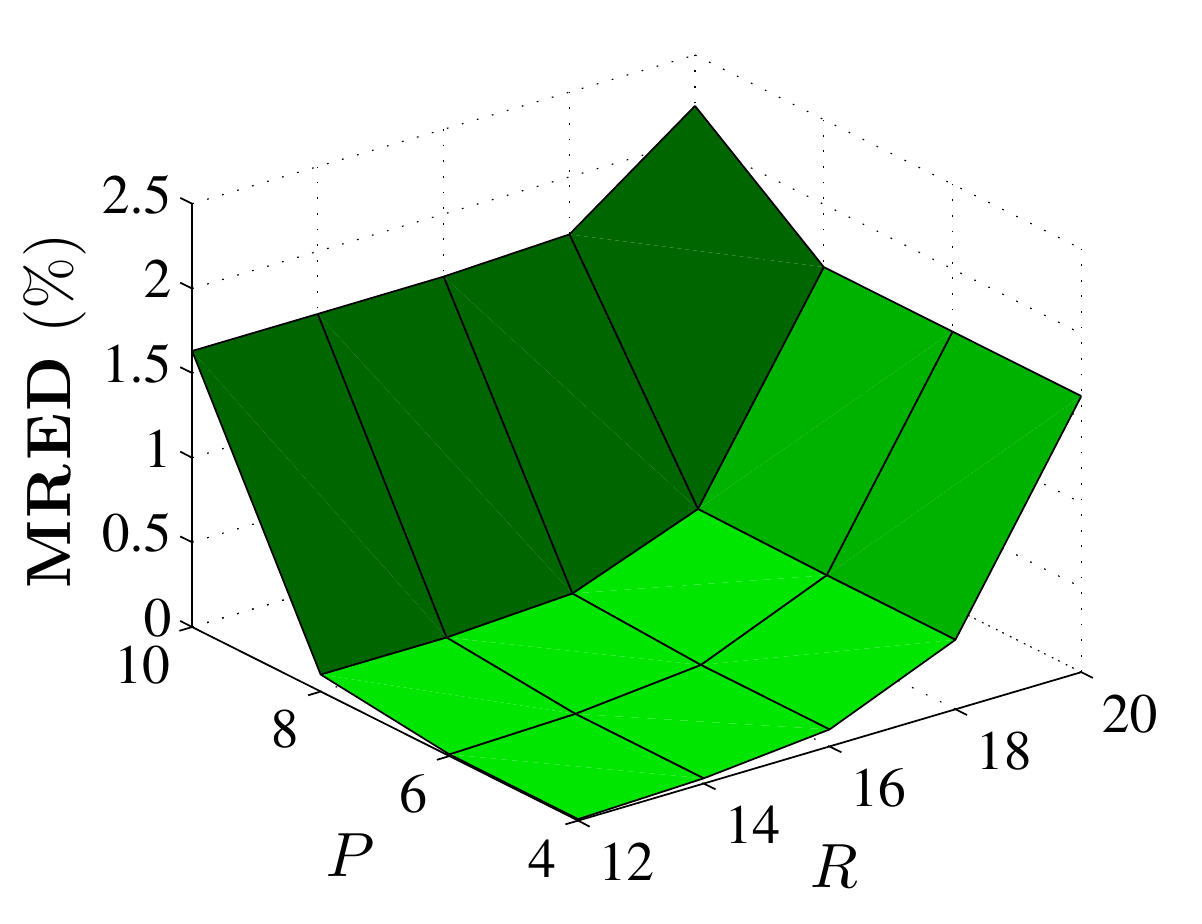}} \hspace{8pt} %
\subfloat[\label{fig_pred32}]{\includegraphics[width=0.47\textwidth]{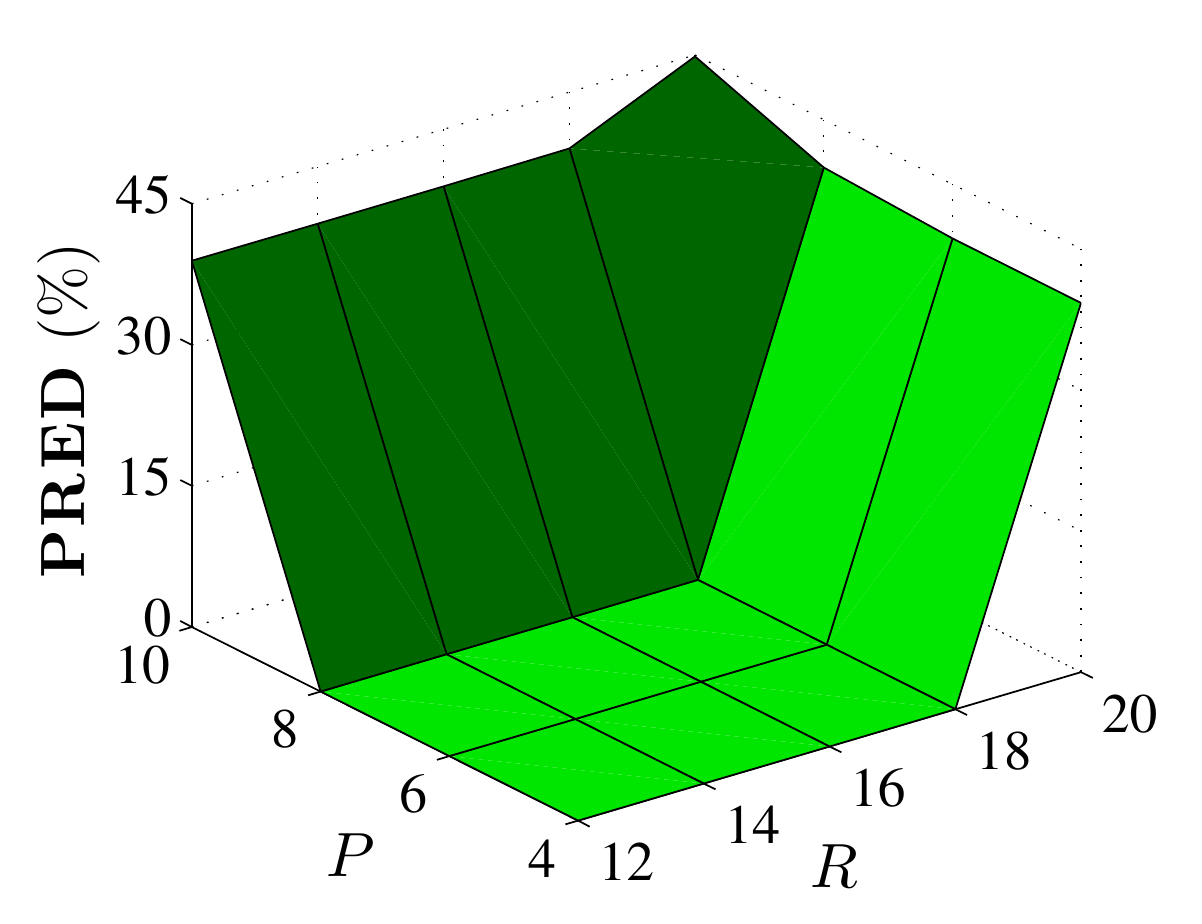}}\\[-10pt]
\subfloat[\label{fig_pon32}]{\includegraphics[width=0.47\textwidth]{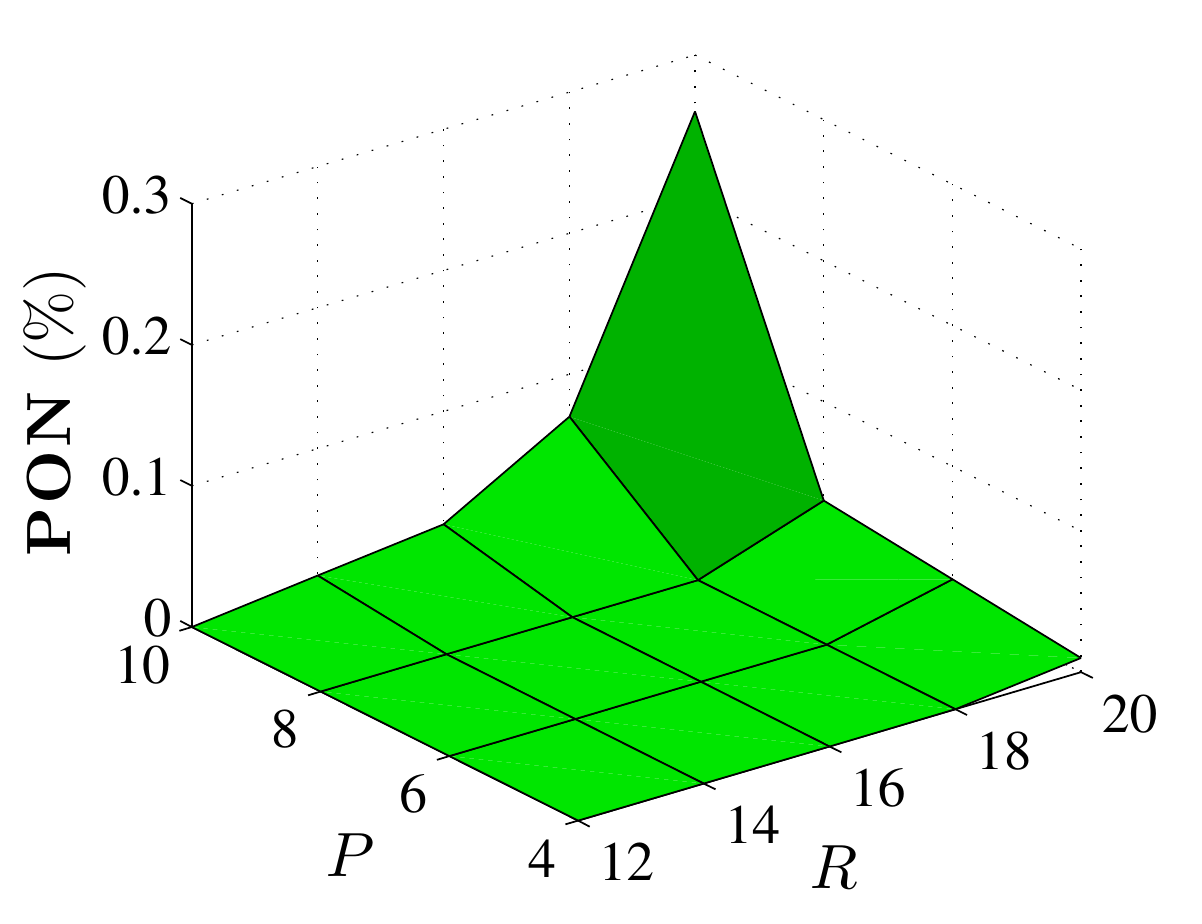}}\hspace{8pt} 
\subfloat[\label{fig_pun32}]{\includegraphics[width=0.47\textwidth]{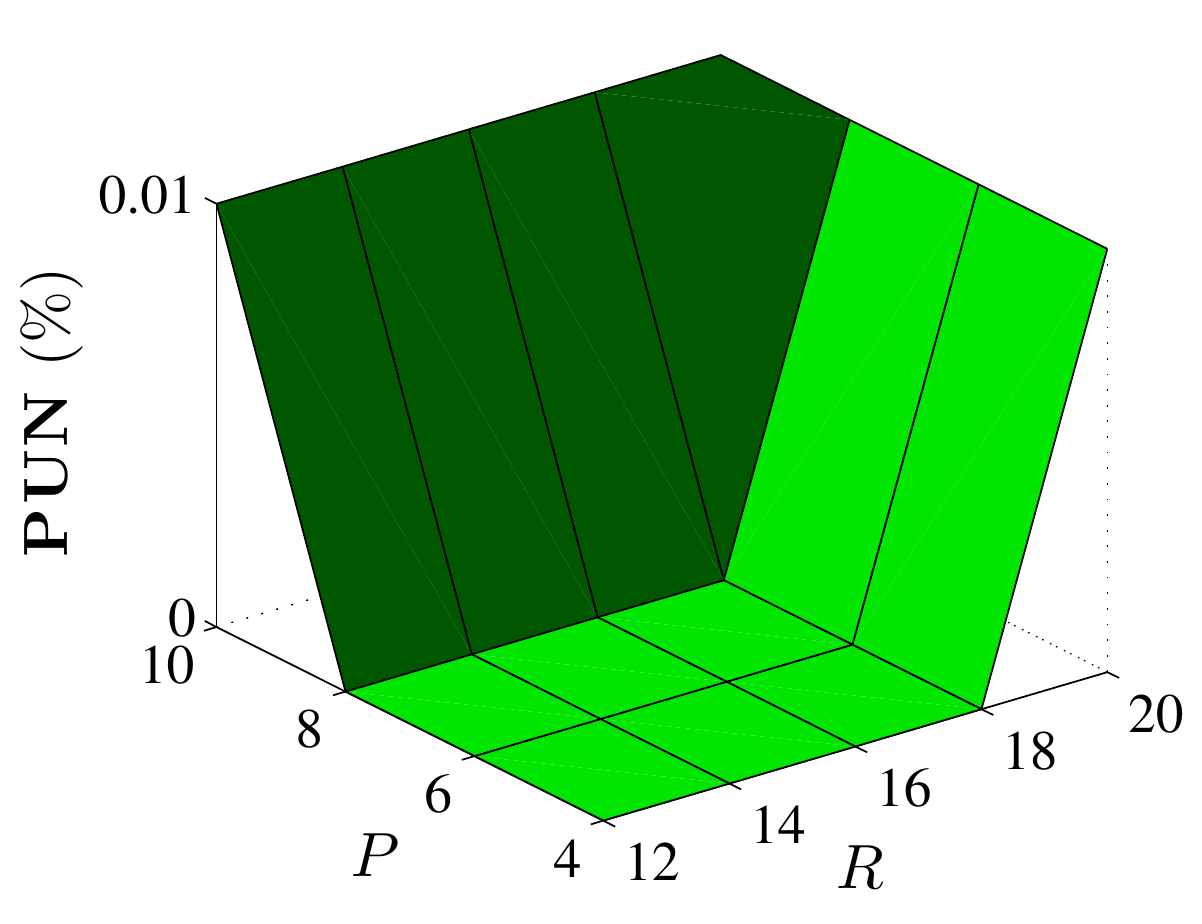}}%
\caption[Error Variation of the Approximate Perforation-\&-Rounding Single-Precision Floating-Point Multiplier]{Variation of error metrics for single-precision AxFPU with respect to the
approximation configuration $P$ (perforation) and $R$ (rounding):
\textbf{(a)} MRED, 
\textbf{(b)} PRED$_2$, 
\textbf{(c)} PON, 
\textbf{(d)} PUN.}%
\label{fig_fp32}
\end{figure}

For the accuracy evaluation,
we use half and single floating-point precision,
i.e., we employ the $16$-bit and $32$-bit AxFPU multipliers,
labeled as AxFPU16 and AxFPU32, respectively.
Similar to the error analysis of AxFXU,
we consider a uniform distribution over the normal floating-point numbers (to cover all the data range)  
and employ $200$K different input pairs. 
Again, 
narrower input distributions, 
e.g., only small floating-point numbers, 
would lead to more biased error values, 
which could be handled 
with a limited set of approximation configurations.    
Figure \ref{fig_fp16} and Figure \ref{fig_fp32} 
present the variation of the error metrics 
for AxFPU16 and AxFPU32,
respectively.
To explore the error scalability, 
we combine different values for $P$ and $R$, 
tailored to the floating-point bit-width,
i.e., for the single-precision AxFPU32, 
we set larger $P$ and $R$ values. 

The derived results show that the range of MRED is $0.05\%$--$3.33\%$ for AxFPU16 
(see Figure \ref{fig_fpmred16}), 
and 
$0.01\%$--$2.20\%$ for AxFPU32 
(see Figure \ref{fig_fpmred32}), 
which are typical mean error values for approximate arithmetic units.
By examining the impact of rounding for fixed perforation values, 
we notice that the MRED of AxFPU16 grows sharply when we set $R=6$, 
i.e., when the bit-width of each partial product is halved. 
Specifically, 
considering perforation values $P<3$, 
there is 
an average \raisebox{0.8pt}{$\scriptstyle\sim$}$4 \times$ increase of MRED 
when moving from $R=4$ to $R=6$.
This threshold is bigger for AxFPU32 
(i.e., $R=20$), 
considering that 
the mantissa bit-width is $23$ rather than $10$.
Moreover, as expected, the error is highly affected by perforation, 
which omits entire partial products. 
For instance, 
the MRED of AxFPU16 explodes from $0.81\%$ to $3.25\%$, 
when discarding $P=4$ rather than $P=3$ partial products.
Multipliers with large bit-widths are favored by more aggressive perforation, 
and as a result, 
AxFPU32 delivers relatively small error, 
even though several partial products are perforated, 
e.g., for $P=10$ the MRED is $1.63\%$ and raises up to $2.20\%$, depending on the rounding configuration.

Regarding the PRED$_2$ metric, 
presented in Figure \ref{fig_pred16} and Figure \ref{fig_pred32},
both multipliers exhibit near-zero values for the smallest examined pairs of configurations, 
which, however, 
offer remarkable resource savings.
For example, 
the PRED$_2$ of AxFPU16 is $0.02\%$ 
in exchange for 
$P=2$ perforated partial products and 
$R=4$ rounding value.
Similarly, 
the PRED$_2$ of AxFPU32 
is $0.01\%$ for $P=8$ and $R=18$, 
i.e., values that deliver significant area reduction
considering the number of the remaining partial products 
and their bit-width.
On the other hand, 
more aggressive approximations produce larger PRED$_2$ values.
We note that these values are decreased  
for input distributions involving bigger floating-point numbers, 
which are less sensitive to the applied approximations. 

Finally, 
the PON values of our approximate designs 
lie below $0.43\%$ and $0.26\%$ 
for AxFPU16 (see Figure \ref{fig_pon16}) and 
AxFPU32 (see Figure \ref{fig_pon32}), respectively.
The corresponding values for 
PUN are $0.1\%$ (see Figure \ref{fig_pun16}) and 
$0.01\%$ (see Figure \ref{fig_pun32}),
namely, near-zero possibilities.
Moreover, 
the low-strength approximation configurations 
completely avoid erroneous type of result
(unexpected overflow or underflow)  
delivering zero PON and PUN values.
Overall, only a negligible number of inputs trigger 
such wrong results, 
which otherwise could render AxFPU inefficient.

\subsection{Experimental Results} 

This section reports 
experimental results for our approximate
fixed- and floating-point multipliers. 
We employ relevant state-of-the-art designs for comparison,
analyze the trade-off between 
accuracy and resources,
and also 
discuss the benefits of the 
design-time and runtime variants.

All the designs are implemented in Verilog 
and 
synthesized with the Synopsys Design Compiler tool
and the TSMC 65-nm standard-cell library.
The simulations for the 
functional verification and the power measurements 
are performed with Mentor Graphics QuestaSim.
The nominal supply voltage ($1$V) is used in both synthesis and simulation.
The critical path delay and the area of the circuits
are reported by Synopsys Design Compiler, 
while the power consumption is measured with Synopsys PrimeTime 
after performing gate-level simulation.
We also evaluate the energy consumption,
which is defined as the product of power and delay. 
For our analysis, 
we define the gain of the approximate design as the relative resource reduction from the respective accurate design.

\subsubsection{Comparative State-of-the-Art Evaluation of Fixed-Point Designs}

We implement 
several configurations of our design-time (AxFXU$|_{P,R}$)
and runtime (DyFXU$|_{P,R}$) multipliers,
as well as 
the accurate radix-$4$ multiplier (ACCR4)
and the approximate multipliers of 
\cite{2016_Jiang_IEEEtc, 2017_Liu_IEEEtc, LeonTVLSI, ZervakisTVLSI2016, Schulte1993, 2015_Hashemi_ICCAD, 2017_Zendegani_IEEEtvlsi}.
In total, 
our evaluation is performed 
considering two design scenarios 
regarding the clock constraint,
and involves:
(i) comparison among state-of-the-art designs, 
(ii) comparison between our design-time and runtime designs,
and 
(iii) Pareto analysis considering the error--energy trade-off.
Below, 
we explain the approximation techniques of the literature's designs,
and then we present the experimental results. 

The perforation of $k$ partial products \cite{ZervakisTVLSI2016}
is labeled as PERF$k$.
We note that the approximation configurations of PERF$k$
are covered by our designs with $P=k$ and $R=0$,
i.e., when applying only perforation and not rounding. 
R8ABM1 \cite{2016_Jiang_IEEEtc} uses the radix-$8$ encoding to
generate the partial products,
while calculating an approximate $3A$ product. 
R8ABM1-15 \cite{2016_Jiang_IEEEtc} is the same design,
but it also truncates $15$ bits of the partial products. 
The R4ABM1-$k$ and R4ABM2-$k$ multipliers \cite{2017_Liu_IEEEtc} 
employ an approximate radix-$4$ encoding 
for generating the $k$ LSBs
of the partial product matrix. 
R4ABM1-$k$ applies less approximations than R4ABM2-$k$  \cite{2017_Liu_IEEEtc}. 
RAD$2^k$ \cite{LeonTVLSI},
which is our design presented in Chapter \ref{chapter4},
generates the partial products
based on the accurate radix-$4$ encoding 
and an approximate high-radix-$2^k$ encoding.
TMC$k$ \cite{Schulte1993} truncates the $k$ LSBs 
of the partial product matrix 
and adds correction terms for reducing the error.
DRUM6 \cite{2015_Hashemi_ICCAD} selects a segment of $6$ bits from the multiplication operands,
starting from the leading `$1$'  
and sets the LSB of the truncated
values to `$1$'.
Finally, 
RoBA \cite{2017_Zendegani_IEEEtvlsi}
rounds the operands to the nearest exponent of two and performs
the multiplication in segments using the shift operation.

\begin{table}[!t]
\fontsize{9}{10}\selectfont
\renewcommand{\arraystretch}{1.2}
\setlength{\tabcolsep}{5pt}
\caption[Experimental Results of Approximate Fixed-Point Multipliers on TSMC 65-nm Standard-Cell]{Experimental results of $16$-bit approximate fixed-point multipliers on TSMC 65-nm standard-cell.}
\label{tb_fxuhw}
\centering
\begin{tabular}{l| c | c c| c c| c c}
\hline
\multicolumn{1}{c|}{\multirow{3}{*}{\textbf{Design}}} & & \multicolumn{2}{c|}{\textbf{MIN-Delay}} &
\multicolumn{2}{c|}{\textbf{ISO-Delay}} & \multicolumn{2}{c}{\textbf{Accuracy}} \\ 
 & 
\textbf{Delay} & 
\textbf{Area} & 
\textbf{Energy} & 
\textbf{Area} & 
\textbf{Energy} & 
\textbf{MRED} &
\textbf{PRED$_\mathbf{2}$} \\[-2pt]
& 
(ns) & 
($\upmu$m$^2$) & 
($\upmu$W$\cdot$ns) &  
($\upmu$m$^2$) & 
($\upmu$W$\cdot$ns) & 
$(\%)$ &
$(\%)$ \\
\hline \hline
ACCR4                & $0.75$ & $4153$ & $3749$ & $2925$ & $1407$ & $-$    & $-$     \\
AxFXU$|_{1,2}$       & $0.73$ & $3209$ & $2793$ & $2341$ & $1116$ & $0.06$ & $0.32$ \\
AxFXU$|_{2,4}$       & $0.69$ & $2672$ & $2274$ & $1800$ & $919$  & $0.23$ & $1.21$ \\
AxFXU$|_{3,4}$       & $0.65$ & $2320$ & $1830$ & $1529$ & $816$  & $0.53$ & $3.01$ \\
AxFXU$|_{3,6}$       & $0.63$ & $2017$ & $1487$ & $1264$ & $741$  & $0.78$ & $4.81$ \\
AxFXU$|_{4,4}$       & $0.60$ & $1888$ & $1453$ & $1253$ & $716$  & $1.55$ & $10.38$ \\
AxFXU$|_{4,6}$       & $0.59$ & $1513$ & $1178$ & $1025$ & $642$  & $1.76$ & $12.17$ \\
DyFXU$|_{1,2}$       & $0.75$ & $4259$ & $2880$ & $3065$ & $1278$ & $0.06$ & $0.32$ \\
DyFXU$|_{2,4}$       & $0.75$ & $4259$ & $2382$ & $3065$ & $1105$ & $0.23$ & $1.21$ \\
DyFXU$|_{3,4}$       & $0.75$ & $4259$ & $2206$ & $3065$ & $1002$ & $0.53$ & $3.01$ \\
DyFXU$|_{3,6}$       & $0.75$ & $4259$ & $2148$ & $3065$ & $981$  & $0.78$ & $4.81$ \\
DyFXU$|_{4,4}$       & $0.75$ & $4259$ & $2024$ & $3065$ & $896$  & $1.55$ & $10.38$ \\
DyFXU$|_{4,6}$       & $0.75$ & $4259$ & $1953$ & $3065$ & $802$  & $1.76$ & $12.17$ \\
PERF1 \cite{ZervakisTVLSI2016} 	& $0.75$ & $3355$ & $2880$ & $2547$ & $1202$ & $0.03$ & $0.17$ \\
PERF2 \cite{ZervakisTVLSI2016} 	& $0.72$ & $2927$ & $2473$ & $2214$ & $1083$ & $0.13$ & $0.60$ \\
PERF3 \cite{ZervakisTVLSI2016} 	& $0.66$ & $2892$ & $2342$ & $1874$ & $965$  & $0.44$ & $2.38$ \\
PERF4 \cite{ZervakisTVLSI2016} 	& $0.65$ & $2285$ & $1859$ & $1564$ & $858$  & $1.48$ & $9.75$ \\
R8ABM1 \cite{2016_Jiang_IEEEtc}      	& $0.77$ & $4210$ & $3895$ & $2694$ & $1268$ & $0.15$ & $1.05$ \\
R8ABM1-15 \cite{2016_Jiang_IEEEtc}    & $0.74$ & $2926$ & $2499$ & $1686$ & $890$  & $0.61$ & $2.70$ \\
R4ABM1-14 \cite{2017_Liu_IEEEtc}    	& $0.74$ & $3958$ & $3460$ & $2634$ & $1257$ & $0.12$ & $0.41$ \\
R4ABM1-16 \cite{2017_Liu_IEEEtc}    	& $0.73$ & $3725$ & $3246$ & $2577$ & $1108$ & $0.49$ & $1.49$ \\
R4ABM2-14 \cite{2017_Liu_IEEEtc}    	& $0.73$ & $3732$ & $3393$ & $2609$ & $1128$ & $0.24$ & $0.68$ \\
R4ABM2-16 \cite{2017_Liu_IEEEtc}    	& $0.72$ & $3467$ & $3101$ & $2547$ & $1079$ & $1.18$ & $2.50$ \\
RAD64 \cite{LeonTVLSI}    		& $0.72$ & $3489$ & $3238$ & $2257$ & $1057$ & $0.08$ & $0.42$ \\
RAD256 \cite{LeonTVLSI}   		& $0.69$ & $2769$ & $2410$ & $1911$ & $926$  & $0.28$ & $1.69$ \\
RAD1024 \cite{LeonTVLSI}  		& $0.65$ & $2624$ & $2224$ & $1663$ & $884$  & $0.93$ & $6.74$ \\
TMC8 \cite{Schulte1993} 		& $0.77$ & $3823$ & $3075$ & $2685$ & $1195$ & $0.11$ & $0.59$ \\
TMC15 \cite{Schulte1993} 		& $0.71$ & $2867$ & $2242$ & $1829$ & $928$  & $1.19$ & $2.75$ \\
DRUM6 \cite{2015_Hashemi_ICCAD} 	& $1.07$ & $3993$ & $2298$ & $3167$ & $1404$ & $1.47$ & $28.85$ \\
RoBA \cite{2017_Zendegani_IEEEtvlsi} 	& $0.94$ & $4040$ & $2345$ & $2999$ & $1216$ & $2.66$ & $50.07$ \\
\hline
\end{tabular}
\end{table}

Table \ref{tb_fxuhw} 
presents the experimental results 
for all the multipliers in $16$-bit arithmetic.
We note that the experiments are presented in two flavours
with respect to the clock frequency of the circuits:
\begin{itemize}
    \item[(i)] \underline{MIN-Delay}: the clock constraint of each circuit is set to its critical path delay (high-performance mode). 
    \item[(ii)] \underline{ISO-Delay}: the clock constraint of all the circuits is set to the same relaxed value (low-power mode). 
\end{itemize}

The advantage of AxFXU is its large approximation space,
which provides the flexibility 
to target multiple error levels,
while delivering remarkable resource gains. 
This is justified by the experimental results
showing
that our approximation technique 
delivers the best exploitation of the error imposed.
Namely, 
for similar error values, 
AxFXU outperforms the rest designs 
in terms of resources (delay, area, energy). 
Indicatively,
we mention that 
AxFXU$|_{4,4}$ 
inserts an MRED of $1.55\%$,
while PERF4 and DRUM6 have an MRED of
$1.48\%$ and $1.47\%$, 
respectively.
However, 
in the MIN-Delay scenario,
AxFXU$|_{4,4}$ 
delivers
$17\%$ area and 
$22\%$ energy
gains versus PERF4,
and
$53\%$ area and 
$37\%$ energy
gains versus DRUM6.
Similarly, 
in the ISO-Delay scenario,
the corresponding gains 
are 
$20\%$ and  
$17\%$ versus PERF4,
and 
$60\%$ and  
$49\%$ versus DRUM6.
We notice that 
in the ISO-Delay scenario,
the gains of AxFXU$|_{4,4}$ versus DRUM6 increase 
due to its large critical path delay (i.e., $1.07$ns),
which is close to the relaxed
clock constraint used. 

The radix-$4$ \cite{2017_Liu_IEEEtc} 
and radix-$8$ \cite{2016_Jiang_IEEEtc} 
multipliers
exhibit small error values,
however, 
compared to AxFXU, 
they suffer from increased energy consumption and delay.
Moreover, 
for every
RAD$2^k$ multiplier \cite{LeonTVLSI},
there is an AxFXU$|_{P,R}$ circuit 
that outperforms it in energy consumption, 
while delivering a slightly smaller MRED. 
For instance, 
in MIN-Delay, 
AxFXU$|_{2,4}$ offers $6\%$ better energy than RAD256,
while it attains smaller error ($0.23\%$ versus $0.28\%$).
In general, 
the RAD$2^k$ family of multipliers
features a limited set of approximation configurations,
and thus, 
its error scaling is abrupt,
which does not allow 
to efficiently serve various error constraints. 
Furthermore, 
AxFXU is more efficacious than the partial product perforation \cite{ZervakisTVLSI2016},
as the application of rounding delivers
significant energy savings, 
whereas the error slightly increases. 
Finally, 
DRUM6 \cite{2015_Hashemi_ICCAD}
and RoBA \cite{2017_Zendegani_IEEEtvlsi}
deliver good energy savings, 
however, 
they exhibit significant errors
(considered large for standalone arithmetic circuits).
In particular, 
their large MRED values
($1.47\%$ and $2.66\%$, respectively)  
are also translated to large PRED$_2$ values
($28.85\%$ and $50.07\%$, respectively).
Moreover, 
the critical paths of the DRUM6 and RoBA circuits
are larger
compared to the encoding-based circuits,
because they implement a different multiplication algorithm. 

Next,
we evaluate the runtime-configurable variant.
We remind that,
contrary to AxFXU,
where each approximation configuration is implemented as a new circuit,
there is only one DyFXU circuit.
This circuit implements ACCR4 plus 
some AND gates
to tune the approximation degree at runtime.
Namely, 
DyFXU$|_{P,R}$ 
denotes the DyFXU circuit
that is 
configured via the control signals
to execute the multiplication
with perforation $P$ and rounding $R$.  
The results for various DyFXU$|_{P,R}$
that are 
presented in Table \ref{tb_fxuhw},
are reasonable.
As expected, 
DyFXU
delivers the same critical path delay
and slightly increased area
compared to ACCR4.
Regarding energy consumption,
which has been measured
when DyFXU is operating 
for its $P$ and $R$ configuration,
it is improved compared to ACCR4
due to the logic paths that are not activated. 
Finally,
it is interesting to compare
DyFXU$|_{P,R}$
with its design-time counterpart,
i.e., AxFXU$|_{P,R}$. 
The energy consumption of DyFXU$|_{P,R}$
is increased,
which is reasonable, 
considering the leakage power of the inactive circuit nodes 
and the power consumed by 
the extra AND gates, 
as well as the control signals that are given as input to the circuit. 
Indicatively, 
DyFXU$|_{1,2}$ 
consumes
$1.03\times$ 
and 
$1.2\times$
more energy than 
AxFXU$|_{1,2}$,
in the MIN-Delay and ISO-Delay scenarios,
respectively. 
When considering more aggressive approximations, 
DyFXU$|_{4,6}$ 
consumes
$1.7\times$
and 
$1.3\times$
more energy than 
AxFXU$|_{4,6}$,
in the MIN-Delay and ISO-Delay scenarios,
respectively. 
Summarizing,   
in terms of area, 
DyFXU 
does not provide gains,
however, 
the overhead is negligible compared to ACCR4.
In terms of energy, 
it provides gains
and outperforms several design-time multipliers of the literature, 
even though these gains are reduced compared to AxFXU.
In any case,
the negligible area overhead
and the reduced, but also significant, 
energy gains
are achieved
while supporting dynamic configuration of the approximation.

\begin{figure}[!t]
\centering
\includegraphics[width=0.93\textwidth]{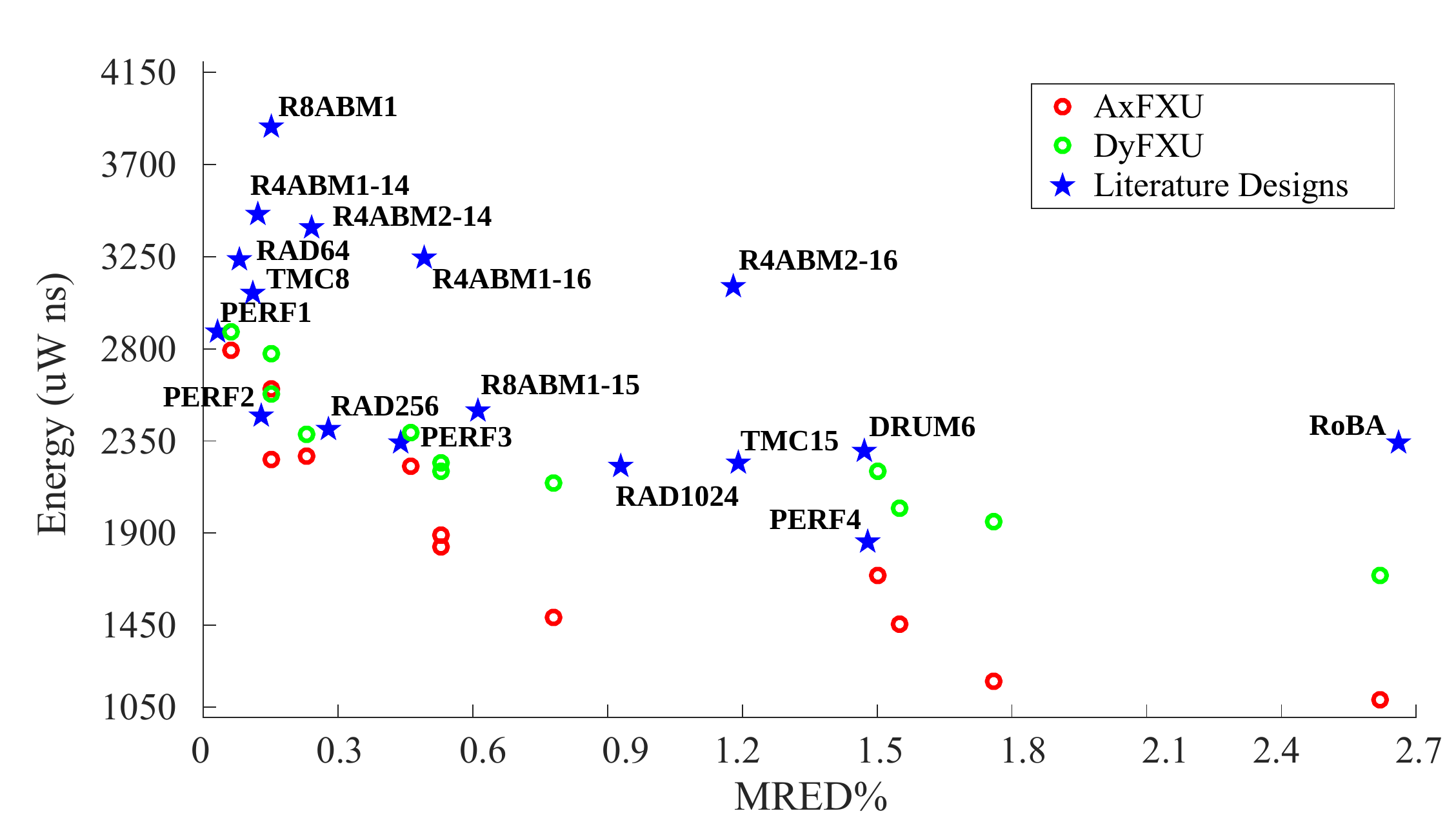}%
\caption[Comparative Pareto Analysis for the Approximate Fixed-Point Multipliers]{Comparative Pareto analysis for approximate fixed-point multipliers considering MRED and energy.}%
\label{fig_prpareto}
\end{figure}

A comprehensive comparison of all the examined designs 
is presented in Figure \ref{fig_prpareto},
where we consider both the MIN-Delay energy and MRED 
in a scatter plot. 
We note that this plot 
includes additional AxFXU and DyFXU designs. 
As shown, 
the Pareto front is formed
exclusively by different configurations of AxFXU,
i.e., our design exhibits the best error--energy trade-off
in this comparison involving several state-of-the-art designs. 
AxFXU$|_{2,2}$
provides significant energy reduction 
in exchange for a very small error.
Similarly, 
AxFXU$|_{2,6}$, AxFXU$|_{3,4}$, and AxFXU$|_{3,6}$
deliver remarkable energy gains
in exchange for MRED values up to $0.78\%$.
Regarding DyFXU, 
even though it has worse energy consumption than AxFXU, 
it retains, 
in almost all cases, 
the Pareto front with regard to the rest approximate multipliers.

\subsubsection{Comparative State-of-the-Art Evaluation of Floating-Point Designs}

The evaluation of 
our approximate floating-point designs,
namely, AxFPU$|_{P,R}$ and DyFPU$|_{P,R}$,
is performed in the following stages:
(i) Pareto analysis of the error--resources trade-off, 
(ii) comparison to state-of-the-art approximate designs,
and 
(iii) efficiency analysis of the runtime variant.

As already discussed, 
the approximation space of our designs is defined by two independent approximation techniques,
and thus, 
it is large. 
It is also obvious that it increases 
as we move on to larger floating-point bit-widths.
Therefore, 
at first, 
we perform a Pareto analysis 
involving the resources (delay, area, energy) and MRED, 
targeting to 
extract the best approximation configurations 
and also 
study the impact of the two approximation techniques in floating-point arithmetic.
Again, we consider half- and single-precision floating-point multiplier,
i.e., AxFPU16 and AxFPU32. 
Our Pareto analysis considers 
approximation configurations that do not produce very large errors. 
Furthermore, 
we stress the tool to perform synthesis 
at the critical path delay of each design, 
and we also configure the clock period of 
the gate-level simulation to be equal to the critical path delay.

\begin{figure}[!t]
\vspace*{-3pt}
\centering
\subfloat[\label{fig_parefp16a}]{\includegraphics[width=0.5\textwidth]{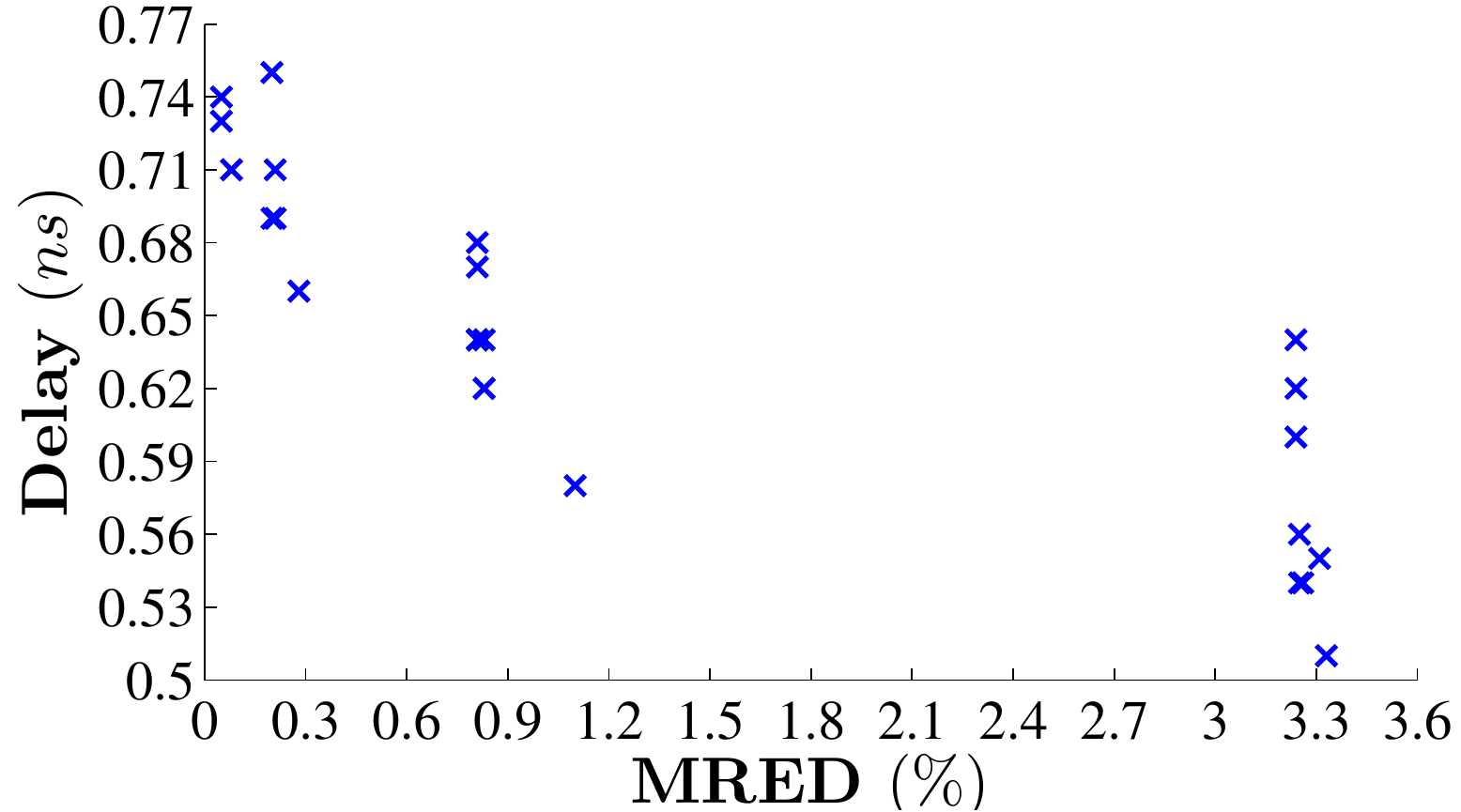}}\\[-10pt]
\subfloat[\label{fig_parefp16b}]{\includegraphics[width=0.5\textwidth]{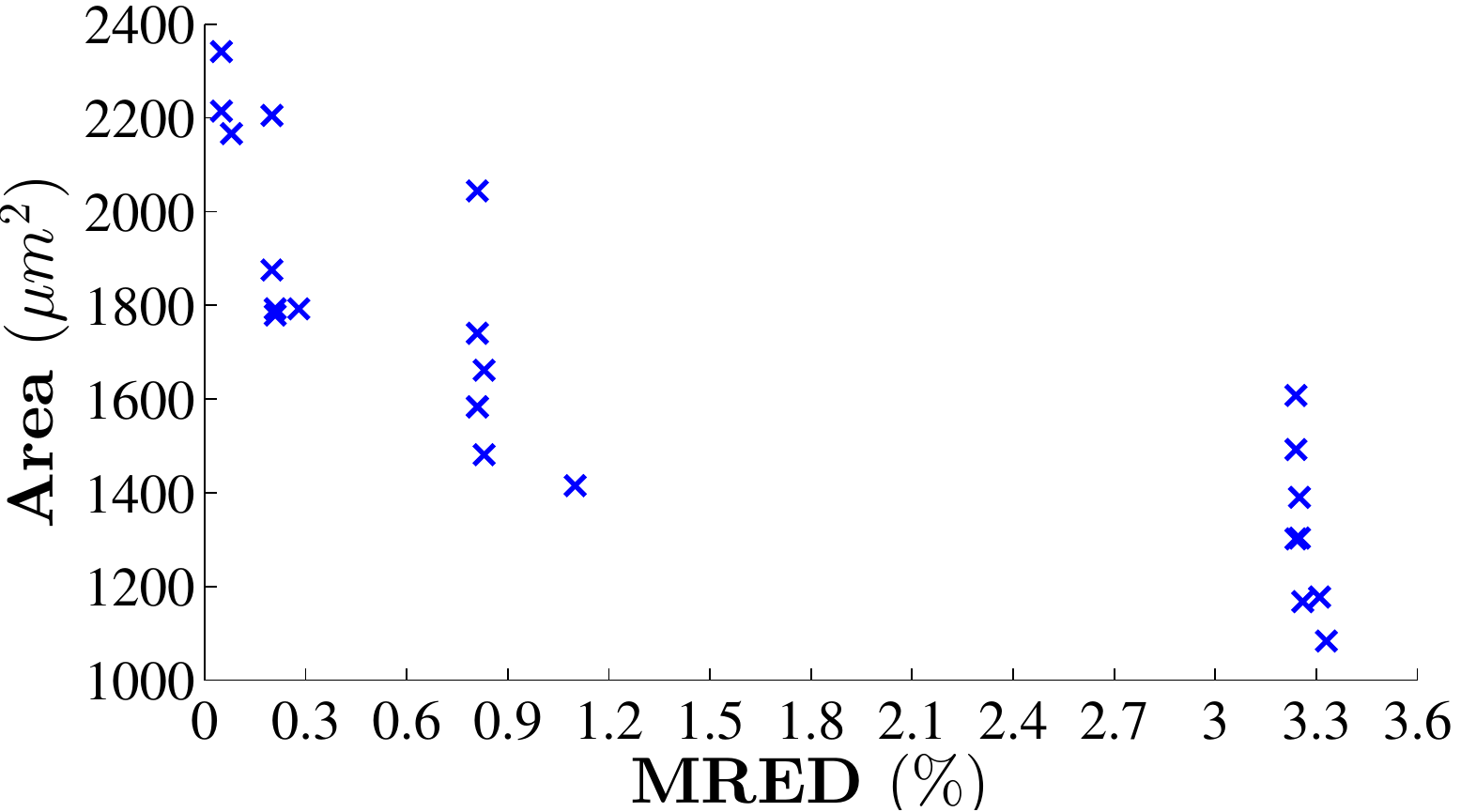}} \hfill
\subfloat[\label{fig_parefp16c}]{\includegraphics[width=0.5\textwidth]{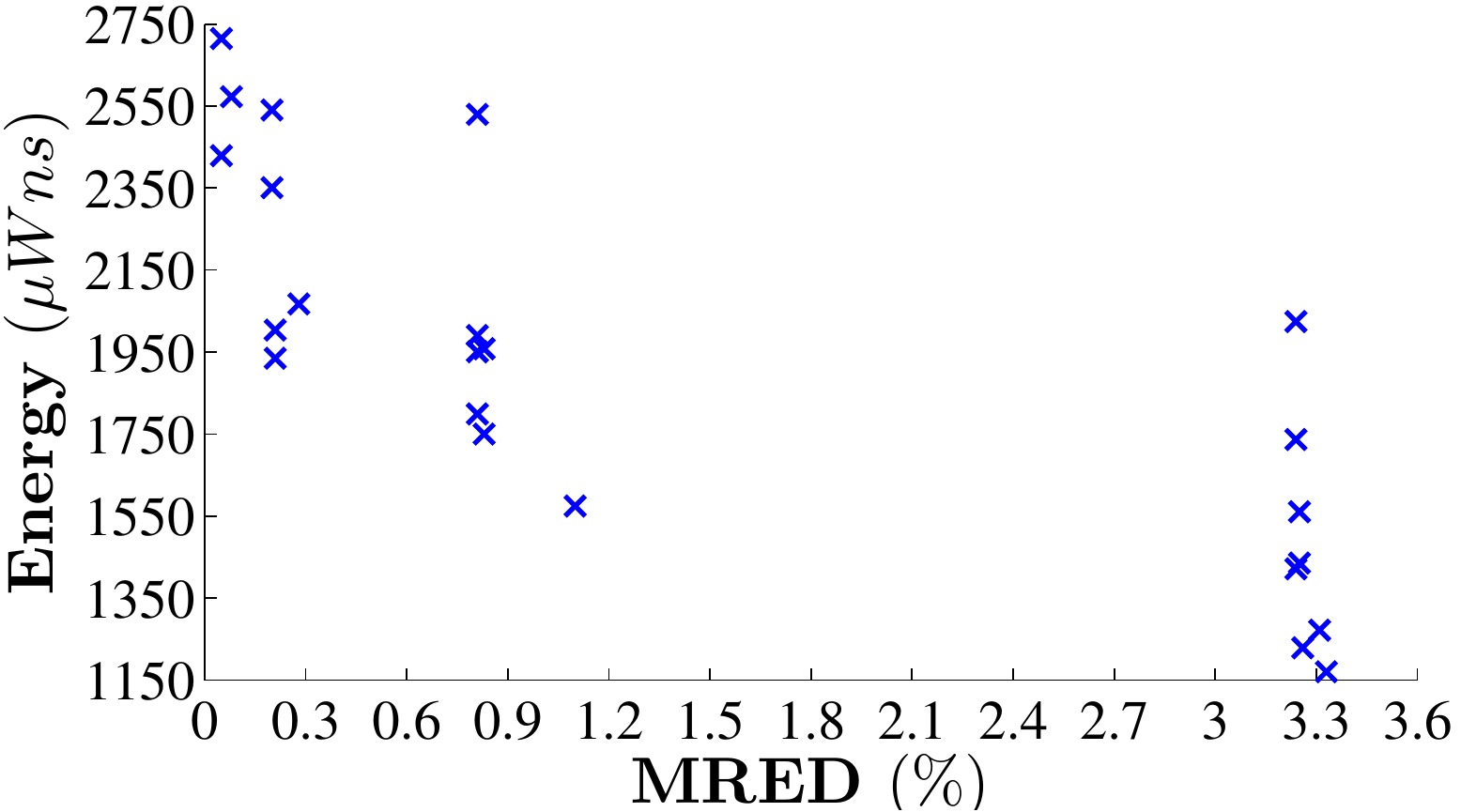}}%
\caption[Pareto Analysis for the Approximate Perforation-\&-Rounding Half-Precision Floating-Point Multipliers]{Pareto analysis for the half-precision AxFPU$|_{P,R}$ considering:
\textbf{(a)} Delay and MRED, 
\textbf{(b)} Area and MRED, 
and
\textbf{(c)} Energy and MRED.}%
\label{fig_parefp16}
\vspace*{-6pt}
\subfloat[\label{fig_parefp32a}]{\includegraphics[width=0.5\textwidth]{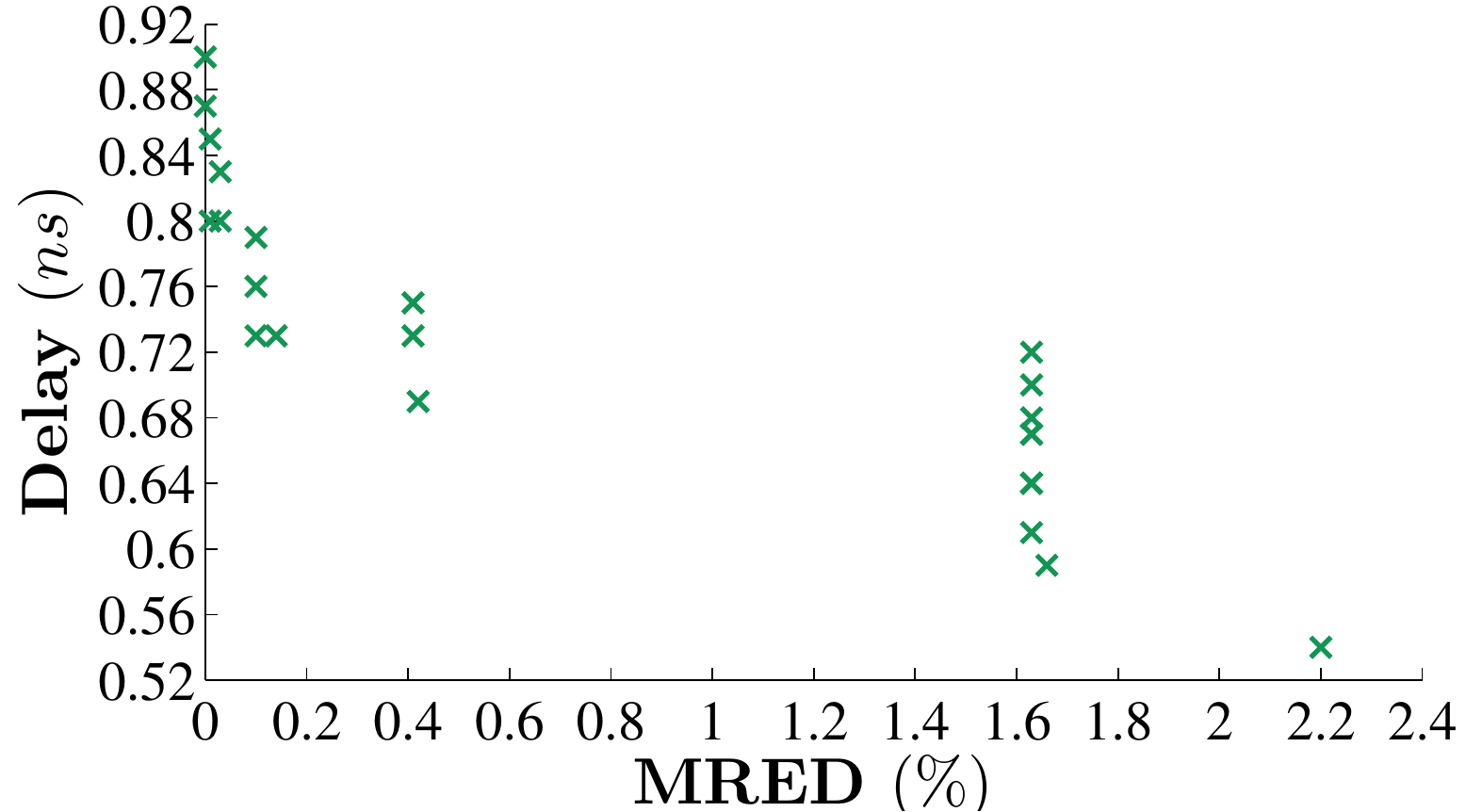}}\\[-10pt]
\subfloat[\label{fig_parefp32b}]{\includegraphics[width=0.5\textwidth]{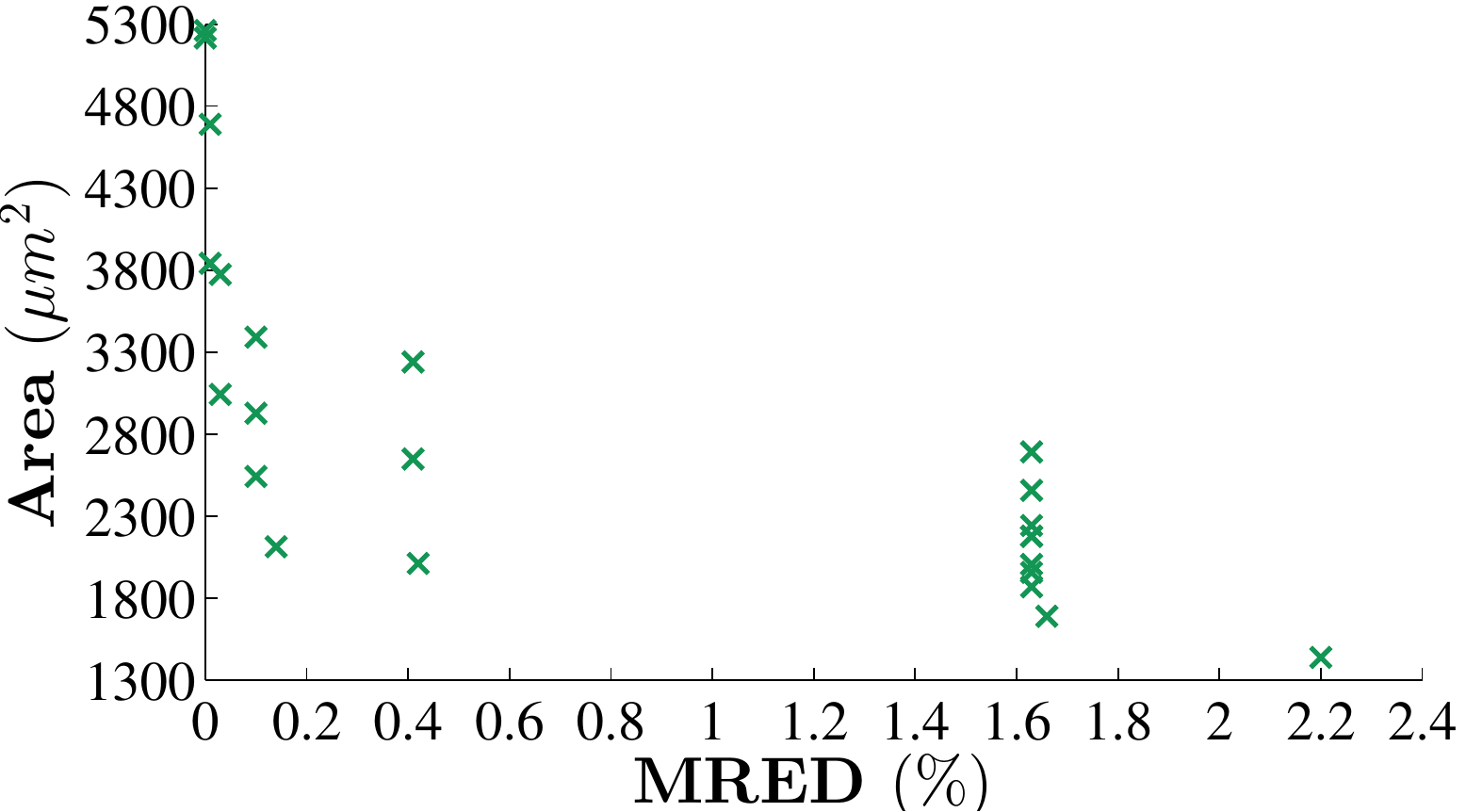}} \hfill
\subfloat[\label{fig_parefp32c}]{\includegraphics[width=0.5\textwidth]{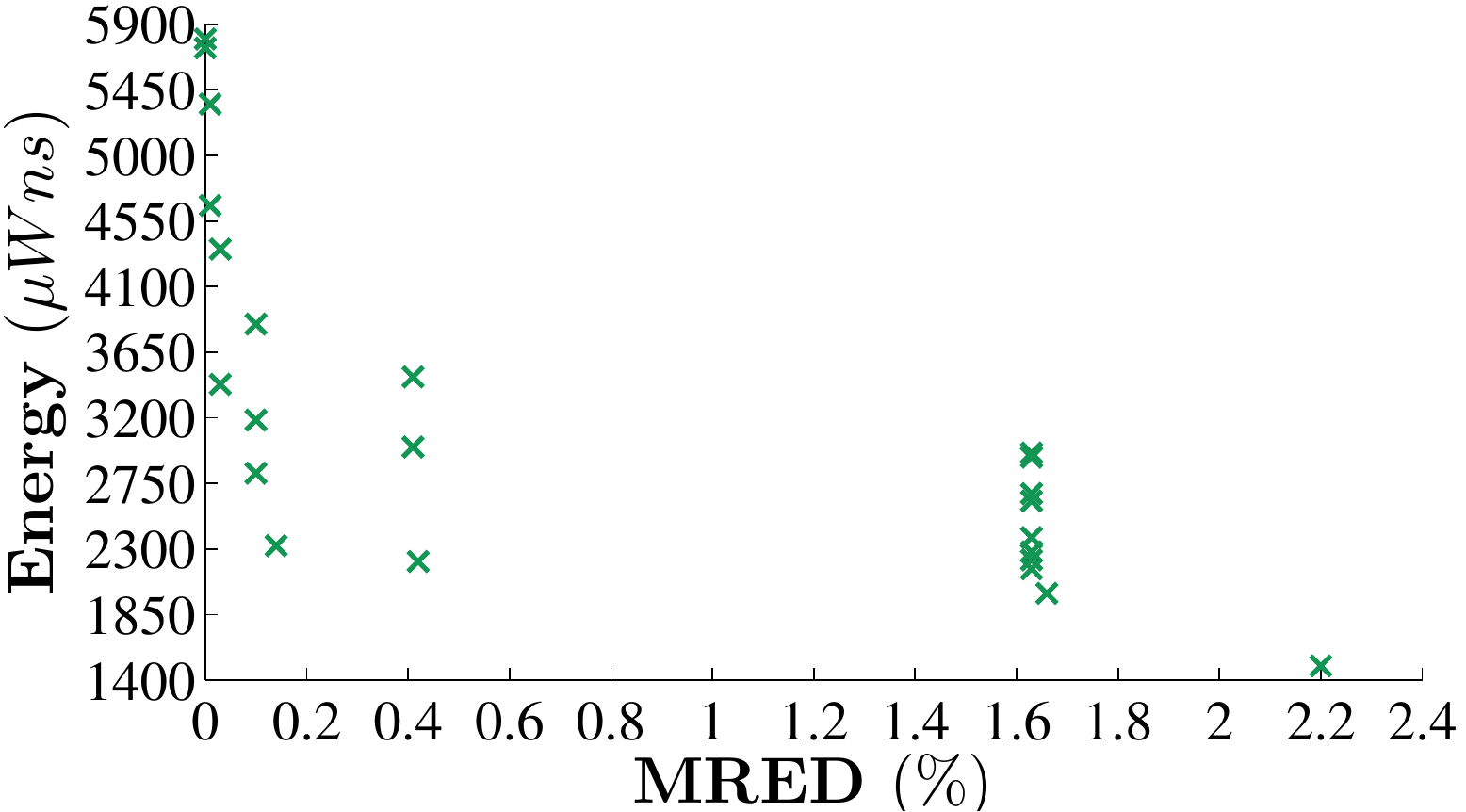}}%
\caption[Pareto Analysis for the Approximate Perforation-\&-Rounding Single-Precision Floating-Point Multipliers]{Pareto analysis for the single-precision AxFPU$|_{P,R}$ considering:
\textbf{(a)} Delay and MRED, 
\textbf{(b)} Area and MRED, 
and
\textbf{(c)} Energy and MRED.}%
\label{fig_parefp32}
\vspace*{-17pt}
\end{figure}

The scatter plots for the Pareto analysis  
of AxFPU16 and AxFPU32 
are presented in Figure \ref{fig_parefp16} and Figure \ref{fig_parefp32}, respectively.
The range of AxFPU16's delay is $0.75$ns--$0.51$ns, 
while the delay range for AxFPU32 is $0.9$ns--$0.54$ns.
Our exploration shows that 
there is a wide range of
area and energy values, 
along with typical MRED,
which constitutes AxFPU as 
a sustainable solution 
for scenarios with diverse area/energy constraints.
According to the results,
rounding has remarkable impact on the accuracy 
for small perforation configurations, 
while for bigger $P$ values, 
MRED almost remains intact.
For instance, 
the MRED of AxFPU16 is stable at \raisebox{0.8pt}{$\scriptstyle\sim$}$3.3\%$ 
for $P=4$ and different values of $R$, 
however, as the rounding aggressiveness increases, 
significant gains are achieved, 
e.g., $25\%$ energy reduction for $R=6$ versus $R=2$.
Regarding perforation, 
the single increment of the $P$ parameter, 
considering stable rounding configuration, 
delivers $5\%$--$12\%$ and $3\%$--$6\%$ energy reduction on average for AxFPU16 and AxFPU32, respectively. 
Furthermore, 
as expected, 
the slower scaling 
of AxFPU32's error due to its larger mantissa bit-width, 
allows more aggressive approximations, 
and as a result, 
its delay, area and energy are similar to those of AxFPU16.
Namely, 
our approximate half- and single-precision multipliers
impose comparable resource demands. 

\begin{figure}[!t]
\centering
\subfloat[\label{fig_fpgains16}]{\includegraphics[width=0.79\textwidth]{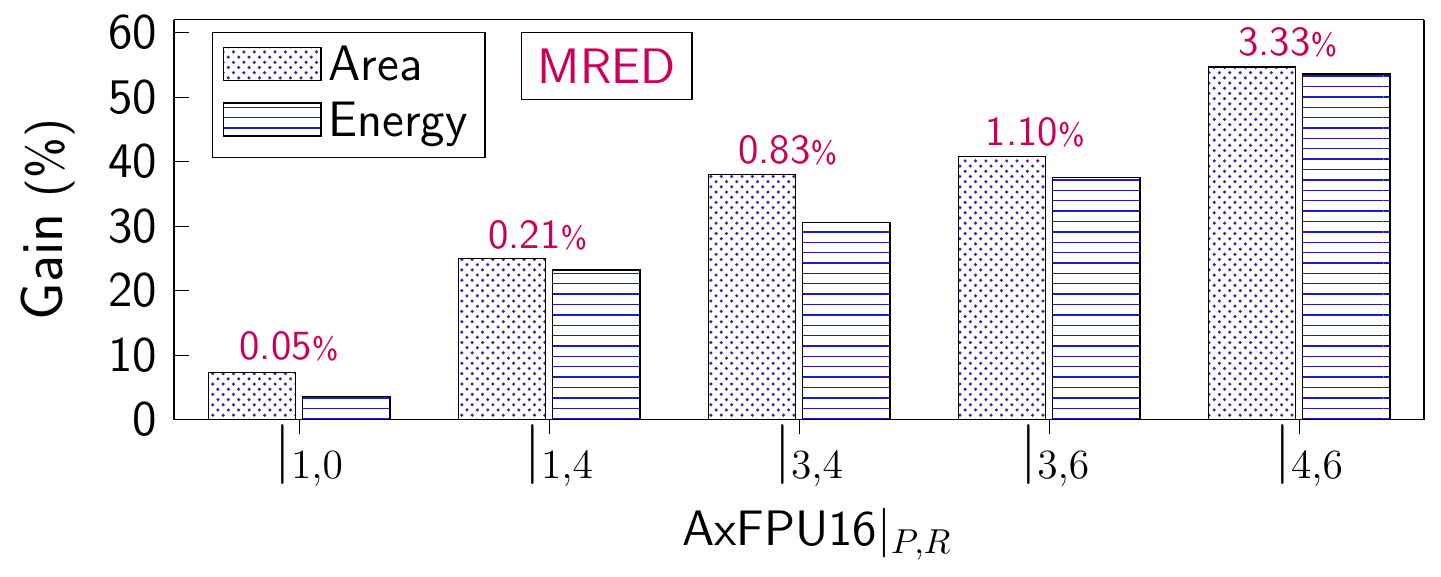}}\\[2pt]
\subfloat[\label{fig_fpgains32}]{\includegraphics[width=0.79\textwidth]{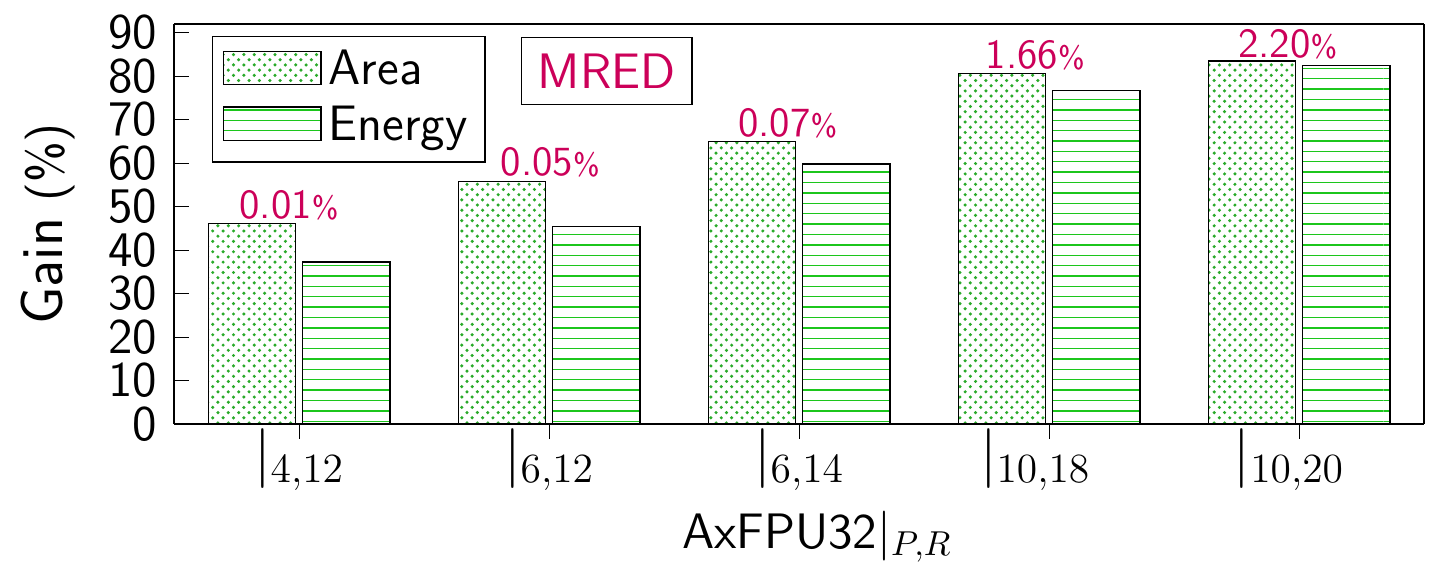}}%
\caption[Area and Energy Gains of the Approximate Perforation-\&-Rounding Floating-Point Multipliers]{Area and energy gains of AxFPU$|_{P,R}$ compared to the accurate floating-point multiplier
for 
\textbf{(a)} half precision 
and 
\textbf{(b)} single precision.}%
\label{fig_fpgains}
\end{figure}

Following our Pareto analysis, 
we examine the total resource gains of the Pareto-front AxFPU designs.  
In Figure \ref{fig_fpgains}, 
we present the area/energy gains of AxFPU
in comparison with the accurate floating-point multiplier,
as well as their MRED values. 
We select designs with varying MRED values, 
i.e., from $0.05\%$ to $3.33\%$ for AxFPU16 and from $0.01\%$ to $2.20\%$ for AxFPU32.
The half-precision AxFPU family 
delivers 
area gains in the range $7.3\%$--$54.6\%$ 
and energy gains in the range $3.6\%$--$53.5\%$.
The corresponding ranges for the single-precision AxFPU family 
are $46.1\%$--$83.4\%$ and $37.2\%$--$82.4\%$.
For both precisions, 
remarkable gains are achieved even for small MRED, 
e.g., $31\%$ and $59\%$ energy reduction for AxFPU16 and AxFPU32, respectively,
in exchange for $0.83\%$ and $0.07\%$ error values. 
Moreover, 
in terms of performance, 
our approximations decrease the critical paths, 
delivering up to $32.9\%$ and $46\%$ delay gain 
in AxFPU16 and AxFPU32, respectively.
According to the results, 
AxFPU32 provides increased gains compared to AxFPU16, even for smaller error values.
This is justified by its larger bit-width, 
which offers more room from approximations 
without significantly affecting the accuracy of the calculations.
We note that larger gains can be achieved for both floating-point precisions  
by increasing the approximation degree.  
However, more aggressive approximations should be accompanied 
by a careful error analysis involving the calculation of the metrics presented in the previous section.

Next, 
we compare AxFPU with the 
approximate floating-point multipliers of \cite{ImaniDAC2017} and \cite{ImaniISLPED2018},
as well as approximate floating-point multipliers 
that employ the designs of 
\cite{2017_Liu_IEEEtc} and \cite{2016_Jiang_IEEEtc}
to calculate the mantissa product. 
CFPU \cite{ImaniDAC2017} does not perform the mantissa multiplication and 
uses directly one of the two input mantissas as output.
On the other hand, 
RMAC \cite{ImaniISLPED2018} replaces the mantissa multiplication with the addition of the input mantissas.
R4ABM \cite{2017_Liu_IEEEtc} and R8ABM \cite{2016_Jiang_IEEEtc} 
are based on the approximate radix-$4$ and radix-$8$ encoding, respectively,
and
they are also included in the evaluation of our fixed-point design.
For this comparison,
we implement the best approximation configuration.
To make a fair comparison, 
we implement the approximation of these designs
in the mantissa multiplication of the accurate floating-point multiplier described
in Section \ref{s5_2}
(the rest components remain intact).
The synthesis of AxFPU, CFPU \cite{ImaniDAC2017}, RMAC \cite{ImaniISLPED2018}, R4ABM \cite{2017_Liu_IEEEtc} and R8ABM \cite{2016_Jiang_IEEEtc} is performed using the same relaxed clock constraint, 
i.e., 
the critical path delay of the accurate floating-point multiplier,
targeting to study their effectiveness, 
but also their sustainability under relaxed design margins.

In Table \ref{tb_fphw}, 
we present the resource gains 
and the accuracy results of the examined floating-point multipliers.
We note that we select AxFPU configurations
with error values that are comparable  
to those of the other designs. 
In terms of accuracy, 
CFPU \cite{ImaniDAC2017} is the most inefficient design,
as it 
delivers around $4\times$--$6\times$ 
larger MRED and 
PRED$_2$ of $70\%$. 
It is important to mention that both
PON and PUN,
i.e., the metrics examining unexpected output concerning overflow and underflow, 
is below $0.5\%$ for all the designs. 
In case of half precision, 
RMAC \cite{ImaniISLPED2018} has slightly better accuracy results than AxFPU, however,
our design provides increased energy gains.
Regarding single precision, 
AxFPU delivers better MRED and PRED$_2$ values than RMAC as well as higher energy gains.
Overall, 
the proposed AxFPU design provides larger energy savings than RMAC \cite{ImaniISLPED2018} for similar errors, 
as well as better energy efficiency than CFPU \cite{ImaniDAC2017}, 
while exhibiting significantly better accuracy.
In terms of critical path delay, 
CFPU \cite{ImaniDAC2017} and RMAC \cite{ImaniISLPED2018} deliver larger gains, 
as they model the mantissa multiplication with less complex operations.
Finally, the gains of R4ABM \cite{2017_Liu_IEEEtc} and R8ABM \cite{2016_Jiang_IEEEtc} are considered small compared to the other designs.
The advantage of these designs is their small error values.
Nevertheless, 
in case there is a constraint of small error values, 
AxFPU with less aggressive approximation 
can provide larger resource gains.
Indicatively, 
we mention that the 
MRED of AxFPU16$|_{3,6}$ is $1.10\%$
and that of AxFPU32$|_{10,18}$ is $1.66\%$. 

\begin{table}[!t]
\fontsize{9}{10}\selectfont
\renewcommand{\arraystretch}{1.2}
\setlength{\tabcolsep}{4.5pt}
\caption[Experimental Results of Approximate Floating-Point Multipliers on TSMC 65-nm Standard-Cell]{Experimental results of approximate half- and single-precision floating-point multipliers on TSMC 65-nm standard-cell.}
\label{tb_fphw}
\centering
\begin{threeparttable}
\begin{tabular}{cl| c c c| c c c c}
\hline
\multicolumn{2}{c|}{\multirow{2}{*}{\textbf{Design}}} &
\textbf{Delay} & 
\textbf{Area} & 
\textbf{Energy} & 
\textbf{MRED} & 
\textbf{PRED$_\mathbf{2}$} & 
\textbf{PON} &
\textbf{PUN} \\[-2pt]
& & $(\%)$\footnotemark\setcounter{footnote}{0} & $(\%)$\footnotemark\setcounter{footnote}{0} & $(\%)$\footnotemark\setcounter{footnote}{0} &  $(\%)$ & $(\%)$ & $(\%)$ & $(\%)$ \\
\hline \hline
\parbox[t]{8mm}{\multirow{5}{*}{\rotatebox[origin=c]{45}{\textbf{\emph{n $\mathbf{=16}$}}}}} 
& AxFPU16$|_{4,6}$              & $32.9$ &  $70.8$ & $67.1$ & $3.33$  & $57.40$ & $0.43$  & $0.10$ \\
& CFPU \cite{ImaniDAC2017}      & $44.7$  &  $71.6$ & $64.6$ & $12.96$ & $70.68$ & $0.46$  & $0.11$ \\
& RMAC \cite{ImaniISLPED2018}   & $42.2$ &  $71.1$ & $63.2$ & $3.16$  & $49.42$ & $0.20$  & $0$ \\
& R4ABM \cite{2017_Liu_IEEEtc}        & $7.8$ & $43.3$ & $39.1$ & $2.12$  & $20.61$ & $0.09$  & $0.01$ \\
& R8ABM \cite{2016_Jiang_IEEEtc}      & $10.1$ & $47.6$ & $41.3$ & $1.56$  & $7.14$ & $0$  & $0$ \\
\hline
\parbox[t]{8mm}{\multirow{5}{*}{\rotatebox[origin=c]{45}{\textbf{\emph{n $\mathbf{=32}$}}}}} 
& AxFPU32$|_{10,20}$            & $46.0$ &  $89.9$ & $87.4$ & $2.20$  & $44.86$ & $0.26$  & $0.01$ \\
& CFPU \cite{ImaniDAC2017}      & $53.9$  &  $86.5$ & $79.7$ & $12.80$ & $70.63$ & $0.05$  & $0.02$ \\
& RMAC \cite{ImaniISLPED2018}   & $50.6$ &  $83.9$ & $78.3$ & $2.92$  & $49.73$ & $0.02$  & $0$ \\
& R4ABM \cite{2017_Liu_IEEEtc}        & $9.1$ & $53.3$ & $50.7$ & $1.44$  & $16.32$ & $0.02$  & $0$ \\
& R8ABM \cite{2016_Jiang_IEEEtc}      & $14.3$ & $59.5$ & $51.1$ & $0.99$  & $5.46$ & $0$  & $0$ \\
\hline
\end{tabular}
\begin{tablenotes}
   \item[1]{\fontsize{7.7}{8.8}\selectfont Refers to $\%$ delay/area/energy gains (relative reduction) in comparison with the accurate design.}
\end{tablenotes}
\end{threeparttable}
\end{table}

Finally, 
we assess the hardware efficiency of DyFPU, 
which is the runtime-configurable variant of AxFPU.
We remind that 
DyFPU$|_{P,R}$ 
denotes the single DyFPU circuit
that is 
configured via the control signals
to execute the multiplication
with perforation $P$ and rounding $R$. 
In contrast, 
AxFPU$|_{P,R}$ is configured at design-time,
i.e., a new circuit is implemented for each 
approximation configuration.
Table \ref{tb_dynres} presents the comparison of the two variants 
by examining the gains compared the accurate design. 
Both DyFPU and AxFPU are synthesized and simulated under tight clock constraints, 
i.e., their critical path delays.
The outcomes from this comparison
are similar to those derived 
from the comparison of the fixed-point designs (AxFXU and DyFXU).
More specifically, 
as expected, 
DyFPU imposes
an area increase of $4.3\%$ and $2.1\%$ for half and single precision, 
respectively.
Regarding energy consumption, 
DyFPU is worse than its design-time counterpart, 
delivering 
\raisebox{0.8pt}{$\scriptstyle\sim$}$1.4 \times$
and \raisebox{0.8pt}{$\scriptstyle\sim$}$1.6 \times$
less gain for remarkable approximation 
in half and single precision, respectively.
However, it is still more energy-efficient than the accurate design, 
providing energy gains up to $49.7\%$, 
while also allowing to dynamically tune the approximation degree
by seamlessly 
changing the values of the $P$ and $R$ parameters. 

In comparison with the runtime-configurable variants of the 
CFPU \cite{ImaniDAC2017} and RMAC \cite{ImaniISLPED2018} designs, 
DyFPU is not designed 
to automatically configure its operation   
to either accurate or approximate mode
based on the mantissa inputs or the approximate mantissa product.
However, it prevails over these two methods in the following factors: 
(i) it supports multiple approximation configurations instead of one fixed configuration,
(ii) it does not insert extra delays (e.g., to check the inputs/outputs or even re-calculate the result using the accurate circuit),
(iii) it exposes the tuning logic (AND gates) to the system,
allowing a high-level policy or framework 
to configure the approximation degree 
with respect to the application's energy and accuracy constraints.  

\begin{table}[!t]
\fontsize{9}{10}\selectfont
\renewcommand{\arraystretch}{1.2}
\setlength{\tabcolsep}{4.1pt}
\caption[Comparison of the Design-Time and Runtime Approximate Floating-Point Multipliers]{Comparison of the design-time and runtime approximate floating-point multipliers.}
\label{tb_dynres}
\centering
\begin{threeparttable}
\begin{tabular}{cc}
\begin{tabular}{cl| c c}
\hline
\multicolumn{2}{c|}{\multirow{2}{*}{\textbf{Design-Time Config.}}} &
\textbf{Area} & 
\textbf{Energy}\\[-2pt]
& & $(\%)$\footnotemark\setcounter{footnote}{0} & $(\%)$\footnotemark\setcounter{footnote}{0} \\
\hline \hline
\parbox[t]{8mm}{\multirow{3}{*}{\rotatebox[origin=c]{45}{\textbf{\emph{n $\mathbf{=16}$}}}}} 
& AxFPU16$|_{1,0}$      &  $7.3$ & $3.6$  \\
& AxFPU16$|_{3,4}$      &  $38.0$ & $30.5$  \\
& AxFPU16$|_{4,6}$      &  $54.6$ & $53.5$  \\
\hline
\parbox[t]{8mm}{\multirow{3}{*}{\rotatebox[origin=c]{45}{\textbf{\emph{n $\mathbf{=32}$}}}}} 
& AxFPU32$|_{4,12}$     &  $46.0$ & $37.2$  \\
& AxFPU32$|_{6,14}$     &  $64.9$ & $59.8$   \\
& AxFPU32$|_{10,18}$    &  $80.5$ & $76.6$  \\
\hline
\end{tabular}
&
\begin{tabular}{cl| c c}
\hline
\multicolumn{2}{c|}{\multirow{2}{*}{\textbf{Runtime Config.}}} &
\textbf{Area} & 
\textbf{Energy}\\[-2pt]
& & $(\%)$\footnotemark\setcounter{footnote}{0} & $(\%)$\footnotemark\setcounter{footnote}{0} \\
\hline \hline
\parbox[t]{8mm}{\multirow{3}{*}{\rotatebox[origin=c]{45}{\textbf{\emph{n $\mathbf{=16}$}}}}} 
& DyFPU16$|_{1,0}$    &  $-4.3$ & $1.1$  \\
& DyFPU16$|_{3,4}$    &  $-4.3$ & $22.5$  \\
& DyFPU16$|_{4,6}$    &  $-4.3$ & $37.2$  \\
\hline
\parbox[t]{8mm}{\multirow{3}{*}{\rotatebox[origin=c]{45}{\textbf{\emph{n $\mathbf{=32}$}}}}} 
& DyFPU32$|_{4,12}$   &  $-2.1$ & $23.1$  \\
& DyFPU32$|_{6,14}$   &  $-2.1$  & $34.1$  \\
& DyFPU32$|_{10,18}$  &  $-2.1$  & $49.7$  \\
\hline
\end{tabular}
\end{tabular}
\begin{tablenotes}
  \item[1]{\fontsize{7.7}{8.8}\selectfont Refers to $\%$ area/energy gains (relative reduction) in comparison with the accurate design.}
\end{tablenotes}
\end{threeparttable}
\end{table}

\section{Conclusion}
\label{s5_4}

In this chapter,
we examined an attractive aspect of Approximate Computing,
i.e., 
the dynamic configuration of the approximation degree. 
This feature is extremely important for
modern embedded systems and circuits,
which need to adapt the accuracy of the calculations
at runtime, 
depending on
the type of the application, 
the input dataset of the application,
and the given energy constraints.
Towards this direction,
we targeted the runtime-configurable arithmetic circuits 
and designed energy-efficient multipliers 
that can change approximation configuration at runtime. 
In particular, 
we employed two orthogonal approximation techniques
for calculating the partial products
and 
generated a large approximation space
that serves different scenarios in terms of accuracy and energy budget. 
The technique of 
partial product perforation 
omits the generation of least significant products,
while partial product rounding
efficiently decreases the bit-width of the remaining ones. 
These two approximation techniques
facilitate dynamic configurability  
when combined with the radix-$4$ operand encoding,
as their application is associated with the input operand bits. 
Therefore,
we added negligible logic overhead in the input of the accurate design,
i.e., $2n$ AND gates, 
where $n$ is the operand bit-width, 
and enabled dynamic configuration via input signals. 
We designed 
both integer/fixed-point and floating-point circuits
integrating our approximation techniques and runtime functionalities.
In terms of accuracy, 
we evaluated our designs 
by performing an in-depth exploration 
involving diverse error metrics for various approximation configurations,
as well as new error metrics that are tailored to floating-point arithmetic. 
The results show that the proposed solution
exhibits dense error scaling,
i.e., it
provides multiple configurations with mean error values 
across the entire range that is considered acceptable (up to $2\%$). 
In terms of circuit resources, 
our design-time variants 
outperform all the examined state-of-the-art multipliers
in both fixed- and floating-point arithmetic.
For similar error values,
the fixed-point AxFXU designs
deliver gains up to 
$60\%$ and $49\%$ 
in area and energy, 
respectively, 
when operating at the same clock frequency. 
Similarly,
the floating-point AxFPU designs
exhibit 
either better accuracy for the same resource gains
or larger resource gains for comparable error values. 
The respective runtime variants
provide negligible area overhead
and smaller energy gains,
however,
they still deliver remarkable gains versus the accurate multiplier
and other state-of-the-art design-time multipliers,
while allowing to seamlessly change approximation configuration at the runtime.
In fixed-point arithmetic,
DyFXU consumes only
up to
$1.2\times$ and $1.7\times$
more energy than 
AxFXU 
for low-strength and more aggressive approximation, respectively. 
Similar behaviour is observed in our floating-point designs, 
where 
DyFPU provides 
\raisebox{0.8pt}{$\scriptstyle\sim$}$1.4 \times$
and \raisebox{0.8pt}{$\scriptstyle\sim$}$1.6 \times$
less energy gains
in half and single precision, respectively. 
\chapter{Cooperative Approximation:
Combination~of~Arithmetic Encodings}
\label{chapter6}

\addtocontents{lof}{\protect\contentsline{chapter}{\protect\numberline{6}Cooperative Approximation: Combination of Arithmetic Encodings}{}{}}
\addtocontents{lot}{\protect\contentsline{chapter}{\protect\numberline{6}Cooperative Approximation: Combination of Arithmetic Encodings}{}{}}

\begin{ChapterAbstract}
The rapid growth of error-resilient applications
with different accuracy constraints and computational demands
creates the need for approximate circuits and systems
that can support multiple approximations. 
Towards this direction,
the designers tend to select approximation techniques 
that do not generate a single approximate design,
but they can provide various design variants with different approximation configuration,
and thus, different accuracy and resource gains. 
In this context,
we combine arithmetic approximation techniques
to create a very large approximation space
for the design of multiplication circuits. 
Besides providing numerous approximation configurations,
we also target to identify the most-efficient design solution
among the well-established arithmetic approximation techniques. 
Our pool of techniques 
consists of high-radix encoding,
partial product perforation and partial product rounding. 
The feasible combinations of these techniques
generate $5$ new design families of approximate multipliers. 
Our extensive design space exploration
shows that the combination of approximation techniques,
called as ``cooperative approximation'',
results in slow error scaling with numerous configurations 
in the typical acceptable range
$\mathit{0}$\%--$\mathit{2}$\%.
This feature creates increased design flexibility,
allowing to efficiently handle workloads and applications
with different constraints. 
The experimental evaluation
is based on a comparative state-of-the-art Pareto analysis 
involving the Pareto-front designs RAD and AxFXU,
namely all the approximate designs presented in the Dissertation,
as well as other designs of the literature.
The results show
that 
the Pareto front is formed exclusively 
by designs with cooperative approximation.
In particular,
the ROUP family of multipliers,
which applies perforation and a new type of rounding,
constitutes the most energy-efficient design alternative,
improving the Pareto front
by up to $\mathit{1.5\times}$--$\mathit{2\times}$.
Furthermore, 
the new Pareto front
has increased resolution 
(i.e., more design configurations)  
due to the large approximation space.  
Finally,
in comparison with state-of-the-art designs, 
the ROUP family 
provides energy gains up to $\mathit{63}$\% for the same error constraint.
\\
This chapter is based on our
\textbf{publication} in \textbf{\cite{LeonDAC}}.
\end{ChapterAbstract}

\newpage

\section{Introduction}

One of the goals of Approximate Computing
is to provide design solutions 
that can handle various accuracy and resource constraints.
This feature is extremely important,
considering that
the error-tolerant applications 
\cite{ChakradharDAC2010, ChippaDAC2013}
impose non-identical demands regarding the quality of results 
and exhibit different error propagation within their  calculations. 
Approximation techniques are applied at software, architecture and hardware layers \cite{2016_Mittal_ACMsrv, 2016_Xu_IEEEdt, 2021_Stanley_ACMsrv}.
To expand the approximation space,  
the designers are examining the simultaneous application
of more than one approximation technique,
either from different or the same layer. 
Although vertical cross-layer approximation techniques have recently emerged \cite{Shafique_2016}, 
the full potential of horizontal approximation, 
i.e., within the same layer of design abstraction, 
still remains an open issue 
for further exploration. 
In this chapter, 
we explore, for the first time, the efficiency of combining arithmetic approximation techniques in the design of  energy-efficient multiplication circuits.  
We focus on combining approximation techniques at the arithmetic level,
as it inherently affects both the dynamic and static power consumption of the underlying circuits. 
Moreover, 
the implementation delivered at this level can be straightforwardly adopted in a vertical cross-layer design approach. 

The work presented in this chapter 
is inspired from the promising results of the AxFXU design (see Chapter \ref{chapter5}),
which combines two orthogonal approximation techniques.
The goal of AxFXU is twofold:
(i) to provide multiple approximation configurations
and enrich the available accuracy options at runtime,
and 
(ii) to apply an approximation scheme that facilitates the integration of dynamic configuration capabilities. 
In this chapter,
we perform an extensive design space exploration on
combining arithmetic approximation techniques.
Contrary to Chapter \ref{chapter5}, 
the goal of this chapter is 
the expansion of the approximation space with diverse sets of approximation configurations
and the 
in-depth exploration of the design space to identify the most efficient solution.
We use the term 
\emph{cooperative approximation}
to state that
two techniques are applied in the design of the approximate circuit. 
As a result of our exploration on cooperative approximation,
we propose 5 families of approximate multipliers,
which combine some of the techniques that have been already examined,
i.e., 
high-radix encoding, partial product perforation, and partial product rounding. 

\begin{figure}[!t]
\centering
\includegraphics[width=0.88\textwidth]{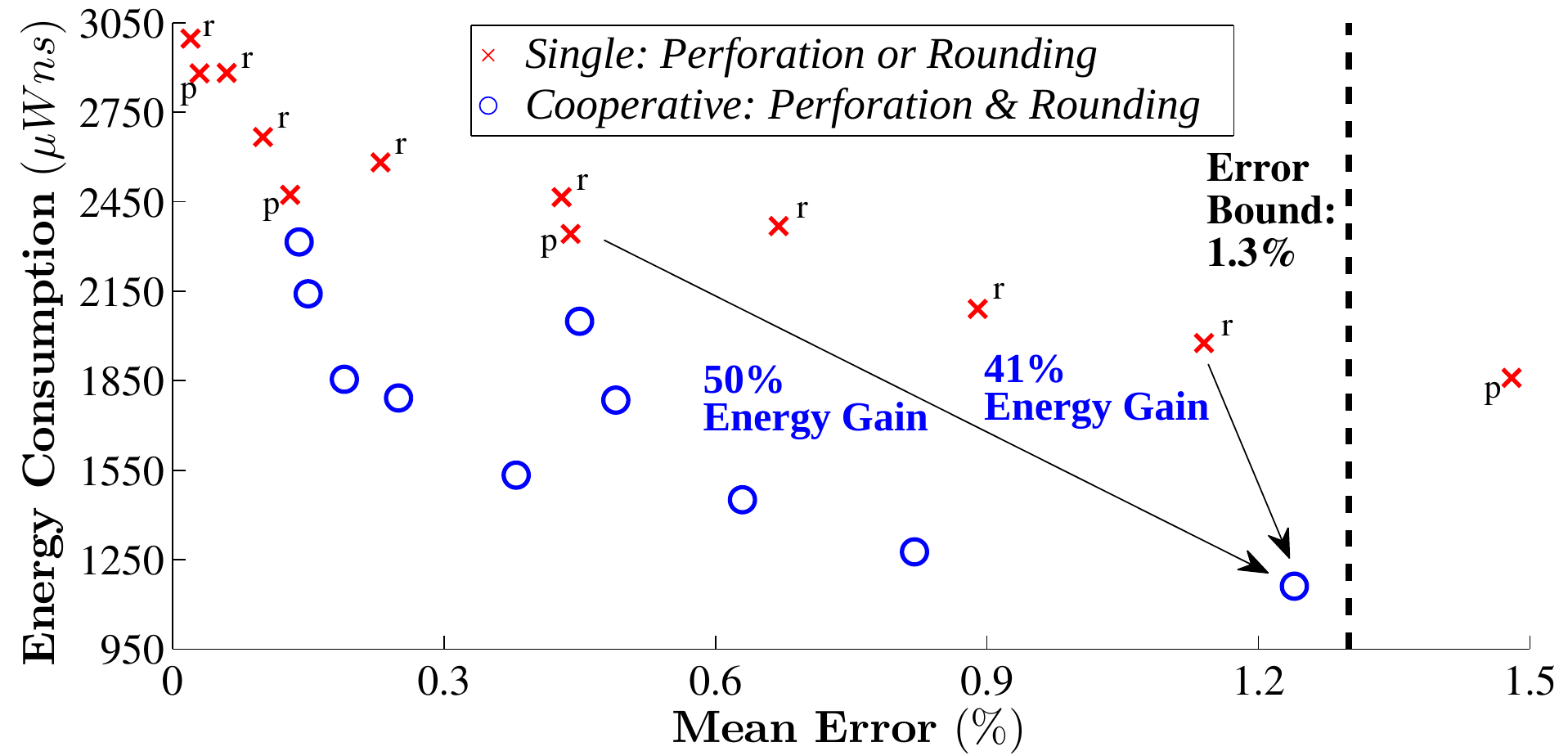}%
\caption[Motivation Plot for Cooperative Arithmetic Approximation]{Motivation plot for cooperative arithmetic approximation.}%
\label{fig_motiv}
\vspace*{-5pt}
\end{figure}

Figure \ref{fig_motiv} summarizes the motivation behind our work by 
demonstrating the benefits of combining approximation techniques rather than using a single one.
This plot illustrates the scatter points 
of partial product perforation,
partial product rounding,
and their combination.
The red points label
the single application of perforation and rounding,
while the blue points label the cooperative perforation \& rounding. 
We note that 
for the cooperative approach,
we employ two perforation configurations 
and combine each one with five rounding configurations. 
As shown, 
the cooperative approximation approach
provides three advantages
versus the single approximation approach (either perforation or rounding): 
(i) it forms the Pareto front, thus, it is considered a better design solution,
(ii) it provides increased Pareto front resolution due to the larger design/approximation space,
and
(iii) it delivers a more energy-efficient circuit for a given error constraint, e.g., for mean error of $1.3\%$, 
it gains $50\%$ and $41\%$ in energy 
versus perforation and rounding, respectively.

The \textbf{contribution} of this chapter is summarized as follows:

\begin{itemize}[]
\item[(i)] We highlight the efficiency of integrating more than one approximation technique in the design of approximate circuits. 
\item[(ii)] We propose 5 new families of approximate multipliers that feature a very large design space 
with differing approximation configurations
that can efficiently handle diverse error constraints. 
\item[(iii)] We reform the state-of-the-art energy/area--error Pareto front
by improving it and also increasing its resolution. 
\item[(iv)] We provide a more energy-efficient design solution than state-of-the-art designs for a given error constraint. 
\end{itemize}

The remainder of this chapter is organized as follows. 
Section \ref{s6_2} classifies the arithmetic approximation techniques and reports representative state-of-the-art works.
Section \ref{s6_3} introduces our approximation techniques
and presents how they can be combined in the design of
multipliers with cooperative approximation. 
In 
Section \ref{s6_4},
we assess the cooperative approximation techniques 
in terms of accuracy and circuit efficiency 
by reporting a comparative state-of-the-art Pareto evaluation,
which includes the previous Pareto-front designs. 
Finally, 
Section \ref{s6_5} draws the conclusions.

\section{Classification of Arithmetic Approximation Techniques}
\label{s6_2}

Arithmetic approximations in circuits have been extensively studied in the past, 
as they deliver significant energy gains at the application/system level. 
Interestingly, 
we focus on techniques that apply approximations
based on the input operands and the bit significance. 
i.e., 
they can be included in the wider range of encoding-based techniques,
and
we attempt to categorize them into the following classes:
(i) pruning, 
(ii) radix encoding, 
(iii) rounding, and
(iv) dynamic scaling.
More specifically, 
we provide a closer look at each class
by describing their basic approximation concept 
and presenting representative state-of-the-art works for multiplication circuits.

\underline{Pruning}: 
This class of techniques aims to reduce the logic by
discarding either bits, terms, or nodes of any arithmetic circuit.
A well-established pruning technique for approximate multipliers 
is the partial product perforation \cite{ZervakisTVLSI2016}. 
In this technique, partial products are omitted, and thus, 
simpler partial product matrices are generated. 
One downside of this technique 
is that the error is exponentially increased as more partial products are excluded.
The literature also provides more general
pruning approaches. 
In \cite{2017_Schlachter_IEEEtvlsi}, 
the authors propose a methodology and tool to automatically trade accuracy for area, power and delay gains. 
Their method applies gate-level pruning to any combinational circuit,
and it is quite effective especially on arithmetic circuits that have a notion of bit significance. 
In the same context, 
probabilistic pruning is proposed in \cite{Lingamneni2013},
where a greedy approach is used to generate approximate circuits. 

\underline{Radix Encoding}:
The techniques that are based on radix encodings,
approximately encode the input operands
to reduce the complexity of the calculations.  
Specifically for multipliers, 
approximate radix encodings result in
less partial product bits or even reductions in the total number of partial products. 
Liu \emph{et al.} \cite{2017_Liu_IEEEtc} designed approximate radix-$4$ encoders by transforming the Karnaugh map of the accurate encoding. 
Moving to radix-$8$ multipliers, Jiang \emph{et al.} \cite{2016_Jiang_IEEEtc} employed an approximate adder for producing the partial product $\pm3A$.
In Chapter \ref{chapter4}, we presented our approximate high-radix encodings \cite{LeonTVLSI}.
In particular, we issue the increased complexity of the conventional accurate high-radix encodings by mapping all the radix values to their nearest of the $4$ largest powers of two. 
This approximation provides 
simpler operand encoders and 
partial product generators.
In the same context,
there are various radix-based approximate multipliers in the literature,
e.g., 
using radix-$4$ \cite{Ratko2020}, radix-$8$ \cite{2020_Waris_IEEEtcasii} and radix-$256$ \cite{Zhu2022} encodings.

\underline{Rounding}:
Both truncation and rounding techniques have been examined in arithmetic circuits.
Truncation simply discards least significant bits
to reduce the bit-width.
In contrast,
rounding performs mapping to a smaller bit-width,
while trying to 
compensate for the error,
e.g., by inserting correction terms.  
These techniques are applied to either
the input operands, intermediate results or the final result. 
Several approximate multipliers are designed 
by applying rounding/truncation 
along with error compensation techniques.
A representative work is the truncated multiplier proposed in \cite{Schulte1993}, 
where the least significant bits of the partial products are vertically discarded and a constant correction term 
is added to reduce the total error. 
In \cite{ChoTVLSI2004},
the authors propose  
an error compensation method 
for the radix-$4$ multiplier 
that outputs products with the same bit-width as the input operands. 
Zhang \emph{et al.} \cite{ZhangIEEEToCaS2018} design an approximate multiplier by dividing the partial product matrix into two segments: 
the main segment,
which is accurately accumulated,
and the truncated segment, 
which is further partitioned into two parts.
The least significant part of the truncated segment is  calculated through a probabilistic approach. 
In Chapter \ref{chapter5},
we introduced partial product rounding,
which rounds the partial products to a smaller bit-width based on low-level optimizations \cite{LeonMicro}. 
As we showed,
the rounding of the partial products is
equivalent to rounding one of the input operands.

\underline{Dynamic Scaling}:
In this class,
the approximation degree is determined 
with respect to the input operands.
In \cite{NarayanamoorthyTVLSI2015}, 
the authors statically capture and multiply bit segments, 
either starting from the most significant bit 
or ending at the least significant bit.
A limitation of this technique is the difficulty in scaling to higher bit-widths, 
and thus, 
its benefits are reduced as the input bit-width grows. 
Based on the varying bit significance, 
Hashemi \emph{et al.} \cite{2015_Hashemi_ICCAD} proposed a more fine-grained input segmentation, using leading one detector circuits to locate the most significant `$1$' in each operand. 
However, 
this dynamic segmentation implies extra circuits for the signed multiplications, 
increasing the total circuit area. 

\section{Design of Multipliers with Cooperative Approximation}
\label{s6_3}

In this section,
we present the proposed combinations of arithmetic approximation techniques.
Firstly, we make a brief introduction
in the examined techniques,
targeting to 
explore the feasibility of all the possible combinations,
and then,
we discuss the technical details of each combination.

\subsection{The Pool of Arithmetic Approximation Techniques}

\begin{figure}[!t]
\vspace*{-15pt}
\centering
\subfloat[\label{fig_tec1}]{\includegraphics[width=0.56\textwidth]{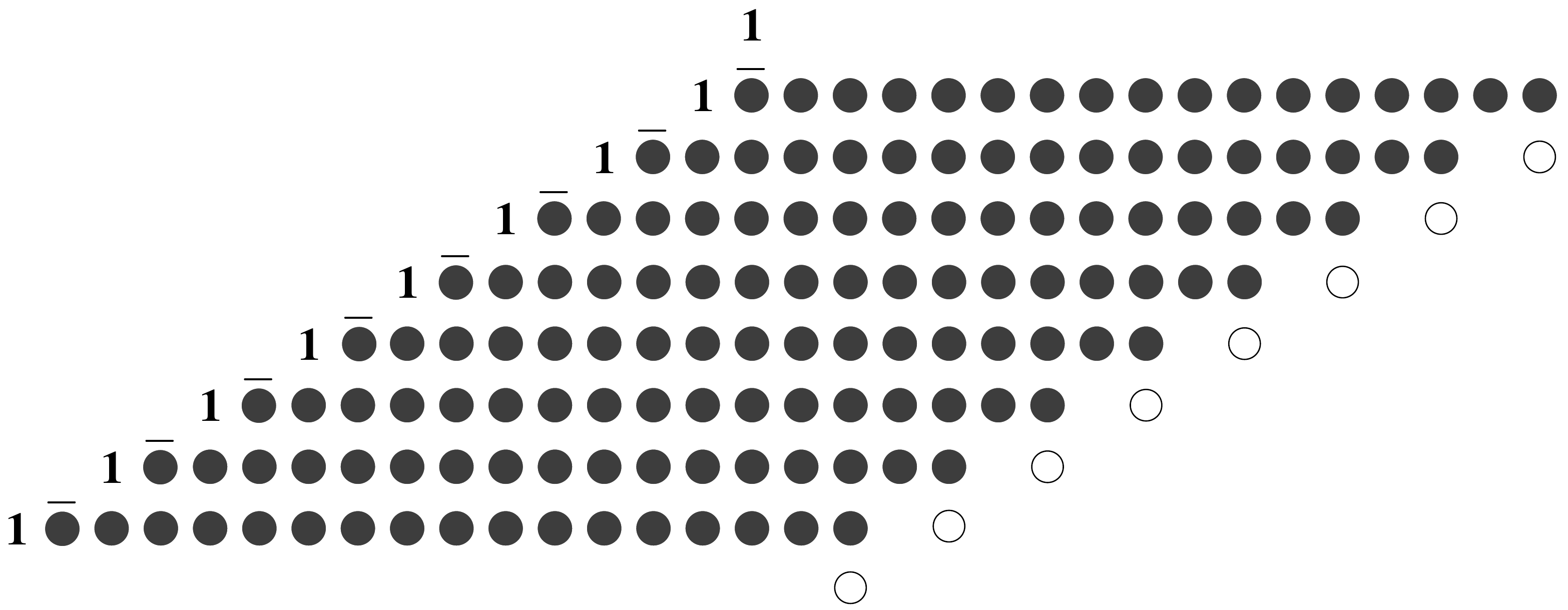}} \hspace*{-47pt} %
\subfloat[\label{fig_tec2}]{\includegraphics[width=0.56\textwidth]{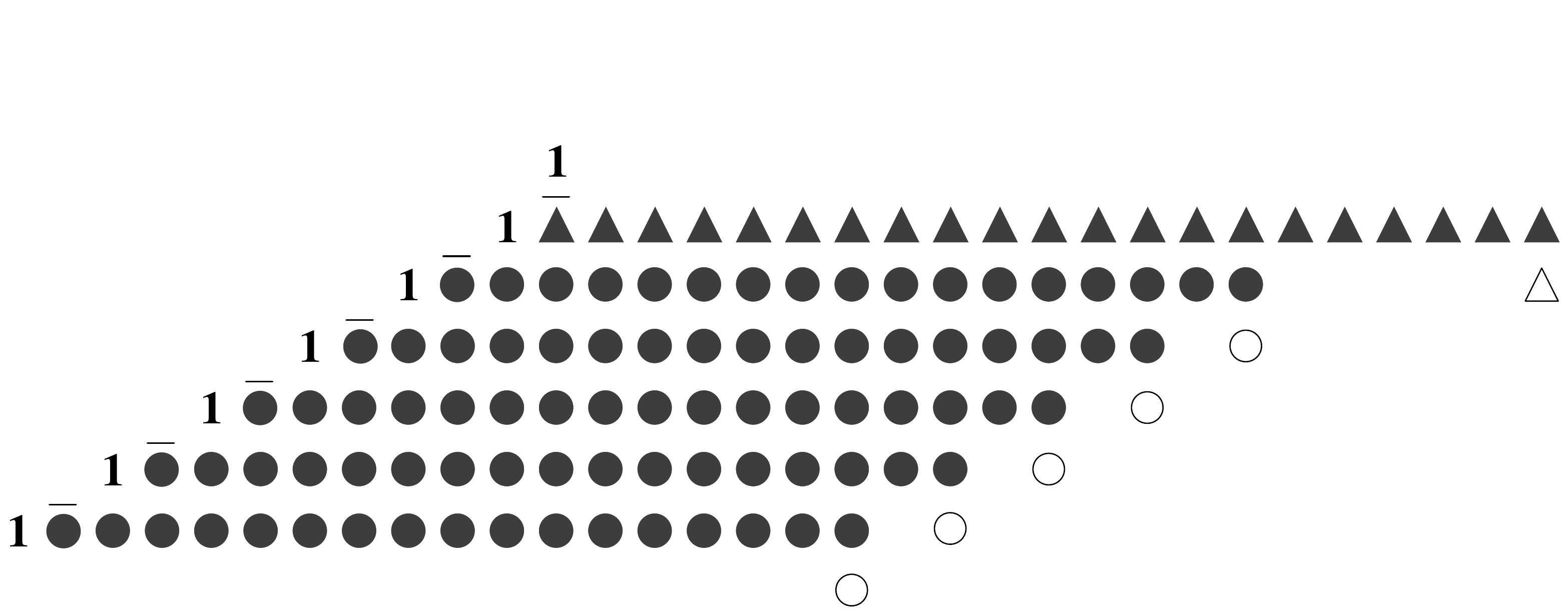}}\\[-10pt]
\subfloat[\label{fig_tec3}]{\includegraphics[width=0.56\textwidth]{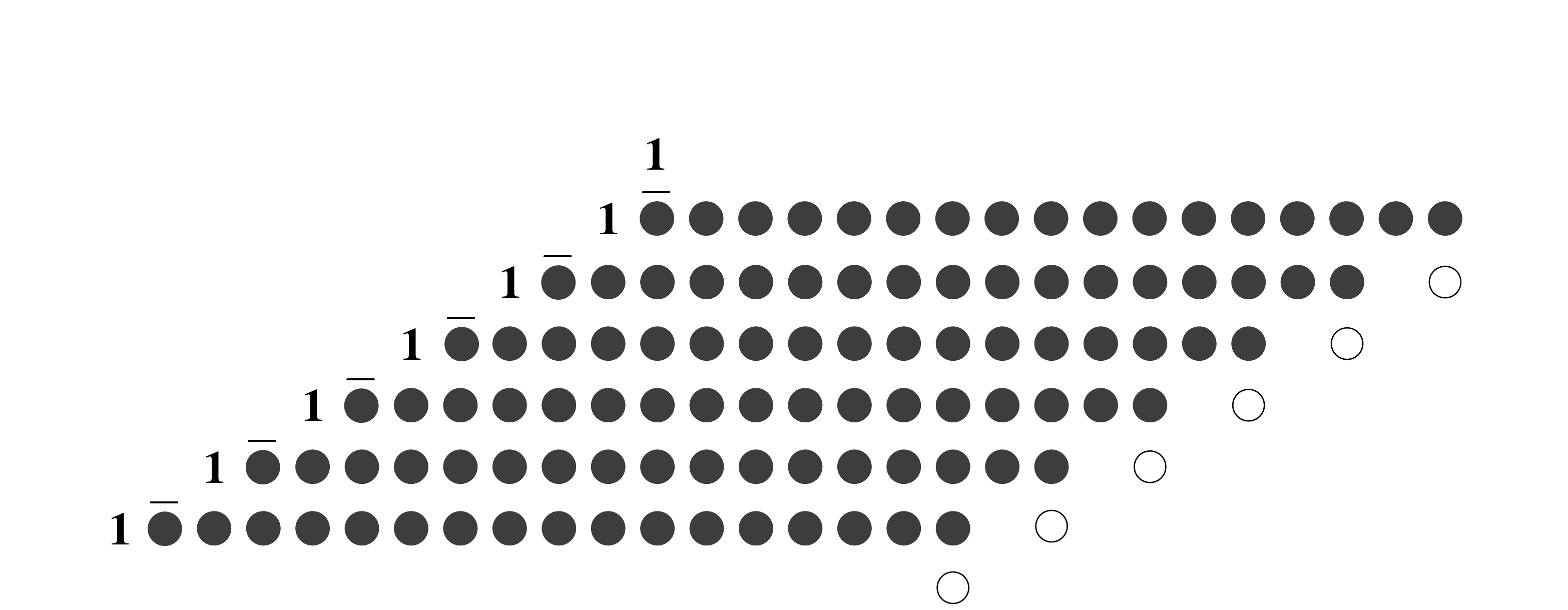}}\hspace*{-47pt} 
\subfloat[\label{fig_tec4}]{\includegraphics[width=0.56\textwidth]{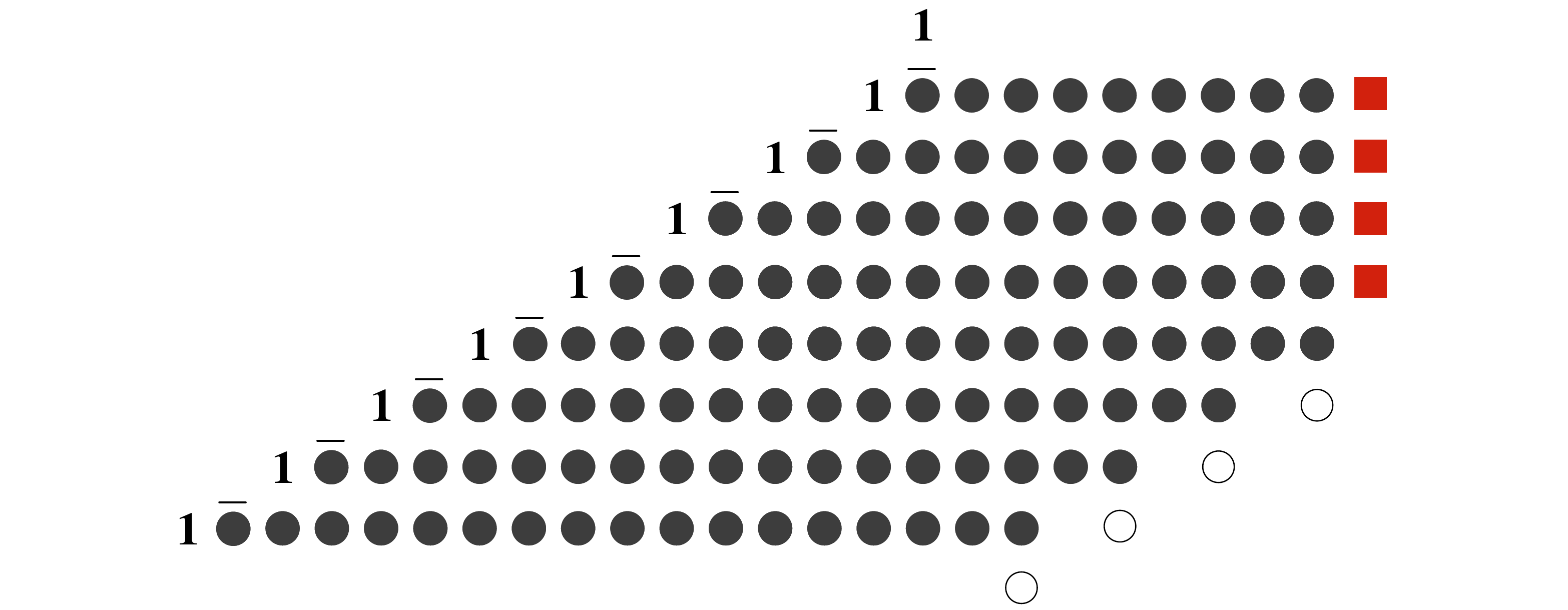}}%
\caption[Partial Product Matrices of Arithmetic Approximation Techniques]{Partial product matrices of $16$-bit multipliers based on:
\textbf{(a)} accurate radix-$4$ encoding (baseline), 
\textbf{(b)} high-radix encoding, 
\textbf{(c)} partial product perforation, 
\textbf{(d)} partial product rounding.}%
\label{fig_coop}
\end{figure}

We consider
the three techniques used in Chapter \ref{chapter4}
and Chapter \ref{chapter5},
i.e., 
high-radix encoding,
partial product perforation,
and partial product rounding.
As baseline, 
we consider the accurate radix-$4$ multiplier 
for the $n$-bit operands $A$ and $B$.
The partial product matrix of the $16$-bit
radix-$4$ multiplier is illustrated in Figure \ref{fig_tec1}.

\underline{High-Radix Encoding}:
This technique is configured with the parameter $k$,
which is an even number belonging in the interval
$[4$, $n-2]$.
The multiplicand $B$ is approximately encoded 
with respect to $k$:
the $n-k+1$ Most Significant Bits (MSBs) are encoded
with the accurate radix-$4$ encoding, 
while the $k$ Least Significant Bits (LSBs) are 
encoded with the approximate high-radix-$2^{k}$ encoding.
This hybrid encoding is then used 
to generate 
$(n-k)/2$ accurate
and $1$ approximate partial product.
Figure \ref{fig_tec2}
illustrates the partial product matrix
of the $16$-bit multiplier
that is based on high-radix encoding with $k=6$.
The least significant partial product (black triangles) is approximately generated from the radix-$64$ encoding,
substituting the three least significant partial products
of the accurate design 
(see Figure \ref{fig_tec1}).

\underline{Partial Product Perforation}:
This technique is configured with the parameter $P$,
which is an integer number belonging in the interval
$[0$, $n/2-1)$.
It is applied
by not generating the $P$ least significant partial products.
In practice,
this technique is equivalent to 
encoding $B$ to 
$B_P = \langle b_{n-1} b_{n-2} \cdots b_{2P-1}\rangle$. 
Figure \ref{fig_tec3}
illustrates the partial product matrix of
the $16$-bit multiplier
that is based on partial product perforation with $P=2$.
As shown,
the matrix has two less partial products
than the accurate design (see Figure \ref{fig_tec1}).

\underline{Partial Product Rounding}:
This technique is configured with the parameter $R$, 
which is an integer number belonging in the interval
$[0$, $n-1)$.
It is applied
by rounding the
partial products to their $R$-th bit.
In practice,
rounding is equivalent to reducing the bit-width of $A$,
e.g.,
the rounding technique of Chapter 
\ref{chapter5}
is equivalent to 
encoding $A$ to 
$A_R = \langle a_{n-1} a_{n-2} \cdots a_R\rangle + a_{R-1}$. 
We note
that in this chapter,
we consider additional methods for rounding,
which reduce the partial products to a tailored bit-width
based on their significance.
In contrast, 
the rounding of Chapter \ref{chapter5}
rounds all the partial products to the same bit-width. 
We call this variant ``asymmetric'' rounding
and use the label ``symmetric''
for the one presented in Chapter \ref{chapter5}. 
Figure \ref{fig_tec4}
illustrates the partial product matrix of
the $16$-bit multiplier
that is based on 
asymmetric rounding. 
As shown,
the four least significant partial products are rounded
to a different bit-width.
The red squares denote the correction terms 
that are added for error compensation. 

The next step is to 
explore all the possible combinations of our approximation techniques.
Table \ref{tb_copos} summarizes the combinations
when considering one encoding per operand.
As shown, there are two combinations,
i.e., rounding \& rounding and perforation \& perforation,
which are not feasible. 
The combination high-radix \& perforation is equivalent to single perforation,
as the high-radix approximate partial product is perforated,
and thus, it can be safely excluded for further examination.
Additionally,
we examine the application of perforation
to all the feasible combinations (bottom part of the table).
The only extra meaningful combination 
is perforation with the double high-radix encoding, 
as the rest ones are equivalent to rounding \& perforation.

\begin{table}[!t]
\fontsize{9}{10}\selectfont
\renewcommand{\arraystretch}{1.2}
\setlength{\tabcolsep}{10pt}
\caption[Combinations of Arithmetic Approximation Techniques]{Combinations of arithmetic approximation techniques.}
\label{tb_copos}
\centering
\begin{tabular}{cc c |c}
\hline
\textbf{1st Operand} & \textbf{2nd Operand} & \textbf{Feasibility} & \textbf{Combination Label} \\
\hline
\hline
high-radix    & high-radix  & \checkmark & DRAD \\
high-radix    & perforation & \checkmark & equivalent to perforation  \\
high-radix    & rounding    & \checkmark & RADR  \\
rounding      & rounding    & \ding{53}  & --  \\
rounding      & perforation & \checkmark & ROUP  \\
perforation   & perforation & \ding{53}  & -- \\
\hline
\end{tabular}\\[8pt]
\setlength{\tabcolsep}{9.1pt}
\begin{tabular}{cc |c}
\hline
\textbf{Extra Combinations} & \textbf{Feasibility} & \textbf{Combination Label} \\
\hline
\hline
high-radix \& high-radix + perforation & \checkmark & DRADP  \\
high-radix \& rounding + perforation   & \checkmark & equivalent to ROUP   \\
rounding \& perforation + perforation  & \checkmark & equivalent to ROUP  \\
\hline
\end{tabular}
\end{table}

\subsection{Combining Arithmetic Approximation Techniques}

Next,
we analyze the selected combinations of 
arithmetic approximations.
More details about the high-radix encoding
and the perforation/rounding 
are provided in Chapter \ref{chapter4} and Chapter \ref{chapter5},
respectively. 

\subsubsection{DRAD: High-Radix \& High-Radix}
Let $m$ and $k$ be the configuration parameters
for encoding $A$ and $B$, respectively. 
The two operands
are transformed 
using the high-radix-$2^m$ and high-radix-$2^k$ encodings
as shown in Eq. \eqref{eq_drada1}--\eqref{eq_dradb3}. 
We note that $B$ is encoded exactly as shown in Chapter \ref{chapter4},
i.e., with the hybrid radix-$4$--radix-$2^k$ encoding, 
while for $A$, we divide it into two words and encode only the least significant with radix-$2^m$.
To create the high-radix digit in $A$,
we add and subtract the term $2^{m-1}a_{m-1}$ in its representation. 

\vspace{-10pt}

\begin{equation}
\hspace{-32pt} \hspace{-50pt} A = -2^{n-1}a_{n-1} + \sum_{i=0}^{n-2} 2^{i}a_{i} = 
A_1 + x^{R2^m}_0  \label{eq_drada1} \\[-10pt]
\end{equation}
\begin{align}
\hspace{-32pt} \text{where } \; \; \; &  A_1 = -2^{n-1}a_{n-1}  + \sum_{i=m-1}^{n-2} 2^{i}a_{i} +  2^{m-1}a_{m-1} \label{eq_drada2} \\[2pt]
&  x^{R2^m}_0 = -2^{m-1}a_{m-1} + 2^{m-2}a_{m-2} + \dots + a_{0} 
\label{eq_drada3} 
\end{align}

\begin{equation}
\hspace{-32pt} \hspace{-53pt} B = -2^{n-1}b_{n-1} + \sum_{i=0}^{n-2} 2^{i}b_{i} = B_1 + y^{R2^k}_0 \label{eq_dradb1} \\[-10pt]
\end{equation}
\begin{align}
\hspace{-32pt} \hspace{24pt} \text{where } \; \; \; &  B_1 = \sum_{j=k/2}^{n/2-1}  4^j y_{j}^{R4}  \text{, } \;\;\;\;\; y_{j}^{R4} = -2 b_{2j+1} + b_{2j} + b_{2j-1} \label{eq_dradb2} \\[2pt]
&  y^{R2^k}_0 = -2^{k-1}b_{k-1} + 2^{k-2} b_{k-2} + \dots + b_{0} 
\label{eq_dradb3} 
\end{align}

Following the approximation approach of the high-radix encoding,
we approximate the high-radix digits of $A$ and $B$, 
i.e., we use 
$\hat{x}^{R2^{m}}_0 \in \{0, \;  \pm 2^{m-4}, \;  \pm 2^{m-3}, \;  \pm 2^{m-2}, \;  \pm 2^{m-1}\}$ 
and 
$\hat{y}^{R2^{k}}_0 \in \{0, \;  \pm 2^{k-4}, \;  \pm 2^{k-3}, \;  \pm 2^{k-2}, \;  \pm 2^{k-1}\}$,
respectively.
Based on these operand encodings,
the multiplication of DRAD$|_{k,m}$
is calculated by Eq. \eqref{eq_drad}.

\begin{equation}
    \text{DRAD}|_{k,m} = A_1 \cdot B_1 + B_1 \cdot \hat{x}^{R2^m}_0 + A \cdot \hat{y}^{R2^k}_0
    \label{eq_drad}\\[2pt]
\end{equation}

\subsubsection{DRADP: High-Radix \& High-Radix + Perforation}
 
When combining the double high-radix encoding with perforation,
we create the DRADP$|_{k,m}$ multiplier,
which perforates the least significant partial product,
as shown in Eq. \eqref{eq_dradp}.

\vspace{-11pt}

\begin{equation}
    \text{DRADP}|_{k,m} = A_1 \cdot B_1 + B_1 \cdot \hat{x}^{R2^m}_0
    \label{eq_dradp}
\end{equation}

\vspace{-4pt}

\subsubsection{RADR: High-Radix \& Rounding}
For this combination,
we consider the approximate high-radix encoding of $B$,
i.e., $B_1 + \hat{y}^{R2^k}_0$,
while $A$ is implicitly encoded via the rounding of the partial products.
To apply asymmetric rounding,
we truncate the $R$ least significant columns
of the partial product sub-matrix that is generated by $A \cdot B_1$.
As a result,
the multiplication of RADR$|_{k,R}$
is calculated by Eq. \eqref{eq_radr}.

\vspace{-6pt}

\begin{equation}
    \text{RADR}|_{k,R} = A \cdot B_1|_{R} + A \cdot \hat{y}^{R2^k}_0\\[2pt]
    \label{eq_radr}
\end{equation}

To compensate for the truncation error,
we insert a correction term
in the place of the truncated MSB of each partial product.
This term is different for each partial product  
and it is affected by the radix-$4$ encoding, 
i.e., it is equal to $one_j + two_j$. 
We remind that the signals $one_j$ and $two_j$ 
are activated if the $j$-th partial product
is equal to $\pm 1A$ or $\pm 2A$, 
respectively 
(see the analysis of Chapter \ref{chapter4}). 
To further improve our rounding,  
we attach a constant `$1$' in the position that the sub-matrix of $A \cdot B_1$
is truncated. 

\subsubsection{ROUP: Rounding \& Perforation}
The symmetric rounding \& perforation combination has been 
already examined in Chapter \ref{chapter5},
where AxFXU$|_{P,R}$ is presented.
Therefore,
in this chapter, 
we combine asymmetric rounding and perforation,
and more specifically,
we employ two different variants of asymmetric rounding.

The first combination is labeled ROUP1$|_{P,R}$
and uses the asymmetric rounding of the RADR design.
In particular,
we perforate the $P$ least significant products,
and then we apply rounding
by truncating the $R$ least significant columns of the partial product matrix
and inserting the correction terms. 
The multiplication of ROUP1 is calculated by Eq. \eqref{eq_roup1}. 

\vspace*{-7pt}

\begin{equation}
\text{ROUP1}|_{P,R} = 
\mathlarger{\sum}_{\substack{j=P}}^{\substack{n\text{/}2-1}} 4^{j} \tilde{P\!P}_j \bigg|_R =
\mathlarger{\sum}_{\substack{j=P}}^{\substack{n\text{/}2-1}} 4^{j} A \cdot  y_{j}^{R4} \bigg|_R
\label{eq_roup1}
\end{equation}

The second combination is labeled ROUP2$|_{P,R}$
and converts the rounding of AxFXU$|_{P,R}$ to asymmetric. 
The multiplication of ROUP2
is performed by accumulating the non-perforated, rounded partial products
as shown in Eq. \eqref{eq_roup2a}--\eqref{eq_roup2b}.

\begin{equation}
\text{ROUP2}|_{P,R} = 
\mathlarger{\sum}_{\substack{j=P}}^{\substack{n\text{/}2-1}} 4^{j} \tilde{P\!P}_j  =
 \mathlarger{\sum}_{\substack{j=P}}^{\substack{n\text{/}2-1}} 4^{j} A_{R_j} \cdot  y_{j}^{R4} 
\label{eq_roup2a}
\end{equation}
\begin{equation}
\hspace{-5pt} \text{where }  \, \;
A_{R_j} = \langle a_{n-1} a_{n-2} \cdots a_{R_{j}}\rangle_{2\text{'s}} + a_{R_{j}-1}
\label{eq_roup2b}    
\end{equation}

Compared to the respective expressions
of AxFXU,
which are given in 
Eq. \eqref{eq_pr1}--\eqref{eq_pr2},
the only difference is that there are $j$ $A_R$ words
(one per partial product)
instead of one.
The rest rounding optimizations are the same,
resulting in integrating
the $a_{R_{j}-1}$ bits with a XOR gate 
in the correction terms,
i.e., 
$(sign_j \oplus a_{R_{j}-1}) \cdot (one_j + two_j)$. 
We note that the notation ROUP2$|_{P,R}$
considers $R=R_P$,
i.e., it shows the rounding configuration of the first non-perforated partial product. 

\subsection{Overview of Cooperative Approximation Techniques}

\begin{figure}[!t]
\vspace*{-10pt}
\centering
\subfloat[\label{fig_tr_drad}]{\includegraphics[width=0.57\textwidth]{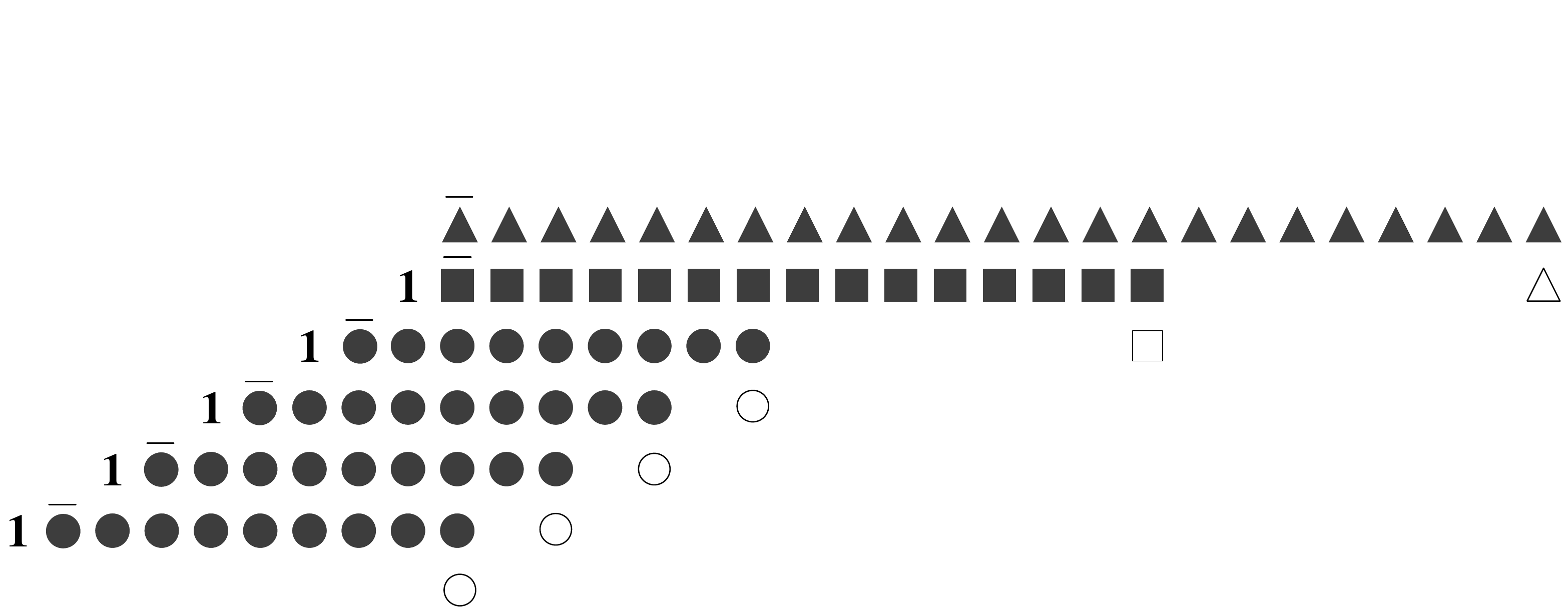}} \hspace*{-43pt} %
\subfloat[\label{fig_tr_dradp}]{\includegraphics[width=0.57\textwidth]{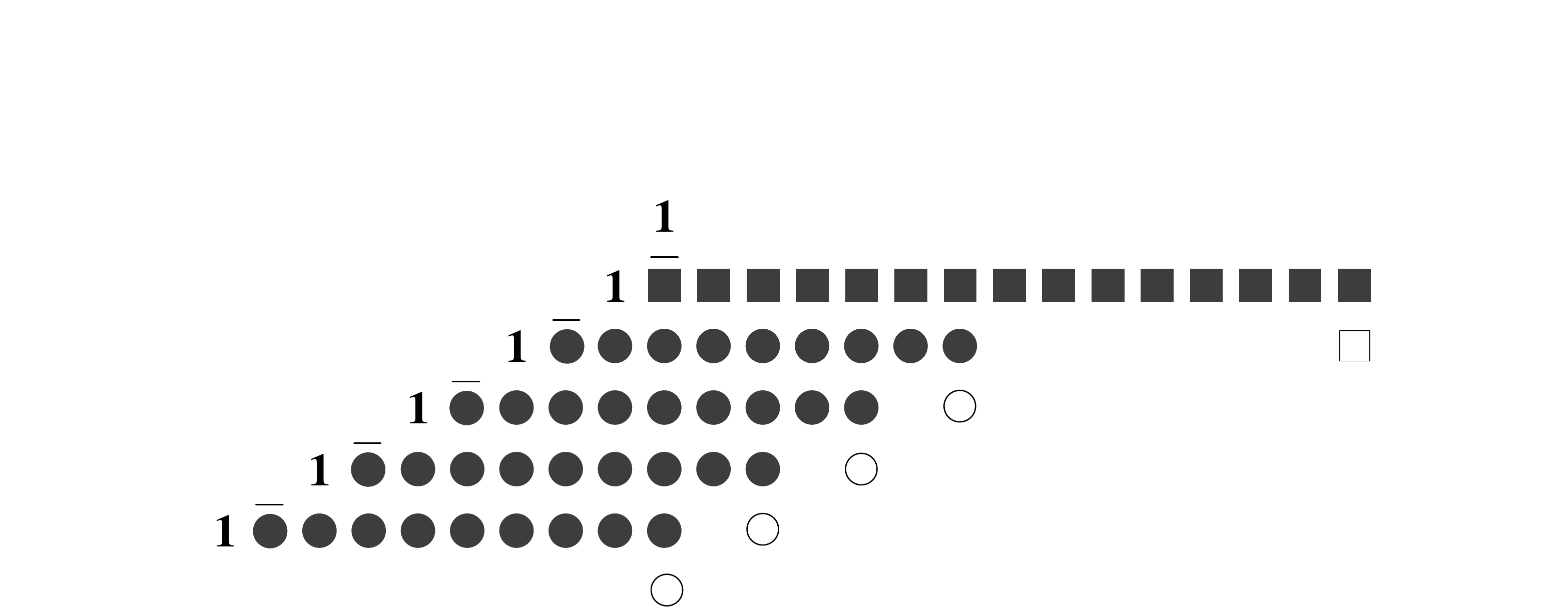}} \\[-10pt]
\subfloat[\label{fig_tr_radr}]{\includegraphics[width=0.57\textwidth]{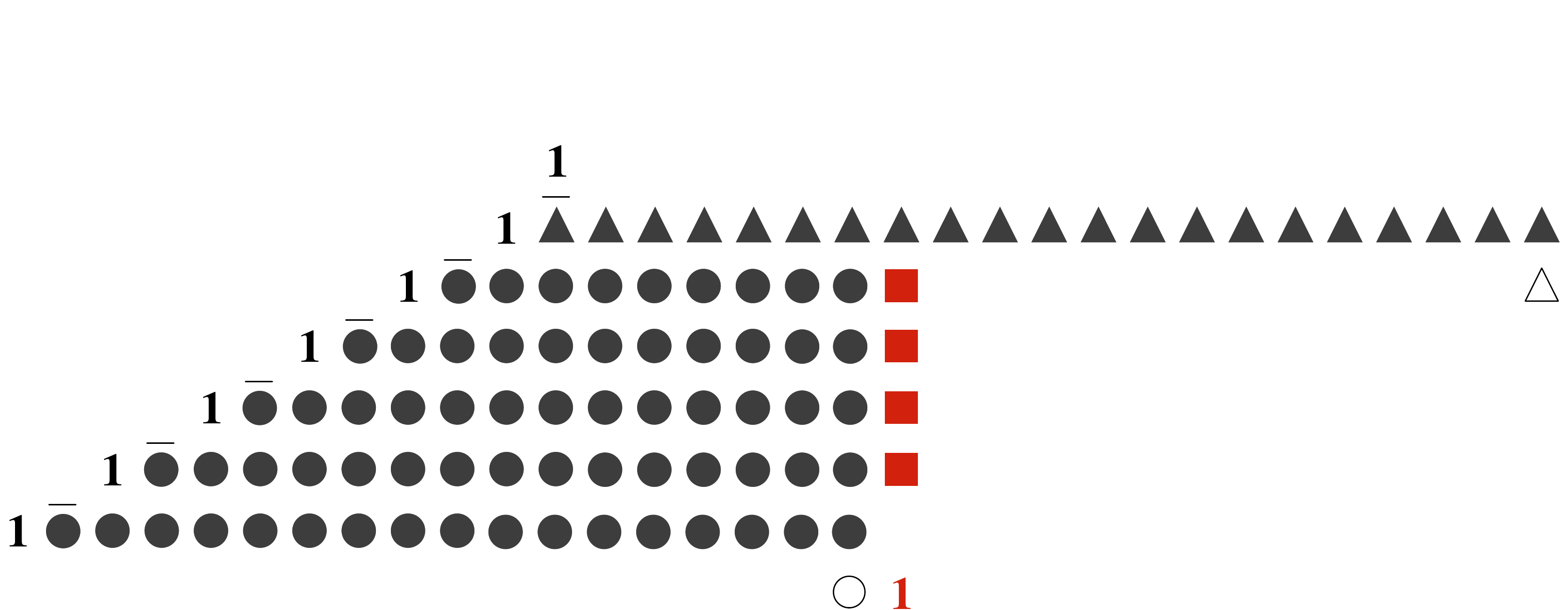}} \\[-25pt]
\subfloat[\label{fig_tr_roup1}]{\includegraphics[width=0.57\textwidth]{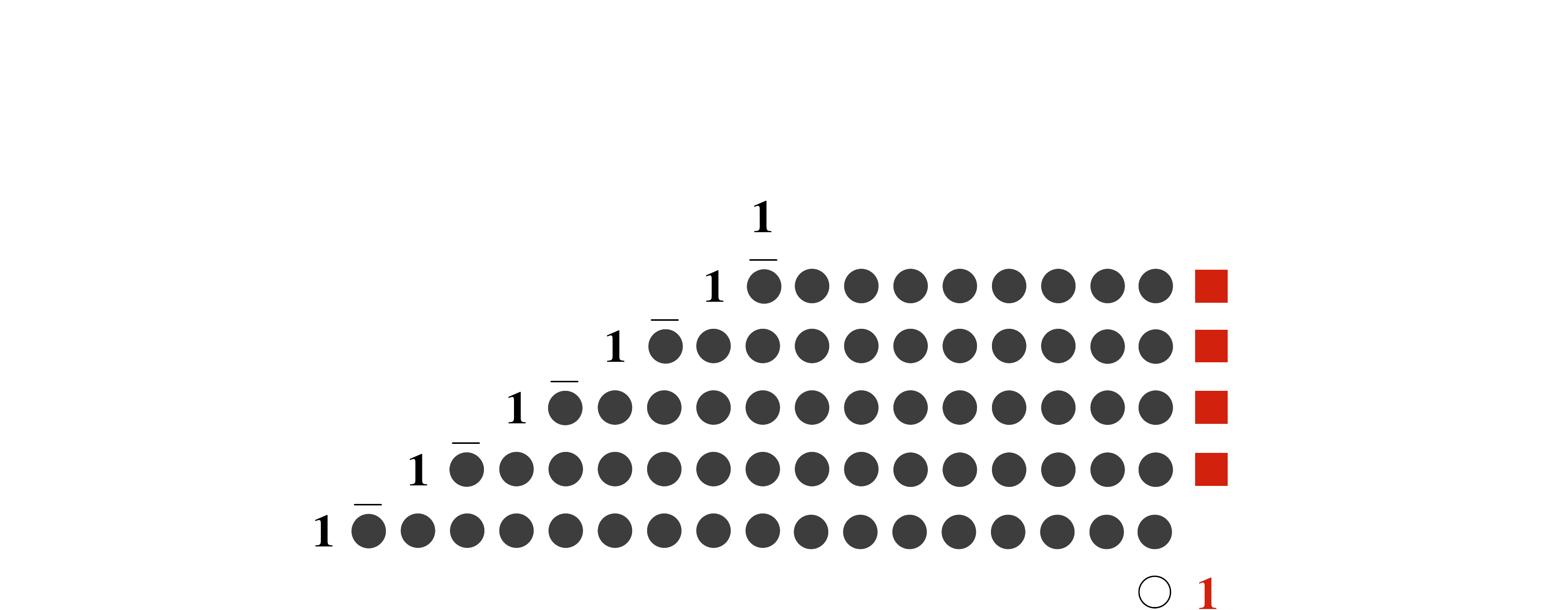}}\hspace*{-43pt} 
\subfloat[\label{fig_tr_roup2}]{\includegraphics[width=0.547\textwidth]{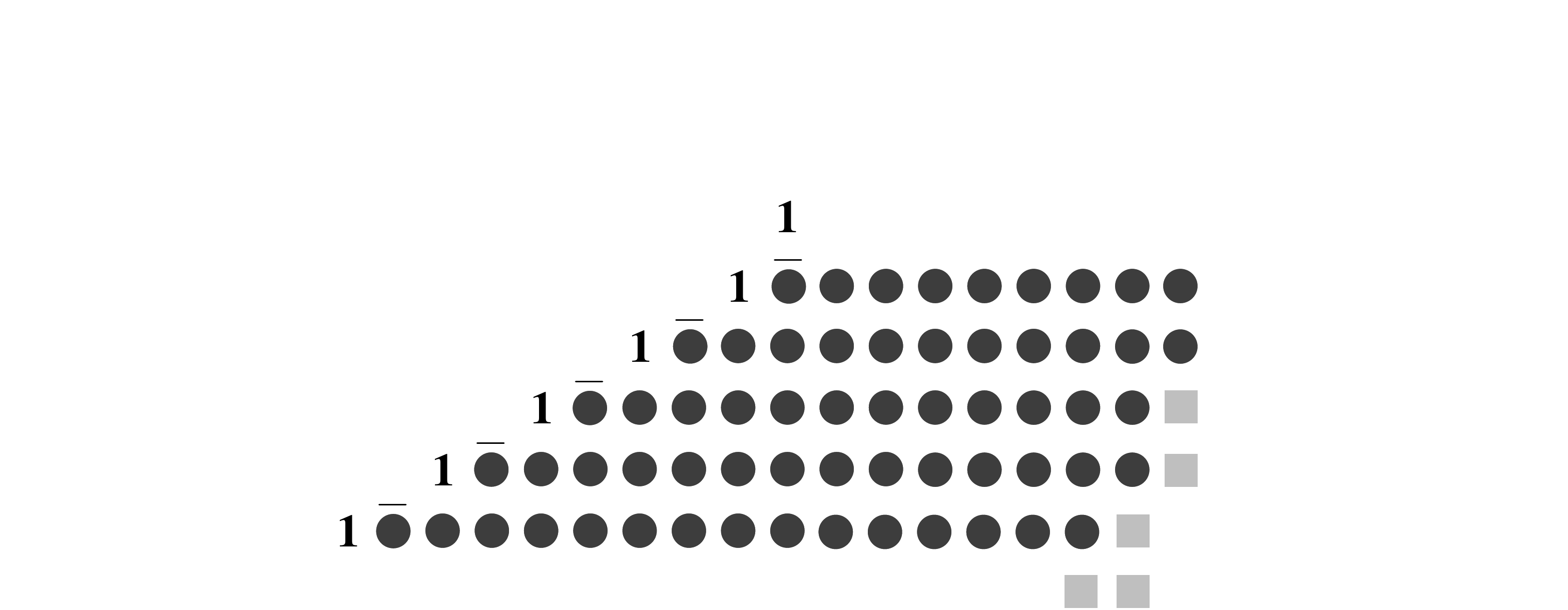}}%
\caption[Partial Product Matrices of Cooperative Approximation Techniques]{Partial product matrices of $16$-bit multipliers based on cooperative approximation:
\textbf{(a)} high-radix \& high-radix (DRAD$|_{8,8}$), 
\textbf{(b)} high-radix \& high-radix + perforation (DRADP$|_{8,8}$), 
\textbf{(c)} high-radix \& asymmetric rounding v.1 (RADR$|_{6,8}$), 
\textbf{(d)} perforation \&  asymmetric rounding v.1   (ROUP1$|_{3,8}$), 
and 
\textbf{(e)} perforation \& asymmetric rounding v.2 (ROUP2$|_{3,10}$).}%
\label{fig_coop2}
\end{figure}

Figure \ref{fig_coop2} illustrates the $16$-bit partial product matrix of each combination with specific configuration.  
As shown, 
depending on the combination,
we achieve a different structure
compared to the accurate matrix of Figure \ref{fig_tec1}. 
Even though 
it is possible to achieve comparable 
horizontal and vertical matrix reduction 
with all combinations, 
we note that
each combination:
(i) requires different logic for implementing the operand encoding and/or partial product generation,
(i) imposes different overheads in the partial product accumulation,
which depends on the bit arrangement within the matrix 
(e.g., for carry propagation penalties),
and 
(iii) has different impact on the accuracy of the results.
For example,
ROUP1$|_{3,8}$ (see Figure \ref{fig_tr_roup1})
and 
ROUP2$|_{3,10}$ (see Figure \ref{fig_tr_roup2})
may deliver similar matrices,
however, 
as we show in the evaluation in the next section,
their results and total efficiency are different.

In the matrix of DRAD,
the triangles denote the product
$A \cdot \hat{y}^{R2^k}_0$ 
and the rectangles denote the product 
$B_1 \cdot \hat{x}^{R2^m}_0$.
The circles are the products from $A_1 \cdot B_1$.
The difference of DRADP is that it perforates 
the product 
$A \cdot \hat{y}^{R2^k}_0$. 
Regarding RADR,
the triangles denote the product 
$A \cdot \hat{y}^{R2^k}_0$,
while the circles are the products from 
$A \cdot B_1|_{R}$.
The red rectangles are the correction terms
that are inserted for error compensation.
The circles in the ROUP1 matrix
are the non-perforated, rounded partial products.
Like in RADR,
the red rectangles are the rounding correction terms. 
Finally,
the matrix of ROUP2 is similar to that of ROUP1,
however, 
in this combination,
the correction terms (gray rectangles) are different
and the matrix is not truncated vertically. 

\section{Evaluation}
\label{s6_4}

In this section,
we evaluate the application
of cooperative approximation
in multiplication circuits.
Like in Chapter \ref{chapter4} and Chapter \ref{chapter5},
we begin with the error analysis of each combination,
and then we 
report comparative experimental results 
involving state-of-the-art works,
as well as all the approximate designs proposed in the Dissertation.

\begin{figure}[!t]
\vspace*{-13pt}
\centering
\subfloat[\label{fig_er_drad}]{\includegraphics[width=0.47\textwidth]{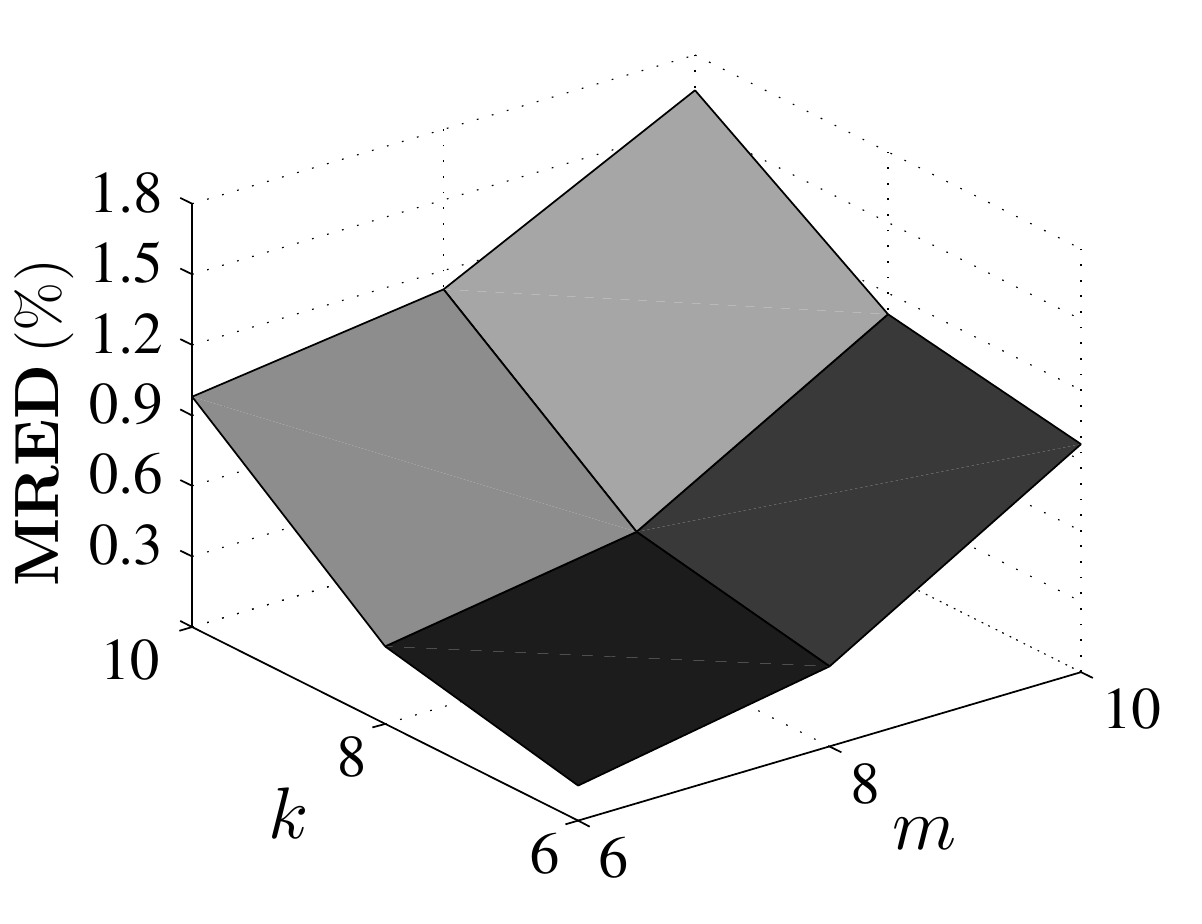}} \hspace{17pt} %
\subfloat[\label{fig_er_dradp}]{\includegraphics[width=0.47\textwidth]{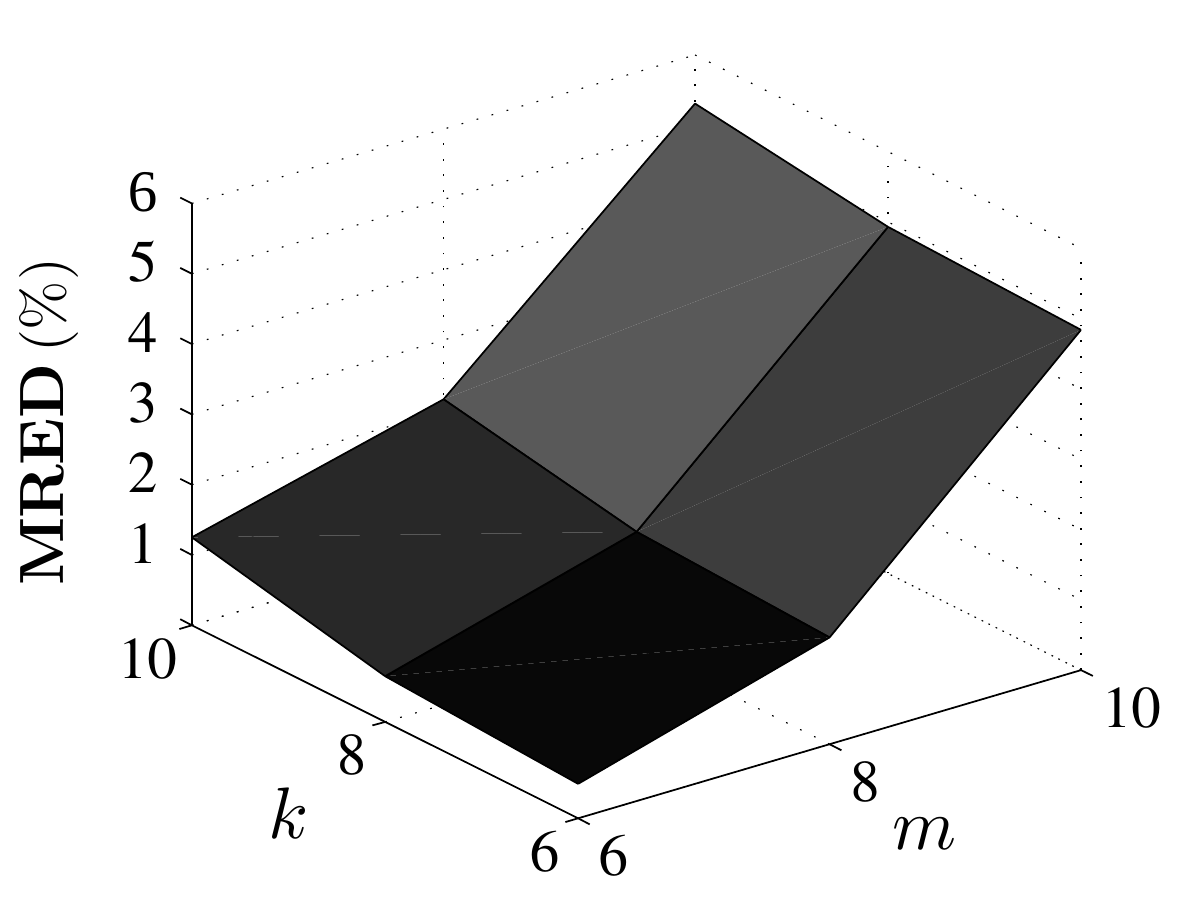}}\\[-20pt]
\subfloat[\label{fig_er_radr}]{\includegraphics[width=0.47\textwidth]{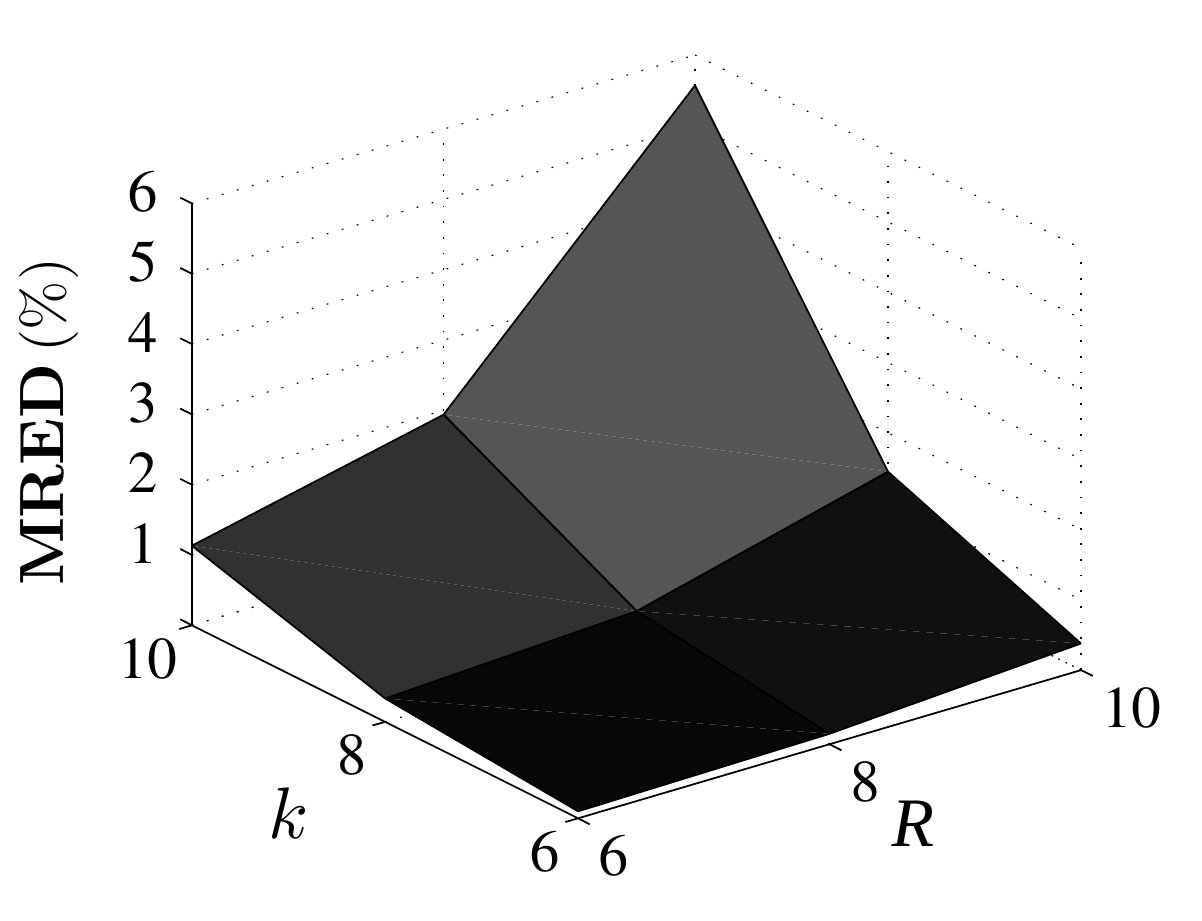}}\\[-27pt]
\subfloat[\label{fig_er_roup1}]{\includegraphics[width=0.47\textwidth]{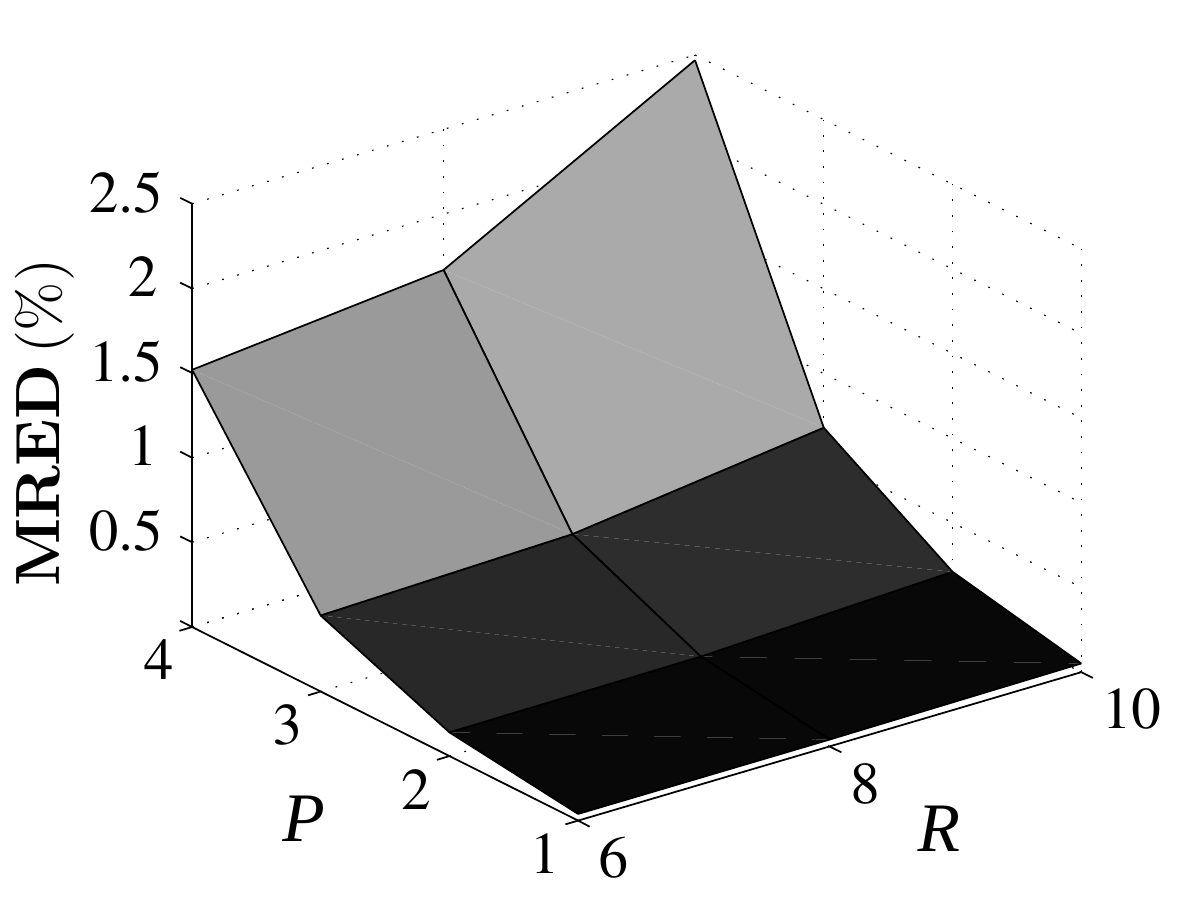}}\hspace{17pt} 
\subfloat[\label{fig_er_roup2}]{\includegraphics[width=0.47\textwidth]{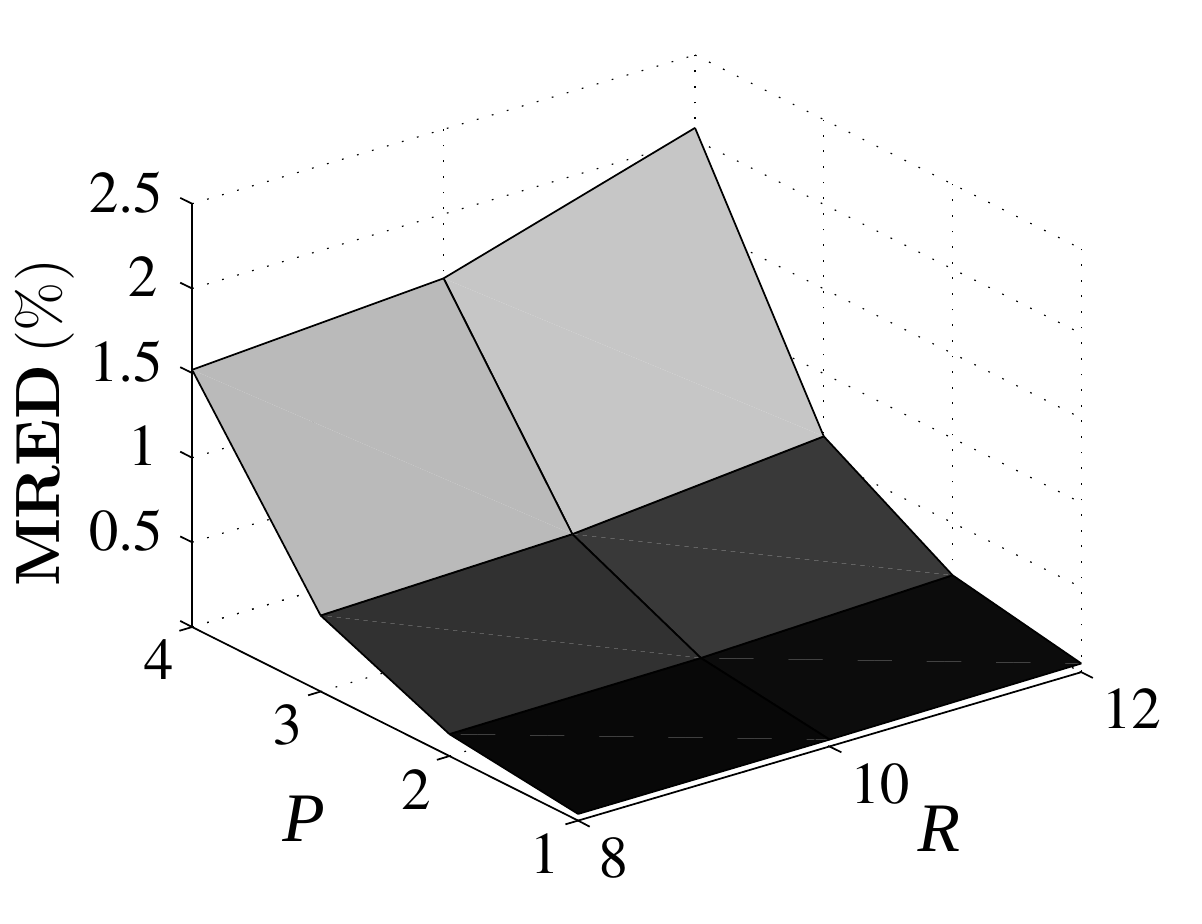}}%
\caption[Error Variation of Cooperative Approximation Techniques]{MRED variation of $16$-bit multipliers based on cooperative approximation:
\textbf{(a)} high-radix \& high-radix (DRAD$|_{k,m}$), 
\textbf{(b)} high-radix \& high-radix + perforation (DRADP$|_{k,m}$), 
\textbf{(c)} high-radix \& asymmetric rounding (RADR$|_{k,R}$),
\textbf{(d)} asymmetric rounding \& perforation v.1 (ROUP1$|_{P,R}$),
and 
\textbf{(e)} asymmetric rounding \& perforation v.2 (ROUP2$|_{P,R}$).}%
\label{fig_coer}
\end{figure}

\subsection{Error Analysis}

For the error analysis,
we rely again on the MRED metric
(see the respective analysis of Chapter \ref{chapter4} and Chapter \ref{chapter5} for more details).
In brief, 
MRED 
is the average of the relative errors 
for a given set of operand pairs.
All the proposed multipliers feature a very large design space, 
as 
they are configured by two independent parameters.
Therefore, for each approximation configuration
we calculate MRED for 
$200$K pairs of operands
that are uniformly distributed over the $16$-bit range.

Figure \ref{fig_coer} 
presents the variation of MRED
with respect to the 
approximation configuration. 
Regarding DRAD,
even though the error range is small,
i.e., $[0.15\%,$ $1.65\%]$, 
it grows rapidly creating blank error segments.
However, 
the error scaling provides increased density
compared to the RAD design of Chapter \ref{chapter4},
which has only three configurations ($k=6,8,10$)
with MRED values $0.08\%$, $0.28\%$ and $0.93\%$. 
As expected,
for the same configurations, 
DRADP exhibits 
larger error values than DRAD, 
starting from $0.49\%$, 
while the rapid error scaling is again observed.
The advantage of RADR is 
that it smooths the rapid error scaling of RAD 
by adding multiple error values between two consecutive RAD configurations, 
especially for the small $k$ values. 
The error range and scaling of ROUP1 and ROUP2
are similar to those of the AxFXU design of Chapter \ref{chapter5}.  
ROUP1
features dense error scaling from $0.04\%$ ($P=1$) to $2.47\%$ ($P=4$) with several intermediate values.
The MRED values of ROUP2 are similar, however, 
its maximum error is smaller ($2.07\%$) compared to that of ROUP1 ($2.47\%$).

\subsection{State-of-the-Art Comparison: Pareto Efficiency Analysis} 

This section includes the experimental evaluation of the
cooperative approximation techniques. 
To provide an overall resource--accuracy Pareto analysis
and extract the most efficient designs proposed in the Dissertation,
we also employ the approximate designs 
of Chapter \ref{chapter4} and Chapter \ref{chapter5}.
We remind that these designs
have already formed the Pareto fronts in comparative evaluations 
including several state-of-the-art designs \cite{2016_Jiang_IEEEtc, 2017_Liu_IEEEtc, 2015_Hashemi_ICCAD, ZervakisTVLSI2016, Schulte1993, 2017_Zendegani_IEEEtvlsi}. 
Table \ref{tb_coopsum} summarizes all our approximate multipliers. 

\begin{table}[!b]
\fontsize{9}{10}\selectfont
\renewcommand{\arraystretch}{1.2}
\setlength{\tabcolsep}{8pt}
\caption[Overview of Dissertation's Approximate Arithmetic Circuits]{Overview of Dissertation's approximate arithmetic circuits.}
\label{tb_coopsum}
\centering
\begin{tabular}{l|c|c} 
\hline
\textbf{Design} & \textbf{Approximation Techniques} &  \textbf{Reference} \\
\hline
\hline
RAD$2^k$ & high-radix-$2^k$ & Chapter \ref{chapter4} \cite{LeonTVLSI} \\
AxFXU|$_{P,R}$ & perforation $P$, symmetric rounding $R$ & Chapter \ref{chapter5} \cite{LeonMicro} \\
DRAD|$_{m,k}$ & high-radix-$2^m$, high-radix-$2^k$ & Chapter \ref{chapter6} \cite{LeonDAC} \\
DRADP|$_{m,k}$ & high-radix-$2^m$, high-radix-$2^k$, perforation $P=1$  & Chapter \ref{chapter6} \cite{LeonDAC} \\
RADR|$_{k,R}$ & high-radix-$2^k$, asymmetric rounding $R$ v.1 & Chapter \ref{chapter6} \cite{LeonDAC} \\
ROUP1|$_{P,R}$ & perforation $P$, asymmetric rounding $R$ v.1 & Chapter \ref{chapter6} \cite{LeonDAC} \\
ROUP2|$_{P,R}$ & perforation $P$, asymmetric rounding $R$ v.2 & Chapter \ref{chapter6} \cite{LeonDAC} \\
\hline
\end{tabular}
\end{table}

All the designs are implemented in Verilog
for $n=16$ multiplication bit-width. 
The synthesis is performed with the Synopsys Design Compiler tool
and the TSMC 65-nm standard-cell library.
The simulations for the 
functional verification and the power measurements 
are performed with Mentor Graphics QuestaSim.
The nominal supply voltage ($1$V) is used in both synthesis and simulation.
The critical path delay and the area of the circuits
are reported by Synopsys Design Compiler, 
while the power consumption is measured with Synopsys PrimeTime 
after performing gate-level simulation.
We also evaluate the energy consumption,
which is defined as the product of power and delay. 
We note that,
like in Chapter \ref{chapter5}, 
we synthesize and simulate the circuits under two different 
design scenarios: 
\begin{itemize}
    \item[(i)] \underline{MIN-Delay}: the clock constraint of each circuit is set to its critical path delay (high-performance mode). 
    \item[(ii)] \underline{ISO-Delay}: the clock constraint of all the circuits is set to the same relaxed value (low-power mode). 
\end{itemize}

\begin{figure}[!t]
\centering
\vspace*{-25pt}
\hspace*{-140pt}\subfloat[\label{fig_ar1}]{\includegraphics[width=0.63\textwidth]{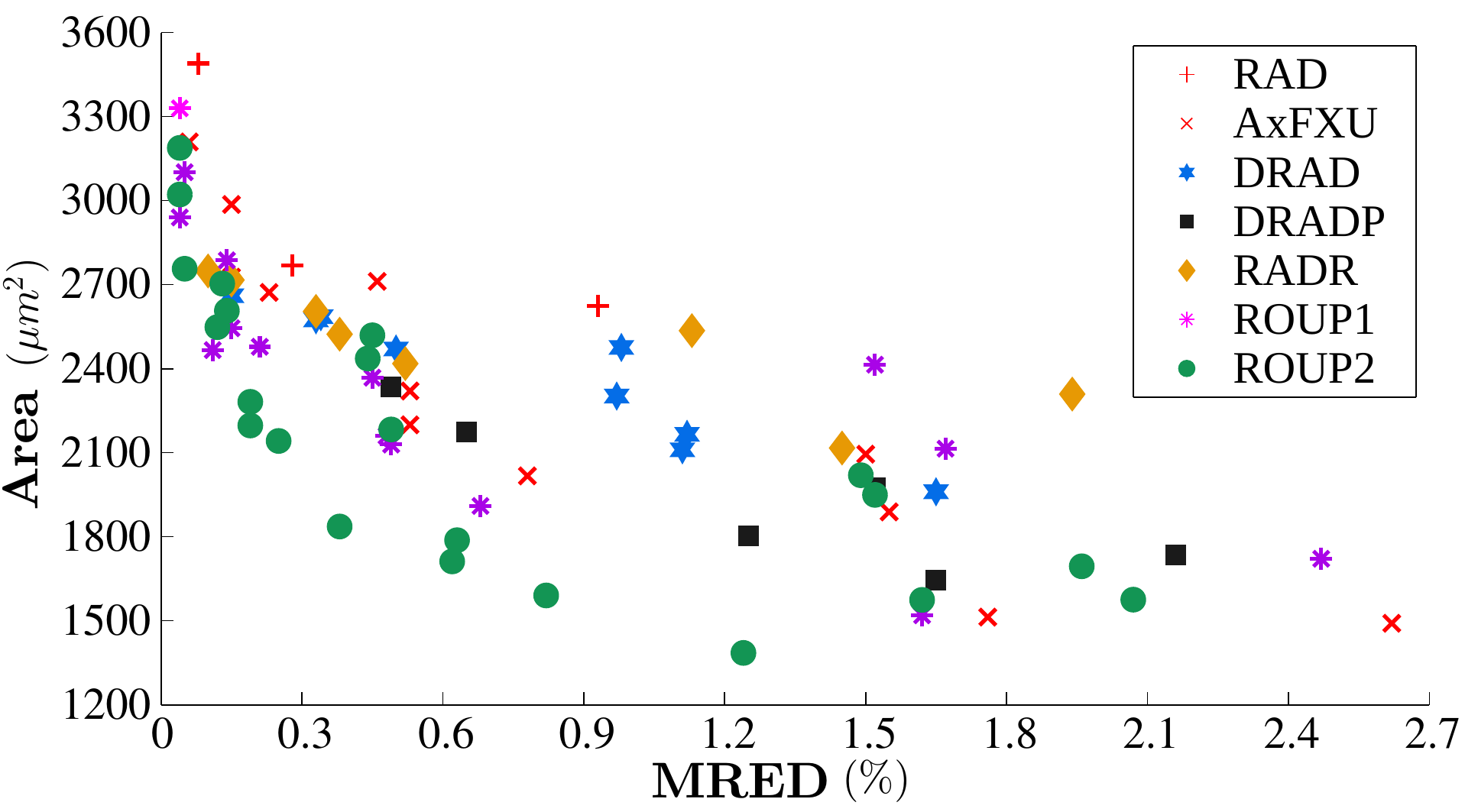}} \\[-23pt]
\hfill\subfloat[\label{fig_en1}]{\includegraphics[width=0.63\textwidth]{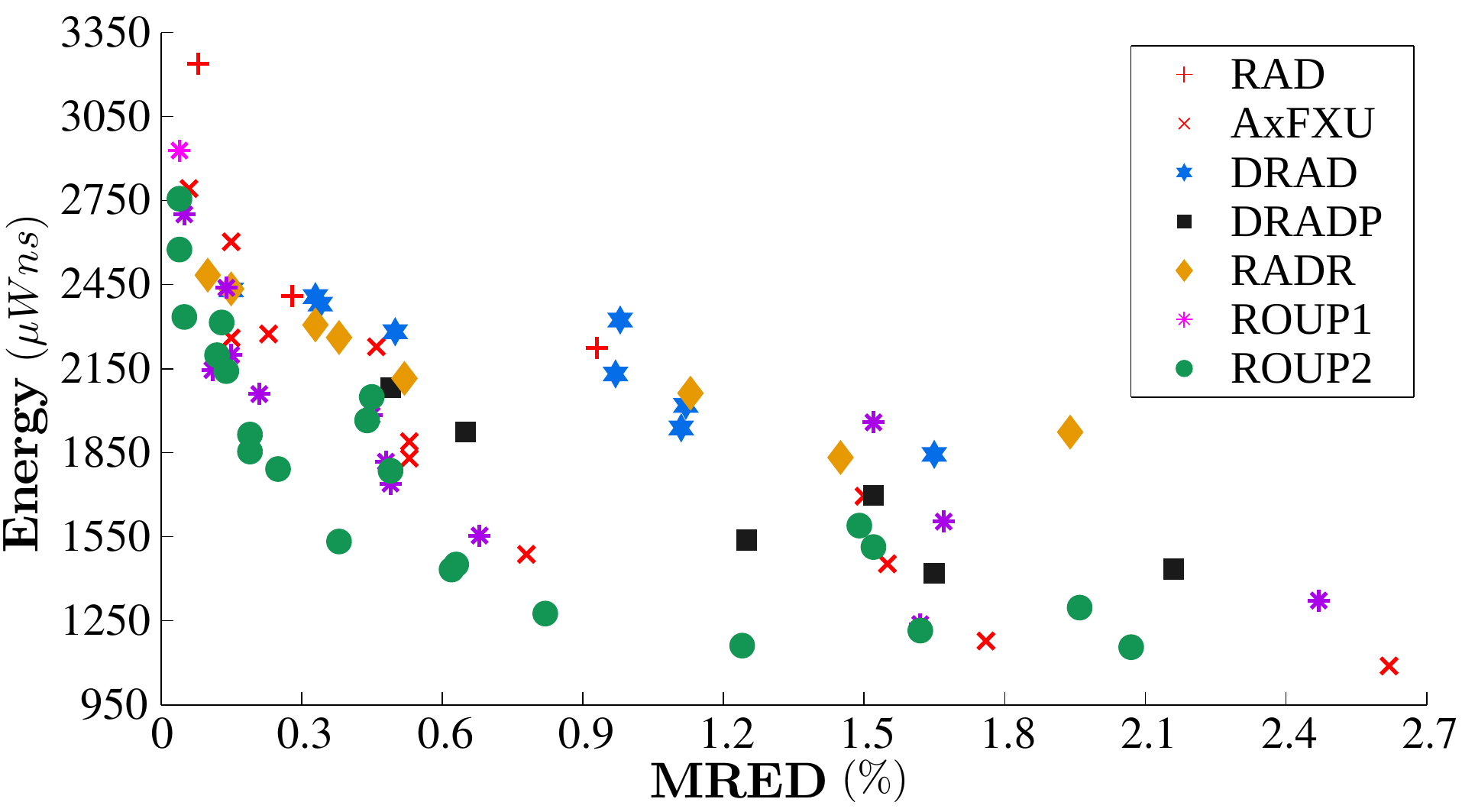}}\\[-29pt]
\hspace*{-140pt}\subfloat[\label{fig_ar2}]{\includegraphics[width=0.63\textwidth]{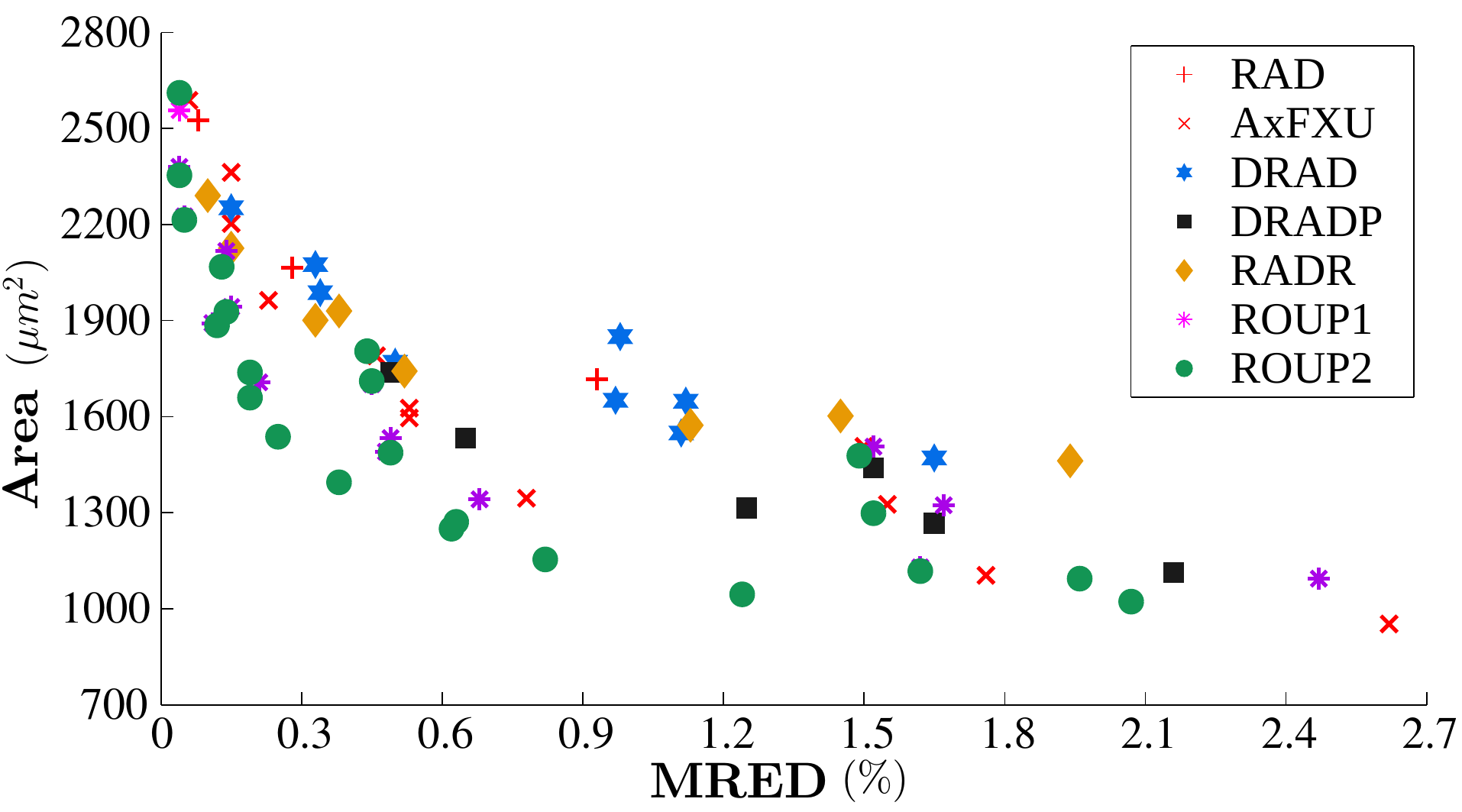}}\\[-23pt]
\hfill\subfloat[\label{fig_en2}]{\includegraphics[width=0.63\textwidth]{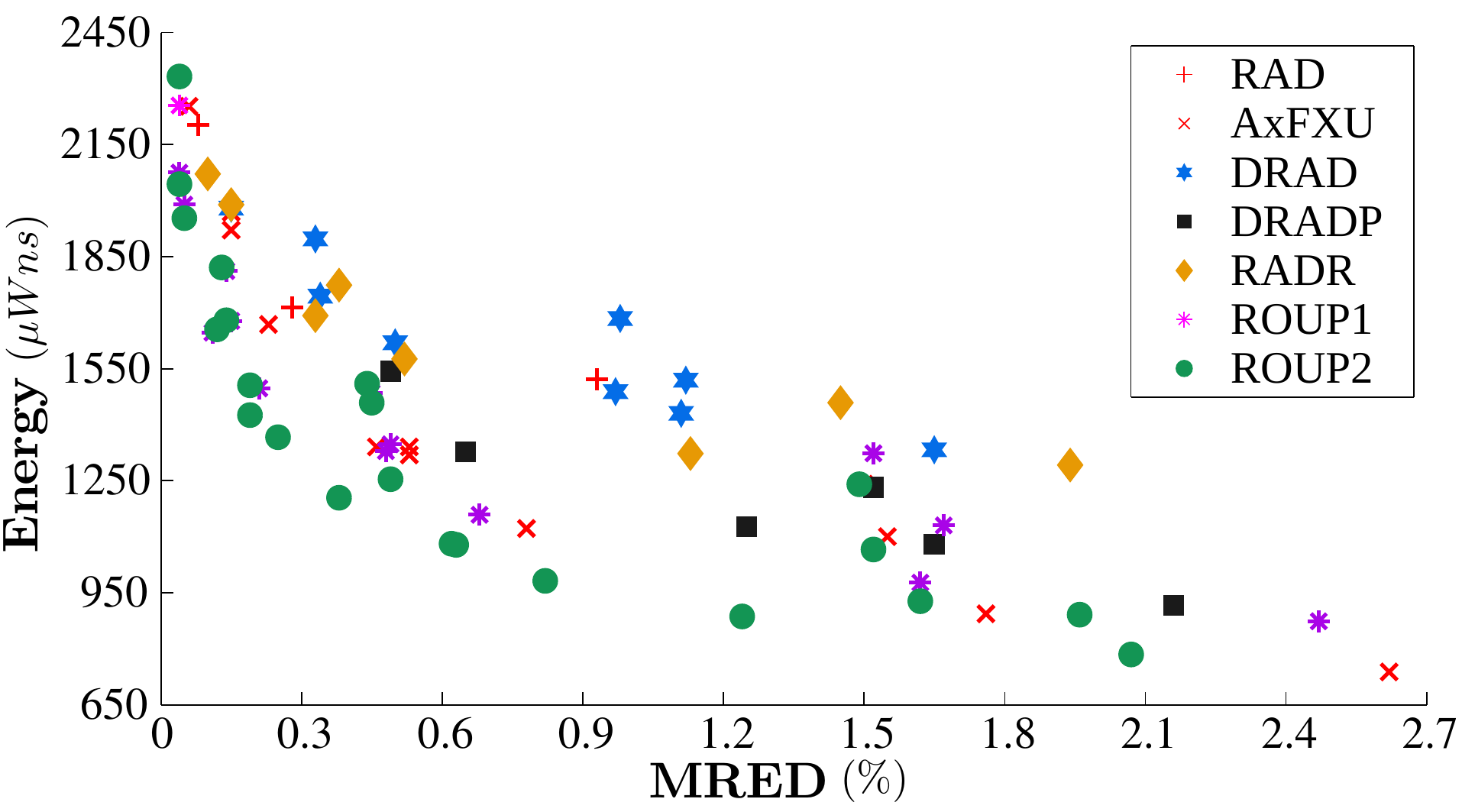}}%
\vspace{-5pt}
\caption[Pareto Analysis for Dissertation's Approximate Arithmetic Circuits]{Pareto analysis for Dissertation's approximate multipliers considering MRED and area/energy under two design scenarios:
\textbf{(a)}, \textbf{(b)} MIN-Delay
and 
\textbf{(c)}, \textbf{(d)} ISO-Delay.}%
\vspace{-28pt}
\label{fig_finpare}
\end{figure}

In Figure \ref{fig_finpare}, 
we present the MRED--area and MRED--energy scatter plots
with all Dissertation's proposed designs. 
Regarding the cooperative high-radix multipliers,
i.e., 
DRAD, DRADP and RADR, 
they increase the resolution of the RAD front, 
and even improve it with DRADP,
which constitutes a better design in most cases.
We note that we present only the RAD multipliers 
for $k=6,8,10$, 
because large error values are produced for $k \geq 12$.
This limitation
of the single high-radix technique 
is surpassed with the cooperative approximation techniques.
Regarding the multipliers implementing cooperative perforation \& rounding,
the asymmetric rounding techniques of ROUP1 and ROUP2
outperform the symmetric rounding of AxFXU,
and thus,
they constitute better design alternatives. 
We conclude that more fine-grained bit-level optimizations,
such as the tailored rounding per partial product 
of ROUP2,
provide better results than generic coarse-grained solutions,
such as the global partial product rounding of AxFXU.
Additionally,
for small error values,
AxFXU is not considered the most energy-efficient design, 
as several configurations of the cooperative approximation techniques provide improved energy consumption.
Nevertheless,
as shown in Chapter \ref{chapter5}, 
the advantage of AxFXU
is that it facilitates dynamic approximation configuration,
which would be more difficult in ROUP1/ROUP2
due to the different rounding per partial product. 

Overall, as shown in both design scenarios, 
the Pareto front is formed exclusively 
by the ROUP2 multipliers.
The ROUP2 design family 
provides the best exploitation of the energy--error trade-off
and further improves
the state-of-the-art Pareto front
(previously held by RAD \cite{LeonTVLSI} and AxFXU \cite{LeonMicro}) 
by up to $1.5\times$--$2\times$.
Moreover,
it is important to mention that
it increases
the resolution of the front,
namely, it 
expands the already-large approximation space even more, 
providing a great variety of design options. 

\begin{figure}[!t]
\centering
\includegraphics[width=0.83\textwidth]{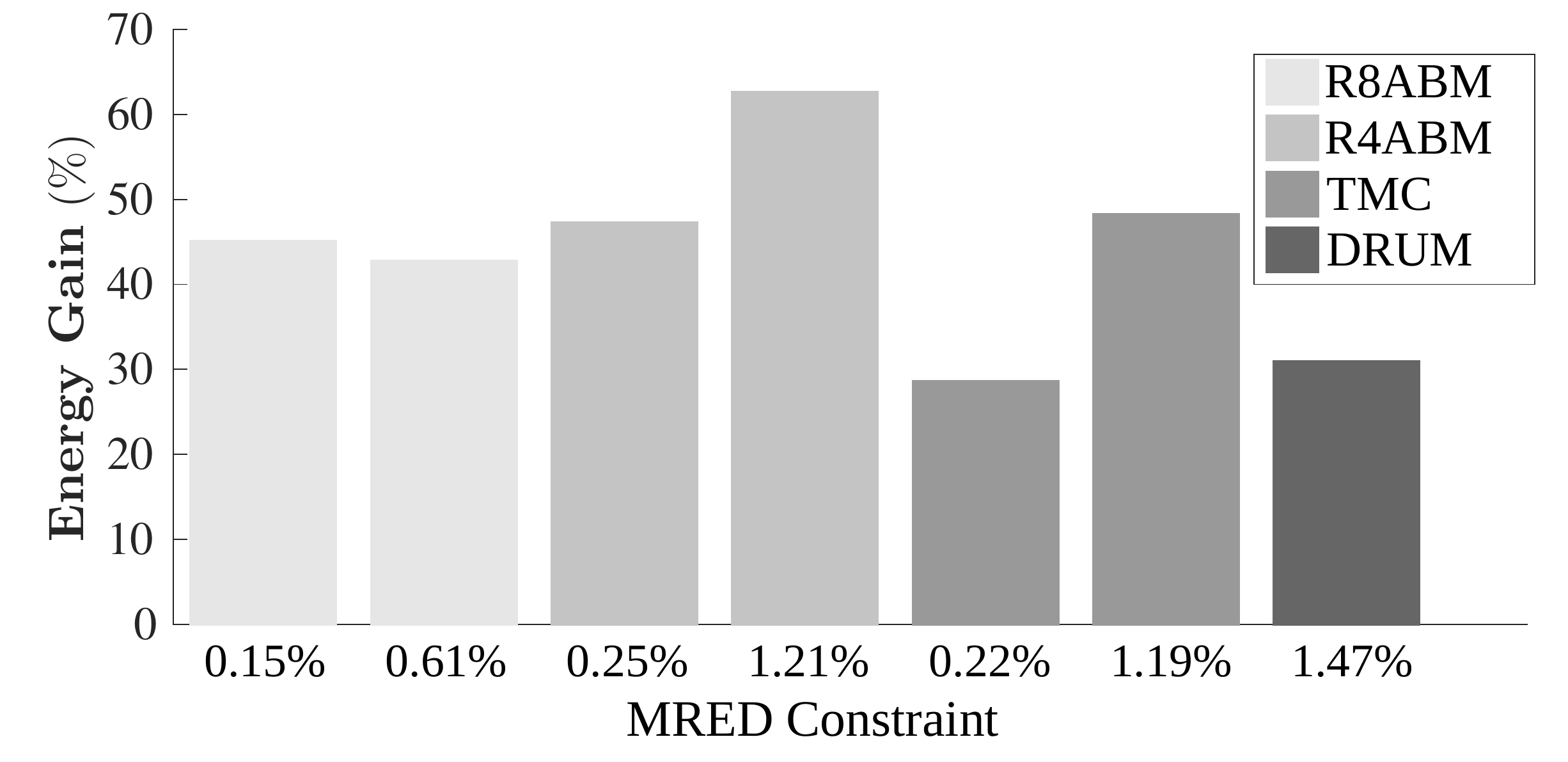}%
\caption[Comparison of Dissertation's Most Efficient Approximate Circuits with State-of-the-Art Designs]{Energy gains of Dissertation's most efficient designs (ROUP2) versus state-of-the-art multipliers
(R8ABM \cite{2016_Jiang_IEEEtc}, R4ABM \cite{2017_Liu_IEEEtc}, TMC \cite{Schulte1993}, DRUM \cite{2015_Hashemi_ICCAD}) 
for the same error constraint.}%
\label{fig_artcomp}
\end{figure}

As a final stage of evaluation,
we compare the new Pareto front 
with other state-of-the-art designs \cite{2016_Jiang_IEEEtc, 2017_Liu_IEEEtc, 2015_Hashemi_ICCAD, Schulte1993}.
In particular,
we 
evaluate the energy consumption of the circuits
for
various MRED constraints in the range $0.15\%$--$1.47\%$.
Figure \ref{fig_artcomp} reports
the energy gains of the ROUP2 multipliers
compared to 
R8ABM \cite{2016_Jiang_IEEEtc}, R4ABM \cite{2017_Liu_IEEEtc}, TMC \cite{Schulte1993}, and DRUM \cite{2015_Hashemi_ICCAD}. 
We
consider
the relative energy reduction from the other design
as the energy gain of ROUP2.  
The large approximation space 
and dense error segments 
of ROUP2
provides increased flexibility,
and thus, 
we select configurations
with MRED equal or slightly smaller than
the MRED constraint. 
This is extremely important,
because
ROUP2 can maximize its gains for a given error constraint.
As expected,
the derived results show that ROUP2 provides significant gains ranging from $29\%$ to $63\%$.

\section{Conclusion}
\label{s6_5}

In this chapter,
we examined the combination of arithmetic approximation techniques, 
targeting to expand the approximation space
and identify the most efficient design solution 
in a comparative state-of-the-art Pareto evaluation. 
We defined the approximation space 
by combining the approximate encodings 
presented in 
Chapter \ref{chapter4} and Chapter \ref{chapter5} 
and as a result,
we propose 5 new families of approximate multipliers.
Each design family
integrates
two approximation techniques
that can be configured independently,
and thus,
it provides numerous approximation configurations
with different accuracy,
namely,  
dense error scaling in the range $0\%$--$2\%$.
The experimental evaluation,
which is performed under two design scenarios (MIN-Delay and ISO-Delay),
reveals that 
the Pareto front is formed exclusively 
by designs applying cooperative approximation techniques,
and more specifically,
by the ROUP2 family. 
The ROUP2 family applies partial product perforation
and asymmetric partial product rounding,
and it improves 
the state-of-the-art Pareto front
that was held by RAD \cite{LeonTVLSI} and AxFXU \cite{LeonMicro}  
by up to $1.5\times$--$2\times$.
Moreover,
besides forming an improved Pareto front,
the cooperative approximations increase its resolution
with their large sets of available configurations. 
Compared to other designs of the literature,
the ROUP2 family 
can efficiently handle all the error bounds 
and provides energy gains up to $63\%$ for the same error constraint.

\chapter{Approximate DSP \& AI Hardware Accelerators}
\label{chapter7}

\addtocontents{lof}{\protect\contentsline{chapter}{\protect\numberline{7}Approximate DSP \& AI Hardware Accelerators}{}{}}
\addtocontents{lot}{\protect\contentsline{chapter}{\protect\numberline{7}Approximate DSP \& AI Hardware Accelerators}{}{}}

\begin{ChapterAbstract}
The worldwide
demand for faster applications with less power consumption
challenges the design of 
Digital Signal Processing (DSP) 
and Artificial Intelligence (AI) 
hardware accelerators. 
As an alternative design strategy,
there is a tendency to exploit 
the error tolerance of DSP/AI workloads
and produce approximate accelerators,
which, however,
improve the power efficiency and/or performance.
Our work
aims at the design and evaluation
of approximate DSP and AI accelerators that integrate our approximate circuits,
i.e., 
RAD, AxFXU/AxFPU, and ROUP.
To provide a great diversity in terms of DSP/AI workloads,
we develop several 
1D/2D signal processors 
and Convolutional Neural Networks (CNNs). 
Both the development and evaluation
are directed by our design methodology,
which includes design space exploration
at both software and hardware level.
The design is not limited only to the use 
of our approximation techniques,
but it also involves
various arithmetic formats,
different algorithms
and state-of-the-art 
parallelization techniques.
The exploration of all these design parameters
targets to provide the 
most efficient approximate variants
of the targeted DSP/AI hardware accelerator. 
Specifically for CNNs,
besides fusing
parallel design techniques with
arithmetic approximations,
we apply fine-grained approximation
without retraining via our MAx-DNN framework. 
Moreover,
we examine networks that are built on floating-point,
fixed-point, and quantized integer arithmetic. 
The evaluation is performed with industrial-strength tools,
i.e., Synopsys Design Compiler and Xilinx Vivado.
In terms of accuracy,
our approximate accelerators
attain mean relative error up to $\mathit{3}$\%
in applications based on multiply-accumulate operations,
typical quality of output image  
in image processing,
and up to $\mathit{5}$\% accuracy loss in CNNs.
We note that the large design space offered by our approximation techniques
can provide,
if necessary,
near-zero accuracy loss. 
In terms of resource gains,
depending on the implementation,
we deliver energy and area gains up to $\mathit{70}$\%.\\
This chapter is based on our
\textbf{publications} in \textbf{\cite{LeonTVLSI, LeonMicro, LeonTECS, LeonDAC, LeonFPL, LeonMOCAST, LeonICECS, LeonLASCAS}}.
\end{ChapterAbstract}

\newpage

\section{Introduction}

In recent years, significant research has been conducted on the development and optimization of applications 
from the Digital Signal Processing (DSP) and Artificial Intelligence (AI) domains.
Such applications
impose strict performance and energy constraints,
which constitute the selection 
of the computing platforms/devices for their deployment
a major open issue \cite{glent}. 
The powerful and compute-intensive DSP/AI algorithms
make the general-purpose Central Processing Units (CPUs)
and the low-end Graphics Processing Units (GPUs) unfit for satisfying the application constraints \cite{fpgagpucpu, georgis}.
As a result, 
alternative computing platforms,
such as the Field-Programmable Gate Arrays (FPGAs)
and the Application-Specific Integrated Circuits (ASICs),
are examined for accelerating DSP/AI workloads.
Both solutions provide high parallelization capabilities
and increased design flexibility.
The FPGAs provide re-configurability and
attractive throughput-per-power ratio \cite{glent}.
The ASICs offer high computational efficiency 
along with low power consumption \cite{asic_vs_fpga},
while allowing more custom implementations. 
In all cases,
the design of optimized circuits 
for DSP/AI workloads
is a key goal for the research community.

AI brings forth
various demanding workloads and algorithms,
including the
Convolutional Neural Networks (CNNs), 
which are 
a class of Artificial Neural Networks (ANNs)
based on deep learning.
CNNs are considered a state-of-the-art AI approach 
to provide high accuracy
in computer vision tasks such as object recognition \cite{Simon} and image classification \cite{Krizh}.
FPGA-based accelerators \cite{Sharma_FPGA, Guo_FPGA, direct_map, Wang_FPGA, lombardi_cnn} have started to take their place as viable and promising solutions, thus,
investing in their efficient implementation forms
an emerging highly valuable design paradigm.
ASIC implementations also provide significant gains in the performance of CNNs \cite{asic1, asic2, asic3, asic4, techno_cnn}, even though they lack of re-configurability.
On the other hand,
classic DSP algorithms are employed
for the implementation of functions for  
image/video/signal processing. 
In this context, 
various FPGA implementations are proposed in the literature,
e.g., 
for image processing \cite{fpga1, fpga2, fpga5}
and telecommunication functions \cite{fpga3, fpga4}.
Correspondingly,
there is significant research on ASIC-based accelerators
for computer vision
\cite{dsp_asic1, dsp_asic2}
and telecommunications \cite{dsp_asic3, dsp_asic4}.

One of the main advantages of FPGA/ASIC design
is the ability
to perform arbitrary bit-level manipulations,
i.e., 
tune the bit-width of the datapath
and optimize the arithmetic optimizations.
Namely,
the designer can select the desired arithmetic representation,
including alternative formats \cite{format1, format2} (also see Chapter \ref{chapter3}),
as well as 
define $n$-bit words without restriction
and use custom arithmetic operators.
All these operations are
seamlessly applied on
FPGA/ASIC technology, 
contrary to the general-purpose GPU/CPU processors.
The design paradigm of Approximate Computing \cite{2016_Mittal_ACMsrv, 2016_Xu_IEEEdt, 2021_Stanley_ACMsrv} 
exploits this flexibility of the FPGA/ASIC design
and 
leverages the error tolerance of DSP/AI applications \cite{ChakradharDAC2010, ChippaDAC2013}
to provide gains in resources (power, energy, area) and/or performance.
The approximation techniques 
that are applied 
in DSP hardware accelerators involve
the use of approximate arithmetic \cite{stratakos},
voltage over-scaling \cite{2010_Kurdahi_IEEEtvlsi}, 
and over-clocking \cite{2013_Shi_FCCM}. 
Obviously,
in FPGA/ASIC design, 
the optimization of the datapath's bit-width
is a typical task, 
non-related to Approximate Computing, 
which may result in accuracy loss to provide 
hardware-friendly designs 
(e.g., due to the use of fixed-point arithmetic or the truncation of the results).
On the other hand,
the design of approximate AI accelerators 
involves application-specific techniques: 
low-bit numerical formats \cite{format2}, 
approximate arithmetic operators \cite{alwann},
weight quantization \cite{quantization}, 
neuron connection pruning \cite{pruning},
and voltage over-scaling \cite{2018_Zhang_DAC}. 

In this chapter,
we focus on \emph{approximate DSP and AI accelerators}
that are 
developed with Hardware Description Language (HDL)
and can be deployed either on FPGA or standard-cell ASIC technology. 
Our main approximation approach lies in
using the Dissertation's approximate multipliers
in DSP and AI hardware architectures. 
The arithmetic components
are key processing units in hardware accelerators,
as they inherently affect the energy efficiency and the performance of the entire application.
Their impact becomes even greater,  
considering that modern accelerators implement parallel architectures consisting of multiple processing units.
The integration of our approximate circuits
in accelerators
is based on a design methodology involving both software- and hardware-level Design Space Exploration (DSE).
Our goal is to
assess the Dissertation's approximate designs in real-world DSP/AI applications,
and in reverse,
explore the error resilience of the applications 
and quantify 
the resource gains of approximate accelerators. 
Moreover,
our approximation approach 
allows to seamlessly
decrease the resources or 
improve the throughput of any given DSP/CNN
without modifying the initial hardware architecture.
Namely,
we improve the efficiency of DSP accelerators
without tuning their underlying arithmetic
(a typical task of the hardware development flow),
and 
specifically for CNNs, 
we provide approximate accelerators
without re-training (a task that may be required in order to reduce the accuracy loss caused by the approximations).
The elimination of additional training/exploration
is considered an advantage,
as proprietary datasets/models may not be available,
and also, 
the training time is usually increased 
due the emulation of the hardware approximations
and 
the use of custom approximate arithmetic operators.

The \textbf{contribution} of this chapter is summarized as follows:

\begin{itemize}[]
\item[(i)] We highlight the benefits of studying and optimizing the arithmetic of DSP and AI hardware accelerators,
while proving the inherent error resilience of their workloads.
\item[(ii)] We propose a methodology for developing  approximate DSP and AI hardware accelerators,
which is based on extensive design space exploration involving arithmetic formats, approximation techniques, and algorithms. 
\item[(iii)] We explore and quantify the resource gains and accuracy of DSP and AI hardware accelerators with approximate arithmetic. 
\end{itemize}

The remainder of this chapter is organized as follows. 
Section \ref{s7_2} introduces our design methodology
and summarizes all the approximate designs, applications and accelerators presented in this chapter. 
Sections \ref{s7_3}--\ref{s7_5} evaluate our approximate designs 
when used in DSP/AI accelerators.
Finally, 
Section \ref{s7_6} draws the conclusions.

\section{Design Methodology}
\label{s7_2}

\begin{figure}[!t]
\vspace*{-5pt}
\centering
\includegraphics[width=0.86\textwidth]{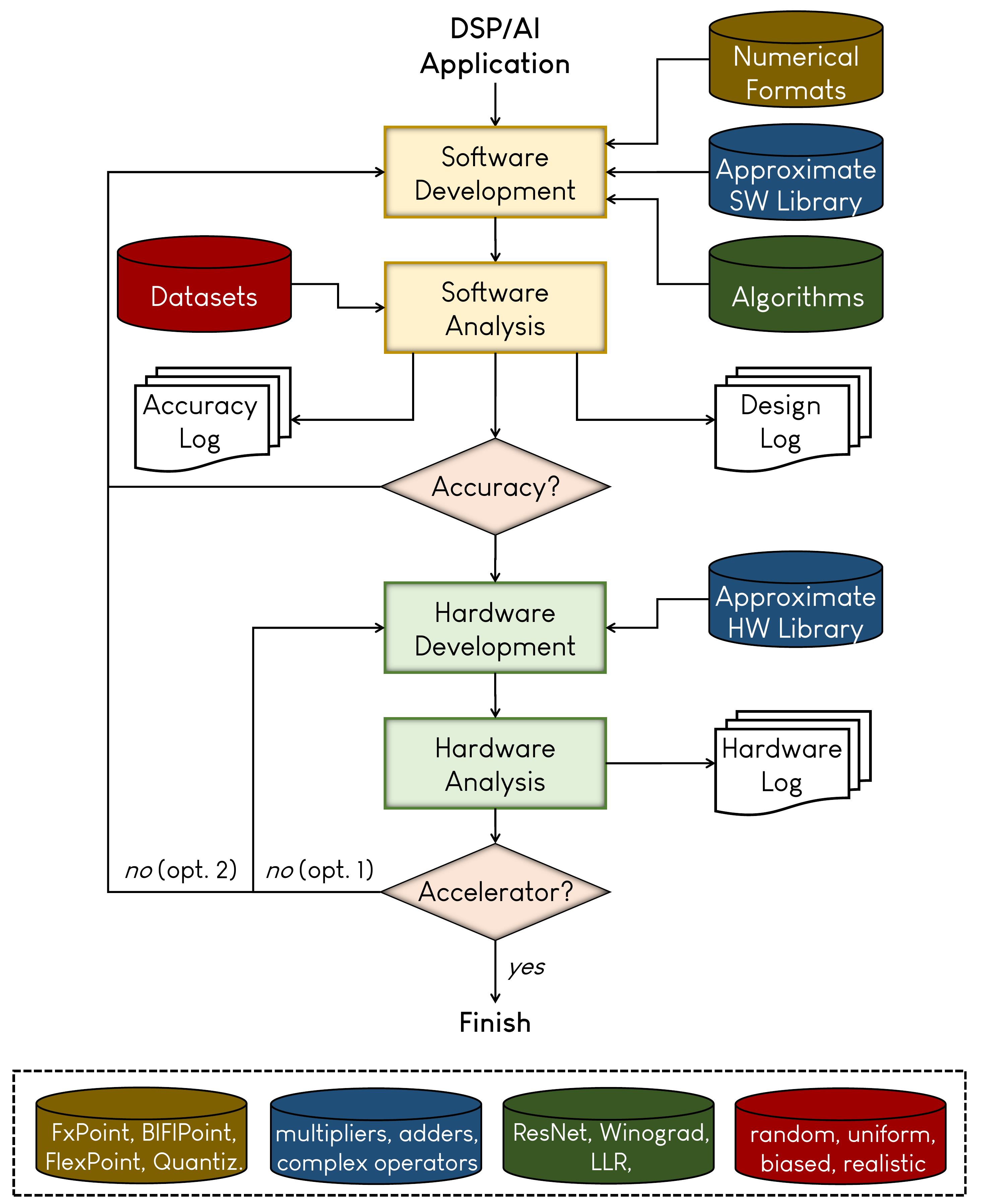}%
\caption[Design Methodology for Approximate DSP and AI Hardware Accelerators]{Design methodology for approximate DSP and AI hardware accelerators.}%
\label{fig_metho}
\vspace*{-4pt}
\end{figure}

The proposed methodology for designing approximate DSP and AI accelerators is illustrated in Figure \ref{fig_metho}.
It consists of two stages:
the exploration at the software level and the development of the accelerator at the hardware level. 
We note that our methodology follows the typical steps
of hardware design.
However, 
it includes 
additional functionalities
(i.e., the use of approximate arithmetic units),
error analysis
(due to the approximations),
and 
multi-level DSE
(involving approximation techniques/configurations, arithmetic formats, and algorithms).
The multi-level DSE of our design methodology is very important, 
because
each combination of approximations, 
arithmetic formats, and algorithms
has a different impact on the accuracy and the hardware efficiency of the accelerator. 

Before analyzing the steps of our methodology,
we make the following clarifications:

\begin{itemize}
    \item \underline{Approximate Design Library}: it includes the software models and the hardware descriptions of the approximate arithmetic units (e.g., RAD \cite{LeonTVLSI}, AxFXU \cite{LeonMicro}, AxFPU \cite{LeonTECS},  and ROUP \cite{LeonDAC}), along with their error analysis and experimental results.
    \item \underline{Arithmetic Pool}: it includes various arithmetic formats, e.g., the conventional fixed- and floating-point, block floating-point \cite{bfl}, Flexpoint \cite{format2}, and quantized fixed-point \cite{quant2}. 
    \item \underline{Algorithmic Pool}: it includes algorithms for the entire application (e.g., the ResNet model \cite{resnet} for CNN or the Log-Likelihood-Ratio (LLR) method \cite{llr} for signal demodulation) and key operations (e.g., the Winograd method \cite{wino} for convolution).
    \item \underline{Dataset Pool}: it includes various application-specific test datasets 
    (e.g., random, Gaussian/uniformly distributed, and realistic data) for evaluating the quality of the results at software level.
\end{itemize}

Firstly,
we develop the application in software
using various arithmetic formats
and algorithms.
Our software models are bit-accurate,
which allows us to study the accuracy 
when using approximate arithmetic formats and/or approximate operators.
To obtain the groundtruth (accurate) results,
we run simulations on the datasets.
Afterwards, 
we start to integrate
approximate arithmetic units
and create approximate model variants of the targeted accelerator. 
For each variant, 
we run simulations
to obtain the approximate results
and evaluate its accuracy with application-specific metrics.
Indicatively, we use metrics such as
the Structural Similarity Index (SSIM)
and Peak Signal-to-Noise Ratio (PSNR) for image processing,
the Bit Error Rate (BER) for telecommunications,  
and the classification accuracy for CNNs.
Furthermore, for each variant,
we report possible bottlenecks in hardware, e.g., regarding parallelization or the implementation of complex operations.
The next step is to examine which approximate variants
satisfy the constrained quality of results,
and then select the most efficient for implementation  
on FPGA/ASIC technology.
In case the accuracy constraints are not met,
we design new approximate variants
by combining different arithmetic, algorithms, and approximate units.

\begin{table}[!b]
\fontsize{9}{10}\selectfont
\renewcommand{\arraystretch}{1.2}
\setlength{\tabcolsep}{7pt}
\caption[Overview of Dissertation's Approximate DSP \& AI Hardware Accelerators]{Overview of Dissertation's approximate DSP \& AI hardware accelerators.}
\label{tb_sumacc}
\centering
\begin{threeparttable}
\begin{tabular}{l|ccc} 
\hline
\textbf{Design} & \textbf{Domain} &  \textbf{Application} & \textbf{Accelerator}\footnotemark\setcounter{footnote}{0} \\
\hline
\hline
RAD    & Digital Signal Processing & Sobel, FIR, MatMul     & ASIC \\
RAD    & Digital Signal Processing & QAM Demodulation       & FPGA \\
RAD    & Artificial Neural Networks  & CNN (Ship Detection)   & FPGA \\
AxFXU  & Digital Signal Processing & Sobel, FIR, MatMul     & ASIC  \\
AxFPU   & Digital Signal Processing & Gaussian Blurring               & ASIC  \\
AxFPU   & Artificial Neural Networks  & CNNs (MNIST, CIFAR-10) & ASIC \\
AxFPU   & Machine Learning  & K-Means Clustering                & --  \\
AxFPU   & Linear Algebra  & LU Decomposition       & -- \\
ROUP     & Artificial Neural Networks  & ResNet-8 (CIFAR-10)    & ASIC \\
\hline
\end{tabular}
\begin{tablenotes}
  \item[1]{\fontsize{7.7}{8.8}\selectfont Implementation details: 
  with HDL, 
  for ASIC (standard-cell) or FPGA (programmable logic).}
\end{tablenotes}
\end{threeparttable}
\end{table}

When the software-level DSE is over,
we continue with the typical 
HDL development. 
We implement the selected approximate variants
integrating our design choices
for arithmetic, algorithms, and approximation techniques.
Finally, 
in the evaluation phase
we examine 
if the accelerator satisfies the constraints for performance and resource utilization. 
In case the results are not the expected ones,
we examine alternative design approaches (e.g., parallelization scheme) 
or even return to the initial software-level DSE to create new approximate variants. 

In the remainder of the chapter,
we present experimental results from DSP/AI hardware accelerators 
employing our approximate multipliers.
More specifically,
based on our methodology,
we design parallel architectures with approximate arithmetic
for 1D/2D signal processing and CNNs.
Each section is dedicated to one
of our approximate designs, 
i.e., 
RAD \cite{LeonTVLSI} 
(presented in Chapter \ref{chapter4}), 
AxFXU \cite{LeonMicro} \& AxFPU \cite{LeonTECS} 
(presented in Chapter \ref{chapter5}), 
and ROUP \cite{LeonDAC} 
(presented in Chapter \ref{chapter6}).
Table \ref{tb_sumacc} summarizes the details for all the applications and accelerators that are presented in the following sections.

\section{Design and Evaluation of Applications with RAD}
\label{s7_3}

In this section,
we employ our approximate high-radix RAD multipliers \cite{LeonTVLSI}
in real-world applications.
From the DSP domain,
we implement 
the Sobel edge detector,
a Finite Impulse Response (FIR) filter,
a matrix multiplication unit,
and a
Quadrature-Amplitude-Modulation (QAM) demodulation filter.
From the AI domain,
we implement a custom 
CNN for ship detection on satellite images.

\subsection{Approximate DSP Accelerators}
\label{s731}

\begin{figure}[!t]
\vspace*{-7pt}
\centering
\subfloat[\label{fig_convo1}]{\includegraphics[width=0.84\textwidth]{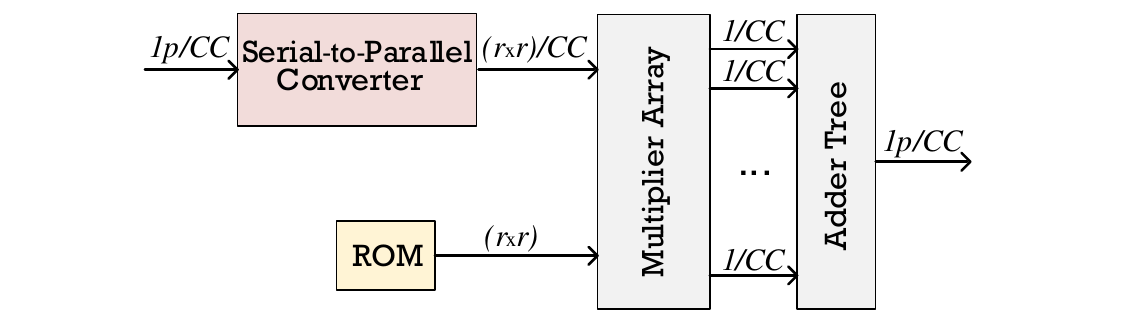}}  \\[-1pt]
\subfloat[\label{fig_convo2}]{\includegraphics[width=0.84\textwidth]{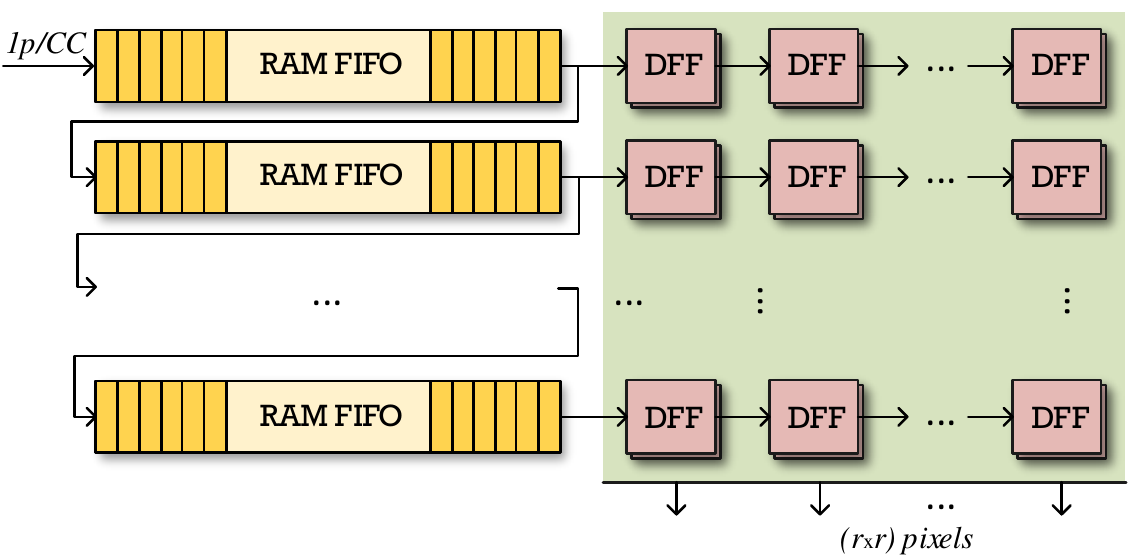}}%
\caption[Hardware Architecture of 2D Convolution]{Hardware architecture of $r \times r$ convolution:
\textbf{(a)} convolution engine
and
\textbf{(b)} serial-to-parallel converter.}%
\label{fig_convo}
\end{figure}

A well-known filter for finding the object boundaries in an image
is the discrete Sobel operator \cite{sobel}. 
The Sobel operator consists of two $3 \times 3$ kernels 
that are applied linearly to the image to compute an approximation of the gradient of the intensity function. 
The two convolutions compute the changes in brightness in the horizontal and vertical orientation.
For the hardware implementation of convolution,
we adopt the ordinary approach \cite{Bailey},
which includes 
a serial-to-parallel converter along with the structure of multipliers and adders,
as illustrated in 
Figure \ref{fig_convo} for generic $r \times r$ kernel.
The convolution engine inputs one pixel per clock cycle
in raster-scan order 
and forwards it
to the serial-to-parallel converter that outputs $r \times r$ pixels per cycle.
The converter consists of $r \times r$ DFFs connected to $r$ FIFOs
in a linear array topology.
In practice, 
the $r \times r$ DFFs slide over the input image in raster-scan order, 
while each FIFO temporarily stores a row of pixels. 
In a pipeline fashion, 
the $r \times r$ pixels 
along with the $r \times r$ kernel weights
are forwarded to $r^2$ multipliers,
and then, 
the multiplication results are added 
to produce the new pixel.
Similar architectures are designed for 
the FIR filter and the matrix multiplication unit.
Regarding FIR,
we employ a $32$-tap low-pass filter with cut-off frequency equal to $20$KHz and $16$-bit coefficients,
while for matrix multiplication
we select $3 \times 3$ tiling. 

\begin{figure}[!t]
\centering
\includegraphics[width=0.91\textwidth]{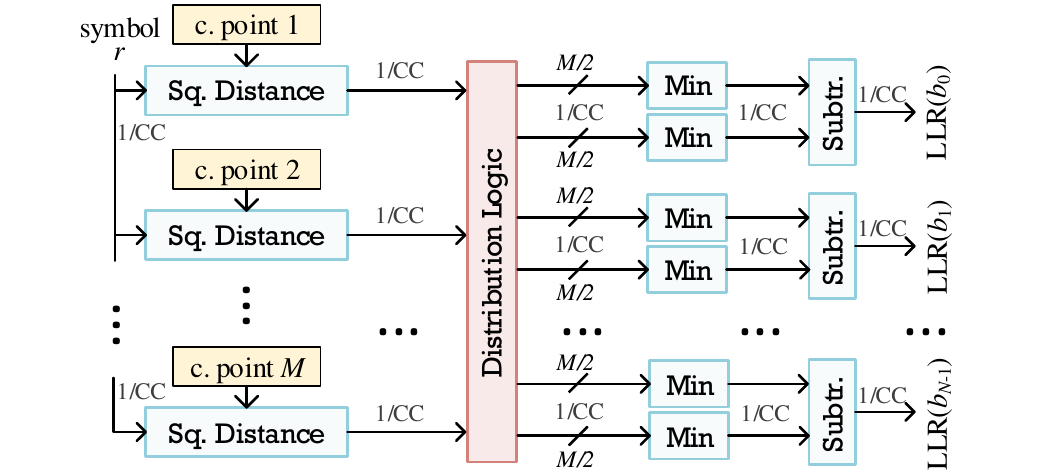}%
\caption[Hardware Architecture of QAM Demodulation]{Hardware architecture of $M$-QAM demodulation.}%
\label{fig_qam}
\vspace*{-5pt}
\end{figure}

Besides the aforementioned hardware architectures,
we design circuits
for QAM demodulation,
i.e., a key digital function in the baseband processing chain (telecommunications).
We consider 
$64$-QAM keying signals
and their corresponding gray-coded constellation map,
where
each constellation point is represented by a $6$-bit
vector. 
We employ the approximate LLR method
\cite{llr}
for predicting the $i$-th bit of the received symbol.
This method avoids 
the expensive logarithmic and exponent calculations
of the exact LLR method
by using only the 
nearest constellation points with 
$i$-th bit equal to $0$ and $1$.
The generic hardware architecture for $M$-QAM demodulation is presented in Figure \ref{fig_qam}.
It processes in parallel the input symbol per constellation point  
and performs on-the-fly computations 
(e.g., square distances, minimum, subtractions).

\subsubsection{Experimental Evaluation}

\begin{figure}[!t]
\centering
\subfloat[\label{fig_s1}]{\includegraphics[width=0.18\textwidth]{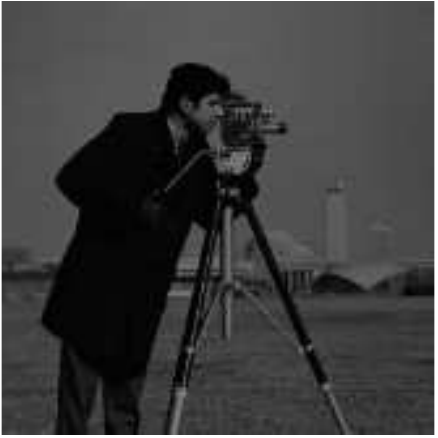}}
\hspace{1pt}
\subfloat[\label{fig_s2}]{\includegraphics[width=0.18\textwidth]{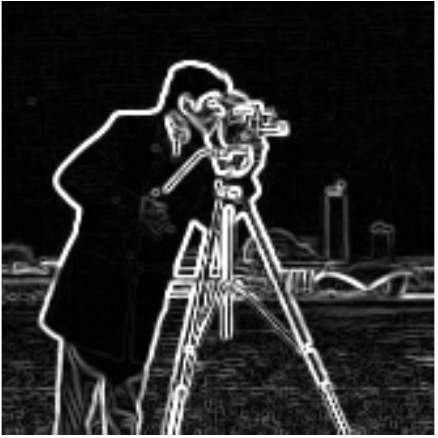}}
\hspace{1pt}
\subfloat[\label{fig_s3}]{\includegraphics[width=0.18\textwidth]{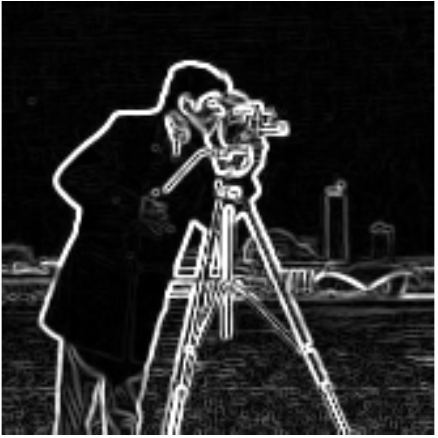}}
\hspace{1pt}
\subfloat[\label{fig_s4}]{\includegraphics[width=0.18\textwidth]{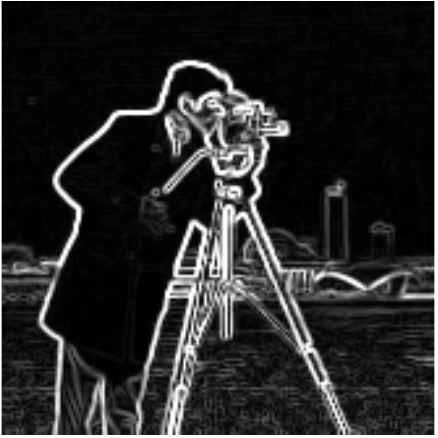}}%
\\[-8pt]
\subfloat[\label{fig_s5}]{\includegraphics[width=0.18\textwidth]{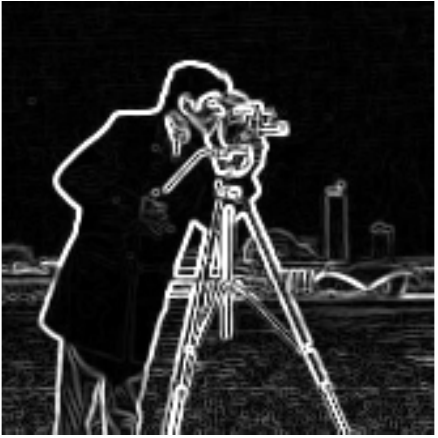}}
\hspace{1pt}
\subfloat[\label{fig_s6}]{\includegraphics[width=0.18\textwidth]{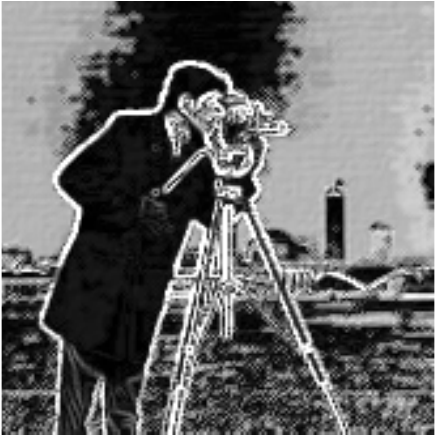}}
\hspace{1pt}
\subfloat[\label{fig_s7}]{\includegraphics[width=0.18\textwidth]{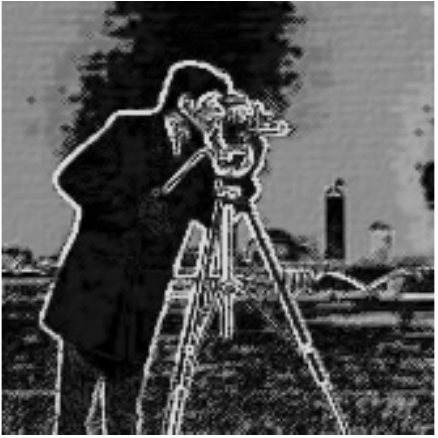}}
\hspace{1pt}
\subfloat[\label{fig_s8}]{\includegraphics[width=0.18\textwidth]{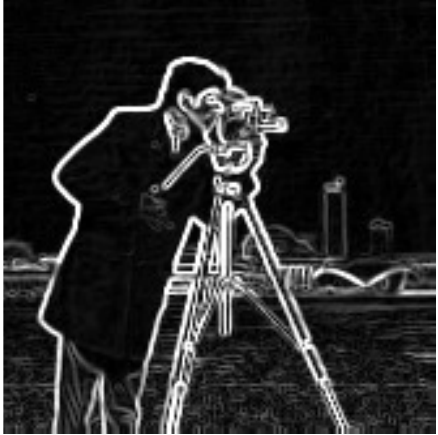}}%
\\[-8pt]
\subfloat[\label{fig_s9}]{\includegraphics[width=0.18\textwidth]{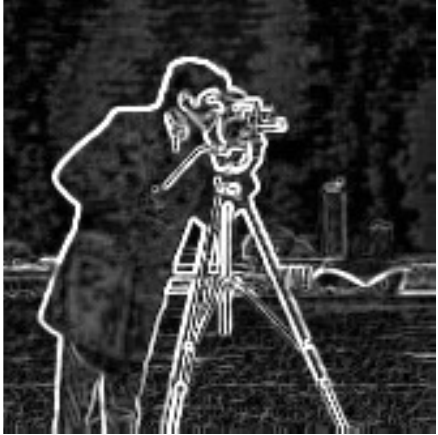}}
\hspace{1pt}
\subfloat[\label{fig_s10}]{\includegraphics[width=0.18\textwidth]{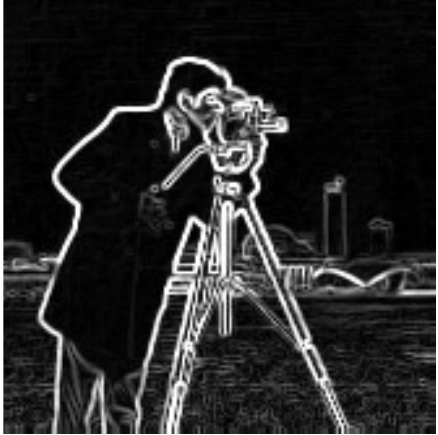}}
\hspace{1pt}
\subfloat[\label{fig_s11}]{\includegraphics[width=0.18\textwidth]{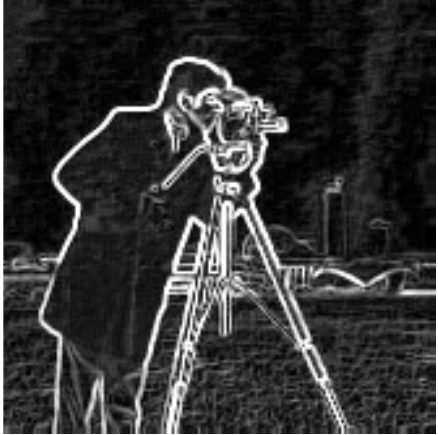}}
\hspace{1pt}
\subfloat[\label{fig_s12}]{\includegraphics[width=0.18\textwidth]{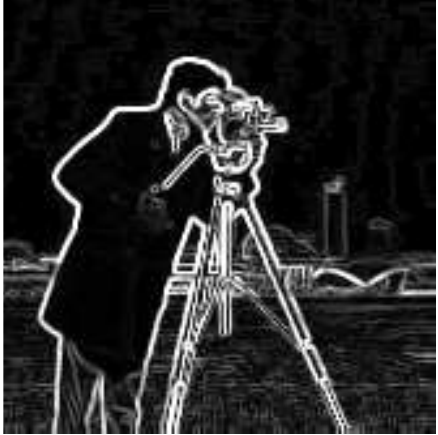}}%
\caption[I/O Images of Approximate Sobel Edge Detector]
{I/O images of approximate Sobel edge detector:
\textbf{(a)} input image 
and output image with
\textbf{(b)} accurate multiplier, 
\textbf{(c)} RAD64, 
\textbf{(d)} RAD256, 
\textbf{(e)} RAD1024, 
\textbf{(f)} R8ABM1 \cite{2016_Jiang_IEEEtc},
\textbf{(g)} R8ABM2-15 \cite{2016_Jiang_IEEEtc},
\textbf{(h)} R4ABM1-14 \cite{2017_Liu_IEEEtc},
\textbf{(i)} R4ABM1-16 \cite{2017_Liu_IEEEtc},
\textbf{(j)} R4ABM2-14 \cite{2017_Liu_IEEEtc},
\textbf{(k)} R4ABM2-16 \cite{2017_Liu_IEEEtc} and
\textbf{(l)} DRUM6 \cite{2015_Hashemi_ICCAD}.}
\label{fig_sob}
\end{figure}

We implement the Sobel edge detector, the FIR filter, and the matrix multiplication 
in Verilog 
and 
synthesize them with Synopsys Design Compiler
and the TSMC 65-nm  standard-cell library,
targeting to ASIC-based approximate DSP accelerators. 
The resource gains and accuracy results
of the accelerators with our RAD multipliers \cite{LeonTVLSI}
are reported in 
Table \ref{tb_sob_hw}.
The Table also includes results for
the multipliers \cite{2016_Jiang_IEEEtc, 2017_Liu_IEEEtc, 2015_Hashemi_ICCAD}
that are employed
in the evaluation of RAD in Chapter \ref{chapter4}.
Regarding accuracy, 
for Sobel,
we use the Correct Edge Ratio (CER),
which measures the number of correct edges detected per total number of edges,
while for the other two accelerators,
we use MRED (as defined in Chapter \ref{chapter4}).
For input data,
we use the Cameraman benchmark image for Sobel
(illustrated in Figure \ref{fig_sob} along with the various output images) 
and $200$K random generated inputs for the FIR filter
and the matrix multiplication.

\begin{table}[!t]
\vspace*{2pt}
\fontsize{9}{10}\selectfont
\renewcommand{\arraystretch}{1.2}
\setlength{\tabcolsep}{11.5pt}
\caption[Experimental Results of Approximate RAD-Based DSP Applications on TSMC 65-nm Standard-Cell]{Experimental results of approximate RAD-based DSP applications on TSMC 65-nm standard-cell.}
\label{tb_sob_hw}
\centering
\begin{threeparttable}
\begin{tabular}{cl| c c c | c}
\hline
& \multicolumn{1}{c|}{\multirow{2}{*}{\textbf{Design}}} &
\textbf{Delay} & 
\textbf{Energy} & 
\textbf{Area} & 
\textbf{\phantom{a}CER\phantom{a}}  \\[-2pt]
& & $(\%)$\footnotemark\setcounter{footnote}{0} & $(\%)$\footnotemark\setcounter{footnote}{0} & $(\%)$\footnotemark\setcounter{footnote}{0} &  $(\%)$ \\
\hline \hline
\parbox[t]{8mm}{\multirow{10}{*}{\rotatebox[origin=c]{45}{\textbf{Sobel}}}} 
& RAD64                         & $1.8$  & $22$   & $20.8$ & $99.98$   \\
& RAD256                        & $6$    & $37.3$ & $33.9$ & $99.96$   \\
& RAD1024                       & $8.4$  & $54.8$ & $46$   & $99.87$  \\
& R8ABM1 \cite{2016_Jiang_IEEEtc}     & $1.1$  & $1.3$  & $1.6$  & $58.90$    \\
& R8ABM2-15 \cite{2016_Jiang_IEEEtc}  & $5.4$  & $9.6$  & $8.4$  & $54.41$   \\
& R4ABM1-14 \cite{2017_Liu_IEEEtc}    & $0.6$  & $3.4$ & $3.8$  & $99.80$   \\
& R4ABM1-16 \cite{2017_Liu_IEEEtc}    & $0.6$  & $6.9$ & $4.7$  & $99.36$   \\
& R4ABM2-14 \cite{2017_Liu_IEEEtc}    & $1.8$  & $7$   & $5$    & $99.11$   \\
& R4ABM2-16 \cite{2017_Liu_IEEEtc}    & $3$    & $7.1$ & $5.1$  & $98.27$   \\
& DRUM6 \cite{2015_Hashemi_ICCAD} & $-7.8$ & $55.3$ & $46.4$ & $98.87$ \\
\end{tabular}
\begin{tabular}{cl| c c c | c}
\hline
& \multicolumn{1}{c|}{\multirow{2}{*}{\textbf{Design}}} &
\textbf{Delay} & 
\textbf{Energy} & 
\textbf{Area} & 
\textbf{MRED}  \\[-2pt]
& & $(\%)$\footnotemark\setcounter{footnote}{0} & $(\%)$\footnotemark\setcounter{footnote}{0} & $(\%)$\footnotemark\setcounter{footnote}{0} &  $(\%)$ \\
\hline \hline
\parbox[t]{8mm}{\multirow{10}{*}{\rotatebox[origin=c]{45}{\textbf{FIR}}}} 
& RAD64                         & $3$     & $6$     & $9.8$ & $0.41$   \\
& RAD256                        & $7.1$   & $17$    & $18.3$ & $0.96$    \\
& RAD1024                       & $11.1$  & $34.1$  & $34.7$ & $3.60$   \\
& R8ABM1 \cite{2016_Jiang_IEEEtc}     & $0$     & $-0.2$  & $-0.1$ & $13.51$     \\
& R8ABM2-15 \cite{2016_Jiang_IEEEtc}  & $6.1$   & $5.3$   & $9.2$ & $33.98$    \\
& R4ABM1-14 \cite{2017_Liu_IEEEtc}    & $0$     & $1.5$   & $2.2$ & $5.28$   \\
& R4ABM1-16 \cite{2017_Liu_IEEEtc}    & $1$     & $3$     & $3.1$ & $23.38$   \\
& R4ABM2-14 \cite{2017_Liu_IEEEtc}    & $1$     & $3.9$   & $4.8$ & $6.10$    \\
& R4ABM2-16 \cite{2017_Liu_IEEEtc}    & $3$     & $4.1$   & $5.5$ & $30.43$  \\
& DRUM6 \cite{2015_Hashemi_ICCAD} & $-32.3$ & $21.4$  & $21.6$ & $9.18$  \\
\end{tabular}
\begin{tabular}{cl| c c c | c}
\hline
& \multicolumn{1}{c|}{\multirow{2}{*}{\textbf{Design}}} &
\textbf{Delay} & 
\textbf{Energy} & 
\textbf{Area} & 
\textbf{MRED}  \\[-2pt]
& & $(\%)$\footnotemark\setcounter{footnote}{0} & $(\%)$\footnotemark\setcounter{footnote}{0} & $(\%)$\footnotemark\setcounter{footnote}{0} &  $(\%)$ \\
\hline \hline
\parbox[t]{8mm}{\multirow{10}{*}{\rotatebox[origin=c]{45}{\textbf{MatMul}}}} 
& RAD64                      & $6.2$  & $7.8$         & $14$ & $0.07$ \\
& RAD256                        & $8$  & $25.8$        & $26.9$ & $0.17$ \\
& RAD1024                       & $11.5$ & $37.7$        & $38.7$ & $0.57$ \\
& R8ABM1 \cite{2016_Jiang_IEEEtc}      & $0$     & $-1.7$      & $0$ & $0.10$ \\
& R8ABM2-15 \cite{2016_Jiang_IEEEtc}    & $5.3$  & $7.1$         & $12.8$ & $0.38$ \\
& R4ABM1-14 \cite{2017_Liu_IEEEtc}     & $0.9$  & $5.8$         & $6.4$ & $0.07$ \\
& R4ABM1-16 \cite{2017_Liu_IEEEtc}     & $0.9$  & $6.6$         & $7.3$ & $0.29$ \\
& R4ABM2-14 \cite{2017_Liu_IEEEtc}    & $3.5$  & $7.1$         & $7.7$ & $0.15$ \\
& R4ABM2-16 \cite{2017_Liu_IEEEtc}    & $3.5$  & $8.9$         & $9.5$ & $0.75$ \\
& DRUM6 \cite{2015_Hashemi_ICCAD}  & $-30.1$ & $32.8$ & $34.9$ & $1.60$ \\
\hline
\end{tabular}
\begin{tablenotes}
  \item[1]{\fontsize{7.7}{8.8}\selectfont Refers to $\%$ resource gains (relative reduction) in comparison with the accurate design.}
\end{tablenotes}
\end{threeparttable}
\vspace*{-10pt}
\end{table}

Starting with the Sobel edge detector, 
the RAD multipliers achieve the highest accuracy by detecting almost all the edges 
(more than $99.87\%$ edges are correctly detected). Moreover, they deliver remarkable resource gains,
i.e., 
up to $54.8\%$ in energy, 
$46.\%$ in area, 
and $8.4\%$ in critical path delay. 
DRUM6 exhibits similar energy gain ($55.3\%$),
however, 
it imposes a delay overhead of $7.8\%$ and detects $1\%$ less edges compared to RAD1024. 
Furthermore, 
it is worth to note that although R8ABM1 and R8ABM2-15 feature small MRED values (see Chapter \ref{chapter4}), 
they decrease the quality of the results in Sobel,
as they exhibit a CER of $58.90\%$ and $54.41\%$, respectively. 

The RAD multipliers also outperform the other designs
in the FIR filter. 
More specifically,
they achieve up to $34.1\%$ energy gain in exchange 
for small a MRED of $3.60\%$.
Similar to Sobel, 
R8ABM1 and R8ABM2-15 introduce large errors,
i.e., MREDs of $13.51\%$ and $33.98\%$, respectively.
In this application, 
DRUM6 achieves significant energy reduction ($21.4\%$) for considerable error ($9.18\%$).
However, 
these values are worse than those of RAD1024, 
which provides $1.6 \times$ larger energy gains
and $2.6 \times$ smaller MRED.

Regarding the matrix multiplication,
all the examined multipliers perform very well in terms of accuracy.
When considering similar error values,
our RAD multipliers achieve larger energy savings than the other designs.
In particular, 
RAD1024 delivers $37.7\%$ energy reduction 
while exhibiting an MRED of $0.57\%$.
R4AMB1-14 and RAD64 feature the smallest MRED ($0.07\%$),
however, RAD64 delivers $1.4 \times$ larger energy gains.
Similarly, 
R4AMB2-14 and RAD256 exhibit similar MRED, however,  
the latter delivers $3.6 \times$ larger energy gains.

\begin{table}[!t]
\fontsize{9}{10}\selectfont
\renewcommand{\arraystretch}{1.2}
\setlength{\tabcolsep}{11.5pt}
\caption[Experimental Results of Approximate RAD-Based $64$-QAM Demodulation on Zynq ZCU106 FPGA]{Experimental results of approximate RAD-based $64$-QAM demodulation on Zynq ZCU106 FPGA.}
\label{tb_qam_hw}
\centering
\begin{threeparttable}
\begin{tabular}{l| c c | c | c}
\hline
\multicolumn{1}{c|}{\multirow{2}{*}{\textbf{Design}}} &
\textbf{LUT} & 
\textbf{Clock} &
\textbf{Throughput} &
\multirow{2}{*}{\setcounter{footnote}{1}\textbf{BER}\footnotemark\setcounter{footnote}{0}} \\[-2pt] 
& $(\%)$\footnotemark\setcounter{footnote}{0} 
& (MHz)
& (MSamples/s)
&  \\
\hline \hline 
Accurate (Fl. Point) & $46$  & $286$ & $286$ & $10^{-1}$ -- $10^{-4}$ \\
Accurate (Fx. Point) & $24$  & $312$ & $312$ & $10^{-1}$ -- $10^{-4}$ \\
RAD64 (Fx. Point) & $16$  & $321$ & $321$ & $10^{-1}$ -- $10^{-4}$ \\
\hline
\end{tabular}
\begin{tablenotes}
  \item[1]{\fontsize{7.7}{8.8}\selectfont Refers to $\%$ resource utilization of ZCU106 ($230400$ LUTs).}
  \item[2]{\fontsize{7.7}{8.8}\selectfont Measured for SNR/symbol values $0$--$14$dB.}  
\end{tablenotes}
\end{threeparttable}
\end{table}

Finally,
we implement the $64$-QAM demodulation
on the Zynq UltraScale+ ZCU106 FPGA
using VHDL
and the 
Xilinx Vivado tool.
Based on our methodology,
we employ 
both $32$-bit floating-point and $16$-bit fixed-point arithmetic.
As accuracy metric,
we use BER.
Table \ref{tb_qam_hw} summarizes the results for the most efficient $64$-QAM circuits.
Fixed-point provides similar BER values with the ``golden'' floating-point model, 
i.e., 
from $10^{-1}$ to $10^{-4}$
for Signal-to-Noise Ratio (SNR) per transmitted symbol in the range $0$--$14$dB.
RAD64 retains 
BER at the same orders of magnitude (the maximum relative error is $1\%$ compared to floating-point).
In terms of resources,
the design using RAD64 provides
$65\%$ LUT reduction and $1.12\times$ speedup versus the floating-point variant, 
and small but still important gains versus the fixed-point variant.
Finally,
according to our design methodology,
the algorithmic pool for QAM demodulation
includes other LLR methods,
such as the exact and the piecewise LLR.
However,
our software/hardware DSE highlights the approximate LLR method as the most efficient one. 
For example,
the approximate LLR method almost matches the BER scaling
of the exact method (for small SNR values, it is the same),
while it provides significantly smaller resource utilization,
i.e., 
\raisebox{0.8pt}{$\scriptstyle\sim$}$3 \times$ less LUTs.

\subsection{Approximate CNN Accelerators}

Following the evaluation in DSP applications,
we employ RAD in CNNs.
In particular,
we develop a parallel CNN architecture 
for detecting ships on satellite images. 
Our CNN model
consists of $4$ convolutional
and $2$ fully connected layers.
In total, 
the convolutional layers 
have $144$ filters 
($32$+$16$+$64$+$32$) with $33$K weights,
while each filter 
includes a ReLU activation function and $4$-to-$1$ max pooling.
The fully connected layers
consist of $50$ neurons
($48$+$2$) with $98$K weights.
The CNN is developed in TensorFlow
and trained with $16$-bit floating-point arithmetic on $128$$\times$$128$ RGB images \cite{kaggleships},
providing a classification accuracy
of $96.8\%$.

For our design,
we consider the typical CNN:
the network consists of convolutional layers, which include filters, which perform multiple convolutions. 
Namely: 
\begin{itemize}
    \item a convolutional layer inputs $M$ image channels, processes them with $N$ filters, and outputs $N$ feature maps (one by each filter).
    \item each one of the $N$ filters inputs $M$ image channels,
    perform $M$ convolutions (one per channel), and outputs $1$ feature map.
\end{itemize}
We apply parallelization at two distinct levels:
(i) at network level, 
where we
deploy multiple parallel convolution engines per filter,
and
(ii) at convolution level,
where 
we parallelize the number of arithmetic operations per $r \times r$ convolution.
The network-level parallelization is supported by a parallel multi-bank memory organization \cite{Bailey},
which allows multiple channels to be accessed concurrently. 
On the other hand,
the convolution-level parallelization is implemented as in Sobel (see Section \ref{s731} and Figure \ref{fig_convo}).
Regarding the arithmetic,
we consider both the conventional fixed- and floating-point formats,
and we also adopt the block floating-point format \cite{bfl}.
Finally, 
for the implementation of the convolutional engine,
besides the ordinary approach used in Sobel,
we employ the Winograd algorithm \cite{wino}. 
Next, we present the details for each one of our design choices.

\begin{figure}[!t]
\centering
\includegraphics[width=0.93\textwidth]{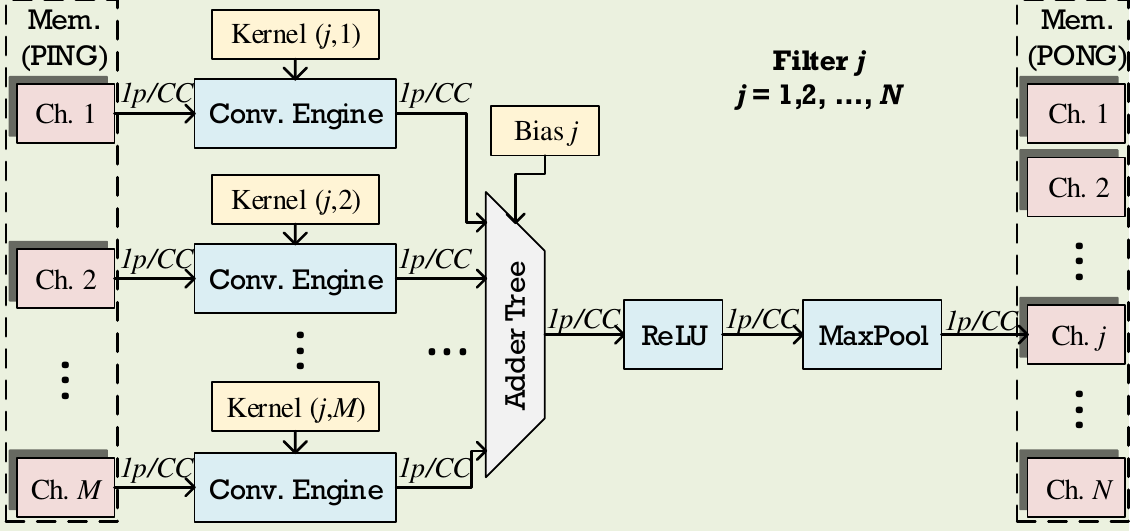}%
\caption[Hardware Architecture of Filter in Convolutional Layers]{Hardware architecture of filter in $M$-input convolutional layers.}%
\label{fig_hwcnn}
\end{figure}

\underline{Network-Level Parallelization}:
The convolution layers are executed serially by 
reusing $E$ convolutional engines in parallel.
Figure \ref{fig_hwcnn} shows a representative setup with a
fully parallelized filter processing all $M$ input channels with $E=M$ engines.
Each engine computes a 2D convolution with a throughput of $1$ pixel per clock cycle,
as shown in Figure \ref{fig_convo}. 
In a pipelined fashion,
the $E$ output pixels per cycle are accumulated in an adder tree along with the filter's bias,
and then,  
they are forwarded through ReLU and MaxPool to the memory storing  
the output channels.
When $E > M$, 
we compute multiple filters in parallel.
The data memories (left and right in Figure \ref{fig_hwcnn}) 
are designed with $1$ bank per channel 
(being reused during CNN steps)
in a ping-pong setup
to interchange between successive convolutional layers.
For each convolution engine,
the corresponding kernel weights are stored in ROM 
and are loaded according to the running filter.
The architecture operates in burst mode
with a convolution's cycle budget being almost equal to the size of the input channel.

\underline{Block Floating-Point}: 
In our floating-point convolution engine,
instead of processing each number as an 
individual exponent-mantissa pair,
we group an entire block of pixels by assuming a common exponent \cite{bfl}.
As a result,
the main operations reduce to fixed-point arithmetic,
i.e., simple mantissa multiplication/addition.
At the start/end of the block processing,
we transform all data (pixels and weights)
from/to the standard floating-point format
to/from block floating-point based on their maximum exponent.
For the transformation to block floating-point
(before starting the processing), 
the mantissas of pixels and weights are shifted according to their max exponent,
which is either detected or given as input.
The transformed mantissas are forwarded to the 
convolutional engine (see Figure \ref{fig_convo}). 
In parallel, 
the block's common exponents (for pixels and weights)
are added and shifted
to be included in the mantissa accumulation
that is performed in the convolutional engine. 
For the transformation to floating-point
(after finishing the processing),
the new max exponent is detected and is used to
normalize the mantissas.
We note that the internal multiplications can be performed by accurate or approximate multipliers,
and also,
another convolution engine can be used (e.g., the Winograd-based one). 

\underline{Winograd Convolution}:
The Winograd 2D convolution algorithm \cite{wino}
with an $r \times r$ kernel computes an $m \times m$ output using 
$(m+r-1) \times (m+r-1)$ image tiles with stride $r-1$.
It requires $(m+r-1) \times (m+r-1)$ 
multiplications to
compute the $m \times m$ output,
while the ordinary convolution 
needs $(m \times m) \cdot (r \times r)$.
The $\{$$m=2$, $r=3$$\}$ configuration is widely used, because it provides resource efficiency and numerical stability.
Compared to the ordinary convolution,  
this Winograd configuration requires $2.25\times$ less element-wise multiplications.
In brief, it 
applies the following steps \cite{wino}:
\begin{enumerate}[label=\roman*), noitemsep]
    \item split the input image into $4 \times 4$ tiles $d_i$ with stride $2$.
    \item transform the $3 \times 3$ kernel $g$ into the $4 \times 4$ kernel 
    $G = AgA^T$, where $A$ is a $4 \times 3$ matrix with elements $0$, $-1/2$, $+1/2$.
    \item transform the image tiles $d_i$ into $4 \times 4$ tiles 
    $D_i = B^T d_i B$, where $B$ is a $4 \times 4$ matrix with elements $0$, $-1$, $+1$.
    \item compute the intermediate outputs $F_i$ via the element-wise multiplication $F_i = D_i \odot G$.
    \item transform the intermediate outputs $F_i$ into $2 \times 2$ matrices $Y_i = C^ T F_i C$, 
    where $C$ is a $4 \times 2$ matrix with elements $0$, $-1$, $+1$. 
\end{enumerate}

\begin{figure}[!t]
\centering
\includegraphics[width=0.88\textwidth]{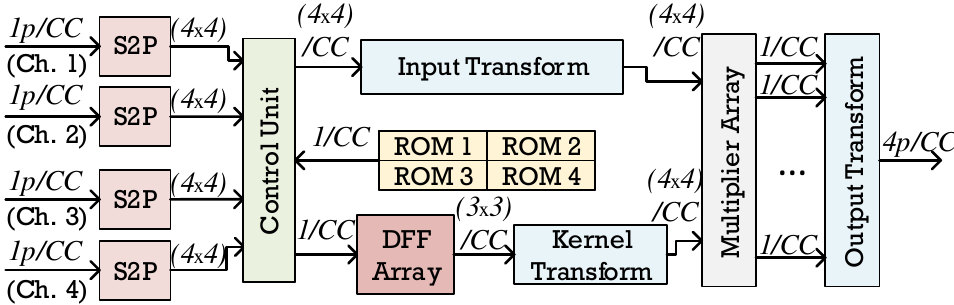}%
\caption[Hardware Architecture of Winograd Convolution]{Hardware architecture of Winograd convolution processing $4$ channels in parallel.}%
\label{fig_hwwino}
\end{figure}

Figure \ref{fig_hwwino} depicts the architecture of our Winograd convolution engine,
which replaces $4$ ordinary convolution engines.
This design exploits the gap created by the aforementioned stride-$2$ steps 
during continuous single-pixel raster-scan streaming,
and it employs one common core for processing $4$ distinct channels
via time multiplexing. 
In the input,
we use $4$ serial-to-parallel converters to
slide a distinct $4 \times 4$ window per channel.
Each $4 \times 4$ window is multiplexed towards the $3$ common processing units
to perform the $D_i$, $F_i$, $Y_i$ calculations
via pipelined multiplications/additions. 
At each clock cycle, 
the corresponding kernel 
is selected from an initialized DFF array and transformed to $G$.
We use fixed input delays and a control unit to synchronize the $4$ streams
and achieve a constant output rate of 
$4$ pixels per clock cycle. 
We note that
the arithmetic operations can be approximated
either using the block floating-point format or approximate arithmetic units.

\subsubsection{Experimental Evaluation}

We develop various CNN variants in parametric VHDL
based on the DSE of our methodology,
and we use 
Xilinx Vivado to implement them 
on the Zynq-7020 FPGA. 
Towards exploiting every specialized resource of the FPGA, 
we implement the entire network with a mixture of accurate and approximate convolution engines.
The key idea is, on one hand, to utilize the DSP blocks by doing `default' multiplications,
while on the other hand, to decrease the LUT utilization via approximations
(e.g., when bound to revert to LUT usage due to lack of DSPs). 
The `default':`approximate' engine ratio depends on the capacity of the FPGA.
For Zynq-7020,
we select the $16$:$16$ ratio, i.e., $16$ typical and $16$ approximate engines,  
for a total of $32\times$ parallel engines 
(which also leads to the full utilization of the available RAMBs).
For the approximations,
we employ our RAD multipliers \cite{LeonTVLSI}, 
as well as all the algorithms and arithmetic representations discussed in this section,
i.e., conventional fixed- and floating-point, block floating-point, and Winograd convolution. 

Table \ref{tb_hat_hw} reports
representative results from our DSE. 
The accurate floating-point CNN achieves a maximum
parallelization of only $8\times$ 
due to increased cost and constraints 
of Xilinx's IP for floating-point calculations,
which relies on DSPs.
By applying the schemes derived by our methodology,
we achieve $4\times$ higher throughput 
with a negligible accuracy loss of $0.1\%$.
More specifically, 
the block floating-point format
delivers $32\times$ parallelization, 
whether with Winograd or not,
delivering up to $730$ Frames Per Second (FPS).
In terms of resource utilization,
as expected,
the increased parallelization
requires more RAMBs,
however,
the use of the RAD256 multiplier
results in not over-utilizing the LUT resources.
Moreover, 
the use of the Winograd algorithm
instead of the ordinary convolution
in the RAD-based block floating-point design
reduces the DSP utilization by $41\%$,
while maintaining almost the same throughput and LUT utilization.
For the accurate fixed-point CNN,
the FPGA resources are automatically balanced, to some extent,
by Vivado. 
The gains provided by our approximate variants 
are 
summarized in the LUT resources (there is also a small increase in throughput).
In particular, 
the mixture of engines 
decreases the LUTs by $7\%$ when using only RAD256
and by $38\%$ when also using Winograd. 
Again, the accuracy is not affected
by either the fixed-point arithmetic or our RAD256 multiplier,
as there is only loss up to $0.2\%$.

\begin{table}[!t]
\fontsize{9}{10}\selectfont
\renewcommand{\arraystretch}{1.2}
\setlength{\tabcolsep}{4.5pt}
\caption[Experimental Results of Approximate RAD-Based Ship-Detection CNN on Zynq-7020 FPGA]{Experimental results of approximate RAD-based Ship-Detection CNN ($128$$\times$$128$$\times$$3$, $132$K param.) on Zynq-7020 FPGA.}
\label{tb_hat_hw}
\centering
\begin{threeparttable}
\begin{tabular}{l| c | c  c  c c | c}
\hline
\multicolumn{1}{c|}{\multirow{2}{*}{\textbf{Design}}} &
\multirow{2}{*}{\textbf{Paral.}} & 
\textbf{LUT} &
\textbf{DSP} &
\textbf{RAMB} &
\textbf{Clock} &
\textbf{Throughput} \\[-2pt] 
& & 
$(\%)$\footnotemark\setcounter{footnote}{0} &
$(\%)$\footnotemark\setcounter{footnote}{0} &
$(\%)$\footnotemark\setcounter{footnote}{0} &
(MHz) &
(FPS) \\
\hline \hline 
Accurate (Fl. Point)     &  $8$        & $37$ & $91$  & $78$ & $125$ & $182$ \\
RAD256 (BFl. Point)      &  $32$       & $65$ & $100$ & $95$ & $125$ & $730$ \\
RAD256 (Win. BFl. Point) &  $8\times4$ & $69$ & $59$  & $95$ & $124$ & $724$ \\
Accurate (Fx. Point)     &  $32$       & $69$ & $58$  & $95$ & $118$ & $689$ \\
RAD256 (Fx. Point)       &  $32$       & $60$ & $54$  & $95$ & $124$ & $724$ \\
RAD256 (Win. Fx. Point)  &  $8\times4$ & $43$ & $58$  & $95$ & $112$ & $654$ \\
\hline
\end{tabular}
\begin{tablenotes}
  \item[1]{\fontsize{7.7}{8.8}\selectfont Refers to $\%$ resource utilization of Zynq-7020 ($53200$ LUTs, $220$ DSPs, $140$ RAMBs).}
  \item[*]{\fontsize{7.7}{8.8}\selectfont The accuracy loss is $0.1\%$--$0.2\%$ among the designs.}  
\end{tablenotes}
\end{threeparttable}
\end{table}

\section{Design and Evaluation of Applications with AxFXU/AxFPU}
\label{s7_4}

In this section,
we evaluate our approximate AxFXU \cite{LeonMicro} and AxFPU \cite{LeonTECS} multipliers
in real-world applications.
We remind that AxFXU is
the fixed-point variant
and AxFPU is
the floating-point variant,
as well as that both variants
combine the perforation and rounding approximation techniques. 
To evaluate AxFXU,
we employ again our hardware architectures for 
the Sobel edge detector,
the FIR filter,
and
matrix multiplication.
Additionally,
we implement a Gaussian blurring filter 
on floating-point arithmetic
to evaluate AxFPU.
From the AI domain,
we develop two floating-point CNNs
on the MNIST and CIFAR-10 datasets.
Finally,
we evaluate the AxFPU design in 
two functions that employ floating-point numbers,
i.e., the K-means clustering and the LU decomposition.

\subsection{Approximate DSP Accelerators}

Starting with AxFXU,
we employ again the fixed-point hardware architectures
for Sobel, FIR, and matrix multiplication. 
The Sobel edge detector uses two $3 \times 3$ convolution kernels,
the FIR filter is low-pass with cut-off frequency of $20$KHz and $16$-bit $32$ coefficients,
while for matrix multiplication
we implement $3 \times 3$ tiling. 

Regarding AxFPU,
we select the Gaussian blurring filter, 
which smooths the image to remove details and noise.
This convolution filter is widely used in image processing,
e.g., before edge detection
to reduce the levels of noise in the image.
We implement the $3 \times 3$ Gaussian blurring
with our convolution engine illustrated in 
Figure \ref{fig_convo}.
We note that
hardware filters such as Gaussian blurring, 
do not require floating-point arithmetic to provide acceptable accuracy results. 
However, 
considering that the optimization of the arithmetic/bit-width
in DSP applications is not a negligible task,
and it is mainly performed by low-level hardware engineers,
we evaluate the scenario where the default floating-point arithmetic is selected.
Moreover, considering that our AxFPU multipliers can be
integrated in an embedded processor,
our evaluation aims to prove their power efficiency
in case the embedded developer does not use integer coefficients for Gaussian blurring.

\subsubsection{Experimental Evaluation}

We develop the DSP kernels  
in Verilog
and 
synthesize them with Synopsys Design Compiler
and the TSMC 65-nm standard-cell library,
targeting to ASIC-based accelerators. 
Like in the evaluation of RAD,
for Sobel 
we use the CER error metric,
which measures the number of correct edges detected per total number of edges,
while we use MRED
for FIR and the matrix multiplication.
For the Gaussian blurring, we use two well-established image processing metrics, i.e., PSNR and SSIM.
As benchmark images,
we use Cameraman for Sobel
and both Cameraman and Lena for Gaussian blurring.
Moreover,
the FIR and matrix multiplication kernels 
take $200$K random generated inputs. 

\begin{figure}
\centering
\subfloat[\label{fig_sobaxfu1}]{\includegraphics[width=0.5\textwidth]{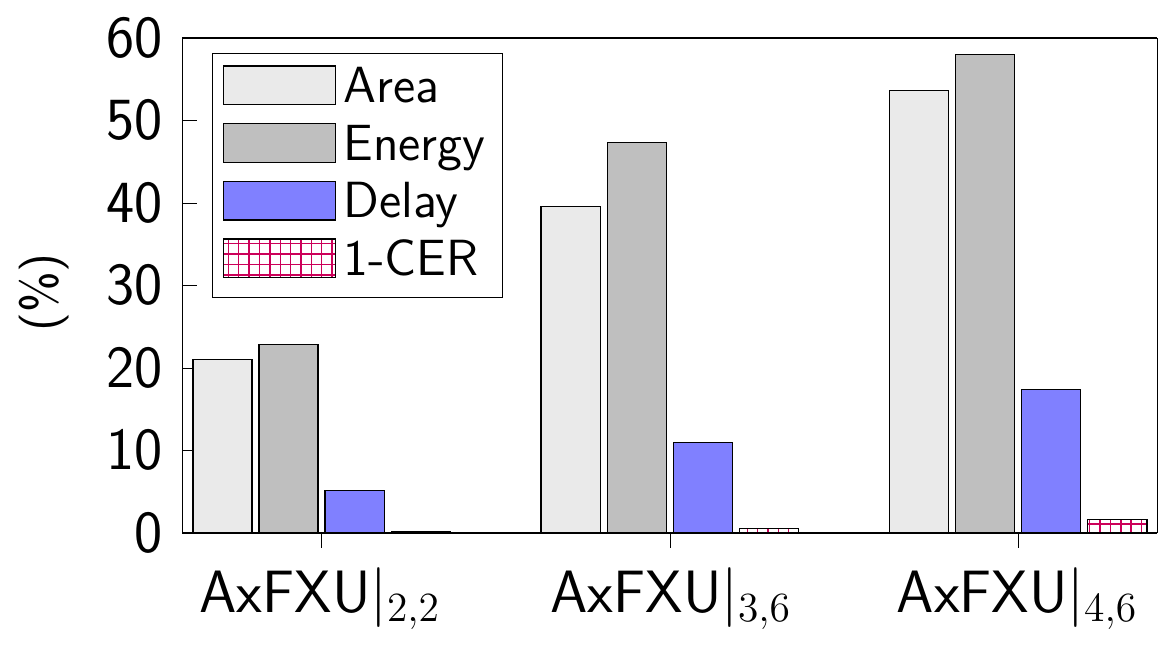}}\\[-10pt]
\subfloat[\label{fig_sobaxfu2}]{\includegraphics[width=0.5\textwidth]{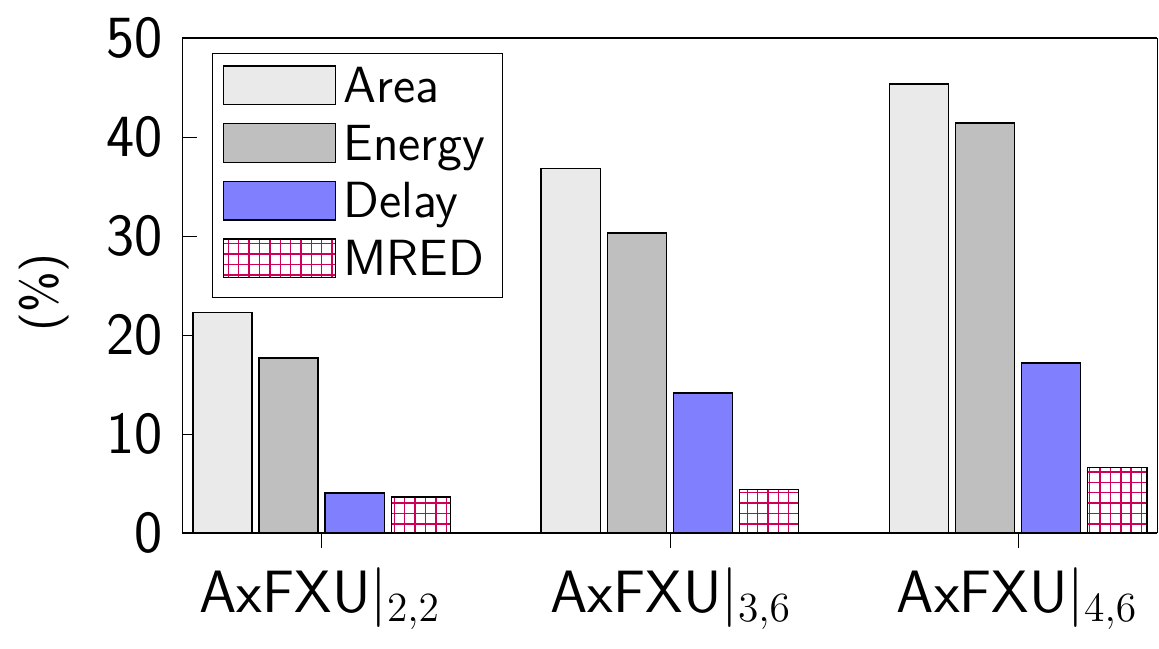}} \hfill
\subfloat[\label{fig_sobaxfu3}]{\includegraphics[width=0.5\textwidth]{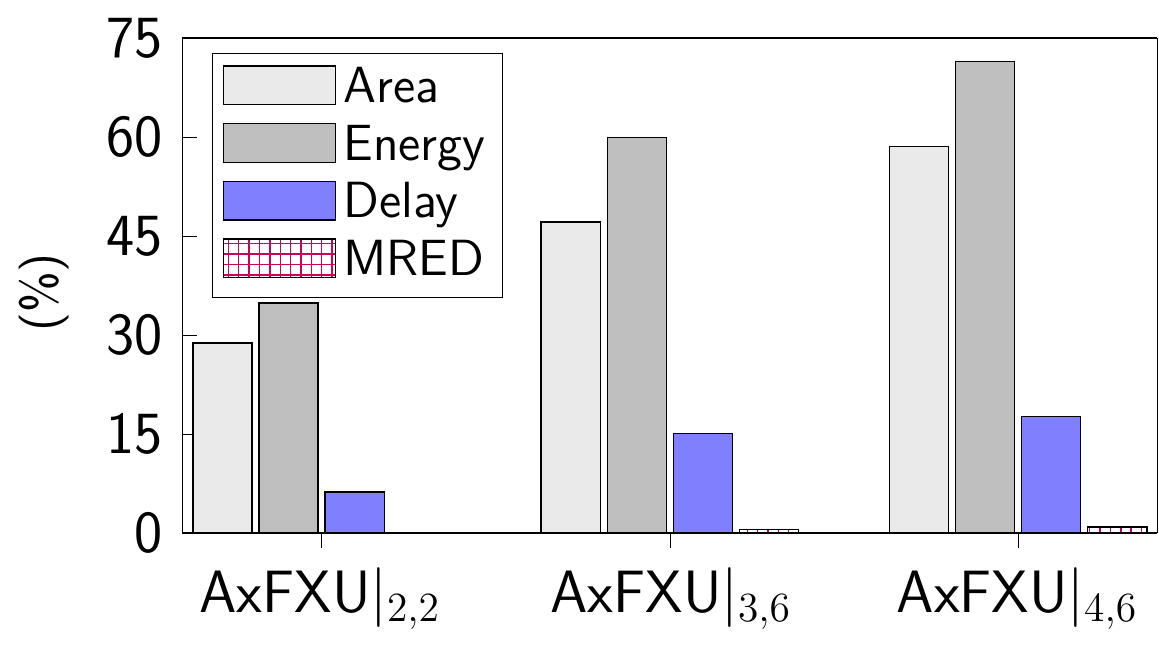}}%
\caption[Experimental Results of Approximate AxFXU-Based DSP Applications on TSMC 65-nm Standard-Cell]{Experimental results of approximate AxFXU-based DSP applications on TSMC 65-nm standard-cell:
resource gains and accuracy results of
\textbf{(a)} Sobel, 
\textbf{(b)} FIR, 
and
\textbf{(c)} MatMul.}%
\label{fig_sobaxfu}
\end{figure}

For evaluating AxFXU,
we employ three configurations with different approximation degree (AxFXU$|_{2,2}$, AxFXU$|_{3,6}$, and AxFXU$|_{4,6}$).
Figure \ref{fig_sobaxfu} shows
the delivered reductions in resources along with the application's accuracy metric.
Regarding Sobel,
as shown in \ref{fig_sobaxfu1},
CER is more than $98.5\%$ in all configurations,
like in the RAD-based Sobel accelerators (see Table \ref{tb_sob_hw}).
However, 
the use of AxFXU provides
increased resource gains, 
i.e., up to 
$54\%$ in area, $58\%$ in energy, and $18\%$ in delay.
Especially for delay,
AxFXU provides $2.3\times$ smaller critical path versus RAD.
Similar behavior is observed for the other two applications.
The important outcome from this exploration is that, 
again, 
the use of our approximate multipliers 
does not damage the accuracy of the application.
Only FIR outputs larger error, 
however, 
this is normal,
as every output depends on the previous ones, 
and the error is propagated to the next calculations. 
Specifically, 
the MRED is $3.66\%$--$6.65\%$,
but 
it can be handled using less aggressive approximation configurations.

\begin{table}[!t]
\fontsize{9}{10}\selectfont
\renewcommand{\arraystretch}{1.2}
\setlength{\tabcolsep}{9pt}
\caption[Experimental Results of Approximate AxFPU-Based Gaussian Blurring on TSMC 65-nm Standard-Cell]{Experimental results of approximate AxFPU-based Gaussian Blurring on TSMC 65-nm standard-cell.}
\label{tb_blur_hw}
\centering
\begin{threeparttable}
\begin{tabular}{l| c c | c c c c}
\hline
\multicolumn{1}{c|}{\multirow{2}{*}{\textbf{Design}}} &
\textbf{Area} & 
\textbf{Power} &
\setcounter{footnote}{1}\textbf{PSNR}\footnotemark\setcounter{footnote}{1}  &
\textbf{SSIM}\footnotemark\setcounter{footnote}{2}  &
\textbf{PSNR}\footnotemark\setcounter{footnote}{2}  &
\textbf{SSIM}\footnotemark\setcounter{footnote}{0} 
\\[-2pt] 
& $(\%)$\footnotemark\setcounter{footnote}{0} 
& $(\%)$\footnotemark\setcounter{footnote}{0} 
& (dB)
&  -- 
& (dB)
&  --\\
\hline
\hline
AxFPU16$|_{1,0}$ & $5.4$  & $2.6$  & $\infty$ & $1$    & $\infty$ & $1$    \\
AxFPU16$|_{3,4}$ & $28.4$ & $22.9$ & $59.22$  & $0.99$ & $55.17$  & $0.99$ \\
AxFPU16$|_{4,6}$ & $41.9$ & $40.3$ & $50.86$  & $0.98$ & $48.20$  & $0.87$ \\
\hline
AxFPU32$|_{4,12}$  & $34.5$  & $27.9$ & $\infty$ & $1$    & $\infty$ & $1$    \\
AxFPU32$|_{6,14}$  & $48.7$  & $44.8$ & $\infty$ & $1$    & $\infty$ & $1$    \\
AxFPU32$|_{10,18}$ & $60.4$  & $57.4$ & $54.46$  & $0.99$ & $52.99$  & $0.95$ \\
\hline
\end{tabular}
\begin{tablenotes}
  \item[1]{\fontsize{7.7}{8.8}\selectfont Refers to $\%$ resource gains (relative reduction) in comparison with the accurate design.}
  \item[2,3]{\fontsize{7.7}{8.8}\selectfont Refers to Lena and Cameraman benchmark images, respectively.}  
\end{tablenotes}
\end{threeparttable}
\end{table}

\begin{figure}[!t]
\centering
\subfloat[\label{fig_len0}]{\includegraphics[width=0.175\textwidth]{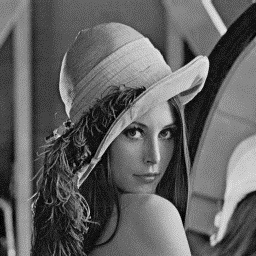}}
\hspace{1pt}
\subfloat[\label{fig_len1}]{\includegraphics[width=0.175\textwidth]{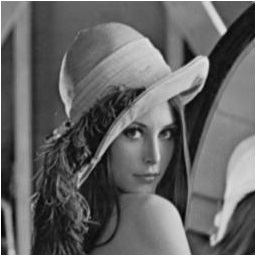}}
\hspace{1pt}
\subfloat[\label{fig_len2}]{\includegraphics[width=0.175\textwidth]{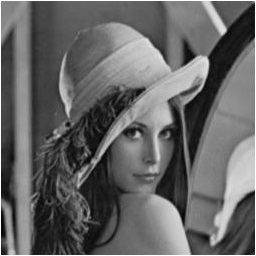}}
\hspace{1pt}
\subfloat[\label{fig_len3}]{\includegraphics[width=0.175\textwidth]{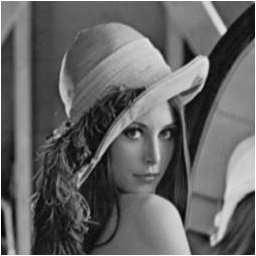}}
\hspace{1pt}
\subfloat[\label{fig_len4}]{\includegraphics[width=0.175\textwidth]{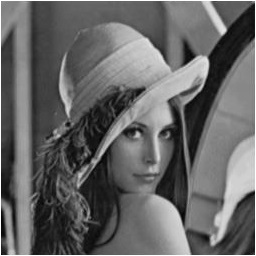}}%
\\
\subfloat[\label{fig_len5}]{\includegraphics[width=0.175\textwidth]{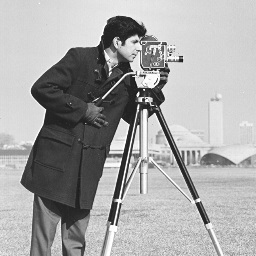}}
\hspace{1pt}
\subfloat[\label{fig_len6}]{\includegraphics[width=0.175\textwidth]{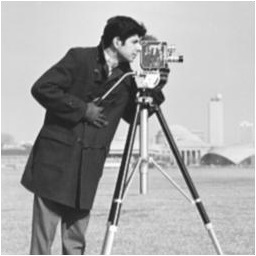}}
\hspace{1pt}
\subfloat[\label{fig_len7}]{\includegraphics[width=0.175\textwidth]{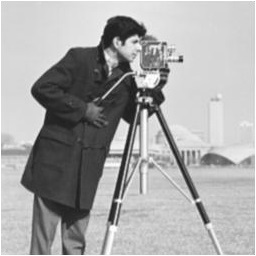}}
\hspace{1pt}
\subfloat[\label{fig_len8}]{\includegraphics[width=0.175\textwidth]{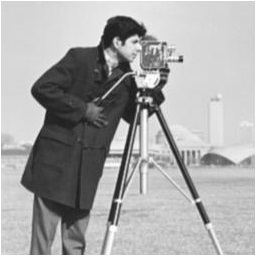}}
\hspace{1pt}
\subfloat[\label{fig_len9}]{\includegraphics[width=0.175\textwidth]{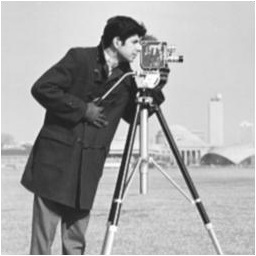}}%
\caption[I/O Images of Approximate Gaussian Blurring]
{I/O images of approximate Gaussian blurring:
\textbf{(a)},\textbf{(f)} input images,
and output image with
\textbf{(b)},\textbf{(g)} accurate multiplier, 
\textbf{(c)} AxFPU16${|_{1,0}}$,
\textbf{(d)} AxFPU16${|_{3,4}}$,
\textbf{(e)} AxFPU16${|_{4,6}}$,
\textbf{(h)} AxFPU32${|_{4,12}}$,
\textbf{(i)} AxFPU32${|_{6,14}}$,
\textbf{(j)} AxFPU32${|_{10,18}}$.}
\label{fig_lenes}
\end{figure}

Regarding the AxFPU-based Gaussian blurring,
we employ Pareto-front configurations 
with diverse approximation degree and error sensitivity.
Our goal is to assess the smoothing's accuracy over different approximation levels.
Table \ref{tb_blur_hw} 
reports the resource gains
and the accuracy metrics for the benchmark images.
The results show that the quality of the calculations 
is barely affected by the approximations, which, at the same time, deliver significant resource gains in the convolution engine.
More specifically, the designs exhibit more than $50$dB PSNR on average, i.e., acceptable PSNR values for lossy image processing, and near $1$ SSIM values, while the less approximation-aggressive designs produce the ``golden'' output.
For instance, AxFPU32, which has already proven to be less sensitive to approximations (see Chapter \ref{chapter5}), produces the ``golden'' output for both images, while delivering power gains up to $45\%$.
The gains reach $58\%$ in exchange for a PSNR of $53.7$dB and a SSIM of $0.97$ on average. 
Finally, Figure \ref{fig_lenes} presents a visual inspection of the images produced using the approximate AxFPU multipliers, indicating a very small difference compared to the image produced with accurate computations.

\subsection{Approximate CNN Accelerators}
\label{s742}

Next, we assess the resource efficiency and accuracy of CNNs
that
replace the multiplications in their the convolutional layers with AxFPU multipliers. 
The first CNN, which is trained and evaluated on the MNIST dataset \cite{mnist}, consists of one convolutional layer with $12$ filters,
followed by another one with $24$ filters. 
The second CNN, which is trained and evaluated on the CIFAR-10 dataset \cite{cifar}, consists of two convolutional layers with $32$ filters and two convolutional layers with $64$ filters. 
Both CNNs use ReLU as activation function 
and rely on two fully-connected layers to generate the $10$-value output vectors. 
The CNN models  
are generated with TensorFlow in two flavors, i.e., with $16$- and $32$-bit floating-point arithmetic, in order to evaluate our half- and single-precision AxFPU multipliers.
The training is performed on
$80\%$ of the datasets, generating $32$-bit floating-point weights and biases.
The $16$-bit floating-point models are created by applying post-training quantization on the $32$-bit models.
Finally, 
we remind that  
we do not aim to generate the most efficient CNN networks 
(e.g., for the selected models/datasets, 
fixed-point or quantized integer arithmetic may be sufficient).
Our goal is to evaluate our approximations 
in floating-point CNN workloads, 
provide hardware efficiency without structural modifications, 
and quantify the actual resource gains. 

\subsubsection{Experimental Evaluation}

To estimate the power consumption 
of the approximate CNN accelerators,
we adopt the model used in  
\cite{alwann}.
Namely,
we calculate
the power of each layer based on 
the number of its multiplications 
and the power of the multiplication circuits,
which is experimentally measured.
To estimate the total power consumption,
we accumulate the individual powers of all the layers.
Moreover, 
we perform similar calculations
to estimate the area of the CNN accelerators.
Targeting to standard-cell ASIC technology,
the AxFPU multipliers are implemented 
on the TSMC 65-nm library
using industrial-strength tools, 
i.e.,
Synopsys Design Compiler for synthesis
and
Synopsys PrimeTime for measuring power.

Table \ref{tb_cnn_tecs} reports the results for the two CNNs using the AxFPU designs.
In terms of accuracy, the small configurations of AxFPU do not affect at all the classification results, especially for the MNIST dataset,
while for CIFAR-10,
the accuracy loss starts growing faster.
For more aggressive approximations,
i.e., 
$P=4$, $R=6$ in AxFPU16
and
$P=11$, $R=19$ in AxFPU32,
the accuracy loss for MNIST is $0\%$, 
while it is small for CIFAR-10 
(in the range $1.3\%$--$5.4\%$).
This accuracy loss is traded for significant gains in area, 
ranging in $33\%$--$67.1\%$ and $30.9\%$--$63.4\%$
for $16$- and $32$-bit floating-point, respectively, 
and similar gains in power (up to \raisebox{0.8pt}{$\scriptstyle\sim$}$64\%$).
Finally, 
according to our exploration,
more aggressive approximation configurations are not recommended,
as the accuracy loss explodes,
while the extra gains are negligible. 
 
\begin{table}[!t]
\vspace*{8pt}
\fontsize{9}{10}\selectfont
\renewcommand{\arraystretch}{1.2}
\setlength{\tabcolsep}{6.5pt}
\caption[Experimental Results of Approximate AxFPU-Based CNNs on TSMC 65-nm Standard-Cell]{Experimental results of approximate AxFPU-based CNNs on TSMC 65-nm standard-cell.}
\label{tb_cnn_tecs}
\centering
\begin{threeparttable}
\begin{tabular}{l| c c c | c c c}
\hline
\multicolumn{1}{c|}{\multirow{3}{*}{\textbf{Design}}} & \multicolumn{3}{c|}{\textbf{2$\mathbf{\times}$ ConvLayers on MNIST}} & \multicolumn{3}{c}{\textbf{4$\mathbf{\times}$ ConvLayers  on CIFAR-10}}
\\
& \textbf{Area} & 
\textbf{Power} &
\textbf{Accuracy} &
\textbf{Area} & 
\textbf{Power} &
\textbf{Accuracy} 
\\[-2pt] 
& $(\%)$\footnotemark\setcounter{footnote}{0} 
& $(\%)$\footnotemark\setcounter{footnote}{1} 
& $(\%)$\footnotemark\setcounter{footnote}{0} 
& $(\%)$\footnotemark\setcounter{footnote}{0} 
& $(\%)$\footnotemark\setcounter{footnote}{1} 
& $(\%)$\footnotemark\setcounter{footnote}{0} \\
\hline
\hline
AxFPU16$|_{1,0}$    & $5.8$      & $2.8$ & $0$    & $5.4$   & $2.5$ & $0$\\
AxFPU16$|_{3,4}$   & $30.3$      & $24.3$ & $0$    & $28.4$   & $22.1$  & $0$\\
AxFPU16$|_{3,8}$   & $33.0$      & $26.1$ & $0$    & $30.9$   & $23.5$  & $1.8$\\
AxFPU16$|_{4,6}$   & $43.7$  & $42.8$ & $0$    & $40.9$   & $38.8$  & $2.5$\\
AxFPU16$|_{5,4}$   & $53.5$      & $49.9$ & $2.7$  & $50.2$   & $45.2$   & $19.8$\\
\hline
AxFPU32$|_{4,12}$  & $36.8$      & $29.8$ & $0$    & $34.5$      & $26.9$ & $0$\\
AxFPU32$|_{8,16}$  & $59.3$      & $54.7$ & $0$    & $55.6$      & $49.6$ & $0$\\
AxFPU32$|_{10,18}$ & $64.4$      & $61.2$ & $0$    & $60.4$      & $55.5$  & $1.3$\\
AxFPU32$|_{11,19}$ & $67.1$      & $63.8$ & $0$    & $63.4$      & $57.6$  & $5.4$\\
AxFPU32$|_{12,4}$  & $69.6$      & $65.5$ & $5.5$  & $65.1$      & $61.2$  & $12.5$\\
\hline
\end{tabular}
\begin{tablenotes}
  \item[1]{\fontsize{7.7}{8.8}\selectfont Refers to $\%$ resource gains (relative reduction) in comparison with the accurate design.}
  \item[2]{\fontsize{7.7}{8.8}\selectfont Refers to $\%$ accuracy loss.}  
\end{tablenotes}
\end{threeparttable}
\end{table}

\subsection{Approximate Clustering and Linear Algebra}
  
In this section, 
we evaluate the use of AxFPU in Rodinia 3.1 benchmarks \cite{rodinia}.
Targeting to explore different application domains than the classic DSP/AI ones,
we employ two benchmarks from machine learning and linear algebra,
i.e., domains that require floating-point arithmetic and high precision, 
and we examine the accuracy results when using AxFPU.
The first benchmark is K-means clustering \cite{kmeans}, which is an unsupervised learning algorithm used extensively in the data mining domain to classify unlabeled data into clusters.
Moreover,
we use AxFPU in the LU decomposition \cite{strang}, which is a typical task of the numerical analysis and linear algebra domains.
For these two benchmarks, 
we do not provide hardware implementations and  
results. 
Nevertheless, 
the accuracy evaluation is extremely important,
considering that such benchmarks are usually implemented in software
and they can be executed in processors integrating AxFPU
in their floating-point units. 
For our experiments, 
we develop a generic C/C++ function for AxFPU that 
takes as input the approximation configuration parameters.
This function is called in the Rodinia programs
instead of the default accurate floating-point multiplication.

\begin{figure}[!t]
\vspace*{-5pt}
\centering
\subfloat[\label{fig_lid1}]{\includegraphics[width=0.74\textwidth]{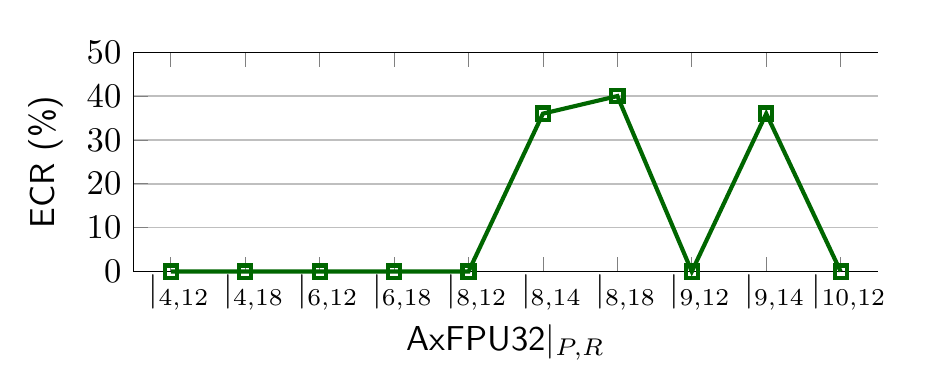}}\\[-5pt]
\subfloat[\label{fig_lid2}]{\includegraphics[width=0.74\textwidth]{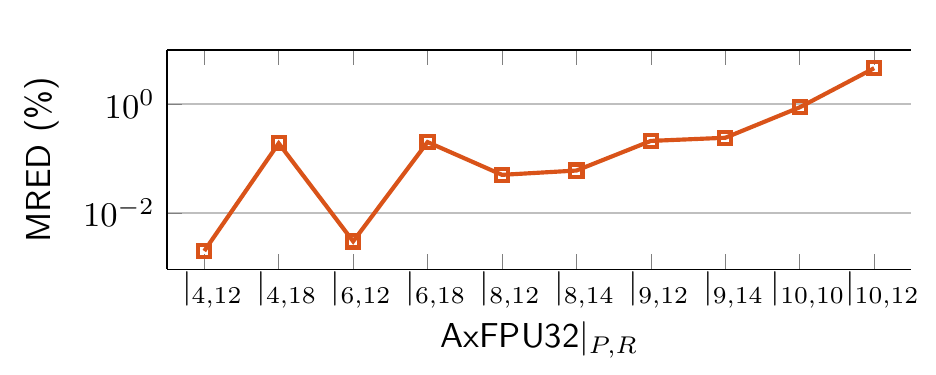}}%
\caption[Accuracy Variation of Approximate AxFPU-Based K-Means Clustering and LU Decomposition]{Accuracy variation of approximate AxFPU-based
\textbf{(a)} K-means clustering
and 
\textbf{(b)} LU decomposition.}%
\label{fig_lud}
\end{figure}

For K-means clustering, 
we consider the accurate $32$-bit floating-point model as baseline,
and examine the Erroneous Classification Ratio (ECR),
i.e., 
the number of wrong classifications per the total number of classifications.
As shown in Figure \ref{fig_lid1}, 
AxFPU achieves the classification results of the ``golden'' model,
even when applying approximations that deliver significant resource gains 
(e.g., AxFPU32$|_{8,12}$ provides $68\%$ area gains versus the accurate multiplier). 
According to the derived results, 
for large perforation values, i.e., $P \geq 8$, 
the impact of rounding increases for $R \geq 14$,
which results in very high ECR. 
Nevertheless, the resource gains and the classification accuracy of the ``golden'' model,
which are provided by either AxFPU32$|_{8,12}$ or AxFPU32$|_{9,12}$, 
establish AxFPU as an efficient solution for such floating-point calculations.
For the LU decomposition, 
we consider MRED as accuracy metric
and the accurate $32$-bit floating-point model as baseline.
Figure \ref{fig_lid2} presents the experimental results for a large matrix size ($2048\times2048$).
Again, the proposed AxFPU achieves good accuracy results, as MRED is smaller than $0.1\%$ for several configurations that deliver significant hardware gains.

\section{Design and Evaluation of Applications with ROUP}
\label{s7_5}

In this section,
we employ our state-of-the-art ROUP multipliers \cite{LeonTVLSI}
in CNNs.
In particular,
we examine the interplay of fine-grained error resilience of CNNs 
in collaboration with arithmetic approximation,
targeting to achieve
higher energy efficiency. 
The large approximation space offered by the ROUP multipliers
allows us to 
systematically explore their fine-grained distribution across the network according to different approaches.
For evaluating our approximations, 
we employ the ResNet-8 CNN on the CIFAR-10 dataset.

\subsection{Fine-Grained Approximate CNN Accelerators}
The approximate CNNs 
presented in Sections \ref{s7_3}-\ref{s7_4}
execute all their multiplications
with the same approximate multiplier.
Namely, each approximate CNN variant employs a single multiplier. 
In contrast, the current section
examines 
the use of different approximate multipliers
within the same network.
More explicitly, 
the approximate multipliers are distributed across the network
based on three approaches corresponding to the 
abstraction levels of the CNN architecture 
(convolutional layers $\rightarrow$ filters $\rightarrow$ convolution engines). 
Moreover,
compared to our previous work on approximate CNN kernels, 
which are built on  
$16$/$32$-bit fixed- and floating-point arithmetic, 
we adopt quantized network models
($8$-bit unsigned integer arithmetic).

To perform an extensive DSE 
regarding the distribution of approximate multipliers within the CNN,
we employ 
ALWANN \cite{alwann},
which is a state-of-the-art framework 
for generating approximate CNN hardware accelerators without retraining.
Towards larger design/approximation space,
we extend it with our approximation approaches
and the ROUP multiplication library.
The proposed framework is called \emph{MAx-DNN}
and 
assigns different ROUP multipliers
either
in each convolutional layer, 
filter, 
or convolution engine (kernel). 
On the other hand, 
ALWANN inserts the same multipliers 
in each convolutional layer 
(multipliers may differ among layers).
In the next paragraphs,
we make a brief introduction to ALWANN,
and then we introduce our approximation approaches.

\underline{ALWANN \cite{alwann}}:
The framework takes as input 
the trained (frozen) network model in protobuf format,
a library of approximate multipliers,
and
architecture constraints for the hardware accelerator (e.g., pipelined or power-gated mode, number of approximate units).
It implements  
accurate addition and approximate multiplication, 
as well as one approximation type per convolutional layer.
Moreover,
to improve the accuracy without re-training, 
the network weights are tuned/updated
according to the multipliers' properties. 
The approximate networks,
labeled as AxNNs,
are modified versions of the initial model and
satisfy the user constraints for the architecture of the accelerator.

To enable the ALWANN functionalities,
the TensorFlow framework is extended 
to support approximate quantized layers.
In particular,
a new operator is created 
that replaces 
the default \texttt{QuantizedConv2D} layers 
with \texttt{AxConv2D} layers.
This operator allows 
to specify which approximate multipliers to employ via the  \texttt{AxMult(str)} parameter
(a C/C++ model of the multipliers is necessary),
and optionally, to use the weight tuning feature via \texttt{AxTune(bool)}.
The frozen model is processed by the TensorFlow's tool for graph transform,
which inserts the \texttt{AxConv2D} layers,
and then, 
the Pareto-optimal AxNNs are extracted by the NSGA-II algorithm.

\begin{figure}[!t]
\vspace*{12pt}
\centering
\includegraphics[width=0.885\textwidth]{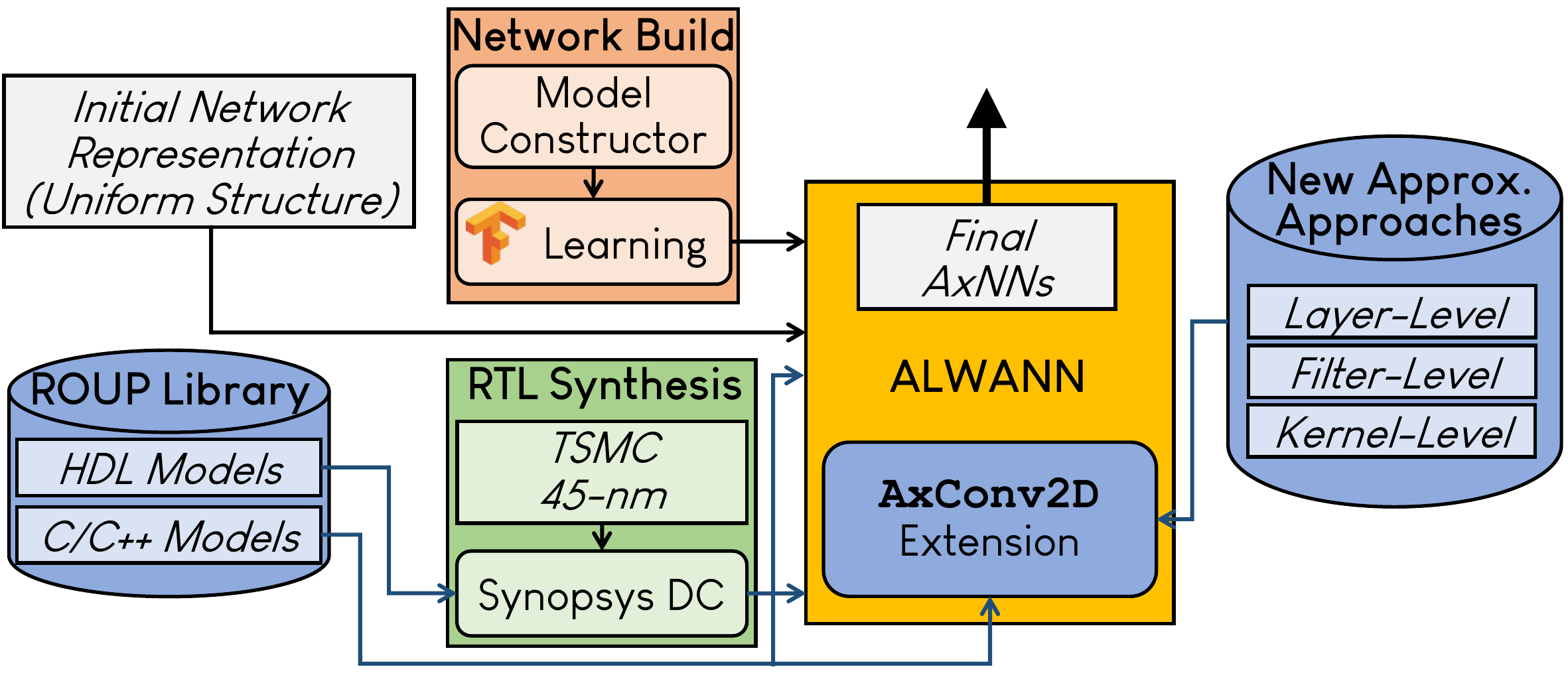}%
\caption[The MAx-DNN Framework for Fine-Grained CNN Approximation]{The  toolflow and architecture of the MAx-DNN framework (extension of ALWANN \cite{alwann}) that applies fine-grained CNN approximation. Max-DNN distributes the state-of-the-art ROUP multipliers across the network based on differing approximation approaches.}%
\label{fig_maxdnn}
\end{figure}

In our work,
we use ALWANN's infrastructure  
and introduce 
new approximation approaches
regarding the distribution of the approximate multiplications across the network.
The toolflow and architecture of MAx-DNN
is illustrated in Figure \ref{fig_maxdnn}.
Compared to ALWANN, 
we modify the \texttt{AxConv2D} TensorFlow operator to support our approximation approaches at different CNN abstraction levels, i.e., 
layer, filter, convolution engine, 
and we also employ the state-of-the-art ROUP library of approximate multipliers. 
Below,
we discuss our approximation approaches, targeting to explore the approximation space of the CNNs 
and identify the best approximation opportunities within the network. 

\begin{figure}[!t]
\centering
\subfloat[\label{fig_llam}]{\includegraphics[width=0.45\textwidth]{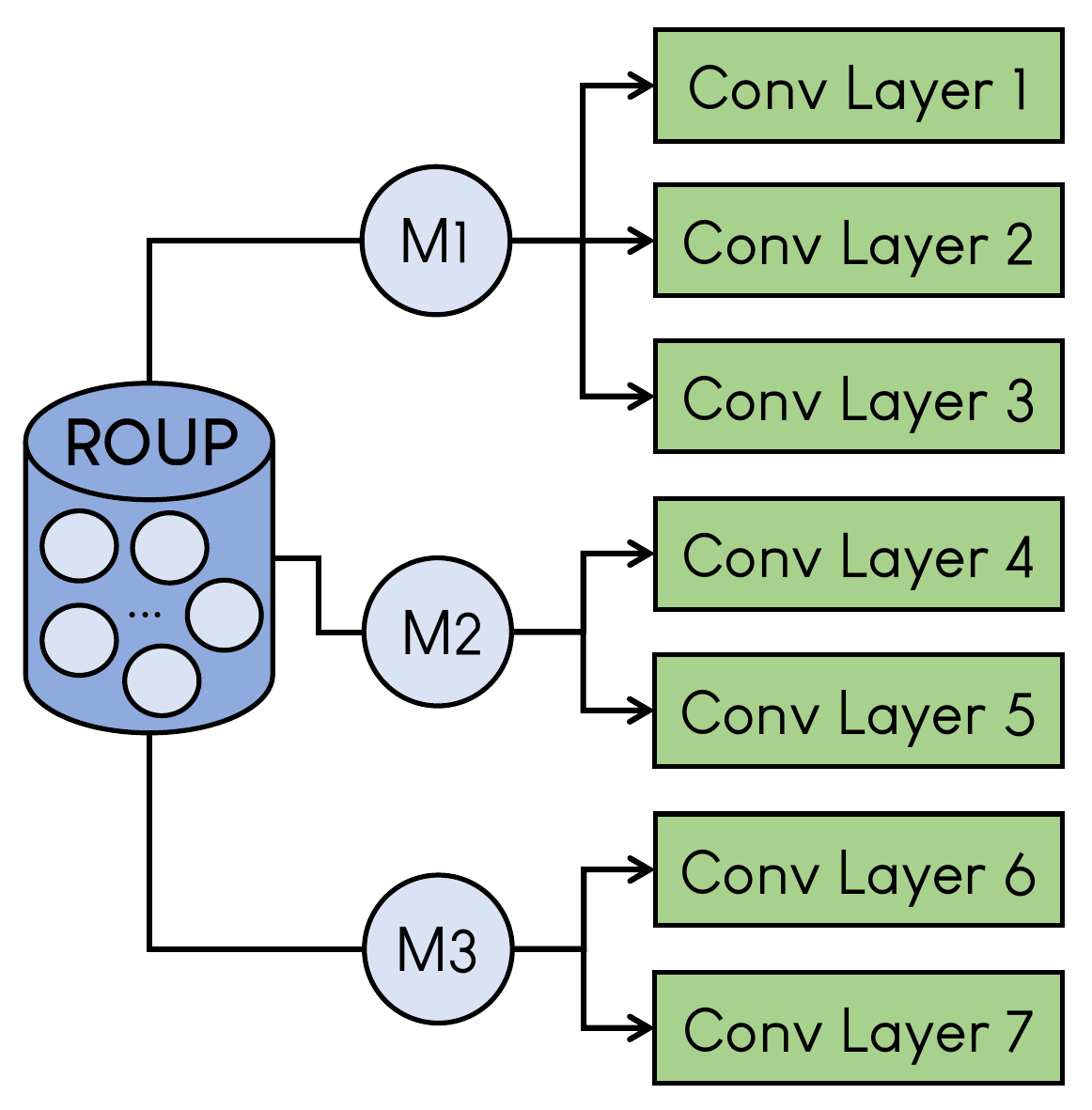}} \\ 
\subfloat[\label{fig_flam}]{\includegraphics[width=0.45\textwidth]{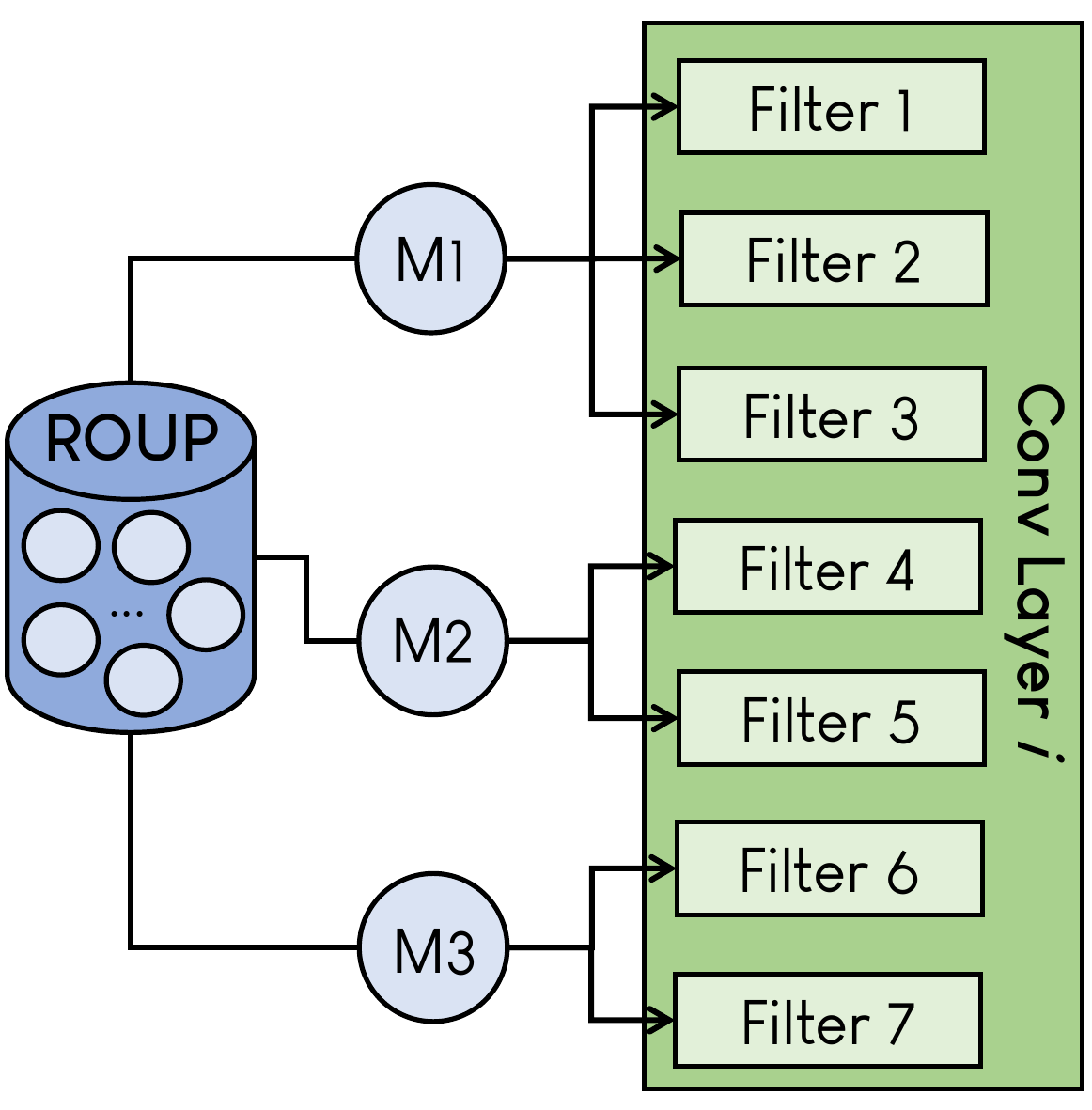}} \hspace{24pt}
\subfloat[\label{fig_klam}]{\includegraphics[width=0.45\textwidth]{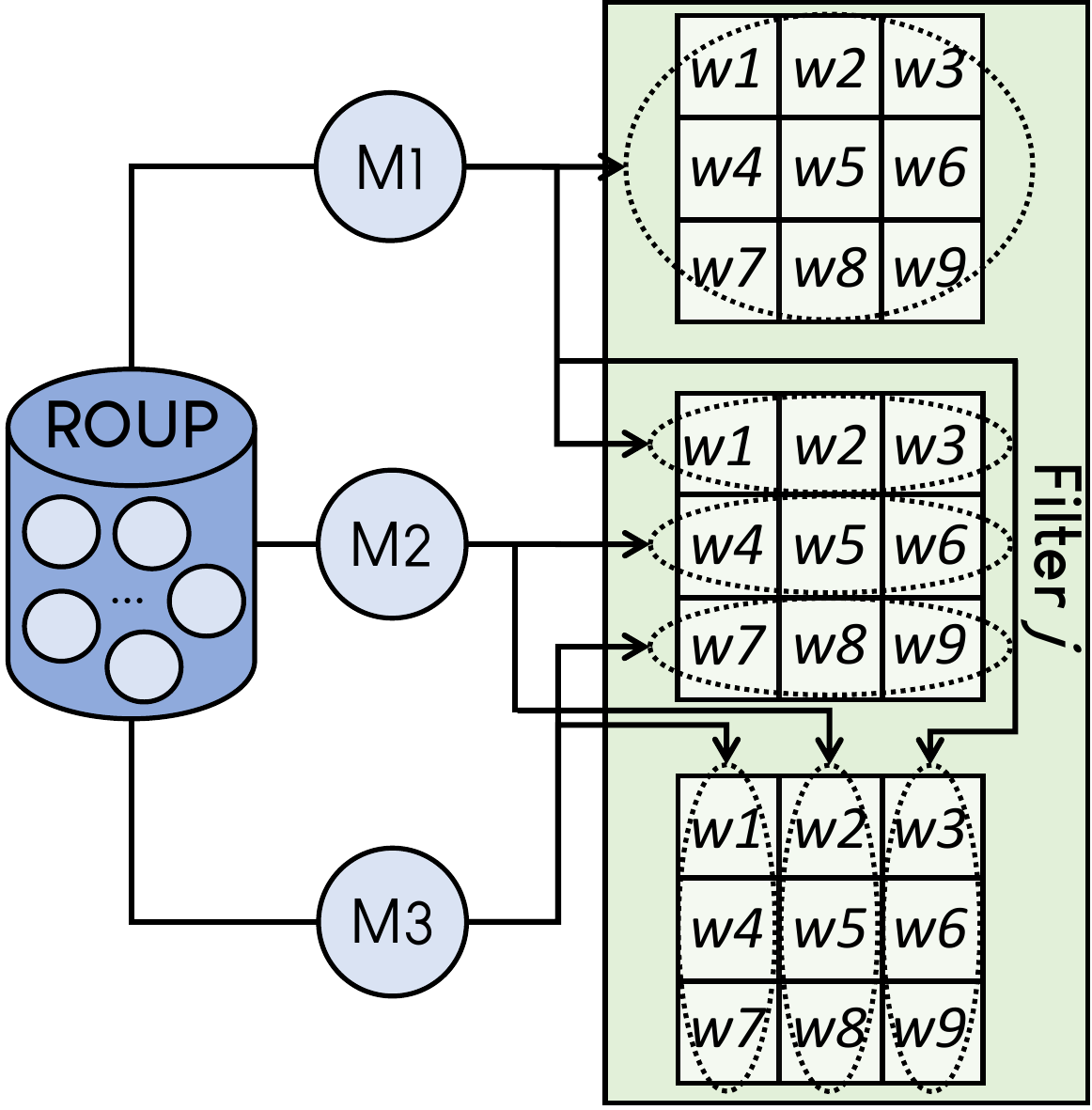}}%
\caption[Fine-Grained Non-Uniform CNN Approximation Approaches]{Fine-grained non-uniform CNN approximation of the MAx-DNN framework at
\textbf{(a)} layer level, 
\textbf{(b)} filter level, 
and
\textbf{(c)} kernel level.}%
\label{fig_maxappro}
\end{figure}

\underline{Layer-Level  Approximate Multiplication}: 
The first approach,
illustrated in Figure \ref{fig_llam}, 
aims to create approximate convolutional layers,
in which all the multiplication operations
are executed with the same approximate multiplier.
This approach is the straightforward one,
where the approximations are distributed across the network
by assigning an approximate multiplier,
either the same (uniform distribution)
or different (non-uniform distribution),
to each convolutional layer.
In MAx-DNN, 
we create various layer-level approximate variants
by using different ROUP configurations
among the convolutional layers.

\underline{Filter-Level Approximate Multiplication}: 
Our second approximation approach,
illustrated in Figure \ref{fig_flam}, 
creates different approximate filters
within each convolutional layer of the CNN.
Namely,
we use different ROUP multipliers
in each filter of the layer,
contrary to the first approach,
where all the filters of a layer have the same ROUP multiplier.
MAx-DNN implements this approach 
by creating groups of filters 
and assigning them ROUP multipliers. 

\underline{Kernel-Level Approximate Multiplication}: 
In the third approximation approach,
we proceed deeper in the CNN architecture 
and examine separately the multiplications
of each convolution engine. 
As a result, 
the convolutions of each filter  
are performed with different ROUP multipliers.
Figure \ref{fig_klam} illustrates the three proposed flavors of this approach:
the channel flavor,
where all
the multiplications of the convolution kernel are performed with the same multiplier,
and the 
row/column flavor,
where 
different multipliers are employed for each kernel's row/column.


\subsubsection{Experimental Evaluation}

For evaluating the ROUP-based MAx-DNN framework,
we employ the ResNet-8 CNN \cite{resnet} and the open-source ALWANN framework \cite{alwann}. 
ResNet-8 is trained with quantization on the CIFAR-10 dataset \cite{cifar}, 
achieving $83\%$ classification accuracy.
The energy consumption of the 
approximate CNN accelerators
is estimated 
with the model used in Section \ref{s742},
which is also used in ALWANN.
According to this model,
the energy consumption is estimated 
based on the energy of the multipliers.
Targeting to standard-cell ASIC technology,
we implement the ROUP 
multipliers 
on the TSMC 45-nm library
using industrial-strength tools, 
i.e.,
Synopsys Design Compiler for synthesis
and
Synopsys PrimeTime for measuring power.
Next, 
we calculate
the energy of each convolutional layer based on 
the number of multiplications and the energy of the multiplication circuits.
To estimate the energy of the entire CNN,
we accumulate the individual energies of the layers.
Our evaluation is performed in the following three stages:
(i) study of CNN layer's error resilience,
(ii) exploration of the CNN's approximation space,
and
(iii) comparison to the original ALWANN approximation framework and its approximate multipliers.

\begin{figure}[!t]
\centering
\subfloat[\label{fig_maxeval1}]{\includegraphics[width=0.44\textwidth]{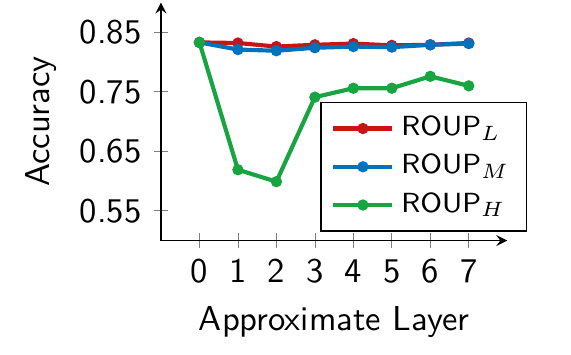}} \hspace{5pt}
\subfloat[\label{fig_maxeval2}]{\includegraphics[width=0.44\textwidth]{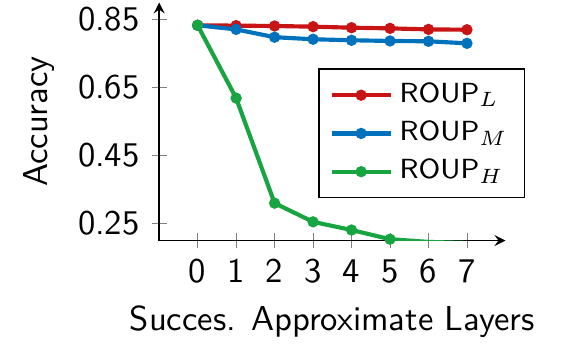}}%
\caption[Scaling of ResNet-8 Accuracy for Different Approximate Convolutional Layers]{The scaling of ResNet-8 accuracy with respect to the layers approximated: 
\textbf{(a)} only the $m$-th layer (e.g., only the $5$th), 
and 
\textbf{(b)} the first $m$ layers (e.g., layers $1$ to $5$). 
``Layer=0'' denotes the baseline quantized model.}%
\label{fig_maxscale}
\end{figure}

At first, we examine the error sensitivity of the convolutional layers,
in an effort to 
understand which layers are offered for approximations.
For this experiment,
we pick three ROUP multipliers with different approximation strength,
i.e., ``low'', ``medium'', and ``high'',
labeled as ROUP$_L$, ROUP$_M$, and ROUP$_H$, respectively. 
Figure \ref{fig_maxeval1} illustrates the accuracy scaling
when using these multipliers 
only in the $m$-th convolutional layer
($m = 1, 2, \dots7$).
Regardless of the approximation strength,
it is shown that 
approximating one of the first layers
results in remarkable accuracy loss
($m = 0$ shows the baseline model with $83\%$ accuracy).
This is more evident in the ROUP$_H$ configurations,
where significant computation errors are inserted.
In this case, 
when approximating one of the layers $4$--$7$, 
the accuracy loss is decreased
and stabilized around $8\%$. 
Figure \ref{fig_maxeval2},
depicts the accuracy scaling
when approximating the first $m$ layers.
As expected,
the accuracy loss is increased
with respect to the number of approximate layers,
however,
we again notice 
the error resilience of the last layers.
Specifically,
the accuracy loss is slowing down 
when extending the approximation after the $4$-th layer.
Another important outcome from this exploration
is the negligible accuracy loss of the ROUP$_L$ and ROUP$_M$ configurations,
regardless of which and how many layers are approximated.
In exchange for such small accuracy loss, 
these multipliers 
provide increased energy gains compared to their accurate design
(around $10\%$ and $20\%$, respectively). 

\begin{figure}[!t]
\centering
\includegraphics[width=0.845\textwidth]{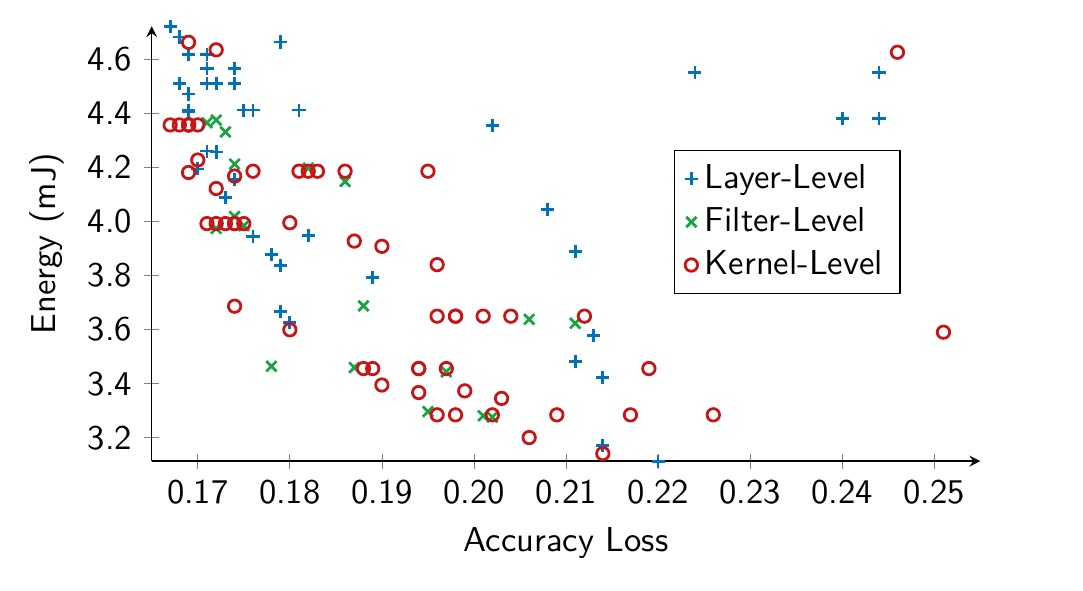}%
\caption[Pareto Analysis for the Approximate MAx-DNN-Based ResNet-8 CNNs]{Pareto analysis for the
approximate MAx-DNN-based ResNet-8 CNNs
(with ROUP multipliers)
considering energy consumption
and accuracy loss.}%
\label{fig_maxpare}
\end{figure}

Subsequently,
we perform an extensive design space exploration on the proposed approaches,
involving various ROUP multipliers,
to extract their most prominent
configurations in terms of accuracy and energy.
For each approach,
several multiplication replacements and combinations are examined. 
In Figure \ref{fig_maxpare}, 
we present
all the configurations that deliver at least $75\%$ classification accuracy, 
i.e., up to $8\%$ accuracy loss compared to the baseline quantized model. 
As shown in the upper left segment of the plot, 
multiple configurations 
deliver negligible accuracy loss compared to the quantized model,
which ranges from $0.02\%$ to $1\%$,
while 
improving the energy efficiency due to using the ROUP multipliers.
Therefore,
our approaches can satisfy near-zero accuracy loss
with more energy-efficient computing.
Regarding the efficiency of each approach,
the Pareto front is formed almost exclusively from the kernel- and filter-level configurations. 
For the same accuracy loss,
these configurations
provide better energy
than the straightforward layer-level approach,
which has to sacrifice a large amount of accuracy,
i.e., more than $40\%$,
to deliver this energy efficiency.

Table \ref{tb_maxalwann} compares the 
Pareto-front configurations
with ALWANN networks employing the EvoApprox8b multipliers \cite{2017_Mrazek_DATE}.
The proposed designs deliver better accuracy,
as the average loss of the
EvoApprox8b configurations
is \raisebox{0.8pt}{$\scriptstyle\sim$}$23\%$, 
while
in terms of energy,
they provide $2\times$ gains.
This comparison highlights the advantage 
of studying the approximations
at a lower CNN abstraction level, i.e., filter or kernel.
Namely, 
the non-uniform fine-grained use of multipliers with different approximation degree
outperforms the straightforward approximation, 
which either selects the same approximate multiplier for the entire network
or assigns an approximate multiplier per layer.
At the same time, 
the use of numerous approximate multipliers
based on different approximation approaches significantly expands the design space.
As a result, 
there is increased design flexibility,
which allows to efficiently handle  various accuracy constraints.

\begin{table}[!t]
\fontsize{9}{10}\selectfont
\renewcommand{\arraystretch}{1.2}
\setlength{\tabcolsep}{11pt}
\caption[Experimental Results of Approximate ROUP-Based ResNet-8 CNNs on TSMC 45-nm Standard-Cell]{Experimental results of approximate ROUP-based ResNet-8 CNNs on TSMC 45-nm standard-cell.}
\label{tb_maxalwann}
\centering
\begin{threeparttable}
\begin{tabular}{c@{\hskip 40pt} l | c  c}
\hline
\multicolumn{2}{c|}{\multirow{2}{*}{\textbf{Approximation Approach \& Configuration}}} &
\textbf{Energy} & 
\textbf{Accuracy} \\[-2pt] 
& & 
$(\%)$\footnotemark\setcounter{footnote}{1} &
$(\%)$\footnotemark\setcounter{footnote}{0}  \\
\hline \hline 
\parbox[t]{12mm}{\multirow{8}{*}{\rotatebox[origin=c]{40}{MAx-DNN/ROUP}}} & FLAM-3clas.\_2\_1\_1 & $49$ & $18$\\
& FLAM-3clas.\_2\_2\_1 & $52$ & $20$\\
& KLAM-chan.\_1\_0\_1  & $46$ & $17$\\
& KLAM-chan.\_2\_0\_2  & $53$ & $21$\\
& KLAM-chan.\_1\_1\_2  & $50$ & $19$\\
& KLAM-chan.\_2\_1\_2  & $54$ & $21$\\
& KLAM-row\_2\_1\_1    & $50$ & $19$\\
& KLAM-row\_2\_1\_2    & $52$ & $20$\\
\parbox[t]{12mm}{\multirow{5}{*}{\rotatebox[origin=c]{40}{ALWANN \cite{alwann,2017_Mrazek_DATE}}}} & Evo\_mul8\_2AC   & $23$ & $20$\\
& Evo\_mul8u\_2HH  & $23$ & $23$\\
& Evo\_mul8u\_NGR  & $32$ & $23$\\
& Evo\_mul8u\_ZFB  & $39$ & $23$\\
& Evo\_mul8u\_7C1  & $20$ & $24$\\ 
\hline
\end{tabular}
\begin{tablenotes}
  \item[1]{\fontsize{7.7}{8.8}\selectfont Refers to $\%$ energy gain (relative reduction) in comparison with the accurate design.}
  \item[2]{\fontsize{7.7}{8.8}\selectfont Refers to total accuracy loss (the baseline quantized model already has $17\%$ loss).}  
\end{tablenotes}
\end{threeparttable}
\end{table}

\section{Conclusion}
\label{s7_6}

In this chapter,
we developed various approximate DSP and AI
ASIC/FPGA-based
accelerators
that integrate 
the Dissertation's approximate circuits.
Our goal was 
twofold:
(i) to evaluate our approximate multipliers in real-world error-resilient applications,
and 
(ii) to quantify the resource gains and examine the accuracy of approximate DSP and AI hardware accelerators.
To facilitate the evaluation,
as well as examine various design scenarios and combinations, 
we proposed a software/hardware design methodology,
which is based on design space exploration
involving arithmetic, algorithms and approximation techniques/configurations. 
Based on our methodology,
we designed several approximate variants 
of accelerators for
1D/2D signal processing and CNNs.
Regarding the HDL development of the accelerators,
we employed various state-of-the-art 
parallelization techniques that 
were combined with our arithmetic approximations.
The most important experimental results
are summarized in Table \ref{tb_concacc}.
In terms of quality of results,
the approximations
attain small accuracy loss,
which is tunable 
due to the various approximation configurations
that are offered by our design families.
More explicitly,
we achieve 
typical values for 
the quality metrics of image processing (convolutions) 
and small errors 
in applications involving 
multiply-accumulate operations 
(signal filtering, matrix multiplication and LU decomposition).
Our approximations also provide promising accuracy results
in applications/algorithms from  other domains,
such as telecommunications (QAM demodulation)
and machine learning (K-means algorithm).
Moreover,
our approximate CNNs
deliver small accuracy loss
in various arithmetic formats
($16$-bit fixed-point, 
$32$/$16$-bit floating-point,
$8$-bit quantized integer),
which can be tuned even to provide the baseline accuracy. 
At the accelerator level,
we achieve remarkable resource gains
on both ASIC and FPGA. 
Depending the application,
we achieve power/energy gains in the range $30\%$--$70\%$ on ASIC technology, 
as well as valuable logic reduction,
i.e., up to $65\%$ in FPGA's LUTs
and up to $60\%$-$70\%$ in ASIC's logic-cell area.

\begin{table}[!t]
\fontsize{9}{10}\selectfont
\renewcommand{\arraystretch}{1.2}
\setlength{\tabcolsep}{1.1pt}
\caption[Summarized Results of Dissertation's Approximate DSP \& AI Hardware Accelerators]{Summarized results of Dissertation's approximate DSP \& AI hardware accelerators.}
\label{tb_concacc}
\centering
\begin{tabular}{l|cccc} 
\hline
\multicolumn{1}{c|}{\textbf{Application}} & \textbf{Approximation} &   \textbf{Accelerator} & \textbf{Gain} & \textbf{Accuracy} \\
\hline
\hline
Sobel Edge Detector & RAD, AxFXU & 65-nm ASIC & $58\%$ energy &  $\text{CER} > 98.5\%$  \\
FIR Filter & RAD, AxFXU & 65-nm ASIC & $34\%$ energy & $\text{MRED} = 3.6\%$  \\
Matrix Multiplication & RAD, AxFXU & 65-nm ASIC & $71\%$ energy & $\text{MRED} = 0.9\%$  \\
QAM Demodulation & RAD & ZCU106 FPGA & $65\%$ LUTs & $\text{BER} \in [10^{\text{-}1},10^{\text{-}4}]$ \\
ShipDetect CNN & RAD & Zynq-7020 FPGA & $38\%$ LUTs & $0.2\%$ loss \\
Gaussian Blurring & AxFPU16 & 65-nm ASIC & $40\%$ power & $\text{PSNR} = 51$dB  \\
Gaussian Blurring & AxFPU32 & 65-nm ASIC & $57\%$ power & $\text{PSNR} = 55$dB\\
CNN/MNIST & AxFPU16 & 65-nm ASIC & $49\%$ power & $2.7\%$ loss \\
CNN/MNIST & AxFPU32 & 65-nm ASIC & $66\%$ power & $5.5\%$ loss \\
CNN/CIFAR-10 & AxFPU16 & 65-nm ASIC & $39\%$ power & $2.5\%$ loss \\
CNN/CIFAR-10 & AxFPU32 & 65-nm ASIC & $58\%$ power & $5.4\%$ loss \\
K-Means Clustering & AxFPU32 & -- & -- & $\text{ECR} = 0\%$  \\
LU Decomposition & AxFPU32 & -- & -- & $\text{MRED} \in [0.1, 0.8]\%$ \\
ResNet-8/CIFAR-10 & ROUP & 45-nm ASIC & $54\%$ energy & $4\%$ loss \\
\hline
\end{tabular}
\end{table}
\partabstract{\begin{center}{\normalsize\textbf{Prologue}}\\[2pt]
\end{center}
{Dissertation's Part \ref{part2}
focuses on higher design abstraction layers
and
proposes design methodologies
for the efficient mapping and acceleration
of DSP and AI algorithms on space-grade FPGAs and embedded heterogeneous VPUs.
Chapter \ref{chapter8} regards the new European space-grade FPGAs, 
which insert several bottlenecks towards efficient implementation, 
mainly 
because the tool is new and the performance is lower 
than that of commercial FPGAs.
We show that  
the systematic development and exploration of all the tool settings can provide sufficient acceleration 
of high-performance DSP algorithms. 
Chapter \ref{chapter9} regards the multi-core VPUs,
which are characterized by 
increased SoC complexity
and decreased performance due to their limited power.
We show that 
the systematic development along with high- and low-level embedded design techniques can provide sufficient acceleration of demanding DSP and AI algorithms. 

The work presented 
in Part \ref{part2} 
was mainly performed in the context of research activities
of the European Space Agency (ESA).
Namely, 
it is product of groups of people
that took part in the respective
activities and
publications. 
The
contribution of the Dissertation's author is 
as follows.
For Chapter \ref{chapter8},
he contributed to the 
methodology, 
FPGA benchmarking,
tool exploration, 
hardware/software development for FPGA execution,
and
experimental analysis/evaluation.
For Chapter \ref{chapter9},
he contributed to the
methodology, 
VPU development/coding, 
and 
experimental evaluation/analysis.

\textbf{Acknowledgements:}
The author of the Ph.D. Dissertation
would like to thank 
Prof. Dimitrios Soudris
and Dr. George Lentaris,
who are the 
Scientific Officer 
and Project Manager \& Chief
of the ESA projects, respectively.}}

\part{Design Methodologies for~Embedded~Computing}
\label{part2}

\chapter{DSP Acceleration with New Space-Grade FPGA Devices~\&~Tools}
\label{chapter8}

\addtocontents{lof}{\protect\contentsline{chapter}{\protect\numberline{8}DSP Acceleration with New Space-Grade FPGA Devices \& Tools}{}{}}
\addtocontents{lot}{\protect\contentsline{chapter}{\protect\numberline{8}DSP Acceleration with New Space-Grade FPGA Devices \& Tools}{}{}}

\begin{ChapterAbstract}
The advent of space applications with increased computational requirements  
pushes the space industry to consider specialized processors
for high-performance on-board data processing.
The excellent performance-per-Watt ratio
of modern Field-Programmable Gate Arrays (FPGAs)
establishes them as a promising device solution,
which outperforms the general-purpose Central Processing Units (CPUs).
Especially for demanding Digital Signal Processing (DSP) algorithms,
the FPGAs offer increased parallelization
(to provide high processing throughput)
and interfacing capabilities
(to handle various sensors and provide high I/O rate). 
In a relatively limited market of space-grade FPGAs,
the new European FPGA family of NanoXplore
offers such novel
radiation-hardened solutions.
Nevertheless, 
the full exploitation
of these new space-grade FPGAs,
considering also that they lack in performance 
compared to their commercial counterparts, 
requires a systematic design approach.
Towards the efficient mapping
and acceleration of high-performance DSP algorithms
on new space-grade FPGAs, 
this chapter devises and applies 
a methodology to thoroughly 
examine the implementation.
We focus on NG-Large, i.e., 
the largest 65nm radiation-hardened SRAM FPGA of NanoXplore,  
and we accelerate demanding computer vision kernels
for feature detection and depth extraction.
Our design approach comprises a number of customized 
steps that
perform exhaustive exploration of all the tool settings
and generate a variety of results,
which are directly compared to results
from well-established FPGA vendors. 
The experimental evaluation shows that
NG-Large provides sufficient performance,
e.g., 
$\mathit{5}$--$\mathit{10}$ FPS 
for feature detection on MPixel images,
while the resource utilization is balanced
and comparable to that of the other FPGA vendors.\\
This chapter is based on our
\textbf{publication} in \textbf{\cite{LeonACCESS}}.
\end{ChapterAbstract}

\newpage

\section{Introduction}

In the last decade,
the Field-Programmable Gate Arrays (FPGAs)
have been established as state-of-the-art devices
in the semiconductor market. 
Their attractive performance-per-Watt ratio
marks a new era,
in which they are not used only for prototyping or interfacing,
but also for
accelerating demanding workloads \cite{fpga1, fpga3, fpga4, Sharma_FPGA, Guo_FPGA, direct_map, Wang_FPGA, lombardi_cnn, mocast_cnn}.
Modern FPGAs still do not match the power consumption of
Application-Specific Integrated Circuits (ASICs) \cite{asic_vs_fpga},
however, their re-configurability
and high-performance constitute them as 
first-class devices for 
acceleration 
in data centers \cite{kachris}
and embedded systems \cite{glent}. 
The throughput rate of FPGAs
is significantly better than that of the
Central Processing Units (CPUs),
while in terms of power consumption, 
they outperform 
the Graphics Processing Units (GPUs) \cite{fpgagpucpu, georgis, glent, marantos_gpu}. 
The FPGAs are used in various fields
(such as telecommunications, robotics, space)
for accelerating Digital Signal Processing (DSP) functions \cite{fpga1, fpga2, fpga3, fpga4, fpga5, karakasis}, 
e.g., for image/video processing and signal filtering, 
and more recently,
for deploying Artificial Intelligence (AI) \cite{Sharma_FPGA, Guo_FPGA, direct_map, Wang_FPGA, lombardi_cnn, mocast_cnn}, 
e.g., neural networks and machine learning algorithms.

The space industry is one of the communities
seeking for high-performance embedded platforms
to handle the increased computational demands and fast data transfers of modern space applications.
The enhanced performance of FPGAs
facilitates on-board/embedded computing in space, 
e.g., for Earth Observation (EO) and Vision-Based Navigation (VBN) tasks,
decreasing the need for   
downlink transmission of sensor data to 
the ground stations.
As a result,
space-grade and 
Commercial-Off-The-Shelf (COTS) FPGAs
are constantly being evaluated 
\cite{mpsoc, iturbe, csp, ngmed, lentaris_tvideo, odometry,   disparity, single_multi_rovers}
as on-board accelerators
and framing processors.
When radiation, thermal and vibration resilience 
are of utmost importance,
space-grade FPGAs 
are used instead of 
their COTS counterparts
to achieve increased reliability. 
The literature also includes numerous works with space avionics co-processing architectures that include FPGAs 
\cite{gpu_fpga, gpu_fpga_vpu, hpcb1}.
In terms of algorithms,
besides
accelerating DSP,
the FPGAs are used 
for implementing 
data transcoding for instruments/sensors 
(e.g., via SpaceWire/SpaceFibre \cite{spw_ffp})
and data compression \cite{lucana_shyloc, paschalis}.

The space-grade FPGAs
are either Radiation Hardened (RH) or Radiation Tolerant (RT).
Their main difference is that RH FPGAs
are radiation hardened by design,
i.e., they are fabricated on special technology 
to endure in radiation environments,
while RT FPGAs are fabricated to 
operate under lower radiation values
(usually along with fault-tolerant techniques to ensure reliable operation).
Nevertheless,
the space-grade FPGAs come with some penalties.
Besides their increased cost,
they are slower
and offer decreased design flexibility (e.g., in terms of tool options, vendor IPs)
than their COTS counterparts \cite{cots, sg_vs_cots}.
In any case, 
they still outperform RH CPUs
such as the PowerPC-based RAD750 ($12$W@$200$MHz) \cite{rad750}
and 
the LEON2-based AT697F ($1$W@$100$MHz) \cite{at697f}.
Currently,
the market offers few space-grade FPGAs,
from which the most well-established 
are categorized per vendor as follows:
\begin{itemize}[wide=3pt, leftmargin=*]
    \item 
    \underline{Xilinx \cite{xilinx}}:
    Virtex-4QV (SRAM, $90$nm),
    Virtex-5QV (SRAM, $65$nm),
    RT Kintex UltraScale (SRAM, $20$nm).
    \item 
    \underline{Microsemi \cite{microsemi}}:
    RTSX-SU (anti-fuse, $250$nm),
    RT ProASIC3 (flash, $130$nm),
    RTAX-S/SL (anti-fuse, $150$nm).
    RTG4 (flash, $65$nm), 
    RT PolarFire (SONOS, $28$nm).
    \item 
    \underline{Atmel \cite{atmel}}:
    AT40K (SRAM, $350$nm), 
    ATF280 (SRAM, $180$nm).
\end{itemize}

All the above FPGAs vary with respect to fabrication technology,
capacity, 
performance capabilities and radiation resilience.  
Very recently,
there are new promising additions
in this limited pool of space-grade FPGAs.
\underline{NanoXplore \cite{nx}} 
provides the first European RH high-density FPGAs \cite{joel, joel2}, 
also known as \emph{BRAVE}, i.e., 
Big Re-programmable Array for Versatile Environments.
The BRAVE FPGAs 
have attracted the interest of the space community 
and
are expected to play a key role in upcoming space missions,
especially in Europe.
This new space-grade FPGA family includes
low-end and high-end RH SRAM-based chips,
i.e., 
NG-Medium ($65$nm), NG-Large ($65$nm) and NG-Ultra ($28$nm), 
as well as
software tools 
for end-to-end development
and seamless re-configuration.
NG-Medium is the first FPGA of the BRAVE series
and was introduced in 2016 along with the initial tool versions.
The BRAVE FPGAs integrate all
the traditional FPGA programmable logic resources,
i.e., 
Look-Up-Tables (LUTs), 
D Flip-Flops (DFFs), 
Carry Units (CYs), 
Digital Signal Processors (DSPs), 
RAM Blocks (RAMBs),
and
Register Files (RFs). 
Additionally, 
NG-Large and NG-Ultra integrate 
the single-core ARM Cortex-R5 
and quad-core ARM Cortex-R52
processors, respectively, 
with the latter implementing the DAHLIA System-on-Chip (SoC).
Furthermore, 
the BRAVE FPGAs include features that are essential for on-board embedded computing in space,
such as the SpaceWire interface for fast I/O data transfers 
and memory scrubbing to ensure continuous error-free functionality.
The BRAVE FPGAs can be configured via multiple interfaces 
(JTAG, SPI/flash, SpaceWire).

The efficient utilization of a new FPGA family
such as BRAVE,
as well as the full exploitation of the associated software tools, 
require a systematic and disciplined approach. 
Moreover,
the developer needs to surpass the bottlenecks
and limitations of new tools 
to provide sufficient resource utilization and/or performance for compute-intensive DSP. 
The development becomes even more challenging
due to the special features of the space-grade FPGAs
compared to the commercial ones (e.g., lower performance).
For these reasons,
the European Space Agency (ESA)
is supporting a set of research
activities\footnote{\fontsize{7.7}{8.8}\selectfont\textbf{ESA QUEENS1/QUEENS2/QUEENS3:}\\4000119331/17/NL/PS, 4000128041/19/NL/AR/va, 4000134874/21/NL/AR/va},
which involve 
hardware benchmarking on the BRAVE chips/boards
and
testing of the BRAVE software tools.
These activities aim to improve the new space-grade 
devices and tools 
and evaluate the suitability of the BRAVE solution 
as on-board data processor.
In the context of these activities,
we develop a methodology
to support the design on the BRAVE FPGAs 
and evaluate the devices/tools
compared to solutions provided by
well-established FPGA vendors. 
We note that
even though our methodology is applied
in space-grade FPGAs,
it can be adopted
to aid the design on any FPGA platform
or perform benchmarking/evaluation among different FPGAs,
either space-grade or COTS. 
Moreover, 
our goal is not to
design new parallel DSP architectures. 
In contrast,
we employ DSP hardware kernels 
mainly from the Computer Vision (CV) field, 
which were developed in past ESA activities 
by Lentaris \emph{et al.} 
\cite{lentaris_tvideo, odometry, disparity,  single_multi_rovers}\footnote{\fontsize{7.7}{8.8}\selectfont Special thanks to
Dr. G. Lentaris \emph{et al.} from NTUA for providing the DSP hardware kernels.},  
and we accelerate them on the BRAVE technology.
More explicitly,
in a systematic manner,
we deploy these high-performance kernels 
on NanoXplore's FPGAs  
and other FPGAs of the market with similar architecture (mentioned as ``3rd-party''),
and we assess the software tools and the underlying hardware.

The \textbf{contribution} of this chapter is summarized as follows:

\begin{itemize}[]
\item[(i)] We highlight the bottlenecks that arise when immigrating to new FPGA technologies,
as well as the significance of the systematic exploration of  
 the tool settings and device capabilities towards improved resource utilization and performance. 
\item[(ii)] We propose a methodology for the efficient 
deployment of high-performance DSP kernels
(developed on well-established technologies) 
on the programmable logic of new FPGAs. 
\item[(iii)] We propose a methodology for the assessment and testing of   
FPGA tools and devices.
\item[(iv)] We evaluate NanoXplore's new space-grade FPGAs as candidate on-board accelerators for future space missions
by testing all the relevant features
and comparing them to well-established competitor devices. 
\end{itemize}

The remainder of this chapter is organized as follows. 
Section \ref{s8_2} overviews the market's space-grade FPGAs
and presents the new BRAVE devices and tools.
Section \ref{s8_3} introduces our design and assessment methodology.
Section \ref{s8_4} presents the CV kernels
and reports issues and tool bugs that arose 
during their implementation with the early tool versions.
Section \ref{s8_5} conducts the experimental evaluation,
which includes various FPGA results and comparisons. 
Section \ref{s8_6} demonstrates the execution of the CV kernels on the BRAVE hardware.
Finally, 
Section \ref{s8_7} draws the conclusions.

\section{Background}
\label{s8_2}

\subsection{The Landscape of Space-Grade FPGAs}

Table \ref{tb_fpgas} presents an overview of the most prominent space-grade FPGAs,
including the BRAVE ones.
The Atmel FPGAs are omitted here
due to their limited on-chip resources.
We note that Xilinx
has recently announced a new space-grade FPGA,
named XQR Versal \cite{versal},
which includes additional features such as
ARM processors and AI engines. 
The examined FPGAs can be categorized in three classes,
with each class including chips 
with similar resources but from different vendors, 
e.g., \{Virtex-5QV, RTG4, NG-Large\}. 
BRAVE is the only FPGA family that offers 
hard embedded processors,
facilitating hardware/software co-design.
In terms of radiation resilience,
similar to Virtex-5QV,
the advantage of BRAVE is their RH SRAM-based chips.
More specifically, 
they are fabricated with $12$-transistor ($12$-T) configuration memory cells
outperforming the simple SRAM cells and
competing with anti-fuse and flash technologies.
This is extremely important, considering that,
for example,
Virtex-4QV would require full Triple Modular Redundancy (TMR) and configuration memory scrubbing to achieve reliable operation.

\begin{table}[!t]
\fontsize{9}{10}\selectfont
\renewcommand{\arraystretch}{1.2}
\caption[Overview of Market's Space-Grade FPGAs]{Overview of market's space-grade FPGAs.}
\label{tb_fpgas}
\centering
\begin{threeparttable}
\begin{tabular*}{\textwidth}{@{\extracolsep{\fill}}l
@{\hspace{0.12cm}}l
@{\hspace{0.17cm}}|
@{\hspace{0.17cm}}c
@{\hspace{0.12cm}}c
@{\hspace{0.12cm}}c
@{\hspace{0.12cm}}c
@{\extracolsep{\fill}}}
\hline 
\multicolumn{1}{c}{\textbf{Vendor}} & \hspace{10pt}\textbf{FPGA} & \textbf{Rad. Resilience} & \textbf{Technology} & \setcounter{footnote}{0}\textbf{TID / SEL}\footnotemark\setcounter{footnote}{1} & \textbf{Processor}\\
\hline
\hline
\parbox[t]{12mm}{\multirow{3}{*}{\rotatebox[origin=c]{38}{Xilinx}}} 
& Virtex-4QV    & Tolerant  & SRAM, $90$nm        & $300$ / $125$  & --          \\
& Virtex-5QV    & Hard      & SRAM, $65$nm        & $1000$ / $125$ & --          \\
& RT Kintex US  & Tolerant  & SRAM, $20$nm        & $100$ / $80$   & --          \\   
\parbox[t]{12mm}{\multirow{3}{*}{\rotatebox[origin=c]{38}{Microsemi}}} 
& RTAX-S/SL     & Tolerant  & anti-fuse, $150$nm  & $300$ / $117$  & --          \\
& RTG4          & Tolerant  & flash, $65$nm       & $100$ / $103$  & --          \\
& RT PolarFire  & Tolerant  & SONOS, $28$nm       & $100$ / $80$   & --          \\
\parbox[t]{12mm}{\multirow{3}{*}{\rotatebox[origin=c]{38}{NanoXplore}}} 
& NG-Medium     & Hard      & SRAM, $65$nm        & $100$ / $60$   & --          \\
& NG-Large      & Hard      & SRAM, $65$nm        & $100$ / $60$   &  Cortex-R5  \\
& NG-Ultra      & Hard      & SRAM, $28$nm        & $50$ / $60$    &  Cortex-R52 \\
\hline
\end{tabular*}
\vspace{8pt}
\begin{tabular*}{\textwidth}{@{\extracolsep{\fill}}l
@{\hspace{0.12cm}}l
@{\hspace{0.17cm}}|
@{\hspace{0.17cm}}c
@{\hspace{0.12cm}}c
@{\hspace{0.12cm}}c
@{\hspace{0.12cm}}c
@{\hspace{0.12cm}}c
@{\extracolsep{\fill}}}
\hline 
\multicolumn{1}{c}{\textbf{Vendor}} & \hspace{10pt}\textbf{FPGA} & \textbf{Logic Cells} & \textbf{Total RAM} & \textbf{DSPs} & \textbf{User I/Os} & \textbf{SERDES}\footnotemark\setcounter{footnote}{0}\\
\hline
\hline
\parbox[t]{12mm}{\multirow{3}{*}{\rotatebox[origin=c]{38}{Xilinx}}} 
& Virtex-4QV    & $55$--$200$K  & $4.1$--$9.9$Mb  & $32$--$192$  & $640$--$960$  & --              \\
& Virtex-5QV    & $131$K        & $12.3$Mb        & $320$        & $836$         &  $18$@$4.3$Gbps    \\
& RT Kintex US  & $726$K        & $38$Mb          & $2.7$K       & $620$         &  $32$@$12.5$Gbps     \\   
\parbox[t]{12mm}{\multirow{3}{*}{\rotatebox[origin=c]{38}{Microsemi}}} 
& RTAX-S/SL     & $2$--$40$K    & $0.05$--$0.5$Mb & $0$--$120$   & $198$--$840$  & --              \\
& RTG4          & $151$K        & $5$Mb           & $462$        & $720$         &  $24$@$3.1$Gbps     \\
& RT PolarFire  & $481$K        & $33$Mb          & $1.4$K       & $584$         & 24@$10.3$Gbps      \\
\parbox[t]{12mm}{\multirow{3}{*}{\rotatebox[origin=c]{38}{NanoXplore}}} 
& NG-Medium     & $34$K         & $2.9$Mb         & $112$        & $566$         & --              \\
& NG-Large      & $137$K        & $10.1$Mb        & $384$        & $684$         & $24$@$6.3$Gbps      \\
& NG-Ultra      & $536$K        & $34$Mb          & $1.3$K       & $744$         & $32$@$12.5$Gbps   \\
\hline
\end{tabular*}
\begin{tablenotes}
  \item[1]{\fontsize{7.7}{8.8}\selectfont Total ionizing dose in Krad (Si) and single-event latch-up immunity to linear energy transfer in MeV-cm$^2$/mg.}
  \item[2]{\fontsize{7.7}{8.8}\selectfont High-performance serialization/deserialization transceivers.}
\end{tablenotes}
\end{threeparttable}
\end{table}

Several of these FPGAs
have been used in space missions
\cite{exomars, arase, xilrov, nasa_mars20}.
More details about their use in past, present and future missions
can be found in our publication in \cite{LeonACCESS}.
At research level,
the space-grade FPGAs
are being evaluated for 
the implementation of compute-intensive functions and novel space applications, 
while they are also examined as part of co-processing architectures \cite{gpu_fpga, gpu_fpga_vpu, hpcb1}.
Next,
we present some representative research works
involving space-grade FPGAs.

In \cite{4qv_compre}, 
the CCSDS 1.2.3 standard for compressing hyperspectral images is implemented on Virtex-4QV,
achieving real-time compression for sensors such as AVIRIS ($680$$\times$$512$$\times$$224$ image size),
while utilizing $1$/$3$ of the chip resources 
and consuming limited power. 
Similarly,
the Virtex-5QV and RTG4 FPGAs
are used for the implementation of the 
SHyLoC 2.0 CCSDS 121 and 123 lossless compression
standards \cite{lucana_shyloc},
providing up to $138$ and $81$ MSamples/s, respectively, for the AVIRIS sensor.
In \cite{kranitis_dpu},
the authors implement a single-chip payload data processing unit on the Virtex-5QV FPGA,
which integrates both the instrument system supervisor and data processing functions.
The proposed architecture supports self-configuration management and mitigation techniques to provide fault-tolerance.
In \cite{space_kintex},
the new SCCC-X telemetry transmitter,
which is an extension of the CCSDS 131.2-B-1 standard, 
is implemented and evaluated on RT Kintex UltraScale and RTG4,
delivering more than $450$ MSym/s and $250$ MSym/s, respectively.
Very recently,
radiation-tolerant FPGA-based platforms for AI applications have gained momentum. 
In \cite{obdp_kintex},
the authors propose 
a deep learning architecture 
for RT Kintex UltraScale,
which is based on Xilinx's deep learning processing unit
and
the TMR MicroBlaze subsystem. 
The authors of \cite{obdp_polarfire} 
use the VectorBlox software development kit
to deploy AI models on a matrix processor implemented on the RT PolarFire FPGA.

The literature includes several works with the BRAVE FPGAs.
In \cite{lucana_brave}, 
the re-configuration capabilities of NG-Medium via the SpaceWire interface are evaluated,
while in \cite{ngmed},
benchmarking results for NG-Medium are reported,
and re-configuration scenarios with different algorithms are examined.
Moreover, 
NG-Medium is used for the implementation of
dense stereo vision algorithms \cite{single_multi_rovers}.
Regarding NG-Large, 
preliminary benchmarking results 
are provided in \cite{obdp_leon}.
In \cite{lucana_shyloc},
both NG-Medium and NG-Large FPGAs
are evaluated
for hyperspectral image compression.

\subsection{The NanoXplore Space-Grade FPGAs and Tools}

\subsubsection{The NG-Large FPGA}
\begin{figure}[!t]
\centering
\includegraphics[width=0.86\textwidth]{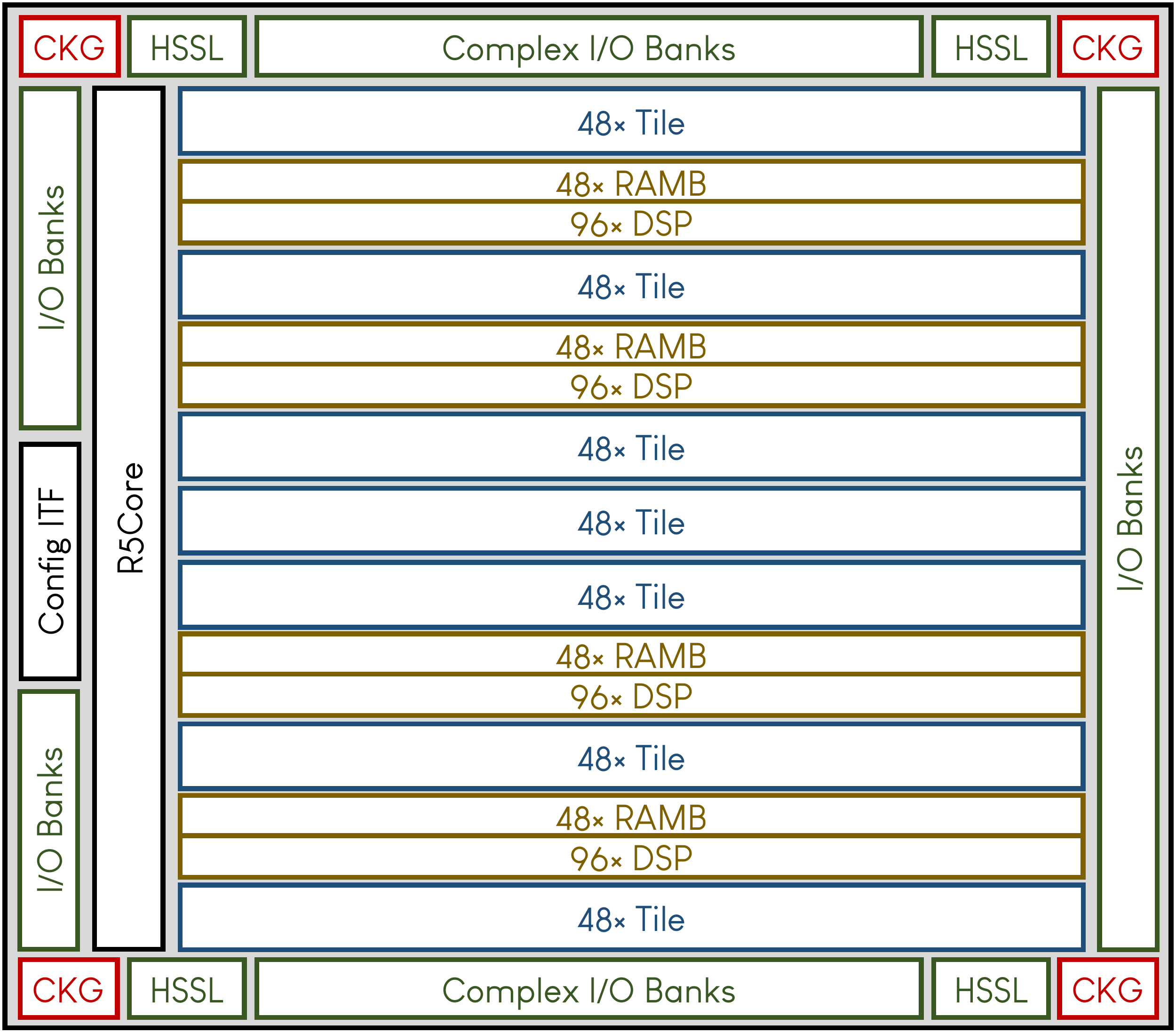}%
\caption[Fabric Architecture of NanoXplore's Space-Grade NG-Large FPGA]{The fabric architecture of NanoXplore's space-grade NG-Large FPGA \cite{nx}.}%
\label{fig_nglar}
\vspace*{-5pt}
\end{figure}

In this Dissertation,
we focus on NanoXplore's NG-Large FPGA,
i.e., the second chip of the BRAVE series,
thus,
we analyze its architectural details and most important features.
NG-Large is a radiation-hardened-by-design SRAM-based FPGA
that is manufactured on the 65nm STM C65-Space process technology \cite{joel}.
Below, we analyze its fabric architecture,
which is illustrated in Figure \ref{fig_nglar}.

The NG-Large die features $7$ rows of $48$ Tiles,
with a single Tile consisting of 
$384$ $4$-input LUTs, 
$384$ DFFs,
$96$ $1$-bit CYs,
$24$ X-LUTs,
and
$2$ $64$$\times$$16$-bit RFs.
Each Tile includes $384$ Functional Elements (FEs),
with a single FE integrating $1$ LUT--DFF pair
and additional logic for CYs, X-LUTs and RFs.
The CYs of NG-Large combine $1$ LUT with carry propagation logic
to support up to $96$-bit carry chain.
To allow the implementation of up to $16$-input logic functions,
$4$ LUTs drive the inputs of another LUT (called X-LUT)
without routing through the interconnect network.
The RF of NG-Large is a synchronous dual-port SRAM
with read-only and write-only ports
and optional output register.
Furthermore,
NG-Large features 
$4$ rows of $48$ $48$-Kbit RAMBs
and
$4$ rows of $96$ DSPs.
Each RAMB is a true dual-port SRAM with optional output register 
and supports multiple memory configurations,
i.e., 
$48$K$\times1$-bit, 
$24$K$\times2$-bit, 
$12$K$\times4$-bit, 
$6$K$\times8$-bit, 
$4$K$\times12$-bit and 
$2$K$\times24$-bit,
as well as Error Detection and Correction (EDAC) configuration.
A single DSP includes a $19\times24$ multiplier,
a $56$-bit Arithmetic Logic Unit (ALU), 
an $18$-bit pre-adder, 
as well as pipeline registers.
The DSPs operate either in signed or unsigned mode,
and can cascade up to $96$ blocks.
Finally,
NG-Large includes $4$ Clock Generators (CKGs),
one at each die corner.
The CKG block includes $1$ Phase-Locked Loop (PLL)
and $10$ Waveform Generators (WFGs), i.e., frequency dividers.
Moreover, it has $4$ High Speed Serial Links (HSSLs) of $6$ lanes,
providing up to $6.25$ Gbps data rate.

Overall, the total resources of NG-Large are summarized as:
$137088$ LUTs,
$129024$ DFFs,
$32256$ CYs,
$384$ DSPs,
$192$ RAMBs,
$672$ RFs,
$4$ PLLs.
Compared to NG-Medium,
NG-Large is \raisebox{0.8pt}{$\scriptstyle\sim$}$4\times$ larger 
(NG-Medium has $34272$ LUTs, $112$ DSPs, $56$ RAMBs) \cite{joel}.
Correspondingly,
NG-Large is \raisebox{0.8pt}{$\scriptstyle\sim$}$4\times$ smaller than NG-Ultra.

\subsubsection{The NXmap Software Tool}

NXmap is the design suite provided by NanoXplore to support 
all classical FPGA development stages
except for simulation, 
which is currently performed with 3rd-party tools (ModelSim/QuestaSim).
The tool supports both Verilog and VHDL hardware description languages 
and includes
functions for timing analysis and power estimation.
It is divided in two components: 
(i) the graphical interface, 
which allows the user to compile existing projects, view the floorplan and inspect the implemented design,
(ii) the Python wrapper, 
which allows the user to build projects by compiling Python scripts with the desired tool settings and functionalities.

The Python wrapper of NXmap supports the Python syntax and structures, 
and it provides a plethora of NanoXplore routines/modules for each stage of the FPGA development flow.
These Python routines can be categorized as follows: 
\begin{itemize}[noitemsep]
    \item Project-related: e.g., \texttt{createProject}, \texttt{addFiles}
    \item Tool-related: e.g., \texttt{setOptions}
    \item Mapping-related: e.g., \texttt{createRegion}, \texttt{addRAMLocation}
    \item Stage-related: e.g., \texttt{synthesize}, \texttt{generateBitstream}
    \item Board-related: e.g., \texttt{addPads}, \texttt{addBanks}
    \item Timing-related: e.g., \texttt{createClock}, \texttt{addFalsePath}
\end{itemize}

The NXmap routines take numerous arguments as input,
providing a wide range of functionalities.
NXmap also offers routines for 
monitoring the design flow
and 
defining the verbosity of the reports.
A Python code snippet demonstrating some basic NXmap functionalities is attached in Code \ref{nxpython}.

\vspace*{11pt}

\begin{lstlisting}[ 
language=Python, 
deletekeywords={print}, 
caption={Example of Python script for building an FPGA project in NanoXplore's NXmap tool.}, label={nxpython}]
# project creation
project = createProject(my_directory) 
project.setVariantName('NG-LARGE')
project.setTopCellName('top_module')
project.addFiles(['top_module.vhd', 'design1.vhd', 'design2.vhd'])

# general tool settings
project.setOptions(['MappingEffort': 'Medium',
					'RoutingEffort': 'High',
					'DisableDSPRegisters': 'Yes',
					'DefaultROMMapping': 'LUT', 
					'ManageUnconnectedOutputs': 'Ground'])

# custom mapping targets
project.addMappingDirective('getModels(.*mult.*)', 'MUL', 'DSP')
project.addMappingDirective('getModels(add_9u_9u)', 'ADD', 'CY')

# custom mapping locations
project.addDSPLocation('*mult_L267*', 'CGB[1x8]:L')
project.addRAMLocation('*RAM_INST0*', 'CGB[48x20]')

# clock constraint
project.createClock('getClockNet(clk)', 'clk', 40000)          

# I/O signals and pads pairing
project.addPad('clk', {'location':'IO_B18D02P', 
                       'standard':'LVCMOS', 'drive':'2mA'}) 
# project save
project.save('project.nym')  

# synthesis
project.synthesize()                  
project.save('synthesis_netlist.vhd')

# place 
project.place()                  
project.save('place_netlist.vhd')

# route
project.route()                   
project.save('route_netlist.vhd')

# reports
project.reportPorts()
project.reportInstances()

# static timing analysis
analyzer = project.createAnalyzer()
analyzer.launch()

# bitstream generation
project.generateBitstream('bitstream.nxb')
\end{lstlisting}

\vspace*{-2pt}

\section{Design \& Assessment Methodology}
\label{s8_3}
In this section,
we introduce a methodology 
for deploying DSP kernels on BRAVE FPGAs,
as well as for assessing 
the capabilities of the tools and chips.
Our methodology 
regards the three main stages
of the typical FPGA design flow,
i.e., synthesis, place \& route (implementation), 
and bitstream generation \& configuration.
Currently, 
the methodology is 
executed manually by the developer, 
however, some segments, such as the exploration of the tool settings via Python scripting, could be performed in an automatic fashion.
Moreover, 
this methodology is generic,
namely it can be adopted to 
test/evaluate other FPGAs
or facilitate the development with new devices and tools.

\subsection{Synthesis of the Design}

From the tool assessment perspective, 
our synthesis-related methodology aims to:
\begin{itemize}[]
\item[(i)] Test the correct functionality of all the tool settings, attributes, and strategies 
by examining if they apply the expected functionality
and 
if the output results of the post-synthesis simulation remain correct.
\item[(ii)] Evaluate the quality of results for different tool configurations
by comparing the resource utilization of the respective synthesis netlists.
\item[(iii)] Evaluate the capability of the tool's synthesizer to efficiently map the input design
on the FPGA blocks
by examining the resource utilization compared to the expected (theoretical) one. 
\item[(iv)] Rate the resource utilization 
via systematic comparisons to 3rd-party devices/tools
that are already established in the market. 
\end{itemize}

From the efficient DSP deployment perspective,
given a kernel developed in Hardware Description Language (HDL),
our methodology aims to:
\begin{itemize}[]
\item[(i)] Quantify the expected resource utilization of the kernel 
based on well-established state-of-the-art 3rd-party devices/tools.
\item[(ii)] Extract the tool configurations that result
in the most efficient synthesis netlists.
\item[(iii)] Resolve the issues
that arise from 
the tool's ``immaturity''
(recently released tool)
or because the kernel was initially developed in other vendor's technology.  
\end{itemize}

\begin{figure}[!t]
\vspace*{-7pt}
\centering
\includegraphics[width=0.85\textwidth]{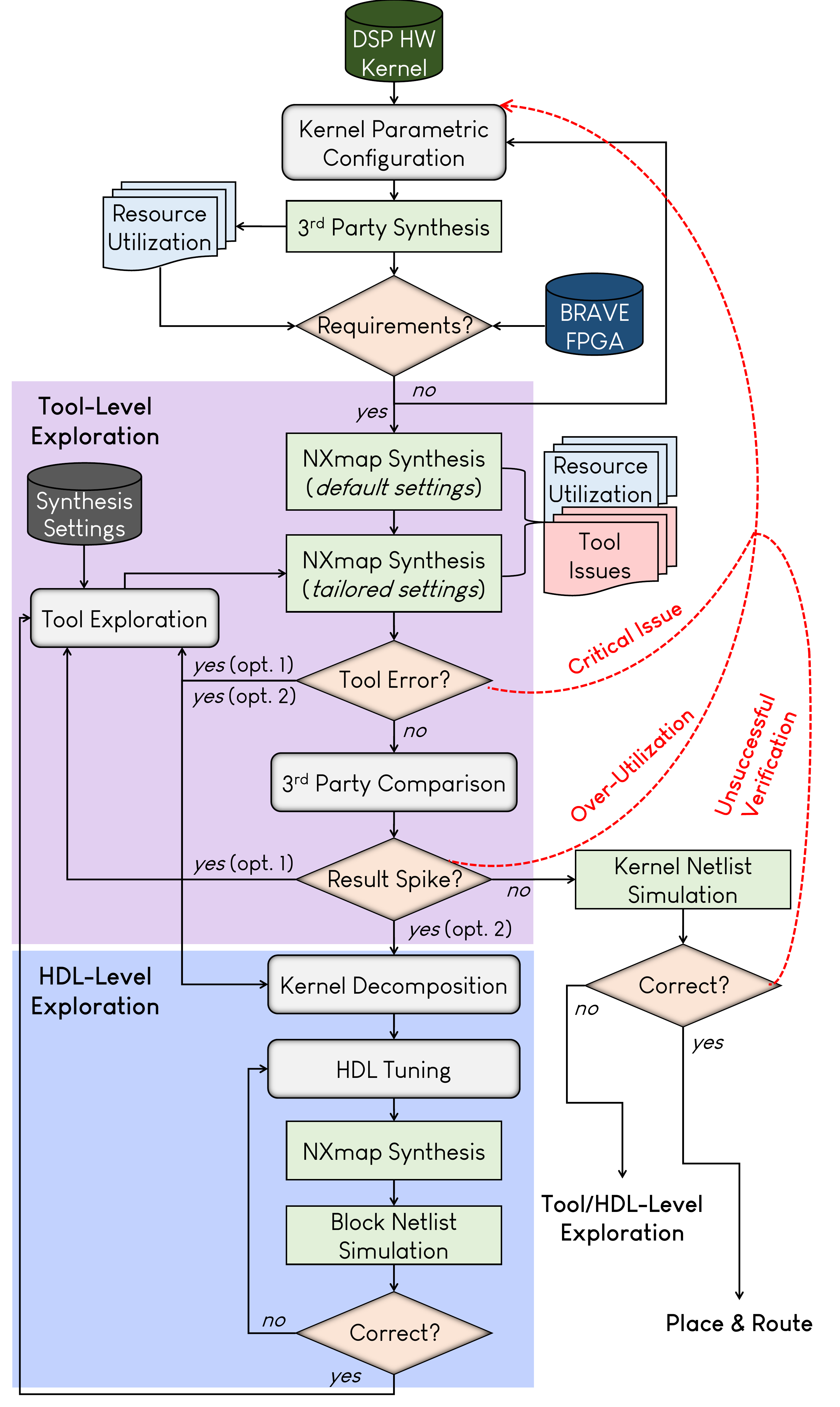}%
\vspace*{-4pt}
\caption[Methodology for the Synthesis of DSP Kernels on the New EU Space-Grade FPGAs]{Methodology for the synthesis of DSP kernels on the new BRAVE FPGAs.}%
\label{fig_synth}
\vspace{-35pt}
\end{figure}

The part of our methodology 
targeting to synthesis
is illustrated in Figure \ref{fig_synth}.
It is divided into three main phases:
the parametric configuration of the DSP kernels,
the tool-level exploration,
and the HDL-level exploration. 
Initially,
we adapt the algorithmic parameters of the DSP kernel
with respect to the features of the targeted BRAVE FPGA
(e.g., resources, architecture of FPGA blocks),  
and we run syntheses on 3rd-party tools.
To be more specific,
we configure parameters such as 
the size of the input image,
the size of the convolution masks,
the data bit-width,
the accuracy of the calculations,
and
the parallelization factor. 
For example,
the bigger BRAVE devices can handle 
larger input images
and/or increased parallelization. 
The next two phases
aim to generate an efficient error-free synthesis netlist.

In the phase of the tool-level exploration,
we perform a preliminary
synthesis with the default tool settings
to retrieve the ``default'' reports 
and detect potential issues.
This step is also considered 
as test for the synthesizer's capability  
to automatically balance the resource utilization 
and generate an error-free netlist.
Next,
we explore all the
available synthesis-related settings
and assess their capability to 
produce a synthesis netlist
according to the developer's choices and preferences.
Indicatively, 
we test tool settings
regarding
the mapping effort of the synthesizer, 
the mapping target of the arithmetic/memory components,
the DSP utilization ratio,
the register duplication,
and 
the style of the finite-state machine encodings.
The settings are applied
in both standalone and
combinatorial fashion. 
In case of error or unexpected tool behaviour,
we report the issue 
(to be resolved, if possible, via an alternative tool configuration or HDL coding).
The resource utilization of the most prominent
synthesis netlists
is compared with that 
of the 3rd-party tools. 
In case spikes are observed,
we apply different tool settings
or we move on to the next phase of our methodology.

In the phase of the HDL-level exploration,
we recursively decompose the kernel
to smaller building blocks 
and test them individually. 
This step is very important in our methodology,
as it provides an in-depth investigation 
of various optimization issues and/or
errors, which, otherwise, 
would be very difficult to be detected in the entire kernel.
This low-level exploration 
uses standard template-based coding and attributes/directives to express memories, 
finite-state machines, and arithmetic components,
as well as vendor-specific HDL templates
(i.e., from NanoXplore in the present case). 
In this phase,
for every new or modified building block,  
we perform synthesis
and functional verification via post-synthesis simulation.
If we identify a type of HDL coding that generates
improved results
or resolves one of the arisen issues,
we adopt it 
and return to the tool-level exploration
to examine the entire kernel (also in comparison with the 3rd-party tools). 

In case neither tool-level nor HDL-level
exploration
can provide a solution
in issues such as 
critical tool error,
resource over-utilization,
and unsuccessful verification,
our methodology 
includes feedback loops (red dashed lines in Figure \ref{fig_synth}).
The developer
can use them 
to return to the first phase
to 
re-customize the algorithmic parameters,
and then perform the tool-level and HDL-level exploration
with 
a new kernel configuration.

\subsection{Placement \& Routing of the Design}

\begin{figure}[!t]
\vspace*{-8pt}
\centering
\includegraphics[width=0.85\textwidth]{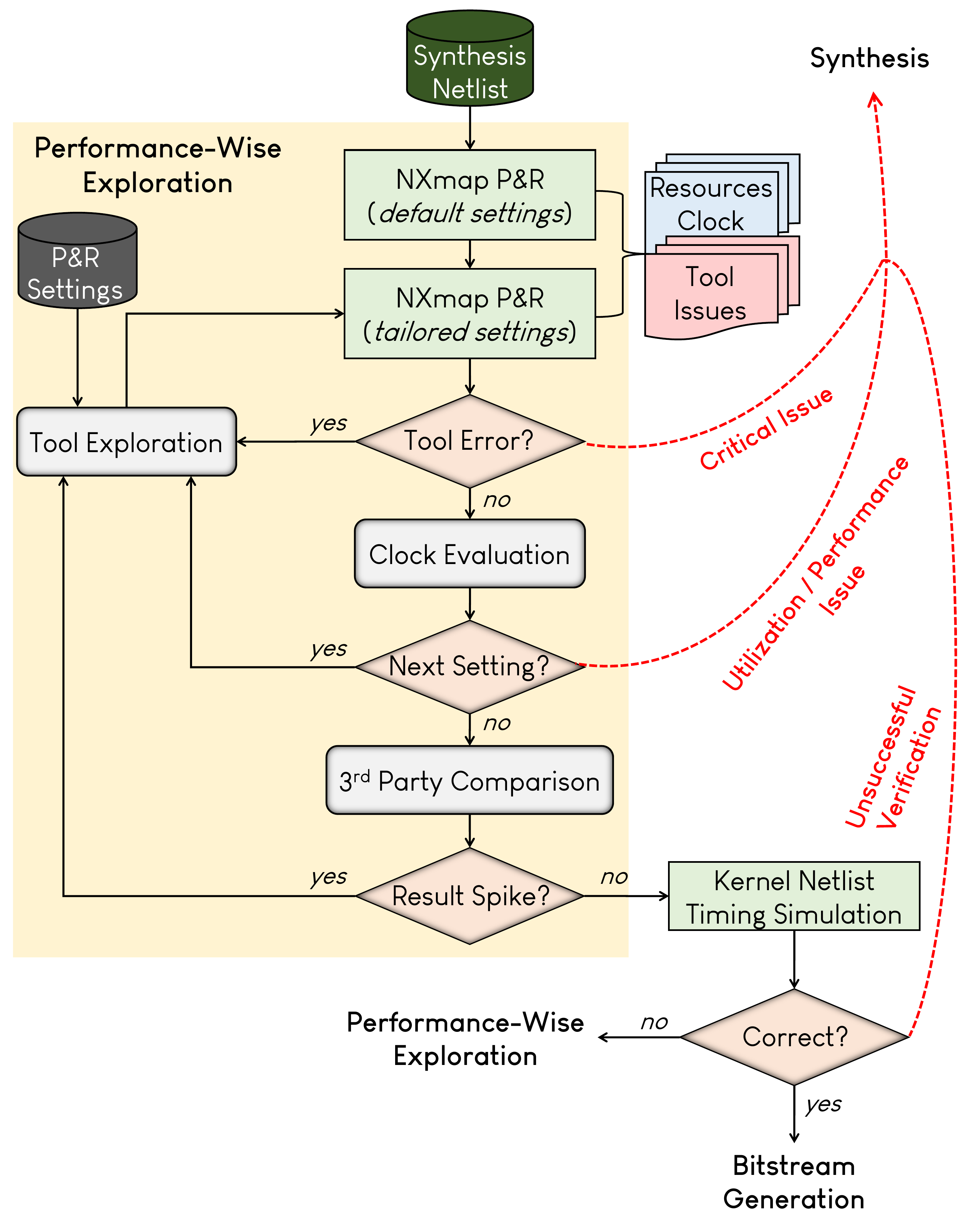}%
\caption[Methodology for the Placement \& Routing of DSP Kernels on the New EU Space-Grade FPGAs]{Methodology for placing \& routing DSP kernels on the new BRAVE FPGAs.}%
\label{fig_pnr}
\vspace*{-5pt}
\end{figure}

A similar methodology is developed for the place \& route (implementation) stage.
This methodology,
illustrated in Figure \ref{fig_pnr},
inputs the error-free synthesis netlist 
of the DSP kernel
and performs a performance-wise tool exploration,
targeting to provide the best possible clock frequency.
Like in synthesis,
we perform a preliminary run with the default settings,
and then start the exploration with tailored tool configurations.
In our exploration,
we employ all the settings related to the implementation,
i.e., placement/routing and physical constraints.

In this phase,
at first, 
we perform location-specific placements 
by specifying mapping regions on the FPGA floorplan,
either at fine-grained (LUTs, DFFs)
or coarse-grained (Tiles/FEs, DSPs, RAMBs) level.
In parallel, 
we examine 
the efficiency of the more general placement settings, e.g., the placement effort.  
Regarding the routing constraints,
we stress the implementation towards performance and increased routing congestion.
Namely, 
we employ 
all the available timing constraints 
(e.g., clock constraint, timing driven option, options for setting false path or max delay) 
and routing settings 
(e.g., router effort, router mode).
We note that we combine these timing/routing settings 
with different placement constraints.
For every combination of settings, 
our performance-wise tool exploration examines the Static Timing Analysis (STA) reports
to identify the most efficacious tool settings. 

Our methodology involves
systematic comparisons with the 3rd-party tools
and 
functional and timing verification via post-place and post-route netlist simulations,
as well as floorplan inspection
for verification purposes.
Moreover,
we include feedback loops (red dashed lines in Figure \ref{fig_pnr})
to return to synthesis
if this stage cannot provide solution to issues such as
critical tool error 
or unsuccessful verification.
As synthesis and place \& route are tightly coupled,
i.e., 
different synthesis netlists may lead to different place \& route results,
we can also return to synthesis and generate a new netlist 
in case of issues with performance or resource utilization.

\begin{figure}[!t]
\vspace*{-7pt}
\centering
\includegraphics[width=0.85\textwidth]{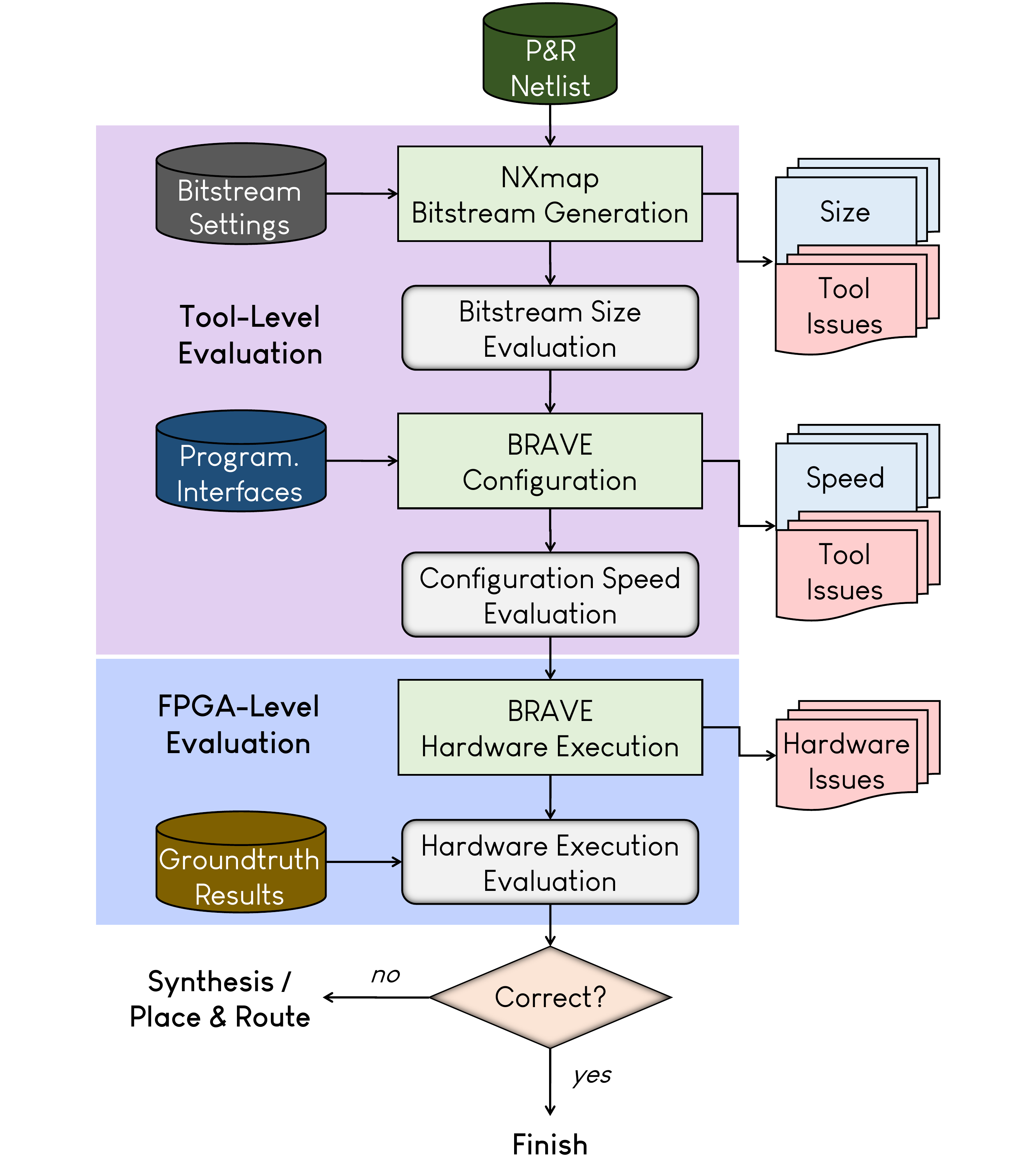}%
\caption[Methodology for the Bitstream Generation and Hardware Execution of DSP Kernels on the New EU Space-Grade FPGAs]{Methodology for the bitstream generation and hardware execution of DSP kernels on the new BRAVE FPGAs.}%
\label{fig_bitt}
\vspace*{-7pt}
\end{figure}

\subsection{Bitstream Generation and Hardware Execution}

The final step of our methodology is to examine the bitstream generation, the FPGA configuration/programming, and the actual hardware execution. 
For these purposes,
we employ the methodology illustrated in Figure \ref{fig_bitt}.
More specifically,
for different relevant tool settings,
we examine 
the correct functionality of the bitstream,
the bitstream size,
and 
the programming speeds via the available configuration interfaces (e.g., JTAG, SpaceWire, EPROM).
The examination of the bitstream size and the programming speed
is essential,
considering that an application may require multiple bitstreams,
e.g., a space mission storing on-board different bitstreams/algorithms and interchanging between them.  
Regarding the chip configuration,
we also examine 
the correct functionality of the FPGA after multiple successive re-configurations, 
as in real-world scenarios (like in space missions),
it may be required to reprogram the FPGA several times,
even in a very short period of time. 
Finally,
we perform hardware executions
to verify the correct implementation of the DSP kernels
on such new chips like the BRAVE FPGAs.
The validation of the results is performed by  
comparing the FPGA outputs with ground-truth data obtained from behavioral or post-place-\&-route netlist simulations.

\section{Porting of Computer Vision Kernels 
on the NG-Large FPGA}
\label{s8_4}

\subsection{CV Kernels for Feature Detection and Depth Extraction}
Based on our methodology,
we accelerate CV hardware kernels from past ESA activities 
\cite{lentaris_tvideo, odometry, disparity,  single_multi_rovers}
on the space-grade NG-Large FPGA. 
In particular,
we employ HDL kernels for 
feature extraction (image edges and corners),
i.e., Canny Edge Detector \cite{lentaris_tvideo} and Harris Corner Detector \cite{odometry}, 
as well as 
stereo matching (depth extraction in 3D
scene reconstruction),
i.e., GAD-Disparity \cite{disparity} and Spacesweep \cite{single_multi_rovers}.
These kernels impose increased  requirements in calculations and memory resources,
stressing the new tools and devices towards efficient and high-performance deployment.
Their development was performed in parametric VHDL code,
and thus,
at compile time, we can change various
parameters, e.g., the input image size, the datapath bit-width, or certain parallelization factors,
as required by our methodology (see Figure \ref{fig_synth}).
Next, we present in brief one kernel from each CV class:
\begin{itemize}[wide =1pt, leftmargin = *]
\item 
\underline{Harris Corner Detector \cite{odometry}}:
The kernel inputs a grayscale image and outputs a set of ``corners'',
i.e., the coordinates 
of the most salient features depicted in the image. 
The image is divided in horizontal stripes, 
which are 
processed successively in the FPGA by resource reusing. 
Functionally, in a loop over all pixels, 
the kernel employs Gaussian-smoothed products of image derivatives to define an
auto-correlation matrix, whose eigenvalues capture the principal intensity changes in the examined point’s neighborhood.
Corners are detected on pixels whose ``cornerness''  is sufficiently high 
and exceeds that of its neighboring pixels in a $3\times3$ region 
(via non-maximum suppression). 
All these operations are implemented via a succession of deep, fine-grained, pixel-based pipelines connecting memories that store intermediate results.
\item 
\underline{GAD-Disparity \cite{disparity}}:
The kernel inputs a pair of stereo images and outputs a two-dimensional disparity map, 
which can be transformed to depth map 
via simple calculations involving the focal length and baseline of the camera. 
The images are divided in horizontal stripes, 
which are processed in the FPGA by resource reusing. 
In an iterative fashion, by implementing an
outer loop over all examined disparities and an inner loop over all pixels, the kernel matches all
$7\times7$ image blocks between images 
by minimizing a 
Gauss-aggregated sum of absolute differences. 
The on-chip memory is mostly used 
to store the pixels 
and their corresponding intermediate aggregated values
that are continuously compared/updated.
On the other hand,
the on-chip logic is mostly used to calculate the Gauss-aggregated values.
\end{itemize}

\subsection{Implementation Details and Issues}
\label{s842}

The initial porting of the CV kernels on NG-Large
generated many tool issues and unoptimized results.
More specifically,
functional errors appeared 
due to the HDL description of the kernels,
which was based on other FPGA technologies.
Moreover, 
the tool reported
unbalanced or increased resource utilization and low clock frequency
due to its ``immaturity'' (newly released tool).
Next,
we report some representative issues that 
appeared during the implementation.  
However, we note that 
most of them have been resolved
either by 
newer tool versions
or via custom tool configuration and alternative HDL coding. 

One of our first issues  
was with the HDL description of the dual-port memory banks. 
The memory core of NG-Large's RAMB and RF
is not read-first,
thus, 
we had to 
add registers to pipeline the write data/address,
and also perform the write operation on the falling edge of the clock.
This issue was not reported by the tool,
however,
after investigation based on our methodology,
we managed to isolate the corresponding components  
and detect the memory problem
via post-synthesis simulation. 
Another incidence 
occurred regarding the recognition of the ROM memories.
The tool always mapped 
our ROM arrays 
(declared as \texttt{constant} in VHDL)
onto the LUTs, even when specifying a different mapping target, i.e., RAMB/RF.
Therefore, 
we used the VHDL
\texttt{signal}
in the description of our ROMs.
This issue was not functional,
however, it did not allow us
to deliver the desired resource optimizations,
while it was difficult to detect.  
Moreover,
we note that the tool could not implement very large multiplications onto the DSPs via cascading,
thus,
in certain cases,
we had to modify the HDL code and split the multipliers into smaller ones
(or alternatively map them onto CYs). 

Regarding the mapping of the CV kernels on the underlying hardware,
we observed several inefficient tool choices. 
Indicatively,
we mention that the tool did not employ the internal registers
of the DSPs and RAMBs 
and used the DFFs of the FEs
(e.g., to implement the registers before a multiplication or after the memory output).
In addition,
the tool did not use all the available memory configurations to 
efficiently map our RAM components onto the RAMBs,
resulting in large fragmentation.
Finally,
we note that the tool did not exploit the CY at all
(e.g., for additions),
resulting in DSP over-utilization and decreased performance. 

Nevertheless,
we note again that all the functional issues have been resolved
and the CV kernels are successfully implemented,
while in terms of mapping efficiency,
the tool is continuously improving.
In any case,
we report indicative issues that arose during the implementation
to highlight the difficulty 
of using new tools for FPGA design,
as well as
to indicate the significance of accompanying 
the development with a methodology.

\section{Evaluation}
\label{s8_5}

\subsection{Experimental Setup}
The experimental evaluation is conducted 
at two levels: 
(i) at tool level, 
where we evaluate the NXmap settings and examine the resource utilization of the CV kernels,
and 
(ii) at hardware level,
where we evaluate the chip's maximum frequency, the throughput of the CV kernels, and the configuration speeds/sizes of the bitstreams. 
Overall,
we report various experimental results for the acceleration of the CV kernels on 
NG-Large,
and we also directly compare them to the results obtained by well-established FPGA tools and devices (mentioned as ``3rd-party'').

From NanoXplore,
besides NG-Large, which is our targeted FPGA device,
we also employ its predecessor NG-Medium for comparison purposes.
Regarding the 3rd-party FPGAs,
we employ Xilinx Virtex-5QV XQR5VFX130
and Microsemi RTG4 RT4G150
(see Table \ref{tb_fpgas}),
but we also use Cyclone III EP3CLS150 of Intel \cite{intelfpga}.
Even though 
the latter is not a space-grade chip,
it provides similar characteristics with NG-Large
(i.e., $65$nm SRAM-based, $150$K LUT4s, $150$K DFFs, $666$ $9$-Kbit RAMBs, $320$ DSPs)
and Intel is considered a well-established vendor in the FPGA market.  
In terms of tools,
we use NanoXplore NXmap v3.5.0.4
and the 3rd-party development suites
(Xilinx Vivado, Intel Quartus, Microsemi Libero). 

The functional kernel verification is performed via post-synthesis simulations, post-place-\&-route simulations,
and actual hardware execution 
on realistic datasets. 
We use
images depicting rocky Martian terrains or satellites
for Harris and Canny,  
and 
synthetic stereo images depicting a rover’s view on Martian terrain
for GAD-Disparity and Spacesweep.
As discussed in our methodology,
we compare the outputs
with the groundtruth results of the behavioral simulations.

\begin{figure}[!t]
\vspace{-15pt}
\centering
\subfloat[\label{fig_canny}]{\includegraphics[width=0.5\textwidth]{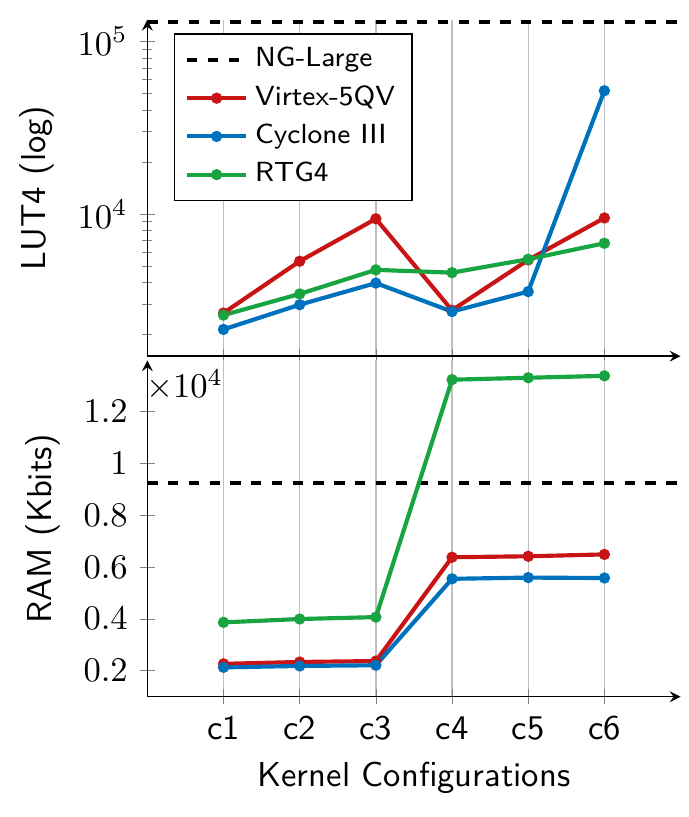}} \hspace{-3pt} %
\subfloat[\label{fig_harris}]{\includegraphics[width=0.5\textwidth]{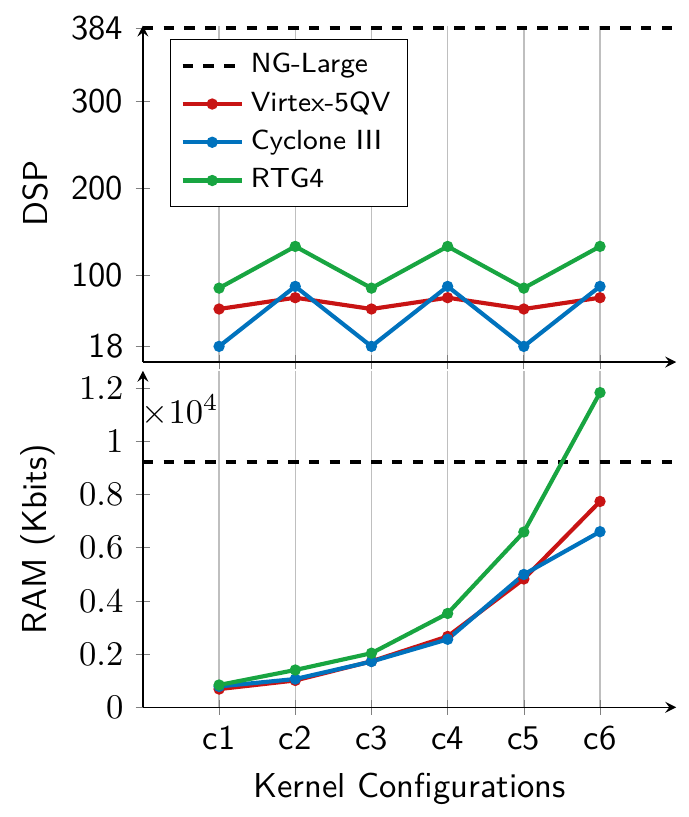}}\\[-7pt]
\subfloat[\label{fig_disparity}]{\includegraphics[width=0.5\textwidth]{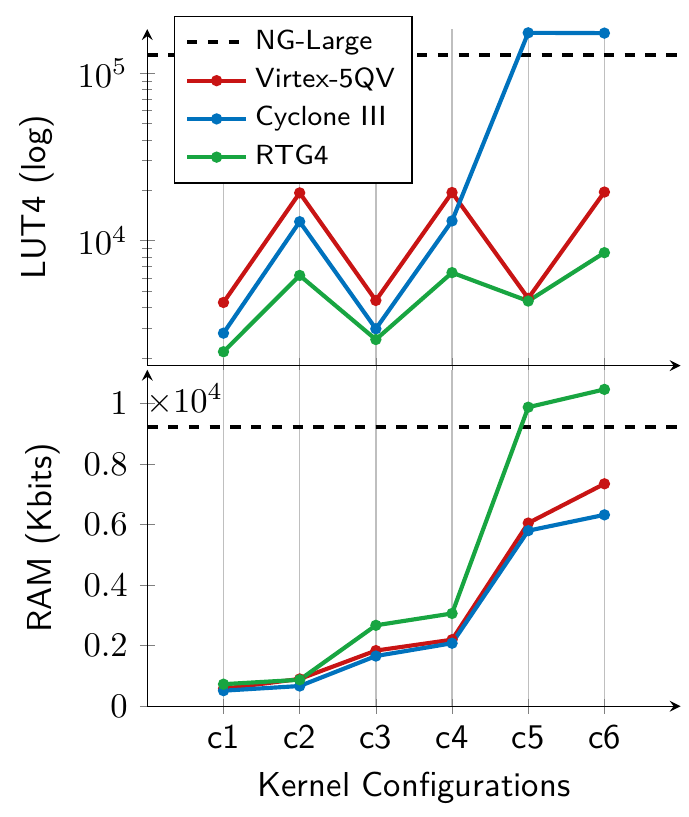}}\hspace{-3pt} 
\subfloat[\label{fig_sweep}]{\includegraphics[width=0.5\textwidth]{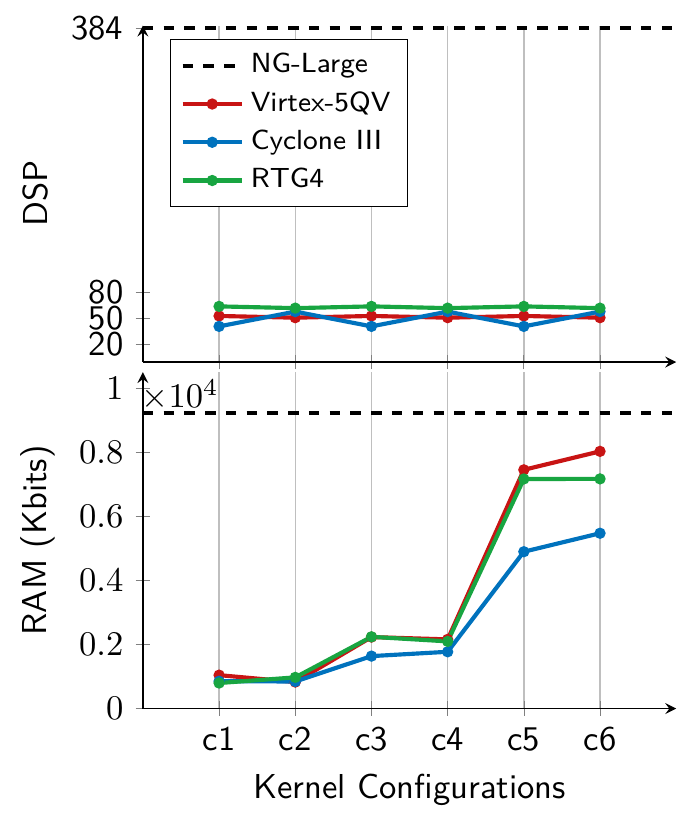}}%
\vspace{-5pt}
\caption[Exploration on the Synthesis of Computer Vision Kernels with Market's FPGA Tools]{Design space exploration on FPGA vendors: synthesis results for
\textbf{(a)} Canny Edge Detector \cite{lentaris_tvideo}, 
\textbf{(b)} Harris Corner Detector \cite{odometry}, 
\textbf{(c)} GAD-Disparity \cite{disparity}, 
\textbf{(d)} Spacesweep \cite{single_multi_rovers}.}%
\label{fig_3rd}
\vspace*{-6pt}
\end{figure}

\subsection{Design Space Exploration on Market's FPGA Vendors}

Before porting the CV kernels on NG-Large,
we perform a design space exploration
on 3rd-party tools/devices, 
as indicated by our methodology (see Figure \ref{fig_synth}).
The goal is to determine the most suitable 
kernel configuration according to NG-Large's capacity,
but also,
to extract baseline results for evaluation purposes. 

Figure \ref{fig_3rd} illustrates the resource utilization 
from the synthesis of $6$ different kernel configurations
with the 3rd-party tools. 
These diverse configurations impose different constraints 
in calculations, datapath and input sizes,
in an effort to stress the tools in both logic and memory. 
We note that
Virtex-5QV has $6$-input LUT
while the other FPGAs have $4$-input LUTs.
To make a fair comparison,
we rely on our experiments
and statistically use LUT4 = $1.5\times$LUT6 for Virtex-5QV. 
Moreover,
as each FPGA has different RAMB size,
we present the total RAM in Kbits. 
As explained, even though our goal is not to compare the 3rd-party tools,
we perform similar scaling in all FPGA vendors.
This is quite evident in the scaling of the RAM resources
of all CV kernels.
The LUT and DSP utilization are tightly coupled,
and they also depend on the strategy of each vendor
regarding the mapping of the arithmetic/logic operations.
Nevertheless,
we observe some spikes,
e.g., 
in the LUT utilization of Cyclone III
for Canny (Figure \ref{fig_canny})
and Harris (Figure \ref{fig_harris}), 
and the RAM of RTGA
for Canny (Figure \ref{fig_canny}) and Disparity (Figure \ref{fig_disparity}).

Following our 3rd-party exploration, 
we configure 
the CV kernels as shown in Table \ref{tb_bravekern}
to port them on NG-Large.
For example, 
Harris inputs a $1024\times1024$ $8$-bit image,
partitioned in $1024\times32$ pixel stripes 
(image divided into $32$ such blocks), 
performs convolutions with $7\times7$ $14$-bit masks,
and outputs $32$-bit corners.

\subsection{Experimental Results}

\subsubsection{Evaluation of NXmap Tool}

In this section,
we evaluate the 
implementation of the CV kernels on NG-Large 
and
present results
from our exploration
in the NXmap synthesis and place \& route stages.
We remind that for the default runs
we do not modify any tool settings,
while for the tailored runs
we apply custom tool settings.
Moreover,
we note that our 
tailored tool configurations
were more important in earlier versions of NXmap,
which provided unoptimized results 
(see Section \ref{s842}). 
In the examined tool version,
the default results
have been greatly improved compared 
to previous versions. 

The upper/lower part
of Table \ref{tb_nxsynth} presents the 
NG-Large utilization of synthesis
with the
default/tailored NXmap
settings.
Towards more balanced resource utilization,
as indicated by our methodology 
(see Figure \ref{fig_synth}), 
we share the arithmetic operations between DSPs and CYs using the corresponding NXmap routines.
For example,
the majority of Harris multiplications are
mapped onto CYs instead of DSPs
in the default synthesis.
As a result,
in the tailored synthesis,
we achieve a balanced utilization 
i.e., 
from $43\%$ CY and $8\%$ DSP 
to $23\%$ CY and $22\%$ DSP.
For Disparity,
the default settings deliver the same resources
with our custom settings that balance the DSP and CY utilization.
For Spacesweep,
the tool maps by default the small memories onto RFs, 
thus,
it is capable of making decisions with respect to the memory size.
However,
we force the tool to use RAMBs
even for these small memories,
because according to our experiments,
it results in better clock frequency.

\begin{table}[!t]
\fontsize{9}{10}\selectfont
\renewcommand{\arraystretch}{1.2}
\setlength{\tabcolsep}{-1pt}
\caption[Configuration of Computer Vision Kernel's Algorithmic Parameters]{Final configuration of computer vision kernel's algorithmic parameters.}
\label{tb_bravekern}
\centering
\begin{tabular*}{\columnwidth}{@{}l @{\extracolsep{\fill}} *{5}{c} @{}}
\hline
\multicolumn{1}{c}{\multirow{3}{*}{\textbf{Kernel}}}   & \multicolumn{3}{c}{\textbf{Data}} & \multicolumn{2}{c}{\textbf{Mask}}
\\
\cmidrule(lr){2-4} \cmidrule(lr){5-6}
  & \textbf{Img Size} & \textbf{Img Partition} & \textbf{I/O Bits} & \textbf{Size} & \textbf{Bits}
  \\
\hline
\hline
Canny Edge Detector \cite{lentaris_tvideo}    
& $1024\times1024$    & --     & $8/4$      & $3\times3$     & $8\times3$  \\
Harris Corner Detector \cite{odometry}     
& $1024\times1024$    & $1024\times32$ & $8$/$32$   & $7\times7$     & $8\times14$  \\
GAD-Disparity Stereo Vision \cite{disparity}
& $1024\times1024$    & $1024\times32$ & $8$/$10$   & $7\times7$     & $8\times7$  \\
Spacesweep Stereo Vision \cite{single_multi_rovers}
& $1024\times1024$    & $1024\times16$ & $8$/$32$   & $13\times13$   & $8\times8$   \\
\hline
\end{tabular*}
\end{table}

\begin{table}[!t]
\fontsize{9}{10}\selectfont
\renewcommand{\arraystretch}{1.2}
\setlength{\tabcolsep}{5.2pt}
\caption[Synthesis' Resource Utilization of Computer Vision Kernels on NG-Large FPGA]{Synthesis' resource utilization of computer vision kernels on NG-Large FPGA.}
\label{tb_nxsynth}
\centering
\begin{threeparttable}
\begin{tabular}{l c c c c c c}
\hline
\multicolumn{1}{c}{\multirow{3}{*}{\textbf{Kernel}}}  & \multicolumn{6}{c}{\textbf{\setcounter{footnote}{0}\textbf{Default Tool Settings}\footnotemark}}\\ 
\cmidrule(lr){2-7}
 &
\textbf{LUT} &
\textbf{DFF} &
\textbf{CY}  &
\textbf{DSP} &
\textbf{RF}  &
\textbf{RAMB} \\
\hline \hline 
Canny         
& 1845 (2\%)   & 2348 (2\%)   & 1167 (4\%)    & 2  (1\%)      & 0  (0\%)     & 177 (93\%)      \\
Harris
& 6210 (5\%)    & 16398 (13\%)   & 13794 (43\%)  & 27 (8\%)      & 0 (0\%)      & 69  (36\%)     \\
Disparity
& 1000 (1\%)   & 3628 (3\%)   & 4548 (15\%)   & 4  (2\%)      & 0 (0\%)      & 85 (45\%)       \\
Spacesweep
& 5500 (5\%)    & 10222 (8\%)   & 6277 (20\%)   & 50 (14\%)      & 8 (2\%)      & 74  (39\%)      \\
\hline
\end{tabular}\vspace{6pt}
\setlength{\tabcolsep}{5.5pt}
\begin{tabular}{l c c c c c c}
\hline
\multicolumn{1}{c}{\multirow{3}{*}{\textbf{Kernel}}}  & \multicolumn{6}{c}{\textbf{\setcounter{footnote}{1}\textbf{Tailored Tool Settings}\footnotemark}}\\ 
\cmidrule(lr){2-7}
 &
\textbf{LUT} &
\textbf{DFF} &
\textbf{CY}  &
\textbf{DSP} &
\textbf{RF}  &
\textbf{RAMB} \\
\hline \hline 
Canny & 1845 (2\%)   & 2299 (2\%)    & 1086 (4\%)    & 4  (2\%)      & 0  (0\%)     & 177  (93\%)     \\
Harris & 6110 (5\%)   & 15304 (12\%)  & 7112 (23\%)   & 81 (22\%)     & 0 (0\%)      & 69 (36\%)      \\
Disparity & 1000 (1\%)   & 3628  (3\%)   & 4548 (15\%)   & 4  (2\%)     & 0  (0\%)     & 85  (45\%)     \\
Spacesweep & 5499  (5\%)  & 10222 (8\%)   & 6277 (20\%)    & 50  (14\%)    & 0 (0\%)       & 79 (42\%)      \\
\hline
\end{tabular}
\begin{tablenotes}
  \item[*]{\fontsize{7.7}{8.8}\selectfont $\%$: utilization of NG-Large ($137$K LUTs, $129$K DFFs, $32$K CYs, $384$ DSPs, $672$ RFs, $192$ RAMBs).}
  \item[1]{\fontsize{7.7}{8.8}\selectfont Previous NXmap tool versions did not utilize CYs by default, and we had to share the arithmetic operations between DSPs and CYs to balance the utilization for our tailored synthesis.}
  \item[2]{\fontsize{7.7}{8.8}\selectfont Tailored tool settings selected towards the best possible clock frequency (see Table \ref{tb_nximpl}).}
\end{tablenotes}
\end{threeparttable}
\end{table}

Before proceeding to place \& route, 
we examine how our mapping choices
affect the clock frequency. 
Canny and Harris achieve better
clock frequency
when using the DSP for their multiplications. 
Therefore,
as in Harris, 
we  map the few multipliers of Canny onto the DSPs.
On the other hand, 
Disparity and Spacesweep achieve better timing when the default mapping is used, 
as the custom mapping 
decreases the clock frequency by $5\%$.
As a result,
we do not adopt the custom mapping in these kernels 
and tune only the settings of place \& route.

Next,
we proceed with the place \& route stage,
where we apply our performance-wise tool exploration
(see Figure \ref{fig_pnr}).
The upper/lower part
of Table \ref{tb_nximpl} presents the 
NG-Large utilization of place \& route
with the 
default/tailored NXmap
settings.
Firstly,
we observe that
the memory and arithmetic resources,
i.e., 
RFs/RAMBs and CYs/DSPs, 
remain equal to those reported by synthesis. 
Moreover, 
the variations in resources for tuning
the place \& route settings
(\texttt{DensityEffort}, 
\texttt{CongestionEffort}, 
\texttt{PolishingEffort},  
\texttt{RoutingEffort},
\texttt{BypassingEffort}) 
are negligible for all the kernels, i.e., $\pm 10$ LUTs. 
In terms of performance, 
all the kernels achieve better clock frequency
by modifying some of the default settings.
We combine different options 
for the aforementioned tool settings 
and report those that provide the maximum frequency according to NXmap's STA.
For Canny, 
the \texttt{PolishingEffort} setting 
is set to ``high'' rather than ``medium'', 
providing an increase of $2.7$MHz. 
For Harris, 
the \texttt{PolishingEffort} setting 
is set to ``low'' rather than ``medium'', 
giving an increase of $9.5$MHz. 
For Disparity, 
the \texttt{PolishingEffort} setting 
is set to ``low'' rather than ``medium'' 
and the \texttt{DensityEffort} to ``medium'' rather than ``low'', 
delivering an increase of $2.6$MHz. 
For Spacesweep, 
the \texttt{CongestionEffort} setting 
is set to ``medium'' rather than ``high'', 
delivering an increase of $0.7$MHz. 
These gains in the clock frequency may be considered limited,
but as we show in the next section,
they are still important
towards improving the 
throughput of the CV kernels.  
Moreover,
we note again
that our tool exploration 
provided significant gains in the initial tool versions,
i.e., up to $3 \times$ frequency improvement. 

\begin{table}[!t]
\fontsize{9}{10}\selectfont
\renewcommand{\arraystretch}{1.2}
\setlength{\tabcolsep}{5.2pt}
\caption[Implementation's Resource Utilization of Computer Vision Kernels on NG-Large FPGA]{Implementation's resource utilization of computer vision kernels on NG-Large FPGA.}
\label{tb_nximpl}
\centering
\begin{threeparttable}
\begin{tabular}{l c c c c c c}
\hline
\multicolumn{1}{c}{\multirow{3}{*}{\textbf{Kernel}}}  & \multicolumn{6}{c}{\textbf{Default Tool Settings}}\\ 
\cmidrule(lr){2-7}
 &
\textbf{LUT} &
\textbf{DFF} &
\textbf{CY}  &
\textbf{DSP} &
\textbf{RAMB}  &
\textbf{MHz} \\
\hline \hline 
Canny & 1844 (2\%)    & 2412 (2\%)   & 1167 (4\%)    & 2 (1\%)      & 177 (93\%)    & 35 \\    
Harris & 6205  (5\%)   & 16516  (13\%) & 13794  (43\%)  & 27  (8\%)     & 69  (36\%)    & 31        \\
Disparity     & 994  (1\%)    & 3664  (3\%)  & 4548  (15\%)  & 4   (2\%)    & 85 (45\%)      & 47  \\
Spacesweep    & 5493  (5\%)  & 10280  (8\%)  & 6277 (20\%)   & 50  (14\%)    & 74  (39\%)     & 51  \\
\hline
\end{tabular}\vspace{6pt}
\setlength{\tabcolsep}{5.5pt}
\begin{tabular}{l c c c c c c}
\hline
\multicolumn{1}{c}{\multirow{3}{*}{\textbf{Kernel}}}  & \multicolumn{6}{c}{\textbf{Tailored Tool Settings}}\\ 
\cmidrule(lr){2-7}
 &
\textbf{LUT} &
\textbf{DFF} &
\textbf{CY}  &
\textbf{DSP} &
\textbf{RAMB}  &
\textbf{MHz} \\
\hline \hline 
Canny & 1843 (2\%)   & 2349 (2\%)    & 1086 (4\%)    & 4 (2\%)      & 177 (93\%)    & 38 \\ 
Harris & 6105 (5\%)    & 15413 (13\%)   & 7112 (23\%)   & 81 (22\%)     & 69 (36\%)     & 40        \\
Disparity     & 998 (1\%)    & 3672 (3\%)    & 4548 (15\%)   & 4 (2\%)      & 85 (45\%)     & 50\\
Spacesweep    & 5458  (5\%)   & 10276 (8\%)  & 6277 (20\%)    & 50  (14\%)     & 79 (42\%)     & 52 \\
\hline
\end{tabular}
\begin{tablenotes}
  \item[*]{\fontsize{7.7}{8.8}\selectfont $\%$: utilization of NG-Large ($137$K LUTs, $129$K DFFs, $32$K CYs, $384$ DSPs, $192$ RAMBs).}
\end{tablenotes}
\end{threeparttable}
\end{table}

Overall, 
with respect to the reported resource utilization, 
we conclude that the NXmap tool
maps
the CV kernels as expected.
Considering that these kernels impose 
diverse memory requirements,
the tool correctly employs
several of the available RAMB 
configurations, 
e.g.,  
$24$K$\times2$ and $12$K$\times4$.
As a result, 
reasonable RAMB utilization is derived for $1024$-pixel-wide images,
ranging from $36\%$ (Harris) to $93\%$ (Canny).
Canny utilizes almost all the on-chip RAM resources,
as it receives the entire image and operates in burst mode,
contrary to the other kernels that input image stripes.
Regarding the arithmetic and logic components of the kernels,
NXmap successfully recognizes, reports, and maps 
all the arithmetic operators, finite-state machines, and logic functions. 

\subsubsection{Evaluation of NG-Large's Performance, Power and Configuration}

Table \ref{tb_nxperf} reports the performance results for the kernels
(clock frequency, latency for one input frame and throughput).
The throughput metric excludes I/O and refers only to processing,
i.e., the single execution of the kernel.
For the feature detection kernels,
we measure the Frames Per Second (FPS),
while for the depth extraction kernels, 
we employ the MPixel Disparities per Second (MPDS),
which is a metric combining both resolution and performance. 
The results show that NG-Large provides sufficient performance 
for $1$-MPixel image,
taking into account the requirements of the corresponding VBN space applications.
The time required for a complete reconstruction using Disparity and Spacesweep 
could improve the conventional depth extraction of Mars rovers
by $1$ order of magnitude (in terms of resolution and speed).
We note that 
in the tested configuration, 
Spacesweep examines 3$\times$ more depth levels than Disparity ($300$ versus $100$), 
and thus,
it provides much higher accuracy.
Furthermore, 
given that most VBN pipelines require $1$--$10$ FPS, 
we conclude that the throughput of Canny and Harris,
which is
$5.3$ FPS and $10$ FPS, respectively,
leaves enough room for the complementary components of an algorithmic chain to finish on time.
Regarding the power consumption of NG-Large,
we report a detailed analysis in \cite{LeonACCESS}.
In brief, 
the static power
(when the FPGA is not programmed) is $2$W,
while 
the dynamic power varies, as it depends on the 
resource utilization 
(e.g., it is \raisebox{0.8pt}{$\scriptstyle\sim$}$200$mW when utilizing $10$K FEs or $200$ DSPs, and \raisebox{0.8pt}{$\scriptstyle\sim$}$40$mW when utilizing $150$ RAMBs).
Based on our power measurements and analysis
in \cite{LeonACCESS},
we report a power estimation in Table \ref{tb_nxperf},
which, however,
regards only the resource utilization
and 
does not take into account the 
I/Os,
interconnections, 
and switching activity.
When considering these parameters,
the power consumption of NG-Large
lies around the typical FPGA values
for such high-performance DSP workloads. 



In Table \ref{tb_nxconf},
we report the bitstream size of each kernel
and the time required for NG-Large to be programmed via JTAG  operating at $8$MHz.
According to the results,
the configuration time of NG-Large is almost proportional to the bitstream size.
In particular,
the JTAG interface
programs the chip with 
a rate of
$452$ KB/s for Canny, 
$389$ KB/s for Harris, 
$381$ KB/s for Disparity 
and $399$ KB/s for Spacesweep. 
The bitstream of Canny is around $1$MB larger than that of the other kernels, 
which is due to its $93\%$ RAMB utilization,
and thus, 
it requires more time to be configured.
The configuration times can be greatly improved
by using the SpaceWire interface,
which can operate up to $400$ Mbps.
Furthermore,
we observe that
the size of the NG-Large bitstream is not fixed,
like in other 3rd-party FPGAs,
where the bitstream size is pre-determined 
regardless of the design size and complexity.
This is an advantage of NG-Large, 
considering that
in space missions
various bitstreams are stored on-board.
These bitstreams implement 
either different kernels (in case of algorithmic pipelines)
or multiple variants of the same kernel 
(e.g., performance-wise and accuracy-wise variants). 
In this context, 
our work in \cite{ngmed}
discusses adaptive scenarios in space applications
that require 
multiple bitstreams to be configured. 
The same work also 
evaluates the reconfiguration capabilities
of NanoXplore's NG-Medium 
based on the expected reconfiguration rates of the space applications. 

\begin{table}[!t]
\fontsize{9}{10}\selectfont
\renewcommand{\arraystretch}{1.2}
\setlength{\tabcolsep}{12.5pt}
\caption[Performance and Power of Computer Vision Kernels on NG-Large FPGA]{Performance and power of computer vision kernels on NG-Large FPGA.}
\label{tb_nxperf}  
\centering
\begin{threeparttable}
\begin{tabular}{l|ccccc} 
\hline
\multicolumn{1}{c|}{\multirow{2}{*}{\textbf{Kernel}}} & \textbf{Clock} &   \textbf{Latency} & \setcounter{footnote}{0}\textbf{Throughput}\footnotemark & \textbf{Power}\footnotemark\\[-1pt]
& (MHz) & (s/frame) & (FPS) | (MPDS) & (W)\\
\hline\hline
Canny &        
$38$ & $0.10$  & $10$    & $2.3$ \\
Harris &
$40$ & $0.19$  & $5.3$   &  $2.5$  \\
Disparity &
$50$ & $6.7$   & $18$   &  $2.3$  \\ 
Spacesweep &
$52$ & $10.8$  & $29$   &  $2.4$\\
\hline
\end{tabular}
\begin{tablenotes}
  \item[1]{\fontsize{7.7}{8.8}\selectfont Excluding I/O: 
  FPS for Harris and Canny, MPDS for Disparity and Spacesweep.}
  \item[2]{\fontsize{7.7}{8.8}\selectfont Estimation based on the static FPGA power and the dynamic resource power (without taking into account the I/Os, interconnections and switching activity).}
\end{tablenotes}
\end{threeparttable}
\end{table}
\begin{table}[!t]
\fontsize{9}{10}\selectfont
\renewcommand{\arraystretch}{1.2}
\setlength{\tabcolsep}{5.5pt}
\caption[Results from the Configuration of the Computer Vision Kernels on NG-Large FPGA]{Results from the configuration of the computer vision kernels on NG-Large FPGA.}
\label{tb_nxconf}  
\centering
\begin{threeparttable}
\begin{tabular}{l|ccc} 
\hline
\multicolumn{1}{c|}{\multirow{2}{*}{\textbf{Kernel}}} & \textbf{Bitstream Size} &  \setcounter{footnote}{0}\textbf{Configuration Time}\footnotemark & \textbf{Configuration Rate} \\[-1pt]
& (KB) & (s) & (KB/s) \\
\hline\hline
Canny &   $2669$ & $5.9$    & $452$  \\
Harris & $1751$ & $4.5$     & $389$  \\
Disparity & $1563$ & $4.1$  & $381$ \\
Spacesweep & $1719$ & $4.3$ & $399$  \\ 
\hline
\end{tabular}
\begin{tablenotes}
  \item[1]{\fontsize{7.7}{8.8}\selectfont FPGA configured via JTAG using Intel Core i7-4500U@1.80GHz$\times$4, 8GB RAM.}
\end{tablenotes}
\end{threeparttable}
\end{table}

\subsection{Comparative FPGA Evaluation}
In this section,
we compare the implementations 
of the CV algorithms on NG-Large
with those on other FPGAs (Virtex-5QV, Cyclone III, RTG4, NG-Medium).
Our goal is to evaluate the BRAVE tools and devices by comparing them to well-established and more mature solutions in the space domain
that exist in the market for much longer.
Moreover, we compare NG-Large with NG-Medium in order to evaluate the evolution of the space-grade BRAVE FPGAs. 

\subsubsection{Comparison to 3rd-Party FPGAs}

For Canny, 
NXmap provides LUT utilization that is comparable to that of the 3rd-party tools,
i.e., LUTs are increased by only $6$\%.
However, 
we note that 
if we also consider the route-thru LUTs that are consumed
due to CY utilization, 
the LUTs are increased by $48$\%.
The DFF resources are increased by almost $50\%$ in NG-Large.
This huge increment is due to
not utilizing the internal registers of RAMBs
(given that the RAMB utilization is $93\%$ and our memories include registers).
In terms of DSPs,
NXmap utilizes the same resources with the 3rd-party tools,
while
it is more efficient in the utilization of the memory resources
(considering the RAMB size difference between devices).
Overall,
we consider the Canny resources as reasonable,  
taking into account that it almost fills up
all the RAMBs, and thus, 
the tool is stressed to provide efficient mapping for the rest resources. 
For Harris, 
NXmap provides a good LUT utilization, i.e., $3.2\times$ less LUT versus the average 3rd-party value. 
When considering the pass-thru LUTs for the CY utilization, the total number of LUTs increases,
but it is still comparable with the 3rd-party value.
Moreover, 
the LUT utilization should be examined along with the number of employed DSPs, 
for which NXmap utilizes $1.5\times$ less.
Regarding the memory resources, 
NXmap delivers $56\%$ less RAMBs,
while 
the total RAMB Kbits consumed are less than the average 3rd-party value. 
Similar results are observed for the other two CV kernels. 
For Disparity, 
NXmap provides promising LUT utilization, 
as it employs a small number,
even when considering the CY resources. 
Regarding RAM resources, 
NXmap is below the 3rd-party value in both blocks and Kbits. 
For Spacesweep,
NXmap provides small LUT utilization,
which is better by $52\%$ compared to the other tools,
and almost the same  
when considering the CYs. 
The DFF utilization is also better by $5\%$, 
while 
the DSP and RAMB utilization is excellent, 
as NXmap outperforms the average 3rd-party values by $20\%$ and $30\%$, respectively. 

\subsubsection{Comparison to the NG-Medium FPGA}

Next, 
we compare NG-Large against its predecessor, 
i.e., NG-Medium,
to evaluate the progress of BRAVE devices 
and examine if NG-Large provides significant advantage due to being $4\times$ bigger in resources.
We implement the same kernels on NG-Medium, 
initially configured as shown in Table \ref{tb_bravekern}, 
and we apply the final tailored tool settings that provided the best frequency on NG-Large.
We note that we do not change the algorithm of the kernels, 
namely,
the same HDL sources are implemented in both devices.
Only in case of resource over-utilization or unexpected tool issue in NG-Medium,
we modify 
either the kernel parameters 
(e.g., use smaller input image) 
or the tool settings 
(e.g., apply different mapping).
Table \ref{tb_bravecomp} reports the resource utilization in both BRAVE FPGAs.
As expected,
the memory resources of NG-Medium force us to modify
the parameters in all kernels.
More specifically,
in Harris and Disparity, 
we decrease the height of the input image stripe
by $4\times$ and $2\times$, respectively,
while in Canny and Spacesweep,
we decrease the size of the entire image by $4\times$ (from $1024\times1024$ to $512\times512$).
The derived results show that, 
even though the designs of NG-Large regard inputs with larger size,
its resource utilization percentage
is significantly better,
leaving room for implementing complementary HDL components, 
increasing the parallelization,
or serving even bigger input images.

\begin{figure}[!t]
\centering
\includegraphics[width=0.87\textwidth]{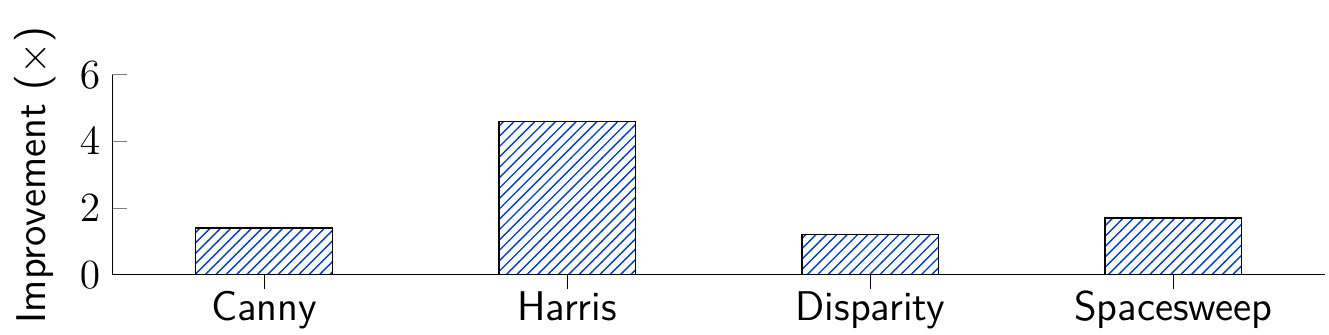}%
\caption[Latency Improvement in Computer Vision Kernels
by NG-Large]{Latency improvement in computer vision kernels
by NG-Large compared to NG-Medium for the same algorithm (no HDL modification, e.g., for increased parallelization).}%
\label{fig_larspd}
\end{figure}

In Figure \ref{fig_larspd},
we report the latency improvement in NG-Large 
compared to NG-Medium.
Harris in NG-Large is better by $4.6\times$,
as the clock frequency is increased by \raisebox{0.8pt}{$\scriptstyle\sim$}$4\times$,
and NG-Medium has to process $4\times$ more (smaller though) image stripes.
Canny achieves better clock frequency and execution time in NG-Medium,
but for the $1/4$ of NG-Large's input image.
To serve an 1024$\times$1024 image, 
it will have to process each $1/4$ of the image, 
send the partitioned edge map back to the CPU,
and at the end,
bring all the partitioned edge maps back to the FPGA. 
These extra steps will add an overhead of \raisebox{0.8pt}{$\scriptstyle\sim$}$60$ms in the good scenario, 
i.e., when using the SpaceWire interface at $100$ Mbps for data transmission.
Thus,
the total performance improvement in NG-Large is $1.4\times$.
Regarding Disparity,
both devices achieve almost the same clock frequency,
i.e., around $50$MHz,
however,
NG-Medium has to process $2\times$ more (smaller though) image stripes.
This differentiation in the partition of the input image is translated to 
$1.2\times$ performance improvement in NG-Large.
For Spacesweep,
considering that 
NG-Medium has to process smaller input image to avoid resource over-utilization,
and also with a clock frequency decreased by $1.7\times$,
NG-Large delivers around $1.7\times$ better performance.
In any case,
we note again that we implement the same algorithms on both devices,
and the performance in NG-Large can be greatly improved 
by exploiting its bigger capacity
and increasing the parallelization,
e.g., implement parallel arithmetic operators or process stripes in parallel.

Concluding,
in comparison with the 3rd-party FPGAs,
NG-Large exhibits comparable,
or even better in several cases,  
resource utilization
(depending on the kernel and the FPGA block).
Considering that NXmap is a newly released tool
and that we have seen continuous improvement in every new tool version,
the results are expected to improve in the future.
Compared to NG-Medium,
NG-Large provides increased flexibility due to its bigger capacity, 
while the former
is stressed (or is unable) to implement CV kernels
for typical image sizes, e.g., 1 MPixel.
NG-Large delivers better performance 
that
can be further improved
by exploiting its resources to increase the parallelization.
Furthermore, 
NG-Large leaves a significant amount of resources
that can be used for implementing other complementary HDL components in the case of algorithmic pipelines,
as shown in the system-level evaluation 
of our publication in \cite{LeonACCESS}. 

\begin{table}[!t]
\fontsize{9}{10}\selectfont
\renewcommand{\arraystretch}{1.2}
\setlength{\tabcolsep}{0pt}
\caption[Comparison of NanoXplore's FPGAs for the Implementation of Computer Vision Kernels]{Comparison of NanoXplore's FPGAs for the implementation of computer vision kernels.}
\label{tb_bravecomp}
\centering
\begin{threeparttable}
\begin{tabular*}{\textwidth}{
@{\extracolsep{\fill}}l
@{\hspace{15pt}}l
@{\hspace{5pt}}|
@{\hspace{0.00cm}}c
@{\hspace{6pt}}|
@{\hspace{0.00cm}}c
@{\hspace{0.00cm}}c
@{\hspace{0.00cm}}c
@{\hspace{0.00cm}}c
@{\hspace{0.00cm}}c
@{\hspace{0.00cm}}c
@{\extracolsep{\fill}}}
\hline
\multicolumn{1}{c}{\setcounter{footnote}{0}\textbf{Kernel}\footnotemark} & \hspace{10pt}\textbf{FPGA} & 
\setcounter{footnote}{1}\hspace{4pt}\textbf{NG-Med. vs. NG-Lar.}\footnotemark &
\textbf{LUT} &
\textbf{DFF} &
\textbf{CY}  &
\textbf{DSP} &
\textbf{RAMB}  &
\textbf{MHz}  \\
\hline \hline
\parbox[t]{8mm}{\multirow{2}{*}{\rotatebox[origin=c]{30}{Canny}}} &
NG-Large & 4$\times$ smaller input image & 2\% & 2\%  & 4\% & 1\%  & 93\% & 38  \\
& NG-Medium & (0.25MP vs. 1MP) & 7\% & 8\% & 13\% & 5\% & 90\% & 50  \\

\parbox[t]{8mm}{\multirow{2}{*}{\rotatebox[origin=c]{30}{Harris}}} &
NG-Large  & 4$\times$ smaller img. partition & 5\% & 13\%  & 23\% & 22\%  & 36\% & 40   \\
& NG-Medium & (1K$\times$8 vs. 1K$\times$32) & 19\% & 38\% & 73\% & 100\% & 63\% & 11 \\

\parbox[t]{8mm}{\multirow{2}{*}{\rotatebox[origin=c]{30}{Disparity}}} &
NG-Large & 2$\times$ smaller img. partition & 1\% & 3\%  & 15\% & 2\%  & 45\% & 50   \\
& NG-Medium & (1K$\times$16 vs. 1K$\times$32) & 4\% & 10\% & 18\% & 41\% & 86\% & 52  \\

\parbox[t]{8mm}{\multirow{2}{*}{\rotatebox[origin=c]{30}{Ssweep}}} &
NG-Large & 4$\times$ smaller input image & 5\% & 8\%  & 20\% & 14\%  & 42\% & 52   \\
 & NG-Medium & (0.25MP vs. 1MP) & 22\% & 32\% & 78\% & 45\% & 95\% & 30 \\
\hline
\end{tabular*}
\begin{tablenotes}
  \item[1]{\fontsize{7.7}{8.8}\selectfont The same algorithms are implemented in both FPGAs (no HDL modification).}
  \item[2]{\fontsize{7.7}{8.8}\selectfont NG-Large is 4$\times$ bigger than NG-Medium in resources.}
\end{tablenotes}
\end{threeparttable}
\end{table}

\section{Demonstration of the Computer Vision Kernels}
\label{s8_6}

To evaluate 
and demonstrate 
the actual hardware execution of the CV kernels on NG-Large,
we develop a hardware/software architecture
that is based on the serial UART communication.
Our goal is
to transmit and receive I/O data to/from NG-Large
and validate the results,
and not to provide the optimal I/O and system throughput
(which can be achieved via the high-performance SpaceWire interface).
For this purpose,
we develop software in our host-PC for I/O data handling and demonstration,
as well as hardware in NG-Large for I/O data handling and kernel control.

\subsection{Development of CPU--FPGA Communication}

\begin{figure}[!t]
\vspace*{8pt}
\centering
\hspace{9pt}\includegraphics[width=0.99\textwidth]{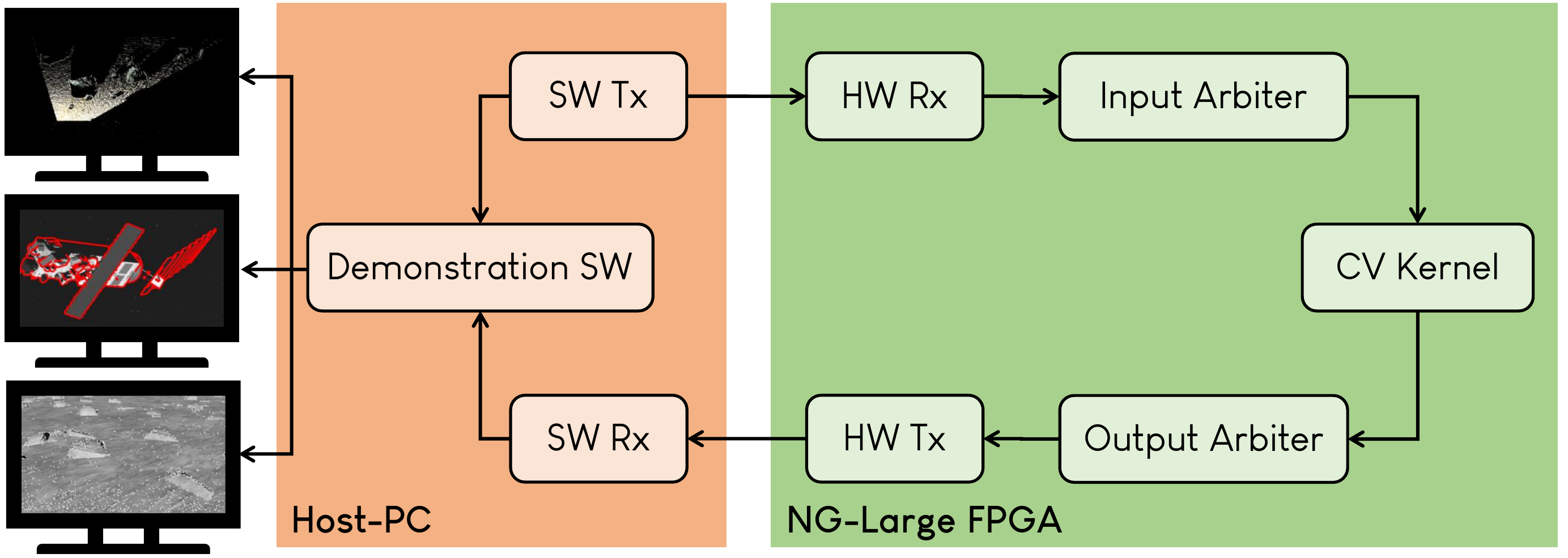}%
\caption[Hardware/Software Architecture for the Execution of Computer Vision Kernels on NG-Large]{Hardware/Software UART-based architecture for the execution of the computer vision kernels on NG-Large (testing and demonstration).}%
\label{fig_uart}
\end{figure}

To establish a communication for data transfers between the host-PC and NG-Large, 
we adopt the infrastructure of Figure \ref{fig_uart}.
In the host-PC side, 
we develop C functions using the GCC compiler,
which are tailored to the
data sizes and bit-widths
of each kernel.
In brief, the host-PC functions perform the following  operations:
\begin{itemize}[]
    \item open the USB port for the serial communication.
    \item specify the settings for the communication, e.g., baud rate, timeouts, blocking or non-blocking mode.
    \item read the input data (stored in the host-PC) and send them to NG-Large by applying the appropriate bit manipulations and writing the USB port.
    \item receive the output data from the NG-Large execution by reading the USB port.
    \item write the output data to files for comparison with the groundtruth data and demonstration.
\end{itemize}
In the FPGA side, 
we implement a hardware architecture for receiving/transmitting 
data from/to the host-PC,
as well as for applying the required data encoding/decoding 
and controlling the kernel's I/Os. 
The development is performed in pure VHDL,
and the architecture is parametric in terms of baud rate and global clock frequency.
The UART modules are implemented based
on the protocol specifications: 
the Rx/Tx component serially receives/transmits the data,
while the tick generator 
synchronizes the UART logic with respect to the baud rate.
The arbiters for controlling the 
kernel I/Os,
i.e.,
the glue logic between UART and the kernel,
are based on the kernel's architecture 
(e.g., processing in image stripes or in burst mode)
and I/O signals.
In general,
the input arbiter 
collects the data from the Rx component, 
packs them accordingly, 
and feeds the kernel when it is ready to start the processing.
Correspondingly, 
the output arbiter 
retrieves the kernel's outputs, 
decouples them to packets, 
and forwards them to the Tx component 
when the latter is ready to transmit them to the host-PC. 

\begin{figure}[!t]
\vspace*{13pt}
\centering
\subfloat[\label{fig_acanny}]{\includegraphics[width=0.95\textwidth]{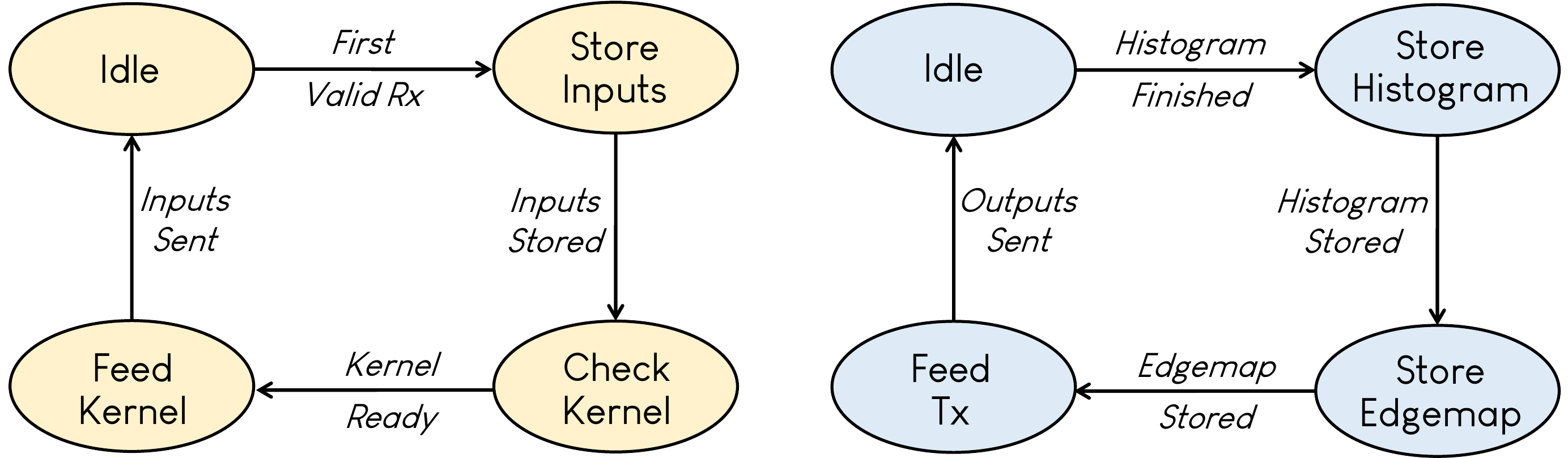}}\\
\subfloat[\label{fig_aharris}]{\includegraphics[width=0.95\textwidth]{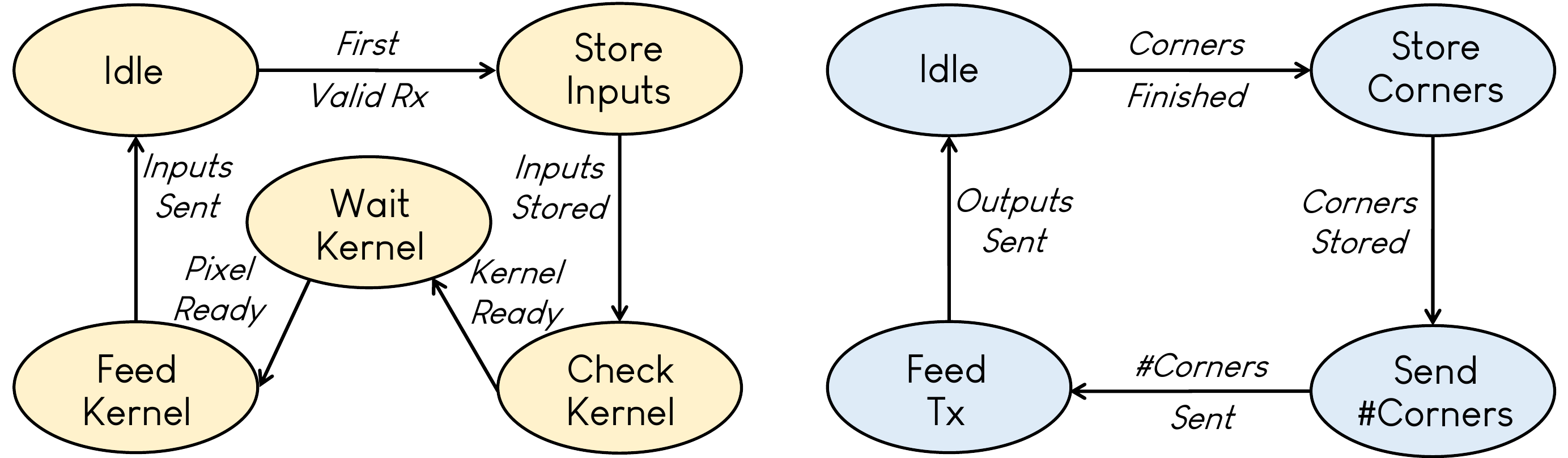}}
\caption[Arbiters
for Handling the I/O Data of Computer Vision Kernels on NG-Large]{Input and output arbiter
for handling the I/O data of
\textbf{(a)} Canny
and
\textbf{(b)} Harris.
The corresponding arbiters for Disparity and Spacesweep are similar to those of Harris.}%
\label{fig_arbi}
\end{figure}

Figure \ref{fig_arbi} illustrates 
the main finite-state machines
of Canny's and Harris' arbiters.
The arbiters of Disparity and Spacesweep
are similar to those of Harris.
At first,
Canny receives the two thresholds
required for the hysteresis thresholding task,
and then it receives the entire image. 
The input arbiter remains in the ``Idle'' state 
until the flag signal for valid data reception 
is activated
for the first time 
by the Rx component,
which means that
the first threshold is successfully received.
This threshold is stored in a register
and the arbiter moves to the state ``Store Inputs''. 
It remains there 
until the second threshold 
and all the image pixels are received and stored in register and memory bank,
respectively.
Next, in the state ``Check Kernel'', 
the arbiter examines if Canny 
is ready to receive the input data. 
When that happens, 
the controller moves to the 
``Feed Kernel'' state,
where it sends the 
thresholds and pixels to Canny 
and 
activates all the signals
required for starting the processing of a new image. 
Upon finishing this process, 
it returns to the ``Idle'' state, 
waiting for the next image.
Correspondingly,
the output arbiter remains in the ``Idle'' state until Canny activates 
its output signals indicating that
both the histogram and the edge map
have been calculated.
When that happens,
the arbiter starts to store the histogram and edgemap in memory banks,
before sending them to the Tx component.
This data storing is required,
as the Tx component is not always ready
to transmit the outputs to the host-PC.
Therefore,
the output memory banks are read
when Tx is ready for transmission.
When all the outputs are sent to Tx,
the arbiter returns to its ``Idle'' state 
to wait for the next Canny outputs. 
In the same context,
we design the arbiters of Harris,
which, however, transfer and process
the image in stripes.
Moreover,
before starting the processing of a new image,
we feed Harris with a new threshold,
which is updated in the software 
based on the previous frame's corners.
Another difference is that Harris
produces $32$-bit output data, 
while the UART communication is designed for
$8$-bit data.
Therefore, 
we transmit the outputs of Harris in bytes, 
and perform the required bit manipulations
in the software
receiving the data from the FPGA.

\subsection{Real-Time Processing and Visualization}

\hypersetup{urlcolor=blue}

We assume the real-world
scenario of 
space missions,
where NG-Large is equipped on-board. 
The camera captures random images during the traversal of the rover,
and NG-Large detects features on them 
using Canny and Harris,
i.e., edges and corners,
respectively.
These features are useful
in VBN pipelines
for pose tracking \cite{lentaris_tvideo}, 
rover localization \cite{odometry},
and other space applications. 
On the other hand,
Disparity and Spacesweep
are used
to extract the depth of the scene
when the camera 
captures stereo images 
(e.g., every $20$cm)
\cite{disparity, single_multi_rovers}. 

\begin{figure}[!t]
\centering
\subfloat[\label{fig_dharris}]{\includegraphics[width=0.75\textwidth]{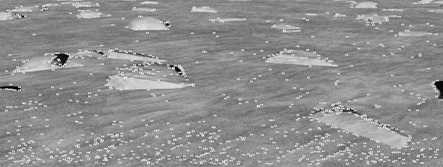}}\\
\subfloat[\label{fig_ddispar}]{\includegraphics[width=0.75\textwidth]{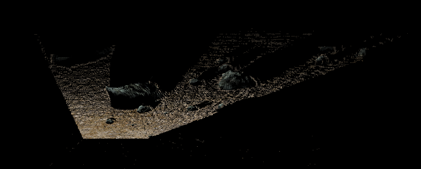}}
\caption[Visualization of the Execution of Computer Vision Kernels on NG-Large]{Visualization of results from the execution of computer vision kernels on NG-Large:
\textbf{(a)} Harris Corner Detector \cite{odometry}
and
\textbf{(b)} GAD-Disparity \cite{disparity}.}%
\label{fig_demos}
\end{figure}

For demonstration purposes,
we visualize
the output results of the CV kernels.
For the
feature detection kernels
(Canny and Harris),
we use both 
real-time camera input
and stored images,
and we plot the edges and corners on them.
For the depth extraction kernels
(Disparity and Spacesweep), 
we use $20$ synthetic stereo images depicting the Martian terrain 
and visualize the 
data in 2D and 3D. 
Figure \ref{fig_demos} illustrates 
output images from the execution of the kernels on NG-Large. 
Our demonstration video is also
available online\footnote{\fontsize{7.7}{8.8}\selectfont Demo available on YouTube: \textbf{\url{https://youtu.be/q8NKV4rpcY4}}.}.
As explained in the previous section,
we aim only to demonstrate the correct hardware execution
and not evaluate the system throughput. 
The serial UART communication does not
provide increased I/O rate,
and thus,
the total system throughput is low.
Considering the processing throughput of the CV kernels (see Table \ref{tb_nxperf}),
the use of the high-performance SpaceWire
interface (e.g., at $100$ Mbps) would provide sufficient system throughput. 

\hypersetup{urlcolor=black}

\section{Conclusion}
\label{s8_7}

In this chapter,
we examined the acceleration capabilities of the new European space-grade BRAVE FPGAs.
In particular,
we employed high-performance CV kernels
that were developed in other FPGA devices/tools,
and we systematically ported them on   NanoXplore's NG-Large FPGA.
Towards exploiting all the available software tools settings,
as well as for surpassing the issues and inefficient results
of such new devices/tools,
we proposed a three-stage 
development \& assessment methodology.
The methodology
regards the typical stages of the FPGA design flow,
i.e.,
synthesis, 
place \& route (implementation),
and bitstream generation \& hardware execution,
and it can be used for testing and development on new FPGAs.
In our work,
we used it
to efficiently port the high-performance CV kernels on NG-Large 
and evaluate this new space-grade chip
as on-board processor for space missions. 
Our experimental evaluation
examined the resource utilization, performance, power consumption,
bitstream size, and configuration time,
which are all taken into account 
in space mission scenarios.
It also included comparisons with well-established FPGA devices/tools. 
Besides measuring the flexibility and efficiency of the
new NXmap tool,
which is continuously evolving and improving,
we showed that NG-Large could achieve feature extraction 
with a throughput of up to $10$ FPS 
and depth extraction with a latency of around $10$s.
Overall,
NG-Large could implement high-performance and complicated algorithms 
with sufficient hardware metrics,
i.e., resource utilization, throughput, and power consumption,
which were all shown to be competitive/comparable 
with the 3rd-party FPGA designs. 
\chapter{DSP \& AI Acceleration on Heterogeneous Multi-Core SoCs}
\label{chapter9}

\addtocontents{lof}{\protect\contentsline{chapter}{\protect\numberline{9}DSP \& AI Acceleration on Heterogeneous Multi-Core SoCs}{}{}}
\addtocontents{lot}{\protect\contentsline{chapter}{\protect\numberline{9}DSP \& AI Acceleration on Heterogeneous Multi-Core SoCs}{}{}}

\begin{ChapterAbstract}

The increased computational demands
of modern applications from domains such as Artificial Intelligence (AI)
and Digital Signal Processing (DSP)
challenges their deployment at the edge.
At the same time,
embedded systems are presented 
with tight energy constraints,
which limit their processing capabilities
compared to the high-performance computers of the cloud. Heterogeneous System-on-Chip (SoC) processors emerge
as an attractive hardware solution, 
however, they still require sophisticated development to
provide efficient implementations.
In this chapter,
we exploit the heterogeneity of
the multi-core Vision Processing Units (VPUs),
which are low-power SoCs with increased diversity in processors and memories,
to accelerate demanding DSP and AI algorithms. 
Towards the efficient utilization
of such heterogeneous and complex SoCs,
we employ a design methodology
and various high- and low-level techniques.
Our implementations include
custom DSP kernels,
as well as 
a Computer Vision (CV) pipeline
and a demanding Deep Neural Network (DNN),
which both perform pose estimation in space.
The individual kernels 
are accelerated by $\mathit{10\times}$--$\mathit{20\times}$
on the Myriad 2 VPU,
while
for the CV pipeline,
we provide a speedup of $\mathit{8.5\times}$--$\mathit{12\times}$.
The throughput of this pipeline for MPixel frames
is up to $\mathit{5}$ FPS,
while the power consumption lies between 
$\mathit{0.8}$W and $\mathit{1.1}$W.
The deployment of the DNN on Myriad X for resampled MPixel frames
provides a throughput of $\mathit{2.7}$ FPS
and a maximum power consumption of $\mathit{2}$W.
Moreover,
we directly compare the VPUs
to other embedded devices.
According to our analysis, 
the Myriad VPUs exhibit significantly better power efficiency,
i.e., $\mathit{5\times}$ versus the Jetson Nano GPU and $\mathit{4\times}$ versus the Zynq FPGA.
When examining the performance-per-Watt ratio of the devices,
the VPUs provide comparable results, 
and in some cases, even better:
e.g., 
for the CV pipeline,
Myriad 2 trades a $\mathit{3\times}$ loss in speed
for a $\mathit{4\times}$ gain in mean power consumption.\\
This chapter is based on our
\textbf{publications} in \textbf{\cite{LeonTECSm, LeonICECSm, LeonM2SOC, LeonELSI}}.
\end{ChapterAbstract}

\newpage 

\section{Introduction}
The last decade is characterized by a rapid growth of 
powerful Artificial Intelligence (AI) and 
complex Digital Signal Processing (DSP) algorithms.
This algorithmic evolution 
along with the large increment in sensor data
have significantly affected
computing systems at the edge. 
The worldwide demand for speed
challenges the integration of 
AI/DSP functionalities in novel applications, 
especially at the power-constrained embedded systems.
Heterogeneous System-on-Chip (SoC) processors 
emerge as a promising solution \cite{heterog},
offering increased programming flexibility
and diversity in terms of processors and storage. 
Besides the well-established SoCs
integrating Central Processing Units (CPUs)
and Graphics Processing Units (GPUs),
a novel class of heterogeneous SoCs has recently appeared,
namely
the Vision Processing Unit (VPU) \cite{myriad_vpu, paralos}.
Compared to the CPU--GPU SoCs,
the VPUs provide better power efficiency.
For example, 
Nvidia's Jetson Nano GPU consumes $10$W/$5$W,
while Intel's Myriad VPUs consume $1$W--$2$W.
Compared to the Field-Programmable Gate Arrays (FPGAs),
the VPUs offer improved programmability, 
smaller development time/effort, 
and lower power consumption.
Moreover,
the VPUs can handle classic DSP workloads,
contrary to AI-specific processors,
such as Google's Tensor Processing Units (TPUs). 
In general, 
the VPUs integrate a variety of processors,
e.g., general-purpose cores,
hardware filters, 
vector cores,  
and neural network accelerators. 
The VPU SoCs excel in low-power imaging applications,
covering domains such as robotics, automotive, and space.

Intel manufactures the Myriad VPUs \cite{intelvpu},
i.e., Myriad 2 (28nm) and Myriad X (16nm), 
which
are prominent and well-established processors
for embedded DSP \& AI applications. 
By offering
numerous heterogeneous processing cores, 
these VPUs enable the
efficient parallelization of demanding algorithms. 
A single VPU chip integrates 
multiple hardware I/O peripherals,
a variety of hardcoded low-power imaging filters,
general-purpose processors,
acceleration vector cores, 
and AI acceleration engine(s).
A similar heterogeneity is provided in storage,
for which the VPU SoCs
offer a
memory hierarchy consisting of 
DRAM, scratchpad, and cache memories.
It is evident
that the full utilization and exploitation
of such complex heterogeneous SoCs
requires a meticulous and systematic development approach.
Furthermore,
considering that these SoCs are build 
towards more ``low-power''
than ``high-performance'',
the need for efficient mapping and deployment 
becomes even more crucial.

As explained in Chapter \ref{chapter8}, 
space is one of the communities
searching for alternative processing platforms 
to comply with the tight constraints of on-board processing.
Besides the FPGAs,
the enhanced performance of modern low-power edge devices 
can efficiently serve tasks of  
Earth Observation (EO) and Vision-Based Navigation (VBN).
The heterogeneity of platforms such as the VPUs,
allows for 
improved adaptability to various mission scenarios
and seamless in-flight re-programmability.
To further improve 
the performance and Size, Weight and Power (SWaP),
as well as additional 
costs (e.g., development effort),
the space industry is studying 
mixed-criticality architectures 
\cite{architectures, criticality2, criticality1, vlsisoc_cots},
i.e., the integration of both 
space-grade and Commercial-Off-The-Shelf (COTS) components.
The use of COTS components 
in Low Earth Orbit (LEO) missions and CubeSats 
relies on 
the partial shielding provided by Earth’s magnetosphere
and/or the short mission lifetime,
which limit the damage or unavailability of electronics
due to radiation. 
In this context,
FPGAs \cite{mpsoc, lentaris_tvideo, iturbe, csp}, 
GPUs \cite{gpu4s, gpu_kosm, gpu_fpga, gpu_fpga_vpu, nano1} and 
VPUs \cite{fsat_myriad, navarro, star_m2, ubo0100, nasa_mx}
are evaluated as accelerators,
while they are also subjected to radiation tests,
such as the Myriad 2 VPU \cite{ai_space}.
A second challenge for the space industry 
is the wider adoption of AI,
which is currently limited to 
offline/ground data processing
and not on-board processing,
mostly due to insufficient computational power
and increased memory footprint,
as well as 
qualification issues when deployed in orbit
\cite{ai_space}.

The promising features of \emph{Myriad VPUs}
has attracted the interest of 
the European Space Agency (ESA),
which 
is thoroughly involved in the safari of COTS embedded devices. 
Towards the use of VPUs in space,
ESA is supporting research
activities\footnote{\fontsize{7.7}{8.8}\selectfont\textbf{ESA LEOTOME:}
4000126083/18/NL/FE\\\fontsize{7.7}{8.8}\selectfont\textbf{ESA HPCB:} 4000126129/18/NL/AF}
that
evaluate the overall performance of the Myriad SoCs,
including their integration in high-performance compute boards and mixed-criticality architectures for space avionics.
These activities aim to assess the suitability of the VPUs as COTS parts 
of the on-board computer,
mainly for low-power DSP/AI acceleration.
In the context of these activities,
we propose a methodology
to support the development on the Myriad VPUs
and accelerate demanding DSP/AI workloads.
Our goal is to 
surpass the bottlenecks of the resource-constrained embedded computing,
unlock the full potential of VPU's
heterogeneity,
and thus,
efficiently deploy complex algorithms.
Our application domain is space,
however, 
both our methodology and development techniques
are generic,
i.e., they can be used as design paradigm
to provide DSP/AI acceleration
on the heterogeneous multi-core VPUs. 
Regarding development,
we apply various high-level parallelization and partitioning techniques,
while at lower level,
we optimize the memories
and the acceleration cores. 
In terms of algorithms,
at first,
we implement custom DSP and AI kernels on Myriad 2,
targeting to demonstrate the VPU's capabilities
and evaluate embedded design techniques for parallelization and optimization.  
Afterwards,
we accelerate a sophisticated 5-stage Computer Vision (CV) pipeline
(developed by Lourakis and Zabulis \cite{lour_zab}\footnote{\fontsize{7.7}{8.8}\selectfont Special thanks to
Dr. M. Lourakis \& Dr. X. Zabulis from FORTH for providing the initial CV SW.})
on Myriad 2. 
Finally,
we accelerate a compute-intensive Deep Neural Network (DNN) \cite{urso}
(that was not developed for embedded systems)
on the AI acceleration engine of Myriad X. 

The \textbf{contribution} of this chapter is summarized as follows:

\begin{itemize}[]
\item[(i)] We highlight the great diversity of heterogeneous SoCs in terms of processors and storage,
as well as the significance of the meticulous exploitation
of the heterogeneity 
towards improved performance. 
\item[(ii)] We propose a methodology for the optimized
mapping and acceleration 
of demanding DSP and AI algorithms on heterogeneous multi-core VPUs,
while we demonstrate several high- and low-level implementation techniques for embedded systems.
\item[(iii)] We report comparative experimental results for embedded CPUs, VPUs, GPUs and FPGAs, and we discuss the trade-offs of each processor. 
\item[(iv)] We test and evaluate Intel's Myriad VPUs
as candidate on-board COTS accelerators for future space missions. 
\end{itemize}

The remainder of this chapter is organized as follows. 
Section \ref{s9_2} overviews the market's embedded platforms
and presents the Myriad VPUs. 
Section \ref{s9_3} introduces our design methodology.
Sections \ref{s9_4}--\ref{s9_6}
discuss the implementation details 
of the DSP and AI algorithms. 
Section \ref{s9_7} reports various experimental results. 
Finally, 
Section \ref{s9_8} draws the conclusions.

\section{Background}
\label{s9_2}

\subsection{The Landscape of Embedded Devices}

Table \ref{tb_embeds}
summarizes well-established embedded devices of the market,
including the Myriad VPUs.
We report platforms
with power consumption lying 
within the same order of magnitude
(e.g., there are other Nvidia Jetson devices consuming more power). 
As shown,
Intel's VPUs integrate LEON general-purpose processors
along with the RTEMS real-time operating system.
On the other hand,
Nvidia's GPUs and Google's TPUs feature ARM processors
and are equipped with Linux-based operating systems.
In terms of accelerators,
the VPUs have more heterogeneity 
than the other devices,
offering multiple
Streaming Hybrid Architecture Vector Engine (SHAVE)
cores
and hardware filters,
while Myriad X also integrates a neural engine. 
The GPUs are based on the Compute Unified Device Architecture (CUDA) cores for acceleration,
whereas the key processing unit of TPUs
is a systolic array with multipliers and accumulators.
The AI performance of these devices varies 
and shows that the TPUs are the winners,
however, these metrics are theoretical
and should also be examined along the AI accuracy
and the actual power consumption.
Moreover,
a disadvantage of TPUs is that they 
are AI-specific accelerators, 
and thus,
they do not provide acceleration cores for classic DSP and CV workloads.
Regarding power,
the VPUs are the most efficient solution,
as they consume up to $2$W. 
The TPU chip consumes $0.5$W per Tera Operation Per Second (TOPS),
however, 
the power of the entire system is 
\raisebox{0.8pt}{$\scriptstyle\sim$}$5$W.
Especially for AI,
the VPUs and the GPUs support various development frameworks, 
e.g., TensorFlow and PyTorch,
while the TPUs rely only on TensorFlow Lite.
Finally,
both the VPUs and TPUs
are produced as USB accelerators that can be hosted on PC or single-board computer 
(e.g., Raspberry Pi). 

\begin{table}[!t]
\fontsize{9}{10}\selectfont
\renewcommand{\arraystretch}{1.2}
\setlength{\tabcolsep}{0pt}
\caption[Overview of Market's Embedded Devices]{Overview of market's embedded devices.}
\label{tb_embeds}
\centering
\begin{threeparttable}
\begin{tabular*}{\textwidth}{@{\hspace{2pt}}
c@{\hspace{4pt}}
l@{\hspace{4pt}}|@{\hspace{4pt}}
c@{\hspace{0.00cm}}
c@{\hspace{2pt}}}
\hline 
\textbf{Vendor} & \hspace{6pt} \textbf{Device} & \textbf{CPU} & \textbf{Accelerators}  \\
\hline
\hline
\parbox[t]{8mm}{\multirow{2}{*}{\rotatebox[origin=c]{38}{Intel}}} 
& Myriad 2         & LEON4 ($\times 2$)  & $12$-Core SHAVE, HW Filters   \\
& Myriad X         & LEON4 ($\times 2$)  & $16$-Core SHAVE, AI Engine, HW Filters  \\
\parbox[t]{8mm}{\multirow{2}{*}{\rotatebox[origin=c]{38}{Nvidia}}} 
& Jetson Nano      & Cortex-A57             & $128$-Core Maxwell GPU       \\
& Jetson TX2       & Cortex-A57, Denver 2   & $256$-Core Pascal GPU          \\
\parbox[t]{8mm}{\multirow{2}{*}{\rotatebox[origin=c]{38}{Google}}} 
& Coral Mini & Cortex-A35             & $64$$\times $$64$-Array\setcounter{footnote}{0}\footnotemark\phantom{i}Edge TPU          \\
& Coral      & Cortex-A53, Cortex-M4F & $64$$\times $$64$-Array\setcounter{footnote}{0}\footnotemark\phantom{i}Edge TPU  \\
\hline
\end{tabular*}
\vspace*{8pt}
\begin{tabular*}{\textwidth}{@{\hspace{2pt}}
c@{\hspace{4pt}}
l@{\hspace{4pt}}|@{\hspace{4pt}}
c@{\hspace{11pt}}
c@{\hspace{11pt}}
c@{\hspace{11pt}}
c@{\hspace{0pt}}}
\hline 
\textbf{Vendor} & \hspace{6pt} \textbf{Device} & \textbf{OS} & \textbf{Max Clock} & \textbf{AI Performance}\setcounter{footnote}{1}\footnotemark & \textbf{Power}\setcounter{footnote}{2}\footnotemark  \\
\hline
\hline
\parbox[t]{8mm}{\multirow{2}{*}{\rotatebox[origin=c]{38}{Intel}}} 
& Myriad 2       &  RTEMS & $600$MHz & $100$ GFLOPS (fp16) & $1$W   \\
& Myriad X       &  RTEMS & $700$MHz & $1$ TFLOPS (fp16) & $2$W \\
\parbox[t]{8mm}{\multirow{2}{*}{\rotatebox[origin=c]{38}{Nvidia}}} 
& Jetson Nano    &  Linux4Tegra & $922$MHz & $472$ GFLOPS (fp16) & $10$W / $5$W      \\
& Jetson TX2     &  Linux4Tegra & $1122$MHz & $1.3$ TFLOPS (fp16) & $15$W / $7.5$W         \\
\parbox[t]{8mm}{\multirow{2}{*}{\rotatebox[origin=c]{38}{Google}}} 
& Coral Mini     &  Mendel Linux & $500$MHz & $4$ TOPS (int8) & $5$W        \\
& Coral          &  Mendel Linux & $500$MHz & $4$ TOPS (int8) & $5$W \\
\hline
\end{tabular*}
\begin{tablenotes}
  \item[1]{\fontsize{7.7}{8.8}\selectfont Estimation based on AI performance and testing (the array dimensions have not been revealed).}
  \item[2]{\fontsize{7.7}{8.8}\selectfont Reported officially by the vendors.}
  \item[3]{\fontsize{7.7}{8.8}\selectfont According to our in-house measurements.}
\end{tablenotes}
\end{threeparttable}
\end{table}

These devices are used for accelerating compute-intensive DSP and AI workloads at the edge.
Because of their increased programming complexity 
and performance issues
compared to their desktop and data center counterparts,
they have received significant research attention.
More specifically, 
the literature 
includes several methodologies and techniques
for the development on these devices 
and provides a plethora of benchmarking/evaluation results. 
There are also relevant works that propose new frameworks/libraries
and co-processing embedded architectures. 
As our application domain is space,
we report related work on the aforementioned devices.

The Myriad VPUs are systematically evaluated
by the space community. 
Furano \emph{et al.} \cite{ai_space} 
report results from the radiation tests on Myriad 2,
involving various functional tests   
and characterization of all SoC's memories. 
Their results show that 
Myriad 2 remains fully functional
after being exposed to a total ionizing dose of 49 krad(Si).
Agarwal \emph{et al.} \cite{star_m2}
use Myriad 2 
to implement a star identification neural network. 
Their experimental evaluation shows that 
Myriad 2 provides sufficient performance,
while it consumes around $1$W 
and retains $99\%$ accuracy.
Myriad 2 is also integrated in custom boards and co-processing architectures.
This VPU is equipped on-board in the
$\mathrm{\Phi}$-{S}at-1 CubeSat mission of ESA
as DNN demonstrator for EO 
\cite{fsat_myriad}.
Furthermore, 
it is integrated in 
Ubotica's CogniSat platform \cite{ubo0100}, 
which is an 
AI inference and CV engine that exposes 
Myriad 2 to the payload developer.
In the same context, 
Myriad 2 is the main accelerator 
of the HPCB platform \cite{navarro}, 
which is a payload data processor board 
provided by Gobham Gaisler.
Interestingly,
the HPCB includes three Myriad 2 SoCs
to provide fault tolerance and increased performance. 
For the successor of Myriad 2,
i.e., Myriad X, 
the authors of \cite{nasa_mx} 
report results from the
deployment of neural networks classifiers. 
The networks are 
trained on Mars imagery from the Reconnaissance Orbiter and Curiosity rover,
and the average inference time is $16$-$20$ms,
while the power consumption lies around $1.9$W.

Regarding the use of embedded GPUs in the space domain, 
Kosmidis \emph{et al.} \cite{gpu4s}
examine their applicability
from both the software and hardware perspectives. 
In particular,
they analyze the algorithms and workloads of space applications
to identify their suitability for GPUs,
and they also perform benchmarking on GPUs. 
In \cite{gpu_kosm},
the authors evaluate two graphics-based computing methodologies
(OpenGL 2.0 and Brook Auto)
for safety-critical systems. 
Their main benchmark is an application modeling a 
VBN scenario where the aircraft performs rendez-vouz
with an object. 
Moreover, 
the literature includes various FPGA--GPU 
co-processing architectures.
The hybrid FPGA-GPU architecture of \cite{gpu_fpga}
employs 
 Nvidia's TX2/TX2i GPU as main accelerator.
The heterogeneous architecture of \cite{gpu_fpga_vpu}
integrates the AMD SoC (CPU--GPU) for acceleration,
and optionally, a VPU for AI deployment.
The authors of \cite{nano1} 
use Nvidia's Jetson Nano GPU
to accelerate neural networks for object detection
along with image compression techniques. 
Finally,
the edge TPUs
are used in several terrestrial applications \cite{tpu1, tpu2},
however, they are still not adopted in the space domain
(ESA is working towards this direction \cite{vlsisoc_tpu}\setcounter{footnote}{2}\footnote{\fontsize{7.7}{8.8}\selectfont\textbf{ESA CAIRS21:}
4000135491/21/NL/GLC/ov}). 
Very recently, 
a CubeSat-sized co-processor with three TPUs was introduced \cite{tpu_spacecube}, 
supporting 
various operation modes 
(high-performance, fault-tolerant, low-power). 

\subsection{The Intel VPUs and Tools}
\label{s922}

\begin{figure}[!t]
\vspace*{-4pt}
\centering
\includegraphics[width=0.94\textwidth]{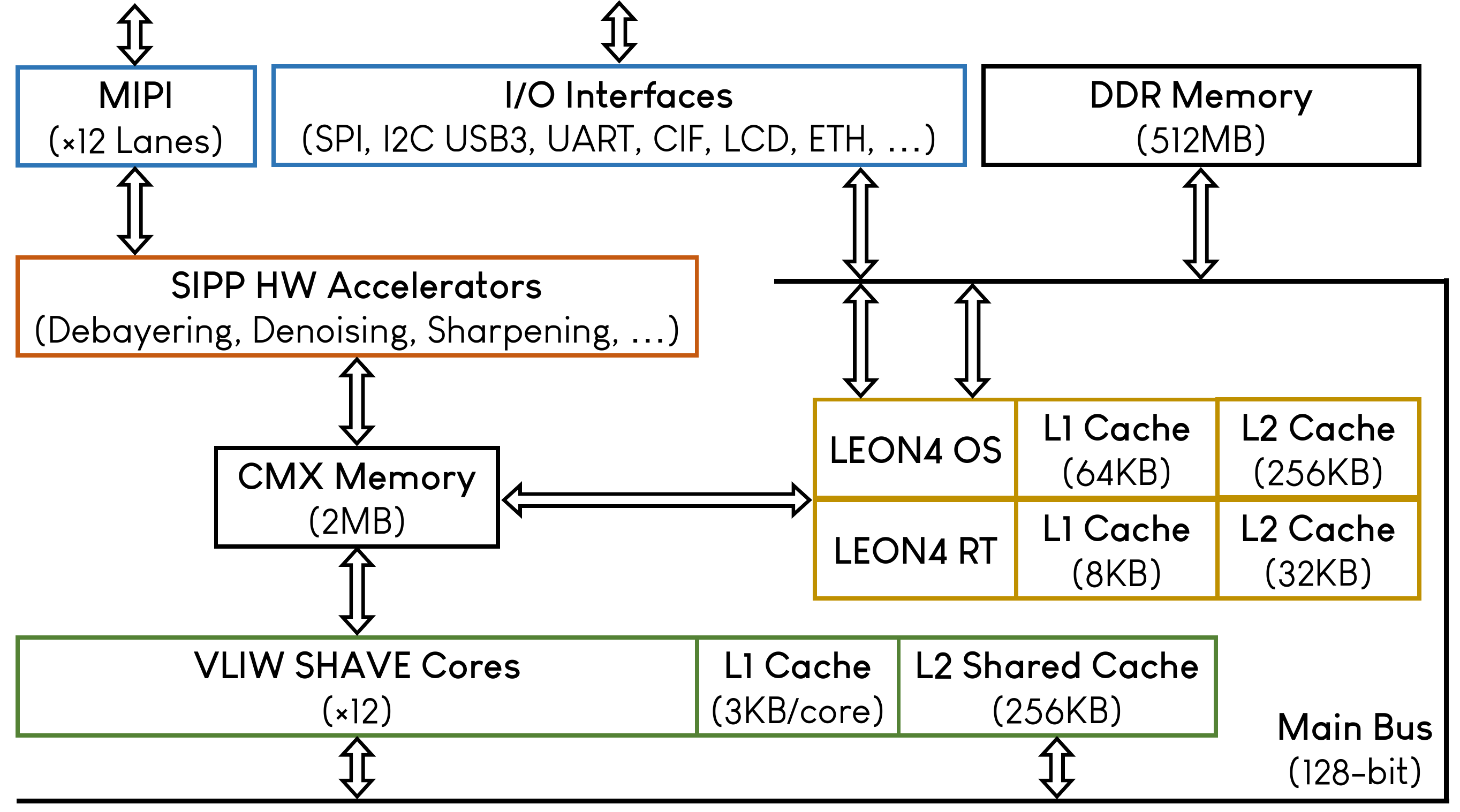}%
\caption[Fabric Architecture of Intel's Myriad 2 VPU]{The fabric architecture of Intel's Myriad 2 VPU  \cite{intelvpu}.}%
\label{fig_m2_new}
\vspace*{10pt}
\includegraphics[width=0.94\textwidth]{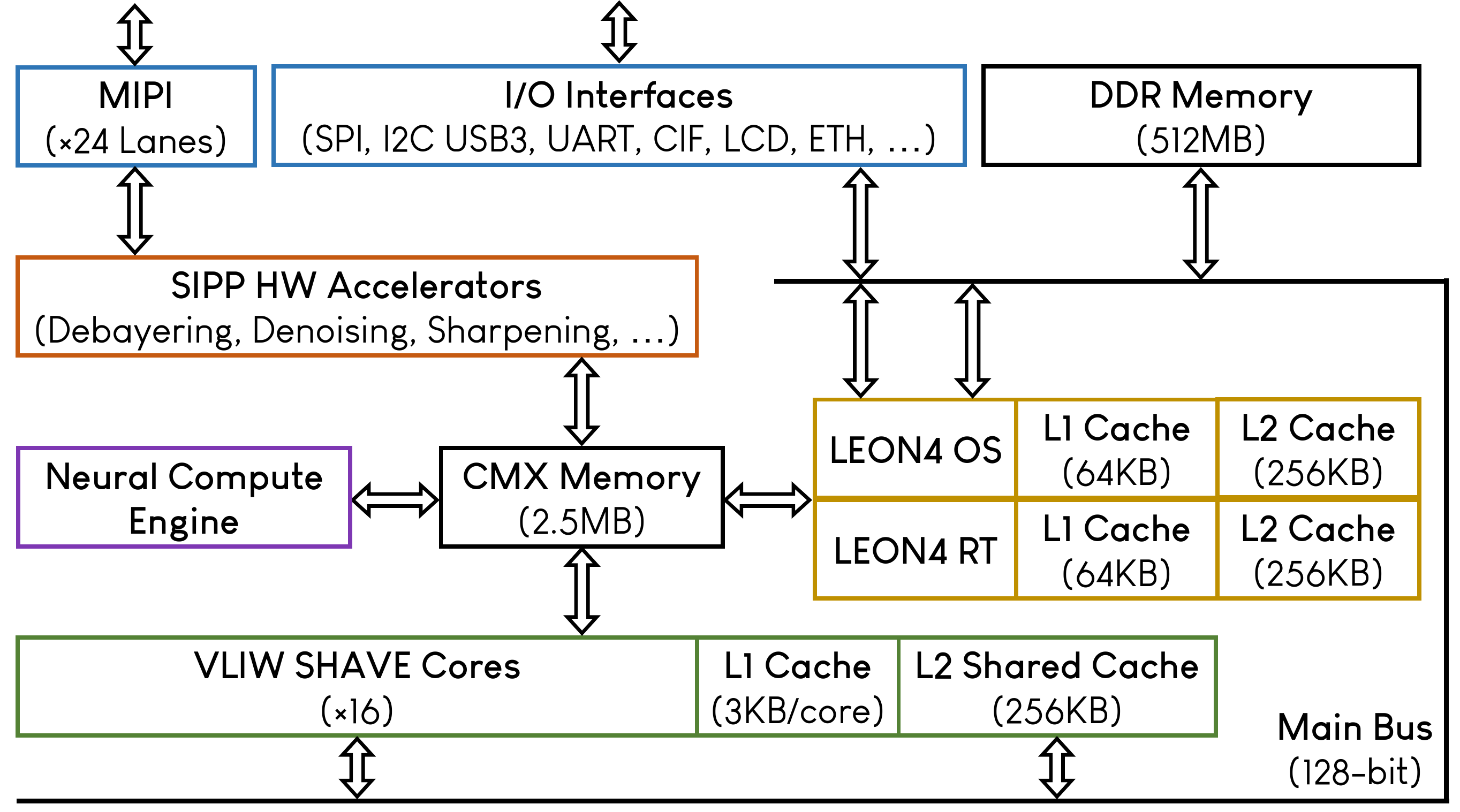}%
\caption[Fabric Architecture of Intel's Myriad X VPU]{The fabric architecture of Intel's Myriad X VPU \cite{intelvpu}.}%
\label{fig_mx_new}
\vspace*{-6pt}
\end{figure}

The Myriad family of VPUs 
offers heterogeneous multi-core SoCs 
for mobile/embedded applications. 
Besides the space domain,
the Myriad VPUs are used
for implementing 
DNNs \cite{myriad_lazdac, myriad_laztcad, movidius_ncs, movidius_ncs2},
machine learning algorithms 
(e.g., SVM classifiers \cite{marantos}),
and 
CV functions (e.g., stereo vision \cite{myriad_cv}).
The fabric architectures of Myriad 2 and Myriad X are illustrated in 
Figure \ref{fig_m2_new}
and 
Figure \ref{fig_mx_new},
respectively. 
Next, we discuss the SoC details
and present the associated software tools. 

The Myriad SoCs integrate
multiple I/O peripherals
and different types of processors.
All these components 
are connected to a multi-ported high-bandwidth shared memory hierarchy.
Regarding general-purpose CPUs, 
the SoCs include 
two LEON4 processors 
that implement the 32-bit RISC SPARCv8 architecture.
The first processor is called LEON OS (LOS)
and primarily handles the external communication, 
i.e., it controls peripherals such as UART, SPI, ETH, and USB3.
Moreover, 
LOS runs the RTEMS real-time operating system.
The second core is called LEON RT (LRT) 
and manages the media devices such as camera sensors and HDMI, while it also
controls all the imaging interfaces (MIPI, LCD, CIF).
The key processing units are the SHAVE cores, 
which are controlled at high level by the LEON processors. 
These cores are $128$-bit 
Very-Long-Instruction-Word (VLIW) processors 
that are suitable for executing 
the bulk of compute-intensive tasks,
offering not only core parallelization,
but also room for low-level custom optimization. 
Each
SHAVE supports 
Single-Instruction-Multiple-Data
(SIMD) instructions on various data types, 
i.e., $16$/$32$-bit floating-point and $8$/$16$/$32$-bit integer. 
The Myriad VPUs are also equipped with
hardware imaging accelerators, 
which are called Streaming Image Processing Pipeline (SIPP) filters. 
These specialized
accelerators are configurable up to a certain degree and provide low-power implementation for
numerous image processing kernels.

Regarding memory hierarchy,
the VPUs provide on-chip DDR DRAM (global memory), 
which is accessed by the processors through a single DDR controller. 
Furthermore,  
the SoCs have a small SRAM memory,
called Connection Matrix (CMX), 
which is primarily used by the acceleration cores
as scratchpad memory.
Each SHAVE core has 
preferential ports into a $128$KB slice of the CMX memory, whereas the remaining storage can be exploited for other purposes. 
For the communication between DDR and CMX,
the hardcoded DMA engine is used,
providing
high-bandwidth data transfers in either direction. 
The processors come along with their cache memories.
More specifically, 
each LEON processor has
both L2 and L1 caches, 
while the SHAVEs share a
common L2 and have a dedicated L1. 

The fabrication process of Myriad 2
is $28$nm HPC+/HPC/HPM. 
This VPU offers $12$ SHAVEs and 
its clock frequency can be configured up to $600$MHz. 
In terms of memories, 
it has on-chip a $512$MB DDR
and a $2$MB CMX.
LOS has $64$KB L1 and $256$KB L2 caches,
whereas 
LRT has $8$KB L1 and $32$KB L2 caches.
SHAVEs
share a common $256$KB L2 cache, 
in addition to a $3$K L1 per core
($1$KB for data and $2$KB for instructions). 
The fabrication process of Myriad X
is $16$nm FFC.  
This next-generation VPU features a dedicated on-chip accelerator, called Neural Computer
Engine (NCE), for inferencing DNNs.
Besides the AI accelerator,
the main differences compared to Myriad 2
are the addition of $4$ more SHAVE cores,
the increment of the CMX capacity by $0.5$MB,
and the increment of the clock frequency to $700$MHz.
There are also more MIPI lanes and slightly  bigger caches for LEONs. 

To implement custom CV and DNN applications on the
Myriad VPUs,
the developer uses the Myriad Development Kit (MDK).
This programming suite integrates  
GCC toolchain for the LEON processors
and
a C/C++ compiler with extensive intrinsic support
for SHAVEs, 
while offering numerous C/C++ libraries (e.g., for DMA transactions, SHAVE initialization, I/O data handling, power measurement).
Moreover, MDK provides
debugger, simulator, trace profiler,
and libraries with CV kernels. 
In essence, 
the Myriad VPUs are programmed via an LLVM-based vectorizing C/C++ compiler,
allowing the generation of assembly code, 
which is on a par
with hand-optimized assembly.
Towards improved assembly code, 
the developer can also write 
Myriad-friendly C/C++ code, 
e.g., 
explicitly apply the vectorization
to leverage the automatic SIMD calculations.

\begin{figure}[!t]
\vspace*{-1pt}
\centering
\subfloat[\label{fig_ncsa}]{\includegraphics[width=0.73\textwidth]{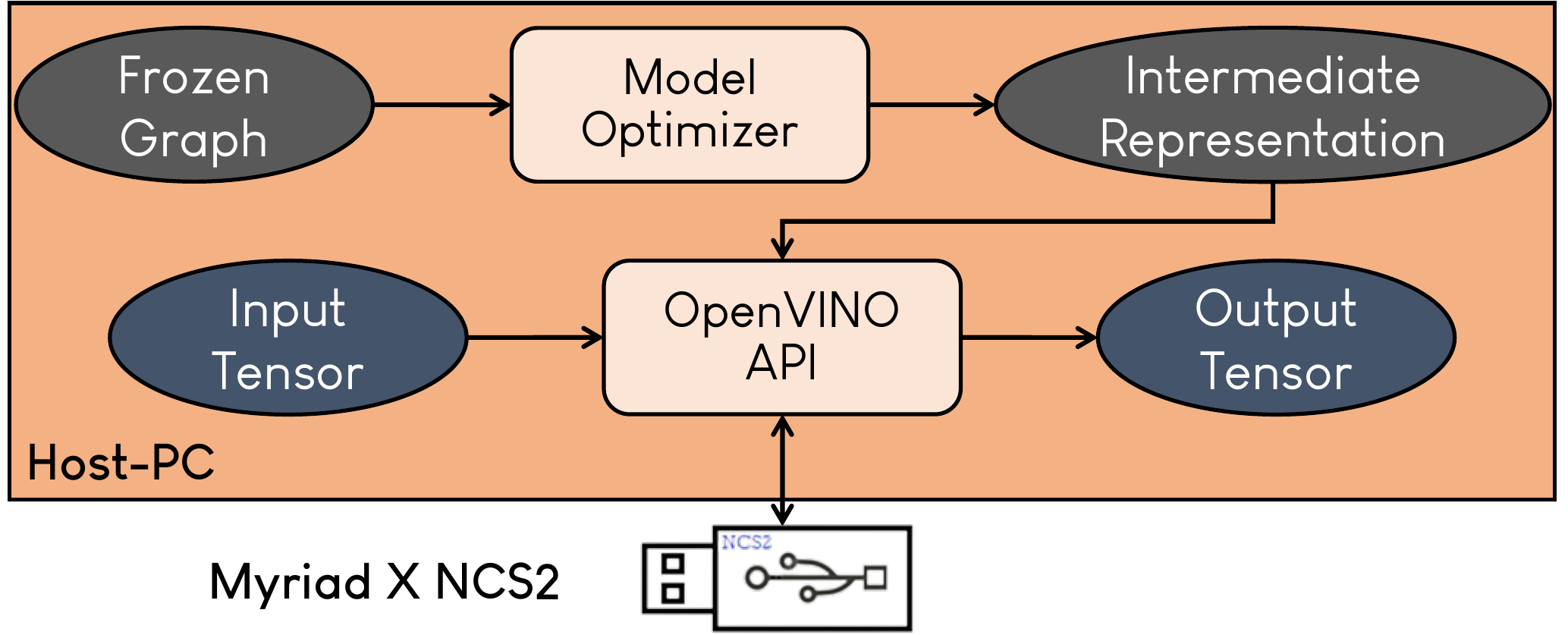}}\\
\vspace{-6pt}
\subfloat[\label{fig_ncsb}]{\includegraphics[width=0.73\textwidth]{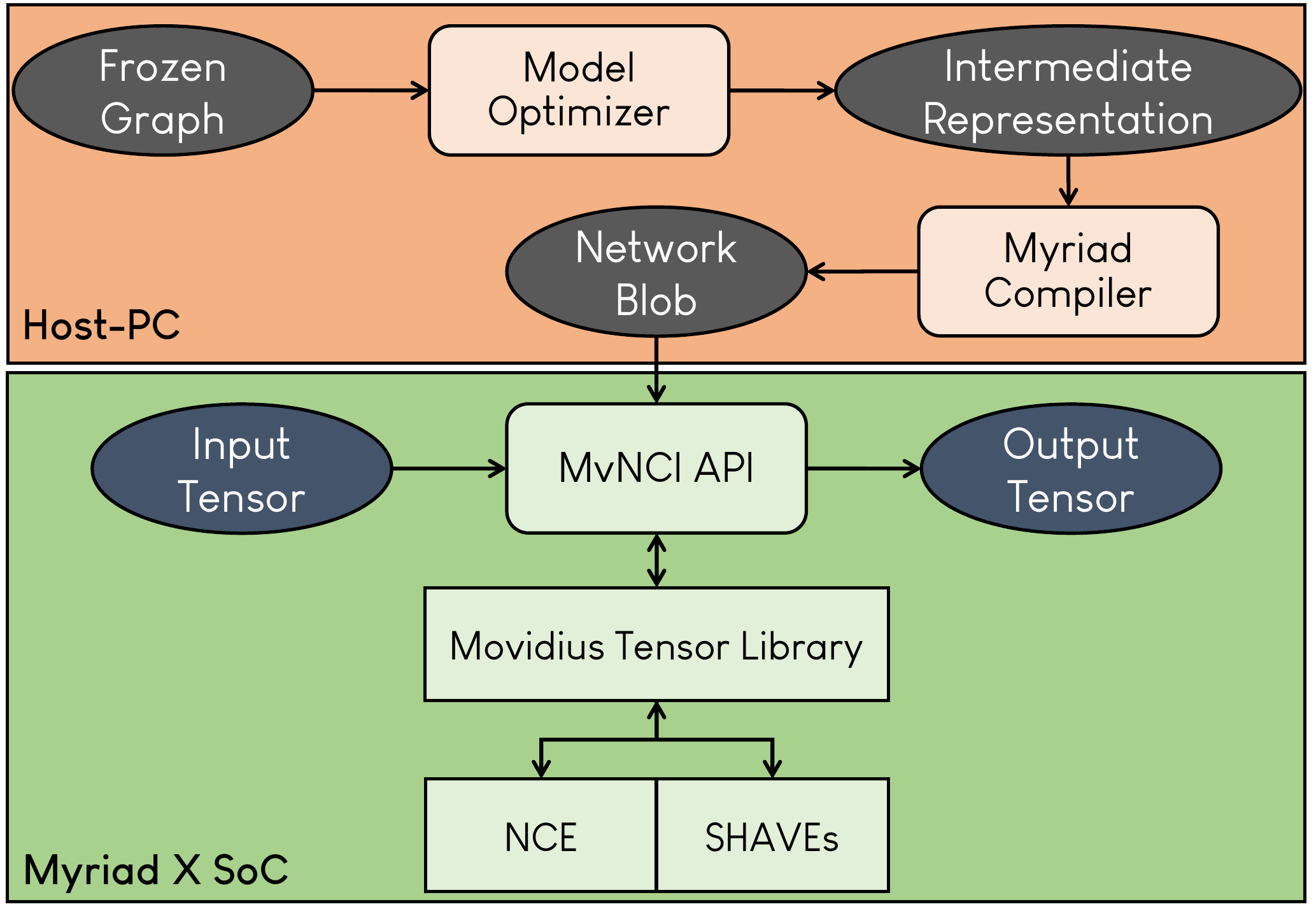}}
\caption[Deployment of Deep Neural Networks on Myriad X via OpenVINO]{Deployment of DNNs on Myriad X via OpenVINO on 
\textbf{(a)} Neural Computer Stick 2
and
\textbf{(b)} Myriad X SoC.}%
\label{fig_ncs2}
\vspace*{-5pt}
\end{figure}

To deploy DNN models
of well-known frameworks 
(e.g., TensorFlow) 
on Myriad X,
the developer uses 
Intel's OpenVINO toolkit \cite{openvino}.
Figure \ref{fig_ncs2}
shows the deployment procedure 
via OpenVINO with respect to the targeted platform:
(i) the USB accelerator integrating Myriad X,
which is called Neural Computer Stick 2 (NCS2) 
and 
(ii) the Myriad X SoC. 
In the first case (Figure \ref{fig_ncsa}),
OpenVINO inputs the frozen graph of the network
and
generates its intermediate representation with  
the model optimizer.
The intermediate representation consists of 
an XML file for the network topology 
and a binary file for the weights and biases.
These files 
are deployed on NCS2 
using the 
OpenVINO C/C++ or Python API. 
In the second case (Figure \ref{fig_ncsb}), 
OpenVINO converts the intermediate representation
to a binary programming file,
which is 
loaded on NCE via the mvNCI API.
The same API is also used 
to feed NCE with input data and receive its outputs in the form of tensors.

\vspace*{-2pt}

\section{Design Methodology}
\label{s9_3}

The efficient utilization of very heterogeneous SoCs, 
such as Myriad 2 and Myriad X,
requires a methodical design approach.
To exploit the full potential of heterogeneity, 
the diverse algorithmic functions of the application
should be mapped to 
the most suitable processing units,
while the development should be 
customized to the underlying hardware.
Towards this direction,
we propose a design methodology
for efficiently partitioning, scheduling and mapping 
any DSP/AI application that consists of multiple algorithmic functions.

We divide our methodology in two branches. 
The first branch regards the implementation of an algorithm from the DSP domain,
e.g., the CV pipeline of \cite{lour_zab}.
The second branch regards the deployment of an AI network, 
e.g., the DNN of \cite{urso}.
The methodology for the AI application
refers only to the Myriad X SoC and the NCS2 USB,
which include NCE,
i.e., the hardcoded AI accelerator.
In case the targeted platform is Myriad 2
or more custom implementations are desired,
the developer can adopt the methodology for the DSP application.

Regarding the DSP application,
we begin by developing all the algorithmic functions
or porting their pure C/C++ code (developed in other platforms)
on the general-purpose LEON CPU.
Upon all the functions are successfully compiled and executed,  
we profile and analyze them
to derive
the execution time, 
the requirements in 
memory, I/O and arithmetic,
and even the programming complexity. 
Based on this analysis,
and according to our design goal,
e.g., 
more ``low-power'' or more ``high-performance'',
we partition the entire algorithm on the SoC,
i.e., 
we determine the mapping target for each algorithmic function.
Moreover,
based on our analysis
as well as 
the SoC's micro-architecture
and the libraries/frameworks provided by the vendor,
we develop the utility software,
e.g., 
for handling the I/Os,
managing the memory allocation, 
and parallelizing the tasks of each function.
Afterwards,
we begin the implementation of each function
at both high and low level.
At high level,
we apply
the most efficient parallelization and mapping scheme 
to the SoC cores
and organize our software  
in terms of time and memory allocation.
At low level,
we accelerate each function by
rearranging its operations to facilitate pipelining 
and maximize the memory reuse,
we apply word-length optimization to further minimize buffering,
and 
we perform parallelization via vectorization and/or data decomposition techniques.
The next step is to integrate all the functions to the system 
and explore all coding parameters 
to fine-tune the implementation for the given problem/dataset.
In case the design constraints are not satisfied,
we use our feedback loop to return either 
to the initial partitioning and scheduling of the entire algorithm
or the high- and low-level implementation.
All these methodology steps are summarized as follows.

\underline{Methodology for DSP Application}:
\begin{enumerate}[wide = 6pt, itemsep=-3pt, leftmargin =18.8pt]
    \item Development/Porting of the application/algorithm on the LEON core.
    \item Profiling and analysis of each algorithmic function with realistic dataset.
    \item Partitioning and scheduling of the entire algorithm on the Myriad SoC.
    \item Development of utility software for I/O handling, memory management, and low-level optimizations.
    \item High-level parallelization of each algorithmic function to the SHAVE cores.
    \item Low-level implementation and optimization in the SHAVE cores.
    \item System integration of each algorithmic function.
    \item Testing and tuning with application-specific datasets.
    \item Return to step 3 or 5 until the constraints are met and the system is optimized.
\end{enumerate}

Regarding the AI application,
we follow steps similar to those proposed by OpenVINO \cite{openvino}.
The training and optimization of the network
with the AI framework (e.g., TensorFlow, PyTorch)
are out of the Dissertation's context. 
OpenVINO supports tuning and optimization on the network frozen graph
via the model optimizer.
One of the provided methods
is the fusing of the linear operations,
e.g., 
the multiplications and additions can be merged into a single multiply-add instance
or fused to convolutional and fully-connected layers.
The tool also offers other methods,
such as
grouped convolution fusing
and
network pruning.
Moreover,
if required, 
the developer can create custom operations in OpenCL. 
All these methodology steps for the AI deployment on Myriad X are summarized as follows. 

\underline{Methodology for AI Application}:
\begin{enumerate}[wide = 6pt, itemsep=-3pt, leftmargin =18.8pt]
    \item Generation of the DNN model in the AI development framework. 
    \item Optimization and development of custom network operations in OpenVINO.
    \item Creation of the network programming file in OpenVINO.
    \item Inference run on Myriad X and analysis of the results.
    \item Return to step 1 or 2 until the constraints are met and the inference is optimized.
\end{enumerate}

\section{Implementation of DSP \& AI Applications on the Myriad 2 VPU}
\label{s9_4}

In this section,
we present the implementation of three custom kernels on Myriad 2.
In particular,
we develop in C/C++ 
an image binning kernel,
floating-point convolutions,
and a Convolutional Neural Network (CNN)\footnote{\fontsize{7.7}{8.8}\selectfont Special thanks to E. Petrongonas for coding on the CNN and Myriad 2.} 
for detecting ships on satellite images \cite{kaggleships}.
This section acts as introductory in the VPU development
and aims to 
introduce the Myriad computing paradigm
and the SoC's functionalities,
as well as 
demonstrate our programming approach.

\subsection{Development of Custom DSP and CNN Kernels}

In image processing, 
binning is the procedure of extracting a single pixel
from an image region (cluster of pixels).
This task, 
although results in loss of information, 
reduces the amount of data to be transferred and processed
(e.g., a $4$-MPixel sensor image is transformed to $1$-MPixel).
For our kernel, 
we adopt $2\times2$ Averaging Binning,
namely,
we assume $2\times2$ regions with stride $2$. 
The new pixels are calculated as the mean values of the regions' pixels. 
For our convolution kernels,
we employ single-precision floating-point masks of different sizes
($3\times3$, $7\times7$, $13\times13$),
targeting to stress the VPU
with this fundamental DSP operation.

\begin{figure}[!t]
\centering
\includegraphics[width=0.70\textwidth]{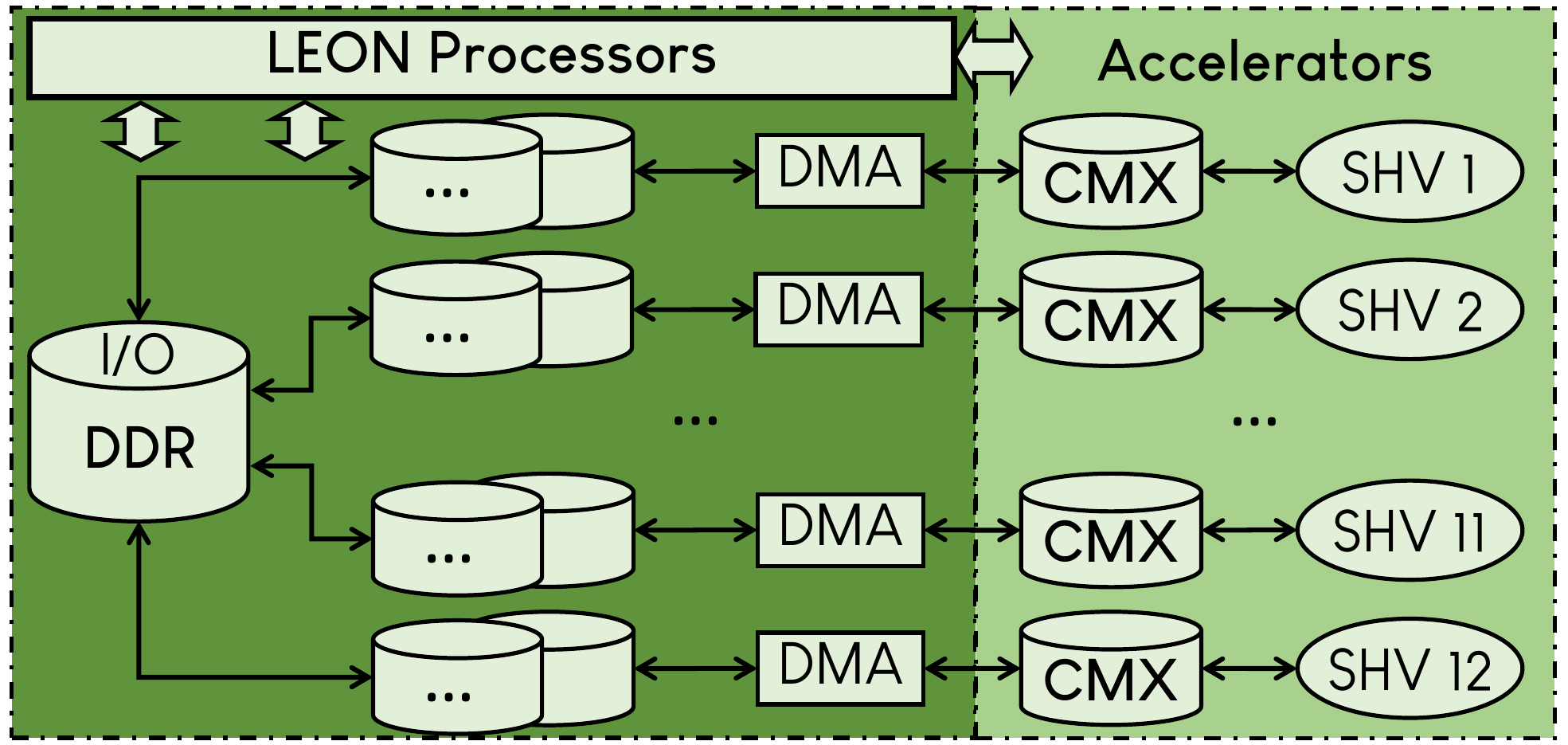}%
\caption[High-Level Parallelization of Image Processing Workload in Myriad 2]{High-level parallelization of image processing workload in the Myriad 2 VPU.}%
\label{fig_shv}
\vspace*{-4pt}
\end{figure}

In the implementation of Averaging Binning and Floating-Point Convolution,
the LEON processor initializes the SHAVEs cores
and performs all the necessary high-level tasks 
for the core parallelization.
Our design choices are the same for both kernels.
The input and output images are stored in DDR (global memory),
while we perform DMA transactions
to transfer image slices in the CMX (scratchpad/working memory),
which is accessed by the SHAVE acceleration cores.
The high-level parallelization procedure is illustrated in Figure \ref{fig_shv}.
Regarding general implementation details in SHAVEs, 
we do not use additional working buffers,
i.e., 
the processing is performed in-place (in the input buffer),
and we enable the caches. 
For Averaging Binning,
we employ $2048$$\times$$2048$ $8$-bit input images,
which are transformed to $1024$$\times$$1024$ $8$-bit output images. 
We divide the input image into $36$ stripes, 
i.e., $35$ stripes of size $2048$$\times$$58$ and $1$ smaller stripe of size $2038$$\times$$18$,
and we assign $3$ stripes to each SHAVE to process them successively.
For the Floating-Point Convolution,
we consider $1024$$\times$$1024$ $8$-bit input images and zero padding.
Our design is parametric in terms of the mask size,
allowing to change masks at compile time.
The masks are stored in 
CMX to be accessed directly by SHAVEs. 
Regarding parallelization,  
we divide the image into $24$ stripes, 
i.e., $22$ of size $1024$$\times$$43$ and $2$ of size $1024$$\times$$39$, 
and we assign $2$ stripes to each SHAVE.
The SHAVE processing uses vectorized operations as indicated by the generated assembly code.
 
Our CNN model is trained in TensorFlow with 
$128$$\times$$128$$\times$$3$ images and $32$-bit floating-point weights and biases.
The accuracy is $96.8\%$ for binary classification 
(ship detected or not).
The network consists of four convolutional layers and two fully-connected layers,
while the total number of weights is $131$K. 
For our custom implementation in Myriad 2, 
the $32$-bit floating-point weights and biases
are converted to $16$-bit floating-point
using the corresponding routine of the MDK suite. 
The implementation is based 
on a custom parametric inference engine for $128\times128\times3$ input tensors,
which is mapped onto SHAVEs
and 
supports all the layers of our 
Ship Detection CNN. 
In our case,
we develop the CNN accelerator 
for $1024\times1024\times3$ $16$-bit input images.
Considering that the inference engine is built for smaller tensors,
we employ a function running 
on LEON that 
divides the input image into $64$ $128\times128\times3$ patches.
Subsequently, 
it successively stores them 
in the engine's input buffer 
and orders the SHAVEs to start the patch processing.
In terms of memory utilization,
we store
the images, weights, and working buffers in DDR. 

\section{Porting of Computer Vision Pipeline on the Myriad 2 VPU}
\label{s9_5}

Based on our design methodology,
we accelerate on Myriad 2 
a vision-based pose tracking algorithm \cite{lour_zab},
which is representative of spacecraft proximity operations. 
In particular,
it inputs a sequence of high-definition images
and continuously performs rendering, feature detection, feature matching, 
and data fitting via robust regression.
The output is the 6D pose of the recorded satellite (Envisat) 
during a hypothetical maneuver in an Active Debris Removal mission \cite{lentaris_tvideo}.
This sophisticated $5$-stage CV pipeline
exhibits increased diversity in terms of computations and memory,
thus, its porting on a low-power SoC is challenging.
Next,
we present the functions of the CV algorithm and 
discuss details about their implementation.

\subsection{The CV Algorithm for Satellite Pose Tracking}
The pose tracking algorithm of \cite{lour_zab}
assumes small motion between successive frames and 
uses a model-based approach to estimate 
the object’s pose relative to the camera.
Frame after frame,
it evolves an initial pose by continuously 
rendering a depth map from the object's mesh model, 
detecting edges on it,
and then matching them to the edges detected on the input image.
These matches are used to refine the current pose
i.e., to perform data fitting
via least median of squares regression 
followed by iteratively reweighted least squares.

\begin{figure}[!t]
\vspace*{9pt}
\centering
\includegraphics[width=0.83\textwidth]{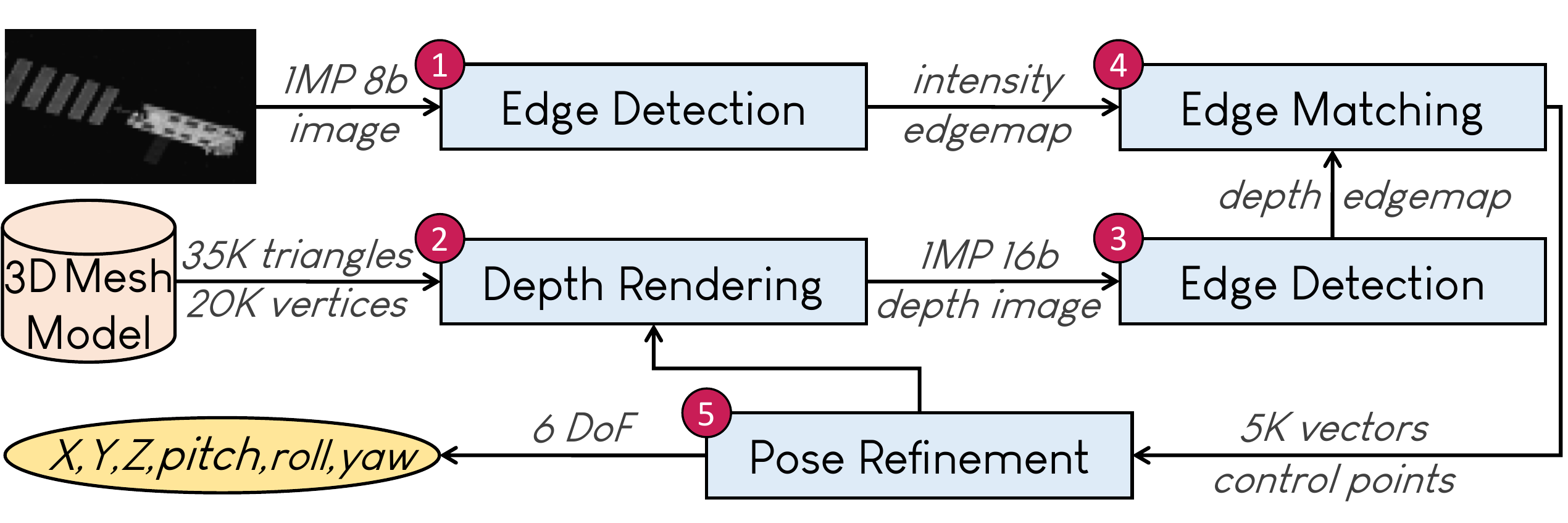}%
\caption[The 5-Stage Computer Vision Pipeline for Satellite Pose Tracking]{The 5-stage computer vision pipeline for satellite pose tracking \cite{lour_zab}.}%
\label{fig_louralgo}
\end{figure}

The key functions and dataflow of the algorithm are presented in Figure \ref{fig_louralgo}.
The algorithmic functions are summarized as follows: 
(i)   edge detection on the input/intensity image, 
(ii)  depth map rendering, 
(iii) edge detection on the rendered/depth image, 
(iv)  perpendicular edge matching, 
and 
(v)   pose refinement.
Regarding I/O,
the algorithm inputs $1024$$\times$$1024$ $8$-bit grayscale images
and outputs a $6$$\times$$1$ floating-point vector corresponding to the pose $\langle x,y,z,pitch,roll,yaw \rangle$.
Moreover, 
it employs the object's mesh model
(the Envisat satellite in our case),
which
consists of $20$K vertices and $35$K triangles.

For Edge Detection,
the algorithm uses the well-known method of Canny \cite{canny} 
for both the intensity and depth images.
This detector performs the following tasks:
(i) Sobel convolution to compute image gradients 
(magnitude and direction),
(ii) non-maximum suppression to remove spurious edges,
(iii) hysteresis thresholding on the magnitudes to identify strong edges and suppress the weak ones.
The hysteresis thresholding task 
initially retains all the mid-strength edges,
but then it traces recursively their spatial connections 
to keep only those related to strong edges. 
This procedure is based on two thresholds,
which are calculated based 
on the median in the histogram of gradients.

For Depth Rendering, 
the algorithm uses a triangle 3D mesh model 
and the current 6D pose 
to generate 
an image,
whose pixels encode the distance 
between the camera and the nearest point on
the model’s surface.
The projection of the triangles 
on the image is performed via rasterization, i.e., by projecting their vertices, then using bounding box traversal to determine the pixels residing inside the projected triangles and, finally, calculating the distance of the model's 
triangles from each pixel. 
When multiple triangles project on the same pixel, 
the algorithm retains the projection that is closest to the camera.

Perpendicular Edge Matching 
finds the correspondences between intensity and depth edges. 
For each edge of the depth map, 
it searches along the gradient direction in the intensity map
until an intensity edge is found or a maximum distance is covered.
Finally, Pose Refinement utilizes the set of control points
(matches plus spatial information)
to determine the change of 6D pose between the two edge maps
representing the previous and current frames.
The change is determined in a robust regression framework that mostly involves linear algebra operations, 
e.g., SVD and QR decompositions.
In an incremental fashion,
the change is used to 
update the pose estimation. 

\subsection{Partitioning and Scheduling}
\label{s9ps}

The most central design choice 
when partitioning a CV algorithm in Myriad 2 
is whether to accelerate each function in the 
SHAVE subsystem, 
execute it on LEON, 
or assign it to a hardware filter. 
The main criteria to select the mapping targets
are:
(a) performance and power constraints, 
(b) library dependencies,
(c) parallelization amenability,
and
(d) memory access patterns.
Regarding criterion (a),
the use of SHAVEs provides better performance 
than the hardware filters,
however,
the latter attain better power efficiency.
Criterion (b)
regards
the software complexity.
For instance, 
it is preferable for the
functions depending on software libraries 
(e.g., for linear algebra)
to run on the general-purpose LEON processor
in case of decreased support and availability in SHAVEs.
Criterion (c)
relates to the efficiency of the parallelization.
For example,
it may be preferable to execute an inherently sequential algorithm
on LEON rather than parallelize it to SHAVEs.
Finally,
criterion (d) refers to the selection and the configuration of the available memories
according to the access patterns
(global or scratchpad, use of cache or not).

\begin{figure}[!t]
\vspace*{7pt}
\centering
\hspace*{-6pt}
\includegraphics[width=1.02\textwidth]{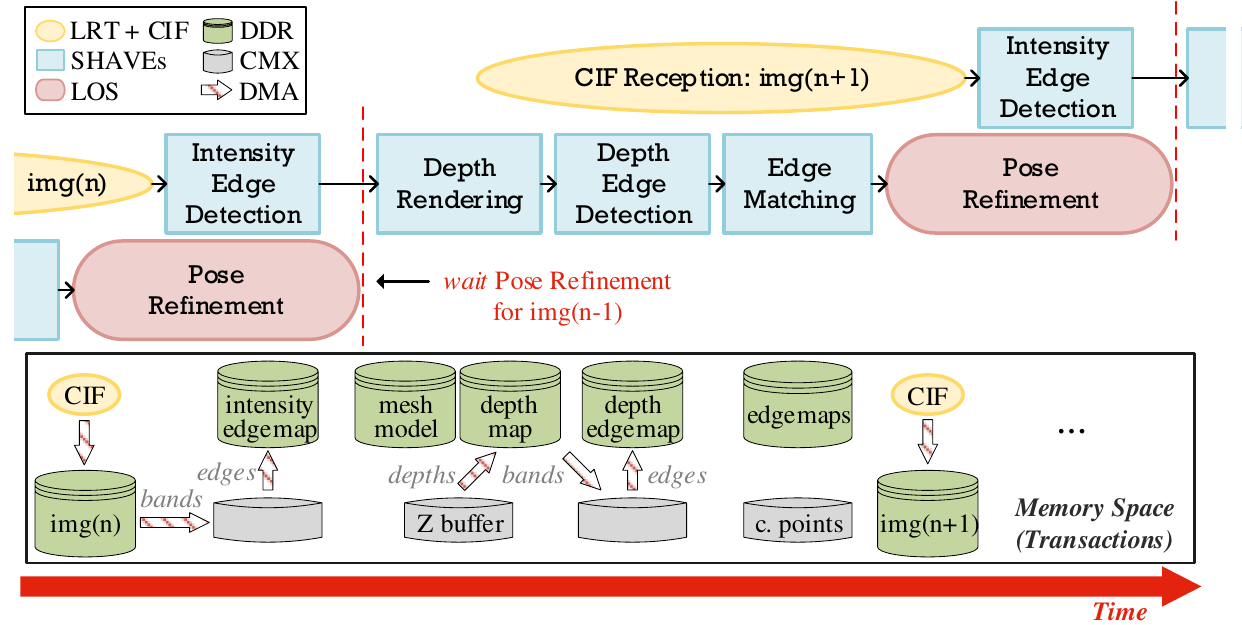}%
\caption[Partitioning, Scheduling and Memory Transactions of the Computer Vision Pipeline in Myriad 2]{Partitioning, scheduling and memory transactions of the computer vision pipeline in Myriad 2.}%
\label{fig_schedul}
\end{figure}

The profiling and analysis of the CV pipeline
based on our methodology and the above criteria 
results in the partitioning and scheduling shown in Figure \ref{fig_schedul}.
First, 
we avoid using the SoC's hardware filters, 
because we target maximum performance gain. 
Second, 
the profiling on LEON shows the very slow execution of the compute-intensive functions,
i.e., Edge Detection and Depth Rendering,
thus, 
we identify all the parallelization opportunities in their algorithmic nature
and accelerate them on SHAVEs.
Even though the Edge Matching function 
does not include demanding computations 
(it applies only image scanning and comparisons),
we also map it onto SHAVEs.
Otherwise, 
LEON would have to scan a $1024$$\times$$1024$ image, 
resulting in increased execution time.
We note that
we sequentially assign each function to all $12$ SHAVEs
instead of splitting them among the three functions.
The latter would
achieve questionable performance gains 
at the cost of significantly increased development effort
for handling split memory and function synchronization.
Regarding Pose Refinement,
we assign it to LEON 
due to its dependencies
on the BLAS/LAPACK libraries.
For high-level scheduling, 
we follow the dependencies shown in Figure \ref{fig_louralgo},
while we maximize the function-level parallelization between SHAVEs and LEON.
More specifically, we start detecting edges on a new input image
even while LEON is still processing the previous frame to output its pose.
When both are finished, we start the execution of 
Depth Rendering for the current frame. 
Finally, 
we consider 
that Myriad 2
receives the input frames
from its Camera Interface (CIF).
In our scheduling,
the reception of each new frame is performed 
in parallel to the main processing by the hardware CIF peripheral of the SoC.

\subsection{Development of Utility Software}
The purpose of developing a utility software is to 
increase the productivity and enable sophisticated parallel programming over MDK, i.e., the 
VPU development tool,
which demands low-level coding when targeting custom implementations.
For example, extra development is required
when multiple diverse functions must be assigned to SHAVEs
for non-embarrassingly parallel execution.
Our utility software 
creates an abstraction layer
for handling the I/O peripherals,
memory management,
task scheduling,
and inter-process communication.
It includes a set of lightweight, standalone, and transparent C/C++ libraries,
which, 
like the other MDK components,
are included at compile-time and alleviate the coding/testing effort from the main development.
Without this software,
the implementation of the 5-function CV pipeline would not be efficient (or even feasible at all).  
Overall,
our software modules extend MDK by
introducing new mechanisms,
improving the performance of the existing ones,
and 
providing automatic/transparent device configuration.
More details
about our utility software for the Myriad VPUs 
can be found in our publications in \cite{LeonTECSm, paralos}.

\begin{figure}[!t]
\centering
\subfloat[\label{fig_util1}]{\includegraphics[width=0.47\textwidth]{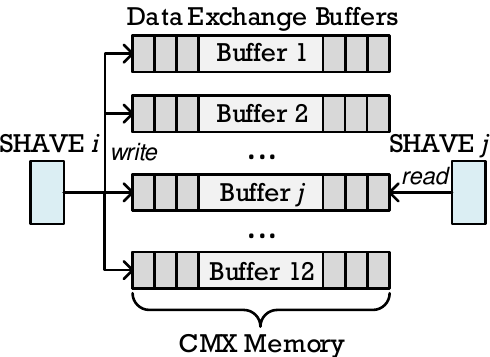}}\\[-9pt] 
\subfloat[\label{fig_util2}]{\includegraphics[width=0.47\textwidth]{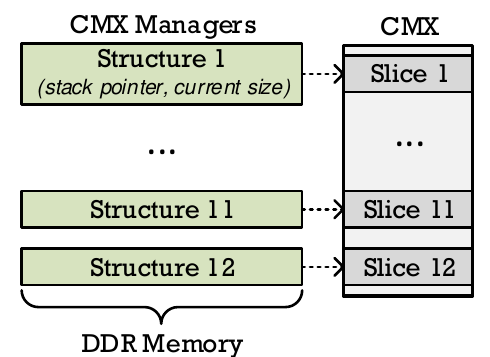}}\hfill
\subfloat[\label{fig_util3}]{\includegraphics[width=0.47\textwidth]{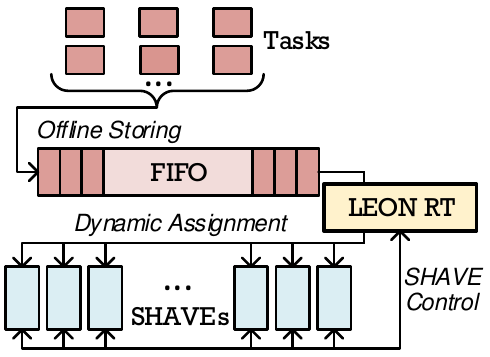}}
\caption[Custom Utility Mechanisms for Development Support in Myriad 2]{Custom utility mechanisms for development support in Myriad 2:
\textbf{(a)} inter-process communication, 
\textbf{(b)} memory management, 
and
\textbf{(c)} dynamic task assignment.}%
\label{fig_paralos}
\end{figure}

Figure \ref{fig_paralos}
presents the basic mechanisms of our utility software. 
The first mechanism,
shown in Figure \ref{fig_util1}, 
provides inter-process communication
using
hardware mutexes and a variable size buffer per SHAVE,
which is placed in the shared CMX memory slice.
For data exchange, the SHAVEs read and write the buffers.
The second mechanism,
shown in Figure \ref{fig_util2}, 
pre-loads the code of each function in DDR 
and allocates the requested CMX memory at runtime,
i.e., during the execution of the function,
based on a pointer technique.
The third mechanism,
shown in Figure \ref{fig_util3}, 
aims to reduce the idle core time that is caused by the static task assignment.
For embarrassingly parallel workloads,
the developer 
creates a task pool by splitting the workload into small (independent) tasks,
which are inserted in a FIFO struct and assigned to SHAVEs at runtime. 
Namely, 
the first task of the FIFO is assigned to the first available SHAVE,
and hence, 
each SHAVE is immediately assigned a new task
upon completing its previous one.

\subsection{Parallelization and Low-Level Optimization}

This section reports the implementation details on SHAVEs
in two interdependent stages: 
(i) core parallelization, 
(ii) low-level mapping and optimization within each core. 
All the functions
employ our memory allocation mechanism
in order to use
the common scratchpad CMX memory.
More specifically, 
before the execution of each function, 
all the necessary working buffers are allocated,
and correspondingly,
after the execution
they are freed.  

\subsubsection{Canny Edge Detection}
To provide improved workload balancing 
among the $12$ SHAVEs, 
we split the input image in half, 
divide each half-image into $12$ slightly overlapping stripes,
and assign each one 
to a distinct SHAVE. 
In total,
each SHAVE processes two $1024$$\times$$43$-pixel stripes
from remote parts of the image,
which have decreased correlation in terms of content.

The initial porting of Canny Edge Detection 
utilized a
large memory, e.g., at least $4$ buffers 
for the images and the derivatives, as well as extra 
$32$-bit integer buffers for the histogram calculation 
and the execution of hysteresis thresholding. 
In our custom embedded implementation,
we adopt in-place processing 
with reduced buffering 
to best fit in the $2$MB CMX memory. 
That is, we interleave the loops for calculating the gradients 
and performing non-maximum suppression, 
such that with extensive memory reuse,
we rely only on the input image buffer 
and two small buffers of size $1024$$\times$$3$ 
to slide all $3$$\times$$3$ kernels.  
Furthermore, 
we calculate the histogram of each stripe
inside this convolution loop 
(while generating the gradients).
Another low-level knob
regards
the optimization 
of the buffers
via word-length tuning and data type adaptation
to work with the smallest required data size.
Besides CMX, this customization exploits 
the capability of the VPU 
to process 
various data types
(e.g., $16$-bit integers instead of $32$-bit). 
Upon the loop completion, 
the image buffer stores all local maxima 
and the hysteresis thresholding task is executed.
This recursive procedure 
re-labels the
weak edges neighboring any strong edge.
The tracing continues until all connecting paths are followed. 
In our implementation, each SHAVE executes a local hysteresis in its own stripe 
and afterwards, 
it exchanges the strong edges of the stripe borders with its neighboring SHAVEs 
using our inter-process communication mechanism.
After updating their border edges, 
SHAVEs execute another iteration of hysteresis 
specifically for those pixels.
The last (low-complexity)
task is executed on LEON
and
calculates the entire histogram of the gradients 
by accumulating the $24$ stripe histograms 
(returned to DDR from CMX).
This process also determines the hysteresis thresholds for the next frame.

\subsubsection{Depth Rendering}
For this function, 
we divide the image into horizontal $R$-row stripes,
which are assigned to distinct SHAVEs
to be rendered almost independently from each other.
That is,
we create a pool of $B=1024/R$ individual tasks,
where $B$ determines
the CMX memory resources per SHAVE and the DMA transactions
(each rendered stripe is sent to DDR).
The full utilization of CMX requires $B=18$ stripes, 
however, our exploration shows that $B=32$ improves the execution.
Namely, 
smaller stripes adapt better to 
image content and distribute 
the workload more fairly to SHAVEs,
even though
larger $B$ increases the  
repetitions of model reading.
To reduce the idle time per core, 
we employ our mechanism for dynamic task assignment:
each SHAVE is assigned a new task (i.e., stripe to render) 
from the pool at runtime,
immediately upon finishing its previous stripe.

After exploration,
we store the static Envisat's mesh model in the DDR global memory instead of CMX. 
The scratchpad is small and is preferred only for storing working buffers.
Each SHAVE gets direct access to the full model in DDR, and hence,
multiple cores read the same set of memory locations repetitively.
However,
due to their shared L2 cache, 
we measure negligible time penalties 
with this approach (compared to storing the model in CMX).
Regarding the low-level optimization in SHAVEs, 
we reduce the number of buffers
and
customize the data types. 
In particular, 
we do not employ the working buffers used in typical CPU implementations,
and ultimately,
we utilize only 
the Z-buffer containing the rendered image stripe.
Furthermore, 
we successfully enable SIMD operations 
to accelerate amenable functions 
(almost half of total computations),
such as
projection, bounding, and depth calculation.  
In detail, 
we arrange the data using the Clang Intrinsic vectors, 
e.g., \texttt{float4} for defining a vector for 4 float numbers,
and we call the corresponding MDK routines 
to calculate the dot/cross products (\texttt{mvuDot}, \texttt{mvuCross}) 
and find the min/max numbers (\texttt{mvuMax}, \texttt{mvuMin}). 
Moreover, we transform extra multiplications and additions
to dot and cross products to expand our parallel/vectorized operations. 

\subsubsection{Perpendicular Edge Matching}
This function searches for matches between the intensity and depth edgemaps.
We divide the image into $12$ independent stripes
and assign each stripe to a distinct SHAVE.
After exploration,
we opt to store both the depth and intensity edge maps in DDR, 
while we enable the L1 (including data) and L2 caches of SHAVEs.  
Considering that Edge Matching 
relies on simple data comparisons (fast computations),
the processing time on SHAVEs cannot mask the overhead of
transferring the edges to CMX via the DMA engine.
We store the output matches in the uncached CMX,   
which is directly accessed by LEON 
(to start the execution of the next function, i.e., Pose Refinement),
in order to avoid disturbing the caching mechanism fetching the input edges.
The race conditions to the shared output buffer are resolved 
by fine grained locking using the SoC's hardware mutexes.

\section{Inference of Deep Neural Network on the Myriad X VPU} 
\label{s9_6}

In this section,
we present the inference of a demanding DNN on Myriad X,
and more specifically,
on the NCS2 USB accelerator via OpenVINO\footnote{\fontsize{7.7}{8.8}\selectfont Special thanks to P. Minaidis for coding on the DNN and NCS2.}. 
In particular, 
we accelerate 
a DNN of ResNet backbone,
namely UrsoNet \cite{urso},
which estimates the satellite's pose.
To surpass the limited embedded computation power 
(compared to high-performance computers and host machines)
and 
comply with the real-time constraints, 
we deploy a mobile version of UrsoNet.
Moreover, 
to improve the DNN inference, 
we decrease the image resolution,
using various resampling algorithms. 
More details about the resampling 
can be found in our publication in \cite{LeonELSI}.

\subsection{Deployment of DNN for Satellite Pose Estimation}

\begin{figure}[!t]
\vspace*{-2pt}
\centering
\includegraphics[width=0.87\textwidth]{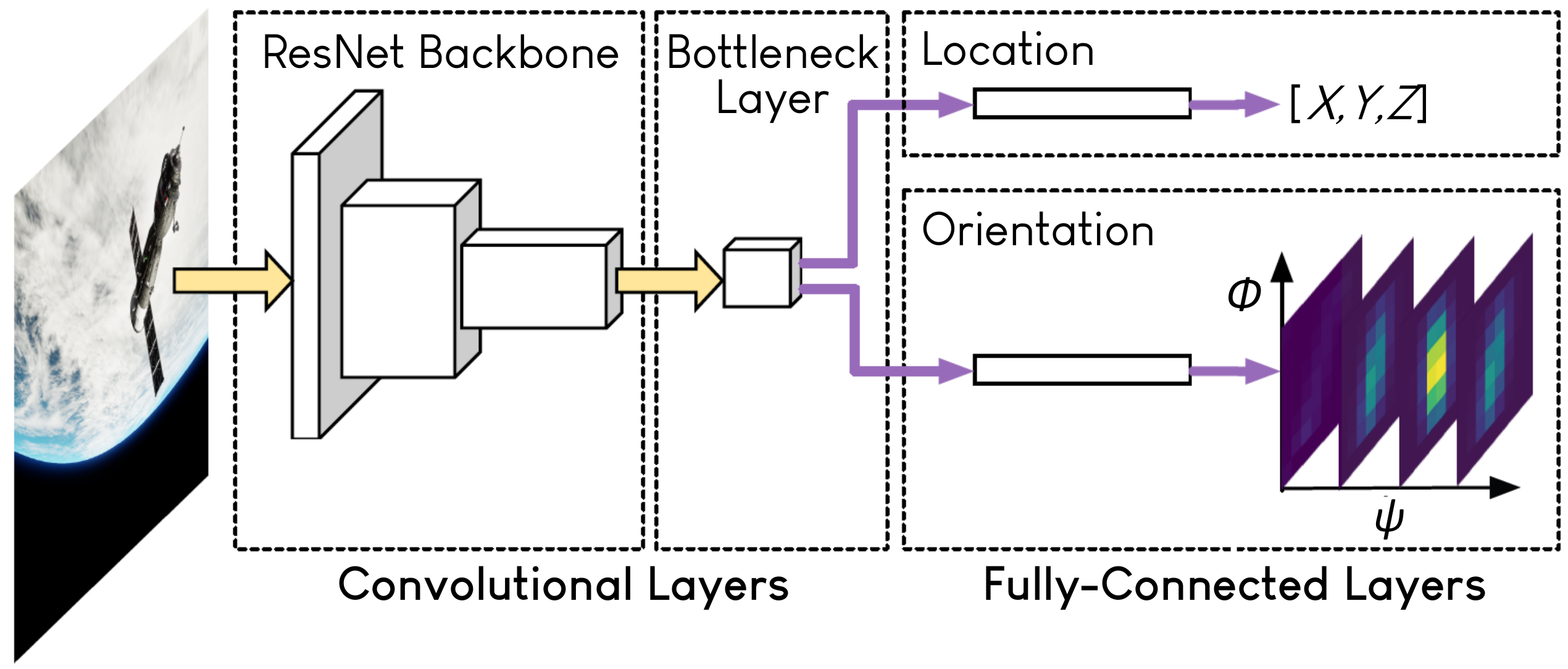}%
\vspace*{-5pt}
\caption[The UrsoNet DNN for Satellite Pose Estimation]{The UrsoNet DNN for satellite pose estimation \cite{urso}.}%
\label{fig_ursoarchi}
\end{figure}

The architecture of the UrsoNet DNN \cite{urso}
is illustrated in Figure \ref{fig_ursoarchi}. 
In comparison with the original ResNet architecture,
the global average pooling layer and the last fully-connected layer
are removed.
In their place,
the designers of UrsoNet insert
a bottleneck layer that consists 
of a $3$$\times$$3$ convolution with stride $2$, 
as well as two fully-connected layers for
calculating 
the satellite's location and orientation.
To generate a mobile network
for deployment on Myriad X,
we adopt the UrsoNet configuration
presented in Table \ref{tb_urso}.
We use the dataset for the Soyuz spacecraft
(``soyuz\_hard'')
and follow the training process of \cite{urso},
but we start from a pre-trained ImageNet model,
perform training for $100$ epoques,
and do not apply data augmentation.
The image resolution is decreased
using different resampling algorithms. 
We consider an $1024$$\times$$1024$$\times$$3$ input image,
which is scaled to $512$$\times$$512$$\times$$3$
for efficient deployment and inference
at the edge. 

\begin{table}[!t]
\fontsize{9}{10}\selectfont
\renewcommand{\arraystretch}{1.2}
\setlength{\tabcolsep}{3.8pt}
\caption[Configuration of the UrsoNet DNN for Deployment on Myriad X VPU (NCS2)]{Configuration of the UrsoNet DNN for deployment on Myriad X VPU (NCS2).}
\label{tb_urso}  
\centering
\begin{tabular}{l c l c} 
\hline
\multicolumn{2}{c}{\textbf{Network}} & \multicolumn{2}{c}{\textbf{Training}}\\
\cmidrule(lr){1-2} \cmidrule(lr){3-4} \multicolumn{1}{c}{\textbf{Parameter}} & \multicolumn{1}{c}{\textbf{Value}} & \multicolumn{1}{c}{\textbf{Parameter}} & \multicolumn{1}{c}{\textbf{Value}} \\
\hline \hline 
 Backbone & ResNet-50 & Pre-Trained Weights & ImageNet\\
 Bottleneck Width & $32$ &  Dataset & ``soyuz\_hard'' \\
 Input Image & $1024$$\times$$1024$$\times$$3$ & Arithmetic & fp32\\
 Resampling & Bilinear/Bicubic/Lanczos &  Epochs & $100$ \\
 Inference Image & $512$$\times$$512$$\times$$3$ & Augmentation & No\\
 Ori./Loc. Resolution & $16$/$16$ & Optimizer & SGD  \\
\hline
\end{tabular}
\vspace*{-3pt}
\end{table}

\begin{figure}[!t]
\vspace*{-2pt}
\centering
\includegraphics[width=0.86\textwidth]{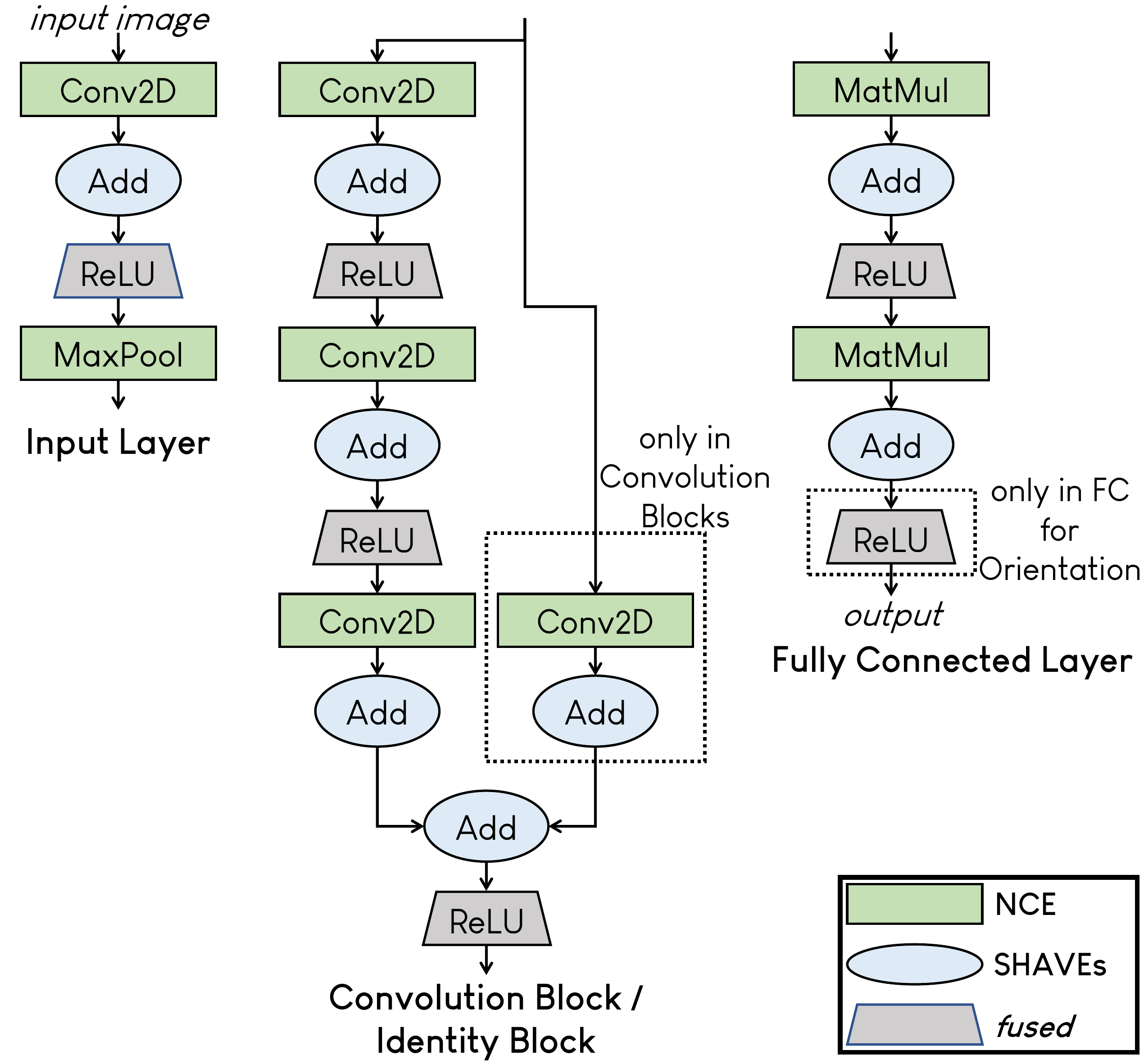}%
\caption[Mapping of the UrsoNet DNN in Myriad X]{Mapping of the UrsoNet DNN in Myriad X.}%
\label{fig_ursomapp}
\vspace*{-7pt}
\end{figure}

To generate the binary network file and deploy it on Myriad X (NCS2 accelerator),
we follow the OpenVINO toolflow discussed in Section \ref{s922}. 
Figure \ref{fig_ursomapp} illustrates the mapping of the basic network blocks of  UrsoNet on Myriad X.
Compared to the original ResNet architecture, 
the batch normalization blocks are replaced by add operations,
which are accelerated on SHAVEs. 
Smaller operations, 
e.g., permutations before the fully-connected layers,
are also executed on SHAVEs. 
The convolutions and matrix multiplications are mapped onto NCE,
while almost all the ReLU activation functions 
are optimized out by OpenVINO,
i.e., fused with other operations 
during the graph transformation stage.

\section{Evaluation}
\label{s9_7}

This section conducts the evaluation of our development on the VPUs.
Section \ref{s971} reports results
from the implementation of the 
DSP and CNN kernels on Myriad 2
(presented in Section \ref{s9_4}).
Section \ref{s972} regards the implementation of the 
CV algorithm for pose tracking on Myriad 2
(presented in Section \ref{s9_5}).
All these algorithms are developed with the MDK tool. 
For their evaluation, 
we consider an FPGA--VPU co-processing
architecture,
in which the FPGA sends/receives the I/O data 
to/from the VPU.
We also employ a host-PC
for validating and demonstrating the results. 
Finally,
Section \ref{s973} evaluates the inference  
of the DNN for pose estimation (presented in Section \ref{s9_6}).
For this experimentation,
we use the OpenVINO tool 
and the Myriad X NCS2 accelerator. 
Table \ref{tb_implvpu}
summarizes all the VPU implementations of the Dissertation 
along with their I/O data.

\subsection{Experimental Results of Custom DSP and CNN Kernels}
\label{s971}

\begin{figure}[!t]
\vspace*{-2pt}
\centering
\includegraphics[width=1\textwidth]{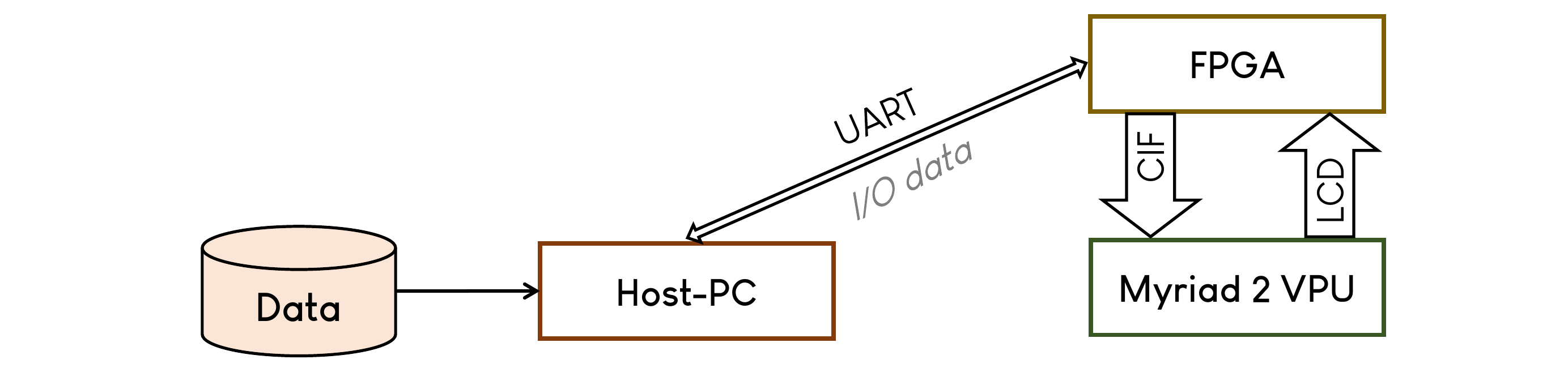}%
\caption[FPGA--VPU Architecture for Accelerating DSPs/CNNs on Myriad 2]{FPGA--VPU co-processing architecture for the acceleration of DSP/CNN kernels on Myriad 2.}%
\label{fig_march2}
\vspace*{-6pt}
\end{figure}

Figure \ref{fig_march2} illustrates our testing setup
for the evaluation of the custom DSP and CNN kernels.
The host-PC is connected with an FPGA,
which transmits/receives the I/O data to/from Myriad 2.
Myriad 2 receives the input data via CIF,
as in our CV pipeline (see Section \ref{s9ps}) 
and sends the output to the FPGA 
via the Liquid Crystal Display (LCD) interface\footnote{\fontsize{7.7}{8.8}\selectfont Special thanks to Prof. D. Reisis \emph{et al.} from NKUA 
for coding on the CIF/LCD interface.}.
More details about our FPGA--VPU architecture,
the components/functions of each device that enable the communication, 
and the I/O transfers
are reported 
in our publication in \cite{LeonICECSm}. 

\begin{table}[!t]
\fontsize{9}{10}\selectfont
\renewcommand{\arraystretch}{1.2}
\setlength{\tabcolsep}{4pt}
\caption[Overview of Dissertation's DSP \& AI VPU Accelerators]{Overview of Dissertation's DSP \& AI VPU accelerators.}
\label{tb_implvpu}  
\centering
\begin{tabular}{l|cccc} 
\hline
\multicolumn{1}{c|}{\textbf{Design}} & \textbf{Input Data} &
\textbf{Output Data} & \textbf{VPU} & \textbf{Reference}  \\
\hline\hline
Averaging Binning &   
$4$-MPixel, $8$b  &  $1$-MPixel, $8$b   & Myriad 2 & Sec. \ref{s9_4}, \ref{s971}\\

Fl. Point Convolution &   
$1$-MPixel, $8$b  &  $1$-MPixel, $8$b  & Myriad 2 & Sec. \ref{s9_4}, \ref{s971} \\

ShipDetect CNN &   
$1$-MPixel$\times3$, $16$b &  $64$$\times$$1$, $32$b  & Myriad 2 & Sec. \ref{s9_4}, \ref{s971} \\

Canny Edge Detection & 
$1$-MPixel, $8/16$b  &  $1$K--$5$K, $5$b & Myriad 2 & Sec. \ref{s9_5}, \ref{s972} \\

Depth Rendering & 
$6$$\times$$1$, $32$b  &  $1$-MPixel, $16$b  & Myriad 2   & Sec. \ref{s9_5}, \ref{s972} \\

Edge Matching & 
$1$-MPixel$\times2$, $8$\&$16$b &   $1$K--$5$K, $8$b  & Myriad 2 & Sec. \ref{s9_5}, \ref{s972} \\

UrsoNet DNN & $1$-MPixel$\times3$, $8$b  &  $4099$$\times$$1$, $32$b  & Myriad X   & Sec. \ref{s9_6}, \ref{s973} \\
\hline
\end{tabular}
\end{table}

\subsubsection{Acceleration of Kernels}

\begin{figure}[!t]
\centering
\subfloat[\label{fig_bn1}]{\includegraphics[width=0.5\textwidth]{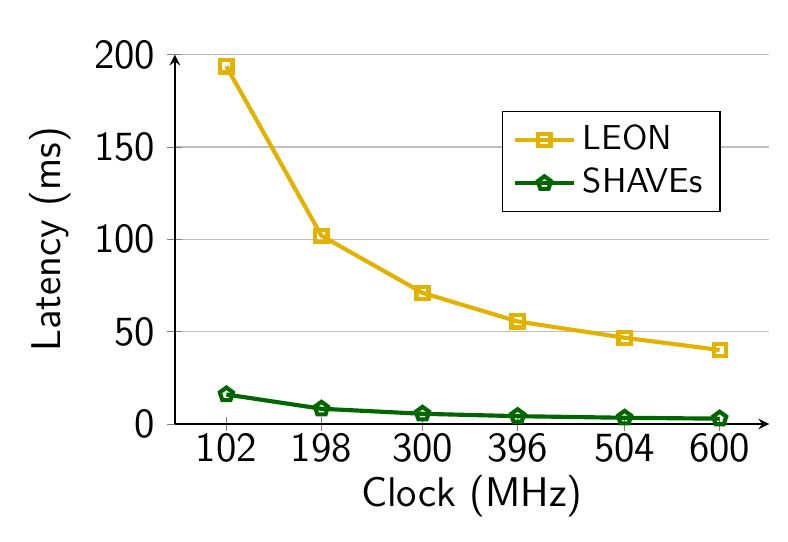}}\\[-2pt] 
\hspace{-5pt}\subfloat[\label{fig_bn2}]{\includegraphics[width=0.5\textwidth]{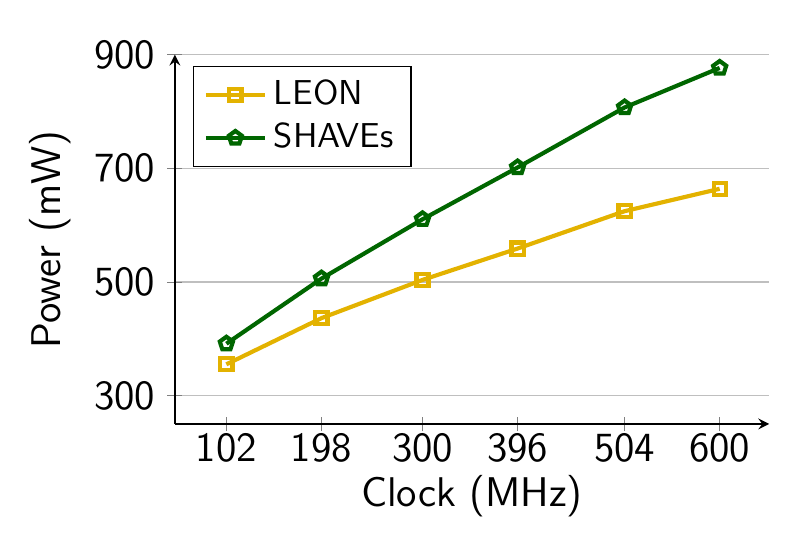}}\hspace{1pt}
\subfloat[\label{fig_bn3}]{\includegraphics[width=0.5\textwidth]{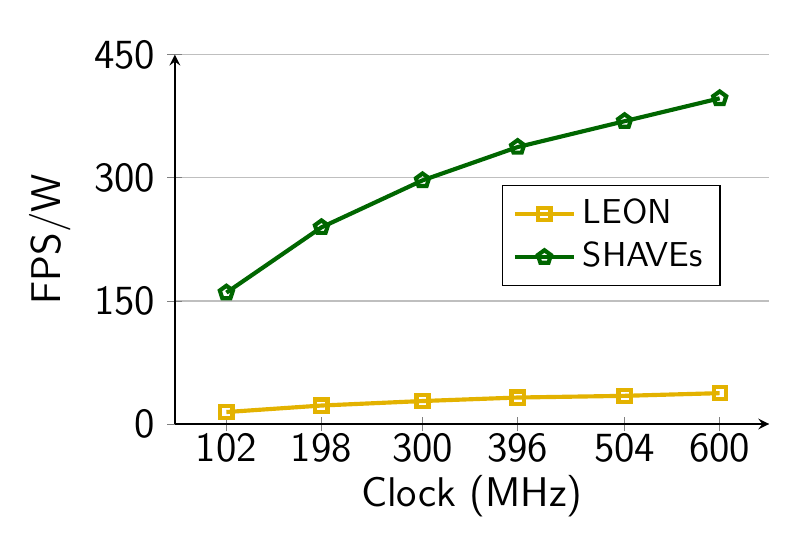}}
\caption[Experimental Results of Averaging Binning on Myriad 2]{Experimental results from the implementation of Averaging Binning on Myriad 2 with respect to the clock frequency:
\textbf{(a)} latency, 
\textbf{(b)} power consumption, 
and
\textbf{(c)} FPS-per-Watt.}%
\label{fig_binn}
\end{figure}

Firstly,
we assess the acceleration of our kernels 
using the general-purpose LEON processor as baseline.
For Averaging Binning, 
we achieve $14\times$ speedup,
which mainly comes from the parallelization to $12$ SHAVEs
(LEON has to scan the entire $4$-MPixel image, 
resulting in significant delay).
Depending on the size of the convolution mask,
we achieve up to $75\times$ speedup
for Floating-Point Convolution. 
Compared to Averaging Binning,
the speedup is larger,
because the convolutions require more demanding computations.
Finally,
the Ship Detection CNN
is implemented only on SHAVEs,
i.e., where the custom inference engine is mapped, 
however,
considering the performance of convolutions,
and given that LEON does not support $16$-bit floating-point
(thus, it must execute the $32$-bit model),
the speedup is expected to be more than $2$ orders of magnitude.

Figure \ref{fig_binn} illustrates the scaling of latency,
power consumption, and performance-per-Watt
of Averaging Binning
for different clock frequencies.
Regarding latency,
it is almost linear to the clock frequency 
whether Averaging Binning is executed on LEON or the $12$ SHAVEs.
The SHAVE implementation is \raisebox{0.8pt}{$\scriptstyle\sim$}$13\times$ faster
than the respective LEON implementation in all experiments. In terms of power, 
as expected, LEON provides $1.1\times$--$1.3\times$ smaller consumption,
however, 
SHAVEs still deliver low power, 
i.e., up to $900$mW.
Moreover,
the power consumption of the SHAVE implementation increases faster than that of LEON.
When considering both Frames Per Second (FPS)
and power,
LEON attains negligible scaling,
while the FPS-per-Watt of SHAVEs increase almost linearly. 
Similar results are derived
for the Floating-Point Convolution,
i.e., up to $900$mW
and $58\times$ speedup.

\subsubsection{System Evaluation}

We also evaluate the system performance
involving both I/O and processing
based on the FPGA--VPU co-processing architecture of Figure \ref{fig_march2},
where the data transfers are performed via the CIF and LCD interfaces. 
In this analysis, the clock frequency of CIF and LCD is configured at $50$MHz,
which guarantees error-free data transfers according to our experiments presented in \cite{LeonICECSm}.
The evaluation regards 
two distinct scenarios concerning the execution order of the I/O handling and processing tasks: 
\begin{enumerate}[ wide = 1pt, leftmargin = *]
    \item[1)] \underline{Unmasked I/O}: 
    assuming serial I/O--processing,
    the VPU receives the input frame from the FPGA, performs the processing, and transmits the output data to the FPGA.
    \item[2)] \underline{Masked I/O}: 
    assuming pipelined I/O--processing and streaming input, 
    the VPU performs in parallel two processes: 
    (i) buffering of output frame $n-1$, CIF reception and buffering of input frame $n+1$, LCD transmission of output frame $n-1$, 
    and  
    (ii) processing of frame $n$.
    In that case, the one LEON processor executes the first process, i.e., the I/O handling,
    and the other takes over the second process,
    i.e., it 
    manages the processing performed by SHAVEs.
 \end{enumerate}
 
The performance results are presented in Table \ref{tb_micecs}.
Both I/O interfaces are operating at $50$MHz and as expected, 
they transmit an $1$-MPixel image in \raisebox{0.8pt}{$\scriptstyle\sim$}$21$ms.
In the Unmasked I/O mode, 
the total throughput ranges between $9$--$20$ FPS for kernels 
with small processing time. 
To implement the Masked I/O mode, 
the I/O data are buffered to an allocated DDR space for data integrity reasons 
(copying an $1$-MPixel image requires \raisebox{0.8pt}{$\scriptstyle\sim$}$42$ms).
As a result, 
the latency of a single frame increases considerably.
Even so, the kernels featuring excessive processing time can benefit
from this masking technique 
and improve their throughput by $1.1\times$--$1.3\times$.
This effect is shown
in the Floating-Point Convolution with a $13\times13$ mask
and the Ship Detection CNN.
In contrast, 
kernels with small processing time suffer 
a throughput decrease when applying masking, 
and the developer must be cautious with respect to the selected mode of operation.
This is evident in Averaging Binning, 
which has only $3$ms computation latency
and
the buffering of its $4$-MPixel input image adds considerable timing overhead.

\begin{table}[!t]
\fontsize{8.95}{10}\selectfont
\renewcommand{\arraystretch}{1.2}
\setlength{\tabcolsep}{3.9pt}
\caption[Experimental Results of Custom DSP \& CNN Kernels on Myriad 2 VPU.]{Experimental results of custom DSP \& CNN kernels on Myriad 2 VPU.}
\label{tb_micecs}
\centering
\begin{threeparttable}
\begin{tabular}{l c c c c c c c}
\hline
\multicolumn{1}{c}{\multirow{4}{*}{\textbf{Kernel}}} 
& 
\multicolumn{3}{c}{\textbf{Function Latency}\setcounter{footnote}{0}\footnotemark} &
\multicolumn{2}{c}{\textbf{Sys-Unmasked I/O}\setcounter{footnote}{1}\footnotemark} &
\multicolumn{2}{c}{\textbf{Sys-Masked I/O}\setcounter{footnote}{2}\footnotemark}\\ 
 \cmidrule(lr){2-4} \cmidrule(lr){5-6} \cmidrule(lr){7-8}
 &
\textbf{CIF} &
\textbf{VPU} &
\textbf{LCD}  &
\textbf{Latency} &
\textbf{Throughput} &
\textbf{Latency} & 
\textbf{Throughput} \\[-1pt]
& (ms) & (ms) & (ms) & (ms) & (FPS) & (ms) & (FPS)\\
\hline \hline 
Avg Binning            & 85	& 3	& 21 & 109 & 9.1  & 906 & 3.2   \\
$3$$\times$$3$ Conv.   & 21 & 8 & 21 & 50 & 20  & 336 & 8    \\
$7$$\times$$7$ Conv.   & 21 & 29 & 21 & 71 & 14.1  & 336 & 8    \\
$13$$\times$$13$ Conv.  & 21 & 114 & 21 & 156 & 6.4  & 336 & 8   \\
ShipDetect              & 63 & 658  & \raisebox{0.8pt}{$\scriptstyle\sim$}0 & 721 & 1.4  & 1505 & 1.5   \\
\hline
\end{tabular}
\begin{tablenotes}
  \item[1]{\fontsize{7.7}{8.8}\selectfont Refers to the input reception via CIF, the processing on VPU, and the output transmission via LCD.}
  \item[2]{\fontsize{7.7}{8.8}\selectfont Regards serial I/O--processing $\rightarrow$ \textbf{Throughput} = 1/(CIF\_Time + VPU\_Time + LCD\_Time).}
  \item[3]{\fontsize{7.7}{8.8}\selectfont Regards pipelined I/O--processing $\rightarrow$ \textbf{Throughput} =     1/($\max\{$VPU\_Time, LCD\_Buffering\_Time + CIF\_Time + CIF\_Buffering Time + LCD\_Time$\}$).}
\end{tablenotes}
\end{threeparttable}
\end{table}

\subsubsection{Comparison to Embedded Devices}

Finally,
we compare our VPU accelerators with other embedded devices.
For the same CNN model,
Myriad 2
provides \raisebox{0.8pt}{$\scriptstyle\sim$}$2.5\times$ less FPS-per-Watt
than the Zynq-7020 FPGA \cite{LeonICECS},
however, 
the latter utilizes almost all the chip resources
and exhibits $4\times$ larger power consumption.
Compared to the Jetson Nano GPU \cite{LeonICECS},
the VPU delivers \raisebox{0.8pt}{$\scriptstyle\sim$}4$\times$ better 
FPS-per-Watt
for the CNN.
For Averaging Binning,
we achieve \raisebox{0.8pt}{$\scriptstyle\sim$}3$\times$ better throughput than 
a typical Zynq FPGA implementation with one binning pipeline in programmable logic 
(one input pixel per clock cycle),
also due to the slower DMA engines of Zynq.

\subsection{Experimental Results of CV Pipeline}
\label{s972}

\hypersetup{urlcolor=blue}

\begin{figure}[!t]
\centering
\includegraphics[width=1\textwidth]{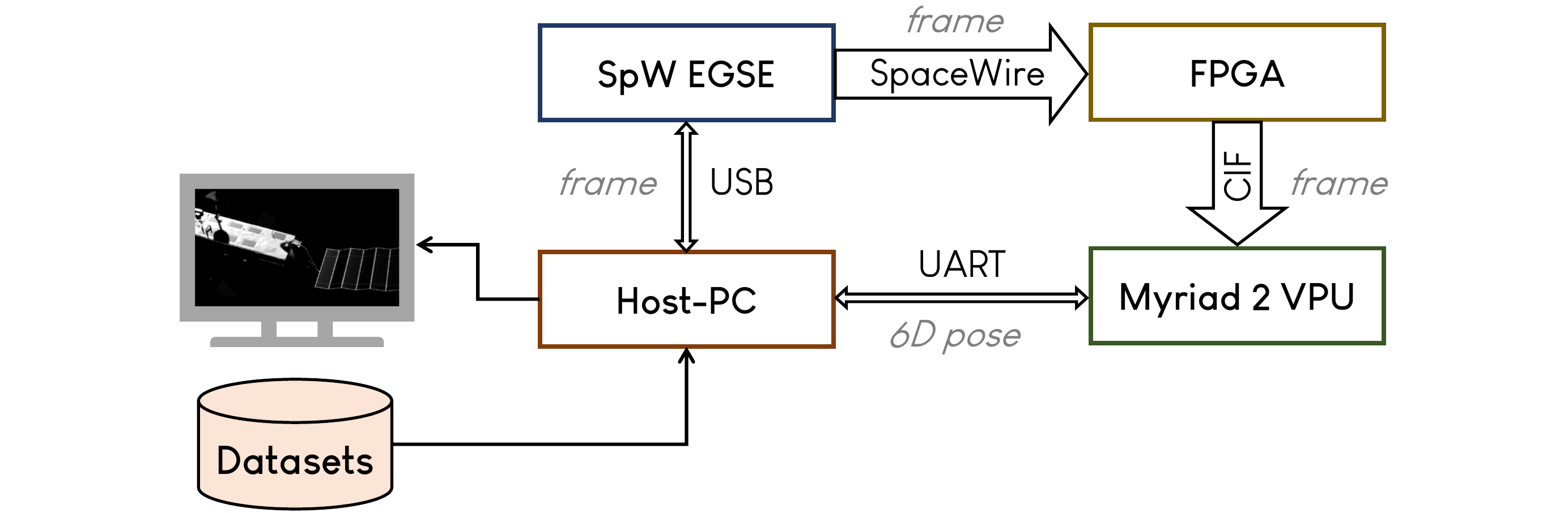}%
\caption[FPGA--VPU Architecture for Accelerating the Computer Vision Pipeline on Myriad 2]{FPGA--VPU co-processing architecture for the acceleration of the computer vision pipeline on Myriad 2.}%
\label{fig_march1}
\end{figure}

Figure \ref{fig_march1} illustrates our testing setup 
for the evaluation of the CV pipeline.
In this setup, 
we employ EGSE \cite{egse},
which comprises an interface unit 
for the real-time simulation of the high-bandwidth SpaceWire link
(i.e., a spacecraft communication network for data transmission). 
The host-PC configures EGSE,
which feeds the input data to the FPGA.
The FPGA transmits the data to Myriad 2 via CIF operating at $5$MHz.
For validation and demonstration\footnote{\fontsize{7.7}{8.8}\selectfont Demo available on YouTube: \textbf{\url{https://youtu.be/9wDLm56zsss}}.} purposes,
we transfer the outputs of the CV pipeline (6D poses)
back to the host-PC.
More details about the FPGA--VPU architecture,
the configuration of SpaceWire,
and
the I/O transfers
are reported 
in our publication in \cite{LeonTECSm}. 

\begin{figure}[!t]
\vspace*{-8pt}
\centering
\subfloat[\label{fig_dtt1}]{\includegraphics[width=0.235\textwidth]{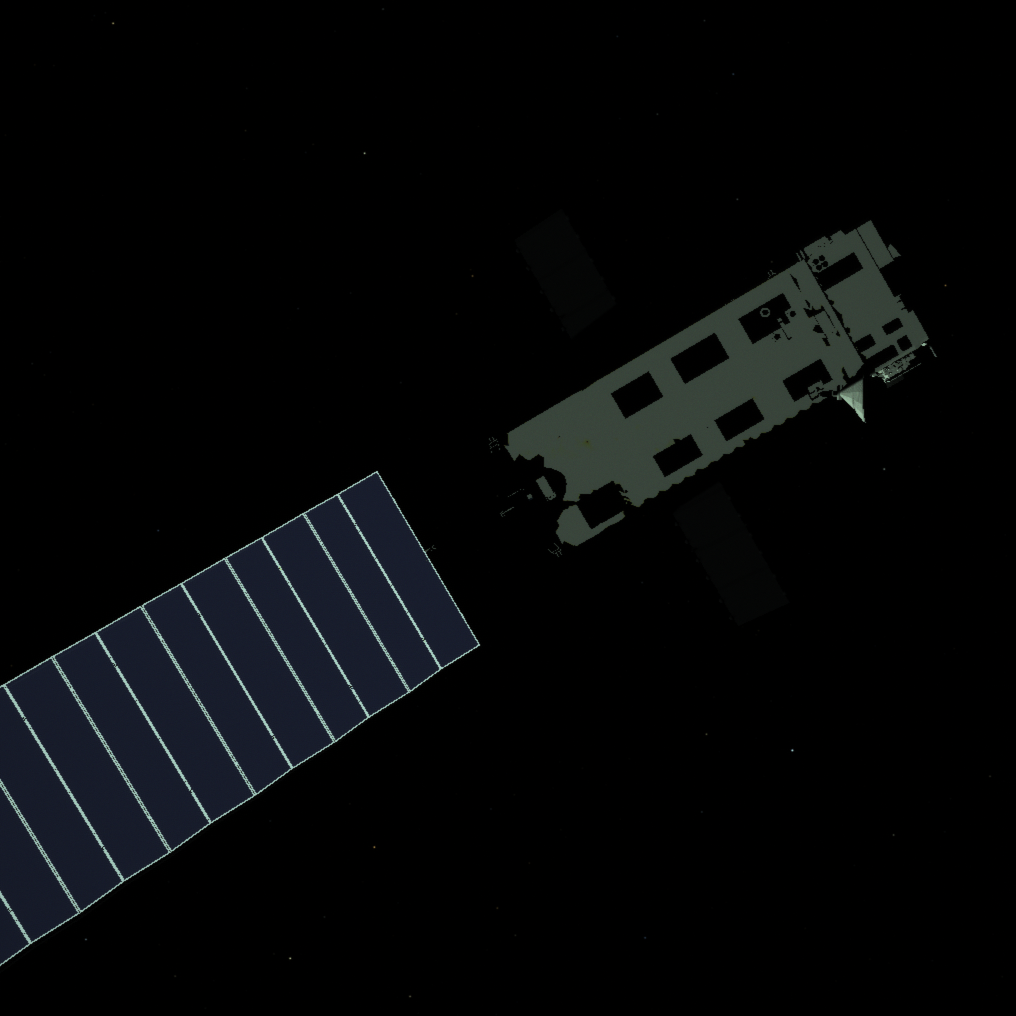}} \hspace*{4pt}
\subfloat[\label{fig_dtt2}]{\includegraphics[width=0.235\textwidth]{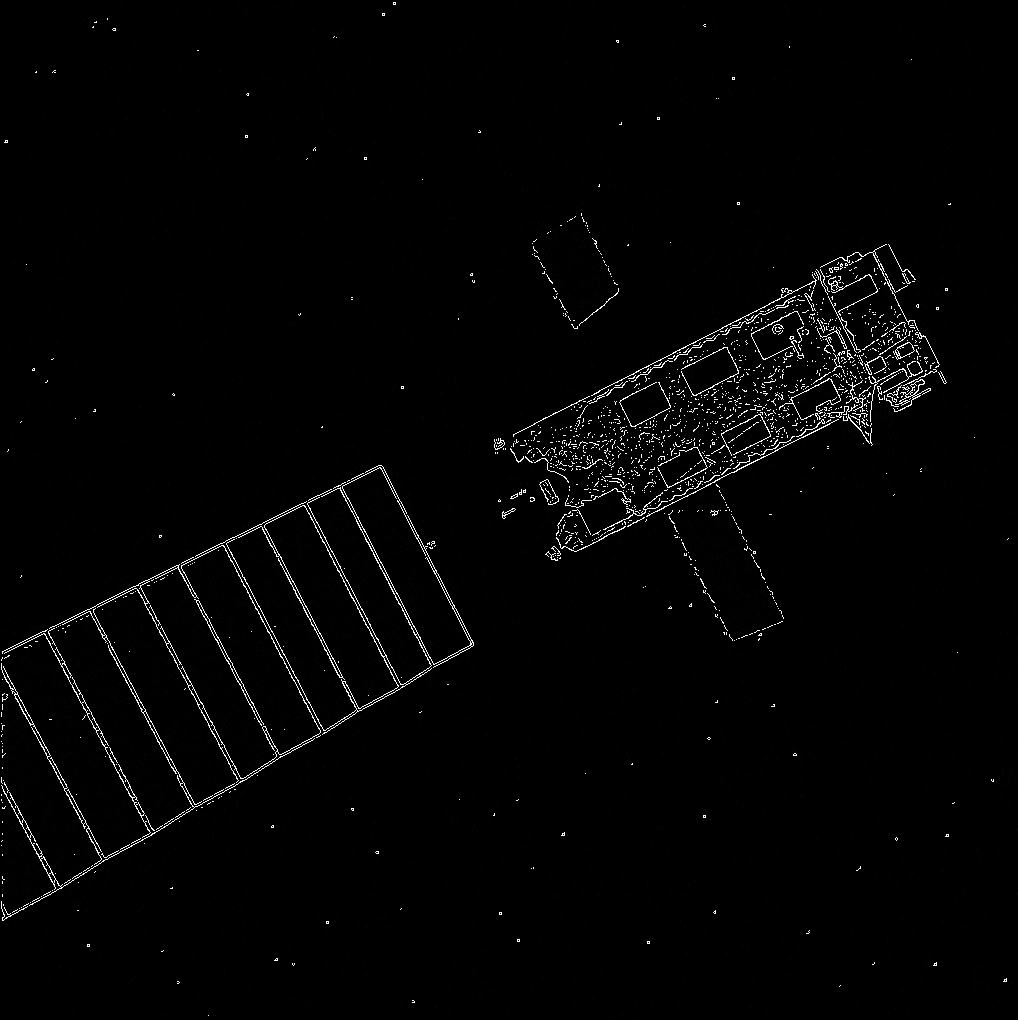}} \hspace*{4pt}
\subfloat[\label{fig_dtt3}]{\includegraphics[width=0.235\textwidth]{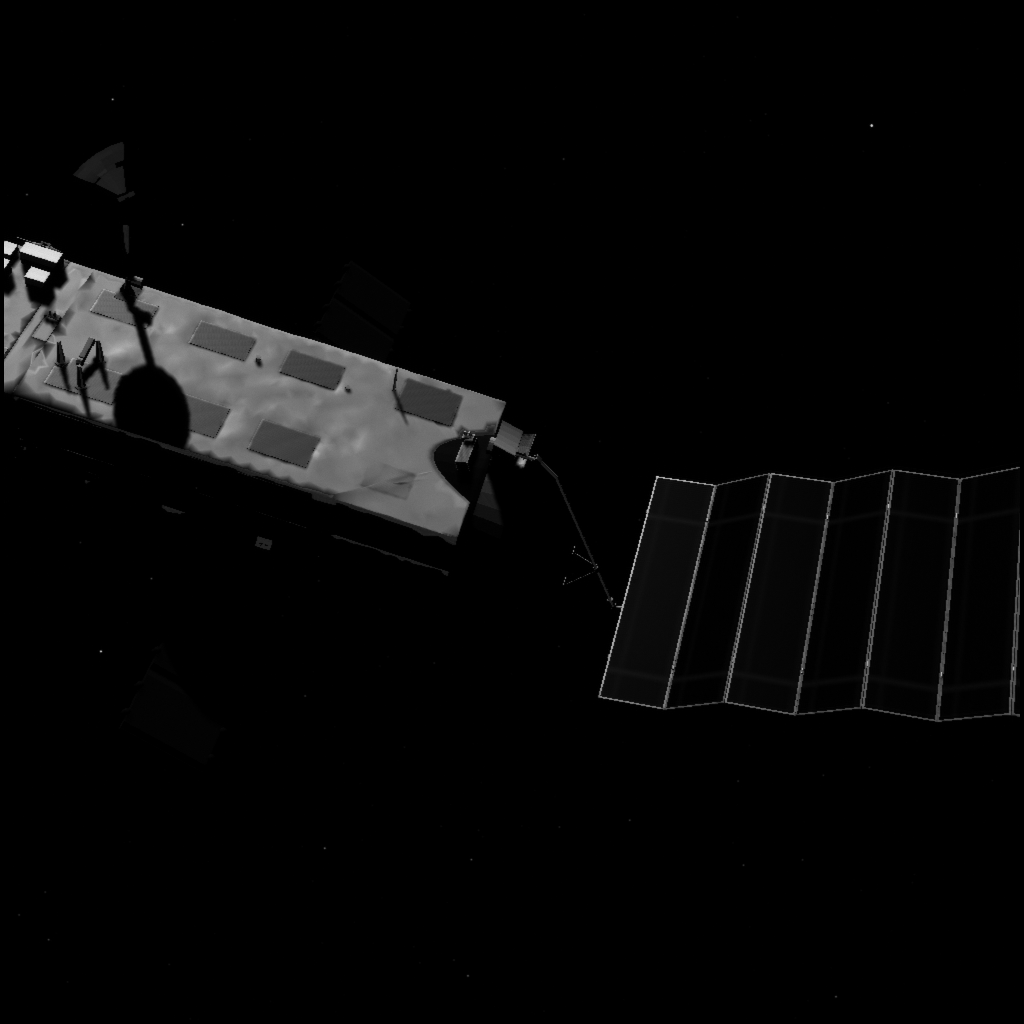}} \hspace*{4pt}
\subfloat[\label{fig_dtt4}]{\includegraphics[width=0.235\textwidth]{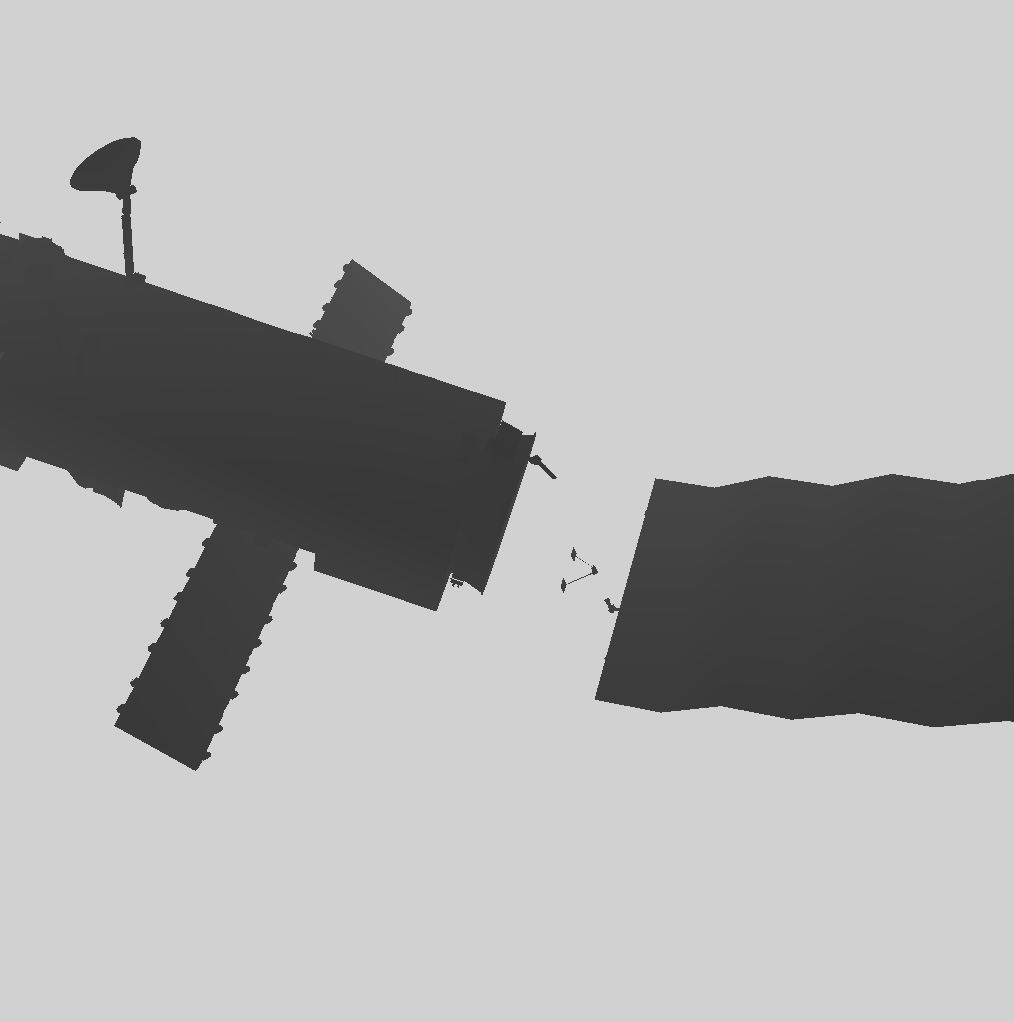}}
\caption[I/O Data of the Computer Vision Pipeline's Functions Accelerated on Myriad 2]{I/O data of main functions in the computer vision pipeline ($30$m--$20$m Envisat sequence):
\textbf{(a)},\textbf{(c)} input, 
\textbf{(b)} output of Edge Detection, 
and
\textbf{(d)} output of Depth Rendering.}%
\label{fig_datas}
\end{figure}

We test our system with a synthetic dataset\footnote{\scriptsize Special thanks to D. Gonzalez-Arjona \emph{et al.} from GMV 
for providing the test dataset of Envisat.} 
that includes two sequences of 
$1000$ images ($1024$$\times$$1024$ $8$-bit grayscale pixels)
realistically simulating the motion of Envisat.
The nature of the data is similar to that of actual rendezvous images
and suffices to evaluate the system performance.
In particular, 
the first frame sequence depicts a satellite rotation out of plane at \raisebox{0.8pt}{$\scriptstyle\sim$}$50$m away from the camera,
while the second one depicts
a satellite approach from $30$m to $20$m while tumbling.
Figure \ref{fig_datas} illustrates frames from the $30$m--$20$m dataset, 
together with outputs from the basic functions of the CV pipeline.

\hypersetup{urlcolor=black}

\subsubsection{Acceleration of Functions}

\begin{figure}[!t]
\vspace*{-5pt}
\centering
\subfloat[\label{fig_spd1}]{\includegraphics[width=0.495\textwidth]{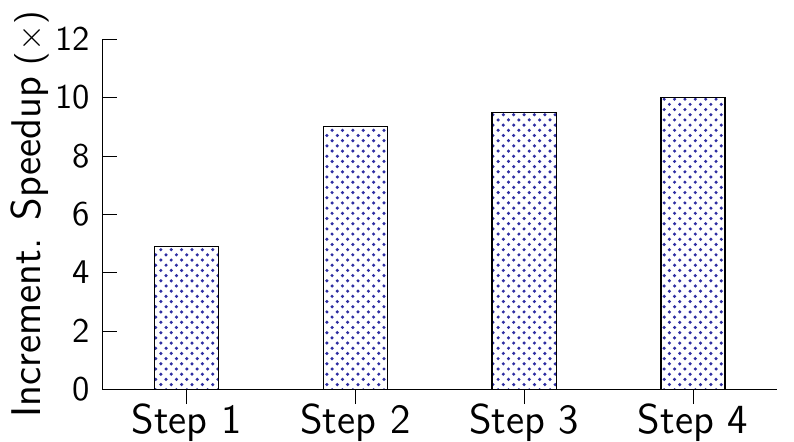}}\\[-8pt] 
\subfloat[\label{fig_spd2}]{\includegraphics[width=0.495\textwidth]{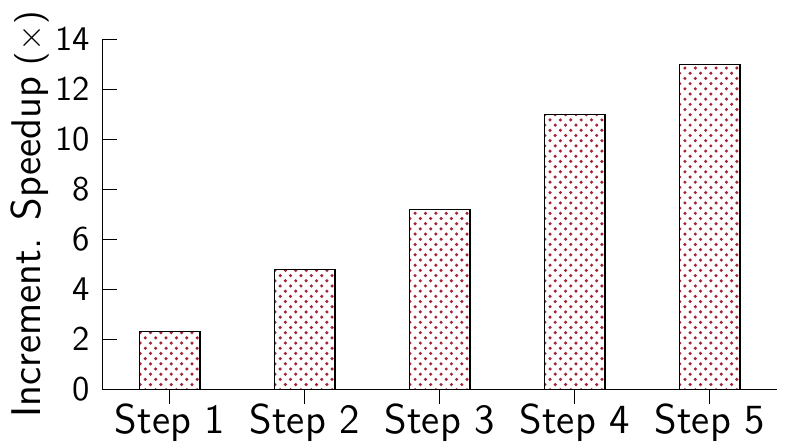}}\hspace{-2pt}
\subfloat[\label{fig_spd3}]{\includegraphics[width=0.495\textwidth]{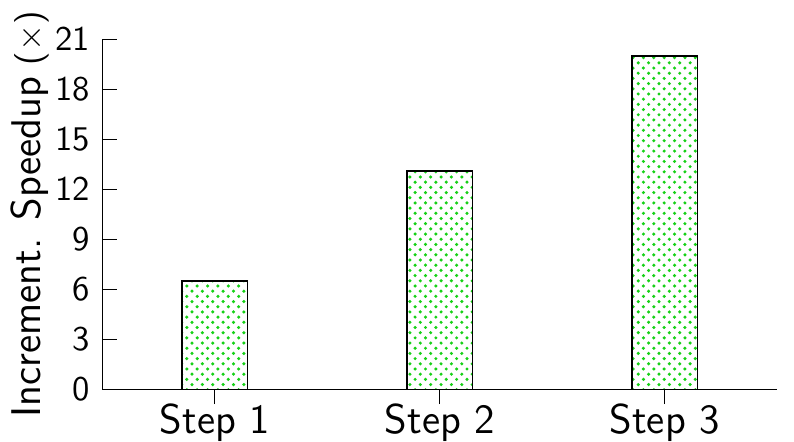}}
\caption[Incremental Acceleration of the Computer Vision Functions on Myriad 2]
{Incremental acceleration of the computer vision functions
with respect to their major implementation steps on Myriad 2:
\textbf{(a)} Edge Detection\setcounter{footnote}{0}\footnotemark,
\textbf{(b)} Depth Rendering\setcounter{footnote}{1}\footnotemark,
and
\textbf{(c)} Edge Matching\setcounter{footnote}{2}\footnotemark.\\
\hspace*{1pt} 
\setcounter{footnote}{0}\footnotemark
\hspace*{-0.7pt} 
{\fontsize{7.7}{8.8}\selectfont 
Step 1: SHAVE porting,
Step 2: improved buffering and in-place computations,
Step 3: loop merging, 
\\ \hspace*{6.5pt} Step 4:   improved task partition.}\\
\hspace*{1pt} 
\setcounter{footnote}{1}\footnotemark
\hspace*{-1.65pt} 
{\fontsize{7.7}{8.8}\selectfont 
Step 1: SHAVE porting,
Step 2: dynamic task assignment,
Step 3: improved task partition,
Step 4: \\ \hspace*{6.5pt}  SIMD  computations,
Step 5: optimized cache configuration.}\\
\hspace*{1pt} 
\setcounter{footnote}{2}\footnotemark
\hspace*{-0.7pt} 
{\fontsize{7.7}{8.8}\selectfont 
Step 1: SHAVE porting,
Step 2: shared L2 and instructions at L1,
Step 3: shared L2 per 2 cores and \\ \hspace*{6.5pt} instructions \& data at L1.}}%
\label{fig_m2speeds}
\end{figure}

At first,
we evaluate the implementation of each function of the CV pipeline.
In particular,
targeting to assess our development methodology
and embedded implementation techniques,
we gradually study the speedup factor. 
Figure \ref{fig_m2speeds} 
presents how each major implementation step contributes 
in the speedup 
for the $30$m--$20$m Envisat dataset.
Namely, 
the speedup of the last step is the final speedup achieved.
Below,
we analyze the implementation steps and the speedup of the CV functions accelerated on SHAVEs. 

The parallelization of Edge Detection to $12$ SHAVEs 
delivers only $4.9\times$ speedup, 
regardless of the pixel bit-width.
The re-design of the function using improved buffering 
and in-place computations 
almost doubles the speedup to $9\times$.
By performing the thresholding before hysteresis 
and by merging loops (histogram calculation during thresholding), 
the speedup is increased to $9.5\times$.
Dividing the image to $24$ stripes rather than $12$ 
adds negligible overhead due to extra DMA transactions, 
however, 
it improves the workload balancing among SHAVEs 
during 
the content-dependent hysteresis task 
and results in the final speedup of $10\times$.

The parallelization of Depth Rendering to $12$ SHAVEs
delivers only 2.3$\times$ speedup when using static task assignment. 
We note that the performance of this function varies among frames, 
as it is highly dependent on the image content.
By employing dynamic task assignment, 
the idle time per SHAVE core 
significantly reduces,
and thus,
the speedup is increased to $4.8\times$. 
Moreover, 
the image partition into more stripes
results in
more fine-grained workload balancing 
and provides an additional speedup boost to $7.2\times$.
Subsequently, 
the exploitation of the SIMD capabilities of SHAVEs
enlarges the speedup by $9\times$--$12\times$.
Finally, 
the read-only nature of the data allows cache optimizations that decrease the execution time even further, to about $119$ms--$212$ms.
In total,
depending on Envisat's orientation in the image,
the Depth Rendering functions attains
a speedup of $10\times$--$16\times$.

Out of all functions,
the configuration of the memory hierarchy has the highest impact on
the performance of Edge Matching.
Its parallelization to $12$ SHAVEs 
provides a remarkable speedup of $6.5\times$,
however, our cache configurations result in even larger speedup.
In particular, 
the employment of the shared L2 and L1 cache for instructions increases it to $13.1\times$. 
By enabling a shared L2 cache of $16$KB per two SHAVEs
and using the dedicated L1 also for data, 
the speedup increases to $20\times$.

\subsubsection{System Evaluation}

Table \ref{tb_mtecs} reports the main experimental results
from the implementation of the CV pipeline in Myriad 2.
It includes results 
from the profiling on LEON and the final SHAVE implementation
for the $30$m--$20$m Envisat dataset. 
Similar results are derived
for the $50$m Envisat dataset.
Nevertheless, 
this dataset 
imposes slightly less computation demands
(the workload is content-dependent),
as the distance of the satellite 
from the camera
is larger,
and thus, its size is smaller. 

The execution times in Table \ref{tb_mtecs}
vary during the test sequence,
because the algorithmic workload is content-dependent,
i.e., 
it is affected by the apparent size of Envisat in the image. 
In addition, 
the efficiency of the task parallelization depends 
on the orientation of Envisat 
due to the mapping of rendering areas to SHAVEs.
We note that the image reception via CIF operating at $5$MHz
requires \raisebox{0.8pt}{$\scriptstyle\sim$}$210$ms, 
however, 
in our final implementation
it is entirely masked by 
the processing of the previous frame 
(see Figure \ref{fig_schedul}).
Hence, the system speedup increases to $8.5\times$--$12\times$ 
(Table \ref{tb_mtecs} does not take into account the I/Os).
Overall, the achieved throughput is $2.6$--$3.8$ FPS 
for the $20$m--$30$m sequence ($325$ms per frame on average),
while for 
the $50$m sequence, 
we achieve a throughput of $3.8$--$4.9$ FPS ($235$ms per frame on average).

\begin{table}[!t]
\fontsize{8.6}{9.5}\selectfont
\renewcommand{\arraystretch}{1.2}
\setlength{\tabcolsep}{1.5pt}
\caption[Experimental Results of the Computer Vision Pipeline on Myriad 2 VPU]{Experimental results of the computer vision pipeline on Myriad 2 VPU.}
\label{tb_mtecs}
\centering
\begin{threeparttable}
\centering
\begin{tabular}{l c c c c c c c c}
\hline
\multicolumn{1}{c}{\multirow{4}{*}{\textbf{Function}}}  & \multicolumn{2}{c}{\textbf{LEON}} & \multicolumn{4}{c}{\textbf{SHAVEs}} &
\multicolumn{2}{c}{\textbf{Gains}}\\ 
\cmidrule(lr){2-3} \cmidrule(lr){4-7} \cmidrule(lr){8-9}
 &
\textbf{Latency} &
\textbf{Memory} &
\textbf{Latency}  &
\textbf{Memory} &
\textbf{Power}  &
\textbf{Accuracy} 
& \textbf{Latency} & \textbf{Memory}\\[-1pt]
& (ms) & (MB) & (ms) & (MB) & (mW) & ($\%$) & ($\times$) & ($\%$)\\
\hline \hline 
I. Edge Det.     & 370--375   & 6.1 & 36--37   & 2.3 & 771 & 99 & 10    & 62.9 
\\
D. Edge Det.     & 375--380   & 7.1 & 39--40   & 3.8 & 771 & 99 & 10    & 47.3  
\\
Depth Rend.     & 1900--2100 & 6.4 & 119--212 & 3.3 & 980 & 100 & 10--16 & 48.6 
\\
Edge Match.       & 100--120   & 3 & 5--6     & 3 & 810 & 100 & 20    & 0 
\\
Pose Refin.     & 100--130   & 6.3 & 100--130\setcounter{footnote}{0}\footnotemark        & 6.3 & 644 & Fig. \ref{fig_alignerr} & 1    & 0 \\
\hline
\multicolumn{1}{c}{\textbf{Pipeline}} & 2845--3105   & -- & 263--388\setcounter{footnote}{1}\footnotemark  & --  & 898 & Fig. \ref{fig_alignerr}  & 8--11\setcounter{footnote}{2}\footnotemark   & --\\
\hline
\end{tabular}
\begin{tablenotes}
  \item[1]{\fontsize{7.7}{8.8}\selectfont Pose Refinement was not accelerated on SHAVEs.}
  \item[2]{\fontsize{7.7}{8.8}\selectfont Intensity Edge Detection is not added in the total latency (masked by Pose Refinement).}
  \item[3]{\fontsize{7.7}{8.8}\selectfont Without I/O. With image reception via CIF (\raisebox{0.8pt}{$\scriptstyle\sim$}$210$ms),
  the speedup increases to $8.5\times$--$12\times$ (CIF is masked).}
\end{tablenotes}
\end{threeparttable}
\end{table}

Regarding power consumption,
the MDK routines report $0.7$W--$1$W 
(see Table \ref{tb_mtecs})
for the execution of each individual function on SHAVEs,  
with Depth Rendering being the most power-hungry.  
In comparison with the porting on LEON,
these values are slightly increased by $1.2\times$--$1.4\times$, 
as LEON consumes $600$mW--$700$mW depending on the function.
On average, when the entire CV pipeline operates, 
we get $0.9$W.
For the execution of the full system, 
i.e., when
including the image reception via CIF, 
the power consumption lies between $0.8$W and $1.1$W.
We note that our board measurements 
via an external multimeter report a power consumption of $1.2$W. 

In terms of memory, 
the integration of the entire CV pipeline along with the CIF module results in small utilization. 
The total memory (data and instructions) is around $20$MB, 
i.e., smaller than the $10\%$ of the available DDR resources. 
Furthermore, 
by storing the instruction code in DDR, 
there are no timing penalties for the execution on SHAVEs, 
and CMX is used only for data and internal structures.
Additionally, 
due to the applied optimizations, i.e., improved buffering and in-place processing, 
we achieve significant memory gains: 
$55\%$ on average for the two Edge Detection functions and $49\%$ 
for Depth Rendering (see Table \ref{tb_mtecs}).

\begin{figure}[!t]
\vspace*{-10pt}
\centering
\subfloat[\label{fig_er30}]{\includegraphics[width=0.41\textwidth]{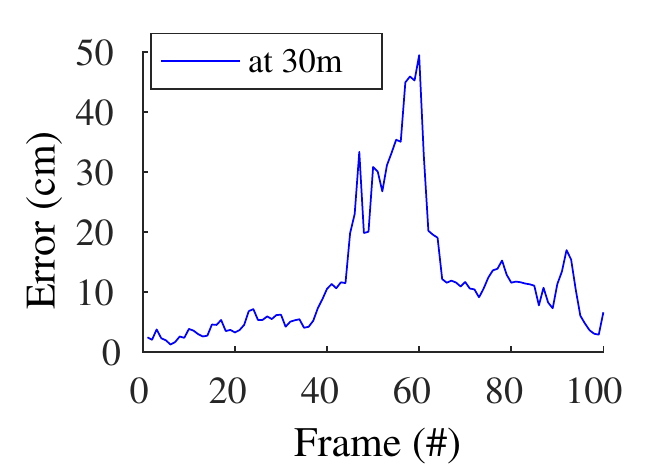}} \hspace*{14pt}
\subfloat[\label{fig_er50}]{\includegraphics[width=0.41\textwidth]{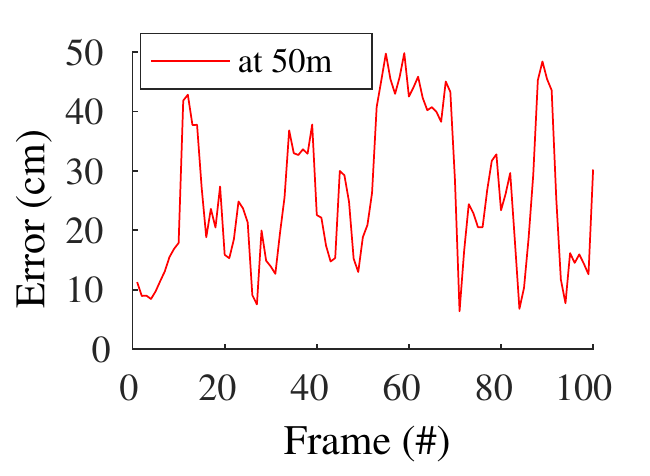}}
\caption[Alignment Error of Envisat's Pose Computed by Myriad 2]{Alignment error of Envisat's pose computed by the computer vision pipeline in Myriad 2 at 
\textbf{(a)} $30$m 
and
\textbf{(b)} $50$m camera distance.}
\label{fig_alignerr}
\end{figure}

In terms of individual accuracy, 
the functions generate the same sets of results 
as the initial software. 
A small exception is in Edge Detection,
which misses $1\%$ of the edges when executing 
slightly fewer hysteresis recursions at the stripe borders (compile-time configurable).  
We also note that
for Pose Refinement, 
we port on LEON a relatively old version of BLAS/LAPACK 
(v3.2.1, without guaranteed compatibility to the initial software).
Figure \ref{fig_alignerr} shows the alignment error of Envisat's pose,
as it is defined in \cite{lentaris_tvideo}. 
The error versus Envisat's distance is in the area of $1\%$ 
and tracking is lost only in few specific frames. 
In general,
the behavior of pose tracking on Myriad 2 is similar 
to that of the CPU and FPGA implementations \cite{lentaris_tvideo}.

Finally, in Table \ref{tb_volred}, we evaluate the efficacy of the proposed architecture as an edge processor by examining the data volume reduction. 
Interestingly, we achieve the very important goal of data reduction at the edge, 
which relieves several potential bottlenecks 
concerning the storage, the network's bandwidth and energy consumption, as well as the I/O throughput.
In particular, 
considering the input and output data, 
i.e., $1024$$\times$$1024$ $8$-bit images and $6$$\times$$1$ $32$-bit vectors, respectively,
as well as the latency of the input reception and processing, 
we achieve a data reduction factor of $10^5$ bps.

\begin{table}[!t]
\fontsize{9}{10}\selectfont
\renewcommand{\arraystretch}{1.2}
\setlength{\tabcolsep}{7pt}
\caption[Data Volume Reduction in the Implementation of the Computer Vision Pipeline in Myriad 2 VPU]{Data volume reduction in the implementation of the computer vision pipeline in Myriad 2.}
\label{tb_volred}  
\centering
\begin{tabular}{l c | l c} 
\hline
\multicolumn{2}{c|}{\textbf{VPU Input}} & \multicolumn{2}{c}{\textbf{VPU Output}} \\
\hline \hline 
Data Type & $1024$$\times$$1024$ $8$-bit frames & Data Type & $6$$\times$$1$ $32$-bit vectors \\
Data Volume & $8$MB per frame & Data Volume & $192$b per vector \\
Reception Time & \raisebox{0.8pt}{$\scriptstyle\sim$}$210$ms & Processing Time & \raisebox{0.8pt}{$\scriptstyle\sim$}$325$ms | \raisebox{0.8pt}{$\scriptstyle\sim$}$235$ms \\
Reception Rate & \raisebox{0.8pt}{$\scriptstyle\sim$}$38$ Mbps & Processing Rate & \raisebox{0.8pt}{$\scriptstyle\sim$}$590$ bps | \raisebox{0.8pt}{$\scriptstyle\sim$}$817$ bps\\
\hline
\end{tabular}
\end{table}

\subsubsection{Comparison to Embedded Devices} 
Finally, 
we provide an overall 
evaluation of the Myriad 2 VPU,
including comparisons to other embedded devices. 
When comparing individually the implementation of the functions
to other ARM-based or GPU-based mobile devices,
we verify the power efficiency of Myriad 2. 
For Canny Edge Detection,
we achieve $1$ order of magnitude faster execution than ARM Cortex-A9
(\raisebox{0.8pt}{$\scriptstyle\sim$}$326$ms per MPixel image, single-threaded at $667$MHz) \cite{lentaris_tvideo},
and approximately $3\times$ better performance-per-Watt than Nvidia's Jetson TK1 GPU
(\raisebox{0.8pt}{$\scriptstyle\sim$}$12$ms per MPixel image, $192$ CUDA cores, but with $10$W) \cite{canny_cmp1}.
For Depth Rendering, 
we achieve \raisebox{0.8pt}{$\scriptstyle\sim$}$7\times$ faster execution than ARM Cortex-A9 
(\raisebox{0.8pt}{$\scriptstyle\sim$}$869$ms per $30$m-distant-image, single-threaded at $667$MHz)
\cite{lentaris_tvideo}
and $5$$\times$--$11$$\times$ 
better power consumption than high-end mobile GPUs
(given their nominal values, see Table \ref{tb_embeds}). 
Regarding system performance, 
our efficient SoC utilization provides tenfold acceleration versus LEON4@600MHz. 
The achieved FPS approach $5$ 
and can be further improved via customization at algorithmic level, e.g., smaller rendering, 
which is generally considered as sufficient for VBN in space \cite{glent}.
When comparing to Zynq-7000
implementing the same algorithm \cite{lentaris_tvideo}, 
the FPGA processes approximately $3\times$ more FPS than Myriad 2, in total. 
For specific functions, i.e., Edge Detection or Depth Rendering,
Zynq also achieves $6\times$--$14\times$ faster execution. 

Regarding the power consumption, the $0.8$W--$1.1$W of Myriad 2 
is significantly lower than that of other high-performance processors, 
with only few CPUs consuming  \raisebox{0.8pt}{$\scriptstyle\sim$}$1$W 
by operating at significantly lower clock frequency. 
Based on our methodology,
we exploit all the capabilities of the tools and SoC
to deliver
a very high performance-per-Watt ratio in Myriad 2,
which matches the notable FPGA results measured  
in \cite{glent, lentaris_tvideo} for such tasks.
Compared to Zynq \cite{lentaris_tvideo}, 
Myriad 2 achieves the same, or slightly better, performance-per-Watt
by trading approximately $3\times$ speed for $4\times$ mean power.
Additionally,
it has a more stable consumption than
the $2$W--$9$W variation of the FPGA, 
allowing for simpler electronics design.

Another advantage of Myriad 2 over the FPGA
is that it utilizes less than $10\%$
of the available memory space,
allowing to implement additional functions 
or alternative algorithms for the current functions.
In contrast, 
the same FPGA implementation utilizes $28\%$--$77\%$ of the chip \cite{lentaris_tvideo},
thus, it would
require dynamic reconfiguration to change functionality.
In terms of accuracy, 
the Myriad 2 processor 
can meet exactly the same arithmetic requirements as an ordinary CPU
during the entire algorithm execution.
This can be viewed as an advantage in cases where the original 
algorithm,
which is usually developed on CPU, 
must remain $100\%$ 
intact with respect to its intermediate and output values.

\subsection{Experimental Results of DNN Kernel}
\label{s973}

\subsubsection{Acceleration of UrsoNet} 
In this section,
we report the results
from the deployment of the UrsoNet DNN on Myriad X.
The latency of the pre-processing stage,
which involves image resampling
to transform the $1024$$\times$$1024$$\times$$3$
input image to   $512$$\times$$512$$\times$$3$,
is $1$ms--$20$ms, depending on the algorithm.
A single DNN inference (synchronous execution) 
on the 512$\times$512$\times$3 image requires around 373ms,
which is 5$\times$ faster than inferencing on the initial image. 
At the same time, 
the mean Location Error (LOCE) is $1.68$m--$1.89$m 
and the mean Orientation Error (ORIE) is $25.08^\circ$--$27.89^\circ$,
namely both remain in the range of the original DNN without resampling,
where LOCE is $1.78$m and ORIE is $27.83^\circ$.
In all cases, 
as shown in \cite{urso},
more fine-grained training,
which is not the purpose of our work, 
can provide better accuracy results for the same performance.
We remind that we apply different training settings
(e.g., another pre-trained model, less epochs)
and architectural modifications
(e.g., decreased resolution for location and orientation, and smaller bottleneck layer). 
Overall,
the pre-processing stage for image resampling,
which is parallelized to $16$ SHAVEs,
adds very small latency,
however, it facilitates the inference stage. 
As a result,
the entire synchronous execution
sustains a throughput of \raisebox{0.8pt}{$\scriptstyle\sim$}$2.7$ FPS.
The throughput is expected to increase for asynchronous inferencing,
while
further speedup can be achieved by
inferencing on smaller tensor, e.g., $192$$\times$$256$$\times$$3$.
In this case, 
Myriad X delivers $15$ FPS,
whereas the accuracy loss in not significant, 
as LOCE lies in the range $1.9$m--$2$m. 


\subsubsection{Comparison to Embedded Devices} 

Following the analysis of the experimental results on Myriad X,
we provide a comparison to other embedded devices
that are also considered for space avionics.
Table \ref{tb_ursocomp} presents the results
for inferencing on VPU, CPU, and GPU.
For this experiment,
we use input images that are scaled to
$512$$\times$$640$$\times$$3$.
The CPU is the ARM Cortex-A57
of Nvidia's Jetson Nano board,
while the GPU is the 128-core Maxwell of the same board.
For Jetson Nano,
we consider its two power modes:
the ``low-power'' at $5$W
and the ``high-performance'' at $10$W.
To make a fair comparison,
we consider the NCS2 VPU
hosted on a single-board computer (i.e., a Raspberry Pi 3), 
and thus,
the total power consumption for inferencing is $5$W.
In terms of latency,
the 4-core Cortex-A57 operating at $1.4$GHz is relatively slow,
thus, NCS2 delivers a $6\times$ speedup.
Compared to the GPU,
NCS2 achieves a slight improvement of $1.3\times$,
but with 
half of the GPU's power consumption. 
When considering both
throughput and power,
NCS2 provides
\raisebox{0.8pt}{$\scriptstyle\sim$}$1.7\times$ more FPS-per-Watt than GPU
and
\raisebox{0.8pt}{$\scriptstyle\sim$}$8.5\times$ more FPS-per-Watt than CPU.
Finally,
we report comparison results for the original ResNet-50 network
(224$\times$224$\times$3 image),
which is the backbone of UrsoNet.
For ResNet-50,
the Jetson Nano GPU  
provides increased throughput versus Pi 3 + NCS2,
i.e., 1.3$\times$--1.9$\times$ more FPS,
however,
when also considering FPS-per-Watt,
the VPU inference outperforms the GPU
by 1.1$\times$--1.5$\times$.

\begin{table}[!t]
\fontsize{9}{10}\selectfont
\renewcommand{\arraystretch}{1.2}
\setlength{\tabcolsep}{2pt}
\caption[Experimental Results of the UrsoNet DNN on Embedded Devices]{Experimental results of the UrsoNet DNN on embedded devices ($512$$\times$$640$$\times$$3$ input tensor).}
\label{tb_ursocomp}  
\centering
\begin{threeparttable}
\begin{tabular}{l  | c c c | c }
\hline
\multicolumn{1}{c|}{\multirow{2}{*}{\textbf{Device}}} 
& \textbf{Latency} & \textbf{Throughput} & \textbf{Power} & \multirow{2}{*}{\textbf{FPS-per-Watt}}  \\[-1pt]
&  (ms) & (FPS) & (W) &  \\
\hline
\hline 
Intel Myriad X VPU (NCS2 + Pi 3)\setcounter{footnote}{0}\footnotemark
& 588   & 1.7 & 5  & 0.34 \\ 
\multirow{2}{*}{ARM Cortex-A57 CPU (Jetson Nano)\setcounter{footnote}{1}\footnotemark} 
& 2830  & 0.4  & 10 & 0.04  \\ 
& 7519  & 0.1  & 5  & 0.02  \\ 
\multirow{2}{*}{Nvidia Maxwell GPU (Jetson  Nano)\setcounter{footnote}{2}\footnotemark}    
& 761   & 1.3  & 10 & 0.13  \\ 
& 958   & 1    & 5  & 0.2  \\ 
\hline
\end{tabular}
\begin{tablenotes}
  \item[1]{\fontsize{7.7}{8.8}\selectfont NCE \& 16-core SHAVE @700MHz. NCS2 (2W) hosted on a Raspberry Pi 3 (3W).}
  \item[2]{\fontsize{7.7}{8.8}\selectfont 10W mode: 4-core @1.4GHz. 5W mode: 2-core @918MHz.}
  \item[3]{\fontsize{7.7}{8.8}\selectfont 10W mode: 128-core @921MHz. 5W mode: 128-core @614MHz.}
\end{tablenotes}
\end{threeparttable}
\end{table}

\section{Conclusion}
\label{s9_8}

In this chapter,
we accelerated various DSP and AI algorithms
on the multi-core Myriad VPUs.
These SoCs are characterized 
by extremely low power consumption 
(i.e., \raisebox{0.8pt}{$\scriptstyle\sim$}$1$W--$2$W)
and increased heterogeneity in terms of processors and memories.
As shown in practice,
the efficient utilization of such an 
heterogeneous SoC architecture
requires a design methodology 
involving
algorithmic analysis, 
multi-level parallelization,
and extensive low-level optimization and tuning,
especially when having 
to deploy multiple diverse software kernels.
Based on our methodology,
at first,
we
implemented custom kernels,
i.e., averaging image binning, floating-point convolutions,
and a CNN for detecting ships on satellite images.
Afterwards,
we accelerated 
a sophisticated 5-stage CV pipeline for tracking the pose of the Envisat satellite,
which includes kernels such as Canny edge detection, depth rendering and perpendicular edge matching.
Moreover,
we deployed the demanding UrsoNet DNN of ResNet-50 backbone
for estimating the pose of the Soyuz spacecraft. 
For the implementations,
we successfully applied various
high- and low-level techniques
such as dynamic task scheduling, 
SIMD operations, improved buffering, 
variable tuning, and optimized memory configurations. 
Our experimental evaluation on Myriad 2
shows that for individual kernels,
we provided $10\times$--$20\times$ speedup
compared to the general-purpose LEON4 CPU.
At system-level,
the entire CV pipeline for pose tracking
on MPixel images
provided a speedup of $8.5\times$--$12\times$,
while sustaining a throughput of
$2.6$--$4.9$ FPS 
with only $0.8$W--$1.1$W.
Regarding Myriad X,
the inference of UrsoNet on resampled MPixel images
delivered a throughput of $2.7$ FPS within the power envelope of $2$W. 
Finally,
compared to other embedded devices,
the Myriad VPUs provide significantly better power efficiency,
i.e., $5\times$ versus the Jetson Nano GPU and $4\times$ versus the Zynq FPGA.
In terms of performance,
they are outperformed by the Zynq FPGA,
however,
when considering the performance-per-Watt ratio,
they provide the same or even better results. 
Indicatively, 
for the CV pipeline,
Myriad 2 trades a $3\times$ loss in speed
for a $4\times$ gain in mean power consumption.

\bookmarksetup{startatroot}
\addtocontents{toc}{\protect\vspace{12pt}}

\chapter{Conclusion}
\label{chapter10}

\section{Summary of Main Contributions}

The goal of the current Ph.D. Dissertation was to design and evaluate DSP and AI accelerators that fulfill the requirements of modern computing systems.
Towards reaching this goal, 
we adopted various design approaches from different layers of the computing stack.
Starting from the bottom to the top design layer,
our work involved
arithmetic circuits, hardware accelerators, FPGA implementations
and implementations on embedded SoCs. 

At circuit level,
we exploited the promising design paradigm of Approximate Computing
to propose new approximation techniques for energy-efficient multipliers,
which are key processing units in DSP/AI hardware accelerators.
In this context,
we examined several aspects of Approximate Computing,
including low-level optimizations, hybrid encodings, runtime configurability, and cooperative approximation. 
Our approximate multiplication circuits
provide a very large approximation space and slow error scaling in the typical acceptable error segment,
while they outperform several state-of-the-art designs. 
All the proposed circuits were efficiently integrated in bigger
DSP/AI hardware 
accelerators
based on our design methodology. 

Next,
we provided acceleration
and efficient mapping of DSP algorithms
on the new European space-grade FPGAs,
which require special treatment due to several factors
(e.g., new tools, lower performance than commercial FPGAs).
Our development was accompanied by a design methodology
targeting to highlight
all the acceleration and mapping opportunities,
and also aid us to
surpass issues that arose
either by the new tool or due to HDL porting on a different FPGA vendor.
The final resource utilization of high-performance 
algorithms for feature detection and stereo vision 
was comparable to that of well-established FPGAs,
while the performance was sufficient for space applications.

Finally,
we provided acceleration and efficient mapping of DSP/AI algorithms
on the multi-core VPUs.
These SoCs
are very heterogeneous 
in terms of processors and memories,
however,
several challenges need to be addressed
to deliver increased acceleration.
To offer very low power consumption,
the VPUs have sacrificed computational power,
while the SoC's complexity
requires systematic study
to efficiently map and schedule compute-intensive algorithms.
Therefore, 
we proposed a
design methodology
and several high- and low-level implementation techniques
to accelerate 
classic DSP kernels,
as well as a sophisticated CV pipeline
and a DNN algorithm
(both
for satellite pose estimation).  

Overall,
the \textbf{key contributions} of the Ph.D. Dissertation
are summarized as follows:

\begin{itemize}[wide =3pt, leftmargin =*, labelsep=3mm]

\item Extensive and up-to-date survey in the field of Approximate Computing,
which reviews and classifies software and hardware approximation techniques 
(\textbf{Chapter \ref{chapter2}}). 

\item Low-level optimizations in the alternative DLSB numerical format 
(\textbf{Chapter \ref{chapter3}}).  

\item New arithmetic approximation techniques, which generate the RAD, AxFXU/ AxFPU and ROUP families of approximate multipliers 
(\textbf{Chapters \ref{chapter4}--\ref{chapter6}}).

\item Improvement ($3$ times) of the
state-of-the-art 
energy-error Pareto front 
of approximate multipliers
(\textbf{Chapters \ref{chapter4}--\ref{chapter6}}).

\item Seamless runtime configuration of the approximation in the DyFXU/DyFPU approximate multipliers 
(\textbf{Chapter \ref{chapter5}}).  

\item The design approach of ``cooperative approximation'', which
combines various orthogonal approximation techniques to provide a very large design (approximation) space
(\textbf{Chapter \ref{chapter6}}).  

\item Methodology for the development of approximate DSP and AI hardware accelerators, either for ASIC or FPGA, 
which is based on extensive design space exploration with differing approximations, algorithms, arithmetic formats, and hardware design techniques 
(\textbf{Chapter \ref{chapter7}}).

\item Experimental results for various approximate DSP and AI hardware accelerators, including kernels for 1D/2D signal processing and CNNs
(\textbf{Chapter \ref{chapter7}}).

\item Methodology for the mapping and acceleration of
high-performance DSP kernels on the new space-grade BRAVE FPGAs and tools
(\textbf{Chapter \ref{chapter8}}).

\item Experimental results from the implementation of various DSP kernels (feature detection and stereo vision) on space-grade FPGAs of the market 
(\textbf{Chapter \ref{chapter8}}).

\item Methodology for the mapping and acceleration of
high-performance DSP and AI algorithms on the heterogeneous multi-core Myriad VPUs
(\textbf{Chapter \ref{chapter9}}).

\item High-level and low-level techniques for the partitioning, scheduling, and mapping of demanding algorithms 
on the heterogeneous multi-core Myriad VPUs
(\textbf{Chapter \ref{chapter9}}).

\item Experimental results from the implementation of various DSP and AI kernels (convolution, image classification, pose tracking) on the heterogeneous multi-core Myriad VPUs 
(\textbf{Chapter \ref{chapter9}}).

\item Evaluation of NanoXplore's space-grade BRAVE FPGAs
and Intel's COTS Myriad VPUs
as candidate on-board processors for space missions
(\textbf{Chapters \ref{chapter8}--\ref{chapter9}}).

\end{itemize}

\section{Future Work}
Regarding the work of \textbf{Part \ref{part1}},
Approximate Computing is applied
in all the layers of the typical computing stack.
As a result, 
research is conducted 
at the transistor layer,
circuits,
hardware accelerators,
micro-architecture, 
runtime systems,
compilers,
and 
programming languages. 
In this Ph.D. Dissertation,
the proposed approximation techniques
are applied at the logic level 
and the design of arithmetic circuits,
which are then integrated in hardware accelerators.
However,
to exploit the full potential of Approximate Computing,
our arithmetic approximation techniques can be combined
with approximations inserted 
in other layers,
namely, 
apply cross-layer approximation.
Other future extensions
include the integration of our approximate circuits
in processors
(e.g., in open-source GPUs or RISC-V CPUs), 
the system-level application 
of our runtime approximation configuration
(e.g., to provide tuning with respect to the desired accuracy constraints), 
and the automatic selection of
the best approximation configurations 
for a given application
(which is currently performed manually). 

Regarding the work of 
\textbf{Part \ref{part2}},
segments of our design methodologies can be performed in an automatic fashion,
e.g., the exploration of the tool settings in the FPGA development.
Our methodology about the programmable logic of the space-grade FPGAs
can be extended to take into account
the ARM processor of the SoCs,
while an open issue is the successful 
integration of high-performance I/O links for data transmission.
Our work in the VPUs
can be extended 
towards the 
synergistic application of both classic CV and AI algorithms,
in order to increase the system robustness,
e.g., execute both the CV pipeline and the UrsoNet DNN for
pose estimation/tracking. 
Moreover, 
we are already developing 
similar methodologies
and performing benchmarking on 
other embedded platforms such as the TPUs. 
Finally,
as
our research activities with the FPGAs and VPUs
target on-board computing in space,
we are also working
on equipping COTS devices
(e.g., Zynq and Myriad)
with fault-tolerant mitigation techniques,
in an effort to increase their reliability 
without affecting the DSP/AI performance.

{
\bibliographystyle{IEEEtran}
\bibliography{REFs}
}
\addtocontents{toc}{\protect\vspace{12pt}}

\chapter*{Publications}
\addcontentsline{toc}{chapter}{Publications}
\chaptermark{Publications}

\section*{Journals}
\begin{enumerate}[leftmargin=11.3mm,labelwidth=*,labelsep=15pt]

\item[J11.]%
\textbf{\underline{V. Leon}}, M. A. Hanif, G. Armeniakos, X. Jiao, M. Shafique, K. Pekmestzi, and D. Soudris, 
{``Approximate Computing Survey, Part II: Application-Specific \& Architectural Approximation Techniques and Applications''},
\textbf{\em ACM Computing Surveys}, under review \newline
DOI: 
\href{https://doi.org/10.48550/arXiv.2307.11128}{\color{blue}{10.48550/arXiv.2307.11128}}

\item[J10.]%
\textbf{\underline{V. Leon}}, M. A. Hanif, G. Armeniakos, X. Jiao, M. Shafique, K. Pekmestzi, and D. Soudris, 
{``Approximate Computing Survey, Part I: Terminology and Software \& Hardware Approximation Techniques''},
\textbf{\em ACM Computing Surveys}, under review \newline
DOI: 
\href{https://doi.org/10.48550/arXiv.2307.11124}{\color{blue}{10.48550/arXiv.2307.11124}}

\item[J9.]%
\textbf{\underline{V. Leon}}, P. Minaidis, G. Lentaris, and D. Soudris, 
{``Accelerating AI and Computer Vision for Satellite Pose Estimation on the Intel Myriad X Embedded SoC''},
\textbf{\em Elsevier Microprocessors and Microsystems}, vol. 103, 2023\newline
DOI: 
\href{https://doi.org/10.1016/j.micpro.2023.104947}{\color{blue}{10.1016/j.micpro.2023.104947}}

\item[J8.]%
I. Stratakos, \textbf{\underline{V. Leon}}, G. Armeniakos, G. Lentaris, and D. Soudris, 
{``Design Space Exploration on High-Order QAM Demodulation Circuits: Algorithms, Arithmetic and Approximation Techniques''},
\textbf{\em MDPI Electronics}, vol. 11, no. 1, 2022\newline
DOI: 
\href{https://www.mdpi.com/2079-9292/11/1/39}{\color{blue}{10.3390/electronics11010039}}

\item[J7.]%
\textbf{\underline{V. Leon}}, I. Stamoulias, G. Lentaris, D. Soudris, D. Gonzalez-Arjona, R. Domingo, D. Merodio Codinachs, and I. Conway, 
{``Development and Testing on the European Space-Grade BRAVE FPGAs: Evaluation of NG-Large Using High-Performance DSP Benchmarks''},
\textbf{\em IEEE Access}, vol. 9, 2021\newline
DOI: 
\href{https://ieeexplore.ieee.org/abstract/document/9543692}{\color{blue}{10.1109/ACCESS.2021.3114502}}

\item[J6.]%
\textbf{\underline{V. Leon}}, T. Paparouni, E. Petrongonas, D. Soudris, and K. Pekmestzi, 
{``Improving Power of DSP and CNN Hardware Accelerators using Approximate Floating-Point Multipliers''}, 
\textbf{\em ACM Transactions on Embedded Computing Systems}, vol. 20, no. 5, 2021\newline
DOI: 
\href{https://dl.acm.org/doi/abs/10.1145/3448980}{\color{blue}{10.1145/3448980}}

\item[J5.]%
\textbf{\underline{V. Leon}}, G. Lentaris, E. Petrongonas, D. Soudris, G. Furano, A. Tavoularis, and D. Moloney, 
{``Improving Performance-Power-Programmability in Space Avionics with Edge Devices: VBN on Myriad2 SoC''}, 
\textbf{\em ACM Transactions on Embedded Computing Systems}, vol. 20, no. 3, 2021\newline
DOI: 
\href{https://dl.acm.org/doi/abs/10.1145/3440885}{\color{blue}{10.1145/3440885}}

\item[J4.]%
\textbf{\underline{V. Leon}}, S. Mouselinos, K. Koliogeorgi, S. Xydis, D. Soudris, and K. Pekmestzi, 
{``A TensorFlow Extension Framework for Optimized Generation of Hardware CNN Inference Engines''}, 
\textbf{\em MDPI Technologies}, vol. 8, no. 1,  2020\newline
DOI: 
\href{https://www.mdpi.com/2227-7080/8/1/6}{\color{blue}{10.3390/technologies8010006}}

\item[J3.]%
\textbf{\underline{V. Leon}}, S. Xydis, D. Soudris and K. Pekmestzi, 
{``Energy-Efficient VLSI Implementation of Multipliers with Double LSB Operands''},
\textbf{\emph{IET Circuits, Devices} \& \emph{Systems}}, vol. 13, no. 6, 2019\newline
DOI: 
\href{https://digital-library.theiet.org/content/journals/10.1049/iet-cds.2018.5039}{\color{blue}{10.1049/iet-cds.2018.5039}}

\item[J2.]%
\textbf{\underline{V. Leon}}, G. Zervakis, S. Xydis, D. Soudris and K. Pekmestzi, 
{``Walking through the Energy-Error Pareto Frontier of Approximate Multipliers''},
\textbf{\em IEEE Micro}, vol. 38, no. 4,  2018\newline
DOI: 
\href{https://ieeexplore.ieee.org/abstract/document/8430622}{\color{blue}{10.1109/MM.2018.043191124}}

\item[J1.]%
\textbf{\underline{V. Leon}}, G. Zervakis, D. Soudris and K. Pekmestzi, 
{``Approximate Hybrid High Radix Encoding for Energy Efficient Inexact Multipliers''},
\textbf{\em IEEE Transactions on Very Large Scale Integration (VLSI) Systems}, vol. 26, no. 3,  2018\newline
DOI: 
\href{https://ieeexplore.ieee.org/abstract/document/8105832}{\color{blue}{10.1109/TVLSI.2017.2767858}}

\end{enumerate}

\section*{Conferences}
\begin{enumerate}[leftmargin=11.3mm,labelwidth=*,labelsep=15pt]

\item[C14.]%
\textbf{\underline{V. Leon}}, G. Lentaris, D. Soudris, S. Vellas, and M. Bernou,
{``Towards Employing FPGA and ASIP Acceleration to Enable Onboard AI/ML in Space Applications''},
\textbf{\em IFIP/IEEE International Conference on Very Large Scale Integration (VLSI-SoC)}, 2022 \newline
DOI: 
\href{https://ieeexplore.ieee.org/document/9939566}{\color{blue}{10.1109/VLSI-SoC54400.2022.9939566}}

\item[C13.]%
\textbf{\underline{V. Leon}}, E.-A. Papatheofanous, G. Lentaris, C. Bezaitis, N. Mastorakis, G. Bampilis, D. Reisis, and D. Soudris, 
{``Combining Fault Tolerance Techniques and COTS SoC Accelerators for Payload Processing in Space''},
\textbf{\em IFIP/IEEE International Conference on Very Large Scale Integration (VLSI-SoC)}, 2022 \newline
DOI: 
\href{https://ieeexplore.ieee.org/document/9939621}{\color{blue}{10.1109/VLSI-SoC54400.2022.9939621}}

\item[C12.]%
\textbf{\underline{V. Leon}}, K. Pekmestzi, and D. Soudris, 
{``Systematic Embedded Development and Implementation Techniques on Intel Myriad VPUs''},
\textbf{\em IFIP/IEEE International Conference on Very Large Scale Integration (VLSI-SoC)}, 2022 \newline
DOI: 
\href{https://ieeexplore.ieee.org/document/9939592}{\color{blue}{10.1109/VLSI-SoC54400.2022.9939592}}

\item[C11.]%
\textbf{\underline{V. Leon}}, G. Makris, S. Xydis, K. Pekmestzi, and D. Soudris, 
{``MAx-DNN: Multi-Level Arithmetic Approximation for Energy-Efficient DNN Hardware Accelerators''},
\textbf{\em IEEE Latin American Symposium on Circuits and Systems (LASCAS)}, 2022\newline
DOI: 
\href{https://ieeexplore.ieee.org/document/9789055}{\color{blue}{10.1109/LASCAS53948.2022.9789055}}

\item[C10.]%
\textbf{\underline{V. Leon}}, C. Bezaitis, G. Lentaris, D. Soudris, D. Reisis, E.-A. Papatheofanous, A. Kyriakos, A. Dunne, A. Samuelsson, and D. Steenari, 
{``FPGA \& VPU Co-Processing in Space Applications: Development and Testing with DSP/AI Benchmarks''}, 
\textbf{\em IEEE International Conference on Electronics, Circuits and Systems (ICECS)}, 2021\newline
DOI: 
\href{https://ieeexplore.ieee.org/abstract/document/9665462}{\color{blue}{10.1109/ICECS53924.2021.9665462}}

\item[C9.]%
\textbf{\underline{V. Leon}}, K. Pekmestzi, and D. Soudris, 
{``Exploiting the Potential of Approximate Arithmetic in DSP \& AI Hardware Accelerators''},
\textbf{\em International Conference on Field Programmable Logic and Applications (FPL)}, 2021\newline
DOI: 
\href{https://ieeexplore.ieee.org/abstract/document/9556418}{\color{blue}{10.1109/FPL53798.2021.00049}}

\item[C8.]%
E. Petrongonas, \textbf{\underline{V. Leon}}, G. Lentaris, and D. Soudris, 
{``ParalOS: A Scheduling \& Memory Management Framework for Heterogeneous VPUs''},
\textbf{\em Euromicro Conference on Digital System Design (DSD)}, 2021\newline
DOI: 
\href{https://ieeexplore.ieee.org/abstract/document/9556499}{\color{blue}{10.1109/DSD53832.2021.00043}}

\item[C7.]%
\textbf{\underline{V. Leon}}, I. Stratakos, G. Armeniakos, G. Lentaris, and D. Soudris, 
{``ApproxQAM: High-Order QAM Demodulation Circuits with Approximate Arithmetic''}, 
\textbf{\em International Conference on Modern Circuits and Systems Technologies (MOCAST)}, 2021\newline
DOI: 
\href{https://ieeexplore.ieee.org/abstract/document/9493421}{\color{blue}{10.1109/MOCAST52088.2021.9493421}}

\item[C6.]%
\textbf{\underline{V. Leon}}, I. Stamoulias, G. Lentaris, D. Soudris, R. Domingo, M. Verdugo, D. Gonzalez-Arjona, D. Merodio Codinachs, and I. Conway, 
{``Systematic Evaluation of the European NG-LARGE FPGA \& EDA Tools for On-Board Processing''},
\textbf{\em ESA/CNES/DLR European Workshop on On-Board Data Processing (OBDP)}, 2021 \newline
DOI: 
\href{https://zenodo.org/record/5521055#.Yf0ArOpBw2w}{\color{blue}{10.5281/zenodo.5521055}}

\item[C5.]%
G. Lentaris, G. Chatzitsompanis, \textbf{\underline{V. Leon}}, K. Pekmestzi, and D. Soudris, 
{``Combining Arithmetic Approximation Techniques for Improved CNN Circuit Design''}, 
\textbf{\em IEEE International Conference on Electronics, Circuits and Systems (ICECS)}, 2020\newline
DOI: 
\href{https://ieeexplore.ieee.org/abstract/document/9294869}{\color{blue}{10.1109/ICECS49266.2020.9294869}}

\item[C4.]%
\textbf{\underline{V. Leon}}, K. Asimakopoulos, S. Xydis, D. Soudris, and K. Pekmestzi, 
{``Cooperative Arithmetic-Aware Approximation Techniques for Energy-Efficient Multipliers''},
\textbf{\em ACM/IEEE Design Automation Conference (DAC)},  2019\newline
DOI: 
\href{https://dl.acm.org/doi/abs/10.1145/3316781.3317793}{\color{blue}{10.1145/3316781.3317793}}

\item[C3.]%
S. Mouselinos, \textbf{\underline{V. Leon}}, S. Xydis, D. Soudris, and K. Pekmestzi,
{``TF2FPGA: A Framework for Projecting and Accelerating TensorFlow CNNs on FPGA Platforms''}, 
\textbf{\em International Conference on Modern Circuits and Systems Technologies (MOCAST)}, 2019\newline
DOI: 
\href{https://ieeexplore.ieee.org/abstract/document/8741940}{\color{blue}{10.1109/MOCAST.2019.8741940}}

\item[C2.]%
K. Maragos, \textbf{\underline{V. Leon}}, G. Lentaris, D. Soudris, D. Gonzalez-Arjona, R. Domin-go, A. Pastor, D. Merodio Codinachs, and I. Conway,
{``Evaluation Methodology and Reconfiguration Tests on the New European NG-MEDIUM FPGA''},
\textbf{\em NASA/ESA Conference on Adaptive Hardware and Systems (AHS)}, 2018\newline
DOI: 
\href{https://ieeexplore.ieee.org/abstract/document/8541492}{\color{blue}{10.1109/AHS.2018.8541492}}

\item[C1.]%
C. Marantos, N. Karavalakis, \textbf{\underline{V. Leon}}, V. Tsoutsouras, K. Pekmestzi, and D. Soudris,
{``Efficient Support Vector Machines Implementation on Intel/Movidius Myriad 2''},
\textbf{\em International Conference on Modern Circuits and Systems Technologies (MOCAST)}, 2018\newline
DOI: \href{https://ieeexplore.ieee.org/abstract/document/8376630}{\color{blue}{10.1109/MOCAST.2018.8376630}}

\end{enumerate}


\chapter*{Curriculum Vitae}
\addcontentsline{toc}{chapter}{Curriculum Vitae}
\chaptermark{Curriculum Vitae}

\textbf{Vasileios (Vasilis) Leon} received the Diploma (M.Eng.) degree from the University of Patras, Department of Computer Engineering and Informatics in 2016. During 2015-2016, he worked as DSP Engineer (started as internship) for Think Silicon S.A., now an Applied Materials company. In 2016, he was admitted to the Ph.D. program of the National Technical University of Athens, School of Electrical and Computer Engineering, where he joined the Microprocessors and Digital Systems Laboratory to pursue his Doctoral Dissertation in the research field of digital integrated circuits and embedded systems. Since 2017, he has been working as R\&D Hardware/Embedded Engineer in technical projects of the European Space Agency (ESA), which involve embedded computing platforms (radiation-hardened and COTS FPGAs, Myriad VPU SoCs, Edge TPU SoMs), high-performance architectures for on-board data processing, benchmarking \& testing of tools/devices, and DSP/AI hardware acceleration. In these projects as well as other research activities, he has been junior, senior, and leader in teams of researchers and engineers, belonging in consortia with companies and universities. 
To date, he has participated in the following ESA projects:\\[-15pt]

\begin{itemize}[wide = 0pt, leftmargin =*, labelsep=3mm, noitemsep]
\item \underline{QUEENS3}: {``Quality Assessment of the New European Ultra BRAVE FPGA Software Tools''}, ESA num. 4000134874/21/NL/AR/va, 2021--2023\\[-8pt]
\item \underline{CAIRS21}: {``COTS AI Accelerators in
Mixed-Criticality High-Performance Avionics for Reconfigurable
Satellites: TPU versus Prominent Embedded Devices, Mitigation
Techniques, SW Frameworks and AI/ML Model Upload''}, ESA num. 4000135491/21/NL/GLC/ov, 2021--2022\\[-8pt]
\item \underline{HPCB}: {``FPGA Accelerated DSP Payload Data Processor Board''}, ESA num. 4000126129/18/NL/AF, 2019--2021\\[-8pt]
\item \underline{QUEENS2}: {``Quality Assessment of the New European Large BRAVE FPGA Software Tools''}, ESA num. 4000128041/19/NL/AR/va, 2019--2020\\[-8pt]
\item \underline{LEOTOME}: {``Demonstration of Visual Based Navigation Algorithms on Myriad2 Processor''}, ESA num. 4000126083/18/NL/FE, 2019\\[-8pt]
\item \underline{QUEENS1}: {``Quality Assessment of the New European Medium BRAVE FPGA Software Tools''}, ESA num. 4000119331/17/NL/PS, 2017--2018
\end{itemize}

During his Ph.D. studies,
Vasileios Leon 
published 25 scientific papers 
in the research fields of Hardware Engineering (digital circuits \& accelerators, computer arithmetic, approximate computing, FPGAs, SoCs) and Space Engineering (on-board data processing, embedded accelerators, mixed-criticality co-processing architectures). 
The papers were published 
in the proceedings of 14 international conferences
(including the Design Automation Conference -- DAC) 
and 11 high-indexed journals
(including IEEE and ACM Transactions).
He was also a reviewer in various IEEE journals and conferences. 
Moreover,
he co-supervised 17 undergraduate Diploma (M.Eng.) theses
and offered teaching assistance in the laboratories of the undergraduate courses
``Microprocessors'',
``Digital VLSI Systems'',
and
``Introduction to VLSI''.\\

\textbf{Research Interests:} 
digital IC design, hardware accelerators, FPGAs, embedded systems, SoCs, approximate computing, computer arithmetic, digital signal processing, computer vision, on-board data processing, space systems.\\

\textbf{Professional Web Profiles}\\[4pt]
\begin{tabular}{l l}
\raisebox{-.12\height}{\includegraphics[scale=0.009]{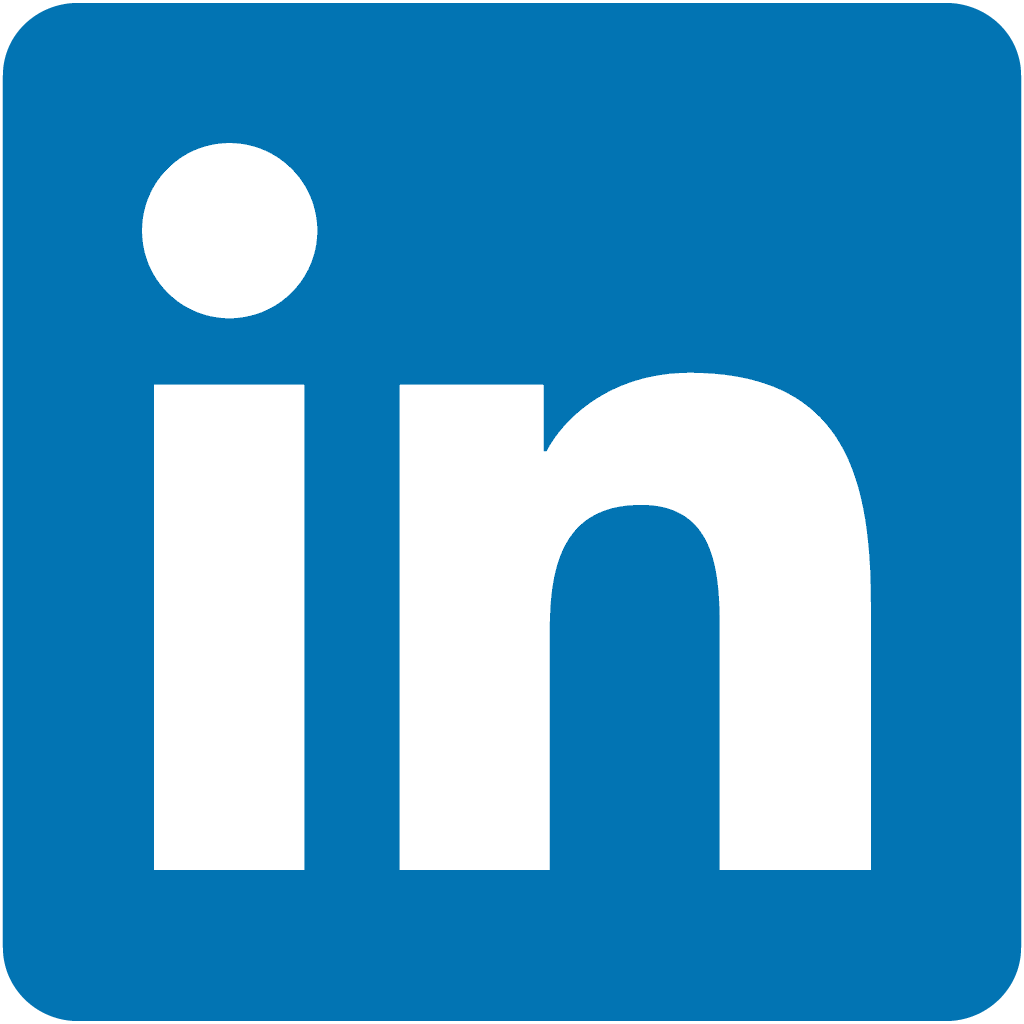}} & \hspace{-10pt} \href{https://www.linkedin.com/in/vasilisleon}{\color{myblue}{linkedin.com/vasileiosleon}}
\end{tabular}\\[4pt]
\begin{tabular}{l l}
\raisebox{-.12\height}{\includegraphics[scale=0.041]{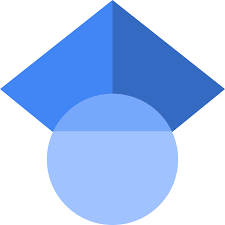}} & \hspace{-10pt} \href{https://scholar.google.com/citations?hl=en&user=9DABzNsAAAAJ}{\color{myblue2}{scholar.google.com/vasileiosleon}}
\end{tabular}\\[4pt]
\begin{tabular}{l l}
\raisebox{-.12\height}{\includegraphics[scale=0.041]{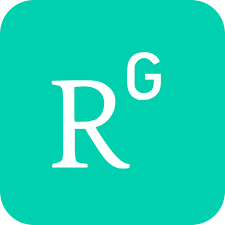}} & \hspace{-10pt} \href{https://www.researchgate.net/profile/Vasileios_Leon}{\color{mygreen}{researchgate.net/vasileiosleon}}
\end{tabular}

\blankpage

\newgeometry{marginparwidth=0.0cm,
left=44.82771pt, right=44.82771pt, 
top=12mm, bottom=12mm} 

\thispagestyle{empty}


\begin{flushright}

\newcolumntype{?}{!{\vrule width 2.3pt}}

\begin{tabular}
{>{\raggedleft\arraybackslash}p{8.7cm}%
   >{\raggedleft\arraybackslash}p{1.8cm} ?
  }
  \Xhline{6\arrayrulewidth}
  \phantom{a} & \\[-6pt]
\textsc{\fontsize{9.5}{11}\selectfont National Technical University of Athens} &  \hspace*{-10pt} \multirow{4}{*}{\raisebox{-.45\height}{\includegraphics[scale=0.035]{MISCs/ntua_logo.png}}}\\[0pt] 
\textsc{\fontsize{9.5}{11}\selectfont School of Electrical and Computer Engineering} & \\[0pt] 
\textsc{\fontsize{9.5}{11}\selectfont Division of Computer Science} & \\[0pt] 
\textsc{\fontsize{9.5}{11}\selectfont Microprocessors and Digital Systems Laboratory} & \\[0pt] 
{\fontsize{9.5}{11}\selectfont 9 Heroon Polytechniou, Zografou Campus} & \\[0pt] 
 {\fontsize{9.5}{11}\selectfont 15780 Athens, Greece} & \\[0pt] 
{\fontsize{9.5}{11}\selectfont \textbf{Ph.D. Dissertation, Vasileios Leon}} & \\

\end{tabular}
\end{flushright}


 \vfill

\begin{tabular*}{\textwidth}{cc}
     \Xhline{12\arrayrulewidth}
\end{tabular*}

\end{document}